%% file: tesi.tex
\documentclass[15pt]{report_cast}	  
\usepackage{latexsym,epsf,rotate}
\usepackage[english]{babel}

\usepackage{makeidx}   
\usepackage{graphicx}  

\newcommand{\be}{\begin{equation}}
\newcommand{\ee}{\end{equation}}
\newcommand{\ben}{\begin{eqnarray}}
\newcommand{\een}{\end{eqnarray}}

\newcommand{\nn}{\nonumber}

\newcommand{\ra}{\rangle}
\newcommand{\la}{\langle}

\font\Bbb =msbm10  scaled \magstephalf
\def\id{{\hbox{\Bbb I}}}
\renewcommand{\theenumi}{\alph{enumi}}
\newcommand{\ia}{\'{\i}}
\newcommand{\nd}{\noindent}
\newcommand{\ket}[1]{|{#1}\rangle}

\def\fh{h}

\def\fu{u}

\begin{document}

\title{
\vspace{-5cm}
\hspace{0cm}\epsffile{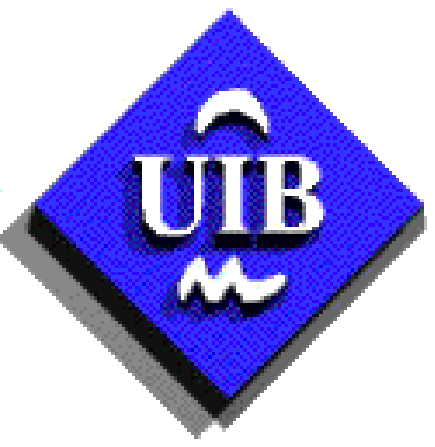} \\
{\Large Departament de F\ia sica}\\ 
{\Large Universitat de les Illes Balears}\\
\vspace{1.5cm}
Characterization of Quantum Entangled States and Information Measures\\
\vspace{3cm}
{\large PhD Thesis}}
\author{Josep Batle-Vallespir}
\date{January 2006}

\maketitle 



\include{enblanc}

\include{dedicatoria}

\include{enblanc}
\include{Shakespeare}

\pagenumbering{roman}

\include{foreword}          
\include{Acknowledgements}  

\tableofcontents

\pagenumbering{arabic}

\include{introductionTesi}



\include{introduction_novel_Tesi}

\include{quantumentanglement}
								    								   


\include{conclusionsTesi}
\include{Appendices}

\include{references} 

\pagestyle{empty}

\include{cvEnglish}

\end{document}

%% file: enblanc.tex
\thispagestyle{empty}
\begin{center}
\begin{minipage}{0.9\textwidth}
 \vspace{7cm} 
\begin{list}{}{\leftmargin=2cm \rightmargin=0cm}\item[]
\,\, 
\end{list}   

\end{minipage}
\end{center}

%% file: dedicatoria.tex
\thispagestyle{empty}
\begin{center}
\begin{minipage}{0.9\textwidth}
 \vspace{7cm} 
\begin{list}{}{\leftmargin=2cm \rightmargin=0cm}\item[]
\emph{\Large{Dedicat a la meva mare, a qui dec el que som; a la 
mem\`oria de mon pare, al cel sia; a la meva germana, per la seva fortalesa davant la vida; 
a na Maria Margalida, en Miquel i n'Andreu Mart\ia, alegries de la meva vida, i 
a n'Ant\`onia, per cada segon}} 
\end{list}   

\end{minipage}
\end{center}

%% file: Shakespeare.tex
\thispagestyle{empty}
\begin{center}
\begin{minipage}{0.9\textwidth}
 \vspace{7cm} 
\begin{list}{}{\leftmargin=2cm \rightmargin=0cm}\item[]
\emph{\Large{\rightline{If you can look into the seeds of time,}}} 

\emph{\Large{\rightline{And say which grain will grow, and which will not,}}} 

\emph{\Large{\rightline{Speak then to me.}}}\newline 

\hspace{0.1cm}
\Large{\rightline{W. Shakespeare, {\it Macbeth}, I, 3.}}
\end{list}   

\end{minipage}
\end{center}

%% file: foreword.tex
\chapter*{foreword}
\addcontentsline{toc}{chapter}{Foreword}

 The present Thesis covers the subject of the characterization of entangled states 
 by recourse to the so called entropic measures, as well as the description of 
 entanglement related to several issues in quantum mechanics, such as the 
 speed of a quantum evolution or the exciting connections existing between 
 quantum entanglement and quantum phase transitions, that is, transitions that 
 occur at zero temperature.\newline 

 This work is divided in four parts, namely, {\it I Introduction}, {\it II Quantum Entanglement}, 
 {\it III The role of quantum entanglement in different physical scenarios}, 
 and {\it IV Conclusions}. 
 At the end of it we include an Appendix with several historical remarks and technical 
 details. The first introductory part consists in turn of three subsections: i) a historical 
 review of what is undestood by Quantum Information Theory (QIT), ii) a brief description of 
 quantum computation, and finally iii) an account on quantum communication. The first part 
 dealing with the roots of information theory and its connection with physics has been 
 included for the sake of completeness, mainly for historical reasons. One believes that 
 a modern subject such as QIT, which is highly diverse and transverse, deserved 
 a few lines so that the reader can realize the importance of the evolution towards 
 the quantum domain of concepts such as (reversible) computability and the physical nature 
 of information, as well as the analyses of fundamental arguments that questioned the 
 completeness of the quantum theory, which in turn motivated and gave rise to the concept 
 of quantum entanglement. 
 The other two subsections on quantum computation and quantum communication 
 are reviewed only because they offer brand new and exciting proposals, which are of 
 common interest to any physicist. However, these former concepts are not present in 
 the description of this Thesis, therefore one can skip them with no loss of continuity.\newline 

 The second part entitled {\it Quantum Entanglement} describes the problem of detecting 
 entanglement, added to the question of characterizing it. The third part covers 
 the role of quantum entanglement in different contexts of quantum mechanics, 
 and finally the Conclusions review some of the most important ideas exposed in the 
 present work.

%% file: Acknowledgements.tex
\chapter*{acknowledgements}
\addcontentsline{toc}{chapter}{Acknowledgements}   

La present tesi \'es el resultat d'anys de recerca que comen\c c\`a amb l'estudi de 
la f\ia sica d'acceleradors al CERN i de les propietats dels agregats at\`omics (Xesca!), 
i que s'ha ha anat acostant amb el temps a l'entanglement o {\it entrella\c cament}. Pel cam\ia, 
romanen la caracteritzaci\'o d'estats amb entanglement amb l'ajut de les mesures d'informaci\'o 
entr\`opiques, altrament conegudes com $q$-entropies (com ara la de R\'enyi i Tsallis), 
l'estudi dels estats entrella\c cats relacionat amb diversos aspectes de la mec\`anica 
qu\`antica, fins arribar a l'estudi de la din\`amica de trancisions de fase qu\`antiques ($T=0$), 
tot emprant una mesura de l'entrella\c cament --la puresa, introdu\"\i da pel grup T-11 
de la divisi\'o te\`orica de Los Alamos-- molt convenient per a l'estudi de 
sistemes de molts cossos.

Aquest treball ha estat fruit d'un esfor\c c personal que ha sorgit de la interacci\'o 
amb diverses persones. Per aix\`o vull fer pal\`es el meu m\'es sincer agra\"\i ment a la meva 
directora de tesi, na Montserrat, pel seu suport, consell cient\ia fic i molta comprensi\'o 
en moments dif\ia cils; a n'\'Angel Plastino fill, co-director d'aquesta tesi, 
amb qui he gaudit de fer feina durant tots aquests anys, i que m'ha fet descobrir 
tantes coses pel que fa aquesta estranya teoria qu\`antica; i a n'\'Angel Plastino pare, 
persona extraordin\`aria amb qui he fru\"\i t de fer feina i de parlar de f\ia sica en tots 
els seus \`ambits possibles. Tamb\'e vull agrair el tracte exquisit que he tengut 
l'oportunitat de gaudir amb en Manuel de Llano, amb qui col$\cdot$labor\`arem sovint 
en afers de superconductivitat en cuprats. Cap a M\`exic va una forta abra\c cada.

Tamb\'e vull donar les gr\`acies als meus col$\cdot$legues i companys de batalletes David Salgado 
i Enrique Rico, amb qui he coincidit en diverses reunions sobre informaci\'o qu\`antica, 
inexplicablement gaireb\'e sempre a It\`alia. Quines coses! A mi no m'agradava el formatge 
i ara m'encanta. Tendr\`a It\`alia la culpa? Tamb\'e vull agrair en Gerardo Ortiz i en 
Rolando Somma per la seva hospitalitat --i la de tota la col\`onia argentina 
de Los Alamos-- durant la meva estada en aquest centre tan important de recerca. Tamb\'e agraesc 
el companyerisme d'en Juanjo Cerd\`a i d'en Pep Mulet, als quals s'han anat apuntant molts 
altres f\ia sics de la famosa Ag\`encia EFE, molts dels quals ja han arribat 
(o estan apunt de fer-ho) de les seves aventures postdoctorals. No voldria 
deixar-me tots els referees (sobretot els bons, sembla que als altres no els 
arrib\`a el pernil que els vaig enviar..) dels nostres articles, que els acceptaren per b\'e 
o per mal. Culpau-los a ells en tot cas.

Tamb\'e deman disculpes per avan\c cat si l'angl\`es emprat no \'es del tot correcte. 
I ja que estem posats, vull fer una petita 
cr\ia tica, si se'm permet: m'agradaria que aquesta 
universitat que tan estim, la nostra universitat, l'\'unica que tenim a les Illes Balears, 
defin\ia s d'una vegada quin model d'universitat vol esser i projectar a l'exterior. 
Com diuen en bon 
mallorqu\ia : ``qui molt abra\c ca, poc estreny" (quien mucho abarca, poco aprieta). I pel que fa 
el tema dels greuges comparatius dins del Departament, millor no parlar-ne. M'agradaria veure 
que en un futur l'estat espanyol fes un gir copernic\`a pel que fa a la seva pol\ia tica de 
recerca, principalment en la figura del {\it becari}, que l'\'unic que fa \'es sembrar 
frustracions arreu i f\ia sics en l'atur o, com a mal menor, en l'educaci\'o secund\`aria per poder 
subsistir.\newline

\nd Palma de Mallorca, gener de 2006.
\pagebreak

The present Thesis summarizes several years of research that started with the physics 
of accelerators at CERN and the properties of atomic clusters (Xesca!), and progressively 
approached the path to {\it quantum entanglement}. In the way to it, there remained the 
characterization of quantum entangled states by recourse to information 
measures --also known as $q$-entropies-- such as R\'enyi's or Tsallis', and the study 
of entangled states in connection with several aspects of quantum mechanics. Finally, 
we studied also the dynamics of quantum phase transitions ($T=0$) by employing a suitable 
entanglement measure, namely, the purity measure introduced by the T-11 group 
of the theoretical division at Los Alamos, specially convenient in the study of 
many-body systems.

This work has been possible due to the interaction with several people. That is why I want 
to express my most sincere gratitude to my advisor, Montserrat,	for her support, scientific 
advice and kind understanding in difficult moments; to \'Angel Plastino Junior, 
co-director of the present Thesis, who has taught me so many things regarding this strange 
quantum theory; and to \'Angel Plastino Senior, extraordinary person with whom I have 
enjoyed working with and discussing about all possible areas of physics. I would also like to 
thank the opporunity of working with Manuel de Llano, with whom we have worked on cuprate 
superconductivity. I kind hug is sent to Mexico!

Also, I want to thank my pals David Salgado and Enrique Rico for several 
discussions and adventures, with whom I have coincided 
several times in quantum information meetings, inexplicably nearly always in Italy. 
The facts of life! I did not like cheese in the past and now I love it. Should I blame 
Italy for it? As well, I want to acknowledge Gerardo Ortiz and Rolando Somma (and all 
the Argentinian scientific colony of Los Alamos) for their kind hospitality. I also 
acknowledge the companionship of Juanjo Cerd\`a and Pep Mulet, and of the other fellows 
at the Ag\`encia EFE, some of them just arrived from their post-doctoral adeventures of 
about to do so. And last but not least, I shall not forget all the referees (specially 
the favourable ones, apparently the others did not receive their gift..) of our 
articles, which were accepted anyhow. Put the blame on them.

I also apologize in advance for the English redaction of this Thesis. And now a little 
bit of criticism. I would like this my beloved University, the only one we have in 
the Balearic Islands, to state what sort of University it wants to be. There is a 
saying in Catalan language that reads ``qui molt abra\c ca, poc estreny" (``Don't spread 
yourself too thin"). I would like to see that someday the Spanish Government 
makes a Copernican turn with respect to its research policy, at least in physics, 
specially in what implies the figure of the ``becario".\newline
					   
\nd Palma de Mallorca, January 2006.

%% file: introductionTesi.tex
\part{Introduction}

The characterization of quantum entangled states and their properties range from the 
description of quantum entanglement itself to the description of the states of quantum 
systems, where this quantum correlation can be present. The description of entangled 
states requires a two-step procedure: the {\bf i) detection of entanglement}, combined with the 
{\bf ii) characterization of entanglement}. 

In the present Thesis we expose the tools employed 
in the characterization of bipartite quantum systems in multiple dimensions (e.g. $2 \times 2$ or 
two-qubit systems, $2 \times 3$, and so forth) by means of entropic or information 
measures. These information measures are described in forthcoming sections of this 
Introduction and more specifically in part II, entitled {\it Quantum Entanglement}, and also 
in the concomitant Chapters where this item is discussed in more detail. The relevance 
of this information-theoretical description resides in the fact that the entropic framework 
offers a highly intuitive and physical meaning to what entanglement represents in a bipartite 
quantum system: the entropy of any of its subsystems cannot be larger that the total entropy 
of the system. These relations, known as {\it entropic inequalities}, applied to the field 
of quantum information theory, constitute the subject of a novel study in this Thesis. 
Mathematically, it is a necessary condition 
for the discrimination of an entangled state: if a state does not possess quantum correlations, therefore 
it fulfils the entropic inequalities. However, the converse is not true, which means 
that one can encounter entangled states that look ``classical". By generating random 
mixed states of bipartite systems in different dimensions, we obtain the volume of 
states that comply with the entropic criteria, and compare it with several other criteria. 
Also in this Thesis, we reveal the interesting connection existing between a particular 
class of entangled states, the maximally entangled mixed states (MEMS), and the violation 
of the aforementioned entropic inequalities. However, previous to the detection of entanglement, 
in this Thesis we refute the fact that 
the maximum entropy principle applied to the inference of states ``fakes" entanglement. That is, 
we show that a proper combination of maximization of entropy followed by minimization of 
entanglement leads to a correct description of the entanglement present in an inferred 
state, contrary to what was believed.

With respect to what implies the characterization of entangled states, we perform a 
Monte Carlo procedure in the exploration of the structure of the simplest quantum 
system that exhibits entanglement, the two-qubit system. Already in these systems, 
the space of mixed states to explore has got 15 dimensions and it is highly anisotropic. 
By generating mixed states of two qubit states according to different measures present in the 
literature, we can observe how entanglement is distributed in this space using 
the so called participation ratio --as sort of degree of mixture-- as a probe. In clear 
connection with quantum computation, quantum gates acting on pure or mixed states act 
as entanglers: quantum gates represent the mathematical abstraction of a physical 
process of interaction. It has then been of interest to study how entanglement is 
distributed when a two-qubit gate acts on an arbitrary pure or mixed state.

Regarding the characterization of entanglement, first of all we review the measures 
of entanglement used to date. We point out that there exist obscure points 
in the current definition of entanglement, which is based in a preferred tensor product partition 
of the Hilbert space of the physical system under study. This is the starting point for 
employing a measure of entanglement introduced by the theoretical group of Los Alamos, 
the so called {\it purity} measure. This measure does not present problems when dealing 
with identical particles, nor with the total number of them. These features make it specially 
suitable for studying quantum entanglement in condensed matter systems. More specifically, 
here we describe the connection that exists between entanglement and quantum phase transitions, 
that is, transitions that occur at zero temperature. Our contribution, which is an extension of 
the work done at Los Alamos, deals with a more specific feature of quantum phase transitions, namely, 
their dynamics. We study the dynamical evolution of the $XY$ anisotropic model in a transverse 
magnetic field, which will reveal that entanglement can be regarded as a property that characterizes 
the overall system. Furthermore, we shall see that entanglement can present non-ergodic 
features, on equal footing with a clear physical magnitude such as the $z$-magnetization. 
In a different scenario, we also show that entanglement can speed up the evolution of 
a quantum state in a very special way: in general terms, an entangled state evolves to its 
first orthogonal faster than an unentangled one. This is also true for the extended usual 
measure of entanglement for indistinguishable particles only in the case of bosons, but not 
for fermions.\newline

The present Thesis has been conceived to be read in a continuous way. In the Introduction (part I) we 
recall several ideas which are of basic nature if one wants to grasp the origins 
of a highly diverse subject such as quantum information theory, in an attempt to stress the 
fact that information has its roots not in abstract mathematical ideas, but in deep physical grounds. The 
tools employed and their description in the study of the characterization of entangled states 
are given in part II, while every Chapter of part III is almost self-contained: it exposes, 
when necessary, the tools that are needed in the description of entanglement in different 
contexts. The Conclusions (part IV) review the most important results obtained and a final 
Appendix contains technical details regarding procedures and concepts constantly referred to 
throughout the present contribution.

\chapter{Quantum information theory: a crossroads of different disciplines}

Quantum information theory (QIT) is a fast developing science that has been 
built upon several branches of different scientific disciplines. The goals of QIT 
are at the intersection of those of quantum mechanics and information theory, while 
its tools combine those of these two theories. Behind what we 
call quantum information (it is yet unclear if quantum computation should be considered 
part of it, at least conceptually) one finds a whole spectrum of researchers working 
not only in the field of theoretical physics, but also mathematicians, computer 
scientists, electronic engineers, experimental physicists and so on. Of course all 
of them have different concerns and interests, and only in very recent years highly 
specialized conferences and meetings do offer a definite frame of reference for 
each one of these researchers. QIT finds room for theoretical physicists interested 
in the foundations of quantum mechanics (theory of measurement, 
quantum Zeno effect, decoherence, interpretation of quantum mechanics, Gleason's 
Theorem, quantum jumps, etc..), Bell's Theorem (Bell inequalities, Bell's Theorem without 
inequalities, etc..), non-locality of Nature (local hidden variables theories, 
quantum entanglement and its description, separability, etc..), information processing 
(quantum cryptography, superdense coding, quantum teleportation, entanglement swapping, 
quantum repeaters, quantum key distribution, etc..); computer scientists and mathematicians 
(quantum computing, quantum algorithms, quantum complexity classes, etc..); both theoretical 
and experimental physicists (quantum circuits, quantum error correction, decoherence-free spaces, 
fault-tolerant quantum computation, nuclear magnetic resonance or NMR quantum computing, 
ion-trap quantum computing, optical lattice quantum computing, solid state quantum 
computing, etc..), and physicist interested in other related abstract features 
(quantum games, quantum random walks, etc..). 

In spite of the clear diversity of interests, there exists a common feature that makes possible 
all previous challenges, which receives the name of {\bf entanglement}. 
Until recent times, in relative terms, fundamental aspects of 
quantum theory were considered a matter of concern to 
epistemologists. While certainly profound questions were debated in the pursue of an 
answer of the ultimate nature of reality, it scarcely seemed possible that they could 
be answered by experiments. The EPR paradox posed by Einstein, Podolsky and Rosen (EPR) in 1935 
\cite{EPR} focused the attention of the physics community on the possible lack of 
completeness of the newborn quantum mechanics. In their famous paper 
they suggested a description of the
world (called ``local realism'') which assigns
an independent and objective reality to the physical properties of the well
separated subsystems of a compound system.
Then EPR applied the criterion of local realism to predictions 
associated with an entangled state, a state that cannot be described solely in 
terms of the properties of its subsystems, to conclude that
quantum mechanics is incomplete.
EPR criticism was the source of many discussions concerning fundamental
differences between quantum and classical description of nature. Schr\"odinger 
\cite{Schro}, regarding the EPR paradox, did not see a conflict with quantum mechanics. 
Instead, he defined that non-locality or ``Verschr\"ankung" (German word 
for entanglement) should be {\it the} characteristic feature of quantum mechanics. 
The link between Information Theory and entanglement was
first considered by him, when he wrote that ``Thus one disposes provisionally 
(until the entanglement is resolved by actual
observation) of only a common description of the two in that
space of higher dimension. This is the reason that knowledge of the
individual systems can decline to the scantiest, even to zero, while that of the 
combined system remains continually maximal\footnote{See Appendix A.}. 
Best possible knowledge of a whole
does not include best possible knowledge of its parts -- and that is
what keeps coming back to haunt us" \cite{Schro}. Schr\"odinger thus identified 
a profound non-classical relation between the 
information that an entangled state gives about the whole system and the corresponding 
information that is given to us about the subsystems. The most significant progress 
toward the resolution of this ``academic" EPR problem was
made by Bell \cite{Bell1,Bell2} in the 60s who proved that the local realism implies constraints
on the predictions of spin correlations in the form of inequalities
(called Bell's inequalities) which can be violated by quantum mechanical
predictions for the system. Experiments were carried out and confirmed Schr\"odinger's 
argument (see forthcoming section on Bell inequalities for more details). 

With time physicists have recognized the possibilities that 
entanglement, which is seen as a fundamental characteristic of Nature, can unfold 
at the technological stage, providing a new framework for developing faster 
computing (quantum computation) or impossible tasks in classical physics such 
as absolutely secure communication (quantum cryptography) or teleportation.

\section{The roots of information theory and computer science}

This section is devoted to the basic features of the origin of information theory 
and computer science. The motivation for doing so become clear as the theory 
of quantum information and computation borrow original ideas from these 
disciplines and translate them into the realm of quantum mechanics. Information in 
a technically defined sense was first introduced in statistics by R. A. Fisher in 
1925 in his work on the theory of estimation. The properties of Fisher's 
definition of information became a fundamental part of the so called statistical 
theory of estimation. Shannon and Wiener, indepently, published in 1948 works describing 
logarithmic measures of information for use in communication theory, which induced 
to consider information theory as synonymous with communication theory \cite{Kullback}. 
As a matter of fact, information theory formulates a communication system as a stochastic process. 
Formally, information theory is a branch of the mathematical theory of probability and mathematical 
statistics. As such, it can be applied in a wide variety of fields. Information 
theory is relevant to statistical inference, provides a unification of known 
results, and leads to natural generalizations \cite{Kullback}. In spirit and concepts, 
information theory has its mathematical roots in the concept of disorder or 
entropy in thermodynamics and statistical mechanics, as we shall see.

It is obvious 
that if one deals with a quantum information theory, it is because there exists a 
classical counterpart. As we shall see in future Chapters, quantum mechanics 
provide a framework where several mechanisms and concepts that appear in classical 
information are improved beyond what was thought to be an impossible barrier to 
overcome, while others simply did not exist. However, this is not so evident in the case 
of computer science. Historically, the first results in the mathematical theory of theoretical 
computer science appeared before the discipline of computer science existed; in fact, 
even before the existence of electronic computers. Shortly after G\"odel proved his 
famous incompleteness theorem, there appeared several papers that drew a distinction 
between computable and non-computable functions. But of course one needed a mathematical 
definition of what was understood as ``computable", each author giving a different 
one\footnote{In his pioneer paper, Turing says: ``The computable numbers may be described 
briefly as the real numbers whose expressions as a decimal are calculable by finite means" 
\cite{Turing}.}. 
But in the end they all resulted in the same class of computable functions. This led 
to the proposal of what is known now as the Church-Turing thesis, named after Alonzo 
Church and Alan Turing. This thesis says that any function that is computable by any means, 
can be computed by a Turing machine. Once ``computability" is defined, the task is 
to classify different problems into complexity classes. But this part will be explained in 
Chapter 2. The bridge existing between this discipline (computer science) and quantum 
information theory can only occur when the principles of quantum mechanics are observed. 
This will certainly take place in the 
near future, if it is not already the case, when the speed of computation will become 
limited by the quantum effects 
that appear in the miniaturization of the basic electronic devices (e.g. the basic 
logical gates). Only when one realizes 
that {\it information is physical..}\footnote{.. {\it but slippery}. Quote attributed 
to R. Landauer.}, information processing, where computing is included, ought to obey 
the laws of quantum mechanics. 

Therefore let us recall the origin of the basic concepts of information theory and 
computer science, so that we could gain more insight into the quantum counterpart.

\subsection{Claude Shannon and the information measure $H$}

In 1948 Shannon published {\it A Mathematical Theory of Communication}. 
This work focuses on the problem of how to best encode the information a sender 
wants to transmit. In this fundamental work he used tools in probability theory, 
which were in their nascent stages of being applied to communication 
theory at that time. His theory for the first time considered communication as a 
rigorously stated mathematical problem in statistics and gave communications 
engineers a way to determine the capacity of a communication channel in terms 
of the common currency of bits. The word ``bit" is the short expression for 
``binary digit", either 0 or 1, which is the abstract state one can assign to two possible 
outcomes in a binary system\footnote{Nothing else but the Boolean algebra upon which all classical 
computers are based.}. Any text can be coded into a string of bits; for instance,
it is enough to assign to each symbol its ASCII code
number in binary form and append a parity check bit. For example, the word ``quanta" can be 
coded as $11100010\,\, 11101011\,\, 11000011\,\, 11011101\,\, 11101000\,\, 11000011$.
Each bit can be stored physically; in classical computers,
each bit is registered as a charge state of a capacitor
(0=discharged,1=charged). They are distinguishable
macroscopic states and rather robust or stable. They are
not spoiled when they are read in (if carefully done) and
they can be cloned or replicated without any problem.
Information is not only stored; it is usually transmitted
(communication) and sometimes processed (computation). With this description, 
we are just advancing some of the properties that ought to be carefully revisited 
in the quantum counterpart.

The transmission part of the theory is not concerned with the 
meaning (semantics) of the message conveyed, though the complementary wing of 
information theory concerns itself with content through lossy compression of 
messages subject to a fidelity criterion. Shannon developed information entropy 
as a measure for the 
uncertainty in a message while essentially inventing what became known as the 
dominant form of ``information theory". Shannon, advised by von Neumann, gave 
the name ``entropy"  

\begin{equation} \label{H}
H\,=\,-\sum_{n=1}^N p_n\,{\rm log}_2\,p_n
\end{equation}

\noindent to the information content of a given message. $p_n$ stands for the 
probability of the event $n$. Let us describe the situation somewhat in more detail. 
What Shannon conceived was the entropy (\ref{H}) associated to a 
discrete\footnote{The generalization to the continuous variable case is done by changing 
a discrete set of probabilities by a probability distribution, and the sum by an 
integral. In the continuous case, however, one can have and {\it infinite} value for 
the entropy: the accuracy needed to address a specific value of the random variable 
$X$ in the continuum may require infinite precision.} random 
variable $X$ which could take $N$ possible values $\{x_1,...,x_N\}$, with $p_n$ being the 
probability that $X$ takes the value $x_n$. $H$ can then be interpreted as a measure of 
ingorance or uncertainty associated to the probability distribution $\{p_n\}$. It is 
not about the knowledge about the distribution itself, but the capacity of predicting 
the results of an experiment subjected to this distribution. Thus, in his 1948 work, 
Shannon formalised the requirements of an information measure $H(p_1,...,p_N)$ with 
the following criteria:

\begin{itemize}
\item i) $H$ is a continuous function of the $\{p_n\}$.
\item ii) If all probabilities are equal, $p_n=1/n$, then $H$ is a monotonic increasing 
function of $N$.
\item iii) $H$ is {\it objective}, that is,
\begin{equation}
H(p_1,...,p_N)\,=\,H(p_1+p_2,p_3,...,p_N)\,+\,(p_1+p_2)\,
H\bigg(\frac{p_1}{p_1+p_2},\frac{p_2}{p_1+p_2}\bigg).
\end{equation}
\end{itemize}

\noindent This last condition entails that information does not change when one
appropiately manages different chunks of it. The principle that entropy is a measure 
of our ignorance about a given physical system was recognized by Weaver, Shannon and 
Smoluchowski. Boltzmann was also aware of it. On the other hand, the mathematical 
theory of information originally was intended as a theory of communication, as we know. 
The simplest problem it deals with could be the following: given a message, one can represent 
it as a sequence of bits and thus, if the length of the ``word" in $n$, one needs $n$ 
digits to characterize it. The set $E_n$ of all words of length $n$ contains $2^n$ elements, 
so the amount of information needed to characterize one element of its is log$_2$ 
of the number of elements of $E_n=$log$_2N$, with $N=2^n$. Elaborating this argument 
a little bit more, one arrives at the result that the amount of information required 
to describe an element of any set of power $N$ is log$_2N$. Now suppose that 
$E=E_1\cup ... \cup E_k$ of pairwise disjoint sets, with $N_i$ representing the number 
of elements of $E_i$. Let $p_i=N_i/N$, $N=\sum_i N_i$. If one knows that an element 
of $E$ belongs to $E_i$, one needs log$_2N_i$ additional information in order to determine 
it completely. Therefore the average amount of information needed to determine an element 
is $\sum_i (N_i/N)$log$_2N_i$=$\sum_i p_i$log$_2Np_i$=$\sum_i p_i$log$_2p_i$+log$_2N$. In 
consequence the lack of information is just -$\sum_i p_i$log$_2p_i$, or just the entropy 
(\ref{H}).

We do not discuss here the details 
of the measure $H$ here. They shall be discussed employing the von Neumann entropy 
$S(\rho)$. Also, the theorems related to information channels and data compression 
will be discussed when compared to the quantum case. However, we must point out that 
all possible candidates to become information measures or entropies have to follow the 
so called Khinchin axioms \cite{Khinchin}. Two of them are convexity and additivity. 
By relaxing the convexity condition, one may encounter R\'{e}nyi's entropy, while 
relaxing the additivity constraint gives rise to Tsallis' entropy. Both of them are 
parameterized with $q$ real, and recover the usual Shannon entropy (\ref{H}) in the limit 
$q \rightarrow 1$. These features are described in detail in Chapters 4 and 7-9.
The entropy $H$, when applied to an information source, 
could determine the capacity of the channel required to transmit the source as 
encoded binary digits. If the logarithm in the formula (\ref{H}) is taken to base 2, then 
it gives a measure of entropy in bits. Shannon's measure of entropy came to be 
taken as a measure of the information contained in a message, as opposed to the 
portion of the message that is strictly determined (hence predictable) 
by inherent structures, such as redundancy in the structure of languages or the 
statistical properties of a language relating to the frequencies of occurrence 
of different letter or words.

Information entropy as defined by Shannon and added upon by other physicists 
is closely related to thermodynamical entropy. A glance to (\ref{H}) would tell 
any physicist that one is talking about entropy, but as we known, the word ``entropy" 
is used in many contexts: information theory, thermodynamics, statistical mechanics,
etc. Information, entropy, order and disorder are words that are often mixed altogether 
in the same context, adding more confusion. Of course there is a formal analogy 
of the information entropy (\ref{H}) and Boltzmann-Gibbs' $S=k\,$log$W$, which applies 
to microscopic systems. With the extension to quantum mechanics made by von Neumann, 
employing the density matrix formalism, one generalizes the concept of entropy to both 
classical and quantum physics. The final connection between information entropy and 
thermodynamical entropy is encountered by maximizing the former --restricted to 
several constrains-- applied to the context of the latter, that is, the description of 
thermodynamics through the tools of statistical mechanics.

\subsection{Jaynes' principle and the thermodynamical connection: 
information and entropy}

The similarities between the information entropy $H$ and Boltzmann's entropy $S=k\,$log$W$, 
especially between the principle of maximum (informational) entropy (which adopts 
as we shall show an exponential form for the probability density distribution or the density 
matrix) and the Gibbs' factor in statistical mechanics, 
were too much coincidence for E. T. Jaynes. To him, the principle of maximum entropy 
\cite{B91,BAD96}, in the framework of inference of information, was advanced as the basis 
of statistical mechanics \cite{J57}. His idea was to view statistical mechanics as a form 
of inference: information theory provides a constructive criterion 
(the maximum-entropy estimate). If one considered statistical mechanics as 
a form of statistical inference rather than a physical theory, the usual rules 
were justified independently of any physical argument, and in particular 
independently of experimental verification. This was the gist of Jaynes' principle.

According to Jaynes' principle, one
must choose the state yielding the least unbiased description of
the system  compatible with the available data. Either if we are 
dealing with classical or quantum statistical mechanics, the spirit is the same. 
That state is provided by the statistical operator  $\hat \rho_{ME} $ that
maximizes the von Neumann entropy $S \, = \, -Tr (\hat \rho \, \ln
\hat \rho)$ subject to the constraints 
imposed by normalization and the expectation values $\langle \hat A_i \rangle \,= \, Tr
(\hat \rho \hat A_i) $ of the relevant observables $\hat A_i$. The outcome 
of this procedure is a density matrix 
$\hat \rho$ proportional to $e^{-\sum_i \lambda_i \, \hat A_i}$, 
$\lambda_i$ being some suitable Lagrange multipliers.

What Jaynes' prescription provides is the best ``bet" that one can
make on the basis of the available data. Clearly, this available
information may not be enough to predict certain properties
of the system. In such cases, Jaynes' prescription is
bound to ``fail" because of a lack of input information \cite{B91,BAD96}. This point 
is the main subject of discussion in Chapter 1, in connection with the inference of 
states using MaxEnt procedures and entanglement.

Jaynes' principle unties statistical mechanics from physical theories, and 
considers it as a suitable inference procedure. Certainly it is an interpretation 
of statistical mechanics in terms of an extremely simple inference scheme based 
of the expectation value of several observables, e.g. the system Hamiltonian $H$, hence 
the Gibbs' factor $e^{-\beta \langle H \rangle}$, $\beta$ being a Lagrange multiplier. 
However, one still needs to associate or to {\it interpret} the concomitant Lagrange multiplier 
with definite physics quantities ($\beta \equiv 1/ k_BT$). 

Notice that by no means the aim of Jaynes' principle is to justify the grounds of 
statistical mechanics. Aspects of the foundations of statistical mechanics \cite{terHaar55} 
such as the $H$-Theorem or the Ergodic Theorem are questions out of the scope 
of these lines: Jaynes' MaxEnt prescription cannot explain the fact that statistical mechanics 
has been able to predict accurately and successfully the behaviour of physical systems 
under equilibrium conditions. It rather simplifies instead the approach to statistical mechanics, 
and the proof of its success, {\it the proof of the pudding, is in the eating}.

\subsection{Alan Turing and the universal computing machine}

The birth of computer science is associated with the publication 
of the work ``On Computable Numbers" by A. M. Turing in 1936, who is considered the father 
of computer science. He basically posed 
the operating principles (further developed by von Neumann\footnote{He was the first to 
formalize the principles of a ``program-registered calculator" based in the
sequential execution of the programs registered in the
memory of the computer (the von Neumann machine).} in 1945) of 
ordinary computers (birth of the Turing machine). Together with A. Church they 
formulated what is known as the ``Church-Turing hypothesis": every physically 
reasonable model of computation can be efficiently simulated on a universal 
Turing machine \cite{Galindo}. 

A Turing machine is the mechanical translation of what a person does during a 
methodical process of reasoning. Turing also provided convincing arguments
that the capabilities of such a machine would be enough to encompass everything 
that would amount to a recipe, which in modern language is what
we call an algorithm. Turing arrived at this concept in his original paper \cite{Turing} 
in an attempt to answer one of the Hilbert's problem, namely, the problem of decidability: 
Does there exist a definite method by which all mathematical
questions can be decided? In the Turing machine context, he found that the problem 
of determining whether a particular machine will halt on a particular input, or on all 
imputs, known as the Halting problem\footnote{Historically, the Halting problem 
was not a merely academic question. In the early times of the first computers, 
composed by thousands of valves, it was usual that the machine entered an infinite 
loop, which had a time and money expense.}, was undecidable. Besides, the logicician A. 
Church shown that the decidability problem was unsolvable: there cannot be a general 
procedure to decide whether a given statement expresses an arithmetic truth. In 
other words, there will not be any Turing machine capable of deciding the truth 
of an arithmetic statement. In point of fact, undecidability shall lead in 
computer science to the classification of the types of problems that can be 
algorithmically solvable into a series of {\it complexity classes}. 
As we have seen, the birth of the Turing machine, of computer 
science, appeared in a context of a crisis in the foundations of mathematics.  

Let us return to the concept of a Turing machine. More precisely, it consists of:

\begin{itemize}
\item A {\it tape} which is divided into cells, one next to the other. Each cell 
contains a symbol from some finite alphabet. This alphabet contains a special 
{\it blank} symbol (``0") and one or more symbols. The tape is assumed to be arbitrarily 
extendible to the left and right. Cells have not been written before, containing the 
blank symbol.
\item A {\it head} that can read and write symbols on the tape and move left and right.
\item A {\it state register} that stores the state of the Turing machine. The number 
of different states is always finite. There exists a special {\it start state} which 
initializes the register.
\item A transition function that tells the machine what symbol to write, how to move the 
head (``L" or ``-1" for one step left, and ``R" or ``1" for one step right) and what 
its new state will be. If there is no entry in the function then the machine will halt.
\end{itemize}

\noindent According to the previous description, a Turing machine contains finite elements, 
except for the potentially unlimited amount of tape, which translates into unbounded 
capacity of storage space. More intuitively, a Turing machine can be viewed as an 
automaton that moves left-right and reads/writes on an infinite tape when it receives an 
order. More formally, a one-tape Turing machine is defined as a 6-tuple 
$M=(Q,A,\delta,s,F,b)$, where $Q$ is a finite set of states, $A$ is a finite 
set of the tape alphabet, $s\in Q$ is the initial state, $F \subseteq Q$ is the set of 
final or accepting states, $b\in A$ is the blank symbol (``0") and $\delta$ is the 
transition function

\begin{equation}
\delta:\,Q \times A\,\rightarrow\,Q \times A \times \{-1,0,1 \}. 
\end{equation}

\noindent Extensions to $k$-tape Turing machines are straightforward. What 
changes is the definition in the transition function

\begin{equation}
\delta:\,Q \times A^k\,\rightarrow\,Q \times \big( A \times \{-1,0,1 \}\big)^k. 
\end{equation}

\noindent The previous definitions for Turing machines on one tape or several 
tapes belong to the class of {\it deterministic} Turing machines (when the 
transition function has at most one entry for each combination of symbol and 
state). Turing machines are useful models of real (classical) computers. 
Despite their simplicity, Turing machines can be devised
to compute remarkably complicated functions. In
fact, a Turing machine can compute anything that the
most powerful ordinary classical computer can compute, which boils down to 
the aforementioned Church-Turing hypothesis. Why {\it universal} Turing machines? 
The importance of the universal machine 
is clear. We do not need to have an infinity of different machines
doing different jobs. A single one will suffice. The engineering
problem of producing various machines for various
jobs is replaced by the office work of programming
the universal machine to do these jobs. In
summary, a Turing machine is comparable to an algorithm
much as the universal Turing machine is to a programmable
computer \cite{Galindo}.

Now, there also exist a model for {\it probabilistic} Turing machines, 
that is more suitable in order to tackle a generalization to the 
quantum domain. Defining the same a 6-tuple $M=(Q,A,\delta,s,F,b)$, 
the new transition function becomes a {\it transition probability distribution}

\begin{equation} \label{pro}
\delta:\,Q \times A\, \times Q \times A\,\times \{-1,0,1 \} \, \rightarrow\, [0,1]. 
\end{equation}

\noindent The value $\delta(q_1,a_1,q_2,a_2,d)$\footnote{For the sake of 
simplicity, on shall assume this number to be rational.} has to be viewed as the 
probability that when the machine is in state $q_1$ and reads the symbol $a_1$, 
it will write the symbol $a_2$, jump to the state $q_2$ and move the head to 
the direction $d\in \{-1,0,1\}$. Clearly, it is required that for all initial states 
$(q_1,a_1) \in Q \times A$

\begin{equation}
\sum_{{\rm All \,possible\, final \,states}\,(q_2,a_2)} \delta(q_1,a_1,q_2,a_2,d)\,=\,1.
\end{equation}

\noindent Although the basics of quantum computation will be exposed in Chapter 2, 
let us show how a quantum Turing machine would look like. In the same vein as before, 
the 6-tuple $M=(Q,A,\delta,s,F,b)$ is formally the same, although the ``states" have to 
be considered quantically, and we replace probabilities with transition amplitudes. 
Therefore we have a {\it transition amplitude function}

\begin{equation}
\delta:\,Q \times A\, \times Q \times A\,\times \{-1,0,1 \} \, \rightarrow\, \mathcal{C} 
\end{equation}

\noindent that generalizes the probabilistic Turing machine {\ref{pro}}. 
$\delta(q_1,a_1,q_2,a_2,d)$ are now complex amplitudes, that satisfy the 
normalization condition

\begin{equation}
\sum_{{\rm All\, possible\, final\, states}\,(q_2,a_2)} |\delta(q_1,a_1,q_2,a_2,d)|^2\,=\,1.
\end{equation}

\noindent A quantum Turing machine\footnote{A pictorical image of a quantum 
Turing computer would be given by replacing 
the alphabet by Bloch spheres (qubits) in the tape and in the state register. As we shall 
see, a qubit is a coherent superposition of the classical bit states 0 and 1.} 
operates in steps of fixed
duration $T$, and during each step only the processor and
a finite part of the memory unit interact via the cursor.
We stress that a quantum Turing machine, much like a
Turing machine, is a mathematical construction \cite{Galindo}; we shall
present explicit experimental realizations of equivalent
quantum circuits (see Chapter 2). The state of the computation is the state 
of the whole quantum Turing machine (Hilbert space $\cal{H}_{QC}$), 
represented by $|\Psi\rangle$. All the set of instructions are 
encoded in the unitary time evolution $U$ of state $|\Psi\rangle$, such 
that after a number $n$ of computational steps, the state 
of the system has evolved to $|\Psi(n\,T)\rangle = U^n\, |\Psi(0)\rangle$. 

Because of unitarity, the dynamics of a quantum computer, as in the case 
of any closed quantum system, are necessary reversible. Turing machines, 
on the one hand, undergo irreversible changes during computations and until 
the 80s it was widely held that irreversibility was an essential feature of 
computation. Turing machines, like other computers, often throw away information 
about their past, by making a transition into a logical state whose predecessor 
is undefined\footnote{A Turing machine is {\it reversible} if each configuration 
admits a unique precessor.}. However, Bennett \cite{Bennett} proved in 1973 that this should 
not be the case by constructing explicitly a reversible classical model 
computing machine equivalent to Turing's. This so called ``logically" irreversibility 
has an important bearing on the issue of thermodynamics of (classical) computation, 
which is the subject of the next section.
\newline

Summing up, Turing's invention was built on the insight of Kurt G\"odel that 
both numbers and operations on numbers can be treated as symbols in a syntactic 
sense. The way to modern (classical) computers starts with the definition of a Turing 
machine, followed by the basic Boolean operations carried out by logic gates 
inserted in integrated circuits. Nowadays we take for granted that all information 
and programming instructions can be expressed by strings of 0 and 1, 
and that all computations, 
ranging from simple arithmetic operations to proving theorems, can be 
carried out when a set of systematic operations (the program) are applied 
to this string of bits. However, we do not discuss here the technical part 
that deals with actual (classical) computing. Instead, it shall 
be revisited in the description of the quantum counterpart.


\subsection{Thermodynamics of information and classical computation} 

``The digital computer may be thought of as an engine that dissipates energy in order to 
perform mathematical work."\newline
{\it C. H. Bennett} in \cite{Bennett82}.\newline

While I am writing these lines, I can perfectly hear the swing of the laptop. 
And if I place my hand behind it, I shall notice the flow of hot air. This fact simply means 
that my computer (and any classical computer known to date) 
dissipates energy in the form of heat. Although modern computers are several orders 
of magnitude more efficient (energetically) than the first electronic ones, they 
still release vast amounts of energy as compared with $k_B\,T$. Even in the 
hypothetical case where perfectly engineered classical computers should not have to be 
cooled down\footnote{Basic components dissipate energy even when the information 
in them is not being used. The energy in an electric pulse sent from one component 
to another could in principle be saved and reused, but it is easier 
and cheaper to dissipate it. The macroscopic size of the components, however, 
is the basic reason for dissipating energy.} in order to perform basic operations, 
we would end up with a physical thermodynamical minimum limit of heat released 
during calculations. 
This is due to the fact that information is not always stored; most of the time a lot 
of useless information is erased. According to Landauer \cite{Landauer}, an irreversible 
overwriting of a bit causes at least $k_BT$ln2 Joules of heat being 
dissipated\footnote{This statement is often known as {\it Landauer's Principle}.}. 
In other words, 
erasure of a bit of information in an environment at temperature $T$ leads to a 
dissipation of energy no smaller than $k_BT$ln2.

Why is this so? In a sense, everything dates back to the 19th century, 
before quantum mechanics or information theory were conceived. Maxwell found in his 
famous paradox an apparent contradiction with the Second Law of thermodynamics. 
Maxwell considered an intelligent being (which could be a programmed machine), later baptized 
as a Demon, which was able to open and close a gate inside a gas vessel, which is divided 
into two parts. By letting faster molecules pass through the gate only, this Demon 
would increase the temperature of one side of the vessel without work, thus 
violating the Second Law. It was Szilard \cite{Szilard} who refined in 1929 the conceptual
model proposed by Maxwell into what is now known as Szilard's engine. 
This is a box with
movable pistons at either end, and a removable partition in the middle. The walls are
maintained at constant temperature $T$, and a particle\footnote{A classical particle. No 
quantum mechanical setting is yet considered.} collides with the walls. A cycle of the 
engine begins with the Demon partitioning the box and observing which side the particle is on. 
It then	moves the piston towards the empty side up to the partition, removes the partition, and
allows the particle to push the piston back to its starting position, the whole cycle 
being isothermal. At each cycle the engine supplies $k_BT$ln2, violating the Second Law. 
Szilard deduced that, if we do not want to admit that the Second Law has been violated, 
the intervention which establishes the coupling between the measuring apparatus and the 
thermodynamic system must be accompanied by a production of entropy, and he gave the 
explicit form of the ``fundamental amount" $k_B$ln2, which is the entropy associated to a 
dichotomic or binary decision process. 
In other words, the entropy of the Demon must increase, conjecturing that this would be 
a result of the (assumed irreversible) measurement process. 
Szilard not only defined the quantity that it is known today as information, which 
found vast applications with the work of Shannon, but also finds the physical connection 
between thermodynamic entropy and information entropy when he establishes that one has to 
pay (at least) $k_BT$ln2 units of free energy per bit of information gain. 
Szilard's argument was a 
pioneering insight of the physical nature of information, indeed. 
Later on, von Neumann also associated 
entropy decrease with the Demon's knowledge in his 1932 {\it Mathematische Grundlagen der 
Quantenmechanik} \cite{vonNeumann}.

\begin{figure}
\begin{center}
\includegraphics[angle=0,width=0.8\textwidth]{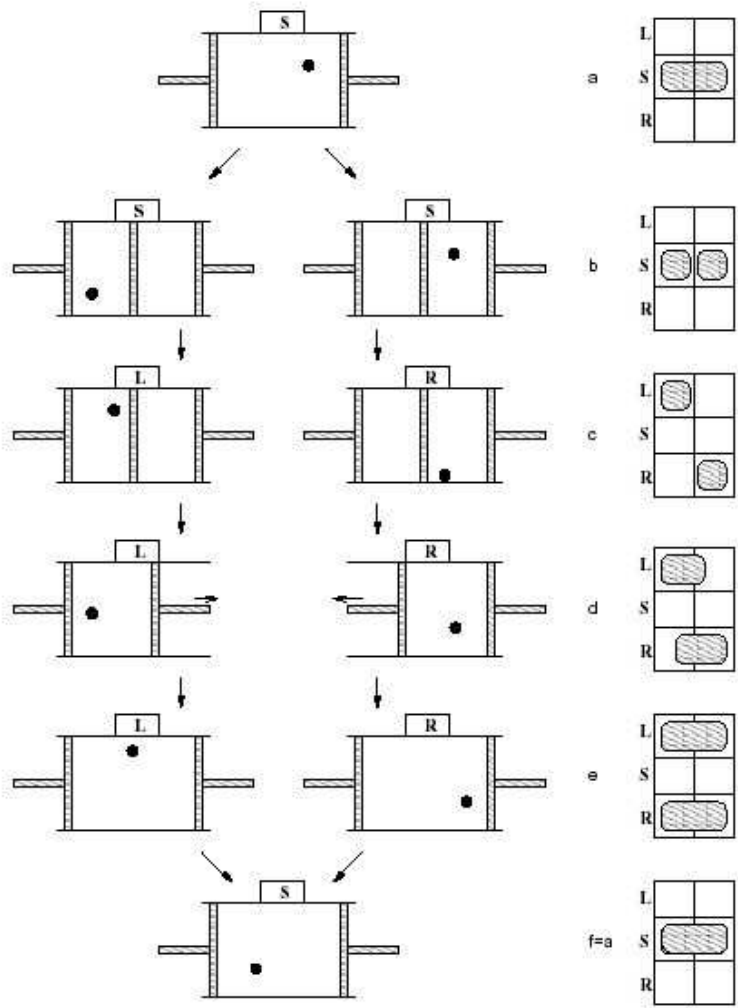} 
\caption{A one molecule Maxwell's Demon apparatus (after \cite{Bennett82}). 
On the right there is a phase diagram with the Demon's coordinates on the vertical 
axis.} 
\label{Szilard}
\end{center}
\end{figure}

The resolution of the paradox would lead to the discovery of a connection between physics 
and the gathering of information. The Demon was finally exorcised in 1982 
by Bennett \cite{Bennett82}. In the meantime, in order to rescue the Second Law, 
many efforts were made involving analyses of the measurement process, such as 
information acquisition via light signals (L. Brillouin in \cite{Brillouin}), which were 
temporary resolutions. Bennett observed that the Demon ``remembers" the information it 
obtains, much as a computer records data in its memory. He then argued that 
erasure of Demon's memory (and here is the link with Landauer's work on computation) 
is the fundamental act that saves the Second Law. Let us follow Bennett's argument 
with the help of Fig.\ref{Szilard}, taken from \cite{Bennett82}, and follow the phase 
space changes of Demon's coordinates through one cycle. In (a) the Demon is in a 
standard state and the particle is anywhere in the box\footnote{Recall that the
entropy of the system is proportional to the phase space volume occupied.}. 
In (b) the partition is inserted and in (c) the Demon makes his measurement. By doing so, 
his state of mind becomes correlated with the state of the particle. Note that the overall 
entropy has not changed. Isothermal expansion
takes place in (e), and the entropy of the particle {\it plus} the Demon increases. 
After expansion the Demon remembers some information, but this is not correlated to 
the particle's position. In the way back to his standard state in (f) we increase its entropy, 
dissipating energy into the environment. If von Neumann had addressed in 1932 the process 
of discarding information in order to bring back the Demon to its initial state, 
he might have discovered what Bennett solved a lot earlier.

Returning to the heat release due to erasure of information, Bennett \cite{Bennett} 
also proved that reversible computation, which avoids erasure of information, 
which in turn is tantamount as avoiding an energy release, was 
possible in principle. Bennett's construction of a reversible Turing machine 
uses in fact three tape Turing machines: input tape, history tape and output 
tape. When we simulate the original machine in the input tape, we store the 
transition rules in the history tape. In this way we obtain reversibility. 
Every time the machine stops, we copy the output from the input tape to the output 
tape, which is empty. Then we compute backwards in order to erase the history time 
for further use. Although we do not give the details, this construction consumes 
considerable memory, being reduced by erasing the history tape recursively, has got 
constant slowdown and increases the space consumed. Nevertheless, Bennett thus 
established that whatever is computable with a Turing machine, it is also 
computable with a reversible Turing machine.
\newline
\newline
{\bf Maxwell's demon revisited: information and measurement 
in the light of quantum mechanics}
\newline
\newline
Until the work of Bennett in 1973 \cite{Bennett}, it was thought that the 
computation process was necessarily irreversible, where energy dissipation 
was associated with information erasure. However he showed that to every 
irreversible computation there exists an equivalent reversible computation. 
But Bennett's work did not addressed concerns realted to quantum effects. 
Certainly processes such as ``measuring" had to be carefully studied 
in the quantum domain.

It was Zurek \cite{Zurek} who careful performed a quantum analysis of Szilard's engine. 
He considered a	particle in an infinite square well potential, with the following 
dimensions: lenght $L$, the classical piston replaced by a finite barrier of 
length $\delta << L$ and height $U >> k_BT$, which is slowly inserted. 
In the quantum version, Zurek shows that the validity of the Second Law is satisfied 
only if the measurement induces an increase of the entropy of the measuring 
apparatus by an amount which has to be greater or equal to the amount of 
information gained \cite{Zurek}. Zurek arrives to this main result 
by observing that the system can at all times be described by its partition function. 
This means that the thermodynamic approximation is indeed valid. A quantum Demon 
explicitly has to reset itself, thus demonstrating that Bennett's conclusion regarding 
the destination of the excess entropy is also correct. 
The measurement of the location of the molecule was of essential nature in the process 
of extracting work in both classical and quantum versions of Szilard's engine.

The fact that erasure of information is a process which costs free energy 
has interesting echoes in quantum information theory. To be more precise, if one 
is able to efficiently erase information, which is tantamount as to saturate 
Landauer's bound $k_BT$ln2, then one can provide a physical interpretation 
\cite{Vedral96,Plenio99} of the so called Holevo bound \cite{Holevo}, which is 
related to the information capacity in quantum channels (Chapter 3). It is 
interesting to see that a bound that is found as a relation satisfied by the 
von Neumann entropy can be interpreted in terms of Landauer's bound $k_BT$ln2.

\section{Foundational and fundamental aspects of quantum mechanics}

``I think I can safely say that nobody today understands quantum mechanics."\newline
{\it R. P. Feynman} in \cite{Feynman65}.\newline

Quantum physics\footnote{There are several excellent books on the history 
of quantum mechanics and the early stages of quantum physics. The reader is 
referred to A. Messiah \cite{M61} and Waerden \cite{Waerden}.} was born in 1900 on Max Planck's 
hypothesis of discretized energy packets or 
{\it quanta} --hence the name {\it quantum}-- as a working hypothesis in order to 
explain the spectrum of a {\it black body}, which put an end to the classical period. 
But it was Einstein in 1905 who became 
the first physicist to apply Max Planck's quantum hypothesis to light (explanation of the 
photoelectric effect). Einstein realized
that the quantum picture can be used to	describe the photoelectric effect. Later on they 
followed the quantization of the energy levels of atoms by Bohr (1913), the famous 
Stern-Gerlach experiment (1922) describing the quantization of the atomic 
systems, the de Broglie 
hypothesis of particles behaving as waves (1924), the first interference 
experiments with electrons carried out by C. J. Davidson and L. H. Germer (1927), and 
the confirmation of the photon theory with the Compton effect (1924). The Bohr 
correspondence principle formulated in 1923, namely, {\it Quantum Theory must approach 
Classical Theory assymptotically in the limit of large quantum numbers} and the subsequent 
Bohr-Sommerfeld quantization rules close the period known as Old Quantum Theory. 
Although the Old Theory undoubtly represented a great step forward, predicting 
a considerable body of experiments from simple rules, it was a rather haphazard mixture 
of classical mechanics and {\it ad hoc} prescriptions.

The physical theory of quantum mechanics (QM) was born by the efforts of men such as M. Born,
P. A. M. Dirac, P. Jordan, W. Pauli, E. Schr\"odinger and W. Heisenberg. The 
founding of QM can be placed between 1923 and 1927 and put an end to the ambiguities 
of the Old Theory. Thereof matrix mechanics and wave mechanics have been proposed 
almost simultaneously: Schr\"odinger's wave formulation and Heisenberg's matrix formulation 
were shown to be equivalent mathematical constructions 
of QM. The transformation theory invented by Dirac unified and generalized Schr\"odinger's and 
Heisenberg's matrix formulation of QM. In this formulation, the state of the quantum 
system encodes the probabilities of its measurable properties or ``observables", which is 
a technical word in QM with a definite meaning. Roughly speaking, QM does not 
assign definite values to observables. Instead, it makes predictions about probability 
distributions of the possible outcomes from measuring an observable.

The problem about quantum mechanics does not lie on its effectivity, but on its interpretation. 
Any attempt to interpret quantum mechanics tries to provide a definite meaning to issues 
such as realism, completeness, local realism and determinism. Historically, 
the understanding of the mathematical structure	of QM went trough various stages. At first, 
Schr\"odinger did not understand the probabilistic nature of the wavefunction of 
the electron. It was Born who proposed the widely accepted interpretation as a probability 
distribution in real space. Also, Einstein had great difficulty in coming to terms with 
QM (section on EPR paradox). Nowadays the Copenhagen 
interpretation\footnote{Born around 1927, while collaborating 
in Copenhagen. They extended the probabilistic interpretation 
of the wavefunction, as proposed by M. Born, in an attempt to 
answer questions which arise as a result of the wave-particle duality, such as 
the measurement problem.} (after Bohr and Heisenberg) of QM 
is the most widely-accepted one, followed by Everett's many worlds interpretation 
\cite{Everett}. Very briefly, 
the Copenhagen assumes two processes influencing the wavefunction, namely, i) its unitary 
evolution according to the Schr\"odinger equation, and ii) the process of measurement. 
As it is well known, the Copenhagen interpretation 
postulates that every measurement induces a discontinuous break in the unitary 
time evolution of the state through the collapse of the	
total wave function onto one of its terms in the state vector 
expansion (uniquely determined by the eigenbasis of 
the measured observable), which selects a single term in 
the superposition as representing the outcome. The nature 
of the collapse is not at all explained, and thus the 
definition of measurement remains unclear. Macroscopic superpositions 
are not a priori forbidden, but never observed since any observation 
would entail a measurementlike interaction. In the words of philosophy, Bohr followed 
the tenets of positivism, that implies that only measurable questions should be discussed 
by scientists.

Some physicists (see Ref. \cite{PeresFuchs}) argue that an interpretation is nothing 
more than a formal equivalence between a given set of rules for processing 
experimental data, thus suggesting that the whole exercise of interpretation is 
unnecessary. It seems that a general consensus has not yet been reached. In the opinion of 
Roger Penrose \cite{Penrose}, who remarks that while the theory agrees incredibly well with 
experiment and while it is of profound mathematical beauty, it ``makes absolute no sense".
\newline

The present status of quantum mechanics is a rather complicated and discussed subject 
(see Ref. \cite{Wheeler,Omnes}). 
The point of view of most physicist is rather pragmatic\footnote{It can be 
expressed in Feynman's famous dictum: ``Shut up and calculate!".}: it is a physical theory with 
a definite mathematical background which finds excellent agreement with experiment. 
In this Chapter we shall 
present the most important results regarding Quantum Mechanics and several issues 
in quantum information theory. We shall not discuss the philosophical implications of 
results such as the interpretation of QM (completely out of the scope of this Thesis) 
which, since the advent of quantum entanglement, has gain considerable attention among 
the physics community.

\subsection{The postulates of quantum mechanics}

In this seccion we are going to provide a brief review of the basic formalism
of quantum mechanics and of its Postulates. Here we closely follow the definitions 
given by C. Cohen {\it et al.} in \cite{Cohen}.

In the mathematical rigorous formulation of quantum mechanics \cite{Mackey}, 
developed by P. A. M. Dirac\footnote{His {\it bra-ket} notation is so extensively 
used that one would not conceive quantum theory without it!.} and J. von Neumann, 
the possible states of a 
quantum system are represented by ``state vectors" (unit vectors) living in a 
complex Hilbert space, usually known in the quantum theory jargon as 
the ``associated Hilbert space" of the system. Observables 
are represented by an Hermitian (or self-adjoint) linear operator acting on the 
state space. Each eigenvector of the operator possesses an eigenstate of an observable, 
which corresponds to the value of the observable in that eigenstate. The operator's 
spectrum can be discrete or continuous. The fundamental role played by complex 
numbers in quantum theory has been found very intriguing by some physicists. Attemps 
to reach a deeper understanding of this aspect of quantum theory have led some 
researchers to consider a quantum formalism based upon quaternions \cite{Adler}. 
However, it seems that the field of complex numbers is enough in order to describe 
quantum phenomena.


The time evolution (time $t$ is {\it not} an observable in quantum mechanics) of a state 
is determined by the Schr\"odinger equation, in which the Hamiltonian $H$, through a unitary 
matrix obtained by complex-exponentiating $H$ times $t$, generates the time evolution of 
that state. The modulus of it describes the evolution of a probability distribution, 
while the phase provides information about interference, hence the name wavefunction 
as being synonymous with quantum state\footnote{In fact this is reminiscent from the 
Old Quantum Theory. It is more correct to employ the technical word {\it quantum 
state}.}. 
Schr\"odinger's equation is completely deterministic, so there is nothing new in this sense 
as compared with classical mechanics.  

Heisenberg uncertainty principle is represented by the fact that two observables 
do not commute. Using Max Born's interpretation, the inner product between two states 
is a probability amplitude (usually a complex number). The possible outcomes 
of a measurement are the eigenvalues of the operator --this explains why 
observables have to be hermitian, i.e., they must be real numbers. The process 
of measurement is not yet understood (it is non-unitary): the system collapses 
from the initial state to one of its eigenstates with a probability given by 
the square of their inner product.  

One can also look at quantum mechanics using Feynman's path integral formulation, 
which is the quantum-mechanical counterpart of the least action principle in 
classical physics.
\newline
\newline
{\bf Postulate 1} 
\newline

``The state of a system is described by a vector in a Hilbert space ${\cal H}$".
\newline

The state of any physical system at time $t$ is defined by specifying a ket 
$|\psi(t)\rangle$ belonging to a state space ${\cal H}$. ${\cal H}$ is a vector 
space, with the concomitant property of linearity. A Hilbert space is complex 
vector space with a scalar product. 
\newline
\newline
{\bf Postulate 2 (principle of spectral decomposition)} 
\newline

``a) Discrete spectrum\newline

The probability	that a measurement of an observable $\hat A$ yields an eigenvalue 
$a_n$ when the system is in a normalized state $|\psi\rangle$ is given by

\begin{equation} 
P(a_n)=\sum_{i=1}^{g_n} |\langle a_n^{(i)}|\psi\rangle|^2,
\end{equation}

\noindent where $|a_n^{(i)}\rangle$ is a normalized eigenvector of $\hat A$ 
associated with $a_n$, and $g_n$ is its degeneracy.

 b) Continuous non-degenerate spectrum\newline

 The probability that a measurement of an observable $\hat A$ yields a value 
 between $\alpha$ and $\alpha$+$d\alpha$ when the system is in a normalized 
 state $|\psi\rangle$ is given by

\begin{equation} 
dP(\alpha)=|\langle \alpha|\psi\rangle|^2 \,d\alpha,
\end{equation}

\noindent where $|\alpha \rangle$ is the eigenvector of $\hat A$ associated with 
the eigenvalue $\alpha$".
\newline
\newline
{\bf Postulate 3} 
\newline

``Physical observables are represented by hermitian operators that act on ket vectors".
\newline

The results of a measurement is given by the eigenvalues of the operator $\hat O$. 
By the spectral decomposition principle we can write this operator as

\begin{equation} 
\hat O = \sum_i o_i  |W_{o_i}\rangle \langle W_{o_i}|,
\end{equation}

\noindent where $|W_{o_i}\rangle$ are the eigenstates of $\hat O$. 

One way of defining a state using an hermitian operator is possible through 
the definition of the {\it density matrix} $\rho$. If the system is found in a 
pure state $|\psi\rangle$, then $\rho=|\psi\rangle \langle \psi|$. Due to 
interaction with the environment, the state of the system is usually found in a 
admixture of states $\rho=\sum_i \lambda_i |\psi_i\rangle \langle \psi_i|$, with 
$0\leq \lambda_i \leq 1$. It has the properties i) Tr($\rho$)=1 and ii) being positive 
for all states $|\phi \rangle$, that is, $\langle \phi|\rho |\phi \rangle \geq 0$, 
where $\rho$ is an hermitian operator, but not an observable. The state 
$\rho$ contains all the information that can be accessed about the system. 
\newline
\newline
{\bf Postulate 4 (reduction postulate, i.e, collapse of the wavefunction)} 
\newline

``The action of a measurement is to project the	state into an eigenstate of 
the observable $\hat O$".
\newline

 Given an eigenvalue $o$ of the observable $\hat O$, the projector $P_o$ onto 
 the subspace expanded by the eigenstates with eigenvalue $o$ is $P_o = \sum_{o_i = o} 
 |\Psi_{o_i}\rangle \langle \Psi_{o_i}|$, where the sum runs over all the 
 eigenvectors sharing the same eigenvalue $o$. 
After the measurement, the state is given by

\begin{equation} 
 |\Psi_{o_i} \rangle = \frac{\hat P_{o_i} |\Psi\rangle}
 {\sqrt{\langle \Psi| \hat P_{o_i} |\Psi\rangle}}.
\end{equation}

\noindent In terms of the density matrix, we have

\begin{equation} 
\rho\prime=|\Psi_{o_i} \rangle \langle \Psi_{o_i}| = 
\frac{\hat P_{o_i} \hat \rho \hat P_{o_i}^{\dagger}}{Tr(\hat\rho \hat P_{o_i})},
\end{equation}

\noindent with the assumptions of i) orthogonality 
Tr($\hat P_{o_i} \hat P_{o_j}^{\dagger}=\delta_{i,j}$) and ii) closure 
$\sum_i \hat P_{o_i} = \hat I$. 

There is a generalization of the concept of measurement	to POVM 
(positive operator-valued measure) where the different measurements are not 
orthogonal (i.e. they are represented by operators $\hat A$,$\hat B$ such that 
Tr($\hat A\hat B$)$\neq 0$). The previous measurement is known as 
a von Neumann or projective measurement \cite{NC00}.
\newline
\newline
{\bf Postulate 5} 
\newline

``The evolution of an isolated quantum system is given by the Schr\"odinger equation

\begin{equation} \label{evolH}
i\,\hbar\,\frac{\partial}{\partial t}\,|\psi(t)\rangle \, =\, \hat H\, |\psi(t)\rangle 
\end{equation}

\noindent where $\hat H$ is the Hamiltonian of the system, the operator related to the 
total energy".
\newline

 There is a formal solution to (\ref{evolH}) given by 
 $|\psi(t)\rangle=\exp(-\frac{i}{\hbar}\int dt\,\hat H)|\psi(0)\rangle$. Because 
 $\hat H$ is hermitian, $U\equiv\exp(-\frac{i}{\hbar}\int dt\,\hat H)$ is a 
 unitary operator. In quantum computation, $U$ is the representation of an algorithm.
 \newline
 \newline

As a final remark, we could postulate also that the state space of a composite 
physical system is given by the tensor product of the state space of its components, 
as opposed to the cartesian product in classical physics. Directly linked to the 
tensor product nature of state space, it emerges the notion of reduced matrices
\footnote{They emerged in the earliest days of quantum mechanics. See Refs. 
\cite{vonNeumann} and \cite{Landau}.}, which in a composite systems are used in order to 
address individual subsystems.

\subsection{The EPR paradox: non-locality and hidden variable theories}

Einstein never liked the implications of quantum theory, 
despite the undeniable success of quantum theory. Einstein's hope was that quantum 
mechanics could be completed by adding various as-yet-undiscovered variables. These 
``hidden" variables, in his opinion, would let us regain a deterministic
description of nature\footnote{His discomfort is clear in 
his celebrated ``God does not play dice".}. The completeness of quantum mechanics 
was attacked by the Einstein-Podolsky-Rosen gedanken experiment \cite{EPR} which 
was intended to show that there have to be {\it hidden variables} in order to avoid 
non-local, instantaneous ``effects at a distance". In the original paper, the 
position-momentum uncertainty relation served as a guideline for their 
argument, although it is most clear to us with the help of D. Bohm \cite{Bohm} 
employing a pair of spin-$\frac{1}{2}$ particles in a singlet state. 

In their paper, EPR argued that any description of nature
should obey the following two properties: 

\begin{itemize}

\item Anything that
happens here and now can influence the result of a measurement 
elsewhere, but only if enough time has elapsed for a 
signal to get there without travelling faster than the speed of light. 

\item The result of any measurement is predetermined. 
In other words a result is fixed even if we do not carry 
out the measurement itself.

\end{itemize}

\noindent EPR then studied what consequences these two conditions would have on 
observations of quantum particles that had previously interacted 
with one another. The conclusion was that such particles 
would have very peculiar properties. In particular, the particles would 
exhibit correlations that lead to contradictions with Heisenberg's
uncertainty principle. Their conclusion was that quantum mechanics 
was an incomplete theory.

The relevance of the EPR paradox was that it motivated a debate in the 
physics community, with the celebrated Schr\"odinger's reply \cite{Schro} introducing 
{\it entanglement} as the characteristic feature of quantum mechanics. As we shall see, 
thirty years later J. Bell \cite{Bell1,Bell2} 
tried to find a way of showing that the notion of hidden variables could remove
the randomness of quantum mechanics. For more than three decades, the EPR 
paradox (or how to make sense of the (presumably) non-local effect one particle's 
measurement has on another\footnote{Interaction of two quantum particles.}) 
was nothing more than a philosophical debate for many 
physicists. Bell's theorem concluded that it is 
impossible to mimic quantum theory with the help of a set of local hidden 
variables. Consequently any classical imitation of quantum mechanics ought 
to be non-local. But this fact does not imply \cite{Terno} the existence of any non-locality 
in quantum theory itself\footnote{Quantum field theory is manifestly local. 
The fact that information is carried by material objects do not allow 
any information to be transmitted faster than the speed of light. This is possible 
because the Lorentz group is a valid symmetry of the physical system under 
consideration (see Ref. \cite{Wein}).}.

\subsection{Testing Nature: John Bell's inequalities}

According to quantum mechanics, the properties of objects are not sharp. They are well 
defined only after measurement. Given two quantum particles that have interacted 
with each other, the possibility of predicting properties without measurement 
on either side led to the EPR paradox. The postulation of unknown random variables, 
``hidden" variables, would restore localism. On the other hand, randomness 
is intrinsic to quantum mechanics.

Bell devised an experiment that
would prove it properties are well-defined or not, an experiment that would 
give one result if quantum mechanics is correct and
another result if hidden variables are needed. Although the concomitant 
theorem is named after John Bell, a number of different inequalities have been 
derived by different authors all termed ``Bell inequalities", and they all purport 
to make the same assumptions about local realism. The most important are Bell's original 
inequality \cite{Bell1,Bell2}, and the Clauser-Horne-Shimony-Holt (CHSH) inequality 
\cite{CHSH69}. Let us recall the original Bell's invitation to his enterprise
\footnote{Speakable and Unspeakable in Quantum Mechanics, pp. 29-31 (Ref. \cite{Bell2}).}.

\vskip 0.5cm
\noindent {\it ``Theoretical physicists live in a classical world, looking out into 
a quantum-mechanical world. The latter we describe only subjectively, in terms of 
procedures and results in our classical domain. (...) Now nobody knows just where 
the boundary between the classical and the quantum domain is situated. (...) More 
plausible to me is that we will find that there is no boundary. The wave functions 
would prove to be a provisional or incomplete description of the quantum-mechanical 
part. It is this possibility, of a homogeneous account of the world, which is for me 
the chief motivation of the study of the so-called ``hidden variable" possibility. 

(...) A second motivation is connected with the statistical character of 
quantum-mechanical predictions. Once the incompleteness of the wave function 
description is suspected, it can be conjectured that random statistical 
fluctuations are determined by the extra ``hidden" variables -- ``hidden" 
because at this stage we can only conjecture their existence and certainly 
cannot control them. 

(...) A third motivation is in the peculiar character of some quantum-mechanical 
predictions, which seem almost to cry out for a hidden variable interpretation. 
This is the famous argument of Einstein, Podolsky and Rosen. (...) We will find, 
in fact, that no local deterministic hidden-variable theory can reproduce all the 
experimental predictions of quantum mechanics. This opens the possibility of bringing 
the question into the experimental domain, by trying to approximate as well as possible 
the idealized situations in which local hidden variables and quantum mechanics 
cannot agree."}
\vskip 0.5cm

Let us follow the development of the Bell's original inequality. With the example advocated 
by Bohm and Aharonov \cite{Bohm}, the EPR argument is the following. Let us consider 
a pair of spin one-half particles in a singlet state, and we place Stern-Gerlach magnets 
in order to measure selected components of the spins ${\bf \sigma}_1$ and 
${\bf \sigma}_2$. If the measurement of the component ${\bf \sigma}_1\cdot {\bf a}$, 
with ${\bf a}$ being some unit vector (observable ${\bf a}$) , yields $+1$, then 
the quantum mechanics says that measurement of the component 
${\bf \sigma}_2\cdot {\bf a}$ must yield $-1$, and vice versa. This is so because 
the two particles are anticorrelated. It is plain that one can predict in advance the 
result of measuring any	chosen component of ${\bf \sigma}_2$, by previously measuring 
the same component of ${\bf \sigma}_1$. 

Now let us construct a classical description of these correlations. Let us 
suppose that there exist a continuous hidden variable $\lambda$\footnote{It makes 
no difference if we have more than one variable, or if they are discrete.}. 
The corresponding 
outcomes of measuring ${\bf \sigma}_1\cdot {\bf a}$ and ${\bf \sigma}_2\cdot {\bf b}$ 
are $A({\bf a},\lambda)=\pm 1$ and $B({\bf b},\lambda)=\pm 1$, respectively. 
The key ingredient 
is that result $B$ for particle 2 is independent of the setting ${\bf a}$, nor $A$ 
on ${\bf b}$, in other words, we address individual particles {\it locally}. Suppose 
that $\rho(\lambda)$ is the probability distribution of $\lambda$ 
(with $\int d\lambda \rho(\lambda) = 1$). If the 
quantum-mechanical expectation value of the product of the two components 
${\bf \sigma}_1\cdot {\bf a}$ and ${\bf \sigma}_2\cdot {\bf b}$ is 

\begin{equation}\label{Bell_1}
\langle \, {\bf \sigma}_1\cdot {\bf a}\, {\bf \sigma}_2\cdot {\bf b} \,	\rangle 
\,=\, -\, {\bf a} \cdot {\bf b},
\end{equation}

\noindent then the hidden variable model would lead to 

\begin{equation}\label{Bell_2}
P({\bf a},{\bf b}) \, = \, \int d\lambda \, \rho(\lambda) \, A({\bf a},\lambda) 
B({\bf b},\lambda).
\end{equation}

\noindent If the hidden variable description has to be correct, then 
result (\ref{Bell_2}) must be equal to (\ref{Bell_1}). 
Now let us impose anticorrelation in this scheme: 
$A({\bf a},\lambda)=-B({\bf a},\lambda)$ and (\ref{Bell_2}) now 
reads 

\begin{equation}
P({\bf a},{\bf b}) \, = \, -\int d\lambda \, \rho(\lambda) \, A({\bf a},\lambda) 
A({\bf b},\lambda).
\end{equation}

\noindent Adding one more unit vector ${\bf c}$, we have

\begin{eqnarray} \label{Bell_3}
P({\bf a},{\bf b}) \, -\, P({\bf a},{\bf c}) \,&=& \, -\int d\lambda \, 
\rho(\lambda) \, [A({\bf a},\lambda)A({\bf b},\lambda)-A({\bf a},\lambda)A({\bf c},\lambda)]\cr
&=& \, \int d\lambda \, \rho(\lambda) \, A({\bf a},\lambda) 
A({\bf b},\lambda) [A({\bf a},\lambda)A({\bf c},\lambda)\,-\,1].
\end{eqnarray}

\noindent Bearing in mind that $A({\bf a},\lambda)=\pm 1$ and $B({\bf b},\lambda)=\pm 1$, 
(\ref{Bell_3}) now reads

\begin{equation}
|P({\bf a},{\bf b}) \, -\, P({\bf a},{\bf c})| \, \leq \, \int d\lambda \, \rho(\lambda) \, 
[1\,-\,A({\bf b},\lambda)A({\bf c},\lambda)]\,=\,1\,+\,P({\bf b},{\bf c}).
\end{equation}

\noindent In a more compact fashion, Bell's original inequality reads

\begin{equation} \label{Bell_4}
1\,+\,P({\bf b},{\bf c}) \, \geq \, |P({\bf a},{\bf b}) \, -\, P({\bf a},{\bf c})|.
\end{equation}

\noindent If we manage to perform an experiment that violates this inequality, 
the local hidden variables theories are not valid. In the case of a singlet state 
$|\psi\rangle=1/\sqrt{2}\,(|01\rangle-|10\rangle)$, the quantum mechanical prediction 
(\ref{Bell_1}) is equal to $-\rm{cos}({\bf a},{\bf b})$, which violates Bell's 
inequality (\ref{Bell_4}) for several ranges of angles. 

In the case of the CHSH inequality \cite{CHSH69}, we can relax the conditions 
$A({\bf a},\lambda)=\pm 1$ and $B({\bf b},\lambda)=\pm 1$ to 
$|A({\bf a},\lambda)|\leq 1$ and $|B({\bf b},\lambda)|\leq 1$. Proceeding as before, 
we thus arrive to 

\begin{equation} \label{CHSH}
|P({\bf a},{\bf b}) \, -\, P({\bf a},{\bf d})| \, + \, 
|P({\bf c},{\bf d}) \, -\, P({\bf c},{\bf b})| \, \leq \, 2.
\end{equation}

\noindent The quantum limit of the CHSH inequality (\ref{CHSH}), that is, the 
right hand side of the inequality is larger by a factor of $\sqrt{2}$. 

As suggested by Bell, these inequalities can be tested experimentally \cite{Aspect}, using 
coincidence counts. Pairs of particles are emitted as a result of a quantum process, and further 
analysed and detected. In practice, to have perfect anticorrelation is difficult to 
obtain. Moreover, the system is always coupled to an environment. Although 
several experiments validate the quantum-mechanical view, the issue is not conclusively 
settled. 
Thanks to the high quality of the crystals used for parametric down conversion 
it is now possible to observe entangled	particles that are separated by a 
distance of almost 10 km. None of these experiments supports the need for 
hidden variables, although we cannot be totally sure because they do not 
detect a big enough fraction of the total flux of photons (detection loopholes). 
An experiment that has no loopholes has not yet been performed \cite{Santos}. 
The ultimate experimental test would not only involve detecting a high proportion 
of entangled particles but also performing measurements so fast (communication loopholes) 
that any mutual faster-than-light influence can be ruled out.

\subsection{Schr\"odinger's Verschr\"ankung: quantum entanglement}

Shortly after Borh's reply to EPR paper on the incompleteness of quantum theory, 
Schr\"odinger published a response to EPR in which he introduced the notion 
of ``entanglement" (or verschr\"ankung, in German) to describe such quantum
correlations. He said that entanglement	was the essence of quantum mechanics 
and that it illustrated the difference between the quantum and classical worlds 
in the most pronounced way. Schr\"odinger realized that the members of an entangled 
collection of objects do not have their own individual quantum states. Only the 
collection as a whole has a well-defined state.

In quantum mechanics we can prepare two particles in such a way that the correlations 
between them cannot be explained classically (the nature of the correlations we are 
interested in does not correspond to the statistics of the particles). Such states 
are called ``entangled" states. As we have seen, Bell recognized this fact and 
conceived a way to test quantum mechanics against local realistic theories. 
With the formulation of Bell inequalities and their experimental violation, 
it seemed that the question of non-locality in quantum mechanics had been settled 
once for all. In recent years we have seen that this conclusion was a bit premature. 
As we shall see in forthcoming Chapters, entanglement in mixed states present special 
features not shown when dealing with pure states, to the point that a mixed state 
$\rho$ does not violate Bell inequalities, but can nevertheless reveal quantum 
mechanical correlations \cite{Werner89}.

Quantum entanglement not only possesses a philosophical motivation, that is, it plays an 
essential role in several counter-intuitive consequences of quantum mechanics \cite{Wheeler}, 
but has got a fundamental physical motivation: the characterization of entanglement 
and entangled states is a challenging problem of quantum mechanics. This physical 
motivation is not only academic, because entanglement can have an applied physical 
motivation as well: entanglement plays an essential role in quantum information 
theory (superdense coding, quantum cryptography, quantum teleportation, etc..) and 
quantum computation. Entanglement, together with quantum parallelism, lies at the 
heart of quantum computing, which finds exciting and brand new applications. Recent work 
has raised the possibility that quantum information 
techniques could be used to synchronize atomic clocks with the help of entanglement \cite{Major}. 
This quantum clock synchronization \cite{Jozsa} requires distribution of entangled singlets 
to the synchronizing parties. The speed-up of quantum evolution of state assisted by 
entanglement has also been proved \cite{GLM03a}. Also, quantum entanglement has shown 
to be a key ingredient in the alignment of distant reference frames \cite{MBaig1,MBaig2}. 
In spite of 100 years of quantum theory with great achievements, we still know very little 
about Nature.

\subsection{Erwin Schr\"odinger's ghost cat}

Schr\"odinger introduced his famous cat in the very same article where entanglement 
was described \cite{Schro}. Schr\"odinger devised his cat experiment in an attempt to 
illustrate the incompleteness of the theory of quantum mechanics when going from 
subatomic to macroscopic systems. Schr\"odinger's legendary cat was doomed to be 
killed by an automatic device triggered by the decay of a radioactive atom. He had 
had trouble with his cat. He thought that it could be both dead and alive. 
A strange superposition of 

\begin{equation}\label{psiSchr}
|\Psi\rangle \,=\, \frac{1}{\sqrt{2}}\,\big(|\rm{excited\,atom},\rm{alive\,cat}\rangle
\,+\, |\rm{non-excited\,atom},\rm{dead\,cat}\rangle \big)
\end{equation}

\noindent was conceived\footnote{The fact of putting the ket symbol $|\rm{object}\rangle$ 
to an object does not automatically convert that object into a quantum one. One can 
consider a quantum property of this macroscopic thing, in our case being ``dead" or 
``alive".}. But the wavefunction (\ref{psiSchr}) showed no such commitment, superposing 
the probabilities. Either the wavefunction (\ref{psiSchr}), as given by the 
Schr\"odinger equation, was not everything, or it was not right. 
The Schr\"odinger's cat puzzle deals with one of the most revolutionary 
elements of quantum mechanics, namely, the superposition principle, mathematically 
founded in the linearity of the Hilbert state space. If $|0\rangle$ and $|1\rangle$ 
are two states, quantum mechanics tells us that $a|0\rangle+b|1\rangle$ is also 
a possible state. Whereas such superpositions of states have been extensively 
verified for microscopic systems, the application of the formalism to macroscopic 
systems appears	to lead immediately to severe clashes with our experience 
of the everyday world. Neither has a book ever observed to be in a superposition 
of macroscopically distinguishable positions, nor seems our Schr\"odinger cat 
that is a superposition	of being alive and dead to bear much resemblance 
to reality as we perceive it. The problem is then how to reconcile the vastness 
of the Hilbert space of possible states with the observation of a comparably few 
``classical" macroscopic states. 

The long standing puzzle of the Schr\"odinger's cat problem has been largely 
resolved in terms of quantum decoherence. Quantum decoherence 
arises from the interaction of a complex object with its internal and external 
environments, and usually results in a fast vanishing of the off-diagonal 
components of its concomitant reduced density matrix. This is of course a 
very rough description of decoherence (see \cite{Schlo} and references therein). 
Decoherence provides a realistic physical modelling and a generalization of 
the quantum measurement process, thus enhancing the 
``black box" view of measurements in the standard Copenhagen interpretation. 

The well-known phenomenon of quantum entanglement had already early in the 
history of quantum mechanics demonstrated that correlations between systems 
can lead to counterintuitive properties of the composite system that cannot be 
composed from the properties of the individual systems. It is the great merit 
of decoherence to have emphasized the ubiquity and essential inescapability of 
system–environment correlations and to have established	a new view on the role 
of such correlations as being a key factor in suggesting an explanation for how 
``classicality"	can emerge from quantum systems. Quite recently \cite{Zaanen}, 
there have been claims that affirm that there is a fundamental limit to how 
long quantum coherence 
can last, showing that spontaneous fluctuations can destroy quantum coherence 
in a time period that depends on the size and temperature of the system. Luckily, 
proposals for quantum computation tend to invoke bits at smaller scales, 
so they are not undermined.

The Schr\"odinger cat is points out the 
paradoxes of playing quantum games with macroscopic objects (probably our 
intuition crashes more with the cat being in a superposition of dead and alive, 
which is obviously a property of animated beings, rather than considering it as 
a macroscopic object). For quantum systems, even at mesoscopic scales, 
decoherence presents a formidable drawback to the maintenance of quantum 
coherence, which is the main drawback in the physical implementation 
of quantum computing. Decoherence typically takes place on extremely 
short time scales. In general, the effect of decoherence 
increases with the size of the system (from microscopic 
to macroscopic scales), but it is important to note 
that there exist, admittedly somewhat exotic, examples 
where the decohering influence of the environment can 
be sufficiently shielded as to lead to mesoscopic and even
macroscopic superpositions, for example, in the case of
superconducting quantum interference devices (SQUIDs)
where superpositions of macroscopic currents become observable. It is in 
these kind of systems\footnote{From time to time there also appear proposals 
with Bose-Einstein condensates in atomic traps.} where ``Schr\"odinger cat" states have 
been reported 
experimentally \cite{SQUID}. In a ring-shaped superconducting device, near absolute 
zero temperature, 
thousands of millions of pairs of electrons can circulate in either a clockwise 
($|C\rangle$) or an anti-clockwise direction ($|A\rangle$) without decaying. 
The system can be represented as 
a potential well with two minima. By exciting the system appropriately, 
one can force the system to change its direction of motion. In a certain 
range of parameters, one encounters the system in a superposition 
$a|C\rangle+b|A\rangle$. If one considers currents of the order of microamps 
or magnetic moments of thousands of millions of Bohr magnetons, as in the 
experiment of \cite{SQUID}, one may think of having something 
``truly macroscopic".


\subsection{John von Neumann and the entropy $S$}

   John von Neumann was a mathematician making a pioneering work 
   (among others) in the fields of quantum mechanics, game theory, and
  computer science. Restricting his contributions to 
  the fields of our interest here, it suffices to say the he contributed to 
   rigorously establish the correct mathematical framework for quantum mechanics 
   in 1932 with his work {\it Mathematische Grundlagen der Quantenmechanik} \cite{vonNeumann}, 
   where the entropy we are about to discuss was introduced. Also, he provided in the 
   same work a theory of measurement, where the usual notion of wave collapse is 
   described as an irreversible process (the so called von Neumann or projective 
   measurement). In the field of computer science, after the conception of the Turing 
   machine, he described 
   the central parts of a physical computer, and most of this structure 
   remains still in the architecture of classical computers\footnote{Classical as 
   opposed to quatum computers, not to ENIAC!.}.
   
   Let us follow the development of the density matrix formalism. The density matrix was 
   introduced, with different motivations, by von Neumann and by 
   Landau \cite{B91}. The motivation that led Landau, in 1927, to introduce 
   the density matrix was the impossibility to describe a subsystem of a composite 
   quantum system by a state vector. On the other hand, von Neumann was led to introduce 
   the density matrix in order to develope both quantum statistical mechanics and a theory of 
   quantum measurements. 
   Ideas and methods drawn from information theory have proved to be useful in
   the study of the probability distributions appearing in quantum mechanics.
   Probabilities in quantum mechanics arise in two different ways. On the one hand,
   we have the probability distribution
   $\tilde p_i = |a_i|^2$, associated with the expansion of a pure quantum state 
   $| \Psi \rangle $
   in a given orthonormal base $| \psi_i \rangle $,

   \be \label{qp2}
   | \Psi \rangle \, = \, \sum_i \, a_i \, | \psi_i \rangle,
   \ee

   \noindent
   where $\sum_i |a_i|^2 = 1$. On the other hand, we have the probabilities $p_i$ appearing 
   when we express the 
   statistical operator $\hat \rho$ as a linear combination of projector operators,

   \be \label{qp1}
   \hat \rho \, = \, \sum_i \, p_i | \phi_i \rangle \langle \phi_i |,
   \ee

   \noindent
   where $\sum_i p_i=1 $, and the states $| \phi_i \rangle$ are not necessarily orthogonal.
   Here the statistical operator $\hat \rho$ describes a mixed quantum state associated
   with an incoherent mixture of states where each (pure) state $|\phi_i \rangle $ appears with
   probability $p_i$. A quantum mechanical statistical operator differs in fundamental ways 
   from a classical
   probability distribution. Nevertheless, the second kind of probability distributions
   described above have some similarities with the standard probability distributions describing
   classical statistical ensembles \cite{SAKU85}. On the contrary, the first kind of probabilities,
   associated with pure states, are essentially quantum mechanical in nature and have no
   classical counterpart. 
   
   The density matrix formalism was developed in order to extend the tools of classical 
   statistical mechanics to the quantum domain. In the classical framework, is it enough 
   to compute the partition function of the system in order to evaluate all possible 
   thermodynamic quantities. The great insight of von Neumann was to introduce an equivalent 
   quantity, the density matrix, in a context of states and operators in a Hilbert space. 
   The knowledge of the statistical density matrix operator would allow us to compute 
   all average quantities (expectation values) in a conceptually similar, but mathematically 
   different fashion. Suppose that we are given a set of wave functions $|\Psi \rangle$ which 
   depend 
   parametrically on a set of quantum numbers $n_1, n_2, ..., n_N$. The natural variable 
   which we have is the amplitude with which a particular wavefunction of the basic set 
   participates in the actual wavefunction of the system. Let us denote the square of this 
   amplitude by $p(n_1, n_2, ..., n_N)$. The goal is to make this quantity $p$ the equivalent 
   of the classical density function in phase space. To do so one verifies that $p$ goes 
   over into the density function in the classical limit and that it has ergodic properties. 
   After checking that $p(n_1, n_2, ..., n_N)$ is a constant of motion, an ergodic assumption 
   for the probabilities $p(n_1, n_2, ..., n_N)$ makes $p$ a function of the energy only. 
   
   After this procedure, one finally arrives to the density matrix formalism when seeking 
   a form where $p(n_1, n_2, ..., n_N)$ is invariant with respect to the representation 
   used. In the form it is written, it will only yield the correct expectation values 
   for quantities which are diagonal with respect to the quantum numbers $n_1, n_2, ..., n_N$. 
   Expectation values of operators which are not diagonal involve the phases of the 
   quantum amplitudes. Suppose we subsume the quantum numbers $n_1, n_2, ..., n_N$ by the 
   sigle index $i$ or $j$. Then our wave function has the form 

   \begin{equation}
   |\Psi \rangle \,=\,\sum_i a_i\, | \psi_i \rangle. 
   \end{equation}

   \noindent If we now look for the expectation value of an operator $B$ which is not 
   diagonal in these wave functions, we find

   \begin{equation} \label{obs}
   \langle B \rangle \,=\,\sum_{i,j} a_i^{*}a_j\, \langle i| B |j\rangle.
   \end{equation}
   
   \noindent The role which was originally reserved for the quantities $|a_i|^2$ is 
   thus taken over by the matrix

   \begin{equation} \label{mrho}
    \langle j|\,\rho \, |i\rangle \,=\,a_j\, a_i^{*}.
   \end{equation} 

   \noindent The matrix (\ref{mrho}) is called the {\it density matrix} of our system. 
   Therefore (\ref{obs}) takes the simple form

   \begin{equation} \label{trc} 
   \langle B \rangle \,=\,{\rm Trace} (\rho \, B).
   \end{equation}

  \noindent The invariance of (\ref{trc}) is now handled with the tools of matrix 
  theory. The final objective is thus accomplished: one finds a mathematical 
  framework where the expectation of quantum operators, as described 
  by matrices, is obtained by tracing the product of the density operator $\hat \rho$ 
  times an operator $\hat B$ (Hilbert scalar product between operators). The 
  matrix formalism is developed here in the statistical mechanics framework, 
  although it applies as well for finite quantum systems, which is usually the case 
  in many Chapters of this Thesis, where the state of 
  the system cannot be described by a pure state, but as a 
  statistical operator $\hat \rho$ of the form (\ref{qp1}). Mathematically, 
  $\hat \rho$ is described by a positive, semidefinite hermitian matrix	with unit 
  trace.
  
  Given the density matrix $\rho$, von Neumann conceived the entropy 	   

  \begin{equation} \label{Sv} 
  S(\rho) \,=\,-{\rm Tr} (\rho \, {\rm ln} \rho),
  \end{equation}

  \noindent which is a proper extension of Shannon's entropy to the quantum case. 
  Needless to say that in order to compute (\ref{Sv}) one has to find a basis in which 
  $\rho$ possesses a diagonal representation. If one deals with statistical mechanics, 
  the entropy $S(\rho)$ times the Boltzmann constant $k_B$ equals the thermodynamical or 
  physical entropy. On the other hand, in the system is finite (finite dimensional matrix 
  representation), entropy (\ref{Sv}) describes the departure of our system from a pure 
  state. In other words, it measures the degree of mixture of our state describing a 
  given finite system. Here we list some properties of the von Neumann entropy 
  \cite{NC00,Wehrl}:

\begin{itemize}
\item 1. $S(\rho)$ is only zero for pure states. On the contrary, $S(\rho)$ is maximal 
and equal to ln$N$ for a maximally mixed state, $N$ being the dimension of the Hilbert 
space.
\item 2. $S(\rho)$ is invariant under changes in the basis of $\rho$, that is, 
$S(\rho)=S(U\,\rho \, U^{\dagger})$, with $U$ a unitary transformation. 
\item 3. $S(\rho)$ is concave, that is, given a collection of positive numbers $\lambda_i$ 
and density operators $\rho_i$, we have

\begin{equation} 
  S\bigg(\sum_{i=1}^k \lambda_i \, \rho_i \bigg) \,\geq\, \sum_{i=1}^k \lambda_i \, S(\rho_i).
\end{equation}

\item 4. $S(\rho)$ is additive. Given two density matrices $\rho_A,\rho_B$ describing different 
systems $A$ and $B$, then $S(\rho_A \otimes \rho_B)=S(\rho_A)+S(\rho_B)$. Instead, 
if $\rho_A,\rho_B$ are the reduced density matrices of the general state $\rho_{AB}$, 
then

\begin{equation} 
 |S(\rho_A)\,-\,S(\rho_B)|\,\leq \, S(\rho_{AB}) \, \leq \, S(\rho_A)\,+\,S(\rho_B).
\end{equation}

\noindent This property is known as subadditivity. While in Shannon's theory 
the entropy of a composite system can never lower the entropy of any of its parts, 
quantically this is not the case. Actually, this can be an indicator of an entangled 
state $\rho_{AB}$.

\item 5. The von Neumann entropy (\ref{Sv}) is strongly subadditive:

\begin{equation} 
S(\rho_{ABC}) \, + \, S(\rho_{B}) \, \leq \, S(\rho_{AB}) \,+\, S(\rho_{BC}).
\end{equation}

\end{itemize} 

The von Neumann entropy is being extensively used in different forms (conditional 
entropies, relative entropies, etc.) in all the framework of quantum information 
theory. This impetus was given by the important work of Schumacher \cite{Schu}, who 
first pointed out the physical interpretation of von Neumann entropy as the measure 
of compression of quantum information in the context of QIT. All measures of entanglement 
are based upon some quantity directly 
related to the von Neumann entropy. Interesting work has been done regarding the application 
of the von Neumann entropy in different physical and mathematical scenarios (see 
for instance Refs. \cite{Dehesa1,Dehesa2}). However, there have appeared in the literature 
several papers dealing with the possible 
inadequacy of the Shannon information, and consequently of the von Neumann 
entropy as an appropriate quantum generalization of Shannon entropy, specially 
pointed out by Brukner and Zeilinger \cite{Brukner1,Brukner2}. The main 
argument is that in classical measurement the Shannon information is a natural 
measure of our ignorance about the properties of a system, whose existence is 
independent of measurement. Conversely, quantum measurement cannot be claimed to 
reveal the properties of a system that existed before the measurement was made. 
This controversy have encouraged some authors \cite{Filipo} to introduce the 
non-additivity property 
of Tsallis' entropy as the main reason for recovering a true quantal information 
measure in the quantum context, claiming that non-local correlations ought to be 
described because of the particularity of Tsallis' entropy. Certainly these entropic 
measures have their application in the context of the detection of entanglement 
(see Part II), but to refuse the utility of Shannon - von Neumann entropy it is a little 
bit risky. In point of fact, all quantities used in the description of quantum 
information processing involve the von Neumann entropy, as we shall see. In a 
different scenario, the von Neumann entropy has proven its validity in the 
statistical mechanics framework. In any case the debate involving the possible 
inadequacy of the Shannon information in the quantum case does not seem to have 
arrived to an end.
\newline
\newline
{\bf A note on Entropy\footnote{Discussion, until {\it the information loss puzzle}, 
taken from ``General properties of entropy", by A. Wehrl. See Ref. \cite{Wehrl}}.}
\newline

Entropy is a crucial concept in thermodynamics, statistical mechanics and (quantum) 
information theory. Its 
sovereign role regarding the behaviour of macroscopic systems was recognized a century 
ago by Clausius, Kelvin, Maxwell, Boltzmann and many others. Therefore it is surprising 
that the main features of it are unknown to many physicists, and many problems remain 
still open, be it in the field of thermostatistics or quantum information theory. 
Traditionally entropy is derived from phenomenological thermodynamical considerations 
based upon the Second Law of thermodynamics. 
The correct definition is only possible in the framework of quantum mechanics, thus 
overcoming the limitation of classical mechanics in the proper definition of entropy.

Admittedly entropy has an exceptional position among other physical quantities. It does 
not show up in the fundamental equations of motion nor in the Schr\"odinger equation. 
It is rather statistical or probabilistic in nature. For instance, entropy can be interpreted 
as a measure of the amount of chaos within a quantum-mechanical mixed state. But by no 
means is an entirely new quantity. The usual concepts of quantum mechanics such as Hilbert 
space, wave function, observables, and density matrices are absolutely sufficient in the 
description of entropy.	Moreover, entropy relates macroscopic with microscopic aspects 
of Nature, and determines the 
behaviour of macroscopic systems in equilibrium, a question that is still puzzling many 
physicists nowadays.

As already mentioned, entropy can be considered as a measure of the amount of disorder, or, 
to what extend a density matrix can be considered as ``mixed". This last meaning (mixedness) 
is actually employed extensively throughout this Thesis. Since entropy can also be regarded 
as a measure of the lack of information about a system, it has been also necessary to comment 
on the relation between physical entropy and information theory. In fact, this is the keypoint 
of quantum information theory, which relies on the grounds of quantum mechanics: the natural 
framework for dealing with information is a physical one.

\begin{itemize}
\item The black hole information loss puzzle.
\end{itemize}

Information entropy may find an ultimate connection with thermodynamical entropy 
in a place where no one should had expected before: a black hole. Thermodynamics 
of black holes \cite{bhole} respect the conservation of energy (when a black hole 
absorbs a mass $m$, the final state of it has augmented its energy by the same amount) 
and angular momentum. But apparently, as J. Wheeler argued, when a chunk of mass 
falls into a black hole, its entropy also disappears, thus violating the Second Law. 
Later on several authors proposed a generalization of the Second Law, namely, that 
the sum of the entropy of the black hole plus the ordinary entropy out of it must 
not decrease \cite{bhole}. Apparently, the entropy of a black hole is proportional to the surface 
of the event horizon, exactly 1/4 of it as measured in Planck units\footnote{The Planck 
length, around $10^{-33}$ cm, is the fundamental unit lenght related to gravity and 
quantum mechanics. Its square is the Planck area.}. So to speak, one bit of information 
is encoded in four Planck areas. But as the black hole evaporates, its mass 
decreases, as well as its surface. The generalization of the Second Law could find an 
explanation to this paradox: the entropy of the outgoing radiation compensates the 
loss of information of the black hole. However one must be careful with entropies. 
The Shannon entropy and the thermodynamical entropy are conceptually equivalent: 
the number of configurations that are counted in the Boltzmann-Gibbs entropy shows 
the quantity of Shannon information that would be needed to carry out any given 
configuration. Apart from the difference in units, which is easily solved by multiplying 
Shannon's entropy (or von Neumann's) by the Boltzmann constant $k_B$ and taking into 
account a factor $ln\,2$, they are different in magnitude. A microchip containing 
a Gigabyte of data possesses a Shannon entropy of $10^{10}$ bits, which is much lesser 
that the concomitant thermodynamical entropy at room temperature (around $10^{23}$ bits). 
This difference is due to the degrees of freedom available in each case. It is plain 
that the matter structure of a piece of silicon will have more configurations 
than an ordered microchip structure, where the ratio of number of atoms required per useful 
bit is immense. Only when the fundamental degrees of freedom are encountered, both 
entropies must be equal. And this is more or less what happens in the black hole 
context.
\newline

Along this line of thought, there recently occurred in the Physics community the solution 
to a controversy dealing with information and black holes. In 1997, Stephen Hawking and 
John Preskill (of the California Institute of Technology) 
made a friendly bet about the so-called ``information paradox" posed by Hawking's work 
on black hole theory. The bet concerned what happens to information that is hidden behind 
the event horizon of a black hole: Is the information destroyed and lost forever, or 
might it in principle be recovered from the radiation that is emitted as the 
black hole evaporates (Hawking radiation)? This question was first raised by Hawking in 
a paper published in 1976. Hawking pointed out that the process of black hole evaporation 
(which he had discovered earlier) could not be reconciled with the principles 
of quantum physics and gravitational physics that were then generally accepted.  
A black hole (formed when a massive star collapses) produces such a strong 
gravitational field that matter or light are sucked in — 
and appear never to escape. Hawking argued that black hole evaporation is fundamentally 
different that ordinary physical processes, that information that falls behind 
the event horizon of a black hole will be lost forever. This, however, violates the so-called 
reversibility requirement of quantum theory: that the end of any process must be 
traceable back to the conditions which created it. According to quantum mechanics, 
although physical processes can transform the information 
that is encoded in a physical system into a form that is inaccessible in practice, 
in principle the information can always be recovered. Possibly not recovered in a very 
accessible form, but anyway information should not be lost. Preskill gambled that, some day, 
a mechanism would be found to allow missing information to be released by a black hole 
as it evaporated. 

At a conference on general relativity and gravitation in Dublin in July 2004, Hawking 
showed that black holes may in fact not form an absolute event horizon 
(a boundary from which nothing can escape). G. 't Hooft, Susskind, and others had 
anticipated before that information is encoded 
in black hole spacetimes in a very subtle way. Rather, black holes may form an apparent 
horizon, thereby reconciling the information paradox as Preskill had predicted. 
Hawking graciously conceded defeat and presented Preskill with his prize: 
something from which "information can be recovered at will" - a new encyclopaedia 
of baseball. However you look at it, information is physical.


\chapter{The language of computer science spoken by quantum mechanics: 
quantum computation}

Interplay between mathematics - computer science and physics seems to be the rule 
in quantum information theory and quantum computation. Structures in mathematics 
are actually deeply rooted in the experiences of the physical world. Examples of 
this observation appear on the recent occasion of the World 
Year of Physics 2005 celebration of Einstein's {\it annus mirabilis}: the principles 
of Geometry can only be tested by experiment, which gave rise to the theory of 
General Relativity\footnote{As we have seen, it was also Einstein's (with Podolsky and Rosen) 
insight into the possible incompleteness of quantum mechanics who triggered Schr\"odinger's 
fundamental response about entanglement, starting the whole thing out.}. In the same 
vein, the type of computation that is based on the laws of classical physics, a 
computation based on a Boolean algebra, leads to absolutely different restrictions on information 
processing than the sort of computation which takes into account the ultimate quantum 
features of Nature (quantum-based computation). The properties of quantum computation 
are not postulated {\it in abstracto} but are deduced entirely from the laws of 
physics \cite{Deutsch}. When quantum effects become important, 
say at the level of single photons and atoms, the purely abstract existing theory 
of computation becomes fundamentally inadequate. Phenomena such as quantum interference, 
quantum parallelism and quantum entanglement can be exploited for 
computation. In few words, quantum computers are huge interferometers that accept 
and an input state in the form of coherent superpositions of many different possible 
inputs, and make them evolve into an output state represented by coherent superpositions 
of many different possible outputs. Computation as such is then understood as a 
sequence of repeated unitary 
transformations, and the time of computation of each one of those is a multiple of the 
finite time $T$ that is necessary to perform a logical action. 

For instance, let us consider the evaluation of a function $f(x)$ at $N$ values. 
To do so, we encode the numbers into states

\begin{eqnarray}
{\rm f}: \, {\bf x} \,& \rightarrow & \, {\rm f}({\bf x})\cr
& &\cr
0\,(|0\rangle) \, & \rightarrow & \, f(0)\,\,\,(|f(0)\rangle)\cr 
1\,(|1\rangle) \, & \rightarrow & \, f(1)\,\,\,(|f(1)\rangle)\cr
&...&\cr
N-1\,(|N-1\rangle) \, & \rightarrow & \, f(N-1)\,\,\,(|f(N-1)\rangle). 
\end{eqnarray}

\noindent Due to linearity of quantum mechanics, we have 

\begin{equation}
\sum |x\rangle \, \rightarrow \, |\Psi\rangle \, \equiv \, \sum |f(x)\rangle.
\end{equation}

By running {\it all} computations at once, we achieve a superposition of all possible 
outputs encoded in the state $|\Psi\rangle$. But the superposition principle 
is not enough. It is certainly of great importance because given an input state 
as $\sum |x\rangle$, the function is evaluated at all $N$ points in a single run. 
One then manages the states to interfere with the help of an algorithm, and a desired 
outcome is obtained with a certain probability. Unitarity, and therefore reversibility, 
is obtained naturally.

In this Chapter we expose the motivation for dealing with quantum mechanics in the 
framework of computing, we describe some of the most important algorithms designed 
to run on a hypothetical 
quantum computer, and finally we review the description of the proposals and existing 
experimental implementations for quantum computing. We shall describe these features 
briefly, because we do not deal with them directly in the present Thesis, but 
nevertheless it is important to grasp their main results in order the have 
a complete perspective of what is going on in current quantum information and 
computation research.

\section{The physical limits of classical computation: the quantal solution. 
Historical background}

In the previous Chapter, we showed the intricacies of classical computing. From 
irreversible Turing machines to (thermodynamical) reversible computation, we saw that 
Bennett established that whatever is computable with a Turing machine, it is also 
computable with a reversible Turing machine. From there on, it could appear that 
everything about improving the capacity of calculation of a computing machine 
was solely a matter of improving technical components, and this could only be achieved 
by miniaturizing more and more the electronic components of classical logical 
gates. Moore's law --the doubling of transistor density on a chip every 18 months-- 
has been describing this ongoing trend for several decades. However, in the 
foreseeable future, each element would shrink to a size at which quantum effects 
become important. Arrived at this point, there exist two possibilities: i) to get 
stuck to the usual Boolean algebra of 0's and 1's, though employing ultimate quantum 
devices such as single electron transistors, or ii) to take advantage of these 
quantum effects in order to perform a new conception of computation, a 
{\it quantum} computation. Clearly the first option represents a short-term answer to 
speeding up computations, but a quantum computer --if ever built-- will definitely 
constitute a long-term solution.

In 1985 David Deutsch had a visionary 
picture of the limitations of 
classical computation when he introduced the universal quantum computer \cite{Deutsch}. 
He suggested that 
a computer made of elements obeying quantum mechanical laws could efficiently 
perform certain computational tasks for which no efficient classical solution was known. 
In his argument, he realizes that the so called Church-Turing hypothesis

\vskip 0.5cm
\noindent {\it ``Every function which would naturally be regarded as computable can be 
computed by the universal Turing machine."}
\vskip 0.5cm

\noindent can be viewed not as a quasi-mathematical conjecture, but as a new physical 
principle. The point is that the word ``naturally" has no precise meaning in a 
mathematical context, for it would be hard to regard a function ``naturally" as 
computable if it could not be computed in Nature. Therefore he introduces a physical version 
of the previous statement in what he calls the Church-Turing principle

\vskip 0.5cm
\noindent {\it ``Every finitely realizable physical system can be perfectly 
simulated by a universal model computing machine operating by finite means."}
\vskip 0.5cm

\noindent In Deutsch's words \cite{Deutsch}, ``The fact that classical physics and the 
classical universal Turing machine do not obey the Church-Turing principle (...) is 
one motivation for seeking a truly quantum model. The more urgent motivation is, of course, 
that classical physics is false". 

As mentioned in the previous Chapter, the advent of quantum computers solves 
the problem of reversibility in a natural way, for quantum evolutions are 
described by unitary --therefore reversible-- transformations. In the way to the 
fully quantum model for computation of Deutsch, several ideas appeared bearing 
the relevance of quantum mechanics and computation. Benioff \cite{Benioff} constructed 
in 1982 a model for computation employing quantum mechanics, but it was 
effectively classical in the sense of Deutsch (the aforementioned Church-Turing principle). 
In this model, at the end of each elementary computational step it did not remained 
any quantum mechanical property such as parallelism or entanglement. In a sense, 
the underlying computations could be perfectly simulated by a Turing machine. It was 
Feynman \cite{Feynman} in the same year who went one step further with his ``universal quantum 
simulator". His model consisted of a lattice of spins with nearest-neighbours 
interactions. Certainly this model contained the important idea of a quantum 
computer being a physical system ``mimicking" another one. Feynman's programming 
consisted of, given the dynamical laws, letting the system evolve from an initial state. But 
one is not able to select arbitrary dynamical laws, certainly. However, Feynman 
hit the nail on the head in the conception of a program being a quantum system being 
evolved in time. In a different vein, Albert \cite{Albert} described the quantum 
counterpart of classical automata. Although Albert's automata were not all general-purpose 
quantum computing machines, they resembled what Deutsch would explain in his 
universal quantum Turing machine description \cite{Deutsch}.\newline

For real purposes, a quantum computer would need several hundreds (and thousands) qubits. 
It is difficult with the present technology to combine the necessary level of control 
over two-level quantum systems with the possibility of mass fabrication. 
At the present time it is not clear whether it will be possible to build practical 
physical devices that can perform coherent quantum computations. However, from 
the point of our concern here, at least it is expected to shed new light 
on the foundations of quantum mechanics.\newline

Since any computational task that is repeatable may be regarded as the simulation 
of one physical process by another, all computer programs are somehow symbolic 
representations of some laws of physics applied to specific processes. Therefore the 
limits of computability coincide with the limits of physics itself.

\section{Qubits, quantum gates and circuits}

The elementary quantity of information is the bit, which usually takes the values 
0 and 1. Any physical realization of a bit needs a system possessing two well-defined 
states (e.g. a charged (1) or discharged (0) capacitor, a pulse in glass fibre, 
the magnetization on a tape, a pit in a compact disc, and so forth..). Also, 
two state systems are used to encode information in quantum systems. The previous 
(classical) states now read $|0\rangle$ and $|1\rangle$. But quantum theory (due 
to linearity) allows a system to be in a coherent superposition of both 
states at the same time. This new feature, which must not be confused with 
{\it probabilistic} bits\footnote{Classically, we have a probability $p$ of being in 
state 0 and $1-p$ of being in 1.}, has no classical counterpart. The state

\begin{equation} \label{qubitQ}
|\Psi\rangle \,=\, \alpha \, |0\rangle \, + \, \beta \, |1\rangle 
\end{equation}

\noindent was coined a ``quantum bit" or a qubit for short by B. Schumacher \cite{Schu} 
in 1995. In the measuring process, state (\ref{qubitQ}) collapses to 
$|0\rangle$ or $|1\rangle$ with probability $\alpha^2$ or $\beta^2$, respectively. 
Such probabilistic behaviour hardly seems to be a good basis for 
processing information. However, as long as we avoid making measurements 
the system will evolve deterministically (the Schr\"odinger equation is deterministic). 
To create a qubit all we need is to isolate a two-level quantum system, which can 
be in the form of the polarization of a photon, the spin of an electron or a nucleous, 
or the left-right flux of Cooper pairs. 

Some tasks in quantum information require the implementation of quantum gates with a very 
high fidelity, that is, performing a logical operation faithfully, being robust 
against the environment \cite{Steane}. This requires all parameters describing the 
physical system on which the quantum computer has to be based to be controlled with 
high precision, which is hard to achieve in practice. Quantum 
gates \cite{Galindo,NC00} constitute 
the quantum generalization of the so called standard logical gates --widely employed 
in usual electronics-- which play a fundamental role in quantum computation 
and other quantum information processes, 
being described by unitary transformations $\hat U$ acting on 
the relevant Hilbert space (usually, that for a multi-qubit system).
\newline
\newline
{\bf One-qubit gates}
\newline

These are the simplest possible gates, transforming one input qubit and into one output 
qubit. Some generalizations of classical logic gates are straightforward. In the case 
of the NOT-gate, its quantum counterpart, the quantum NOT gate is given by the 
unitary evolution (in the basis $|0\rangle$, $|1\rangle$)

\be \label{NOT}
U_{{\rm NOT}} \, = \, \left(
\begin{array}{cc}
 0 & 1 \\
 1 & 0 
\end{array}
\right).
\ee

\noindent While the NOT gate acts of classical states, the quantum NOT gate (\ref{NOT}) 
acts on qubits. Noteworthy, the gate (\ref{NOT}) can be decomposed into twice 
the application of a simpler gate, the square-root-of-not gate $\sqrt{{\rm NOT}}$

\be \label{SQRNOT}
U_{\sqrt{{\rm NOT}}} \, = \, \frac{1}{2}\left(
\begin{array}{cc}
 1+i & 1-i \\
 1-i & 1+i 
\end{array}
\right).
\ee

\noindent such that $U_{\sqrt{{\rm NOT}}}^2=U_{{\rm NOT}}$, which clearly has no classical 
counterpart. Another interesting one-qubit gate is given by the Hadamard gate 
transform $U_H$ ($U_H^2=1$), given by $U_H= \frac{1}{\sqrt{2}}
[\sigma_1 + \sigma_3],$ that acts on the single qubit basis $\{ |0>,\,\, |1> \}$
in the following fashion, $U_H |0> = \frac{1}{\sqrt{2}}[|1> - |0>]$,  $U_H |1>
=\frac{1}{\sqrt{2}}[|0> + |1>]$. The Hadamard gate has no classical counterpart 
either. 
\newline
\newline
{\bf Two-qubit gates}
\newline

These gates act on systems of two qubits. Examples of two-qubit gates are given by 
the controlled-NOT (or exclusive-OR) gate, and the controlled-phase gate 
(in the computational basis $|00\rangle,|01\rangle,|10\rangle,|11\rangle$)

\begin{equation} 
U_{{\rm CNOT}} \, = \, \left(
\begin{array}{cccc}
 1 & 0 & 0 & 0 \\
 0 & 1 & 0 & 0 \\
 0 & 0 & 0 & 1 \\
 0 & 0 & 1 & 0
\end{array}
\right),\,
U_{{\rm CPh(\phi)}} \, = \, \left(
\begin{array}{cccc}
 1 & 0 & 0 & 0 \\
 0 & 1 & 0 & 0 \\
 0 & 0 & 1 & 0 \\
 0 & 0 & 0 & \exp(i\phi)
\end{array}
\right)
\end{equation}

\noindent Another gate is the SWAP gate, that interchanges the states of the two qubits 
\cite{Galindo}. The CNOT gate flips the second qubit if the first one is in state 
$|1\rangle$. Experimental realizations of these gates have been obtain using different 
techniques \cite{NC00}. Extension to multiqubit gates are done in the same vein 
as the CNOT gate, for instance, the Toffoli gate or controlled-CNOT acting on three qubits. 
Also, quantum gates can be viewed as entanglers, because the can entangle or decrease the 
amount of entanglement of the input states. This is the subject of study of forthcoming 
Chapters.\newline

The previous gates can be assembled into a networklike arrangement that enables 
us to perform more complicated quantum operations. In other words, a quantum 
circuit is a computational network composed of interconnected
elementary quantum gates. These circuits perform a ``black box" operation in the form 
of a unitary matrix on the input states. This unitary matrix describing the quantum 
circuit composed of smaller unitarities, that is, the basic one-qubit and two-qubit 
gates. A quantum algorithm, for instance, is designed to run in a quantum circuit 
that perfoms the desired operations on several registers of qubits.
\newline
\newline
{\bf Universal gates for quantum computation}
\newline

Given a desired task to be performed, can it be decomposed into the simplest 
logical operations of all one-qubit and two-qubit gates, or only some 
of them support universality? This question was addressed in the seminal 
1995 paper of D. Deutch, A. Barenco, and A. Ekert \cite{universal}

\vskip 0.5cm
\noindent {\it ``Both the classical and the quantum theory of computation 
admit universal computers. But the ability of the respective universal computers 
to perform any computation that any other machine could perform under the respective 
laws of physics, could only be conjectured (the Church-Turing conjecture). In the 
quantum theory it can be proved \cite{Deutsch}, at least for quantum systems 
of finite volume. This is 
one of the many ways in which the quantum theory of computation has turned 
out to be inherently simpler that its classical predecessor. (...) we concentrate 
on universality for components, and in particular for quantum logical gates."}
\vskip 0.5cm

Barenco showed in \cite{Bar95} that any two-qubit gate ${\bf A}(\phi,\alpha,\theta)$ 
that effects a unitary transformation of the form (in the previous 
computational basis)

\begin{equation} 
A(\phi,\alpha,\theta) \, = \, \left(
\begin{array}{cccc}
 1 & 0 & 0 & 0 \\
 0 & 1 & 0 & 0 \\
 0 & 0 & \exp(i\alpha)\cos\theta & -i\exp(i(\alpha-\phi))\sin\theta \\
 0 & 0 & -i\exp(i(\alpha+\phi))\sin\theta & \exp(i\alpha)\cos\theta
\end{array}
\right)
\end{equation}

\noindent is universal, with $\phi, \alpha$ and $\theta$ being irrational multiples of 
$\pi$ and of each other. He then proves that gates $U$ of the form ${\exp i\hat H}$, with 
$\hat H$ a hermitian operator, are universal. By showing that any two-qubit can can be 
arbitrary close to the form of ${\exp i\hat H}$, he concludes that almost two-qubit 
gates are universal. Therefore, in conjunction with simple single-qubit operations, 
the CNOT gate constitutes a set of gates out of which {\it any quantum gate may be built}.

\section{Quantum algorithms: Grover's and Shor's}

The goals of QIT are at the intersection of those of quantum mechanics 
and information theory, while 
its tools combine those of these two theories. This has proved to 
be very fruitful. A remarkable case is that of the discovery of 
quantum algorithms that outperform classical ones. 
By adding ``quantum" one means that the resources employed by a 
classical algorithm are quantized, such as the passage from bits to 
qubits. Several algorithms have appeared in the literature over the past 
decade, such as the Deutsch-Jozsa algorithm \cite{Deutsch,Jozsa92}, 
Simon's algorithm \cite{Simon}, Grover's algorithm \cite{Grover1,Grover2} or 
Shor's algorithm \cite{Shor}. They usually exploit the coherence of the 
quantum wave function of a quantum register implemented by an array 
of qubits. We focus here on these two last algorithms. One is 
referred to \cite{NC00} for a thorough review.

Grover's algorithm solves the problem of
searching for an element in a list of $N$ unsorted elements. 
Classically one may devise
many strategies to perform this search, but if the elements
in the list are randomly distributed, then we shall
need to make $O(N)$ trials. Grover's quantum
searching algorithm takes advantage of quantum mechanical
properties to perform the search with an
efficiency of order $O(\sqrt{N})$ \cite{Grover1,Grover2}. This algorithm 
is discussed in detail in Chapter 13 in connection with entanglement. 

A more drastic improvement over a classical algorithm (from exponential to 
shortened to polynomial time) due to quantum mechanics is given by 
Shor's algorithm for factorizing large integers. A strong incentive for 
attempts to develop practical quantum computers arises 
from their possible use in the speed-up of factoring very large numbers 
for cryptographic purposes (see Chapter 3). 
While the best classical algorithm known to date requires of the order 
of $e^{ ({\rm ln}L)^{2/3} L^{1/3} }$ steps to factorize a $L$-digit 
number \cite{EJ96}, Shor's requires only of the order of $L^3$ steps. The 
present description is taken from \cite{Brandt}. The goal is the 
following: to factor a number $N$, let us choose a number $x$ at random 
that is coprime with $N$. The we use the quantum computer to calculate 
the order $r$ of $x{\rm mod}N$. In other words, let us find the $r$ 
such that\footnote{If $r$ is even, the greatest common divisor of $x^{r/2} \pm 1$ 
and $N$ is a factor of $N$, which can be obtained with Euclid's algorithm. 
For example, if $N=1295$ and $x=6$, from $6^r=1{\rm mod}1295$ we get 
that $r=4$, $6^{4/2} \pm 1=\{35,37\}$ and $1295=5\times 7 \times 37$.}

\begin{equation}
x^r\,=\, 1 {\rm mod} N.
\end{equation}

\noindent Shor's algorithm calculates the order $r$ quantum-mechanically. 
First a number $q$ having small prime factors is chosen, such that $N^2<q<2N^2$. 
Employing several quantum gates, the qubits of a quantum 
register are manipulated to produce the state

\begin{equation} \label{psi1}
|\psi_1\rangle\,=\, \frac{1}{\sqrt{q}} \sum_{a=0}^{q-1} |a,0\rangle.
\end{equation}

\noindent Next, an additional set of
quantum gates must be used to implement a unitary transformation of the state 
(\ref{psi1}) to produce the state

\begin{equation} \label{psi2}
|\psi_2\rangle\,=\, \frac{1}{\sqrt{q}} \sum_{a=0}^{q-1} |a,x^a{\rm mod}N\rangle.
\end{equation}

\noindent After this, state (\ref{psi2}) is Fourier transformed producing 
the state

\begin{equation} \label{psi3}
|\psi_3\rangle\,=\, \frac{1}{\sqrt{q}} \sum_{m=0}^{q-1} \sum_{a=0}^{q-1}\,e^{i\frac{2\pi am}{q}} 
|m,x^a{\rm mod}N\rangle.
\end{equation}

\noindent Now a measurement is performed on the arguments, obtaining $m=c$, $x^a=x^k$ for 
$0<k<r$. The probability of this particular outcome is given by 

\begin{equation}\label{pShor}
P(c,x^k)\,=\,\bigg|  \frac{1}{\sqrt{q}} \sum_{a=0,x^a=x^k{\rm mod}N}^{q-1}\, 
e^{i\frac{2\pi ac}{q}} \bigg|.
\end{equation}

\noindent This probability is periodic in $c$ with period $q/r$, being sharply peaked 
at $c=pq/r$ for some integer $p$. After few trials, one obtains the period $r$ 
probabilistically. The classical algorithm for checking whether a given number is 
a factor of $N$	is a faster one, so it is not a big deal to multiply large integers. 
The computation must be repeated enough times to determine the peaks in $c$ of the 
probability distribution (\ref{pShor}). In this algorithm, it can be shown that 
entanglement is present in 
(\ref{psi2}), and we exploit a massive quantum parallelism in passing from state 
$|\psi_1\rangle$ to $|\psi_2\rangle$.

Although a quantum factorizer capable of factoring a 250-digit number does not 
presently exist, poof-of-principle experiments with nuclear magnetic resonance 
techniques (NMR) have factored small numbers as the number $15$ \cite{NMR}.

\section{Fault-tolerant quantum computation. Quantum error correction}

Schemes for quantum computation rely heavily on the maintenance of quantum 
superpositions, particularly including those of entangled states. Preventing errors 
in quantum information is an important part of quantum information theory itself 
and a central goal in quantum computing. Since efficient algorithms make use of 
many particle quantum states which are very fragile, 
this will be a key component of any working quantum computing device. Interaction 
with the environment will inevitably cause decoherence. When this occurs, decoherence 
downgrades severely the performance of a quantum computer. For instance, Plenio 
and Knight \cite{PleKnig} have estimated that the amount of decoherence may be at least	
$10^{-6}$ every time a controlled-NOT gate is used. One can imagine the relevance of these 
estimations in the case of factorizing, for example, a 130-digit number by Shor's 
algorithm, which requires billions of gates!

A quantum error correcting code can be regarded as a set of states which can be 
used to store information in a way that errors are able to be detected and corrected 
during a certain task in quantum information. The general scheme is to encode the 
quantum state of a single qubit into particular states of a collection of qubits. 
When decoherence or any other error corrupts the encoded quantum state, the process may 
be reversed by the application of an error-correction procedure. Similarly to the classical 
case, the code is a repetition or redundancy code where information is stored in a 
state within a subspace. One can correct these errors as long as orthogonal states are 
mapped to orthogonal states. Historically, Shor \cite{Shor95} and 
Steane \cite{Steane98} were the first to demonstrate the implementation 
of a quantum error correcting code. They proved possibility of correcting errors in 
quantum computing 
devices\footnote{However, when errors are not independent or when gating errors 
are present, the number of physical qubits required to encode one logical qubit 
increases dramatically..}. In the case of Shor's code, it uses nine physical qubits to
encode one logical qubit and thus stores one qubit of information reliably. It protects 
the logical qubit against single independent errors on the physical qubits. Steane's 
code contains seven qubits. Several authors investigated quantum error correcting codes, 
showing that there are large classes of such codes. Two especially important classes 
are the CSS codes (after Calderbank, Shor and Steane) and their generalization to 
the stabilizer codes. See Ref. \cite{RefCode} for a lucid account of these codes and references 
therein.

Although we do not give the mathematical details of these codes, let us provide a simple 
example. Let us consider a single qubit, for which there may be three possible error 
operations

\begin{eqnarray} \label{errors}
\alpha |0\rangle\,+\,\beta |1\rangle\, & \rightarrow & \, \alpha |1\rangle\,+\,\beta |0\rangle \cr
\alpha |0\rangle\,+\,\beta |1\rangle\, & \rightarrow & \, \alpha |0\rangle\,-\,\beta |1\rangle \cr
\alpha |0\rangle\,+\,\beta |1\rangle\, & \rightarrow & \, \alpha |1\rangle\,-\,\beta |0\rangle 
\end{eqnarray}

\noindent which are called $X$ (bit flip), $Z$ (phase flip) and $Y$ 
(bit flip {\it plus} phase flip). The correction of these errors is achieved 
by means of a 5-qubit code\footnote{Five qubits is the minimum number required.}. 
An initial qubit is encoded in a linear combination of a number of 5-qubit states 
with an even and odd number of 0's. In the case of an error of the type 
(\ref{errors}) occurs, a sequence of operations is performed on 
the encoded qubit as it passes through each stage of the circuit; the operations also
involve an ancillary qubit. At the end of each stage a measurement
is performed on the ancillary qubit. If all the results are +1, there has been no
error. However, if any of these measurement results are -1, then at least one error
has taken place, and the number and type of errors may be read off and corrected.\newline

Another form of correcting errors is to avoid them. Let us give an example of 
error prevention (or error 
avoiding)\footnote{Taken from a talk from the workshop ``Quantum 
Information and Quantum Computation", held at the Abdus Salam International 
Centre for Theoretical Physics, Miramare-Trieste, October 2002.}. Suppose that we have 
a system of $N$ spins 
with Hamiltonian $H_S=\epsilon \sum_{i=1}^{N} \sigma_i^{(z)}$, coupled the environment 
$H_E=\sum_k \omega_k\,b_k^{\dagger} b_k$. The interaction term is given by 
$H_{S-E}=\sum_{i=1}^{N} \sum_k \big( f_{ik}\sigma_i^{(+)}b_k+g_{ik}\sigma_i^{(-)}b_k^{\dagger}+
h_{ik}\sigma_i^{(z)}b_k +{\rm h.c.} \big)$. The first two terms try to flip the spin 
of the particles, while the last one represents a change of phase. Instead of 
addressing individual spins, and in the case that interacting coefficients do not 
depend on the paticle ($f_{ik}=f_k$, $g_{ik}=g_k$, $h_{ik}=h_k$), it is more 
convenient to deal with total spin magnitudes. If this do happens, then the 
interaction terms reads as $H_{S-E}^{\prime}=\sum_k \big( f_k S_k^{(+)}b_k + 
g_k S_k^{(-)}b_k^{\dagger}+h_k S_k^{(z)}b_k +{\rm h.c.}\big)$, 
with $S^{(\mu)}=\sum_{i} \sigma_i^{(\mu)}$, and it is more likely that a singlet state 
remains as such ($S^{(+)}|{\rm singlet}\rangle=0,\,S^{(-)}|{\rm singlet}\rangle=0$). 
In this setting, the number of possibilities of realizing a singlet state is given 
by the combinatorial number $C_{N}^{N/2}\sim N^N/(N/2)^N \sim 2^N$. The number of 
configurations for having a singlet state increases exponentially with the number of 
particles.\newline

Still another novel approach to the problem is given by decoherence-free subspaces. 
A decoherence-free subspace is a state or set of states which is not vulnerable to 
decoherence. See Ref. \cite{Lidar} for a recent review. This approach takes 
advantage of a symmetry in the interaction with the environment in order to store 
information which is invariant under the action of the interaction Hamiltonian. Then 
operations on the system serve as a universal set of gating operations.

\section{Proposals and experimental implementations for quantum computing}

The previous theoretical prescriptions of quantum algorithms have to run on a 
quantum computer, which is based on quantum circuits in a network on basic 
quantum gates. There have appeared several proposals in different 
areas of quantum physics. Yet, these devices are very modest in size 
and the real breakthrough will be to scale them up to 
sizes capable of doing tasks which are not possible with classical
computers. There is a generic foundation for building a
quantum computer. According to \cite{Galindo}, we basically need 

\begin{itemize}
\item (i) any two-level quantum system
\item (ii) interaction between qubits, and
\item (iii) external manipulation of qubits.
\end{itemize}

\noindent The two-level system is used as a qubit and the interaction
between qubits is used to implement the conditional
logic of the quantum logical gates. The system of 
qubits must be accessible from outside, to
read in the input state and read out the output, as well as
during the computation if the quantum algorithm requires
it. David P. DiVincenzo's came up in 2000 with several requirements \cite{10rules} that
any physical setting has to fulfil in quantum computation. They appear 
in the following list:

\begin{itemize}
\item 1. {\it The physical system must support scalability}. This means 
that we must add a reasonable number of qubits without an enormous cost.

\item 2. {\it Initial state preparation}. Quantum registers must be initialized 
(to $|000..00\rangle$, for instance) or reset every time a computation has to be 
performed.

\item 3. {\it Long decoherence times are required}. The ratio $\tau_{Gate}/\tau_{Dec.}$ 
of the time required for gates operations must be considerably greater that the 
typical decoherence time of the system under consideration.

\item 4. {\it A ``universal" set of quantum gates}. The system must be able to 
support one-qubit and two-qubit gates for universal quantum computing \cite{universal}.

\item 5. {\it Readout}. The system must have a qubit-specific measurement capability.

\item 6. {\it The ability to interconvert stationary and flying qubits}. 

\item 7. {\it The ability to faithfully transmit flying qubits between specified 
locations}.

\end{itemize}

How much do we gain with quantum computing over classical computing? There does not 
seem to be a near answer to this question. But at present we know 
that quantum-mechanical tools do offer a framework to speed up 
all information processing tasks: we can speed up exponentially 
the factorization of an $N$-digit integer (Shor's algorithm), we speed up 
moderately (from $O(N)$ to $O(\sqrt{N})$) the location of a random entry in a 
database of $N$ entries	(Grover's algorithm), and some tasks are not sped up 
at all, such as the $n$th iterate of a function $f(f(...f(x)...))$ \cite{10rules}.

There are several settings in which one can fulfil the
aforementioned three requirements. We shall not go into all the
technical details of the experimental proposals below
but instead present the basic physical details of their foundation. 
The scope of the approaches to the implementation of the 
``quantum hardware" are diverse, ranging from atomic physics \cite{trapped}, 
quantum optics \cite{Qoptics} and cavity-QED \cite{QED}, nuclear magnetic resonance 
spectroscopy \cite{nmrQC}, 
superconducting devices \cite{superQC}, and electrons in quantum dots \cite{QDQC}. 
Some of them are proposed, and other are underway, such as optical lattices. These 
proposals are only the tip of the iceberg. As we said, we shall only mention 
the description of the basics of these 
proposals, such as NMR, ion-trap and optical lattices quantum 
computing. One is referred to \cite{NC00} for a thorough review.

\begin {table}[tbp]
\centering
\begin {tabular}{|c|c|c|c|}
Type of hardware &  Qubits needed  & Steps before decoherence & Status \\
\hline
Quantum Cryptography  & 1  & 1  & implemented \\
Entanglement based & 2 & 1 & demonstrated \\ 
quantum cryptography &  &  &  \\ 
Quantum CNOT gate  & 2  & 1  & demonstrated \\
Composition of gates & 2 & 2 & demonstrated \\
Deutsch's algorithm & 2 & 3 & demonstrated \\
Channel capacity doubling & 2 & 2 & imminent \\
Teleportation & 3 & 2 & demonstrated \\
Entanglement swapping & 4 & 1 & demonstrated \\
Repeating station for  & a few  & a few  & theory still\\
quantum cryptography  &   &   & incomplete \\
Quantum simulations  & a few  & a few  & simple demos\\
Grover's algorithm  & 3+  & 6+  & demonstrated\\
with toy data  &   &   & with NMR\\
Ultra-precise frequency  & a few  & a few  & foreseeable\\
standards  &   &   & \\
Entanglement purification  & a few  & a few  & foreseeable\\
Shor's algorithm  & 16+  & hundreds+  & \\
with toy data ...  & ...  & ...  & \\
Quantum factoring engine  & hundreds  & hundreds  & \\
Universal quantum computer  & thousands+  & thousands+  & \\

\end{tabular}
\label {Table X}
\caption{Relevant achievements in the development of quantum computing 
(after \cite{PhysQI}).}
\end{table}

Finally, there have been interesting proposals concerning the use of the so called 
{\it holonomies} --abelian and non-abelian geometric operations depending 
on the degeneracy of the eigenspace of the governing Hamiltonian, the Berry 
phase\footnote{A quantal system in an eigenstate, slowly transported round a circuit C 
by varying parameters {\bf R} in its Hamiltonian $H({\bf R})$, acquires 
a geometrical phase factor $\exp(i\gamma(C))$ known as the {\it Berry phase}, 
in addition 
to the familiar dynamical phase factor.} \cite{Berry} being an abelian case-- 
to implement 
robust quantum gates. The holonomies \cite{ZanGeo99} are acquired when a quantum 
system is driven to undergo some appropriate cyclic evolutions by adiabatically 
changing the controllable parameters in the governing Hamiltonian. 
This is best known as ``holonomic" quantum computation \cite{Duan}, 
where quantum gates are carried out by varying certain parameters, whose 
outcome only depends on geometrical properties of the paths in parameter space 
\cite{ZanGeo99,Pachos99}. This scheme has some built-in fault-tolerant features, 
which might offer practical advantages, such as being resilient to certain types 
of computational errors. The fact that it requires an adiabatic procedure is 
a serious disadvantage, because decoherence takes place in the meantime. However, 
interesting experimental quantum computation schemes with trapped ions have been 
addressed, discussing the pros and cons of this geometric approach \cite{CiracScience}.

\subsection{NMR quantum computing}

Most progress has been made with nuclear magnetic resonance techniques (NMR), 
which has the initial advantage that many of the necessary manipulations of spin-state 
required for quantum computation are rather similar to those carried out routinely 
over recent years, and so many of the basic effects of quantum computing have already
been demonstrated. See Ref. \cite{NMR} for comprehensive survey of the 
proof-of-principle achievements and physical details of NMR quantum computing.

\begin{figure}
\begin{center}
\includegraphics[angle=0,width=0.65\textwidth]{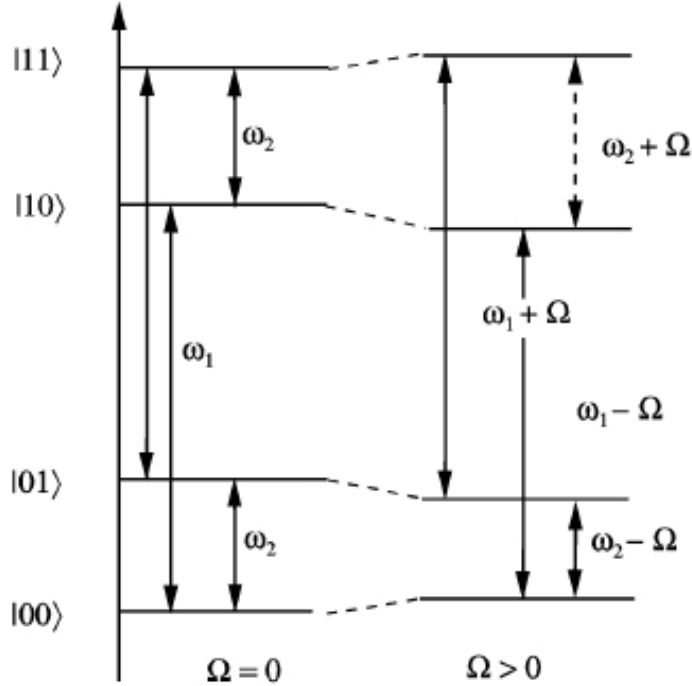} 
\caption{Energies of the basis states with coupling ($\Omega \neq 0$) and without 
coupling ($\Omega=0$). Without coupling, the signal is highly peaked around frequencies 
$\omega_1$ and $\omega_2$. When the coupling between spins of the corresponding NMR 
molecule is on, there appear four peaks due to splitting ($\omega_1\pm \Omega$ 
and $\omega_2\pm \Omega$). With $\omega_2<\omega_1$, this states represent 
the abstract computational ones $|00\rangle$, $|01\rangle$, $|10\rangle$ and $|11\rangle$.} 
\label{NMR}
\end{center}
\end{figure}

The qubits are the spins of atomic nuclei in the molecules constituting	
the liquid. These qubits are extremely well isolated from their environment. 
The nuclear spin orientations in a single molecule form a quantum data 
register. The choice of nuclear spins as qubits has several pros and cons. 
While nuclear spins in a molecule
of a liquid are very robust quantum systems, yielding to 
decoherence times of the order of seconds, long enough for quantum gate operations, 
on the other hand, in a liquid at finite temperature the nuclear spins
form a highly mixed state, which is quite different to the formulation presented so far 
of quantum computing tasks using pure states. However, this formalism is 
modified by using density matrices to describe the mixed states of spins and 
their evolution. One encodes the abstract states $|0\rangle$ and $|1\rangle$ 
in the spin states $|\uparrow\rangle$ and $|\downarrow\rangle$ of the nuclei of the molecule. 
The liquid contains about $10^{23}$ molecules at room temperature and undergoes 
strong random thermal fluctuations. Therefore on can assume that the molecules 
in the solution are in thermal equilibrium at some temperature $T$. In this case, 
the density matrix describing the quantum state of the nuclear spins in each 
single molecule is given by $\rho_{eq.}(T)=e^{-H/k_BT}/Z(T)$, where $H$ is the 
Hamiltonian of the system. The liquid is located in a large magnetic field, 
and each spin can be oriented in either directions of the magnetic field. 
Usually, $H$ is of the form 

\begin{equation}\label{Hnmr}
H\, = \, \frac{1}{2}\hbar \omega_1 \sigma_z^{(1)}\,
+\,\frac{1}{2}\hbar \omega_2 \sigma_z^{(2)}\,+\,\hbar \Omega \sigma_z^{(1)} \sigma_z^{(2)},
\end{equation}

\noindent where the first two terms are of the form of a Zeeman-splitting, and 
the last term represents the dipole-dipole interaction, which is assumed to 
be small. Without interaction, there would only be two frequencies involved, as
shown in Fig.\ref{NMR} (left), but, when it is included, there are four.

The state $\rho_{eq.}(T) \simeq 1/Z(T)\,\big( 1-\omega_1/2k_BT\,\sigma_z^{(1)}-
\omega_2/2k_BT\,\sigma_z^{(2)}+...\big)$ is expanded, and we keep only the first terms. 
This approximation is valid because the system is at room temperature. 
The input to the computer is an ensemble of nuclear spins initially in thermal 
equilibrium. Each spin can be manipulated with resonant radio frequency (rf) pulses, 
and the coupling between neighbouring nuclear spins can be exploited to produce 
quantum logical gates (one and two-qubit gates). By controlling the pulses, one 
can build for instance a CNOT gate. As we have seen, the spins have dipole-dipole interactions, 
and a driving pulse in resonance can tip a spin conditional on the state of 
another spin, thus providing a quantum bus channel. 
A sequence of rf pulses and delays produces a series of quantum logic gates 
connecting the initial state to a desired final state. The liquid consists 
effectively of a statistical ensemble of single-molecule quantum computers, which 
can be described by a density matrix. 

The NMR quantum computers have poor scaling with the number of qubits. 
The measured signal is of the order $\alpha\,N/2^N$, with $\alpha=\hbar \omega/k_BT$ 
and $N$ being the number of molecules. This feature limits NMR quantum
computing to applications requiring only 10 to 20 qubits \cite{Warren}. 
Certainly, NMR technology does not offer a solution to the requirements of 
quantum computation. However, this technology has demonstrated the basic effects 
of quantum computing and quantum information.

Finally, the fact that we deal with mixed states which are weakly entangled (expansion 
of $\rho_{eq.}(T)$ around the maximally mixed state $\frac{1}{N}\hat I$) 
raised an interesting discussion to what extent entanglement is necessary for 
quantum computing \cite{LindPop,Schack}. Even though there is still no general 
answer, it was shown \cite{LindPop} that Shor's algorithm requires entanglement. 
This is indeed the case for Grover's (see Chapter 13).

\subsection{The ion-trap quantum computer}

\begin{figure} 
\begin{center}
\includegraphics[angle=0,width=0.75\textwidth]{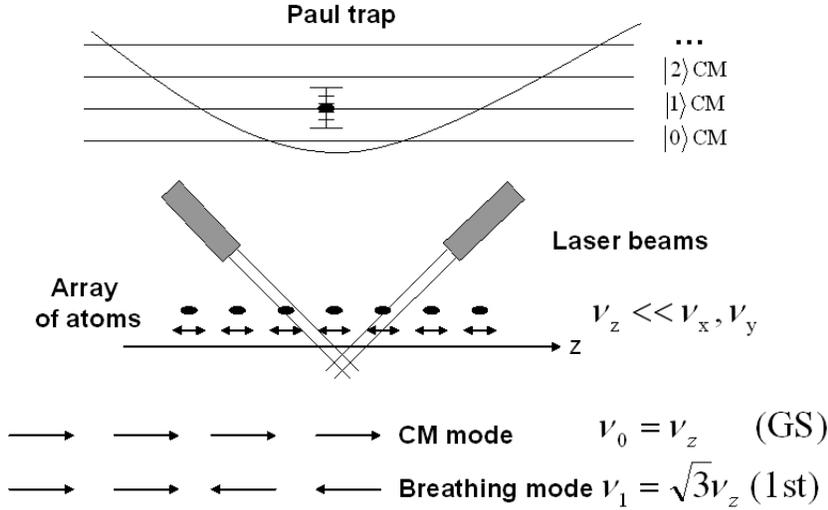} 
\caption{The ion-trap quantum computer confines an array of atoms along the axis 
of a combined electrostatic and radio frequency trap (the Paul trap). Qubits are 
stored in the spin states of the atoms, which can be addressed individually with 
high-precision lasers. The combination of internal degrees of freedom of the 
atom plus the quantized harmonic motion of the whole system of atoms is the basis 
of performing one an two-qubit gates.} 
\label{iontrap}
\end{center}
\end{figure}

In the ion-trap quantum computer \cite{Brandt,trapped} a one-dimensional lattice 
of identical ions is stored and laser cooled in a linear Paul trap, which is a 
radio frequency quadrupole trap. This linear array of ions acts as a quantum 
register. The trap potential strongly confines the ions 
radially about the trap axis, and an electrostatic potential causes the ions to
oscillate along the trap axis in an effective harmonic potential (see Fig.\ref{iontrap}). 
A set of lasers shines the atoms, which are cooled so that they are localized 
along the trap axis, with spacing determined by
Coulomb repulsion and the confining axial potential. The lowest collective 
excitation is the axial centre-of-mass mode (see Fig.\ref{iontrap}). 
Each of the trapped ions acts as a qubit, in which the two pertinent states are 
the electronic ground state and a long-lived excited state. By means of 
coherent interaction of a precisely controlled laser
pulse with any one of the ions in a standing wave configuration, one can
manipulate the ion's electronic state and the quantum state of the collective centre
of mass mode of the oscillator. The centre of mass mode can then be used as a
bus, quantum dynamically connecting the qubits, to implement the necessary
quantum logic gates. The general state of the line of ions comprising the quantum
register is an entangled linear superposition of their states. 
Experimental demonstration of the ion-trap approach began with state
preparation, quantum gates, and measurement for a single trapped ion \cite{Monroe}. 
Since then, a number of experimental and theoretical issues regarding the ion-trap
approach to quantum computation have been explored \cite{NC00}. 

The primary source of decoherence is apparently the heating due to coupling between 
the ions and noise voltages in the trap electrodes. Besides decoherence, the main 
goal of ion-trap set ups is to maintain an equilibrium with loading more and more 
ions in the trap. Also, the speed of an ion trap quantum computer would	
apparently be limited by the frequencies of vibrational modes in the trap. 

\subsection{Quantum computing using optical lattices}

There exists a wide interest in the study of cold atoms confined 
in optical lattices, which are systems with remarkable features for quantum
computing. An optical lattice \cite{Bloch} is essentially an artificial crystal of 
light, which is obtained when a periodic intensity pattern is formed by the 
interference of two or more laser beams. Typically, a periodic 2D 
intensity pattern is formed where two perpendicular standing waves 
interfere. The atoms sit in a regular array at every position of maximum 
brightness, like eggs in an egg-box. An optical lattice is able to trap 
atoms because the electric fields of the lasers induce an electric dipole 
moment in the atom \cite{Bloch}. If the laser frequency is less than 
a specific electronic transition one within the atom, they are then pulled 
towards regions of maximum intensity. Conversely, if the laser frequency is higher, 
the atoms are pushed away. These lattices are ``loaded" with cold bosonic atoms 
which are transferred from a Bose condensate from a magnetic trap to an 
optical lattice. By controlling the confining lasers it is possible to move the 
atoms over precise separations, and even to make them contact with neighbouring atoms, with 
exquisite degree of control. The interesting thing about this set-up is that it is possible 
to convert a weakly interacting Bose gas into a Mott insulator --a strongly interacting 
quantum state. This proposal has been achieved experimentally \cite{Bloch}. When 
the system is in the form of a Mott insulator, each lattice is occupied by a single 
atom or qubit, the interaction energy is zero and there is no phase coherence 
between atoms. This fact converts this set-up in a perfect quantum ``simulator" 
for several models of current study in mesoscopic systems (Hubbard model, Anderson 
model, etc.).

In these systems, a large number of atoms can be trapped in the lattice at very low 
temperatures, providing a large number of qubits. Besides, neutral atoms 
interact weakly with the environment, which boils down to slow decoherence times. 
In an experiment reported in 2004, Bloch and co-workers created 
a Bose Einstein condensate of Rubidium atoms, and about 10,000 them were 
transferred to an optical lattice. Optical lattices pose important experimental 
challenges, such as loading the lattice with one atom per site or being able to measure 
the interaction and tunnelling constants with high accuracy. See Ref. \cite{Jaksch} 
for current proposals for quantum computations with neutral atoms. Very briefly, 
one considers a set of bosonic atoms confined in a periodic lattice at sufficiently 
low temperatures such that the only first Bloch band is occupied. The qubit 
is then stored in the two relevant ground levels $|0\rangle$ and $|1\rangle$ \cite{Jaksch}. 
In \cite{Jaksch}, single qubit gates are realized using lasers and two-qubit gates 
are obtained by displacing the atoms in a particular internal state to a next neighbour 
site.


%

%% file: introduction_novel_Tesi.tex
\chapter{Novel (or improved) aspects in quantum information: 
quantum communication}						       

Entangled states have offered a new perspective on secret communication between 
parties, usually two parties, named Alice (A) and Bob (B)\footnote{In Catalan language 
it would me more appropiate to speak of Ant\`{o}nia and Bernat, for instance, 
without prejudice of any other names (\`{A}gata and Bartomeu, Al\ia cia and 
Bernad\ia, etc). The choice of names does not affect quantum effects. Historically, Alice 
and Bob appeared in late 70s in classical cryptography.}. This Chapter is far 
from being a comprehensive study of the quantum communication features. Our aim is 
to grap some of the applications that are of current research interest in quantum 
communication. For instance, quantum channels are not discussed, though their study 
constitutes a basic means for successful communication. The reader is intended 
to Refs. \cite{NC00,PhysQI} for a comprehensive review of the subject.

\section{Quantum dense coding}

Entangled states permit a completely new way of encoding information, as first suggested 
by Bennett and Wiesner in 1992 \cite{BW93}. Suppose Alice wants to send two bits of 
classical information to Bob. One possibility is to send him two particles with 
information encoded in their polarization states. From the point of view of (classical) 
information theory, two bits is the maximum Alice can send in this way. 
One method of doing this would be to send one of the four Bell states\footnote{Bell states 
are maximally entangled pure states ot two qubits.} $|\Phi^{+}\rangle$, $|\Phi^{-}\rangle$, 
$|\Psi^{+}\rangle$, $|\Psi^{-}\rangle$, with probability 1/4. Suppose on the contrary 
that Alice and Bob share the Bell state	$|\Phi^{+}\rangle$, from a EPR 
source\footnote{Name that receives any physical system that produces entangled 
pairs of particles.}. The local actions that alice can perform (on the polarization 
state of the entangled photons) are given by the transformations $U_0=\hat I \otimes \hat I$, 
$U_1=\hat \sigma_x \otimes \hat I$, $U_2=\hat \sigma_y \otimes \hat I$ and 
$U_3=\hat \sigma_z \otimes \hat I$. Applying the aforementioned transformations 
to $|\Phi^{+}\rangle$, we have $U_0|\Phi^{+}\rangle=|\Phi^{+}\rangle$, 
$U_1|\Phi^{+}\rangle=|\Psi^{+}\rangle$, $U_2|\Phi^{+}\rangle=-i|\Psi^{-}\rangle$ and 
$U_3|\Phi^{+}\rangle=|\Phi^{-}\rangle$. 

Now Alice and Bob take out their EPR pairs, and Alice performs one of these operations 
on her side. She sends her qubit $A$ of pair $|\Phi^{+}\rangle$ to Bob, who performs 
the measurements $\sigma_x^{A}\sigma_x^{B}$ and $\sigma_z^{A}\sigma_z^{B}$ (the superindex 
labels the side where the particle forming the shared entangled state of two qubits 
$|\Phi^{+}\rangle$ comes from). Because all Bell states are eigenstates of these 
operators, with different eigenvalues, these measurements completely identify 
the state. Once Bob obtains his outcome, he can infer which of the 
four local operations Alice used. He has received two bits of information 
despite the fact that Alice only sent him one qubit (a two-level system). This is 
the basis of superdense coding. Quantum dense coding has been achieved experimentally by 
Mattle {\it et al.} from the group of Innsbruck \cite{Mattle}, among other 
laboratories. The experiment relies on the process of spontaneous 
parametric down-conversion in a non-linear crystal, which produces pairs of 
photons with entangled polarizations. Quantum dense coding was the first experimental 
demonstration of the basics of quantum communication.

The maximum ``compression factor", so to speak, for superdense coding in the case 
of pure states was given by Hausladen {\it et al.} in \cite{Hausladen}, where for 
any pure $N$-state entangled state that Alice can send, she communicates log$N$ 
bits of information. As one should expect, the excess from superdense coding is 
exactly equal to the entanglement of the state.

\section{Quantum teleportation}

An even more interesting process in quantum communication is given by quantum 
teleportation. Let us suppose that Alice has an object and wants Bob the have the same 
object she has. In principle she could (classically) sent all information relative 
to that object to Bob, in order to reconstitute it. But this is forbidden in 
quantum mechanics, which prohibits a complete knowledge of the state of any object.

Luckily, there exists another strategy. All one has to do is to guarentee that what Bob 
receives has the same properties as Alice's original, {\it without} knowing 
the properties of the original object, that is, without measurement. This 
was the Bennett's way to avoid the measuring process through teleportation \cite{BBCJPW93}. 
Let us suppose as in the case of dense coding that that Alice and Bob share the Bell 
state $|\Phi_{AB}^{+}\rangle$\footnote{The subscript $AB$ points out that the state 
is shared by both parties.}. Also, on one side, Alice holds an (unknown) qubit state 
$|\Psi\rangle_C = a|0\rangle_C + b|1\rangle_C$ on system $C$. Coefficients $a$ and $b$ need 
not be known, otherwise the teleportation scheme is not valid (a measurement process 
would destroy the quantum information). The nice feature about system $A\oplus B \oplus C$ 
is that the state $|\Phi_{AB}^{+}\rangle |\Psi\rangle_C$ can be written as

\begin{eqnarray} \label{teleport}
 \frac{1}{2\sqrt{2}}\,\bigg( |\Phi^{+}\rangle_{AC}(a|0\rangle_B + b|1\rangle_B) +
 |\Phi^{-}\rangle_{AC}(a|0\rangle_B - b|1\rangle_B)+\cr
 |\Psi^{+}\rangle_{AC}(a|0\rangle_B + b|1\rangle_B)+
 |\Psi^{-}\rangle_{AC}(a|0\rangle_B - b|1\rangle_B) \bigg).
\end{eqnarray}

\noindent Alice then performs a joint Bell-state measurement on the photon se wants 
to teleport and one of the ancilliary photons. This measurement projects the 
other ancillary photon into a quantum state (one of the four in (\ref{teleport})), 
which is uniquely linked to $|\Psi\rangle$ up to some rotations. Alice then 
telephones Bob, telling him the result of her measurement. All Bob has to do is to 
perform the appropriate operation on his qubit and ends up with the state $|\Psi\rangle$.

No faster than speed of light transmission of information takes place, as one might 
wonder. Quantum teleportation does not violate causality because it requires a means 
of classical communication in order to restore the original state. Quantum 
teleportation has been experimentally realized using single photons \cite{MartTel,ZeilTel} 
and nuclear spins as 
qubits (NMR techniques) \cite{KnillTel}. Recently, the group of R. Blatt reported 
the teleportation 
of the quantum state of a trapped calcium ion to another calcium ion \cite{Blatt}, and the group of 
M. D. Barrett reported a similar experiment with beryllium ions \cite{Barrett2004}. 
Ref. \cite{criticTel} offers a critical view of the nuts and bolts of the 
experimental and theoretical status of teleportation. Still, we are far from {\it Star Trek}.

\pagebreak

\section{Quantum cryptography}

``Shaken, not stirred."

\noindent {\it Bond, James Bond}.\newline

Quantum cryptography, also known as quantum key distribution (QKD), exploits the 
principles of quantum mechanics to enable secure distribution of private 
communication. The scenario is the following: we have Alice and Bob who want 
to communicate with each other, and an eavesdropper Eve, who wants to 
``listen" to what they say. In classical cryptography, public key distribution 
is widely used in the Internet in the form of the {\it RSA cryptosystem}, developed 
by\footnote{It is believed, though, that the British intelligence agency 
had discovered it before back in the 60s or 70s.} 
{\bf R}ivest, {\bf S}hamir, and {\bf A}dleman in 1978 \cite{RSA}. Its security 
lies on the difficulty of factorizing large numbers with the available classical 
algorithms, even for supercomputers. It is based on secret key sharing and public key 
distribution. Suppose that Alice has her the message $Y$ encoded in a string of bits (an 
integer number), and so does Bob with $Z$, both of them being private and random. 
$X$ and $N$ are public and random. Alice perfoms $X^Y$mod$N$ on her side, while 
Bob does the same thing with $X^Z$mod$N$. Now they exchange information. 
Alice will have $(X^Z{\rm mod}N)^Y$mod$N$=$(X^{ZY}{\rm mod}N)$ and Bob 
$(X^Y{\rm mod}N)^Z$mod$N$=$(X^{YZ}{\rm mod}N)$, which is clearly the same message. 
These operations are easy to do. For Eve to find the secret message 
$(X^{ZY}{\rm mod}N)$, knowing $X$, $N$, $X^Y$mod$N$ and $X^Z$mod$N$ (all public stuff), 
would require a very long time with a classical algorithm. 

A quantum computer would factorize numbers quickly in the RSA scenario, thus 
revealing secret messages. This is why, among other reasons, quantum cryptography 
is the most mature area of quantum information, both at the theoretical and experimental 
level (it has been demonstrated over distances of tens of kilometres!). 
See Ref. \cite{LPS98} for a comprehensive survey of experimental results. 
However, this does not imply that a classical breaking of the RSA cryptosystem is not 
possible. A classical algorithm --in case of existence-- that 
factorizes large integers in a polynomial number of steps, has not yet been found.

In the context of QKD, using the quantum mechanical properties of information carriers, 
Alice and Bob can generate a truly random sequence of classical bits —- 0's and 1's —- 
which they both know perfectly, unknown by any third party.
If Alice and Bob share a secret key in this way, they can transmit information completely
securely over a public (insecure) channel. They do this by using the One-Time Pad, which 
is the only guaranteed unbreakable code known. If Alice wishes to send the 
secret message 101110 to Bob and they share a secret key 001011, then by bitwise-adding the
message and the key she arrives at the encrypted message 100101 which she sends to Bob. By 
bitwise-adding the secret key to the message, Bob uncovers the original message. It can be shown
that, as long as Alice and Bob use the secret key only once, an eavesdropper Eve can obtain
no information about the message \cite{NC00}. The problem is how to share a secret key with 
someone when one cannot decide on a capable courier to carry it.
There are several protocols in QKD that solve this problem using properties of 
quantum mechanics. The first one is the BB84 protocol, conceived by Bennett and Brassard 
in 1984 \cite{BB84}. Let us encode the information in the polarization state 
of a photon. First Alice chooses randomly one state out of four (four 
non-orthogonal vectors in $C^2$), and sends a qubit to Bob. Then, Bob measures the qubits 
that receives (he selects two operators, say $\hat \sigma_x, \hat \sigma_z$) and 
obtain the outcomes. After this there is a public discussion (classical communication), 
where they want to get rid of the uncorrelated results (both look at the coincidences 
and erase those which are uncorrelated). Finally there comes an authentification, where 
they check that nobody has been listening. They can do so because due to the 
quantum measurement Eve disturbs the system when she tries to obtain information. Also, 
as we shall see, she cannot make perfect copies of states by virtue of the 
non-cloning theorem. Therefore secret communication is based on basic principles 
of quantum mechanics.

\section{The non-cloning theorem and quantum repeaters}

Perfect copies of classical bits of information are 
carried out in everyday life technology. The copy-paste routine of text editors 
is possible because classical information can be copied at will. It took 
considerable long time to realize that this simple procedure is not possible 
in the quantum domain. This is the surprising ``Non-cloning Theorem" due to 
W. K. Wootters and W. H. Zurek \cite{Nclon}. Let us consider the Hilbert 
space $H_n$ of a system having $n$ basis states $\{|a_1\rangle ... |a_n\rangle\}$. 
The action of cloning a given state $|a_1\rangle$ is given by a unitary mapping 
in $H_n \otimes H_n$ that, for any state $|{\bf x}\rangle \in H_n$, results in 
$U(|{\bf x}\rangle |a_1\rangle)=|{\bf x}\rangle |{\bf x}\rangle$. The proof that 
this is not possible is so simple that it can be done in few lines. 
Assume that a quantum copymachine exists and $n>1$. 
Therefore two orthogonal states $|a_1\rangle$ and $|a_2\rangle$ exist. 
The application of the copymachine leads to 
$U(|a_1\rangle |a_1\rangle)=|a_1\rangle |a_1\rangle$ and 
$U(|a_2\rangle |a_1\rangle)=|a_2\rangle |a_2\rangle$. Combining both of them, we have 

\begin{eqnarray} \label{Nclon1}
U\big( \frac{1}{\sqrt{2}}(|a_1\rangle+|a_2\rangle)\, |a_1\rangle \big) \,&=& \,
\big( \frac{1}{\sqrt{2}}(|a_1\rangle+|a_2\rangle)\, \frac{1}{\sqrt{2}}(|a_1\rangle+|a_2\rangle) \big) \cr
&=&\, \frac{1}{2}\,(|a_1\rangle |a_1\rangle+|a_1\rangle |a_2\rangle+|a_2\rangle |a_1\rangle+
|a_2\rangle |a_2\rangle).
\end{eqnarray} 

\noindent Owing to linearity of $U$, we also have

\begin{eqnarray} \label{Nclon2}
U\big( \frac{1}{\sqrt{2}}(|a_1\rangle+|a_2\rangle)\, |a_1\rangle \big) \,&=& \,
\frac{1}{\sqrt{2}}U(|a_1\rangle|a_1\rangle)\, +\, \frac{1}{\sqrt{2}}U(|a_2\rangle|a_1\rangle) \cr
&=&\, \frac{1}{\sqrt{2}}\,(|a_1\rangle |a_1\rangle) \,+\, \frac{1}{\sqrt{2}}\,(|a_2\rangle |a_2\rangle).
\end{eqnarray}

\noindent Formulas (\ref{Nclon1}) and (\ref{Nclon2}) clearly do not coincide. We thus have arrived to a 
contradiction and a quantum state cannot be {\it perfectly} cloned\footnote{We have used the property of 
linearity of a quantum evolution. We could had used unitarity instead.}.

We just have shown that a quantum state cannot be {\it perfectly} cloned, but this fact does not 
mean that imperfect copies of states can be supplied, with a degree of accuracy 
(as measured by so called fidelity $F$) high enough so that 
certain processes of quantum communication are possible.
\newline

The non-cloning theorem has to be taken into account in the design of quantum 
channels. Because of decoherence and absorption by fibres, entangled pair of 
particles (or qubits, photons in this case) can only be maintained through finite lenghts, 
so it is required a number of repetitions, hence {\it quantum repeaters}, 
for a succesful transmission. Amplification of the qubit cannot be done without 
destroying the quantum correlations present \cite{Nclon}. A solution \cite{PhysQI} is 
obtained in a way simmilar to classical communication. The channel is divided 
in $N$ parts, where a combination of fidelity-enhancement, entanglement purification 
and entanglement swapping take place in small ``quantum processors" 
(composed by a few qubits) at every connection point. See Ref. \cite{PhysQI} 
and \cite{Dur} for more details on quantum repeaters.

%% file: quantumentanglement.tex
\part{Quantum Entanglement}

\chapter{Detection of entanglement}

Entanglement is
one of the most fundamental and non-classical features exhibited by quantum
systems \cite{LPS98}, that lies at the basis of some of the most important
processes studied by quantum information theory \cite{Galindo,NC00,LPS98,WC97,W98}
such as quantum cryptographic key distribution \cite{E91}, quantum
teleportation \cite{BBCJPW93}, superdense coding \cite{BW93}, and quantum
computation \cite{EJ96,BDMT98}. It is plain from the fact that entanglement 
is an essential feature for quantum computation or secure quantum 
communication, that one has to be able to develop some procedures (physical 
or purely mathematical in origin) so as to ascertain whether the state $\rho$ 
representing the physical system under consideration is appropriate for 
developing a given non-classical task. Besides, detecting entanglement is a 
way of characterizing the system possessing quantum correlations, a fundamental 
physical property that need not have to find any application whatsoever.

In this chapter we state the so called ``separability problem", which is of 
great importance in quantum information theory, and expose the methods or 
criteria (operational and non-operational) available in order to 
detect quantum entanglement. Unless explicitly stated we consider entanglement 
between two parties, which is the common situation, say, in quantum 
communication protocols. Although some results can be extended to multipartite 
systems easily, some other scenarios are either under current research or 
remain open questions, in the same manner that some problems in the bipartite 
case have not yet been solved.

\section{The separability problem}

 As pointed out before, it is essential to discriminate the states that contain 
classical correlations only. Historically, the violation of Bell's inequalities 
have become equivalent to non-locality or, in our this context, 
to entanglement. For every pure entangled state there is a Bell inequality 
that is violated and, in consequence, from a historic viewpoint, the first 
separability criterion is that of Bell (see \cite{T02} and references therein). 
It is not known, however, whether in the case of many entangled mixed states, 
violations exist: some states, after ``distillation" of entanglement (this is 
done by performing local operations and classical operations (LOCC), that is, 
operations performed on each side independently) eventually violate the 
inequalities, but some others don't. 

The first to point out that an entangled state did not imply violation of 
Bell-type inequalities (that is, they admit a local hidden variable model) 
was Werner \cite{Werner89}, providing himself with a family of mixed states 
(the {\it Werner} $\rho_{W}$) that do no violate the aforementioned 
inequalities. Werner also provided the current mathematical definition for 
separable states: a state of a composite quantum system 
constituted by the two subsystems $A$ and $B$ is called ``entangled" if it can
not be represented as a convex linear combination of product states. In other
words, the density matrix $\rho_{AB} \in {\cal H}_A \otimes {\cal H}_B$ 
represents an entangled state if it cannot be expressed as

\begin{equation} \label{sepa} 
\rho_{AB} \, = \, \sum_k \, p_k \, \rho_{A}^{(k)}
\otimes \rho_{B}^{(k)},
\end{equation}

\noindent with $0\le p_k \le 1$ and $\sum_k p_k =1$. On the contrary, states of
the form (\ref{sepa}) are called separable. The above definition is physically
meaningful because entangled states (unlike separable states) cannot be
prepared locally by acting on each subsystem individually \cite{P93} 
(LOCC operations). An example of a LOCC operation is provided by

\begin{equation} \label{locc} 
\rho\prime \,=\, (U_A\otimes U_B)\,\rho \,(U_A\otimes U_B)^{\dagger},
\end{equation}

\noindent where $U_A$($U_B$) represents a local action (unitary transformation) acting 
on subsystem $A$($B$). Equivalently, a separable state is a state that can be written 
as a mixture of factorizable pure states. 
Apparently this should be the end of the story. If one is able to write a given 
state as a convex combination of product states as in (\ref{sepa}), then that 
state is separable. In practice, though, this is an impossible task because 
there are infinitely 
many ways of decomposing a state $\rho$ (for instance, the pure states constituting 
the alluded mixture need not be orthogonal, what makes it even more arduous). 
Physically, it means that 
the state can be prepared in many ways. Another intriguing feature is that 
a mixture of entangled states is not necessary entangled\footnote{For instance, 
$\rho = \frac{1}{2}
(|0_{A}\rangle\otimes|0_{B}\rangle\langle 0_{A}|\otimes\langle 0_{B}|+
|1_{A}\rangle\otimes|1_{B}\rangle\langle 1_{A}|\otimes\langle 1_{B}|) 
= \frac{1}{2}(|\Phi^{+}\rangle\langle \Phi^{+}|+|\Phi^{-}\rangle\langle 
\Phi^{-}|)$, where $|\Phi^{\pm}\rangle=\frac{1}{\sqrt{2}}
(|0_{A}\rangle\otimes|0_{B}\rangle\pm|1_{A}\rangle\otimes|1_{B}\rangle)$ are 
two maximally entangled pure states (called {\it Bell states}).}. 
On the contrary, the set of separable or unentangled states is convex: 
any linear convex combination of separable states gives another separable state. 
Thus we arrive at the {\it separability problem}: given a state $\rho$ 
describing a quantum system, is it entangled or not?

This (not generally solved) problem can be related, as we shall see, to 
challenging open questions of modern linear algebra: the characterization 
of positive maps. However, we require a criterion which could decide wheter 
a state $\rho$ is entangled or not. Such procedure can be cast following 
a simple algorithm or recipe (a {\it functional} criterion) or abstractly 
(a {\it non-functional} criterion). The next sections give an account 
of such criterions.

\section{Functional criteria: PPT, reduction, majorization and $q$-entropic. 
Inclusion relations among them} 

The development of criteria for entanglement and separability is one aspect of
the current research efforts in quantum information theory that is receiving,
and certainly deserves, considerable attention \cite{T02}. Indeed, much
progress has recently been made in consolidating such a cornerstone of the
theory of quantum entanglement \cite{T02}. 
Before discussing the general mixed case, we must mention that there is a simple 
necessary and sufficient separability criterion for pure states, namely, the 
Schmidt decomposition \cite{T02}. Given a pure state $\rho_{AB} \in 
{\cal H}_A \otimes {\cal H}_B$, $N$ being the dimension of system $A$ and 
$M \ge N$ the one for system $B$, with rank $r\le N$, it can be decomposed as 
a sum of products of orthogonal states

\begin{equation} \label{Schmidt}
|\Psi\rangle \, = \, \sum_{i=1}^r w_i |x_i\rangle \otimes |y_i\rangle,
\end{equation}

\noindent with $w_i>0$ and $\sum_{i=1}^{r} w^{2}_{i}=1$, 
$\langle x_i|x_j\rangle=\langle y_i|y_j\rangle=\delta_{ij}$. The criterion is 
then extremely simple: a the state $|\Psi\rangle$ is separable or unentangled 
iff $r=1$.

Returning to the mixed case, from the formal point of view 
it was shown by the Horodecki family (see \cite{T02} and references therein) 
that a density matrix $\rho \equiv \rho_{AB}$ is
entangled if and only if there exists an entanglement witness
(a hermitian operator $\hat W=\hat W^{\dagger}$) such that

\ben \label{wit} Tr\, \hat W\, \hat \rho &\le& 0, \,\,\,\, {\rm while} \nn \\
 Tr\, \hat W\, \hat \rho_s &\ge& 0,\,\,{\rm for \,\,all\,\,separable\,\,states\,\,\rho_s}.
 \een
\noindent This rather abstract definition (it defines what is called an 
{\it entanglement witness}) exposes the need of more operational 
criteria that are easy to check in an explicit case. Let us briefly sketch the 
situation of the state-of-the-art functional criteria, and later on explain in 
more detail each one of them. A special, but quite important LOCC operational 
separability criterion,
necessary but not sufficient, is provided by the positive partial transpose
(PPT) one. Let $T$ stand for matrix transposition. The PPT requires that  \be
\label{W}  [\hat 1 \otimes \hat T](\rho) \ge 0. \ee

Another operational criterion is called the {\it reduction} criterion, that
is satisfied, for a given state $\rho \equiv \rho_{AB}$, when both \cite{T02}

\ben \label{reducti} \id \otimes \rho_B -\rho &\ge&  0\nn \\  
\rho_A \otimes \id -\rho &\ge& 0  .\een

Intuitively, the distillable entanglement is the maximum asymptotic yield of
singleton states that can be obtained,  via LOCC, from a given mixed state.
Horodecki {\it et al.} \cite{Horo97} demonstrated that any entangled mixed state
of two qubits can be distilled to obtain the singleton. This is not true in
general. There are entangled mixed states of two qutrits, for instance, that 
cannot be distilled, so that they are useless for quantum communication. These 
are the so called {\it bound entangled} states. 
An important fact is that {\it all states that violate
the reduction criterion are distillable}  \cite{Horo99}.

 Another criterion associates PPT to the rank of a matrix. Consider two
subsystems $A$, $B$ whose description is made, respectively, in the Hilbert
spaces ${\cal H}_n$ and ${\cal H}_m$. Focus attention now in the  density
matrix $\rho \equiv \rho_{AB}$ for the associated composite system. If
\begin{enumerate}
\item $\rho$ has PPT, and
\item its rank ${\cal R}$ is such that  ${\cal R} \le {\rm max}(n,m)$,
\end{enumerate}
then, as was proved in \cite{HLVC00}, $\rho$ is separable. 

The entropic 
criteria are also functional separability ones. Still another one
is majorization.
\newline
\newline
{\bf PPT}
\newline

The PPT criterion was suggested by Peres in \cite{Peres}. So far, it has been 
shown to be the strongest criterion providing the closest approximation to the 
set of separable states. Formally, it can be cast in the following way. Let 
$\rho_{AB}$ be a generic state of a bipartite Hilbert space 
${\cal H}={\cal H}_A \otimes {\cal H}_B$, with $N_A$ and $N_B$ the dimensions 
of the concomitant subsystems. If we express $\rho\equiv\rho_{AB}$ in the 
corresponding ortonormal product basis

\begin{equation}
\rho\,=\,\sum_{i,j=1}^{N_A}\sum_{m,n}^{N_B} \langle i,m|\rho|,j,n\rangle \, 
|i,m\rangle\langle j,n|\,=\,\sum_{i,j=1}^{N_A}\sum_{m,n}^{N_B} 
\langle i,m|\rho|,j,n\rangle \, 
|i\rangle_{A}\langle j|\otimes |m\rangle_{B}\langle n|,
\end{equation}

\noindent the partial transpose with respect to $A$ is given by

\begin{equation} \label{PPT}
\rho^{T_A}\,=\,\sum_{i,j=1}^{N_A}\sum_{m,n}^{N_B} 
\langle i,m|\rho|,j,n\rangle \, 
|{\bf j}\rangle_{A}\langle {\bf i}|\otimes |m\rangle_{B}\langle n|.
\end{equation}

\begin{figure}
\begin{center}
\includegraphics[angle=0,width=0.65\textwidth]{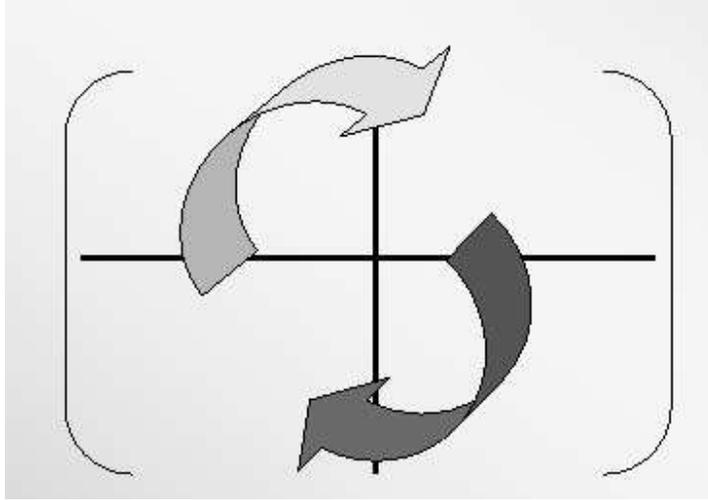}
\caption{Pictorial action of partially transposing a state $\rho$ in the computational 
basis $|00\rangle,|01\rangle,|10\rangle,|11\rangle$. PPT action on subsystem $A$ moves 
the matrix elements as shown is the figure.}
\label{PPTfig}
\end{center}
\end{figure}

\noindent If $\rho$ is separable, then $\rho^{T_A}$
\footnote{Transposition on $B$-side leads to the same conclusion 
($\rho^{T} \ge 0$, and $(\rho^{T_A})^{T_B}=\rho^{T}$).} must have all their 
eigenvalues defined positive (positive operator). The reverse it was 
conjectured to be true. In Fig.\ref{PPTfig} we provide a pictorial image of partial 
trasposition in system of two qubits, expressed in the so called {\it computational basis} 
($|0_A\rangle \otimes |0_B\rangle,|0_A\rangle \otimes |1_B\rangle,|1_A\rangle 
\otimes |0_B\rangle,|1_A\rangle \otimes |1_B\rangle$). 
A state that fullfils $\rho^{T_A} \ge 0$ is called 
PPT, otherwise NPT. If a state $\rho$ is separable, that is, it can be written 
as (\ref{sepa}), then it must possess a $\rho^{T_A} \ge 0$. 
It was shown by the Horodecki family \cite{HoroPPT} 
with use of positive maps, that PPT in systems of dimensions 
$2\times 2$ and $2\times 3$ is not only necessary but sufficient for 
separability. 

The physical significance of the positivity of (\ref{PPT}) is the following. 
Precisely, it's all about producing {\it unphysical} results. Taking the 
partial transposition of 
a separable state $\rho$, which can be written in the form (\ref{sepa}), 
does not produce any new matrix due to the fact that (\ref{sepa}) factorizes, 
and therefore must be a valid state (must remain positive). If applied to 
a state that is entangled, then it can give an unphysical result 
(some negative eigenvalue). Partial transposition is equivalent to time reversal 
on either one of the parties! 
\newline
\newline
{\bf Reduction}
\newline

The reduction criterion certainly resembles PPT in the fact that a certain 
operation is done on one of the subsystems independently. What lies beneath 
is more about positive maps, and will be apparent from the discussion in the 
next section. Quite recently \cite{Hiro03} it was shown that if a 
bipartite quantum state satisfies the reduction criterion for distillability, 
then it satisfies the majorization criterion for separability, therefore 
stablishing a strong link between reduction and the next criterion. 
\newline
 \newline
{\bf Majorization}
\newline

 Let $\{\lambda_i\}$ be the set of eigenvalues of the matrix $\xi_1$
 and $\{\gamma_i\}$ be the set of eigenvalues of the matrix
 $\xi_2$. We assert that the ordered set of eigenvalues  
$\vec \lambda$ of $\xi_1$ {\it majorizes} the ordered set of 
eigenvalues  $\vec \gamma$ of $\xi_2$
(and writes $\vec \lambda \succ \vec \gamma$) when
$\sum_{i=1}^k \,\lambda_i \,\ge\, \sum_{i=1}^k \,\gamma_i$
for all $k$. It has been shown \cite{NK01} that, for all separable
 states $\rho_{AB} \equiv \rho$,

 \ben \vec \lambda_{\rho_A} &\succ&   \vec \lambda_{\rho}, \,\,
{\rm and}
 \nn
 \\
  \vec \lambda_{\rho_B} &\succ&   \vec \lambda_{\rho}.    \een
This last fact can be cast in the following sentence: separable states are 
more disordered globally than locally.

 In point of fact, it is not possible to find a necessary and sufficient
   criterion for separability based
  solely upon the eigenvalue spectra of the three density matrices
  $\rho_{AB}, \rho_A=Tr_B[\rho_{AB}]$, and $\rho_B=Tr_A[\rho_{AB}]$
  associated with a composite system $A\oplus B$ \cite{NK01}. That is why 
both majorization and the $q$-entropic criterions are weaker than PPT.
Besides, there is an intimate relation between this majorization criterion 
and entropic inequalities, as discussed in \cite{T02,VW02}: if a state 
is separable in view of the reduction criterion then it must comply 
with the $q$-entropic criterion. 
\newline
\newline
{\bf $q$-Entropic}
\newline

The separability question has  quite interesting  echoes in information theory
and its associate information measures or entropies. When one deals with a 
classical composite system described by a suitable probability distribution 
defined over the concomitant phase space, the entropy of any of its subsystems 
is always equal or smaller than the entropy characterizing the whole system. 
This is also the case for separable states of a composite quantum
system \cite{NK01,VW02}. In contrast, a subsystem of a quantum system described 
by an entangled state may have an entropy greater than the entropy of
 the whole system. Indeed, the von Neumann entropy of either
 of the subsystems of a bipartite quantum system described (as a whole)
 by a pure state provides a natural measure of the amount of entanglement
 of such state. Thus, a pure state (which has vanishing entropy)
 is entangled if and only if its subsystems have an entropy
 larger than the one associated with the system as a whole.

  Regrettably enough, the situation is more complex when the composite system
  is described by a mixed state. There are
  entangled  mixed states such that the entropy of the complete
  system is smaller than the entropy of one of its subsystems.
  Alas, entangled mixed states such that the entropy of the
  system as a whole is larger than the entropy of either of
  its subsystems do exist as well. Consequently, the classical
  inequalities relating the entropy of the whole system with the
  entropy of its subsystems provide only necessary, but not
  sufficient, conditions for quantum separability.
  There are several entropic (or information) measures that can be
  used in order to implement these  criteria for separability.
  Considerable attention has been paid, in this regard, to the
  $q$-entropies \cite{T02,VW02,HHH96,HH96,CA97,V99,TLB01,TLP01,A02},  
  which incorporate both R\'enyi's \cite{BS93} and Tsallis' 
\cite{T88,LV98,LSP01} families of
  information measures as special instances (both admitting, in turn,
  Shannon's measure as the particular case associated with the limit
  $q\rightarrow 1$). Here we recall the definitions that appear in the 
Introduction.

 The ``$q$-entropies" depend upon the eigenvalues $p_i$ of the density
 matrix $\rho$ of a quantum system through the quantity $\omega_q =
\sum_i p_i^q$. More explicitly, we shall consider either the R\'enyi entropies
\cite{BS93},

  \be 
   S^{(R)}_q \, = \, \frac{1}{1-q} \, \ln \left( \omega_q \right),
  \ee

\noindent
  or the Tsallis' entropies \cite{T88,LV98,LSP01}

  \be 
  S^{(T)}_q \, = \, \frac{1}{q-1}\bigl(1-\omega_q \bigr),
  \ee
\noindent
 which have found many applications in many different fields of Physics.
These entropic measures incorporate the important (because of its relationship
with the standard thermodynamic entropy) instance of the von Neumann measure,
as a particular  limit  ($q\rightarrow 1$) situation

  \be 
  S_1 \, = \,- \, Tr \left( \hat \rho \ln \hat \rho \right).
  \ee

\noindent The concomitant ${\it conditional \, q-entropies}$  are defined as

 \be \label{qurela}
  S^{(T)}_q(A|B) \, = \,
  \frac{S^{(T)}_q(\rho_{AB})-S^{(T)}_q(\rho_B)}{1+(1-q)S^{(T)}_q(\rho_B)}
  \ee

\noindent for the Tsallis case, while its R\'{e}nyi counterpart is

\be 
  S^{(R)}_q(A|B) \, = \, S^{(R)}_q(\rho_{AB})-S^{(R)}_q(\rho_{B}), 
  \ee
  \noindent
  where $\rho_B = Tr_A (\rho_{AB})$ (the conditional $q$-entropy
  $S^{(T)}_q(B|A)$ is defined in a similar way as (\ref{qurela}),
  replacing $\rho_B $ by $\rho_A = Tr_B (\rho_{AB})$). 
The ``classical $q$-entropic inequalities" finally read

  \ben 
  S^{(T,R)}_q(A|B) &\ge & 0, \cr
  S^{(T,R)}_q(B|A) &\ge & 0,
  \een

  \noindent accomplished by all separable states for several $q$-values 
\cite{Horo99}. 

 The early motivation for the studies reported in
  \cite{VW02,HHH96,HH96,CA97,V99,TLB01,TLP01,A02} was
  the development of practical separability criteria for density matrices.
  The discovery by Peres of the partial transpose criteria, which for
  two-qubits and qubit-qutrit systems turned out to be both necessary
  and sufficient, rendered that original motivation somewhat outmoded. 
However, their study provide a more physical insight into the issue of quantum 
separability.
\newline

We have shown that all these criterions obey a chain of implications: if a state complies with 
PPT, it follows that it must satisfy reduction; in turn, majorization is 
satisfied, as well as the entropic inequalities \cite{inclus}. Symbolically, we have 
\newline

\begin{center}
${\it \rho\,\, separable \rightarrow PPT \rightarrow reduction 
\rightarrow majorization \rightarrow q-entropic}$. 
\newline
\end{center}

\noindent None of the implication 
relations can be reversed for a general state, except for systems in 
$2 \times N_B$ dimensions, where reduction is equivalent to PPT. Also, PPT 
provides a necessary and sufficient condition in $2\times 2$ and $2\times 3$ 
dimensions. This implications and other results are exhaustively studied in 
Chapter 7.

\section{Non-functional criteria: the theory of positive maps and entanglement 
witnesses} 

The criterions described in this section are not easy to implement in practice, 
but nevertheless constitute necessary and sufficient means of discriminated 
whether a state $\rho$ possesses entanglement or not. 

There is an interesting connection between
entanglement and the theory of positive maps \cite{AB01}, which requires some 
mathematical definitions regarding positive operators, positive and completely 
positive maps. First of all we recall that any physical action\footnote{Except time-reversal, 
which is in fact used for PPT. By physical action we mean evolution, such that 
{\it positive} probabilities of the state $\rho$ are mandatory.} is represented 
by a positive map. Let us suppose that ${\cal O}_A$ and ${\cal O}_B$ denote 
the set of operators acting on the subsystems of 
${\cal H}={\cal H}_A \otimes {\cal H}_B$, and let us denote by 
$L({\cal O}_A,{\cal O}_B)$ the space of the 
linear maps from ${\cal O}_A$ to ${\cal O}_B$. Then, a map 
$\Lambda\in L({\cal O}_A,{\cal O}_B)$ is said to be positive if it 
maps positive operators in ${\cal O}_A$ into positive operators. 

Completely positive is an extension of the previous maps. A map 
$\Lambda\in L({\cal O}_A,{\cal O}_B)$ is completely positive if the 
extended map

\begin{equation}
\Lambda_x=\Lambda\otimes \id_x:{\cal O}_A\otimes {\cal M}_x \rightarrow
{\cal O}_B\otimes {\cal M}_x
\end{equation}

\noindent is positive for all extensions\footnote{An example of
a completely positive is $\rho\rightarrow A\rho A^\dagger$ where $A$ is an
arbitrary operator} of dimension $x$, where $\id_x$ is the identity map on 
the space ${\cal M}_x$. Now here we have the desired characterisation of 
separable states {\it via} positive maps, bearing in mind that complete 
positivity is not equivalent to positivity. A state is separable iff for any 
positive map \cite{AB01} $\Lambda$

\be \label{sepMap}
(\id \otimes \Lambda) \rho \geq 0
\ee

\noindent holds.

Interpretation of PPT is thus straightforward if we regard the action 
of partial transposition in (\ref{PPT}) as an extension on the total 
transposition $\hat T$ of a state (a positive map that leaves the positive 
eigenvalues of 
$\rho$ untouched): $[\id \otimes \hat T](\rho) \ge 0$ (or its dual 
form $[\hat T \otimes \id](\rho) \ge 0$). Thus, a state $\rho$ that verifies 
PPT (thus separable) has an associated completely positive map. The fact that 
systems in low dimensions can be characterised without much 
difficulty, allow us to ascertain that PPT is a necessary and sufficient 
criterion for $2\times 2$ and $2\times 3$ systems 
($\Lambda:M_2\rightarrow M_2$, $\Lambda:M_3\rightarrow M_2$). However, 
the problem remains open (here we have the link with unsolved challenges of 
modern mathematics) because the full description of separability is 
equivalent to the characterization of the set of all positive maps, which 
is {\it per se} a formidable task.

The reduction criterion can also be reviewed in terms of positive maps. 
In this case, the map under consideration is given by
$\Lambda({\cal O})=$Tr$({\cal O}) \id - {\cal O}$. The eigenvalues of
the resulting operator $\Lambda({\cal O})$ are given 
by $\lambda_i=$Tr$({\cal O}) - o_i$ where $o_i$ are
eigenvalues of $O$. It follows immediately that if ${\cal O} \geq 0$, the map 
is positive ($\lambda_i \ge 0$). Taking the formula (\ref{sepMap}) and 
its dual form $(\Lambda \otimes \id) \rho \geq 0$ for the aforementioned map, 
the following inequalities

\be
\id\otimes \rho_B-\rho\geq0,\quad \rho_A\otimes \id -\rho\geq0\,
\ee

\noindent must be observed by separable states. That is, a separable state 
must remain a physical operator (non-negative eigenvalues) under the action 
of a complete positive map. If not, there is room for entanglement.
\newline

The next non-functional criterion is the one provided by {\it entanglement 
witnesses}, which is, roughly speaking, a kind of Bell inequality. As already 
stated, $\rho$ is entangled if and only if there exists an entanglement witness
(a hermitian operator $\hat W=\hat W^{\dagger}$) such that

\ben \label{witx} Tr\, \hat W\, \hat \rho &\le& 0, \,\,\,\, {\rm while} \nn \\
 Tr\, \hat W\, \hat \rho_s &\ge& 0,\,\,{\rm for \,\,all\,\,separable\,\,states\,\,\rho_s}.
 \een

\noindent As expected from linear algebra, which is the mathematical framework 
of these non-functional criterions, there is a correspondence between the 
two approaches. They are linked together through the 
{\it Jamiolkowski isomorphism} \cite{Jam72}. Here we do not expose the details of the 
relation between them, focusing our attention only in the properties of 
entanglement witnesses. 

\begin{figure}
\begin{center}
\includegraphics[angle=0,width=0.65\textwidth]{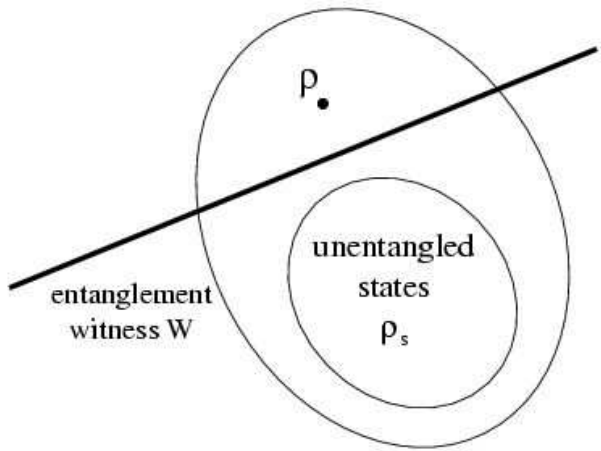}
\caption{Hyperplane $W$ that separates the entangled state $\rho$ from the set of 
separable states ${\cal S}_{sep}$. See text for details.}
\label{witness1}
\end{center}
\end{figure}

As stated, there is a fundamental difference between the set of separable 
or unentangled states ${\cal S}_{sep}$ and the remaining entangled states 
${\cal S}-{\cal S}_{sep}$ (${\cal S}$ being the set of all states): while the 
former is convex, that is, any linear convex 
combination of separable states gives another separable state, we find 
counterexamples to the latter. As a consequence, the mathematical grounds for 
the existence of entanglement witnesses lie 
basically in the fact that ${\cal S}_{sep}$ is convex and compact
\footnote{Compacticity comes from the fact that the set of product states 
$\rho_{A}\otimes \rho_{B}$ is indeed compact, because it is the tensor product 
of two compact sets. Because ${\cal S}_{sep}$ is the convex hull of 
$\rho_{A}\otimes \rho_{B}$ (see (\ref{sepa})), we conclude that it must 
be compact.}. This allow us to introduce the {\it Hahn-Banach} theorem 
\cite{Ban32}: For any convex, compact subset ${\cal S}_{sep}$ of a 
finite Hilbert space ${\cal H}$ and $\rho \not \in {\cal S}_{sep}$, there 
exist a hyperplane $W$ that separates $\rho$ 
from ${\cal S}_{sep}$. Fig.\ref{witness1} illustrates this fact.

\begin{figure}
\begin{center}
\includegraphics[angle=0,width=0.65\textwidth]{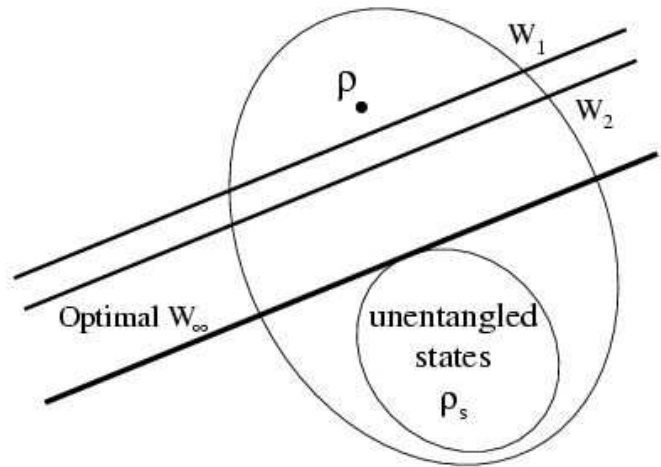}
\caption{There exist different witnesses. Once a ``direction" in the space 
of mixed states is given, the optimal witness $W_{\infty}$ is reached 
($W$ ``tangent" to ${\cal S}_{sep}$). See text for details.}
\label{witness2}
\end{center}
\end{figure}

\begin{figure}
\begin{center}
\includegraphics[angle=0,width=0.65\textwidth]{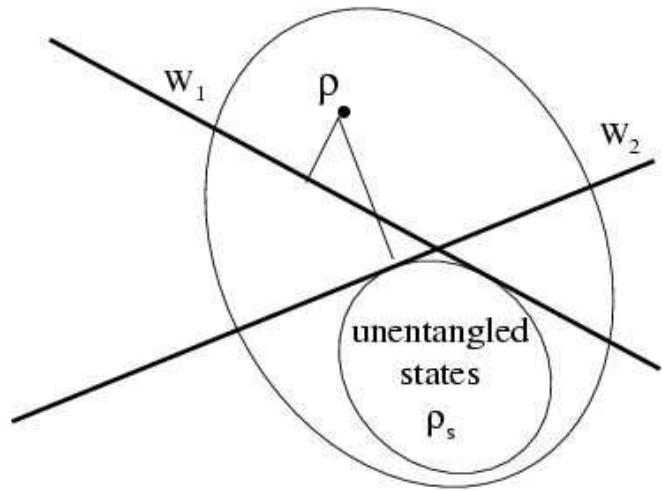}
\caption{Entanglement witnesses $W_1$ and $W_2$ are optimal. The witness 
we seek is obtained after a survey of {\bf all} optimal witnesses. 
See text for details.}
\label{witness3}
\end{center}
\end{figure}

Noting that Tr$(W\rho)$ is nothing but an inner                          
product for operators in ${\cal H}$, it can be regarded as a scalar 
product (indeed Tr$(W)=1$) of two vectors, where the orientation of the 
hyperplane is taken such that separable states always lie on 
the positive side, whereas the entangled ones remain on the 
negative side (\ref{witx}). From Fig.\ref{witness2} we see that parallel 
transports of witness $W$ can be performed until it becomes 
``tangent" to ${\cal S}_{sep}$, which defines an {\it optimal} 
witness $W_{opt}$. By no means this is the end of the story, 
because one then has to perform a minimization over all 
possible optimal $\{$$W_{opt}$$\}$ (see Fig.\ref{witness3}), which is tantamount 
to explore the whole shape of ${\cal S}_{sep}$ and, 
unless it is a polytope, it is an impossible task. Thus, 
the difficulty of the complete characterization of all positive 
maps here is translated into the complexity of ``moulding" 
the egg-type shape of ${\cal S}_{sep}$.
Nevertheless several steps have been done towards a better 
characterization on entanglement using these operator witnesses.

Lots of fruitful results have been obtained with the theory of 
entanglement witnesses, when applied not only to separability, 
but also to the distillability problem. The work \cite{WitRev} 
reviews the achievements of the active group of 
Hannover and Innsbruck. Nevertheless, there is no general 
procedure for obtaining an optimal witness for a given 
arbitrary state $\rho$ yet.

The problem of distillability, that is, given a certain state 
$\rho$, the possibility of ascertaining whether it is 
distillable or not, follows the footsteps of the separability 
problem. One way of solving to problem 
finds its way in a non-operational criteria, which turns out 
to be necessary and sufficient \cite{WitRev}: The state 
$\rho$ is distillable iff there exists 
$|\psi\rangle=a_1|e_1\rangle|f_1\rangle+a_2|e_2\rangle|f_2\rangle$ 
such that $\langle\psi|(\rho^{T_A})^{\otimes n} |\psi\rangle<0$ 
for some n. The operational criteria provide some insight into 
the problem, but still remains open.

\section{Schematics of the set of all states}

Bipartite quantum states are classified into three categories: 
separable states, bound entangled states, and free entangled states. 
It is of great relevance to completely characterize these families 
of states for the full development of quantum information theory. 
We recall that PPT is the strongest operational criterion known 
to date, which provides a necessary condition for separability 
in any bipartite system, being indeed sufficient for cases 
of small dimensionality ($2\times 2$ and $2\times 3$ systems). 
Also, bound entangled states are those which preserve positivity 
(positive eigenvalues) under the action of partial transposition 
(PPT) and cannot be distilled. In the following, we clarify the 
meaning of {\it distillation}.

Thus, in view of the existence of these particular 
states, the set of all states ${\cal S}$ must be described 
either according to the separability problem (through PPT) 
or the distillability problem (PPT + reduction/majorization).

\subsection{Decomposition according to PPT} 

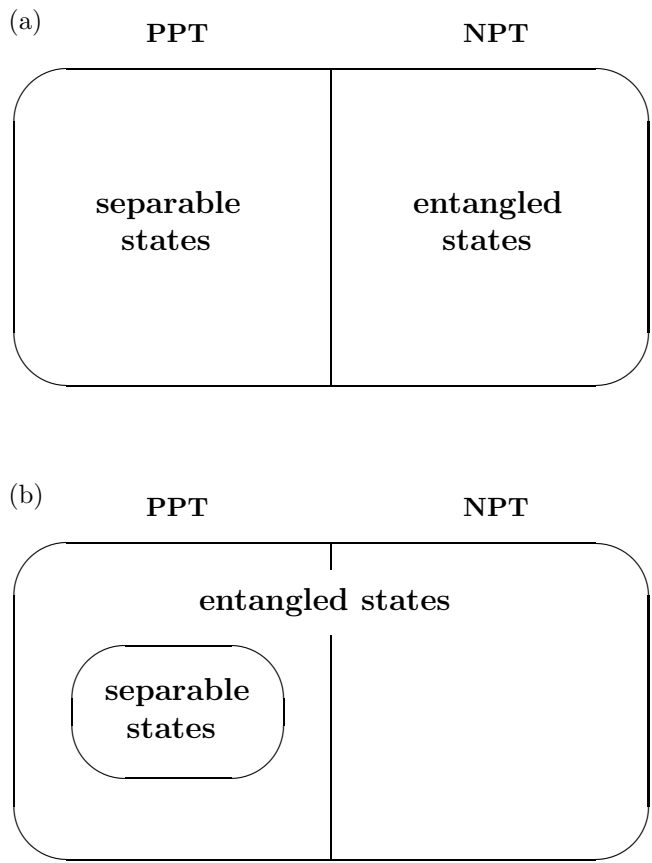
\begin{figure}
 \ \vskip 1cm
\begin{center}
\parbox{200pt}{
\begin{picture}(200,150)
\put(28,175){(a)}
\put(150,100){\oval(240,120)}    
\put(150,40){\line(0,1){120}}    
\put(80,170){\bf PPT}
\put(200,170){\bf NPT}
\put(61,100){\parbox{2.6cm}{\large \bf separable \\ \centerline{states\ 
\hskip1mm  \ \ \ }}}
\put(181,100){\parbox{2.6cm}{\large \bf entangled\\ \centerline{states \ \ \ }}}
\end{picture}}

\vskip1cm

\parbox{200pt}{
\begin{picture}(200,150)
\put(28,175){(b)}
\put(150,100){\oval(240,120)}    
\put(150,40){\line(0,1){85}}    
\put(150,150){\line(0,1){10}}    
\put(92,96){\oval(80,50)}
\put(80,170){\bf PPT}
\put(200,170){\bf NPT}
\put(61,95){\ \parbox{2.4cm}{\large \bf separable \\ \centerline{states \ \ \ }}}
\put(100,135){\parbox{5cm}{\large \bf entangled states }}
\end{picture}}
\caption{a) Decomposition of mixed states according to PPT for $2\otimes 2$ and 
$2\otimes3$ systems and b) for higher dimensions (after \cite{AB01}).}
\label{figHor1}
\end{center}
\end{figure}

If we are interested only in the separability problem, 
the only operational tool that we have at hand in order to describe 
the entanglement properties of the set ${\cal S}$ is 
the positive partial transposition (PPT). While waiting for a 
new general and more restrictive separability criterion, we 
discriminate states according to PPT in Fig.\ref{figHor1}. 
The only clear solution provided by PPT is restricted to 
low dimensions, alas, simple states easily described by positive 
maps or entanglement witnesses. Such states are the two-qubits 
systems ($2 \times 2$) and the qubit-qutrit systems ($2 \times 2$).

\subsection{Decomposition according to distillability. Bound entanglement} 

It is known that the creation of maximally ({\it ergo}, pure) entangled states 
is possible in principle, but the most common situation encountered in practice 
is that those pure states evolve to mixed states due to interactions with 
the environment. This is the norm, for instance, whenever trying to create 
entangled pairs of photons for quantum communication protocols. Thus we are 
naturally led to the idea of {\it distillation}: we must concentrate the 
entanglement present in the mixed state by LOCC operations. Therefore, the 
classification of bipartite states according to their distillability properties 
is an important problem in quantum information theory.

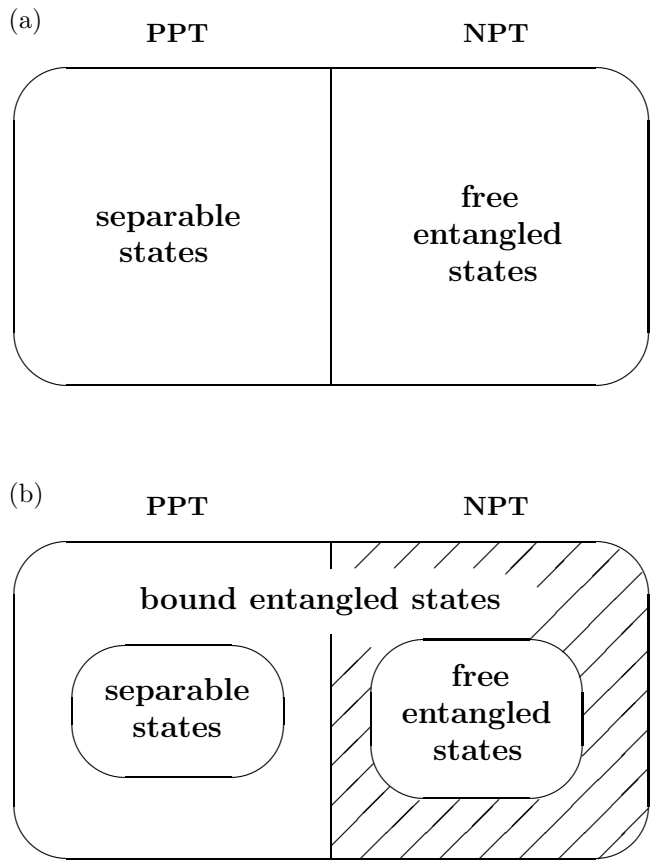
\begin{figure}
\begin{center}
\ \vskip1.5cm
\parbox{200pt}{
\begin{picture}(200,150)
\put(28,175){(a)}
\put(150,100){\oval(240,120)}    
\put(150,40){\line(0,1){120}}    
\put(80,170){\bf PPT}
\put(200,170){\bf NPT}
\put(61,95){\parbox{2.6cm}{\large \bf separable \\ \centerline{states \ \ \ \ }}}
\put(181,102){\parbox{2.6cm}{\large \bf \phantom{l} \ \ \ \ free\\ 
entangled\\ \centerline{states \ \ }}}
\end{picture}}

\vskip1cm

\parbox{200pt}{
\begin{picture}(200,150)
\put(28,175){(b)}
\put(150,100){\oval(240,120)}    
\put(150,40){\line(0,1){85}}    
\put(150,150){\line(0,1){10}}    
\put(92,96){\oval(80,50)}
\put(80,170){\bf PPT}
\put(200,170){\bf NPT}
\put(61,95){\ \parbox{2.4cm}{\large \bf separable \\ \centerline{states\ \ \ }}}
\put(55,135){\parbox{7cm}{\large \bf \ \ \ \ \ bound entangled states }}
\put(205,93){\oval(80,60)}
\put(173,93){\parbox{2.6cm}{\large \bf \ \ \ \ \ free\\  
\phantom{l}entangled\\ \centerline{states \ }}}
%
%
\put(150,105){\line(1,1){15}}
\put(150,90){\line(1,1){15}}
\put(150,75){\line(1,1){15}}
\put(150,60){\line(1,1){16}}
\put(150,45){\line(1,1){22.3}}
\put(161,40){\line(1,1){22.3}}
\put(176,40){\line(1,1){22.5}}
\put(191,40){\line(1,1){22.5}}
\put(206,40){\line(1,1){22.5}}
\put(221,40){\line(1,1){49}}
\put(236,40){\line(1,1){34}}
\put(251,40){\line(1,1){19}}
\put(245,79){\line(1,1){25}}
\put(245,94){\line(1,1){25}}
\put(244,109){\line(1,1){26}}
\put(237.5,118.5){\line(1,1){30}}
\put(225,123){\line(1,1){34}}
\put(228,143){\line(1,1){17}}
\put(216,148){\line(1,1){12}}
\put(204,150){\line(1,1){10}}
\put(189,150){\line(1,1){10}}
\put(174,150){\line(1,1){10}}
\put(159,150){\line(1,1){10}}
\put(147,153){\line(1,1){7}}
\end{picture}}
\caption{a) Decomposition of mixed states according to distillability 
for $2\otimes 2$ and $2\otimes3$ systems and b) for higher dimensions 
(after \cite{AB01}).}
\label{figHor2}
\end{center}
\end{figure}

There is something strange about those mixed states that, being 
already entangled, cannot be distilled: bound entangled states are invariant 
under the operation of partial transposition. Physically, it means that they 
remain physical under the unphysical action of partial transposition 
(time-reversal on one side only). 
Undistillable -- separable and bound entangled -- states have a common property 
when viewed through the glass of the characterized separability criteria: the 
eigenvalue vector of the global system is majorized by that of the local 
system. In other words, the majorization criterion is implied by reduction, 
and both of them constitute sufficient conditions for 
distillability of bipartite quantum states. Thus, whenever any of these 
criterions is violated, the state is distillable (the converse is not true). 
Using these tools, we characterize the set ${\cal S}$ of all states 
according to their distillability features in Fig.\ref{figHor2}. 
We must point out that that PPT and reduction coincide in 
$2 \times N_B$ systems, and particularly in $2\times 2$ and 
$2\times 3$ systems 
-- which are distillable (have got no bound entanglement) -- PPT is also 
a sufficient criterion for separability. 

\chapter{Characterization of entanglement}

Detecting entanglement may not be sufficient. Given a matrix $\rho$ 
representing the state of the system, it is possible that interaction with 
environment destroys the entanglement present in $\rho$. We therefore require 
a means of {\it quantifying} entanglement. This feature, quantification, it is 
decisive whenever we must decide whether the entanglement provided by a physical 
set up (e.g. a certain physical implementation for quantum computation) 
is sufficient or not in order to accomplish with the quantum information related 
tasks. One such dramatic example is provided by nuclear magnetic resonance (NMR) 
computing, as explained throughout this Chapter.

A fundamental question remains still. The characterization of entanglement 
obviously implies a knowledge of the physical meaning of entanglement itself. 
In point of fact entanglement should be regarded as a resource, like energy, 
which underlies most of the striking new applications of the newborn science 
of quantum information. This extremely non-classical feature that is assumed 
to be {\it the} characteristic feature of quantum mechanics -- even more 
than the superposition of possible states of a particle or the impossibility 
of attribution of a priori well defined properties to quantum states 
(problem of measurement) -- looses its merely fundamental aspect in physics 
and finds lots of practical applications impossible to achieve before. 
This is perhaps what is more striking about entanglement, and indeed it is 
hardly possible to find a similar analogue that had undergone a similar 
transition in quantum mechanics. 

However, we have not answered the question regarding the meaning of 
entanglement. As a matter of fact, {\it no one really knows what 
entanglement is}. This is a similar problem encountered whenever trying to 
define the absolute meaning of the word ``energy". In an analogous way, one 
describes the spectra of capacities for what entanglement is able to perform 
instead. This situation is somewhat similar to the historical development of 
Thermodynamics. The familiar entropy appeared in order to clarify the processes 
involving temperature, work and heat, and one had to find links to that 
newborn quantity through specific heat, Joules and calories, that is, computable 
magnitudes. 

In the context of quantum information, entanglement receives several 
definitions depending on the discipline: 

\begin{itemize}
\item for a physicist working in quantum cryptography, entanglement is 
an essential tool for absolute secure communication;
\item for a physicist working in teleportation, entanglement is 
the basic tool for making teleportation of states possible;
\item for a physicist in the field of quantum correlations/Bell inequalities, 
entanglement represents some sort of ``extra" correlation that enables to 
refute local hidden variable theories, which try to describe physical reality 
in local terms;
\item for a computer scientist, entanglement is the basic ingredient for 
building new kinds of algorithms that solve problems exponentially 
faster that classical Turing machines;
\item for a any physicist interested in solving {\it hard} problems, entanglement 
is the 8th Wonder of the World that provides him/her with a tool 
that is able to simulate a given physical system\footnote{Nothing but 
Feynman's original idea that the best way of simulating a quantum physical 
system is using another quantum physical system (in the form of a quantum 
computer).}. 
\end{itemize}

Probably the best way to tackle this precise definition could be the one 
originally provided by Schr\"odinger himself: a state becomes 
entangled when after 
some interaction of its parts, the knowledge of the whole state does not 
include the best knowledge of its parts. This sentence reminds us that 
entanglement between parties, i.e. between certain degrees of freedom of 
the parties, arises in the form of correlations between them {\it after} 
these parties have ``spoken" to each other through some interaction. This 
fact is physically meaningful because particles which do not interact or have 
interacted in the past are not expected to shown any quantal correlation. Of 
course this is not the case for non-interacting particles with some associated 
statistics (e.g. identical fermions or bosons), which clearly possess intrinsic 
correlations, though they are useless for quantum information purposes. 
Thus, the characterization of entanglement takes into account the 
distinguishability of the subsystems, as exposed in detail 
throughout this Chapter.

To characterize entanglement is tantamount to quantify this resource as well. 
Let us take the International System of Units and the meter, wherefrom any 
lenght is described in units of that bar. Any distance is then described in 
terms of meters, and eventually ``distance" and ``meters" become linked. This 
of course does not define length, but {\it describes} it in terms of meters. 
Similarly, when we consider entanglement, one might choose a physical system 
being representative of maximal quantal correlation, and define it as an 
``entanglement ruler". Of course the situation is more involved, but 
intuitively it remains the same. One feature that has to be taken into account 
is the fact that entanglement cannot be enhanced by local operations acting 
on the subsystems individually or by classical communication between 
them (LOCC operations), 
albeit it can be decreased. We revisit this situation when we expose 
several physically motivated measures.

New definitions of entanglement may come through 
the field of relativity \cite{Terno} (bear in mind that 
the theory of entanglement is developed in the framework of non-relativistic 
quantum mechanics) or even from information theoretical aspects \cite{Chef00} 
(entanglement arises whenever there is an incomplete information transfer 
between quantum systems), or some new approximations to the problem. 
For instance, one of the first ones points out the induced tensor product 
partition of the Hilbert space representing a physical system \cite{ZLLl04} and 
the corresponding way in which entanglement is described, 
or relates the available information about the system 
(in the form of Lie algebras) 
with the quantum correlations present \cite{barnum1,barnum3}. The usual 
definitions and the new views of entanglement are exposed in this 
Chapter.

\section{Entanglement for distinguishable particles}

To start with, let us recall the usual definition of entanglement (\ref{sepa}) 
generalized here to an arbitrary number of parties $N$: {\it a state $\rho$ is 
entangled iff it cannot be written in the form}

\begin{equation} \label{sepaN}
\rho \,=\, \sum_{k} p_k \, \rho^{(k)}_1 \otimes ... \otimes \rho^{(k)}_N
\end{equation}

\noindent with $0\le p_k \le 1$ and $\sum_k p_k =1$. It is implicit 
from (\ref{sepaN}) that the $N$ parties of this composite system, whose state 
$\rho$ belongs to the Hilbert space 
${\cal H}={\cal H}_1 \otimes ... \otimes {\cal H}_N$, are {\it distinguishable} 
or, on the contrary, are indeed identical but can be addressed individually 
because the individual wavefunctions do not overlap. Furthermore, for all 
practical purposes encountered so far, it suffices to consider localized 
particles. If we had been given a system of $N$ identical particles, we could 
no longer use (\ref{sepaN}) as a definition for entanglement. Thinking of 
fermions, we should rather had used Slater determinants, so as to take into 
account the antisymmetric features of the associated statistics.

However, the most common situation found in practice is that of a bipartite 
system, where each subsystem is clearly localized. Notice for instance 
quantum communication, when the exchange of information takes place between 
two entities, or the teleportation of a state between two parties. 
Even in this -- apparently simple -- case the general detection of entanglement 
is a highly non-trivial task.

\subsection{Bipartite entanglement}

As mentioned, most of the protocols for quantum communication and related 
tasks deal with two separated physical systems which may or may not become 
entangled or use entangled states to transfer information. The cases of 
entanglement present in pure and mixed bipartite states appear next. 
\newline
\newline
{\bf Pure states and their entanglement}
\newline

\begin{figure}
\begin{center}
\includegraphics[angle=0,width=0.5\textwidth]{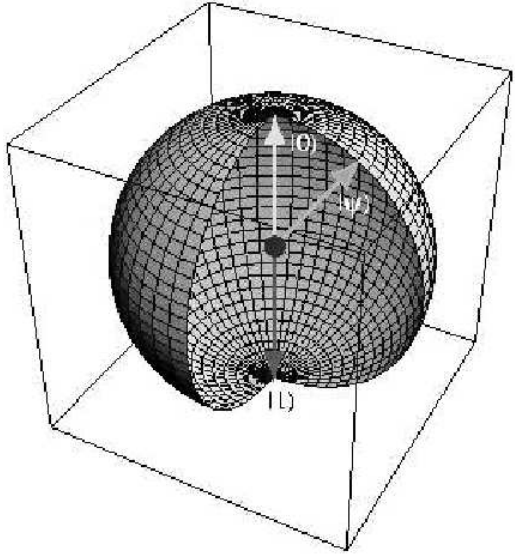}
\caption{Representation of a qubit as a Bloch sphere. Any state of a qubit 
$|\psi\rangle=\cos(\theta)|0\rangle+e^{i\phi}\sin(\theta)|1\rangle$ corresponds 
to a point in the so called Bloch sphere $S^2$. The states $|0\rangle$ and $|1\rangle$ 
are localized in the poles. See text for details.}
\label{Bloch}
\end{center}
\end{figure}

The simplest quantum mechanical systems that exhibit the feature of quantum 
entanglement are bipartite systems composed of two subsystems, each one described 
by a two-dimensional Hilbert space ($2 \times 2$). These systems are 
generically known as two-qubit systems. Let us recall that ``qubit" stands for 
``quantum bit", and constitutes 
the quantum extension of the binary digits $|0\rangle$ and $|1\rangle$ in the 
form of $\alpha|0\rangle+\beta|1\rangle$, with $\alpha,\beta \in {\cal C}^2$ and 
$|\alpha|^2+|\beta|^2=1$. The usual classical bits 0 and 1 can refer to 
voltages in a certain logical gate in a transistor, which appropriately used 
constitute the Boole algebra upon which modern -- though classical -- computers 
base their operations. Even though the aforementioned voltages have a quantal 
origin due to doping in semiconductors, they have a well defined property after 
and during the measure of the state of the gate. On the contrary, a qubit has 
not a well defined state, being a {\it superposition} of two possible states. 
After a measurement is performed, the state of the qubit collapses to either 
$|0\rangle$ or $|1\rangle$ with a definite probability given by $|\alpha|^2$ or 
$|\beta|^2$, respectively. In point of fact, the previous definition of a qubit 
can be rewritten in a more elegant fashion: any state of a qubit 
$|\psi\rangle=\cos(\theta)|0\rangle+e^{i\phi}\sin(\theta)|1\rangle$ corresponds 
to a point in the so called Bloch sphere $S^2$ (see Fig.\ref{Bloch}). 
Also, a particularly convenient representation of such the state 
$|\psi\rangle \langle \psi|$ is given by 

\begin{equation} \label{qubit}
\rho \,=\, \frac{1}{2} \big[ I + {\bf r·\sigma} \big]
\end{equation}

\noindent where $I$ is the $2\times 2$ identity matrix, 
${\bf r} \in {\cal R}^3$ and $\{$$\sigma_n$$\}_{n=1}^{3}$ are 
the Pauli matrices. Gathering two-qubits, we have 
a two-qubit system.

A pure state $\rho_{AB}=|\psi\rangle_{AB} \langle \psi|$ representing 
an arbitrary two-qubits system can be written as 
\begin{equation} \label{pure}
|\psi\rangle_{AB} \,=\, a |0\rangle_A \otimes |0\rangle_B + 
b |0\rangle_A \otimes |1\rangle_B + 
c |1\rangle_A \otimes |0\rangle_B + 
d |1\rangle_A \otimes |1\rangle_B,
\end{equation}
\noindent with $a,b,c,d \in {\cal C}^2$ and $|a|^2+|b|^2+|c|^2+|d|^2=1$. 
From now on we omit the tensor product symbol between kets and 
the subscript referring to which system they belong 
(e.g. $|0\rangle|1\rangle \equiv |0\rangle_A \otimes |1\rangle_B$). 
The definition of entanglement for pure states is particularly simple: after 
Schmidt-decomposing (\ref{pure}) into (\ref{Schmidt}) with rank $k$ and 
coefficients $\{$$w_{i=1..k}$$\}$, entanglement is defined as the Shannon 
entropy of $\{$$w_{i=1..k}$$\}$ squared. Thus, a state is separable 
if $k=1$ (zero entropy).

An alternative way of describing the entanglement of a pure state of two-qubits 
is the following. Consider the total density matrix 
$\rho_{AB}=|\psi\rangle_{AB} \langle \psi|$. By the action of partial tracing 
we eliminate the degrees of freedom of either subsystem and end up with a 
reduced or marginal density matrix (e.g. $\rho_A=$Tr$_B$($\rho_{AB}$)=
$\langle 0_B|\rho_{AB}|0_B\rangle$+$\langle 1_B|\rho_{AB}|1_B\rangle$). The 
entanglement $E$ of $\rho_{AB}$ is then defined as the von Neumann entropy of 
$\rho_A$, $E \equiv S(\rho_A)=-$Tr($\rho_A\,\log\rho_A$). 
In point of fact, it is not difficult to see that $S(\rho_A)=S(\rho_B)$, as 
it should be for symmetry reasons. The logarithm of the von Neumann entropy 
is taken is base 2 ($\log_{2}$) such that maximum entanglement corresponds to 1. 

A very interesting family of bipartite pure states is constituted by the 
so called {\it Bell states}. The Bell states correspond to pure states 
with maximal entanglement and are defined, up to a global phase, as 

\begin{eqnarray} \label{Bell}
|\Phi^{\pm}\rangle &=& \frac{1}{\sqrt{2}}(|00\rangle \pm |11\rangle), \cr
|\Psi^{\pm}\rangle &=& \frac{1}{\sqrt{2}}(|01\rangle \pm |10\rangle),
\end{eqnarray} 

\noindent where $|\Psi^{-}\rangle$ is nothing but the singlet state of 
two $\frac{1}{2}$-spins. These states have lots of theoretical and 
practical uses, as we shall see. They are synonymous with EPR states, named 
after Einstein, Podolsky and Rosen (see Appendix A).                            
As a matter of fact, our ``entanglement ruler" 
could be defined as the quantum correlations contained in the antisymmetric 
singlet state $|\Psi^{-}\rangle$. Indeed, its entanglement $E$ is maximum 
($E(|\Psi^{-}\rangle \langle \Psi^{-}|)=1$).
\newline
\newline
{\bf Mixed states and their entanglement}
\newline

Pure states are difficult to obtain in the laboratory and hard to store for long 
periods of time. Due to the interaction with the environment ${\cal E}$, they 
rapidly spread (decoherence time) into an statistical mixture of different 
available pure states (a mixed state). In a way, it is more natural to think of 
entangled mixed states rather than entangled pure states. Actually, 
if $|\psi_S\rangle$ is our initial pure state of the system at $t=0$, 
the environment described by the state $|{\cal E}\rangle$ dilutes the 
individuality of $|\psi_S\rangle$ into some new state $|\psi_{SE}\rangle$ at 
$t=T$. Mathematically, $|\psi_S\rangle \otimes |{\cal E}\rangle (t=0) 
\rightarrow |\psi_{SE}\rangle (t=T)$, wherefrom if one could trace out the 
degrees of freedom of the environment, we would end up with the state of the 
system at $t=T$ ($\rho_S=$Tr$_{E}$($|\psi_{SE}\rangle \langle \psi_{SE}|$)). 
The characterization of entangled mixed states is also necessary for the study 
of the entanglement properties of {\underline{pure}} states of multipartite systems 
with more than two components. For instance, let us consider a pure state 
$|\psi_{ABC}\rangle$ of a system with three subsystems A, B, and C. If we want 
to know the amount of entanglement present between subsystems say A and B, we 
have to consider the state $\rho_{AB}=$Tr$_C$($|\psi_{ABC}\rangle \langle\psi_{ABC}|$) 
which is, in general, mixed.

The complete characterization of mixed two-qubits states requires $4^2-1=15$ 
real parameters. This is so because $\rho$ is a $4\times 4$ hermitian, positive 
semidefinite matrix ($\rho=\rho^{\dag}$), which implies that $N^2$ real entries 
are needed. The 
requirement of normalization Tr($\rho$)$ = 1$ reduces the number to $N^2-1$. 
This is the most general structure of the space of two-qubits given in the 
so called computational basis ($|00\rangle,|01\rangle,|10\rangle,|11\rangle$). 
Because a mixed state can be prepared in infinitely many ways, some other 
decompositions can be more or less interesting depending on the context:

\begin{itemize}

\item Sometimes it is useful to decompose a given mixed state 
in the form of a superposition of Bell states (Bell states form an 
ortonormal basis). A subclass of these states are named 
{\it Bell diagonal states}, and are written as the convex sum 

\begin{equation} \label{Belldiagonal}
\rho\,=\,p_1|\Psi^{-}\rangle \langle \Psi^{-}|+
p_2|\Psi^{+}\rangle \langle \Psi^{+}|+p_3|\Phi^{+}\rangle \langle \Phi^{+}|+
p_4|\Phi^{-}\rangle \langle \Phi^{-}|. 
\end{equation}

\noindent The entanglement properties of these 
states are easy to describe, as we shall see. They appear quite often in 
quantum teleportation. One such example of Bell diagonal state is given by 
the so called ``Werner states" $\rho_W$. The Werner density matrix reads 

\be
\label{unoW} \rho_W\,=\,x|\Phi^{+}\rangle\langle
\Phi^{+}|\,+\,\frac{1-x}{4}I,\ee 

where $|\Phi^{+}\rangle$ is a
Bell state (maximally entangled). $\rho_W$ is a mixture of a Bell state, 
usually a singlet, with the remaining states. The state (\ref{unoW}) is
separable (unentangled) for the mixing coefficient $x\le 1/3$ \cite{Werner89}. 
For $x > 1/3$ they are entangled and violate the CHSH inequality for 
$x > 1/\sqrt{2}$ \cite{pop94,pop95}.
We see that Werner states are mixtures of noise and a
maximally entangled state, and therefore, for values of
the mixing parameter $x > 1/3$ they are entangled and
exhibit non-classical features \cite{pop94,pop95}. The fact that Werner 
density matrices violate the CHSH Bell inequality (when each of the two 
concomitant subsystems is subjected to a single ideal measurement) for 
$x > 1/\sqrt{2}$, but being entangled, motivated Werner himself to provide 
a hidden variable simulation of these correlations \cite{Werner89}. 
The Werner state (\ref{unoW}) is very popular in the literature. For instance, we have shown 
\cite{PLA2005} that there is a one-to-one correspondence between Werner
states and the Heisenberg anti-ferromagnet thermal states of two spinors (two qubits), 
that is, $\rho(T)=\exp(-H/k_{B}T)/Z(T)$ with $H$ being the Heisenberg Hamiltonian.
 
\item There exists a more pedagogical way of presenting two-mixed states which 
owes much of its simplicity to the use of Pauli matrices. Resembling Eq. 
(\ref{qubit}) for one qubit, any given two-qubits state can decomposed 
in the following way:

\begin{equation} \label{rhoPauli}
\rho\,=\,\frac{1}{4} \big[I\otimes I + {\bf r·\sigma}\otimes I + 
I \otimes {\bf s·\sigma} + 
\sum_{m,n=1}^{3} t_{m,n}\sigma_n\otimes \sigma_m \big]
\end{equation}

\noindent where $I$ is the identity matrix, 
${\bf r,s} \in {\cal R}^3$, $\{$$\sigma_n$$\}_{n=1}^{3}$ are 
the Pauli matrices, and $t_{m,n}=$Tr($\rho\,\sigma_n\otimes \sigma_m$) are 
the coefficients of a real matrix. The state (\ref{rhoPauli}) is written 
in such a way that all non-local terms are contained in the last addend. 
All quantum correlations present in the mixed state appear due to this last 
term, while all the others refer to local addressings. Notice that we still 
need 15 real parameters. This form is specially suitable in the field of 
quantum optics and quantum tomography.

\end{itemize}

Nevertheless, when we study global properties of the set of all states 
${\cal S}$ we use the generic form in the computational basis. 
The quantification of entanglement for mixed states is much more difficult 
than for pure states. Due to this fact, several measures of entanglement 
that recover the usual one for pure states had been advanced. A concise 
study is done in the 
following sections. Also, a thorough exposition of the space of two-qubits 
and entanglement is drawn in Chapter 8.                                       

\subsection{Multipartite entanglement}

Contrary to what may seem an exception, multipartite entanglement is the rule. 
The outcome of many parties in mutual interaction result in a statistical 
matrix $\rho$ that, if entangled, cannot be decomposed as a mixture of product 
states of the individual subsystems. Nevertheless, at present we have only
partial knowledge of the complete picture of multiparticle entanglement. 
For instance, several quantum algorithms 
require two registers, that is, two bunches of qubits, to be entangled and 
this feature is essential for quantum computation. This latter case, however, 
only requires entanglement of multipartite {\it pure} states. A much more 
difficult problem consists in the classification of multipartite {\it mixed} 
states, which is still under current study. 

Historically, the interest in entanglement between more than two parties was 
motivated by the fact that that correlations among more
than two particles present novel and highly nontrivial features not present 
in states of two particles. This fruitful path was opened by the seminal paper 
by Greenberger, Horne, and Zeilinger \cite{GHZ}. Here we present a brief sketch 
of the GHZ experiment. Let us take three well-separated parties $A$, $B$ 
and $C$, each one of 
those having two observables $X$ and $Y$, which can adopt the discrete values 
$\{$$\pm 1$$\}$ 
{\it only}\footnote{In fact, this experiment corresponds to a local 
variable theory}. Let us suppose that we perform three measurements 1, 2 and 3, 
where in each experiment we measure either $X$ ($m_x$) or $Y$ ($m_y$) for 
every party, as arranged in Table 5.1.

 \begin{table}[tbp]
 \centering
\begin {tabular}{|c|c|c|c|c|c|}
       & A & B & C & product \\
\hline
1  & $m_x$ & $m_y$ & $m_y$ & $r_1$   \\
2  & $m_y$ & $m_x$ & $m_y$ & $r_2$   \\
3  & $m_y$ & $m_y$ & $m_x$ & $r_3$   \\
$X-$result  & $m^{A}_{x}m^{A}_{y}m^{A}_{y}=m^{A}_{x}$ & 
$m^{B}_{y}m^{B}_{x}m^{B}_{y}=m^{B}_{x}$ & 
$m^{C}_{y}m^{C}_{y}m^{C}_{x}=m^{C}_{x}$ & $r_{1}r_{2}r_{3}$   \\

\end{tabular}
\label{TableGHZ}
\caption{Scheme for the GHZ experimental setting.}
\end{table}

\noindent The last column reports the product of the outcomes of the 
{\it different} parties. The last row shows the results of the product 
of the outcomes of the {\it individual} measurements 
(note that $(m^{\mu}_y)^2=1 ~\forall \mu$). One then should expect that 
$r_{1}r_{2}r_{3}$ be equal to $m^{A}_{x}m^{B}_{x}m^{C}_{x}$. Let us go 
to the laboratory and prepare three photons in the state
\footnote{First introduced by Mermin, though.}

\begin{equation} \label{GHZ}
|GHZ\rangle \, = \, \frac{1}{\sqrt{2}} (|000\rangle-|111\rangle),
\end{equation}

\noindent where $|0\rangle$ or $|1\rangle$ may denote opposite states of 
polarization of a photon. Suppose that we perform, in analogy 
with the $r_i$ results, the following 
measurements ((\ref{GHZ}) is an eigenstate of these operators)

\begin{eqnarray} \label{eigenGHZ}
\sigma^{A}_{x}\sigma^{B}_{y}\sigma^{C}_{y} \, |GHZ\rangle &=& + 1\,|GHZ\rangle \nonumber \\
\sigma^{A}_{y}\sigma^{B}_{x}\sigma^{C}_{y} \, |GHZ\rangle &=& + 1\,|GHZ\rangle \nonumber \\
\sigma^{A}_{y}\sigma^{B}_{y}\sigma^{C}_{x} \, |GHZ\rangle &=& + 1\,|GHZ\rangle,
\end{eqnarray}

\noindent we therefore expect 
$\sigma^{A}_{x}\sigma^{B}_{x}\sigma^{C}_{x}|GHZ\rangle$ to be 
$(+1)(+1)(+1)=+1$, but if we check the experimental results, we find 
$\sigma^{A}_{x}\sigma^{B}_{x}\sigma^{C}_{x}|GHZ\rangle=-1\,|GHZ\rangle$ instead! The 
contradiction with the ``expected" value expresses the fact that the GHZ state 
(\ref{GHZ}) is an entangled state of three particles. Any separable state 
of the form 
$|\psi^{A}\rangle \otimes |\psi^{B}\rangle \otimes |\psi^{C}\rangle$ would 
comply with the value predicted by a local variable theory. It is a kind of 
Bell theorem without inequalities: superposition exists and properties are not 
sharp (one cannot attribute properties before measurement, that is why 
$r_{1}r_{2}r_{3}$ does not equal $m^{A}_{x}m^{B}_{x}m^{C}_{x}$).

Already in the simplest extension of a bipartite system of two qubits, 
that is, a three qubit system, the characterization of entangled states (pure 
or mixed) is not an easy task. Even the detection of entanglement is not clear. 
However, classes of multipartite pure states are known \cite{Acin} such as

\begin{eqnarray} \label{classes3}
|GHZ\rangle &=& \frac{1}{\sqrt{2}} (|000\rangle-|111\rangle) \nonumber \\
|W\rangle &=& \frac{1}{\sqrt{3}} (|001\rangle+|010\rangle+|100\rangle) \nonumber \\
|B\rangle &=& |\Psi^{+}_{AB}\rangle \otimes |0_{C}\rangle \nonumber \\
|S\rangle &=& |\psi^{A}\rangle \otimes |\psi^{B}\rangle \otimes |\psi^{C}\rangle, 
\end{eqnarray}

\noindent where $|B\rangle$ correspond to {\it biseparable} states and 
$|S\rangle$ belong to the class of {\it product} states ({\it ergo} separable). 
In the class $|B\rangle$ one can encounter for instance that A(BC) and (AB)C 
partitions are separable, while (AC)B is entangled! Another interesting example 
is provided by the complementary state to the so called SHIFTS UPB tripartite 
mixed state introduced in \cite{xino}. Given the set SHIFTS UPB of product states 
$|\psi_i=\{|0,1,+\rangle,|1,+,0\rangle,|+,0,1\rangle,|-,-,-\rangle \}$, with 
$|\pm \rangle=(|0\rangle \pm |1\rangle)/\sqrt{2}$, one defines its complementary state 

\begin{equation} \label{shift}
\rho\,=\,\frac{1}{4} \bigg( 1\,-\,\sum_{i=1}^4 |\psi_i \rangle \langle \psi_i| \bigg).
\end{equation}

\noindent According to \cite{xino}, state (\ref{shift}) has the curious property 
that it is not only two-way PPT, but also two-way separable, which means that 
(\ref{shift}) has got genuine tripartite mixed entanglement.

In the case of pure states of three qubits, a pioneer result \cite{CKW00} in 
the description of entanglement in many systems states the 
``monogamy of entanglement". That is, 
for a certain bipartite entanglement measure $E$ and a given system composed by 
three qubits A, B and C, it is found that

\begin{equation} \label{ineq}
E[\rho_{AB}]\,+\, E[\rho_{AC}] \leq E[\rho_{A(BC)}],
\end{equation}

\noindent where $\rho_{AB}=$Tr$_C$($|\Psi\rangle_{ABC}\langle \Psi|$) and 
$\rho_{A(BC)}=$Tr$_{BC}$($|\Psi\rangle_{ABC}\langle \Psi|$) are the concomitant 
reduced density matrices. The result (\ref{ineq}) is conjectured to hold 
in for arbitary number of qubits. A more detailed account is given in Chapter 11.                                

In general terms, the characterization of multipartite 
entanglement is an open question, at least in the terms of the mathematical 
problem of separability (\ref{sepaN}). When more than two parties are involved, 
the notion of partitions of the system has to be taken into account, which 
induces a certain arbitraricity in a rigorous definition of entanglement.

\subsection{Physically motivated entanglement measures. General properties}

Bearing in mind the aforementioned problem of the lack of a general and 
complete description of the entanglement present in many-body systems, next 
we give an account of several well known bipartite entanglement measures. 
To start with, let us investigate some properties of the already used 
von Neumann entropy (of a subsystem) as a measure of the quantal correlations 
$E(|\Psi\rangle)$ present in a 
pure state $|\Psi\rangle$. One such characteristic is that $E(|\Psi\rangle)$ 
is invariant under local unitary transformations $U=U_A \otimes U_B$. That is, 
$E(|\Psi'\rangle = U|\Psi\rangle) = E(|\Psi\rangle)$. In other words, 
$E(|\Psi\rangle)$ cannot increase under LOCC. Thus, one of the many features 
required by a proposed entanglement measure is that it cannot increase under the 
action of local operations. In point of fact, the fundamental property is that 
entanglement between two systems cannot be increased without quantum
interaction between them. If the systems are spatially separated, then
entanglement between the quantum systems is only allowed to decrease under LOCC. 
A good source on entanglement measures can be found in \cite{Rinton}.

The basic postulates or properties that a reasonable measure of entanglement must 
exhibit can be summarized as follows.

\begin{itemize}

\item 1) For any separable state $\rho_{sep}$ the entanglement measure should be 
null

\begin{equation}
E(\rho_{sep})\,=\,0.
\end{equation}

\item 2) The entanglement of a state $\rho$ must remain invariant under local 
unitary transformations $U=U_A\otimes U_B$, 

\begin{equation}
E(\rho)\,=\,E(U \, \rho \, U^{\dag}).
\end{equation}

\noindent It is equivalent to say that a change in the two local basis 
(associated with each of the two subsystems) upon which states 
$\rho$ are decomposed must not neither augment nor diminish the entanglement. 
Besides, any LOCC operation $\Lambda$ should at most leave the 
entanglement of $\rho$ untouched: $E(\Lambda(\rho))\leq E(\rho)$.

\item 3) Suppose that a LOCC operation action on $\rho$ is capable of 
transforming our initial state into a series of possible final states 
$\{$$\rho_{f}^{(i)}$$\}$ (an ensemble), each one of those with 
probability $p_i$. We then require

\begin{equation}
E(\rho)\,\ge \, \sum_i p_i\, E(\rho_{f}^{(i)}).
\end{equation}

\noindent This last condition is strongly related to the purification procedure 
of extracting pure maximally entangled states out of a noisy state $\rho$. Also, 
we must require a good measure $E$ to be a convex function, 
$E\big(\sum_i p_i\rho_i\big) \leq \sum_i p_i E(\rho_i)$.

\item 4) Once we gather together two entangled non-interacting states $\rho_1$ 
and $\rho_2$ in the form $\rho_1\otimes \rho_2$, we must have that

\begin{equation}
E(\rho_1\otimes \rho_2)\, = \, E(\rho_1) +  E(\rho_2).
\end{equation}

\end{itemize}

\noindent As we can see, most of these conditions form a set of statements that 
linger around the notion that entanglement is something that cannot be created 
or prepared locally, and therefore different local, linear manipulations of states 
can only diminish -- never increase -- the ``quantity" of quantum correlations. 

In the case where $\rho$ corresponds to the density matrix of a pure 
state, all measures must recover the von Neumann entropy of the reduced 
density matrix $S(\rho_A)=-$Tr($\rho_A \log_2 \rho_A$), which complies with 
these demandings. The problem arises in the case of 
mixed states. There are not many measures that observe the aforementioned 
demands. One then can either extend the definition for pure states to mixed 
states $\rho$ by performing a minimization over all possible decompositions of 
$\rho$ (what is known as a convex roof procedure) or introduce some new measure 
based on distances between states (distance-based measures), which can relax 
(not always) some postulate of the previous list.
\newline
\newline
{\bf A. Entanglement of distillation, entanglement cost and 
entanglement of formation}
\newline

\begin{figure}
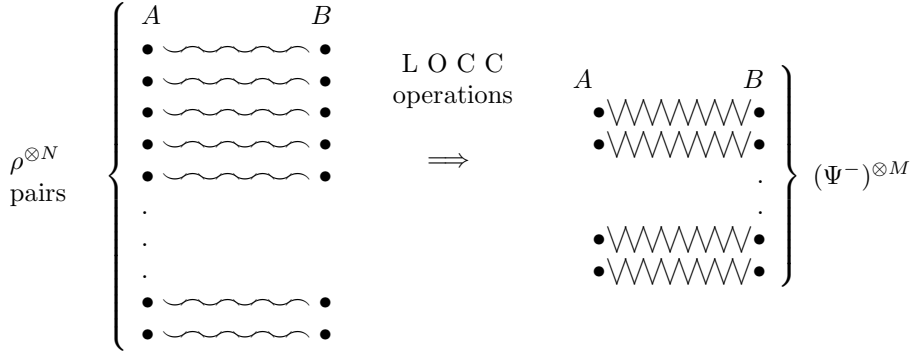

\begin{center}
\def\AB{$A \,\,\,\,\,\,\,\,\,\,\,\,\,\,\,\,\,\,\,\,\,\,\,\,
\,\,\,\,\,\,\,\,\,\, B$}
\def\weakpair{$\bullet \smile\!\!\frown\!\!\smile\!\!\frown
\!\!\smile\!\!\frown\!\!\smile\!\!\frown
\bullet$}
\def\strongpair%
{$\bullet\backslash\!/\!\backslash\!/\!\backslash\!/\!\backslash\!/%
\!\backslash\!/\!\backslash\!/\!\backslash\!/\!\backslash\!/%
\bullet$}
$
\parbox{1.8cm}{$\rho^{\otimes N}$ \\ pairs}
\kern-6mm
\left\{\begin{array}{l}
$\AB$ \\ 
$\weakpair$ \\
$\weakpair$ \\
$\weakpair$ \\
$\weakpair$ \\
$\weakpair$ \\
.\\
.\\
.\\
$\weakpair$ \\
$\weakpair$
\end{array} \right.
\quad
\begin{array}{c}
{\rm  L\ O \ C \ C}\\
{\rm  operations}\\
\ \\
\Longrightarrow
\ \\
\ \\
\ \\
\ \\
\ \\
\end{array}
\quad
\left.\begin{array}{r}
$\AB$ \\ 
$\strongpair$\\
$\strongpair$\\
.\\
.\\
$\strongpair$\\
$\strongpair$\\
\end{array} \right\}
\hskip1mm \parbox{1.8cm}{$(\Psi^{-})^{\otimes M}$}
$
\caption[Distillation of entanglement]{Distillation of entanglement.} 
\label{distil}
\end{center}
\end{figure}

Suppose that two parties $A$ and $B$ share $N$ pairs of pure states 
of the form $\psi=a|00\rangle + b|11\rangle$, with $a,b \in {\cal R}$ and 
$a^2+b^2=1$ without loss of generality. It is known that these states are 
partially entangled, because maximum entanglement occurs for 
$a=b=\frac{1}{\sqrt{2}}$. Imagine that both parties would like to 
convert these entangled states $\psi^{\otimes N}$ to a smaller supply 
of $M$ singlet states $\Psi^{-}$, which are maximally entangled. 
This process is known as {\it distillation} of entanglement\footnote{This problem 
was first addressed by Bennett {\it et al.} in \cite{Bennett96}.}. 
In other words, entanglement is concentrated in fewer pairs. Fig.\ref{distil} 
provides an sketch of the situation. By acting under LOCC operations, one 
may wonder what is the yield $\frac{M}{N}$. This is the starting 
point for defining the {\it entanglement of distillation} or 
distillable entanglement of $\psi$, $D(\psi)$

\begin{equation}
Sup._{LOCC} \lim_{N \to \infty} \frac{M(N)}{N} \equiv D(\psi).
\end{equation}

\noindent For pure states, it can be shown that $D(\psi)=S(\rho_A)$. 
It gives us the number of units of singlets that can be distilled out of 
a given state $\psi$.

The reverse process is called {\it dilution} of entanglement. Starting 
from $M$ singlets, $N$ copies of the state $\psi$ are obtained under 
LOCC operations, defining the {\it entanglement cost} $E_C(\psi)$

\begin{equation}
Sup._{LOCC} \lim_{N \to \infty} \frac{M(N)}{N} \equiv E_C(\psi).
\end{equation}

\noindent Also for pure states, $E_C(\psi)=S(\rho_A)$. Thus, we have 
that both distillation and dilution of entanglement are equal and 
reversible asymptotically. The proof that $D(\psi)$ equals the reduced  
von Neumann entropy $S(\rho_A)$ is not difficult, but tedious. It basically 
consists in expanding the general form of $\psi^{\otimes N}$. After assuming 
that the unit of entanglement is given by the entanglement of 
the singlet state and some asymptotic approximation is made, it is seen 
that $E(\psi) \sim N S(\rho_A)$, for $N \rightarrow \infty$. The formal 
proof of $E_C(\psi)=S(\rho_A)$ is a bit more involved, and can be performed 
with the aid of the theory of generalized measurements. 

It is also important to stress that the above procedure of 
distillation is reversible, but this is not the case for mixed states, 
where both entanglement of distillation and entanglement cost differ from 
each other. This is an important issue: there appears an intrinsic 
irreversibility feature in passing from pure to mixed states: the so called 
{\it bound entangled} states are so thoroughly mixed that cannot be distilled. 
This fact has indeed led to conjecture some connection between the theory 
of entanglement and the Second Law, as discussed in forthcoming sections. 

Now then, how can we define a measure of entanglement for \underline{mixed 
states}? If we agree that the reduced von Neumann entropy 
is a good measure of entanglement for pure
states, it is somewhat natural to extend this definition to mixed states. 
One way to do so is by doing the {\it convex roof}

\begin{equation} \label{roof}
E(\rho)\,=\, inf\, \sum_i p_i E(|\psi\rangle \langle \psi|),
\end{equation}

\noindent with $\sum_i p_i=1$ and $0 \leq p_i \leq 1$. The minimum is taken 
over all possible decompositions of the mixed state $\rho=\sum_i p_i 
|\psi\rangle \langle \psi|$ into pure states (which need not be orthogonal). 
This huge task of optimizing (\ref{roof}) is still an open problem, but luckily 
for us there exists a closed formula for the two-qubits instance, which results  
in the physically motivated measure of entanglement 
provided by the {\it entanglement of formation} $E_F$ \cite{BDSW96,WO98}.
This measure quantifies the resources needed to create
a given entangled state $\rho$. That is, $E_F(\rho)$ is equal to
the asymptotic limit (for large $N$) of the quotient $\frac{M}{N}$, where $M$ 
is the number of singlet states needed to
create $N$ copies of the state $\rho$ when the optimum
procedure based on local operations is employed. This procedure goes in the opposite 
direction to the one sketched in Fig.\ref{distil}. 
The relationship between the entanglement of formation $E_F$ and the 
entanglement cost $E_C$ in the case of mixed states remains 
and open question. That is, it is not known if $E_F=E_C$ in the general 
mixed-state case, as it is for pure states.

The entanglement of formation for two-qubit systems is
given by Wootters' expression \cite{WO98}

\be \label{Ef}
E[\rho] \, = \, h\left( \frac{1+\sqrt{1-C^2}}{2}\right), \ee

\noindent where

\be
h(x) \, = \, -x \log_2 x \, - \, (1-x)\log_2(1-x), \ee

\noindent and the concurrence $C$ is given by

\be \label{concurrence}
C \, = \, max(0,\lambda_1-\lambda_2-\lambda_3-\lambda_4), \ee

\noindent $\lambda_i, \,\,\, (i=1, \ldots 4)$ being the square
roots, in decreasing order, of the eigenvalues of the matrix $\hat
\rho \tilde \rho$, with

\be \label{rhotil} \tilde \rho \, = \, (\sigma_y \otimes \sigma_y)
\rho^{*} (\sigma_y \otimes \sigma_y). \ee

\noindent The above expression has to be evaluated by recourse to
the matrix elements of $\hat \rho$ computed  with respect to the
product basis. $C^2$ can be regarded as a proper measure of entanglement, 
as (\ref{Ef}) is a monotonic increasing function of $C^2$. For pure 
states (\ref{Ef}) reduces to the usual von Neumann entropy. As a matter of 
fact, $C^2=4\,$det$\rho_A$ for pure two-qubits states. The general form 
for a bipartite pure state \cite{Rungta01} 
$|\psi\rangle \in {\cal H}_N \otimes {\cal H}_K$ is ($N \leq K$ is assumed) 
$C(|\psi\rangle)=\sqrt{2[|\langle \psi|\psi\rangle|^2-Tr(\rho_{N}^{2})]}$, 
with $\rho_N=$Tr$_K(|\psi\rangle \langle \psi|)$. Although there is no 
general expression for the concurrence of mixed bipartite states, 
F. Mintert {\it et al.} derive in \cite{concumany} a lower bound 
for the concurrence of mixed bipartite quantum states, valid in arbitrary
dimensions.

In the case of Bell diagonal states $\rho$ (\ref{Belldiagonal}), which appear 
quite often in quantum teleportation and quantum cryptography scenarios, the 
concurrence reads $C=2\lambda_m-1$, where $\lambda_m$ is the maximum 
eigenvalue of $\rho$. Thus, for this family of states, it suffices to say 
that a given state is entangled if $\lambda_m \geq \frac{1}{2}$.
\newline
\newline
{\bf B. Robustness of entanglement}
\newline

A nice measure is provided by the {\it robustness of entanglement} 
\cite{Guifre99}. Given an entangled state $\rho$ and
 separable state $\rho_{s}$, a new density matrix $\rho(s)$
 can be constructed as

 \begin{equation}\label{rhos}
 \rho(s)=\frac{1}{s+1}(\rho+s\rho_s),\quad s\geq0.
 \end{equation}

\noindent The ensuing state $\rho(s)$ can be either separable or 
entangled. What is certain is that there always exits a minimal ${\bf s}$
 corresponding to $\rho_s$ such that $\rho(s)$ is separable. This 
optimal value ${\bf s}$ is called the robustness of $\rho$ with respect to
 $\rho_s$, and expressed as $R(\rho\parallel\rho_s)$. The inferior 
of all optimal values is known as the absolute robustness of $\rho$, 
or simply the {\it robustness of entanglement} 

\begin{equation} \label{rob}
 R(\rho\parallel S)\equiv\min_{\rho_s\in S_{sep}} R(\rho\parallel
 \rho_s),
\end {equation}

\noindent where the whole set of separable states ${\cal S}_{sep}$ is explored. 
This measure is based on a simple physical operation: mixing
with locally prepared states, and certainly it does not increase under 
LOCC operations. It admits the geometrical interpretation of the minimal 
distance of the entangled state $\rho$ to the boundary of the set ${\cal S}_{sep}$.

The robustness of entanglement can be computed analytically for pure states. 
Recalling the Schmidt decomposition 
(\ref{Schmidt}) for a pure state in a system ${\cal H}_N \otimes {\cal H}_M$, 
with $K=$min$(N,M)$ being the rank of the state and coefficients 
$\{$$w_i$$\}_{1}^{K}$, $R(\rho\parallel S)=\big(\sum_{i=1}^{K} w_i \big)^2-1$. 
There is no general expression\footnote{Not exactly true: S. J. Akhtarshenas {\it et al.} 
claim in \cite{Iranian} to have found a complex procedure for 
computing the robustness of entanglement for any mixed state of two qubit systems, 
together with a special parameterization. Their formulas have to be confirmed yet.}, 
to our knowledge, even for the bipartite 
case of two-qubits systems. However (\ref{rob}) is not difficult to compute 
numerically taking advantage of the fact that ${\cal S}$ is a convex and 
compact set.
\newline
\newline
{\bf C. Relative entropy of entanglement and similar measures}
\newline

The relative entropy measure belongs to the class of entanglement measures  
that are based on distances between states. It need not has to be a proper 
metric is the space of all states ${\cal S}$, and in fact it is not, but it can 
be extended to recognized metrics in that space ${\cal S}$. The 
relative entropy of formation is defined as \cite{Vedral97}

\be \label{rel}
E^{rel}[\hat \rho] \, \equiv \, min_{\sigma \in {\cal S}_{sep}} \, D(\rho||\sigma), 
\ee

\noindent where the minimum is taken over all possible states $\sigma$ that are 
unentangled (set ${\cal S}_{sep}$), and $D$ is the quantum version of the 
Kullback-Leibler relative entropy\cite{Kull} 
$S^{rel}(\rho||\sigma)=$Tr$\left( \rho \log \rho- \rho \log \sigma \right)$. 
The measure (\ref{rel}) can be interpreted as the minimum distance from the 
set ${\cal S}$ when one uses $S^{rel}(\rho||\sigma)$ as a ruler. Besides, measure 
(\ref{rel}) acquires a well-defined statistical meaning, in terms of the quantum Sanov's 
Theorem \cite{PhysQI}: the probability of not distinguishing two quantum states $\sigma$ 
and $\rho$ after $n$ measurements is 

\begin{equation}
p(\rho \rightarrow \sigma)\,=\,e^{-n\,S(\sigma||\rho)}.
\end{equation}

\noindent In this context the situation deals with the probability of mistaken an 
entangled state from a disentangled one.

This measure 
possesses some similarities with the robustness of entanglement (\ref{rob}): 
they are both taken as minima of quantities representative of the entangled 
state $\rho$ with respect to ${\cal S}_{sep}$, and the two of them are easy 
to extend to arbitrary number of parties and dimensions. 
However, (\ref{rel}) reduces to the 
von Neumann entropy of entanglement for pure states, while this is not 
the case for (\ref{rob}). The relative entropy fulfils the 
requirement of a ``decent" entanglement measure. There exists only one 
serious drawback, though: it can only be computed analytically 
for Bell diagonal states (\ref{Belldiagonal}). However, convexity of 
${\cal S}_{sep}$ again allows us to numerically compute \cite{BatleCamerino} 
the quantity (\ref{rel}).

Other measures for $D(\rho||\sigma)$ can be used instead of 
$S^{rel}(\rho||\sigma)$. In point of fact, the latter expression is not 
a true metric simply because it is not symmetric. Therefore one could in 
principle use some metric distance to be minimized, such as the Bures or 
Hilbert-Schmidt distances \cite{Vedral97bis} between states, but they do not 
incorporate the von Neumann entropy of entanglement for pure states.

\section{Entanglement for indistinguishable particles}

So far we have delt with entanglement for distinguishable particles, assuming 
that a qubit or a {\it qudit} (coherent superposition of $D$ states) is 
encoded in the degrees of freedom of that particle. Examples of this situation 
could be the spin degrees of freedom of a spin-$s$ particle ($2s+1$), the 
$D$ relevant energy levels of an atom, or the polarizations of a photon. 
Speaking of qubits, we have implicitly supposed that the $N$ parties possess 
identical qubits (encoded in identical particles), but this is not at all the 
general framework. We could easily create an entangled state of two-qubits 
where the two important energy levels are encoded in each one of two perfectly 
distinguishable atoms, say Rubidium and Calcium. In this picture of entanglement, 
the distinction of states as tensor products is perfectly licit. Even in the 
case where the particles (representing the qubits) are identical, this picture 
preserves its validity because the systems are well-located and far apart from 
each other, so that the intrinsic correlations 
-- which are always present -- due to the statistics of identical particles 
is practically negligible for all practical purposes.

Now let us tackle the problem of entanglement between identical particles 
where the overlap of the wavefunctions is neither zero nor negligible. 
This is a typical situation encountered in the proposals of quantum 
gates for quantum computation based on the present semiconductor technology 
(see \cite{Brazil} and references therein). 
It is possible nowadays to confine a well defined number of electrons 
in quantum dots. Even for the simple case of two quantum dots confining 
one electron per site, the global wave 
function $|\Psi\rangle_{AB}$ has to be properly antisymmetrized unless they 
do not interact by any means. As soon as the electrons start 
to ``see" each other, we can no longer describe the concomitant 
Hilbert space ${\cal H}$ as a tensor product of individual subspaces 
${\cal H}_A$,${\cal H}_B$. Let us consider the case of two identical fermions 
(bosons receive similar a treatment) 
in an $N$-dimensional single particle space ${\cal H}_{N}$, where the total 
Hilbert space corresponds to ${\cal A}({\cal H}_{N}\otimes{\cal H}_{N})$ with 
${\cal A}$ being the antisymmetrization operator. Under this conditions, 
an arbitrary state can be written as

\be \label{anti}
|w\rangle=\sum_{i,j=1}^{N}w_{ij}f^{\dagger}_{i}f^{\dagger}_{j}|vacuum\rangle,
\ee

\noindent where $f^{\dagger}_{i}$ are the fermionic creation operators, and 
the antisymmetric matrix $w_{ij}$ fulfils the normalization condition
Tr$(w^* w)=-1/2\,$. The cases of pure and mixed two-fermion states have been 
studied in \cite{identical} and an extension of the {\it concurrence} measure 
(\ref{concurrence}) is given.

\begin{itemize}
\item i) Pure states. Rotating appropriately the state basis of a two-fermion 
state $|w\rangle$ we obtain

\be\label{eqn:slaterDecomp}
|w\rangle=2\sum_{k=1}^{m}r_k f^{\dagger}_{2k} f^{\dagger}_{2k-1}|vacuum\rangle,
\ee
\noindent with $2m\leq N$ and $r_k$ real and positive. Regarding each term 
as an Slater determinant, $|w\rangle$ has a minimal number $m$ of Slater
determinants, in analogy with the Schmidt rank (\ref{Schmidt}). 
In this case ($N=2\times 2=4$) the concurrence $C(|w\rangle)$ is defined as 

\be \label{Cpure}
C(|w\rangle)=\Big|\frac{1}{2}\sum_{i,j,k,l=1}^{4}
\epsilon^{ijkl}w_{ij}w_{kl}\Big|,
\ee 

\noindent where $\epsilon$ is the usual antisymmetric tensor.

\item ii) Mixed states. Let $\rho=\sum_i p_i |w_i\rangle \langle w_i|$ 
be a given mixed two-fermion state. Defining 
$|\widetilde w_i\rangle=\sum_{i,j,k,l=1}^4\epsilon^{ijkl}w_{kl} 
f_{i}^{\dagger} f_{jl}^{\dagger} |vaccum\rangle$, 
$\tilde\rho=\sum_i p_i |\widetilde w_i\rangle \langle \widetilde w_i|$
and $\lambda_i$ as the eigenvalues (real and positive) of $\rho\tilde\rho$
in decreasing order, the concurrence for mixed states is constructed as

\be \label{concmixed}
C(\rho)=\max(0,\lambda_1-\sum_{i=2}^6\lambda_i).
\ee
\noindent Note the similarities between (\ref{concmixed}) and the definition 
(\ref{concurrence}) for distinguishable two-qubits. This last definition is 
obtained as the convex roof of (\ref{Cpure}).
 
\end{itemize}

Antisymmetrization, essential whenever we deal with identical (fermionic or bosonic)
particles, might lead to misunderstandings 
as far as entanglement is concerned. This was clearly pointed out in 
\cite{Guir}. Suppose that we have the state $|R\rangle = |Me$ exposing 
$\rangle \otimes |You$ reading $\rangle$ describing two composite systems. 
From the tensor product structure one judges that this state is separable.
The systems need not be identical, but if they do are, we must antisymmetrize 
the global state even if they are far apart. The ensuing state 
${\cal A} |R\rangle$ possesses no entanglement at all because identical 
particles do not produce entanglement by themselves. On the contrary, 
a state like $\frac{1}{\sqrt{2}} \big[|Me$ exposing 
$\rangle \otimes |You$ reading $\rangle$ - $|Me$ reading 
$\rangle \otimes |You$ exposing $\rangle \big]$, which is entangled 
from the very beginning, must remain non-separable after ${\cal A}$.

However, the correlations that arise from the statistics of the particles 
find interesting applications in quantum information theory. In 
\cite{Omar1,Omar2} it is shown that indistinguishability enforces a 
transfer of entanglement from internal to spatial degrees of freedom 
of a system without interaction, and an entanglement concentration scheme 
which uses only the effects of quantum statistics of indistinguishable 
particles is exposed.

Summing up, the total correlations present in a given state representing the 
concomitant physical system are the result of the sum of statistical 
correlations {\it plus} quantum correlations, which arise from previous 
interactions between the constituents of the system. 
It is difficult to ascertain the feasibility of addressing the latter type 
of correlations in quantum information, at least experimentally. There might 
be some reason for choosing distinguishable qubits to the detriment of 
indistinguishable particles. However, systems of indistinguishable 
particles possess very interesting features as described in the field of 
condensed matter. It is in along this line of reasoning that a novel approach 
to entanglement makes its appearing.

\section{Relativity of entanglement: a new insight}

In the previous case of the problem of indistinguishable particles 
there are some important difficulties in the 
partition of the Hilbert space: when adding a particle $s$ to a system of 
$N$ identical particles, formally one can write the total Hilbert space of 
the system as ${\cal H}_{N+1} \cong {\cal H}_{N} \otimes h_{s}$, where 
$h_{s}$ is Hilbert space of $s$. If the particle which is added is 
identical to the rest of $N$ particles, then this description cannot be possible 
because it does not take into account the statistics of identical particles, 
either bosons or fermions. Instead, one has to recall the Fock space of 
second quantization $\oplus_{N=0}^{\infty} {\cal H}^{(N)}_{F,B}$. It is nothing 
but the occupation number representation: $({\cal C}^2)^{\otimes N}$ for 
fermions and $(h_{\infty})^{\otimes N}$ for bosons, where 
$h_{\infty}$ spans the harmonic oscillator basis. These considerations 
led us to Eq. (\ref{anti}). However, some paradoxes appear. Suppose that 
we have a single particle state which can be either in the states $A$ or $B$, 
similar to two different sites or ``parties". In second quantization, 
it reads as 

\begin{equation} \label{psimode}
|\Psi\rangle \,=\, \frac{1}{\sqrt{2}} (c_{A}^{\dagger} + c_{B}^{\dagger}) 
|vacuum\rangle \,=\, \frac{1}{\sqrt{2}} (|1,0\rangle+|0,1\rangle),
\end{equation}

\noindent where $c_{A,B}^{\dagger}$ are the creation operators of sites $A$ and 
$B$ acting on the vacuum, respectively. This case is illustrative of the fact 
that even in the case of 
a single particle, we can have entanglement between its possible {\it modes}. 
This is what is known as the counterpart of ``particle entanglement", wherefrom 
we now deal with modes. This picture \cite{ZanMode} is in some sense 
complementary to the one presented in the discussion of identical particles. 
Now, the paradox arises when (\ref{psimode}) can be transformed into the 
state $|1,0\rangle$ after a unitary transformation action upon 
$c_{A,B}^{\dagger}$. How is it possible to convert a maximally entangled 
state like (\ref{psimode}) into a separable state like $|1,0\rangle$? 
The solution consists in differentiating 
mode from particle entanglement, pointing out that 
entanglement may not be an absolute quantity. In other words, a state is 
entangled with respect to which picture you use. This fact induces some loose 
sense of arbitrarity in the definition of entanglement.

Due to this last consideration, the formal description of the separability 
problem and, in turn, of entanglement, in terms of (\ref{sepaN}) has 
been questioned. Zanardi {\it et al.} pointed out in \cite{ZLLl04} that 
the partition of a quantum system into subsystems is dictated by the set of
operationally accessible interactions and measurements available by the 
observer: ``Suppose one is given a four-state quantum system\footnote{For 
instance, the two-qubit case.}.
How does one decide whether such a system supports
entanglement or not? In other words, should the given
Hilbert space (${\cal C}^4$) be viewed as bipartite 
($\cong {\cal C}^2  \otimes {\cal C}^2$), or irreducible?" \cite{ZLLl04}. 

This idea arises at the same time that a new theory of entanglement is 
developed in the enlightening work of H. Barnum {\it et al.} from 
the group of Los Alamos \cite{barnum1,barnum3}. This new theory argues that 
entanglement is relative to a distinguished subspace of observables 
(in terms of algebras) rather than a distinguished subsystem decomposition. 
A new measure of entanglement -- the purity $P_{h}$ -- is properly introduced. 
It is remarkable that in this new framework the measure for pure states is 
recovered under certain conditions of the information available to the observer. 
As we shall see next, it also finds an exciting place in the description of 
many-body systems.

\subsection{The Purity measure $P_{h}$. Mathematical grounds}

In this subsection we shall highlight the basics of the mathematical theory of Generalized 
Entanglement (GE) developed by the group of Los Alamos, together with the concomitant 
definition of the so called ``purity" measure $P_{h}$. It will be necessary to 
say a few words on Lie algebras, in order to clarify the tenets of GE. Finally, 
we will establish a link between the purity measure $P_{h}$ and condensed matter 
theory, through the issue of quantum phase transitions (QPTs).

Let us start by recalling the standard framework for entanglement. Throughout this 
Chapter and the previous one, devoted to the detection of entanglement, it has become 
clear that the Hilbert space ${\cal H}$ of a given system may support different 
tensor product decompositions. In fact, the entanglement of state $\rho$ in ${\cal H}$ 
is defined once a preferred subsystem decomposition has been chosen 
(${\cal H} \cong \otimes_i \, {\cal H}_{i}$). One interesting feature of entanglement 
in the usual context is that a pure state $|\Psi\rangle \in {\cal H}$ is entangled 
iff the state of any of its subsystems is described by a mixed state. In point of fact, 
if $|\Psi_{AB}\rangle$ happens to be any of the Bell states, which are maximally entangled, 
it is immediate to obtain that the reduced density matrices 
$\rho_{A,B}=$Tr$_{B,A}|\Psi\rangle_{AB} \langle \Psi|$ represent maximally mixed states, 
and therefore are unentangled states. In other words, entangled pure states look 
mixed to {\bf local} observers. This observation will be of crucial importance in 
the development of a theory of GE.

Usually these decompositions appear naturally 
in the description of quantum information processing features, because when that parties 
are well-separated (in real space), one can treat them as distinguishable quantum subsystems. 
The problem arises in the conflict described previously between particle or mode entanglement, 
and the question of describing the entanglement of more than two particles, specially 
crucial in the case of indistinguishable particles.
\newline
\newline
{\bf Entanglement as an observer-dependent concept}
\newline

The formal theory describing the GE is given in references \cite{barnum1,barnum3}. 
Let us recall here the key concepts.
\newline

\nd i) \underbar{The GE is relative to a subspace of observables $\Omega$ of the quantum system}. 
\newline

Let us start with the description of the usual Lie algebra. A Lie 
algebra $h$ is a vector space endowed with a binary operation 
or map (the commutator operation) satisfying antisymmetry, and the well-known Jacobi 
identity. It is said that an ideal $I$ in a Lie algebra is a subalgebra such that 
for $x \in I$ and arbitrary h $\in h$, $[{\rm h},x] \in I$. A Lie algebra is simple if 
it contains no proper ideals. The algebra $h$ will be semisimple iff it can be written as 
a direct sum of simple Lie algebras. Given a semisimple Lie algebra $h$, a maximal 
commutative subalgebra is known as a Cartan subalgebra $c$ of $h$. If a vector space 
carries a representation of $h$, then it decomposes into orthogonal joint eigenspaces 
$V_{\lambda}$ of the operators in $c$, that is, for each $V_{\lambda}$ there exists a 
set of states $|\psi\rangle$ such that for $x \in c$, $x|\psi\rangle = 
\lambda(x)|\psi\rangle$ ($\lambda$ is called the weight of $V_{\lambda}$). The 
space of operators of $h$ orthogonal to $c$ can be classified as raising ($e_{\mu}$) and 
lowering ($e_{-\mu}$) operators. The set of lowest-weight spaces for all Cartan subalgebras 
is the orbit of any one such state under the Lie group generated by $h$\footnote{Notice that 
the group is obtained by exponentiating the algebra, as immediately seen.}. These states 
are the generalized coherent states \cite{barnum1,barnum3}, which can be represented as 

\begin{equation}
|GCS\rangle\,=\,e^{\sum_{\mu} (\alpha_{\mu}\,e_{\mu}-\overline{\alpha}_{\mu}\,e_{-\mu})} |ref\rangle,
\end{equation}

\noindent with $|ref\rangle$ being an extremal state (in physical terms, the ground state 
of a system Hamiltonian, for instance). These generalized coherent states constitute a natural 
extension of the familiar ones corresponding to the harmonic oscillator.
\newline

\nd ii) \underbar{Extension of the fact that entangled pure states look mixed locally}. 
\newline

The basic idea is to generalize the observation previously made, namely, that pure entangled 
states possess maximally mixed states when we trace over the degrees of freedom of the 
rest of the subsystems (reduced state). In other words, they look entangled as a whole, 
but unentangled locally. Once we are given a pure state and we distinguish a relevant subspace 
of observables $\Omega$, with respect to which the GE is considered, the associated 
reduced state is obtained in this framework by only taking into account the expectation 
values of those $\Omega$-observables. Therefore, we say that a state is generalized unentangled 
relative to $\Omega$ if its reduced state is pure, and vice versa\footnote{With this definition, 
one recovers the usual definition of entanglement for bipartite systems $AB$ provided one has 
access to local observables, that is, 
$\Omega=\{ \hat A \otimes \hat I \oplus \hat I \otimes \hat B\}$.}. 
Mixed unentangled states are addressed in the usual way (convex combinations of unentangled 
pure states). One can grasp more physical insight about generalized unentangled states when 
viewed as the set of states that are unique ground states of a distinguished observable, say a 
Hamiltonian. The rest of extensions of this theory, such as the generalized LOCC operations 
can be found in \cite{barnum1,barnum3}.
\newline
\newline
{\bf The Purity measure $P_{h}$}
\newline

Let us give the definition of the $h$-purity. Let $\{x_i\}$ be a Hermitian and normalized 
orthogonal basis for $h$, that is, $x_i^{\dagger}=x_i$ and Tr($x_i x_j$)$ \infty \,\delta_{ij}$. 
For any pure state $|\Psi\rangle \in {\cal H}$, the purity of $|\Psi\rangle$ relative 
to $h$ is

\begin{equation} \label{purpur}
P_h(|\Psi\rangle)\,=\,\sum_i |\langle \Psi|x_i|\Psi\rangle|^2.
\end{equation}

\noindent This measure is endowed with a clear geometric meaning: $P_h(|\Psi\rangle)$ 
is the (square) distance from 0 of the projection of $|\Psi\rangle \langle \Psi|$ onto 
$h$, provided we employ the usual trace inner-product norm. In the case that 
$h$ is a Lie algebra, the purity measure (\ref{purpur}) is invariant under group transformations, 
$\tilde x_i=g^{\dagger} x_i g$, where $g\in e^{i\,h}$. Let us suppose that we have a preferred 
set of operators $h$, say Hamiltonians for instance. These Hermitian operators may generate 
a Lie group of unitary operators via $h\rightarrow e^{i\,h}$, or observables. 

With this definition of the purity measure $P_{h}$, it is plain that the previous generalized 
coherent states GCS possess maximal purity, that is, $P_{h}(|GHS\rangle)=1$, because they are 
already extremal (extremal projection onto $h$) with respect to 0. Therefore, in the framework 
of GE, the more entangled is a state with respect to some set $h$, the less $h$-purity it 
possesses, and vice versa. Once the mathematical details of the theory of GE are unfold, 
they prove the power of the Lie-algebraic setting. 

This measure recovers\footnote{We do not mean that the purity measure, or some function 
of it, reduces to the {\it reduced} von Neumann entropy for pure states. Conceptually 
they are {\it similar}, but not {\it equal}. Roughly speaking one may think of the 
purity measure as 1 minus the normalized (to 1) reduced von Neumann entropy, once a local set 
of observables is chosen, which is tantamount as partitioning the Hilbert space of the 
system in a preferred way.} the results 
obtained in the traditional framework of bipartite entanglement, provided the right set of 
observables is chosen. The physics behind these 
formulas is that entanglement is relative to the information (in terms of observables 
following an algebra) one has access to. On the one hand, in the case of two-spin 1/2 
particles, having 
access to $\Omega=h=su(2)\oplus su(2)=\{S_x^j,S_y^j,S_z^j,j=1,2\}$ give rise to quantum 
correlations --entanglement-- between the spin degrees of freedom of the two particles, that is, 
a Bell state has zero $P_{h}$. 
On the other hand, if we have access to all correlations, expressed in the form 
$\Omega'=h'=su(4)$, a Bell state will have non-zero $P_{h'}$.

The applications of this purity measure can extend to the condensed matter framework. In Chapter 
14 we describe how the purity may describe a quantum phase transition, once a proper 
algebra of observables has been chosen.

\section{Thermodynamic analogies for entanglement}

This section is mainly devoted to the similarities that some authors have 
established between entanglement and the Laws of Thermodynamics \cite{HOH02}. 
One must say that entanglement is understood as the usual quantum correlation 
existing between definite distinguishable parties. The analogy no longer works 
outside the orthodox view of entanglement exposed in previous sections. 
Along these lines of reasoning, we clarify some aspects previously 
discussed regarding some processes involving manipulations of 
entanglement. 

At first sight, the fact that there is a unique measure of entanglement 
for pure states (the von Neumann entropy of entanglement) shared with its 
thermodynamical counterpart may point out some connexion between the theory 
of quantum entanglement and Thermodynamics. Both of them share the historical 
problem of defining and quantifying the resources in terms of known quantities 
and brand new ones, such as entropy, which had little intuition by the time 
of Carnot. Also, strictly speaking we do not handle entanglement, but states 
that possess entanglement. Starting the analogy, the same situation is 
encountered in Thermodynamics, where work is the addressed quantity. The 
conjunction of work and heat (disordered work) boils down to the First Law, 
the preservation of energy. Irreversibility and reversibility appears in the 
manipulation of entangled states. We learned that the processes of distillation 
and dilution of entanglement for pure states are {\it reversible}: $E_D=E_C$, 
recalling that $E_D$ is the asymptotic yield of singletons drawn out of 
non-maximally pure states, and that $E_C$ is the reverse procedure. However, 
this is not the case for mixed states, where the total entanglement $E$ is 
imperfectly distilled so that $E_D < E_C$. This is due to the fact that 
some entanglement (the so called {\it bound entanglement}) can not be distilled. 
So some ``arrow of time" appears in quantum information when dealing with 
mixed states. It is like there existed two fundamental classes of entanglement, 
namely, pure state -- ordered -- and mixed state  
-- disordered -- entanglement. 
Then the process of extracting useful ``free" entanglement $E_{free}$ is similar 
to the one of drawing useful energy (work) between two heat reservoirs. 
The basic assumption in quantum information that entanglement cannot be created 
or increased by local operations or classical communications (LOCC) finds its 
counterpart (first emphasized by Popescu and Rohrlich in \cite{PR97}) 
in the Second Law, which asserts that the entropy of a system cannot 
decrease in a isolated system. Going one step further along this analogy, 
one could in principle 
regard the free entanglement $E_{free}$ ($\equiv E_D$) as a sort of generic ``free" 
energy $F$ \cite{Slovaca}, and the total entanglement $E$ as an internal 
energy $U$, so that 
$E_{bound} = E - E_{free} \equiv T_x S$ (recall $U = F + TS$). Thus, for 
entangled pure states $T_x S=0$, while some undefined ``temperature" 
$T_x$ exists for entangled mixed states.

Some may argue that the generalization of the above conclusions to the whole 
class of states $\rho$ by means of thermodynamic analogies is somewhat risky, 
precisely because some problems regarding separability remain open still. 
Furthermore, the analogy with Thermodynamics is by far incomplete (e.g. what is 
the equivalent of the temperature $T$?). Perhaps the 
fact that entanglement can be regarded as a resource and that 
certain manipulations of it are irreversible opened a path for developing 
some analogy with energy and the way it is manipulated in physics, hence 
the connection with the First and Second Law. What is certain, however, is that 
the process of changing pure entanglement into mixed 
entanglement is irreversible, which makes the distinction between pure and mixed 
states essential for all practical purposes in quantum information theory.


\part{The role of entanglement in different physical scenarios}


\chapter{The maximum entropy principle and the ``fake" inferred entanglement}

 The inference of entangled quantum states by recourse to the maximum entropy
 principle has been considered in the literature
 \cite{HHH99,BDADK97,RPPC00,R99,AR99}. In particular, the question of how to
estimate in a reliable way  the amount of entanglement of a bipartite
quantum system when only partial, incomplete information about its state
is available was addressed by Horodecki {\it et al.} \cite{HHH99}. Various
strategies have been advanced in order to tackle this problem
\cite{HHH99,RPPC00,R99,AR99,SH00}. Horodecki's question has also been
considered in connection with procedures for the entanglement purification
of unknown quantum states \cite{BCS00}. The motivation behind these lines
of inquiry is the full description of quantum entanglement, which constitutes, 
as we know, the basic resource required to
implement several of the most important processes studied by quantum
information theory \cite{LPS98,WC97,W98}.

If one has enough information it is possible to determine the
amount of entanglement of a quantum system even if the available
information does not allow for a complete knowledge of the
system's state. An interesting example of this situation was 
discussed by Sancho and Huelga in \cite{SH00}, where the minimal
experimental protocol required for determining the entanglement of
a two-qubits pure state from local measurements is exposed.
Another important result is that the 
knowledge of the expectation value of just one observable ({\it
local or not}) does not suffice to determine the entanglement of a
given unknown pure state of two particles. The case in
which the prior information is not sufficient for a complete
determination of the amount of entanglement was further examined
by Horodecki {\it et al.} \cite{HHH99}. These authors did not
restrict their analysis to pure states. They assumed that the
available information consists of the mean values of a given set
of observables $\hat A_i$. Jaynes' maximum entropy (MaxEnt)
principle \cite{B91,BAD96} provides a general inference scheme to
treat this kind of situations. According to Jaynes' principle, one
must choose the state yielding the least unbiased description of
the system  compatible with the available data. That state is
provided by the statistical operator  $\hat \rho_{ME} $ that
maximizes the von Neumann entropy $S \, = \, -Tr (\hat \rho \, \ln
\hat \rho)$ subject to the constraints imposed by normalization
and the expectation values $\langle \hat A_i \rangle \,= \, Tr
(\hat \rho \hat A_i) $ of the relevant observables $\hat A_i$.

Even though Jaynes' principle does provide a very satisfactory
answer in many situations \cite{B91,BAD96}, Horodecki {\it et al.}
\cite{HHH99} showed that the straightforward application of Jaynes'
prescription in its usual form is not always an appropriate strategy
for dealing with entangled states. It was shown in \cite{HHH99} that
the standard implementation of Jaynes' principle may create ``fake"
entanglement. For example, the MaxEnt density matrix may correspond
to an entangled state even if there exist separable states compatible
with the prior information. Since quantum entanglement is, in
many cases, the basic resource needed when processing quantum
information, statistical inference procedures that
overestimate the amount of available entanglement should be handled 
with care. Furthermore, it is  well-known that local operations and
classical communication (LOCC) can never increase the amount of
entanglement between remote systems, but they can make it
decrease. As a consequence, one should often bet on the
decrease of entanglement and not be very ``optimistic" when estimating
the available amount of this resource. The above considerations
suggests that, in order to deal with some situations involving
entanglement, the usual form of Jaynes' prescription needs to be
modified or supplemented in an appropriate way. Various such schemes
have been proposed. Horodecki {\it et al.} \cite{HHH99} proposed a
combined strategy based on a constrained {\bf minimization of entanglement
followed by a maximization of the von Neumann entropy}. Alternatively,
Abe and Rajagopal \cite{AR99} explored the possibility of inferring
entangled states by recourse to a variational principle based
on non-extensive information measures.

So far, the work done in connection with
Horodecki's problem of fake inferred entanglement
focused on that particular case in which the prior information
is given by the mean value of the Bell-CHSH operator
\cite{HHH99,RPPC00,R99,AR99}. In this Chapter we 
explore what happens when
the available prior information consists of the
expectation value of operators exhibiting a more general
form, such as operators diagonal and non diagonal in the Bell basis 
(\ref{Belldiagonal}), while we provide counterexamples 
to the general prescription proposed in \cite{R99} by Rajagopal for
solving the problem of fake entanglement. Finally, for bipartite
systems consisting of two qubits, and assuming that we know the expectation
value {\it b} of the most general operator diagonal in the Bell basis, we
explore the whole set of physical states ${\hat \rho}$ in the $({\it b},E({\it
b}))$-plane, where $E(\hat \rho)$ is the entanglement of formation (\ref{Ef}). 
States that possess the maximum, or the minimum, amount of
entanglement are derived explicitly.

\section{Sketch of Horodecki's and Rajagopal's treatment}

Following Horodecki {\it et al.} let us assume
that the prior (input) information is given by the expectation value
$b$ of the Bell-CHSH observable \cite{CHSH69}

\begin{equation} \label{campana}
\hat B = \sqrt{2} \, \Bigl( \sigma_x\otimes \sigma_x+\sigma_z\otimes \sigma_z \Bigr)
= 2\sqrt{2} \, \Bigl(|\Phi^+\ra\la\Phi^+|-|\Psi^-\ra\la\Psi^-| \Bigr)\,
\end{equation}

\noindent
which is defined in terms of the components of the well-known Bell basis,

\begin{eqnarray} \label{babel}
|\Phi^\pm \ra &=& \frac{1}{\sqrt{2}} \, \Bigl(| 00 \ra\pm | 11 \ra \Bigr), \cr
|\Psi^\pm \ra &=& \frac{1}{\sqrt{2}} \, \Bigl(| 10 \ra\pm | 01 \ra \Bigr).
\end{eqnarray}

\noindent The Bell observable is {\it nonlocal}. In order to
measure the Bell observable one relies upon local
operations and classical communication between the parts (that is,
LOCC operations). It can not be measured without interchange of 
classical information (e.g. a telephone call) between the 
observers \cite{HHH99}.

\noindent
The MaxEnt state obtained by recourse to the standard prescription, when
the sole available information is given by $b=\langle \hat B \rangle $,
is described by the density matrix \cite{HHH99}

\begin{eqnarray} \label{rhoj}
\hat \rho_{ME}(b) &=&
\frac{1}{4}\Bigg[ \left(1+\frac{b}{\sqrt{2}}+\frac{b^2}{8}\right)
|\Phi^+\rangle \langle \Phi^+|
+ \left(1-\frac{b}{\sqrt{2}}+\frac{b^2}{8}\right)
|\Psi^-\rangle \langle \Psi^-|   \nonumber \\
& & + \left(1-\frac{b^2}{8}\right)
\Bigl( |\Psi^+\rangle \langle \Psi^+| +
|\Phi^-\rangle \langle \Phi^-| \Bigr) \Bigg].
\end{eqnarray}

Rajagopal \cite{R99} and Abe and Rajagopal \cite{AR99} showed that the inclusion
of $\sigma^2 = \langle \hat B^2 \rangle$ within the input data set entails important
consequences for the inference of entangled states. The main idea of Rajagopal's
proposal \cite{R99} is to consider the density matrix $\hat \rho_{MS} $ obtained by considering
both mean values $b=\langle \hat B \rangle$ and  $\sigma^2=\langle \hat B^2 \rangle $
as constraints in the MaxEnt prescription, and assuming that the mean value of
$\hat B^2$ adopts the minimum value compatible with the given value of $b$. Rajagopal
proved that $\hat \rho_{MS}$ is {\it separable} if and only if $b <\sqrt{2}$. The
method employed by Rajagopal to characterize the states $\hat \rho_{MS}$ of
minimum-$\sigma^2$ rests heavily on the particular form of the Bell operator. A
different approach is needed if one wants to implement Rajagopal's inference scheme
when the input information consists of the mean value of more general observables.

The operators $\hat B$ and $\hat B^2$ verify the relations

\ben \label{b2b1}
\hat B^2 \, &=& \, 16 |\Phi^+ \rangle \langle \Phi^+ | \, - \, 2\sqrt{2} \hat B \cr
            &=& \, 16 |\Psi^- \rangle \langle \Psi^- | \, + \, 2\sqrt{2} \hat B.
\een

\noindent
It is easy to see, computing the trace of the above equations, that

\be \label{ineq1}
\sigma^2 \, \ge \, 2 \sqrt{2} \,\, |b|,
\ee

\noindent
and, consequently, the minimum value of $\sigma^2$  compatible
with a given value of $b$ is

\be \label{minimobdo}
\sigma^2 = 2 \sqrt{2} |b|.
\ee

\noindent From
  the trace of equation (\ref{b2b1}) it also transpires that  density
matrices with the minimum value of $\sigma^2$ compatible with a given value
of $b$ comply  with

\ben \label{sigmins}
\langle \Phi^+ |\hat \rho | \Phi^+ \rangle \, &=&  \, 0
\,\,\,\,\,\,\,\,\,\,\,\,\,\, ({\rm if} \,\,\, b<0) \cr
\langle \Psi^- |\hat \rho | \Psi^- \rangle \, &=&  \, 0
\,\,\,\,\,\,\,\,\,\,\,\,\,\, ({\rm if} \,\,\, b>0).
\een

\noindent
This means that a state complying with the  minimum uncertainty requirement belongs
to the three dimensional subspace spanned by the vectors
$\{|\Psi^{\pm} \rangle, |\Phi^- \rangle \}$ ($b<0$),
or by the vectors
$\{|\Psi^+ \rangle, |\Phi^{\pm} \rangle, |\Phi^- \rangle \}$ ($b>0$).
For the density matrices defined within this subspaces we have

 \ben \label{muc}
 b \, &=& \, - 2\sqrt{2} \, \langle   \Psi^- |\hat \rho| \Psi^- \rangle
 \,\,\,\,\,\,\,\,\,\,\,\,\,\,\,\,\,\, ({\rm if}\,\,\, b<0) \cr
 b \, &=& \,\,\,\,\,\, 2\sqrt{2} \, \langle   \Phi^+ |\hat \rho| \Phi^+ \rangle
 \,\,\,\,\,\,\,\,\,\,\,\,\,\,\,\,\,\, ({\rm if}\,\,\, b>0).
 \een

 \noindent
 The matrices provided by Rajagopal's scheme are

 \ben \label{rajarho}
 \hat \rho_{MS} \, &=& \, \frac{-b}{2\sqrt{2}}
 |\Psi^- \rangle \langle \Psi^- | \, + \, \frac{1}{2} \left(1+\frac{b}{2\sqrt{2}} \right)
 \left[ |\Psi^+ \rangle \langle \Psi^+ | + |\Phi^- \rangle \langle \Phi^- | \right], \cr
 \hat \rho_{MS} \, &=& \, \frac{b}{2\sqrt{2}}
 |\Phi^+ \rangle \langle \Phi^+ | \, + \, \frac{1}{2} \left(1-\frac{b}{2\sqrt{2}} \right)
 \left[ |\Psi^+ \rangle \langle \Psi^+ | + |\Phi^- \rangle \langle \Phi^- | \right],
 \een

\noindent where the first (second) state corresponds to $b<0$ ($b>0$). 
States that are diagonal in the Bell basis (\ref{babel}) are separable if and
only if they have no eigenvalue larger than $\frac{1}{2}$\footnote{As discused 
in Chapter 5, it is a characteristic feature of Bell diagonal states.}. 	  
Hence, it follows
from equation (\ref{rajarho}) that the states $\hat \rho_{MS}$ are separable if
and only if $|b| < \sqrt{2}$.

Let us now consider general minimum uncertainty states (that is, states
$\hat \rho$ verifying (\ref{minimobdo}) but not necessarily of the MaxEnt
form). Expressing the matrix elements of $\hat \rho$ in the Bell basis
(\ref{babel}), let us equate all the nondiagonal elements to zero and leave
unchanged the diagonal ones. The new density matrix $\hat \rho_D$ thus
obtained has always less entanglement than the original $\hat \rho$.
If the original $\hat \rho$ is such that $ b \, > \, \sqrt{2}$, then the
matrix $\hat \rho_D$ (which is diagonal in the Bell basis) will have one
eigenvalue greater than $1/2$ (see equation (\ref{muc})). Thus, $\hat \rho_D$ is
entangled and so is $\hat \rho$. Summing up, there is no separable density
matrix complying with the minimum-$\sigma^2$ condition (\ref{minimobdo}) and
having $b \, > \, \sqrt{2}$. This means that, for $b \, > \, \sqrt{2}$, the
inference scheme proposed by Rajagopal does not produce ``fake" inferred
entanglement. {\it At least when the input data is related to the Bell observable}
(\ref{campana}), {\it Rajagopal's prescription does not lead to an entangled inferred
state $\hat \rho_{MS}$ if there are separable states compatible with the constraints
$b$ and $\sigma^2$}. This is the main result obtained by Rajagopal \cite{R99,AR99},
although he arrived to it by recourse to a different line of reasoning.

\begin{figure}
\begin{center}
\includegraphics[angle=270,width=0.65\textwidth]{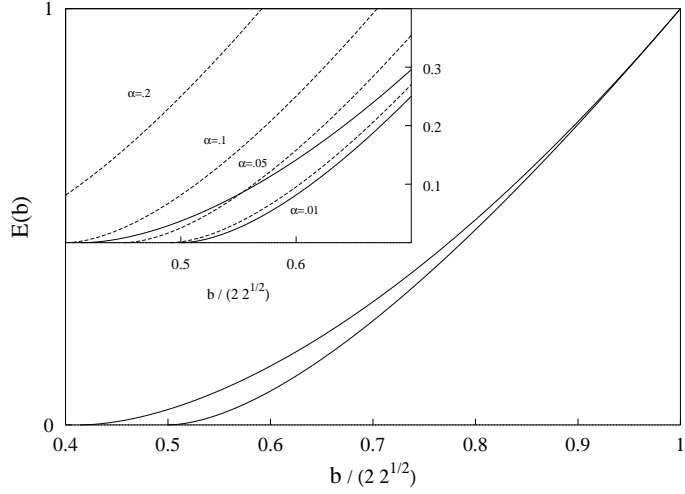}
\caption{The entanglement of formation $E[\hat \rho]$, as a
function of i) the expectation value $b$ of the
Bell operator, ii) the MaxEnt density matrix $\hat
\rho_{ME}$ (Eq. (\ref{rhoj})) (upper solid line),
and iii) the minimum-$\sigma^2$ density matrix
$\hat \rho_{MS}$ (Eq. (\ref{rajarho})) (lower
solid line). The results corresponding to the
density matrix ansatz (\ref{alfalfa}) (dashed
lines) are shown in the inset.} 
\label{fig1fake}
\end{center}
\end{figure}

Fig.\ref{fig1fake} depicts the entanglement of formation as a function of the 
input data $b$ (for $b>0$).
Two types of inferred density matrix are used to compute the
entanglement of formation, namely, (i) the density matrix $\hat
\rho_{ME}$ yielded by the standard MaxEnt procedure (upper solid
line) and (ii) the density matrix $\hat \rho_{MS}$ provided by
Rajagopal's minimum-$\sigma^2$ scheme (lower solid line).

Let us suppose that the ``true" state of the system is described by
a density matrix of the form

\ben \label{alfalfa}
\hat \rho_T(\alpha) & =	&
\left(\frac{b}{2\sqrt{2}}+\alpha \right)|\Phi^+ \rangle \langle \Phi^+ |
 +  \alpha |\Psi^- \rangle \langle \Psi^- |  \cr 
 & & + \frac{1}{2}\left(1 - \frac{b}{2\sqrt{2}} - 2\alpha \right)
\Bigl( |\Phi^- \rangle \langle \Phi^- | +
 |\Psi^+ \rangle \langle \Psi^+ | \Bigr).
\een

\noindent The (``true") density matrices belonging to the above
family are characterized by a parameter $\alpha$ and verify
$Tr(\hat \rho_T \hat B) = b$. We assume that the only knowledge we
have about $\hat \rho_T$ is given by the mean value $b$. From this
piece of data we can determine the inferred matrices $\hat
\rho_{ME}$ and $\hat \rho_{MS}$ provided, respectively, by the
standard MaxEnt and Rajagopal's prescriptions. In the inset of Fig.\ref{fig1fake} 
we can see, together  with the entanglement of formation of both
$\hat \rho_{ME}$ and $\hat \rho_{MS}$, the behaviour (as a
function of $b$) of the entanglement of formation $E[\hat
\rho_T(\alpha)]$, i.e., that of the ``true" state. The ($b,E(b)$)-plane 
depicted in Fig.\ref{fig1fake}, representing input information $b$ versus 
the inferred entanglement $E(b)$, constitutes a useful device for visualizing
the entanglement-related properties of an inference scheme. In
Fig.\ref{fig1fake} we can compare how both the standard MaxEnt scheme, and the  
one advanced by Rajagopal,  behave in the ($b,E(b)$)-plane. The
most noteworthy feature of Fig.\ref{fig1fake} is that (when the input 
information is related to the Bell observable) the results
obtained using the usual MaxEnt method do not seem to differ too
much from those obtained using Rajagopal's prescription.

\section{General non-diagonal and diagonal Bell states. 
Entanglement ``boundaries"}

Let us explore to what extent the conclusions previously reached 
are valid when the available prior information consists on the
expectation values of more general observables. Let us explore first 
what happens when observables non diagonal in
the Bell basis are considered. An interesting
example is provided by the quantum observable associated with the 
hermitian operator

\be \label{boperat}
\hat A \, = \, \kappa \Bigl(|1\rangle \langle 1| \, + \,
 |3\rangle \langle 3|\Bigr)
\, + \, \lambda |2\rangle \langle 2|,
\ee

\noindent
where $\kappa$ and $\lambda$ are real parameters such that

\ben \label{kala}
\kappa \ge 0 \ge \lambda ,
\een

\noindent
and whose eigenvectors
$|i\rangle\,\, (i=1,\ldots 4)$ are

\ben \label{basis}
|1\rangle \, &=& \, \frac{1}{\sqrt{2}} \, \Bigl( |11\rangle \, + \, |00\rangle  \Bigr), \cr
|2\rangle \, &=& \, \frac{1}{\sqrt{2}} \, \Bigl( |11\rangle \, - \, |00\rangle  \Bigr), \cr
|3\rangle \, &=& \, |01 \rangle, \cr
|4\rangle \, &=& \, |10 \rangle.
\een

\noindent {\it It is clear that $\hat A$ is non diagonal in the
Bell basis}. The observable $\hat A$ is nonlocal. 
Consequently, and as far as its nonlocality
properties are concerned, the observable $\hat A$ has the same
status as the Bell-CHSH. Suppose that we know the
expectation value $a$ of $\hat A$, given by

\be \label{valmeb}
a \, = \, Tr (\hat \rho \hat A) \, = \, \kappa
\Bigr( \langle 1 |\hat \rho | 1 \rangle + \langle 3 |\hat \rho | 3 \rangle \Bigl) \, + \,
\lambda \langle 2 |\hat \rho | 2 \rangle.
\ee

\noindent
Following the Rajagopal procedure, we 
incorporate a new constraint associated with the expectation
value of

\be \label{boperatsqu}
\hat A^2 \, = \, \kappa^2  \Bigl(|1\rangle \langle 1| \, + \, |3\rangle \langle 3|\Bigr)
\, + \, \lambda^2  |2\rangle \langle 2|,
\ee

\noindent
which is

\be \label{velmesig}
\sigma^2 \, = \, Tr (\hat \rho \hat A^2) \, = \, \kappa^2
\Bigr( \langle 1 |\hat \rho | 1 \rangle + \langle 3 |\hat \rho | 3 \rangle \Bigl) \, + \,
\lambda^2 \langle 2 |\hat \rho | 2 \rangle,
\ee

\noindent
so that the problem of fake inferred
entanglement can be solved if in order to describe
our system  we adopt a density
matrix
$\hat \rho_{MS}$ complying with two requisites.
First, $\hat \rho_{MS}$ must have the MaxEnt
form corresponding to the constraints associated with the expectation
values of both
$\hat A$ and $\hat A^2$.
Secondly, the expectation value $\sigma^2 $ must adopt
the lowest value compatible with the given value of $a$. Notice that the
mean value
$a=\langle \hat A \rangle$ is the only independent input data. For the sake of
simplicity we are going to restrict our considerations to the case of positive
values of $\langle \hat A \rangle$.

The mean values of $\hat A$ and $\hat A^2 $ are related by

\be \label{sigmabe}
\sigma^2 \,= \, \kappa a \, + \, \lambda (\lambda - \kappa) \langle 2 |\hat \rho | 2 \rangle,
\ee

\noindent
which implies that those mixed states  characterized by
exhibiting  the minimum possible
$\sigma^2$-value  compatible
with a given $a>0$ must verify $\langle 2|\hat \rho | 2 \rangle = 0$.
Consequently, for those
states with minimum $\sigma^2$ we have

\be \label{sigmamin}
\sigma^2 \, = \, \kappa a.
\ee

When we have a single constraint corresponding to the mean value of $\hat A$,
the maximum entropy density matrix is

\be \label{romeia}
\hat \rho_{ME}^I \, = \, \frac{1}{Z} \exp(-\beta \hat A),
\ee

\noindent
where $\beta$ is a Lagrange multiplier and $Z=Tr (\exp(-\beta \hat A))$.
Alternatively, $\hat \rho^I_{ME}$ can be cast as

\be \label{romeib}
\hat \rho_{ME}^I \, = \, \frac{1}{1 + 2w + w^{\lambda/\kappa}}\,
\left[ w \Bigl(|1\rangle \langle 1| \, + \, |3\rangle \langle 3|\Bigr) +
w^{\lambda/\kappa}|2\rangle \langle 2| + |4\rangle \langle 4| \right],
\ee

\noindent
where $w = \exp(-\beta \kappa)$ verifies

\be
\frac{a}{\kappa} \, = \,
\frac{2w + (\lambda/\kappa) w^{\lambda/\kappa}}{1 + 2w + w^{\lambda/\kappa}}.
\ee

\noindent
The maximum entropy statistical operator associated with
the expectation values $a$ and $\sigma^2$ as input information
is

\be \label{roMaxEnta}
\hat \rho_{ME}^{II} \, = \, \frac{1}{Z} \exp(-\beta \hat A -\gamma \hat A^2),
\ee

\noindent
where $\beta $ and $\gamma$ are appropriate Lagrange multipliers and the
partition function $Z$ is given by

\be \label{z2}
Z=Tr (\exp(-\beta \hat A -\gamma \hat A^2)).
\ee

\noindent
The matrix $\hat \rho^{II}_{ME}$ can be expressed explicitly in terms
of the input mean values $a$ and $\sigma^2$,

\ben \label{roMaxEntb}
\hat \rho_{ME}^{II} \, & = & \,
\frac{1}{2} \frac{\sigma^2 - \lambda a}{\kappa(\kappa - \lambda)} \,
\Bigl(|1\rangle \langle 1| \, + \, |3\rangle \langle 3|\Bigr)
\, + \,
\frac{\kappa a - \sigma^2}{\lambda(\kappa - \lambda)} \,
|2\rangle \langle 2| \cr
& & \, + \,
\frac{\sigma^2 - a (\kappa + \lambda) + \lambda \kappa}{\lambda \kappa} \,
|4\rangle \langle 4|.
\een

\noindent
When the further requirement of
a minimum value for $\sigma^2 $ is imposed, the above MaxEnt
density matrix reduces to

\be \label{romasigmin}
\hat \rho_{MS} \, = \,
\frac{a}{2\kappa} \,
\Bigl(|1\rangle \langle 1| \, + \, |3\rangle \langle 3|\Bigr)
\, + \, \left( 1 - \frac{a}{\kappa} \right)\,
|4\rangle \langle 4|.
\ee

\noindent
Since we always have $\kappa \ge a$, the above matrix is positive semidefinite.

Now, in order to find out whether Rajagopal's prescription is
plagued with the problem of fake inferred entanglement (when
applied in connection with the observable $\hat A$), we need to
proceed according to what follows. First, we adopt  a form for the
``true" density matrix describing the system. Second, we assume
that the only available information about the true state consists
on the expectation value of $\hat A$. From this sole piece of data
we obtain, via the inference scheme we are studying, the inferred
density matrix. Finally, we compare the entanglement properties
associated with
 the original, true density matrix with the entanglement properties
exhibited by the inferred one. In particular, we can evaluate on
both matrices an appropriate  quantitative measure of
entanglement. In what follows we assume that the true
state of the system is described by an statistical operator
belonging to the family of density matrices

\be \label{rosepar}
\hat \rho_{S} \, =
\, p |1\rangle \langle 1| \, + \, \alpha |3\rangle \langle 3|
\, + \, (1-p-\alpha ) |4\rangle \langle 4|,
\ee

\noindent
where $p$ and $\alpha $ are real positive parameters verifying

\ben \label{palfa}
0 \le  p  \le  1  \cr
0 \le \alpha \le 1-p.
\een

\noindent Notice that the ``true" density matrices (\ref{rosepar})
that we are trying to infer by recourse to different schemes are
not of the maximum entropy form, nor of the form associated with
any other statistical inference scheme. The expectation values of
$\hat A $ and $\hat A^2 $, evaluated on $\hat \rho_S$ are

\be \label{bsep}
a \, = \, p \kappa \, + \, \alpha \kappa,
\ee

\noindent
and

\be \label{sigsep}
\sigma^2 \, = \, p \kappa^2 \, + \, \alpha \kappa^2.
\ee

\noindent
Suppose we are given the expectation values $a$
and $\sigma^2 $ corresponding to a given state
belonging to the family (\ref{rosepar}) (notice
that, for this family of density matrices, the
mean values $a$ and $\sigma^2$ always verify the
minimum-$\sigma^2$ condition (\ref{sigmamin})). We can
take those mean values as input information and
generate the concomitant inferred density matrix.
That is, we can associate a MaxEnt state to each
member of (\ref{rosepar}). The performance of the
inference scheme can be studied by comparing the
entanglement properties of a member of the
parameterized family (\ref{rosepar}) with those of
the concomitant inferred state. As a first step we
are going to find out, by recourse to Peres' PPT criterion
\footnote{It is necessary and sufficient for 
$2\times 2$ and $2\times 3$ systems.}, whether there
are separable states of the form (\ref{rosepar})
leading to entangled inferred states. Applying PPT to 
the minimum-$\sigma^2$ MaxEnt density matrix $\hat
\rho_{MS}$ (Eq. \ref{romasigmin}) we find that
there is only one eigenvalue of the partial
transpose matrix that may adopt negative values.
This eigenvalue is

\be \label{negati}
\delta \, = \,
- \frac{a}{4\kappa} \, + \, \frac{1}{2} \, - \, \frac{1}{4}
\sqrt{\frac{a}{\kappa} \, \left(10 \frac{a}{\kappa} - 12 \right) + 4}.
\ee

\noindent
Hence, we have

\ben \label{peres} a/ \kappa \, &\le & 8/9 \, \Longleftrightarrow
\, \delta \ge 0,
 \cr a/ \kappa \,  &>&  8/9 \, \Longleftrightarrow \,
\delta   < 0  \een

\noindent Consequently, $\hat \rho_{MS}$ is separable if $a/\kappa
\le 8/9$ and entangled otherwise. Using the Peres' criterion we
can also determine just when the parameterized (true) density
matrix $\hat \rho_S$ is separable. For the considerations that
follow it will prove convenient to rewrite $\hat \rho_S$ in terms
of the expectation value $a= Tr(\hat \rho_S \hat A)$,

\be \label{rose} \hat \rho_S \, = \, \left(\frac{a}{\kappa} -
\alpha \right) |1\rangle \langle 1| \, + \, \alpha |3\rangle
\langle 3| \, + \, \left(1 - \frac{a}{\kappa} \right) |4\rangle
\langle 4|.
 \ee

\noindent
 It is important to stress that the above expression describes
 the same family of mixed states defined by equation
 (\ref{rosepar}). The states $\hat \rho_S$ associated with
 equation (\ref{rose}) still depend on two independent parameters,
 i.e., $\alpha$ and $a/\kappa$. Equation (\ref{rose}) is just a
 re-parameterization of the family (\ref{rosepar}) where, for the
 sake of convenience, we have chosen
 $a/\kappa=Tr(\hat \rho_S \hat A)/\kappa$
  as one of the two relevant parameters. The separability of
  $\hat \rho_S$ is determined by the quantity

\be
Q \, = \,\frac{1}{2} \, - \, \frac{a}{2\kappa} \, + \, \frac{\alpha}{2} \, - \, \frac{1}{2}
\sqrt{2 \left(\frac{a}{\kappa} \right)^2 - \frac{2a}{\kappa} + 1 -2\alpha + 2\alpha^2}.
\ee

\begin{figure}
\begin{center}
\includegraphics[angle=270,width=0.65\textwidth]{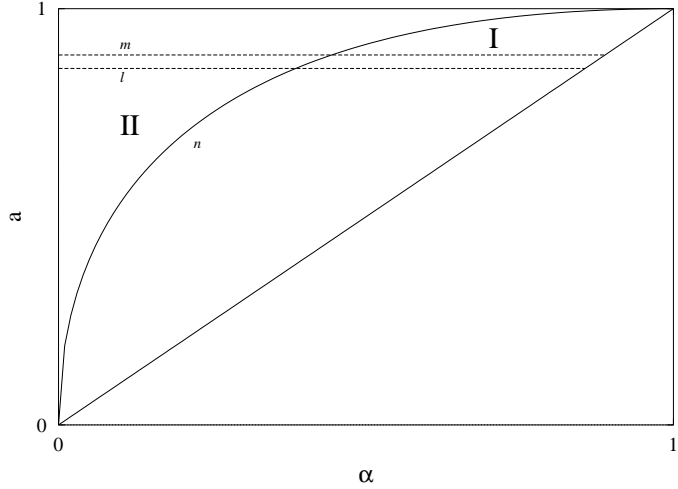}
\caption{Boundaries between the regions corresponding
to separability and entanglement for states
described by the density matrices $\hat
\rho_{ME}^{I}$ (line $l$), $\hat \rho_{MS}$ (line
$m$), and $\hat \rho_S$ (line $n$). The
expressions for the matrices $\hat
\rho_{ME}^{I}$, $\hat \rho_{MS}$, and $\hat
\rho_S$ are given, respectively, by equations
(\ref{romeib}), (\ref{romasigmin}), and
(\ref{rose}).} 
\label{fig2fake}
\end{center}
\end{figure}

\noindent The statistical operator $\hat \rho_S$ is separable if
$Q\ge 0$ and entangled otherwise. The boundaries (in the plane
$(\alpha,a)$) between the separability and the entangled regions
corresponding to (i) the density operators $\hat \rho_S$, (ii) the
standard MaxEnt statistical operators $\hat \rho_{ME}^I$, and
(iii) the minimum-$\sigma^2$ MaxEnt density matrices $\hat
\rho_{MS}$, are depicted in Fig.\ref{fig2fake}, where we take $\kappa=1$ and 
$\lambda=-1$. Notice that only those points with $\alpha <a$ are
physically meaningful, since $(\alpha,a)$ pairs not complying
with that inequality lead to a matrix $\hat \rho_S$ with one
negative eigenvalue. Fig.\ref{fig2fake} is to be interpreted as follows. 
There are three density matrices associated with each point in the
plane $(\alpha,a)$:

\begin{itemize}
\item{(i) The (``true") $\hat \rho_S$ matrix given by
the expression (\ref{rose})}.
\item{(ii) The (inferred) density matrix $\hat \rho_{ME}^I$}, of the
standard MaxEnt form (\ref{romeia}-\ref{romeib}).
\item{(iii) The (inferred) density matrix  $\hat \rho_{MS}$ of  the
minimum-$\sigma^2$ MaxEnt form
(\ref{romasigmin})}.
\end{itemize}

\noindent For all the three aforementioned density matrices
 the  expectation value of $\hat A$ is $a$,
 (that is,
$a = Tr(\hat \rho_{MS} \hat A)=Tr(\hat \rho_{ME}^I)=Tr(\hat
\rho_{S} \hat A)$). The density matrix $\hat \rho_{MS}$ is the one
yielded by Rajagopal's prescription if one tries to infer $\hat
\rho_S$ from the sole knowledge of the expectation value $a = Tr
(\hat \rho_S \hat A) $. The standard MaxEnt procedure, instead,
would lead to $\hat \rho_{ME}^I$. Using the Peres' criterium we
can determine when the inferred density matrix $\hat \rho_{ME}^I$
is entangled. For $\kappa=1$ and $\lambda=-1$ we found that $\hat
\rho_{ME}^I $ is separable when $a\le 0.8564$ and entangled
otherwise. The lines $l$ and $m$ in Fig.\ref{fig2fake} corresponds to 
$a=0.8564 $ and $a=8/9$, respectively. The curve $n$ represents
the equation $Q(a,\alpha)=0$. The density matrices $\hat
\rho_{ME}$ ($\hat \rho_{MS}$) are entangled for points
$(a,\alpha)$ lying above the line $l$ ($m$) and separable
otherwise. On the other hand, the matrices $\hat \rho_S $ are
separable when $(a,\alpha)$ lies below the curve $n$ and entangled
if $(a,\alpha)$ lies above $n$. Of particular interest are the
regions I and II. {\it In region I the (``true") density matrix to
be inferred, $\hat \rho_S$, is separable, while the associated
(``inferred") matrix $\hat \rho_{MS}$, provided by Rajagopal's
inference scheme, is not}. In region II things are quite
different: the inference scheme provides a separable statistical
operator $\hat \rho_{MS}$ while the matrix to be inferred, $\hat
\rho_S$ is entangled. It is clear that {\it the maximum entropy
minimum-$\sigma^2$ inference procedure advanced by Rajagopal 
generates fake entanglement when applied to
states $\hat \rho_S$ associated with points $(a,\alpha)$ belonging
to region I}. Contrary to previous evidence obtained when the
Bell's observable mean value is taken as the prior information, 
we must conclude that the MaxEnt
minimum-$\sigma^2$ scheme does not provide a general solution to
the problem of fake entanglement.

\begin{figure}
\begin{center}
\includegraphics[angle=270,width=0.65\textwidth]{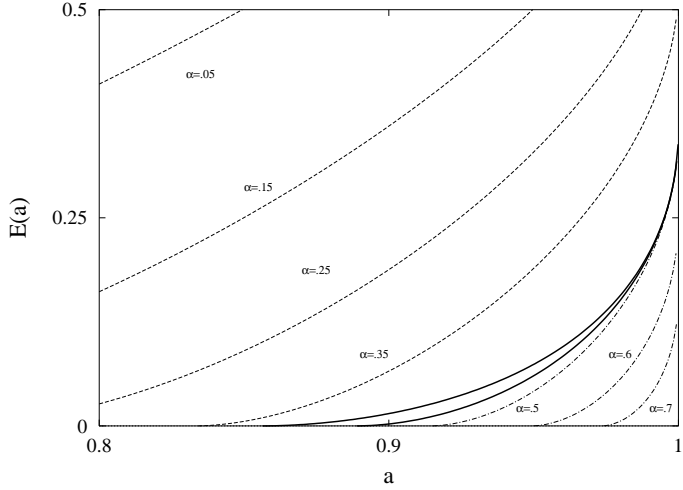}
\caption{The entanglement of formation $E[\hat \rho]$ as a
function the expectation value of the observable $\hat A$ (Eq.
(\ref{boperat})) with $\kappa =1$ and $\lambda=-1$, corresponding
to $\hat \rho_{ME}^{I}$ (upper solid line), to $\hat \rho_{MS}$
(lower solid line), and to $\hat \rho_S$, for the values of
$\alpha$ indicated in the figure (dashed and dot-dashed lines).
The expressions for the matrices $\hat \rho_{ME}^{I}$, $\hat
\rho_{MS}$, and $\hat \rho_S$ are given, respectively, by
equations (\ref{romeib}), (\ref{romasigmin}), and (\ref{rose}).} 
\label{fig3fake}
\end{center}
\end{figure}

 The comparison of the amount of entanglement of formation exhibited by
the states
 $\hat \rho_S$ and $\hat \rho_{MS}$ enables us to study the problem of fake
inferred
 entanglement in a quantitative way. The curves depicted in Fig.\ref{fig3fake} 
display the
 behaviour of $E[\hat \rho_{MS}]$ and $E[\hat \rho_S]$ as a function of
the mean value
 $a$ of the observable $\hat A$ (again, with $\kappa=1$ and $\lambda=-1$).
The upper solid line
corresponds to $E[\hat \rho_{ME}^{I}]$, the lower solid line to $E[\hat
\rho_{MS}]$,
and the dashed and dot-dashed lines to $E[\hat \rho_S]$, for different values of the
parameter $\alpha$.
The results exhibited in							
Fig.\ref{fig3fake} illustrate how, for each given value of the input data $a=Tr(\hat \rho
\hat A)$, the
entanglement of formation $E$ of
 the density operators yielded by both the standard MaxEnt method
($\hat \rho_{ME}^{I}$) and Rajagopal's scheme ($\hat \rho_{MS}$) compare
with the  entanglement
of formation
of the state to be inferred ($\hat \rho_{S}$). It is clear from Fig.\ref{fig3fake} 
that, with regards to
the behaviour of the inferred amount of entanglement as a function of the
input information
(at least when this input data consists of $\langle \hat A \rangle$),
the prescription advanced by Rajagopal does not appreciably
differ  from the
standard MaxEnt result.
In particular, both prescriptions tend to yield the same results
in the limit
$a\rightarrow 1$. We have shown a more detailed account of the extensions to general 
observables in the Bell basis in \cite{JPhysA1}.
\newline
\newline
{\bf Bell diagonal operators and states}
\newline

Let us now slightly generalize the observable $\hat B$ (\ref{campana}), 
but first let us recall	a basic property of the Bell basis \cite{HHH99}. 
For a given 
density matrix $\hat \rho$, let us consider the statistical operator $\hat
\rho^{(Diag)}$ having (in the Bell basis) the same diagonal elements as $\hat
\rho$, and all the non diagonal elements equal to zero. That is, $\hat
\rho_{ii}=\hat \rho^{(Diag)}_{ii}, \,\,(i=1,\ldots,4)$, and
$\rho^{(Diag)}_{ij}=0,\,\,(i\ne j)$. Then, it can be shown that  $ E(\hat
\rho^{(Diag)}) \le E(\hat \rho)$. As a consequence, when seeking
the state of minimum entanglement compatible with a constraint consisting on
the expectation value of an observable diagonal in the Bell basis, we can
restrict our search to the family of states diagonal in that basis. Notice that
this situation does not necessarily hold when dealing with an observable
diagonal in a basis different from Bell's.
 We legitimately focus  attention then on states of
 the general form \newline $ \hat
\rho_{E}=(p_{1},p_{2},p_{3},p_{4});\, 0 \le p_{i} \le 1; \, \sum_{i=1}^{4}
p_{i}=1.$ 

Consider the $\lambda$-family, with $\lambda \in {\cal R}^+$, 
$\hat D =(2 \sqrt{2},-\lambda, \lambda,-2 \sqrt{2})$: 
\ben \label{obs2} \hat D &=& 2 \sqrt{2} \vert 1 \ra \la 1 \vert-
\lambda \vert 2 \ra \la 2 \vert + \lambda \vert 3 \ra \la 3 \vert - 2\sqrt{2}\,
\vert 4\ra \la 4 \vert \cr &=& ( \sqrt{2}+
\frac{\lambda}{2})\sigma_x\otimes\sigma_x + ( \sqrt{2}- \frac{\lambda}{2} )
\sigma_z\otimes\sigma_z.\een This observable is easily seen to violate the
usual  Bell's inequalities \cite{mann}.	As we know, for states $\hat \rho$ diagonal
in the Bell basis the entanglement of formation $E(\hat \rho)$ depends only on
the largest eigenvalue $\lambda_m$ of $\hat \rho$. Furthermore, $E(\hat
\rho)$ is (for $\lambda_m>\frac{1}{2}$) an increasing function of $\lambda_m$. 
Consequently, to
minimize $E(\hat \rho)$ is tantamount to minimize the largest eigenvalue of
$\hat \rho$. Given the prior information $d = \la \hat D \ra$, we
proceed now to infer two states: i) $\hat \rho_{ME}$ and ii) $\hat \rho_{I}$. 

\begin{figure}
\begin{center}
\includegraphics[angle=0,width=0.5\textwidth,clip=true]{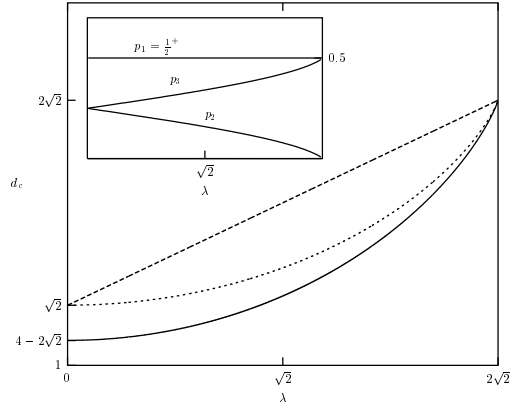}
\caption{The quantity $d_c$ (see text) vs. $\lambda$ (that parameterizes the
family of Bell operators (\ref{obs2})). The long dashed line corresponds to the
minimum possible amount of entanglement ($d_c^{MM}$). The short dashed curve
corresponds to the inference technique here advanced ($d_c^{I}$), the solid one
to Jaynes's MaxEnt approach ($d_c^{ME}$). The two inference schemes are seen to
produce ``fake"  inferred entanglement. Inset: The eigenvalues of $\rho_I$
versus $\lambda$ for $\langle \hat D \rangle = d_c$. Note that $p_1= 1/2^{+}
\equiv 1/2 + \epsilon$.} 
\label{fig1PRA}
\end{center}
\end{figure}

 We note that in order to apply here Horodecki's scheme, a numerical minimization of
entanglement is mandatory. We wish to ascertain just how far we can proceed
with mere ``algebraic" considerations. According to the operating constraints
we have, $p_1-p_4=1/(2\sqrt{2})[d - \lambda (p_3-p_2)]$, instead of
$p_1-p_4=d/(2\sqrt{2})$ as one has for (\ref{campana}). 
Moreover, the inset of Fig.\ref{fig1PRA} shows that i) $p_3>p_2$, and  
ii) $p_1$ is the largest 
eigenvalue\footnote{These facts are evident in the case $\lambda << 1$, 
although they are also true for $\lambda\le 2 \sqrt{2}$.}. 
The idea is then to regard $p_{1}, p_{4}$ as independent variables, 
that together with the constraints give

\ben
p_3\, &=& \, \{d+\lambda-(2\sqrt{2}+\lambda)p_1+(2\sqrt{2}-\lambda)p_4\}/2\lambda , \cr
p_2\, &=& \, \{\lambda-d+(2\sqrt{2}-\lambda)p_1-(2\sqrt{2}+\lambda)p_4\}/2\lambda.
\een

\noindent In view of the above facts i) and ii), it is obvious that one
minimizes entanglement by letting $p_4 \rightarrow 0$. Finally, we maximize the
von Neumann entropy, which uniquely fixes the weight $p_{1}$, and
 yields the desired, final state $\hat \rho_{I}$.
    The present treatment can be regarded as a new inference
procedure, reminiscent\footnote{At first sight it looks almost identical, since 
the same line of reasoning is employed.} of that of Horodecki's,
although it does not coincide with it. It is introduced to show just how 
insidious the problem of overestimating the 
amount of entanglement problem really is.

 For each value of $\lambda$ we have, of course, a different Bell observable and
  confront three (in principle, different) entanglement values, namely,
 {\bf (i)} {\it minimum minimorum} $E_{MM}$,
 {\bf (ii)} the one corresponding to $\hat \rho_{ME}$, namely, $E_{ME}$, and,
  {\bf (iii)} the one corresponding to $\hat \rho_I$, that is, $E_I$,
 which will yield three monotonous $\lambda$-curves when these amounts of
entanglement are plotted as a function of $d$. Let us call $d_{critical} \equiv
d_c$ those particular $d$-values for which $E=0$. Fig.\ref{fig1PRA} depicts, as a function 
of $\lambda$, the three values $d_c^{MM,\,ME,\,I}$. It is easy to see that
$d_c^{MM}= \frac{2\sqrt{2} + \lambda}{2}.$ Clearly, at $\lambda=0$ we have
$d_c^{MM}=d_c^{I}$. The three curves coincide at $\lambda= 2\sqrt{2}$. At
intermediate $\lambda$-values, both $\hat \rho_{ME}$ and $\hat \rho_{I}$ are
afflicted by the problem of overestimation of entanglement. (The inset depicts
$p_1,\,p_2,\,p_3\,\,versus\,\, \lambda$ for $d=d_c$, where $p_1$ is slightly
larger that $1/2$, by an infinitesimal amount $\epsilon$.)  
\newline
\newline
{\bf Entanglement boundaries}
\newline

One would like to know, given some piece of information (constraint), just
which state is the one with the minimum
 possible amount of entanglement. We will study now the whole set of states
 compatible with one a priori known mean value in a search for this desideratum.
  We restrict our consideration to operators diagonal in the
Bell basis. Let ${\hat B} = (B_{1},B_{2},B_{3},B_{4})$ be a general operator 
describing some system's observable, and $\hat \rho=(p_{1},p_{2},p_{3},p_{4})$ a
normalized state with $b=Tr(\hat \rho \hat B)=
p_{1}B_{1}+p_{2}B_{2}+p_{3}B_{3}+p_{4}B_{4}$. Although in this Chapter 
we shall describe only non-degenerate ${\hat B}$, the most general case 
is considered in \cite{PRA}.

For the sake of simplicity we assume the eigenvalues of ${\hat B}$ ordered 
according to 
$B_{1}<B_{2}<B_{3}<B_{4}$, without loss of generality. Let us regard $p_{1}$
and $p_{2}$ as the``true"  unknown weights. Thus, the positivity and
normalization of $\hat \rho$ clearly determine the remaining weights. One
writes then the final state in the following fashion $\hat \rho =
(p_{1},p_{2},p_3,p_4 )$, with 

 \ben
 p_3\,&=&\,\{B_{4}-b+(B_{1}-B_{4})p_{1}+(B_{2}-B_{4})p_{2}/[B_{4}-B_{3}] \}, \cr
 p_4\,&=&\,\{b-B_{3}-(B_{1}-B_{4})p_{1}-(B_{2}-B_{4})p_{2}/[B_{4}-B_{3}] \}.
 \een

\noindent Forcing {\it both} $p_{1}$ and $p_{2}$ to vanish, the remaining weights
$p_{3}$ and $p_{4}$ adopt minimum or maximum values, depending on the value of
$b$. This  defines for the interval $[B_{3},B_{4}]$ low and high entanglement-degree
 states. Repeating this procedure for the remaining 
 $\frac{N(N-1)}{2}=6$ different ways of	selecting two independent weights 
 (out $N=4$), we obtain a whole family of high and low degree of entanglement states.

\begin{figure}
\begin{center}
\includegraphics[angle=270,width=0.8\textwidth,clip=true]{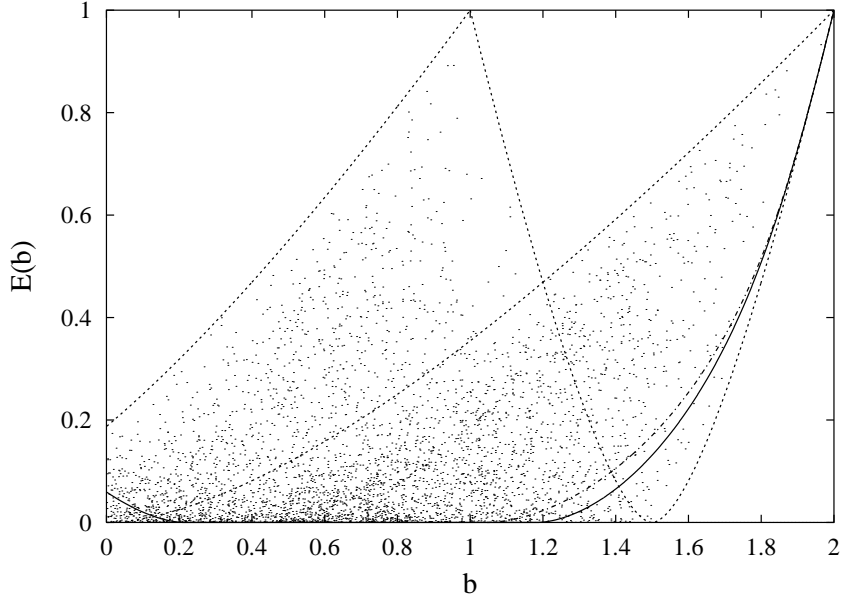}
\caption{States $\hat \rho_{\lambda, \kappa}$ defining the entanglement
boundaries are plotted (dashed lines), together with the back-ground of
accessible states. See text for details and Ref. \cite{PRA}.} 
\label{fig2PRA}
\end{center}
\end{figure}

To be more precise, these states are described by the following family

\begin{equation} \label{concu}
\hat \rho_{\lambda, \kappa} = (\frac{b-\kappa}{\lambda-\kappa},0,
\frac{\lambda-b}{\lambda-\kappa},0)
\end{equation}

\noindent where both $\kappa$ and $\lambda$  ($\kappa < \lambda$) belong to the set
$\{B_{i}\}$. The positions of the non-zero weights should ensure that the
condition $b=Tr(\hat \rho_{\lambda, \kappa} \hat B)$ is fulfilled. We are now
in a condition to ascertain which is the minimum amount of entanglement for a
given observable $\hat B$ or, conversely, which states define the curve of
least possible entanglement in the $({\it b},E({\it b}))$-plane. By evaluation
of the concurrence \cite{WO98} of (\ref{concu}), it is easy to show that the
state $\hat \rho_{\lambda=B_{2}, \kappa=B_{1}}$ defines the least entangled
state for constraints ranging from $b=B_{1}$ up to $b=\frac{B_{1}+B_{2}}{2}$,
where the entanglement of formation keeps vanishing till we reach, for our
constraint, the value $b=\frac{B_{3}+B_{4}}{2}$. Afterwards, the state $\hat
\rho_{\lambda=B_{4}, \kappa=B_{3}}$ is the one that determines the curve of
minimum possible entanglement,  until we reach for $b$ the value $b=B_{4}$.
These features are illustrated in Fig.\ref{fig2PRA} for $\hat B=(2,-2,1,-1)$. 
A reasonable sample of all possible physical states is plotted (randomly) as background.
The above determined boundaries  are also drawn (dashed lines), together with
the amounts of entanglement corresponding to both ${\hat \rho_{ME}}$
(dot-dashed line) and ${\hat \rho_{E}}$ (solid line).

\section{Concluding remarks}

In this Chapter we have exhaustively investigated Horodecki's  
``fake"  inferred entanglement problem, related with the use 
of the maximum entropy principle,
with reference to distinct inference schemes, and advanced a 
new one, reminiscent of Horodecki's, for dealing with more general 
observables than the Bell-CHSH one (\ref{campana}).

  There is no doubt that Jaynes'  MaxEnt principle has to
play an important role in any appropriate scheme
for the inference of entangled quantum states.
Indeed, one of the most remarkable features of
Jaynes' principle is its robustness: usually, when
it seems to fail, the real problem is not the
inadequacy of the MaxEnt principle itself, but
rather that some piece of relevant (prior)
information is not being taken into account. In point of fact 
it was pointed out in \cite{BCS00} that the various 
inference schemes advanced to solve the fake inferred
entanglement problem admit of an interpretation
within the strictures of Jaynes' approach. These
inference prescriptions may be regarded as
implementations of the MaxEnt principle in which
some extra prior information (that may not
consists just of the expectation values of some
observables) is assumed to be known. This is
certainly the case with Rajagopal's MaxEnt
minimum-$\sigma^2$ proposal, which assumes extra
information related to the square of the relevant
observable. However, the results reported here
show that this approach works only in very special
situations.

Besides enabling us to asses the usefulness of the minimum-$\sigma^2$
scheme, we shed some light on the 
entanglement features exhibited by the standard MaxEnt principle
within contexts more general than those previously considered in
the literature \cite{HHH99,RPPC00,R99,AR99}.

Summing up, we conclude that:

\begin{itemize}


\item For arbitrary operators of the type (\ref{boperat}), Rajagopal's 
$\sigma^2$-scheme does not provide a satisfactory solution as in the 
Bell-CHSH case.

\item For a quite general {\it Bell diagonal} observable
 ${\hat B} = (B_{1},B_{2},B_{3},B_{4})$,
  we have studied {\it all} the normalized states $\hat \rho$ (pure and mixed)
such that $Tr [\hat \rho \hat B]= b$ and established just which is the one with
minimum (maximum) amount of entanglement. We have drawn the ``entanglement
boundaries" in the $(b, E(b))$-plane. Once these states have been found, 
the problem of ``fake" inferred entanglement is immediately solved.

\end{itemize}

\chapter{Detection of entanglement at work: hierarchy of 
separability criteria. Volume occupied by the set of unentangled states 
according to different criteria} 

The development of criteria for entanglement and separability is one aspect of
the current research efforts in quantum information theory that is receiving,
and certainly deserves, considerable attention \cite{T02}, and so we devote 
the present Chapter to its study. Indeed, much
progress has recently been made in consolidating such a cornerstone of the
theory of quantum entanglement \cite{T02}. The relevant state-space here is of
a high dimensionality, already 15 dimensions in the simplest instance of
two-qubit systems. The systematic exploration of these spaces can provide us
with valuable insight into some of the theoretical questions extant.

As a matter of fact, important steps have been recently made towards a 
systematic exploration of the space of arbitrary (pure or mixed) states of 
composite quantum systems \cite{ZHS98,Z99,Z01} in order to determine the typical 
features exhibited by these states with regards to the phenomenon of quantum
entanglement \cite{ZHS98,Z99,Z01,MJWK01,IH00,BCPP02a,BCPP02b}.

It is well known that, for a  composite quantum system, a state
described by the density matrix $\rho$ is called ``entangled" if it can not be
represented as a mixture of factorizable pure states. Otherwise, the state is
called separable. The separability question has interesting  echoes in 
information theory 
and its associate information measures or entropies, as adressed in Chapter 4. 
We know that the early motivation for the studies reported in
  \cite{VW02,HHH96,HH96,CA97,V99,TLB01,TLP01,A02} was
  the development of practical separability criteria for density matrices. 
  However, the discovery by Peres of the partial transpose criteria, which for
  two-qubits and qubit-qutrit systems turned out to be both necessary
  and sufficient, rendered that original motivation somewhat outmoded. A crucial 
  fact, it is not possible to find a necessary and sufficient
   criterion for separability based
  solely upon the eigenvalue spectra of the three density matrices
  $\rho_{AB}, \rho_A=Tr_B[\rho_{AB}]$, and $\rho_B=Tr_A[\rho_{AB}]$
  associated with a composite system $A\oplus B$ \cite{NK01}.

  Interesting concepts that revolve around the separability issue have been
developed over the years. A comprehensive account is given in Terhal in \cite{T02}.
Among them we find criteria like the so-called Majorization, Reduction and
Positive Partial Transpose (PPT) (all of them described in Chapter 4), together 
with the concept of distillability. Certainly, quantum entanglement is a 
fundamental aspect of quantum physics that deserves to be investigated in full 
detail from all possible points of view. The chain of implications,
\newline

\begin{center}
${\it \rho\,\, separable \rightarrow PPT \rightarrow reduction 
\rightarrow majorization \rightarrow q-entropic}$
\newline
\end{center}

\noindent and the related inclusion relation, among the different separability 
criteria is certainly a vantage point worth of detailed scrutiny \cite{inclus}. 
   
   It is our purpose here to revisit, with such a goal
   in mind, the separability question by means of an exhaustive Monte Carlo
exploration involving the whole space of pure and mixed states. Such an effort
should shed some light on the inclusion issues that interest us here.
 Concrete numerical evidence will thus be provided on the relations among the
separability criteria. We will then be able to quantify, for a bipartite system
of arbitrary dimension, the proportion (or volume) of states $\rho$ that can be
distilled according to a definite criterion. 
\newline
\newline

Although the complete description of the separability criteria was already given 
in Chapter 4, let us briefly sketch the mathematics of these criteria. 
From a historic viewpoint, the first separability criterion is that
of Bell (see Chapter 1 in the Introduction). For every pure entangled state
there is a Bell inequality that is violated. It is not known, however, whether
in the case of  many entangled mixed states, violations exist. There does exist
a witness for every entangled state though \cite{Horodeckis1996}. It was shown
by Horodecki {\it et al.} that a density matrix $\rho \equiv \rho_{AB}$ is
entangled if and only if there exists an entanglement witness
(a hermitian superoperator\footnote{A {\it superoperator} acts on 
density matrices as ordinary operators act on pure states.} 
$\hat W=\hat W^{\dagger}$) such that

\ben \label{witxx} Tr\, \hat W\,\rho &\le& 0, \,\,\,\, {\rm while} \nn \\
 Tr\, \hat W\,\rho_s &\ge& 0,\,\,{\rm for \,\,all\,\,separable\,\,states\,\,\rho_s}.  \een

Also, an important LOCC operational separability criterion,
necessary but not sufficient, is provided by the positive partial transpose
(PPT) one. Let $T$ stand for matrix transposition. The PPT requires that  \be
\label{Wx}  [\hat 1 \otimes \hat T](\rho) \ge 0. \ee

Another operational criterion is {\it reduction}, that
is satisfied, for a given state $\rho \equiv \rho_{AB}$, when both \cite{T02}

\ben  \hat 1 \otimes \rho_B -\rho &\ge&  0\nn \\  
\rho_A \otimes \hat 1-\rho &\ge& 0  .\een

As we know, the distillable entanglement is the maximum asymptotic yield of
singleton states that can be obtained, via LOCC, from a given mixed state.
It was shown in \cite{Horo97} that any entangled mixed state
of two qubits can be distilled to obtain the singleton. However, there are 
entangled mixed states (in higher dimensions) that cannot be distilled, 
so that they are useless for quantum communication. In our scenario an 
important fact is that all states that violate
the reduction criterion are distillable \cite{Horo99}.

Majorization criterion compares the spectra of two matrices in a special way. 
Let $\{\lambda_i\}$ be the set of eigenvalues of the matrix $\xi_1$
 and $\{\gamma_i\}$ be the set of eigenvalues of the matrix
 $\xi_2$. We assert that the ordered set of eigenvalues  
$\vec \lambda$ of $\xi_1$ {\it majorizes} the ordered set of 
eigenvalues  $\vec \gamma$ of $\xi_2$
(and writes $\vec \lambda \succ \vec \gamma$) when
$\sum_{i=1}^k \,\lambda_i \,\ge\, \sum_{i=1}^k \,\gamma_i$
for all $k$. It has been shown \cite{NK01} that, for all separable
 states $\rho_{AB} \equiv \rho$,

 \ben \vec \lambda_{\rho_A} &\succ&   \vec \lambda_{\rho}, \,\,
{\rm and}
 \nn
 \\
  \vec \lambda_{\rho_B} &\succ&   \vec \lambda_{\rho}.    \een
There is an intimate relation between this majorization criterion and entropic
inequalities, as discussed in \cite{T02,VW02}.

We ommit the description of the entropic criteria \cite{batle}, which will constitute the 
subject of main study in the next Chapter.

\section{Separability probabilities: exploring the whole state space}

We shall perform a systematic numerical  survey of
the properties of arbitrary (pure and mixed) states of a given quantum system
by recourse to an exhaustive exploration of the concomitant state-space ${\cal
S}$. To such an end it is necessary to introduce an appropriate measure $\mu $
on this space. Such a measure is needed to compute volumes within ${\cal S}$,
as well as to determine what is to be understood by a uniform distribution of
states on ${\cal S}$.  The natural measure that we are going to adopt here is
taken from the work of Zyczkowski {\it et al.} \cite{ZHS98,Z99}.
An arbitrary (pure or mixed) state $\rho$ of a quantum system
described by an $N$-dimensional Hilbert space can always be
expressed as the product of three matrices,

\be \label{udot} \rho \, = \, U D[\{\lambda_i\}] U^{\dagger}. \ee

\noindent Here $U$ is an $N\times N$ unitary matrix and
$D[\{\lambda_i\}]$ is an $N\times N$ diagonal matrix whose
diagonal elements are $\{\lambda_1, \ldots, \lambda_N \}$, with $0
\le \lambda_i \le 1$, and $\sum_i \lambda_i = 1$.
The group of unitary matrices $U(N)$ is
endowed with a unique, uniform measure: the Haar measure $\nu$
\cite{PZK98}. On the other hand, the $N$-simplex $\Delta$,
consisting of all the real $N$-uples $\{\lambda_1, \ldots,
\lambda_N \}$ appearing in (\ref{udot}), is a subset of a
$(N-1)$-dimensional hyperplane of ${\cal R}^N$. Consequently, the
standard normalized Lebesgue measure ${\cal L}_{N-1}$ on ${\cal
R}^{N-1}$ provides a natural measure for $\Delta$. The
aforementioned measures on $U(N)$ and $\Delta$ lead then to a
natural measure $\mu $ on the set ${\cal S}$ of all the states of
our quantum system \cite{ZHS98,Z99,PZK98}, namely,

\be \label{memu1}
 \mu = \nu \times {\cal L}_{N-1}.
 \ee

 \noindent
  All our present considerations are based on the assumption
 that the uniform distribution of states of a quantum system
 is the one determined by the measure (\ref{memu1}). Thus, in our
 numerical computations\footnote{The quantities $\mu_i$ computed with a Monte 
 Carlo procedure have an associated error which is on the type 
 $t_{M-1,\alpha/2}\frac{\sigma_x}{\sqrt{M-1}}$, where M is the number 
 of generated states, $t_{M-1,\alpha/2}$ is the value corresponding to 
 the Student distribution with M-1 degrees of freedom, computed with a 
 certain desired accuracy $1-\alpha$, and $\sigma_x$ is the usual computed 
 standard deviation. Therefore, if we seek a result with an error say less than 
 $10^{-3}$ units, we have to generate a number of points M around 
 10 or 100 million. If not stated explicitly, from now on all quantities computed 
 are exact up to the last digit.} we are going to randomly generate
 states according to the measure (\ref{memu1}). In the forthcoming 
 Chapters we will be dealing several times with this measure, which 
 is fully described in Appendix B.

\section{Survey's results}

{\bf The overall scenario}
\newline

\begin{figure}
\begin{center}
\includegraphics[angle=0,width=0.65\textwidth]{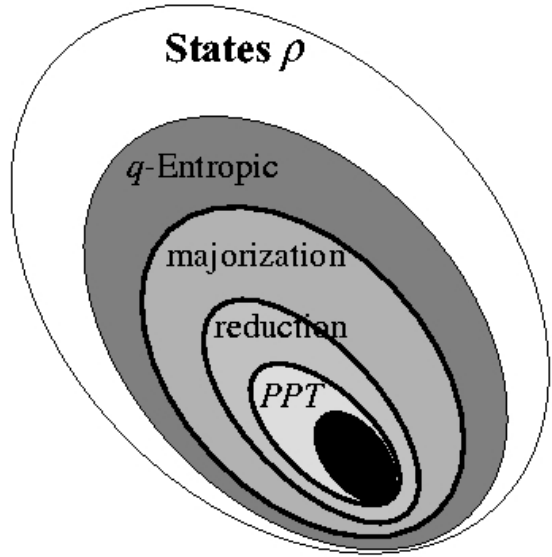}
\caption{Schematics of the inclusion relations among separability
criteria as given by the volume occupied by states $\rho$ for a given dimension
$N$ which fulfil them.} 
\label{fig1inclus}
\end{center}
\end{figure}

An overall picture of the situation we encounter is sketched in Fig.\ref{fig1inclus}.  
The set of all mixed states presents an onion-like shape, as conjectured by 
Terhal \cite{T02}. Which among these states are separable? As reviewed above, 
several criteria are available. We start with the $q$-entropic one. 
By using a definite value of $q$, namely $q = \infty$, and the sign of the
 associated, conditional $q$-entropy, we are able to define a 
 closed sub-region, whose states are supposedly separable. This region has a
definite border, that separates it from the sub-region of states entangled
according to this criterion. What we see now is that, if we use now {\it other}
separability-criteria, the associated sub-regions shrink in a manner prescribed
by the particular  criterion one employs. The shrinking process ends when one
reaches the sub-region defined by the Positive Partial Transpose (PPT)
criterion, which is a necessary and sufficient separability condition for $2
\times 2$ and $2 \times 3$ systems, being only necessary for higher dimensions.

Summing up, the volume of states which are separable according to different
criteria diminish as we use stronger and stronger criteria. There is a first
shrinking stage associated to entropic criteria, from its Von Neumann ($q=1$)
size, as $q$ grows,  to the limit case $q\rightarrow \infty$ \cite{batle}. A
second stage involves majorization, reduction, and, finally positive partial
transpose (PPT) \cite{T02}.
\newline
\newline
{\bf PPT and Reduction}
\newline

\begin{figure}
\begin{center}
\includegraphics[angle=270,width=.65\textwidth]{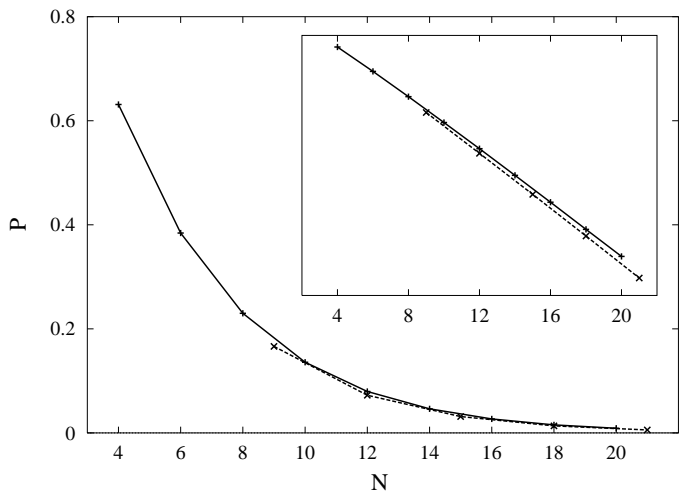}
\caption{Probability of finding a state with 
positive partial
transpose. The solid line corresponds to states with dimension $N=2 \times
N_2$, while the dashed line corresponds to $N=3 \times N_2$ states.  The
difference between these curves can be appreciated in the inset 
(semi-logarithmic plot). Our probabilities  decrease, to a good approximation,
in exponential fashion.} 
\label{fig2inclus}
\end{center}
\end{figure}

We report now on our state-space exploration with regards to the probability of
finding a state with positive partial transpose. The results are depicted in
Fig.\ref{fig2inclus}. The solid line corresponds to states with dimension $N=2 \times N_2$,   
while the dashed line corresponds to $N=3 \times N_2$ states. Note how similar
are the pertinent  values in both cases. The tiny difference between them can
be inspected in the inset (a semi-logarithmic plot). To a good approximation,
our PPT probabilities decrease exponentially. There exist lower bounds to the volume 
of separable states. In \cite{Guifre99}, G. Vidal and R. Tarrach show that the 
probability of finding a separable state in a $n$-party system with $N$-dimensional 
Hilbert space is

\be
P_{sep}\,\geq \, \bigg( \frac{1}{1\,+\,\frac{N}{2}} \bigg)^{(n-1)(N-1)}
\ee

\noindent which is clearly nonzero for finite systems.

\begin{figure}
\begin{center}
\includegraphics[angle=270,width=.65\textwidth]{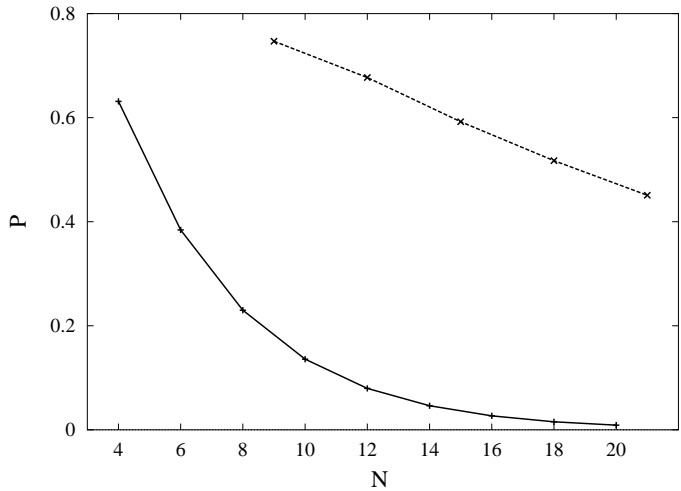}
\caption{Probability of finding a state fulfilling the reduction
criterion for $N=2 \times N_2$ (solid line) and $N=3 \times N_2$ (dashed line). 
PPT and reduction coincide for $N=2 \times N_2$ systems.} 
\label{fig3inclus}
\end{center}
\end{figure}

Fig.\ref{fig3inclus} deals instead with the probability of finding a state which obeys the    
strictures of the reduction criterion, for $N=2 \times N_2$ (solid line) and
$N=3 \times N_2$ (dashed line). As a matter of fact, PPT and reduction coincide
for $N=2 \times N_2$. It is known that if a state satisfies PPT, it
automatically verifies the reduction criterion \cite{T02}. We have
demonstrated that, at least in the  $N=2 \times N_2$-instance, the converse is
also true \cite{inclus}. However, in the $N=3 \times N_2$-case, it is much more likely to
encounter a state that verifies reduction than one that verifies PPT.
\newline
\newline
{\bf Entropic criteria and Majorization}
\newline

We begin with a brief recapitulation of former $q$-entropic results.
 The situation encountered in \cite{JPhysA2} was that the ``best" result
within the framework of the ``classical $q$-entropic inequalities" as a
separability criterion was reached using the limit case $q\rightarrow \infty$,
but considerably less attention was paid to other values of $q$. This was
remedied in \cite{batle}, where the question of $q$-entropic inequalities for
finite $q$-values was extensively discussed. It was there re-confirmed that the
above mentioned limit case does indeed the better job as far as separability
questions are concerned \cite{batle}. For such a reason, this limit $q$-value
is the only one to be employed below. 

\begin{figure}
\begin{center}
\includegraphics[angle=270,width=.65\textwidth]{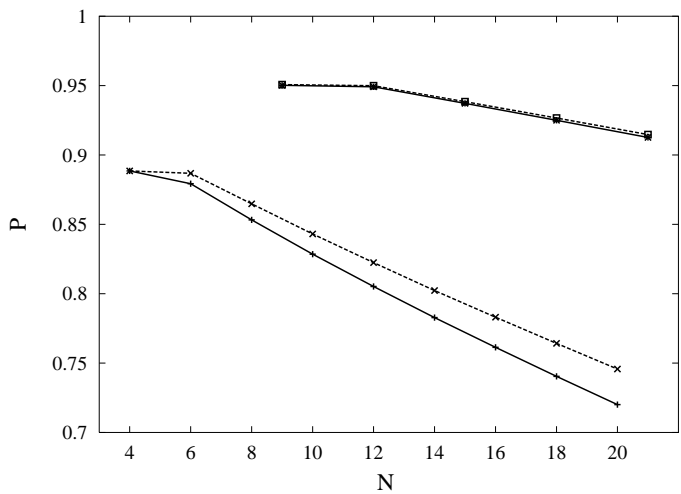}
\caption{Probability of finding a state whose two relative
$q$-entropies are positive for $q \rightarrow \infty$ (dashed curves). The
probability that a state be completely majorized by both of their subsystems is
represented by the solid line. Bottom: curves correspond to states $\rho$ with
$N=2 \times N_2$. Top: $N=3 \times N_2$.} 
\label{fig4inclus}
\end{center}
\end{figure}										 

In Fig.\ref{fig4inclus} we depict the probability of finding a state which, for $q              
\rightarrow \infty$, has its two conditional $q$-entropies positive (dashed
curves). In view of the intimate relation of entropic inequalities with
majorization \cite{T02,VW02}, we also analyze in Fig.\ref{fig4inclus} the probability that a  
state is completely majorized by both of their subsystems (solid line).
 It is shown in \cite{VW02} that, if $\rho_{AB}$ satisfies the reduction 
criterion, its two associated conditional $q$- entropies are non-negative as well.

 In the same work the authors assert that majorization is not implied by the 
 conditional entropy criteria. Our results confirm this assessment. In Fig.\ref{fig4inclus},      
 the lower curves correspond to states $\rho$ with $N=2 \times N_2$,
while the upper curves have $N=3 \times N_2$. Majorization results and
$q$-entropic do coincide for two-qubits systems ($N_1=N_2=2$). More generally,
majorization probabilities are a lower bound to probabilities for conditional
$q$-entropic positivity, an interesting new result, as far as we know. Notice
also that the two approaches yield quite similar results in the 
$N=3 \times N_2$ case.
\newline
\newline
{\bf Comparing more than two criteria together}
\newline

\begin{figure}
\hspace{2cm}
\includegraphics[angle=270,width=.4\textwidth]{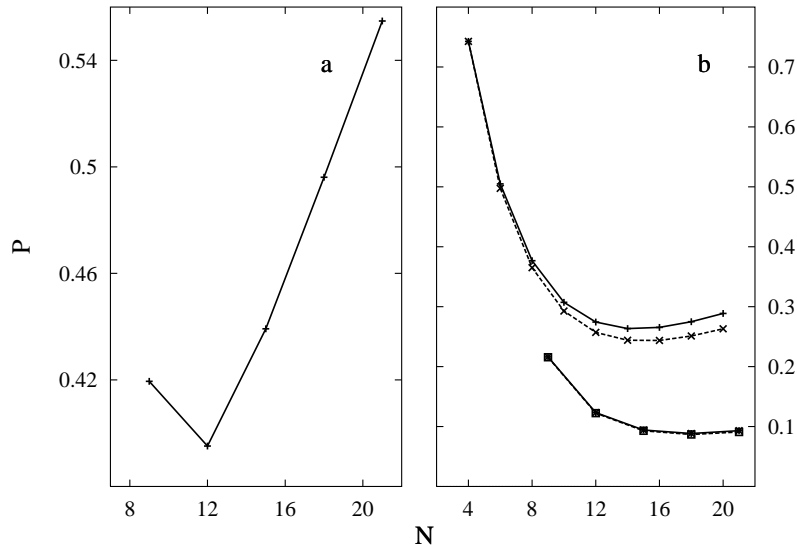}
\caption{a) Probability that the state $\rho$ with $N=3 \times N_2$ 
either has i) a positive partial transpose {\it and} does not violate the 
reduction criterion, or ii) has a non positive partial transpose {\it and} 
violates reduction. In the case $N=2 \times N_2$ the outcome is always unity. 
b) Probability that i) PPT and majorization
(solid line) and, ii)  PPT and the $q$-entropic criterion (dashed line) 
lead to the same conclusion regarding separability. Top: $N=2 \times N_2$. 
Bottom: $N=3 \times N_2$.} 
\label{fig5inclus}
\end{figure}

We compare now the reduction criterion to the PPT one. The former is implied by
the latter but is nonetheless a significant condition since its violation
implies the possibility of recovering entanglement by distillation, which is as
yet unclear for states that violate PPT \cite{VW02}. Fig.\ref{fig5inclus} a) depicts the           
probability that state $\rho$ with $N=3 \times N_2$ either:
\begin{enumerate}
\item has a positive partial transpose and does not violate the reduction
criterion, or \item has a non positive partial transpose and violates
reduction.
\end{enumerate}
Remember that in the case $N=2 \times N_2$, the two criteria always coincide
\cite{T02}. For  $3 \times N_2$ the agreement between the two criteria becomes
better and better as $N_2$ augments.

Of more interest is to compare the relations among PPT, majorization, and the
entropic criteria (Fig.\ref{fig5inclus}b), since it is not yet known how the majorization         
criterion is related to other separability criteria like PPT, undistillability,
and reduction \cite{VW02}. In this vein, Fig.\ref{fig5inclus} b) plots the                        
``coincidence-probability" between, respectively,
\begin{enumerate}
\item  PPT and majorization (solid line), and
\item PPT and
the $q$-entropic criterion (dashed line).
\end{enumerate}
 The  curves on the top correspond to  $N=2 \times N_2$, 
while those at the bottom to
  $N=3 \times N_2$. In this last case the two curves agree with each other
  quite well.

  The conclusion here is that, as $N_2$ augments, the probability of coincidence
among the three criteria, and in particular  between majorization
  and PPT (our main concern),
   rapidly diminishes at first, and stabilizes
  itself afterwards. For two qubits the three criteria do agree with each other
  to a large extent.

\begin{figure}
\hspace{2cm}
\includegraphics[angle=270,width=.4\textwidth]{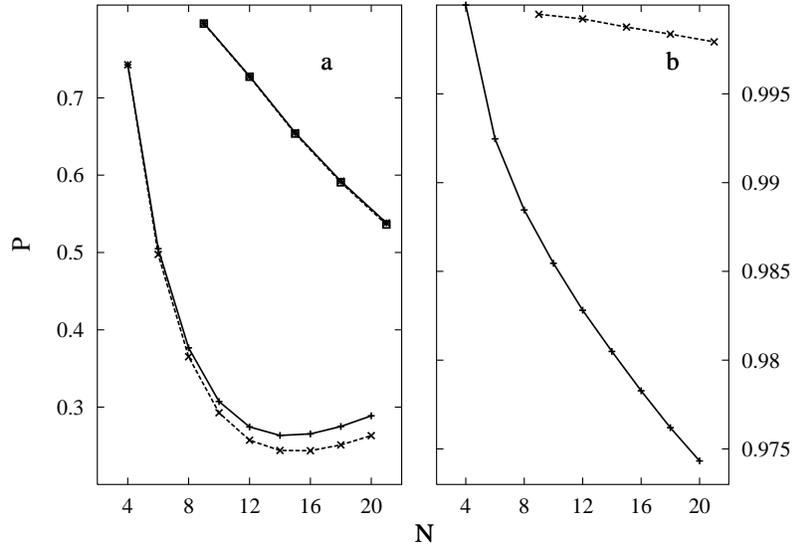}
\caption{a) Probability that
reduction and majorization (solid line) and reduction and the $q$-entropic
criterion (dashed line) yield the same conclusion reagrding separability. Top:
$N=3 \times N_2$. Bottom:  $N=2 \times N_2$ (lower curves). 
b) Probability that a state, for $q \rightarrow \infty$, either i) has
both positive relative $q$-entropies and fulfils majorization, or ii) has a
negative relative $q$-entropy and is majorized by both of their subsystems. The
solid line corresponds to the case $N=2 \times N_2$, while the dashed line
corresponds to $N=3 \times N_2$.} 
\label{fig6inclus}
\end{figure}

Fig.\ref{fig6inclus} a) depicts the probability that, for a given  state $\rho$,                    
\begin {enumerate}
\item reduction and majorization (solid line) and
\item reduction and the $q$-entropic criterion (dashed line)
\end{enumerate}
yield the same conclusion as regards separability. Without PPT in the game, and         
opposite to what we encountered in Fig.\ref{fig5inclus}, we find better coincidence for $N=3        
\times N_2$ systems (top) than for $N=2 \times N_2$ (bottom). The deterioration
of the degree of agreement as $N_2$ grows is similar to that of Fig.\ref{fig5inclus}, though.         

Fig.\ref{fig6inclus}b) represents the probability that a state, for $q \rightarrow \infty$,          
either:
\begin{enumerate}
\item has both positive conditional $q$-entropies and satisfies the majorization 
criterion, or
\item has a negative conditional $q$-entropy and is majorized by both of their
subsystems.
\end{enumerate}
The solid line corresponds to the case $N=2 \times N_2$,  while the dashed
lines corresponds to the $N=3 \times N_2$ instance. These results together with
those of  Figs.\ref{fig4inclus}-\ref{fig5inclus} could be read as implying that 
majorization and the $q$-entropic criteria provide almost the same answer for 
dimensions greater or equal than $N=3 \times N_2$.

\begin{figure}
\begin{center}
\includegraphics[angle=270,width=.65\textwidth]{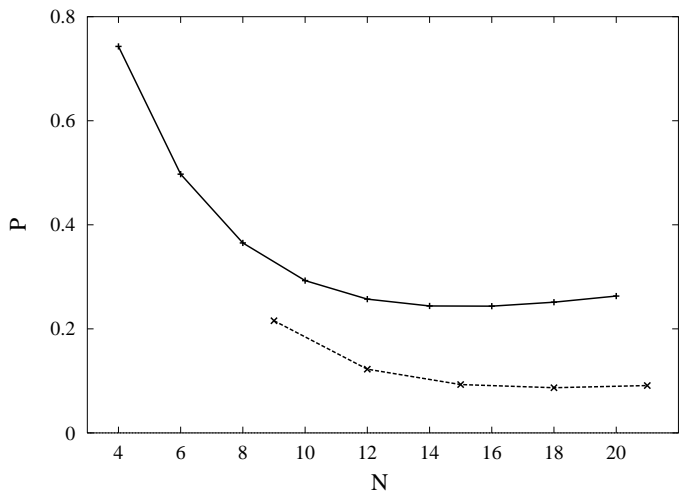}
\caption{Total probability that all criteria considered in the present
work lead to  the same conclusion regarding separability. Probabilities are
plotted as a function of the total dimension $N=N_1 \times N_2$, with $N_1=2$
(solid line) and $N_1=3$ (dashed line).} 
\label{fig7inclus}
\end{center}
\end{figure}	

Finally, in Fig.\ref{fig7inclus} we look for the  probability $P_{agree}$ that all criteria           
considered in the present work do lead to the same conclusion on the
separability issue. $P_{agree}$ is plotted  as a function of the total
dimension $N=N_1 \times N_2$, with $N_1=2$ (solid line) and $N_1=3$ (dashed
line). The agreement is quite good for two qubits, deteriorates first as $N_2$
grows, and rapidly stabilizes itself around a value of 0.26 for $N_1=2$ and of
0.1 for $N_1=3$.
\newline
\newline
{\bf Distilling}
\newline

\begin{figure}
\begin{center}
\includegraphics[angle=270,width=.65\textwidth]{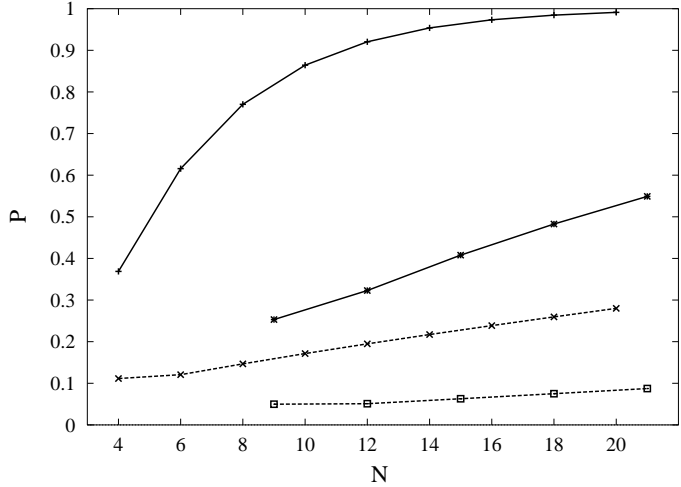}
\caption{Solid line: probability that a state  violates the reduction
criterion. Dashed line: the same for violation of the majorization criterion.
Top: $N=2 \times N_2$. Bottom:  $N=3 \times N_2$. The dashed curve with
crosses represents the case $N=2 \times N_2$, while the one with squares
indicates the $N=3 \times N_2$ instance.} 
\label{fig8inclus}
\end{center}
\end{figure}

Let us at now consider the results plotted in Fig.\ref{fig8inclus}. We ask first 
for the relative number of states that violate the reduction criterion  and are thus
distillable \cite{Horo97} (solid line), and appreciate the fact that, as $N$
grows, so does the probability of finding distillable states.
 On the other hand, the probability of encountering states that violate 
the majorization criterion, represented by dashed lines, is much lower than 
that associated to distillation.

For both criteria, the upper solid line corresponds to the case $N=2 \times
N_2$, and the lower one to $N=3 \times N_2$. The dashed curve with crosses
represents the case $N=2 \times N_2$, while the one with squares indicates the
$N=3 \times N_2$ instance. The dependence with $N_2$ of the dashed curves
(majorization violation) is not so strong as that of the solid ones
(distillability). Our results are lower bounds to the total volume of states 
that can be distilled.

\section{Concluding remarks}

In this Chapter, we have explored the application of 
different separability criteria by recourse to an exhaustive Monte Carlo
exploration involving the pertinent state-space of pure and mixed states. The
corresponding  chain of implications of different criteria is in such a way
numerically elucidated. We have also quantified, for a bipartite system of arbitrary
dimension, the proportion of states $\rho$ that can be  distilled according to
a definite criterion. 

To be more precise, we have performed a systematic numerical 
survey of the space of pure and mixed states of bipartite systems of dimension 
$2 \times N_2$ and $3 \times N_2 $ in
order to investigate the relationships ensuing among different separability
criteria. Our main results are

\begin{itemize}
\item Regarding the line of separability implication, our graph in Fig.\ref{fig1inclus}            
constitutes a confirmation of some propositions in Terhal's work \cite{T02}
\item It is known that if a state satisfies PPT, it
automatically verifies the reduction criterion \cite{T02}. In the present work
we show that in the  $N=2 \times N_2$-instance, the converse is also true. In
the $N=3 \times N_2$-case, it is much more likely to encounter a state that
verifies reduction than one that verifies PPT.
\item We have numerically verified the assertion made in \cite{VW02} that
majorization is not implied by the conditional entropic criteria. Majorization
results and $q$-entropic criteria coincide for two-qubits systems. In general,
majorization probabilities constitutes lower bounds for conditional $q$-entropic
positivity.
\item Regarding the relation between majorization and PPT, the agreement between
the criteria deteriorates as $N_2$ grows.
\item For dimensions  $\ge 3 \times N_2$, as illustrated by Figs.\ref{fig4inclus}-\ref{fig5inclus}, 
majorization and the $q$-entropic criteria provide almost the same answers.
\item Lower bounds to the total volume of states that can be distilled are found.
\end{itemize}

We believe that the results of this numerical exploration shed
some light on the intricacies of the separability issue. 
Indeed, the size of the	volume of separable states would reflect the 
fact important for numerical analysis of entanglement, to
what extent the	separable or entangled states are typical. 
But, according to \cite{AB01}, also there appeared 
a technical motivation: the considerations on volume of separable
states lead of important results concerning the question of relevance 
of entanglement	in quantum computing \cite{nmr}, specifically NMR 
quantum computing. 
These considerations proved to be crucial for analysis of the experimental
implementation of quantum algorithms in high-$T$ systems via NMR methods. 
This is because a generic state used in this approach is the maximally mixed
one with a small admixture of some pure entangled state. 
In \cite{nmr} the sufficient conditions of the above sort were further
developed and it was concluded,	that in all the NMR quantum computing 
experiments performed to date the admixture of the pure state was too small. 
Thus the total state used in these experiments was separable.


\chapter{Conditional $q$-entropies and quantum separability}

Some entangled states of quantum composite systems (in particular, all pure
entangled states) exhibit the notable property of having an entropy smaller
than the entropies of their subsystems. This feature of composite quantum
systems, and its connections with other of their entanglement-related
properties, has been recently investigated by several authors
\cite{T02,VW02,HHH96,HH96,CA97,V99,TLB01,TLP01,A02,AT02,TPA03}. The phenomenon
of entanglement is one of the most fundamental and non-classical features
exhibited by quantum systems \cite{Schro,LPS98}. Quantum entanglement is the
basic resource required to implement several of the most important processes
studied by quantum information theory
\cite{Galindo,NC00,LPS98,WC97,W98,AB01}, such as quantum teleportation
\cite{BBCJPW93}, superdense coding \cite{BW93} and the exciting issue of
quantum computation \cite{NC00}. Due to 
the significance of quantum entanglement, it is important to survey the state
space of composite quantum systems, in order to get a clear picture of the
concomitant entanglement properties, and of the relationships between
entanglement and other relevant features exhibited by the quantum states.
Significant advances have been made  by a program that attempts performing a
systematic exploration of the space of arbitrary (pure or mixed) states of
composite quantum systems \cite{ZHS98,Z99,Z01} in order to determine the
characteristic features shown by these states with regards to the phenomenon of
quantum entanglement \cite{ZHS98,Z99,Z01,MJWK01,IH00,BCPP02a,BCPP02b,batle}.

Separable quantum states share with classical composite systems the following
basic property: the entropy of any of its subsystems is always equal or smaller
than the entropy characterizing the whole system \cite{NK01}. In contrast, as
already mentioned, a subsystem of a quantum system described by an entangled
state may have an entropy greater than the entropy of the whole system, thus
violating the concomitant classical entropic inequalities. This situation holds
for the well known von Neumann entropy, as well as for the more general
$q$-entropic (or $q$-information) measures
\cite{T02,VW02,HHH96,HH96,CA97,V99,TLB01,TLP01,A02,AT02,TPA03}, which
incorporate both R\'enyi's \cite{BS93} and Tsallis' \cite{T88,LV98,LSP01}
families of measures as special instances. These entropic functionals are
characterized by a real parameter $q$.

The alluded to classical entropic inequalities constitute necessary and
sufficient separability criteria for pure states. The situation is, however,
more involved in the case of mixed states. In the latter case we can find
entangled states that do not violate these inequalities. Consequently, the
classical entropic inequalities provide only necessary separability criteria.
As a matter of fact, the main motivation for studying the classical entropic
inequalities (and their violation by some entangled states) is not any more the
development of practical separability criteria. This is the case particularly
since the introduction of the Positive Partial Transposition (PPT) criterion by
Peres \cite{Peres}, and the related results obtained by the Horodecki´s
\cite{Horodeckis1996}. However, {\it the violation of the classical entropic
inequalities is interesting in its own right, because they constitute, from the
perspective of classical physics, a highly counterintuitive property exhibited
by some entangled quantum states}. Moreover, this non-classical feature of
certain entangled states is of a clear and direct information-theoretical
nature.

 The goal of the present Chapter is to investigate further aspects of the
relationship between quantum separability and the violation of the classical
$q$-entropic inequalities.

\section{Features of conditional $q-$entropies of composite quantum systems}

By performing a systematic numerical survey of the
space of pure and mixed states of bipartite systems of any dimension we are about to 
determine, for different values of the entropic parameter $q$, the volume in
state space occupied by those states characterized by positive values of the
conditional $q$-entropies. We pay particular attention to the monotonic
tendency shown by these separability ratios as they evolve with $q$ from finite
values to the limiting case $q\rightarrow \infty$, for any Hilbert space's
dimension. 

\subsection{$q$-Conditional entropies}

 As we have shown in Chapter 4, the ``$q$-entropies" depend upon the 
 eigenvalues $p_i$ of the density matrix $\rho$ of a quantum system through the 
 quantity $\omega_q = \sum_i p_i^q$. More explicitly, we shall consider either 
 the R\'enyi entropies \cite{BS93},

  \be 
   S^{(R)}_q \, = \, \frac{1}{1-q} \, \ln \left( \omega_q \right),
  \ee

\noindent
  or the Tsallis' entropies \cite{T88,LV98,LSP01}

  \be 
  S^{(T)}_q \, = \, \frac{1}{q-1}\bigl(1-\omega_q \bigr),
  \ee
\noindent
 which have found many applications in many different fields of Physics. 
 In the  $q=2-$case, $ S_{q=2}$ is often called {\it the linear 
 entropy} $\mathcal{S}_L$  \cite{MJWK01}. These entropic measures incorporate the 
 important (because of its relationship 
 with the standard thermodynamic entropy) instance of the von Neumann measure,
as a particular  limit  ($q\rightarrow 1$) situation

  \be 
  S_1 \, = \,- \, Tr \left( \hat \rho \ln \hat \rho \right).
  \ee

\noindent

 Tsallis' and R\'enyi's measures are
  related through $
  S^{(T)}_q \, = \,F( S^{(R)}_q),
  $
  \noindent
  where the function $F$ is given by $
  F(x)  =   \left\{ e^{(1-q)x} - 1 \right\}/(1-q).
  $
  \noindent
  As an immediate consequence, for all non vanishing values of $q$, Tsallis' measure
  $ S^{(T)}_q$ is a monotonic increasing function of R\'enyi's
  measure $ S^{(R)}_q $. We will be here rather  more interested in
${\it conditional \, q-entropies}$ than in  total entropies, because of the
former's relation with the issue of quantum separability. Conditional entropic
measures are defined as

 \be \label{qurelaxxx}
  S^{(T)}_q(A|B) \, = \,
  \frac{S^{(T)}_q(\rho_{AB})-S^{(T)}_q(\rho_B)}{1+(1-q)S^{(T)}_q(\rho_B)}
  \ee

\noindent for the Tsallis case, while its R\'{e}nyi counterpart is

\be \label{relarenyi}
  S^{(R)}_q(A|B) \, = \, S^{(R)}_q(\rho_{AB})-S^{(R)}_q(\rho_{B}).
  \ee
  \noindent
  Notice that the denominator in (\ref{qurelaxxx}),
  $
  1+(1-q)S_q \, = \, w_q \, > \, 0
  $
  \noindent
  is always positive. Consequently, as far as the sign of the
  conditional entropy is concerned, the denominator in (\ref{qurelaxxx})
  can be ignored. Now, since Tsallis' entropy is a monotonous
  increasing function of R\'enyi's,
  it is plain that (\ref{qurelaxxx}) has always the same sign as (\ref{relarenyi}). 
  The matrix $\rho_{AB}$ denotes an arbitrary quantum state of the
  composite system $A\otimes B$, not necessarily factorizable nor separable,
  and $\rho_B = Tr_A (\rho_{AB})$ (the conditional $q$-entropy
  $S^{(T)}_q(B|A)$ is defined in a similar way as (\ref{qurelaxxx}),
  replacing $\rho_B $ by $\rho_A = Tr_B (\rho_{AB})$).
     Interest in the conditional $q$-entropy
  (\ref{qurelaxxx}) arises  in view of  its relevance with regards to
   the separability of density matrices describing composite
  quantum systems \cite{TLB01,TLP01}. For separable states,
  we have \cite{VW02}

  \ben \label{qsepar}
  S^{(T)}_q(A|B) &\ge & 0, \cr
  S^{(T)}_q(B|A) &\ge & 0.
  \een

  \noindent
  As already mentioned, there are entangled states (for instance, all
entangled pure states) characterized by negative conditional $q$-entropies.
That is, for some entangled states one (or both) of the inequalities
(\ref{qsepar}) are not verified. Since just the sign of the conditional entropy
is important here, we can either use Tsallis' or R\'{e}nyi's entropy, for
(\ref{qurelaxxx}) and (\ref{relarenyi}) will always share the same sign. In what
follows,  when we  speak of  the positivity of either Tsallis' conditional
entropy (\ref{qurelaxxx}) or of R\'enyi's conditional entropy (\ref{relarenyi}),
we will make reference  to  the ``classical $q$-entropic
 inequalities" issue.

\subsection{Volumes in state space sccupied by states of special entropic
properties.}


The systematic numerical study of pure and mixed states of a bipartite quantum
system of arbitrary dimension $N=N_1 \times N_2$ requires the introduction of
an appropriate measure $\mu $ defined over the corresponding space ${\cal S}$
of general quantum states. Such a measure is necessary in order to compute
volumes within the space ${\cal S}$. The measure we are going to adopt in the
present approach ($\mu = \nu \times {\cal L}_{N-1}$) was introduced by Zyczkowski 
{\it et al.} in several valuable
contributions \cite{ZHS98,Z99}, and is the same employed in Chapter 7 and throughout 
the present one. All our present considerations are based on the assumption
 that the uniform distribution of states of a quantum system
 is the one determined by measure $\mu$ (\ref{memu1}). Thus, in our
 numerical computations we are going to randomly generate
 states according to measure (\ref{memu1}).
 The situation that we shall encounter in next section is the following
 one: the volume in phase space corresponding to those states complying with the
classical $q$-entropic inequalities monotonically  decreases  as the entropic
parameter $q$ increases, adopting its minimum value in the limit case
$q\rightarrow \infty$. In this limit case, the volume  of states with positive
conditional entropies adopts simultaneously: i) its lowest value and also ii)
the one most closely resembling that of the set of states with positive partial
transpose (PPT). The  volume of states with positive conditional $q$-entropies
is, however, even in the limit case $q\rightarrow \infty$, larger than the
volume associated with states with a positive partial transpose. This means
that, regarded as a separability criterion, the classical entropic inequality
with $q=\infty$ is (among the conditional $q$-entropic criteria) the strongest
one, though it is not as strong as the PPT criterion (see Fig.\ref{fig1inclus}). 
In point of fact, it has
been proven that there is no necessary and sufficient criteria for quantum
separability based solely on the eigenvalues of $\rho_{AB}$, $\rho_{A}$, and
$\rho_{B}$. Our main concern here is {\it not} the study of the classical
inequalities {\it qua} separability criteria. Their study is interesting {\it
per se} because it provides us with additional insight into the issue of
quantum separability, on account of their intuitive information-theoretical
nature. We want to survey the state-space in order to obtain a picture, as
detailed as possible, of i) how the signs of the $q$-conditional entropies are
correlated with {\it other} entanglement-related features of  quantum states,
and ii) how these correlations depend both on the value of $q$  and on the
dimensionality of the systems under consideration.

As reported in \cite{JPhysA2}, the volume occupied by states with
positive values of the conditional $q$-entropies decreases with
$q$ in a monotonous fashion as the entropic parameter grows from
finite $q$-values to $q=\infty$. It is to be remarked that some
authors had previously conjectured \cite{TLB01} that the
conditional $q$-entropy $S_q(A|B)[\rho]$, evaluated in each
particular density matrix $\rho$, is a monotonous decreasing
function of $q$. This conjecture implies that it should be enough
to consider the value $q\rightarrow \infty$ in order to decide on
the positivity of the conditional $q$-entropies for all $q$. If
this conjecture were true it would lead, as an immediate
consequence, to the monotonous behaviour (as a function of $q$) of
the volume of states with positive values of the conditional
$q$-entropies.

Alas, one can find several low-rank counterexamples to the monotonicity of the
conditional Tsallis or R\'enyi entropies with $q$ (a particularly interesting
case of non-monotonicity with $q$ of Tsallis' conditional entropies has been
recently discussed by Tsallis, Prato, and Anteneodo in \cite{TPA03}).  A rather
surprising situation ensues:  the volumes associated with positive valued
conditional $q$-entropies behave in a monotonous way {\it in spite of the fact
that the alluded to conjecture is not valid}. One of the aims of the present
effort is precisely to investigate this point in more detail. By recourse to a
Monte Carlo calculation we have determined numerically (both for two-qubits and
qubit-qutrit systems) the proportion of states which behave monotonously as $q$
changes. This involves exploring either the $15$-dimensional space of
two-qubits ($N=4$) or the $35$-dimensional space of one qubit-one qutrit mixed
states. Table 8.1 shows the results for different ranks, dimensions, and
entropies used for the mixed state $\rho$. In each case (that is, for each set
of values for $q$, total Hilbert Space dimension $N=N_1\times N_2$, and rank of
$\rho$) we have randomly generated $10^7$ density matrices. This implies that
the relative numerical error associated with the values reported in Table 8.1 is
less than $10^{-3}$. We consider it remarkable that most of the states have a
conditional entropy that behaves monotonically with $q$, this fact being more
pronounced for the case of the Tsallis entropy. The proportion of these states
diminishes as the rank of the state $\rho$ decreases, regardless of the
dimension and the conditional entropy used. The general trend suggested by
Table 8.1 is that the percentage of states with monotonous conditional
$q$-entropies increases with the total (Hilbert space's) dimension of the
system and, for a given total dimension, increases with the rank of the density
operator. This is fully consistent with the monotonic behaviour (as a function
of $q$) exhibited by the total volume corresponding to states with positive
conditional $q$-entropies.

\begin {table}[tbp]
\centering
\begin {tabular}{|c|c|c|}
       &  Tsallis  & R\'{e}nyi \\
\hline
$2 \times 2. Rank ,\ 4$  & 0.972  & 0.719   \\
$Rank ,\ 3$ & 0.850  & 0.434   \\
$Rank ,\ 2$ & 0.204 & 0.003  \\
$2 \times 3. Rank ,\ 6$  & 0.996  & 0.888   \\
$Rank ,\ 5$ & 0.99  & 0.79   \\
$Rank ,\ 4$ & 0.96 & 0.64  \\
$Rank ,\ 3$ & 0.84 & 0.38  \\
$Rank ,\ 2$ & 0.32 & 0.003  \\

\end{tabular}
\caption{Proportion of states which behave monotonously as $q$
changes.  Both Tsallis' and R\'{e}nyi's conditional entropies, for
two-qubits and one qubit-one qutrit systems, are considered. For a
given dimensionality one is to notice how the system evolves with
the rank of the pertinent state $\rho$.}
\end{table}

\begin{figure}
\begin{center}
\includegraphics[angle=270,width=.75\textwidth]{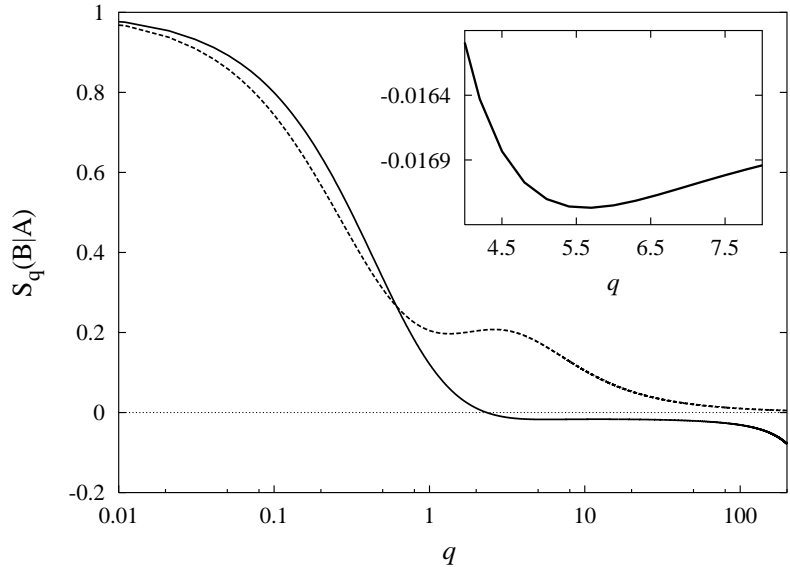}
\caption{Conditional Tsallis entropy $S_q(B|A)$ for two sample states
$\rho$ of a two-qubits system (with rank 4) which do not change in monotonous
fashion when $q$ grows. The dashed line corresponds to a state whose
conditional entropy remains positive for all $q$-values. The solid line
corresponds to a state whose conditional entropy eventually becomes negative
(and, consequently, the state becomes entangled) for large values of $q$. The
inset depicts, for the last case, details of the rather tiny region where
monotonicity is broken. All quantities depicted are dimensionless.} 
\label{fig1Some}
\end{center}
\end{figure}

 Examples of non-monotonous behaviour of the conditional $q$-entropy are
depicted in Fig.\ref{fig1Some}, for a pair of two-qubits states of range four. 
The dashed line corresponds to a state whose conditional entropy, although non-monotonous,
remains always positive. The continuous line refers to an entangled state such
that $S^{(T)}_q(A|B) < 0$ for large enough $q$-values. The $q$-interval in
which the monotonicity of the last state is broken is depicted in the inset.
One gathers form these results that it seems correct to regard $q\rightarrow
\infty$ as the right value to ascertain positivity for a single given state
$\rho$, as was recently suggested by Abe \cite{A02} on the basis of his
analysis of a mono-parametric family of mixed states for multi-qudit systems.

\begin{figure}
\begin{center}
\includegraphics[angle=270,width=.75\textwidth]{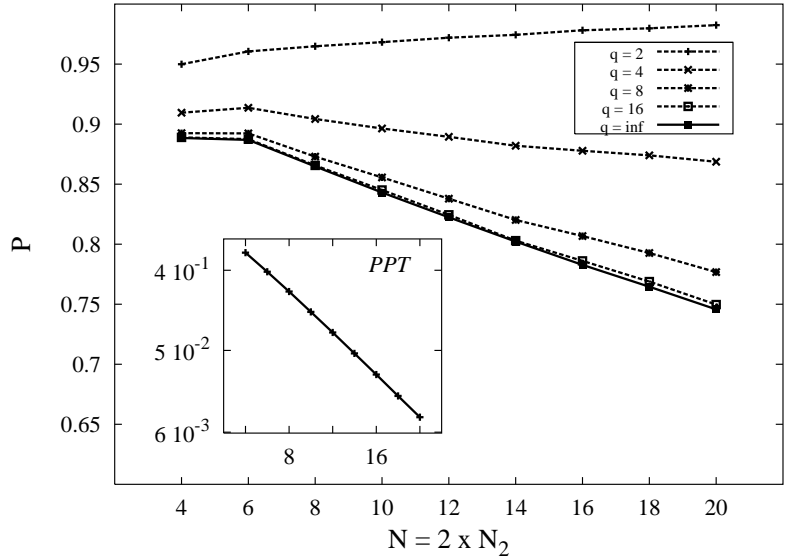}
\caption{Probability of finding a state $\rho$ for systems in $2
\times N_2$ dimensions which, for different values of $q$, has its two
conditional $q$-entropies positive. Different curves are assigned  to various
values of $q$. These curves ``saturate" when the limit $q \rightarrow \infty$
is reached. Also, two regimes of growth with the dimension are to be noticed.
See text for details. The inset depicts the concomitant probabilities for 
PPT. The lines are just a guide for the eye. All quantities
depicted are dimensionless.} 
\label{fig2Some}
\end{center}
\end{figure}

\begin{figure}
\begin{center}
\includegraphics[angle=0,width=0.65\textwidth,clip=true]{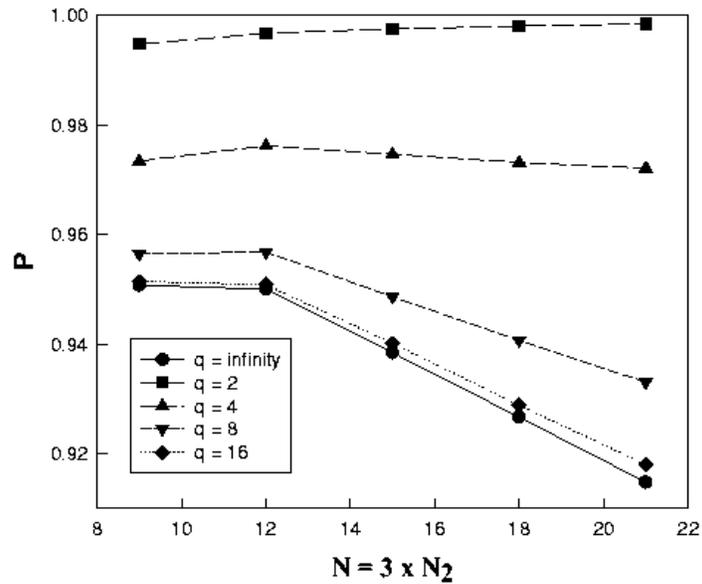}
\caption{Same as in Fig.\ref{fig2Some} for systems of $3 \times N_2$
dimensions. Values of probabilities are higher and the rate of
saturation is different. All quantities depicted are
dimensionless.} 
\label{fig3Some}
\end{center}
\end{figure}

 To further explore the issue of monotonicity we have computed the
fraction of the total state space volume occupied by (that is, the
probabilities of finding) states with positive conditional $q$-entropies (for
both (i) different finite values of $q$ and (ii) $q=\infty$), in the case of
bipartite quantum systems described by Hilbert spaces of increasing
dimensionality \cite{batle}. Let i) $N_1$ and $N_2$ stand for the dimension of the Hilbert
space associated with each subsystem, and ii) $N=N_1\times N_2 $ be the
dimension of the Hilbert space associated with the concomitant composite
system. We have considered two sets of systems: (1) systems with $N_1=2$ and
increasing values of $N_2$, and (2) systems with $N_1=3$ and increasing
dimensionality. The computed probabilities for the first set of systems are
depicted in Fig.\ref{fig2Some}, as a function of the total dimension $N$. 
The case of the second set is depicted in Fig.\ref{fig3Some}. In order to 
obtain each point in Figs. \ref{fig2Some} and \ref{fig3Some} (as well as 
to obtain each of the points appearing in the subsequent
Figures in this section) we have randomly generated $10^7$ density matrices.
This leads to Monte Carlo results with a relative, numerical error less than
$10^{-3}$. In Fig.\ref{fig2Some} one plots different values of the probabilities 
associated with positive conditional $q$-entropy for (a) $q=2,4,8,16,$ 
and $\infty$ and (b) different values of the total dimension $N$ of the system.

 With respect to the behaviour of these probabilities, one is to focus attention
upon two aspects: i) evolution with $q$ for a given $N$ and  ii) evolution with
the dimension for fixed $q$. In the first instance one clearly sees a common
behaviour for all $N$. As $q$ increases, the probabilities of finding states
that comply with the classical entropic inequalities decreases, with different
rates, down to the saturating value corresponding to $q\rightarrow \infty$.
This tendency is universal for any dimension and answers the query about the
monotonicity of the ``$q$-volumes" occupied by states behaving ``classically"
in what regards to  their conditional $q$-entropy. With respect to the second
aspect, i.e., evolution with $N$ for fixed $q$, one sees that for any value of
$q$, and for $N\le 6$, all the curves of Fig.\ref{fig3Some} behave in an approximately 
linear fashion (sure enough, this linear behaviour can not continue for
arbitrarily large values of $N$). There is also a sort of ``transition" in the
behaviour of the probabilities, depending on the value of $q$. For small $q$
values, as the total dimension $N=2 \times N_2$ grows, the conditional
$q$-entropies tend to behave classically: the probabilities of positive
conditional entropies increase in a monotonous way with $N$ and approach 1.
This ``classical behaviour" is ruled out beyond a certain value of $q$, when the
system, as its dimension increases, exhibits the quantum feature given by
negative conditional entropies. This behaviour is more pronounced for higher
$q$-values. Interestingly, these two behaviours seem to be ``separated" by a
certain ``critical" value $q=q{*}$. The probabilities of finding states with
positive conditional $(q=q^{*})$-entropies are (when keeping $N_1$ constant)
rather insensitive to changes in $N_2$. In the case of Fig.\ref{fig1Some} we have 
$q{*}\in[2,4]$.

We pass now to the consideration of systems for which the former qubit is
replaced by a qutrit (Fig.\ref{fig3Some}). This figure exhibits the features already 
encountered in Fig.\ref{fig2Some} (for the same values of $q$). 
For a fixed dimension, all probabilities are monotonous with $q$ and, again, the 
curves exhibit two types of qualitative behaviour. As $q$ grows, one seems to pass 
from one of them to
the other at a certain critical $q=q^{*}$-value. This special $q$-value
discriminates between i) the region where the ``classical" behaviour of the
conditional entropies becomes more important with increasing $N$, from ii) the
region where negative conditional entropies (which can not occur classically)
are predominant for large $N$. In this case, $q^{*}$ lies, as before, between
the values 2 and 4. It is interesting to notice, after glancing at both 
Figs. \ref{fig2Some} and \ref{fig3Some}, that the probabilities of finding states 
with positive conditional
$q$-entropies are not just a function of the {\it total} dimension $N=N_1\times
N_2$, as is the case, with good approximation, for the probability of having a
positive partial transpose (this was already noted in \cite{JPhysA2}). The
probabilities of having positive conditional $q$-entropies depend on the {\it
individual dimensions} ($N_1$ and $N_2$) of both subsystems.

\begin{figure}
\begin{center}
\includegraphics[angle=0,width=0.65\textwidth,clip=true]{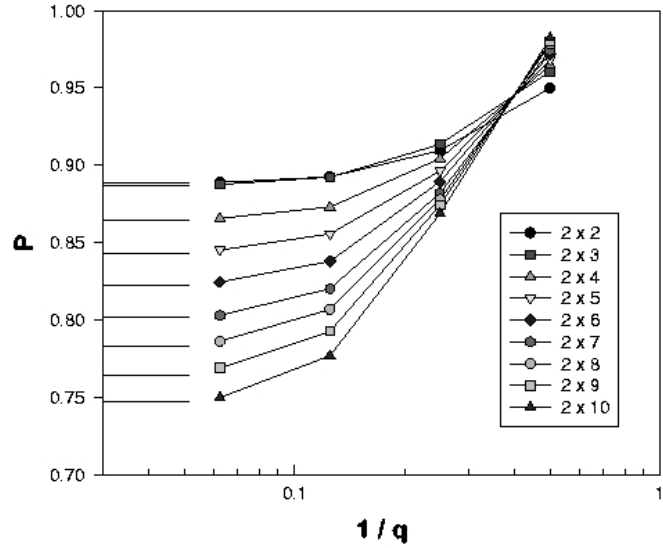}
\caption{Probability of finding a state $\rho$ 
(for systems of $2 \times N_2$ dimensions) which has its two conditional
$q$-entropies of a positive nature vs. $1/q$. Different curves correspond to
different dimensions. The monotonic decreasing behaviour of these probabilities
is apparent. The lines are just a guide for the eye. All quantities depicted
are dimensionless.} 
\label{fig4Some}
\end{center}
\end{figure}

\begin{figure}
\begin{center}
\includegraphics[angle=0,width=0.65\textwidth,clip=true]{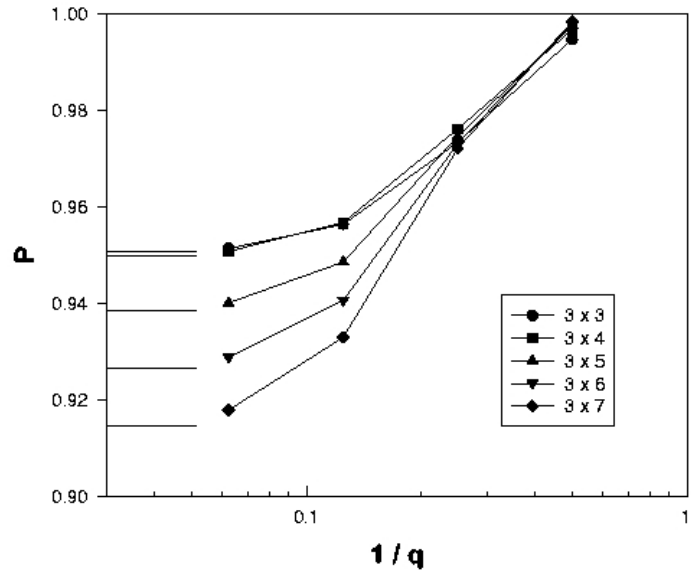}
\caption{Same as in Fig.\ref{fig4Some}, but  for systems in $3
\times N_2$ dimensions.  All quantities depicted are
dimensionless.} 
\label{fig5Some}
\end{center}
\end{figure}

A better insight into the monotonicity issue (how  the probabilities of having
positive conditional entropies change with $q$) is provided by 
Figs. \ref{fig4Some} and  \ref{fig5Some}. In Fig.\ref{fig4Some} we depict, 
for $N=2 \times N_2$ systems, the evolution of those probabilities with $q$, 
for fixed values of the total dimension $N$. A similar
evolution is plotted in Fig.\ref{fig5Some} for $N=3 \times N_2$ systems. The 
curves in these two figures behave in similar fashion. For given values of $N_1$ and
$N_2$, the probabilities decrease in a monotonous way with $q$. On the other
hand, for a fixed $q$-value, the probabilities behave in a monotonous fashion
with $N_2$. Again (as was already mentioned in connection with Figs. \ref{fig2Some} and
\ref{fig3Some}), there seem to be a special q-value, $q^{*}$, such that above $q^{*}$ the
probabilities decrease with $N_2$, while below $q^{*}$, the opposite behaviour
is observed.

 Thus far we have considered specific systems for which one of
the parties has fixed dimension while that of its partner augments. But what if
we consider the case of composite systems with $N_1=N_2=D$ (that is, two-{\it
qudits} systems)?. It was already shown in \cite{JPhysA2}, and as we shall see 
in the forthcoming section, for the case $q=\infty$ that the concomitant probabilities 
of finding states complying with
the classical entropic inequalities (that is, having positive both conditional
q-entropies) exhibit a behaviour that is quite different from the one previously
 discussed. Indeed, the numerical evidence gathered for $q=\infty$ in \cite{JPhysA2}
 suggests that, for an $N_1 \times N_2$-composite system of increasing dimensionality,
 the  probability trends that interest us here are clearly different if
 one considers either (i) increasing dimension for one of the
 subsystems and constant dimension for the other, or (ii)
 increasing dimension for both subsystems. In case (i) we have that,
 for $q=\infty$, the probabilities of finding states with positive conditional
 $q$-entropies diminish as $N$ grows. In the present effort we have
 extended the study of case (i) to finite values of $q$, obtaining
 a similar type of behaviour for $q$-values above a certain
 special value $q^{*}$. In case (ii) the probability of
 finding states complying with the classic entropic
 inequalities steadily grows with $N$ and approaches unity as $N \rightarrow
 \infty$. The reader is referred to the forthcoming Fig.\ref{fig9cond}. The 
 evolution of the probabilities for systems with $N=D \times D$ for
finite  $q$ does  not qualitatively differ from that pertaining to
the limit case $q\rightarrow \infty$. As far as monotonicity is
concerned, these probabilities share the  monotonic behaviour (with
$q$) so far discussed for a fixed dimension.

We will now look at two-qudits systems from the following, different
perspective: instead of considering the probability of states having  positive
conditional entropies for both parties, consider the behaviour, as a function of
the entropic parameter $q$, of the {\it global probability that an arbitrary
state of a two-qudit systems} exhibits either (i) both a positive conditional
$q$-entropy and a positive partial transpose, or, (ii) both a negative
conditional $q$-entropy and a non positive partial transpose. That is, we now
focus attention on the probability that i) Peres' PPT criterion and ii) the
signs of the conditional $q$-entropies (regarded as the basis of a separability
criterion), {\it both} lead to the same answer as far as separability or
entanglement are concerned.
 
\begin{figure}
\begin{center}
\includegraphics[angle=0,width=0.5\textwidth,clip=true]{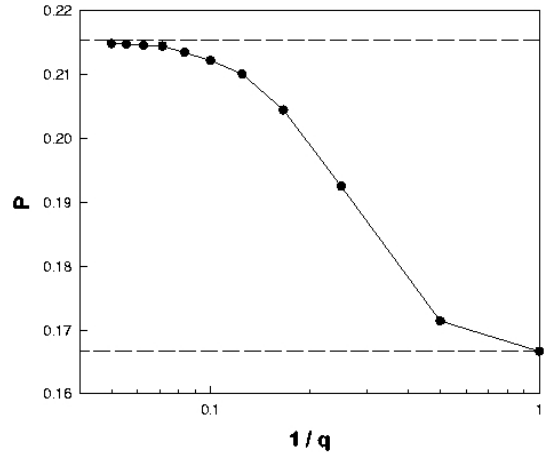}
\caption{Probability (as a function of $q$) of finding a two-qudits
state ($D \times D, D=3$) which is characterized by either i)  positive
conditional $q$-entropy and  positive partial transpose, or ii) a negative
conditional $q$-entropy and a non positive partial transpose. As $q$ grows so
does the degree of agreement with the PPT-criterion,  from the von Neumann
($q=1$) case to the ``best"  $q \rightarrow \infty$ improves. The lines are
just a guide for the eye.  All quantities depicted are dimensionless.} 
\label{fig6Some}
\end{center}
\end{figure}									      

Fig.\ref{fig6Some} illustrates the case $D=3$ ($N=3 \times 3$). We depict there 
the referred to probabilities 
as a function of $1/q$, for values of $q\in [2,20]$. Keeping also in mind the 
results plotted in Fig.\ref{fig5cond} in next section (for $D=2$), we conclude that 
(i) agreement with Peres' criterion becomes larger in all cases as $q$
increases up to $q=\infty$, and (ii) the largest degree of agreement, achieved
in this limit case, rapidly decreases as $D$ augments from its $D=2$-amount
(nearly 75 per cent \cite{JPhysA2}) to the $D=3$-one (Fig. \ref{fig6Some}) of nearly 
22 per cent, and further down to the $D=4$-instance in \cite{JPhysA2} of 4.5 per cent.


\section{Probabilities of finding states with positive conditional $q$-entropies.}
 
  We have see that when one deals with a classical composite system, 
  described by a suitable probability distribution defined over the concomitant
 phase space, the entropy of any of its subsystems is always equal
 or smaller than the entropy characterizing the whole system. This
 is also the case for separable states of a composite quantum
 system \cite{NK01,VW02}. In contrast,
 a subsystem of a quantum system described by an
 entangled state may have an entropy greater than the entropy of
 the whole system. Then we are naturally led to the entropic inequalities, whose 
 violation provides a clear and direct information-theoretical
  manifestation of the phenomenon of entanglement.

  This section is similar to the previous one, but we extend our results 
  using the so called participation ratio,

  \be \label{partrad1}
  R(\hat \rho) \, = \, \frac{1}{Tr(\hat \rho^2)},
  \ee

  \noindent
  and the maximum eigenvalue $\lambda_m$ of $\rho$ as a degree of mixture, 
  $\rho$ being the density matrix describing the state of the system under 
  consideration. Besides, we derive the proportion of states in $2\times 2$ 
  and $2\times 3$ systems (two-qubits and qubit-qutrit systems) which are 
  separable or unentangled vs. $R$.

\begin{figure}
\begin{center}
\includegraphics[angle=270,width=.65\textwidth]{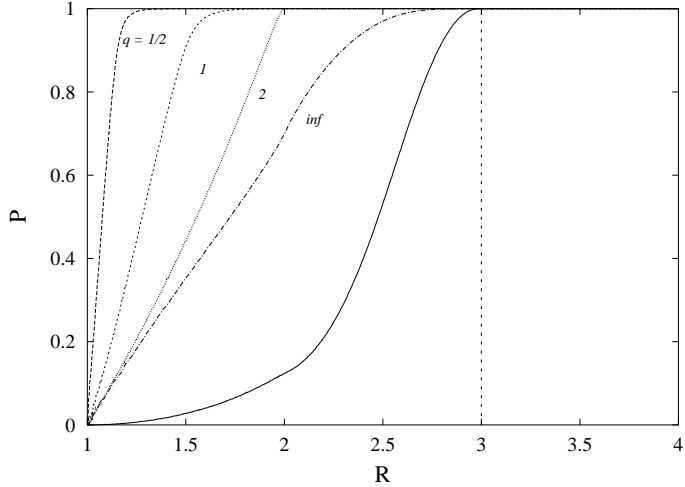}
\caption{Probability of finding (for different values of
$q$) a two-qubits state which, for a given degree of mixture $R=
1/Tr \, (\rho^2)$, has positive relative $q$-entropies. The solid
line corresponds to the probability of finding, for a given degree
of mixture $R= 1/Tr \, (\rho^2)$, a two-qubits state with a
positive partial transpose.} 
\label{fig1cond}
\end{center}
\end{figure}

 We determined numerically, by recourse to the usual Monte Carlo calculation
 and for different values of the entropic parameter $q$,
 the probability of finding a two-qubits state which,
 for a given degree of mixture $R= 1/Tr \, (\rho^2)$,
 has positive conditional $q$-entropies. The results
are depicted in Fig.\ref{fig1cond}. The solid line in Fig.\ref{fig1cond} 
corresponds to the probability of finding, for a given degree of mixture 
$R= 1/Tr \,(\rho^2)$, a two-qubit state with a positive partial transpose.
Since Peres' criterium for separability is necessary and
sufficient, this last probability coincides with the probability
of finding a separable state. We see that, as the value of $q$
increases, the curves associated with the conditional entropies
approaches the curve corresponding to Peres criterium. However,
even in the limit $q \rightarrow \infty$ the entropic curve lies
above the Peres' one by a considerable amount. This means that,
even for $q \rightarrow \infty$, {\it there is a considerable
volume in state space occupied by entangled states complying with
the classical entropic inequalities} (that is, having positive
conditional entropies).

   The probability of finding separable states increases with the
degree of mixture \cite{ZHS98}, as it is evident from the solid
curve in Fig.\ref{fig1cond}. Also, one can appreciate the fact that a similar    
trend is exhibited by the probability of finding, for a given
$q$-value, states with positive conditional $q$-entropies.

\begin{figure}
\begin{center}
\includegraphics[angle=270,width=.65\textwidth]{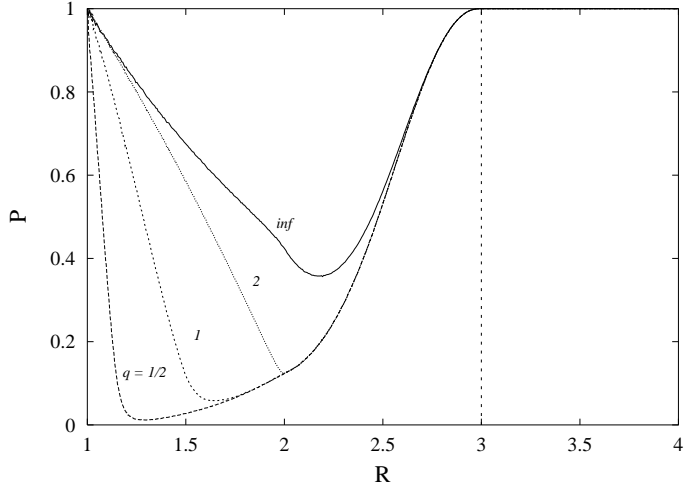}
\caption{Probability of finding (for different values of $q$) a
two-qubits state which, for a given degree of mixture $R= 1/Tr \, (\rho^2)$,
either (i) has both relative $q$-entropies positive, as well as a positive
partial transpose, or (ii) has a negative relative $q$-entropy and a non
positive partial transpose.} 
\label{fig2cond}
\end{center}
\end{figure}

 We have computed numerically the probability (for different
 values of $q$) that a two-qubits state with a given degree of
 mixture be correctly classified, either as entangled or as separable,
 on the basis of the sign of the conditional $q$-entropies. The results
 are plotted in Fig.\ref{fig2cond}. That is, Fig.\ref{fig1cond} depicts the probability of   
 finding (for different values of $q$) a two-qubits state which,
 for a given degree of mixture $R= 1/Tr \, (\rho^2)$, either has
 (i) both conditional $q$-entropies positive, as well as a positive partial transpose,
 or (ii) has a negative conditional $q$-entropy and a non positive partial
 transpose. We see that, for all values of $q>0$,
  this probability is equal to one both for pure
 states ($R=1$) and for states with ($R>3$). The probability
 attains its lowest value $P_m(q)$ at a special value $R_m(q)$ of the
 participation ratio. Both quantities $R_m(q)$ and $P_m(q)$
 exhibit a monotonic increasing behaviour with $q$, adopting their
 maximum values in the limit $q\rightarrow \infty$.

\begin{figure}
\begin{center}
\includegraphics[angle=270,width=.65\textwidth]{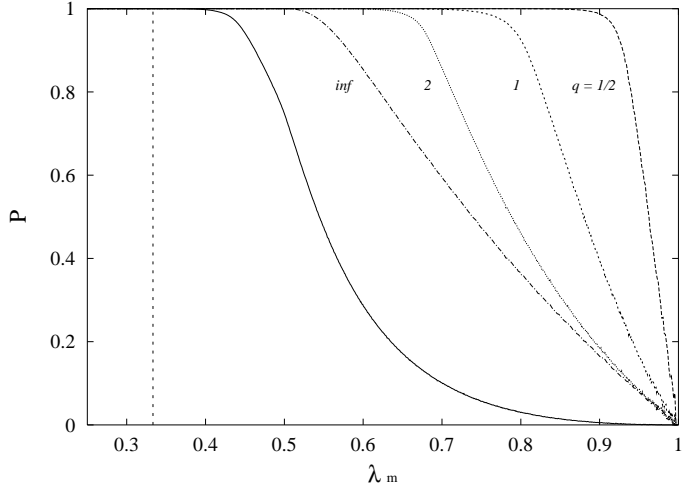}
\caption{Probability of finding (for different values of
$q$) a two-qubits state which, for a given value of the maximum
eigenvalue $\lambda_m$, has positive relative $q$-entropies. The
solid line corresponds to the probability of finding, for a given
value of $\lambda_m$, a two-qubits state with a positive partial
transpose.} 
\label{fig3cond}
\end{center}
\end{figure}

\begin{figure}
\begin{center}
\includegraphics[angle=270,width=.65\textwidth]{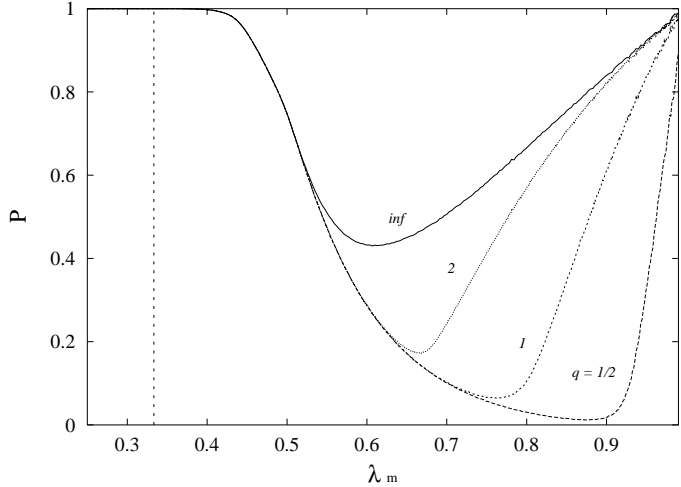}
\caption{Probability of finding (for different values of
$q$) a two-qubits state which, for a given value of the maximum
eigenvalue $\lambda_m$, either (i) has its two relative
$q$-entropies positive, as well as a positive partial transpose,
or (ii) has a negative relative $q$-entropy and a non positive
partial transpose.} 
\label{fig4cond}
\end{center}
\end{figure}

In Fig.\ref{fig1cond} and Fig.\ref{fig2cond} we have used the participation 
ratio $R$ as a measure of mixedness. The quantity $R$ is, essentially, a
$q$-entropy with $q=2$. The $q$-entropies associated with other
values of $q$ are legitimate measures of mixedness as well, and
have already found applications in relation with the study of
entanglement \cite{ZHS98,BCPP02b}. It is interesting to see what
happens, in the present context, when we consider measures of
mixedness based on other values of $q$. Of particular interests is
the limit case $q\rightarrow \infty$ which, as already mentioned,
is related to the largest eigenvalue of the density matrix. The
largest eigenvalue constitutes a legitimate measure of mixture in
its own right: states with larger values of $\lambda_m$ can be
regarded as less mixed. Its extreme values correspond to (i) pure
states (with $\lambda_m =1$) and (ii) the density matrix
$\frac{1}{4}\hat I$ (with $\lambda_m = 1/4$). In Figures \ref{fig3cond}-\ref{fig4cond} 
we have considered (in the horizontal axes) the largest eigenvalue
$\lambda_m$ as a measure of mixedness.
 We computed the probability of finding (for different values
of $q$) a two-qubits state which, for a given value of the maximum
eigenvalue $\lambda_m$, has positive conditional $q$-entropies. The
results are depicted in Fig.\ref{fig3cond}. The solid line corresponds to the 
probability of finding, for a given degree of mixture $R= 1/Tr \,
(\rho^2)$, a two-qubits state with a positive partial transpose.
We see in Fig.\ref{fig3cond} that, for $\lambda<1/3$, the probability of 
finding states verifying the classical entropic inequalities
(i.e., having positive conditional entropies) is, for all $q>0$,
equal to one. This is so because all states whose largest
eigenvalue $\lambda_m$ is less or equal than $1/3$ are separable
\cite{BCPP02b}.

Fig.\ref{fig4cond} depicts the probability of finding (for different values of 
$q$) a two-qubits state which, for a given value of the maximum
eigenvalue $\lambda_m$, either has (i) both conditional $q$-entropies
positive and a positive partial transpose, or (ii) a negative
conditional $q$-entropy and a non positive partial transpose.

  A remarkable aspect of the behaviour of the sign of the
conditional $q$-entropies, which transpires from Figures \ref{fig1cond} and \ref{fig3cond}, 
is that, for any degree of mixture, the volume corresponding to
states with positive conditional $q$-entropies ($q>0$) is a
monotonous decreasing function of $q$. This feature interesting because, for a 
single given state $\rho $, the conditional $q$-entropy is not necessarily 
decreasing in $q$ \cite{VW02}. This means that the positivity of the conditional
entropy of a given state $\rho $ and for a given value $q^{*}$ of
the entropic parameter does not imply the positivity of the
conditional $q$-entropies of that state for all $q<q{*}$. That is,
$q<q^{*}$ does not imply that the family of states exhibiting
positive conditional $q^{*}$-entropies is a subset of the family of
states with positive $q$-entropies. This fact notwithstanding, the
numerical results reported here indicate that for $0<q<q{*}$ the
volume of states with positive $q^{*}$-conditional entropies is
smaller than the volume of states with positive $q$-entropies.
This implies that, among all the $q$-entropic separability
criteria, the one corresponding to the limit case $q\rightarrow
\infty$ is the strongest one, as was suggested by Abe in 
\cite{A02} on the basis of his analysis of a monoparametric family
of mixed states for multi-qudit systems.

\begin{figure}
\begin{center}
\includegraphics[angle=270,width=.65\textwidth]{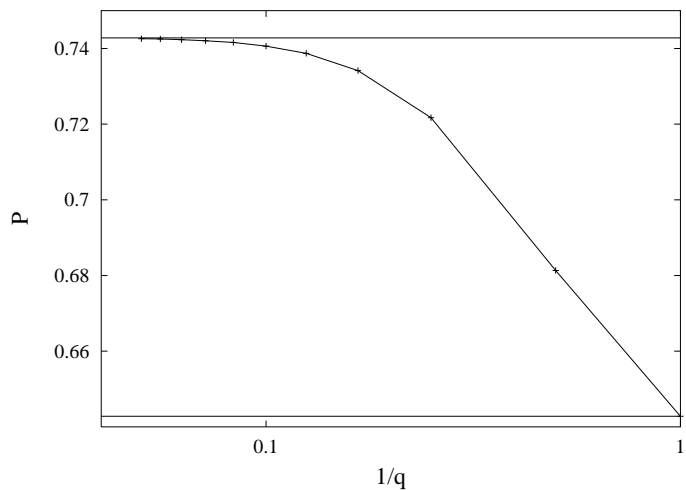}
\caption{Probability (as a function of $q$) of finding a
two-qubits state which either has both  positive relative
$q$-entropies and a positive partial transpose, or has a negative
relative $q$-entropy and a non positive partial transpose.} 
\label{fig5cond}
\end{center}
\end{figure}

It is interesting to see the behaviour, as a function of the
entropic parameter $q$, of the global probability (regardless of
the degree of mixture) that an arbitrary state of a two-qubit
system exhibits simultaneously (i) a positive conditional $q$-entropy
and a positive partial transpose, or (ii) a negative conditional
$q$-entropy and a non positive partial transpose. In order words,
this is the probability that for an arbitrary state the entropic
separability criterium and the Peres' criterium lead to the same
``conclusion" with respect to the separability (or not) of the
state under consideration. In Fig.\ref{fig5cond} we depict this probability as 
a function of $1/q$, for values of $q\in [2,20]$. We see that this
probability is an increasing function of $q$. In the limit
$q\rightarrow \infty $ this probability approaches the value
$0.7428$. On the other hand, for $q=1$ (that is, when we
use the standard logarithmic entropy) the probability is
approximately equal to $0.6428$.

\begin{figure}
\begin{center}
\includegraphics[angle=270,width=.65\textwidth]{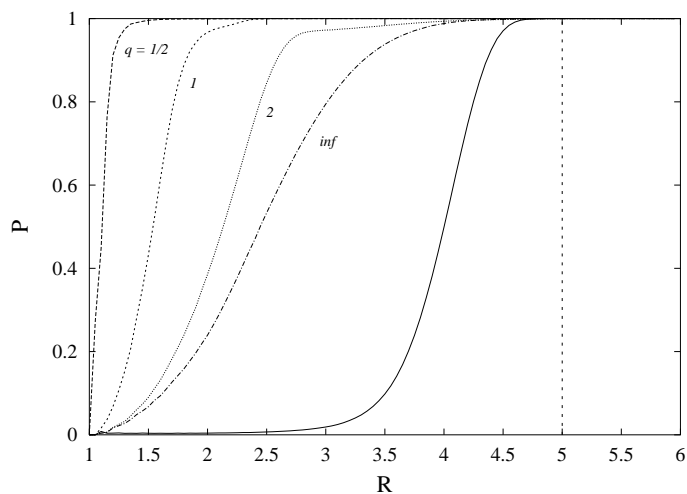}
\caption{Probability of finding (for different values of
$q$) a qubit-qutrit state which, for a given degree of mixture $R=
1/Tr \, (\rho^2)$, has positive relative $q$-entropies. The solid
line corresponds to the probability of finding, for a given degree
of mixture $R= 1/Tr \, (\rho^2)$, a qubit-qutrit state with a
positive partial transpose.} 
\label{fig6cond}
\end{center}
\end{figure}

\begin{figure}
\begin{center}
\includegraphics[angle=270,width=.65\textwidth]{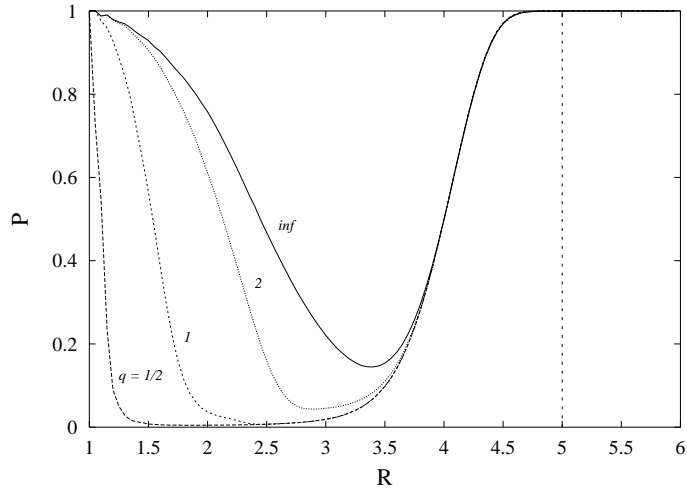}
\caption{Probability of finding a qubit-qutrit state
which, for a given degree of mixture $R= 1/Tr \, (\rho^2)$, and
for different values of $q$, either (i) has its two relative
$q$-entropies positive, as well as a positive partial transpose,
or (ii) has a negative relative $q$-entropy and a non positive
partial transpose.} 
\label{fig7cond}
\end{center}
\end{figure}

\begin{figure}
\begin{center}
\includegraphics[angle=270,width=.65\textwidth]{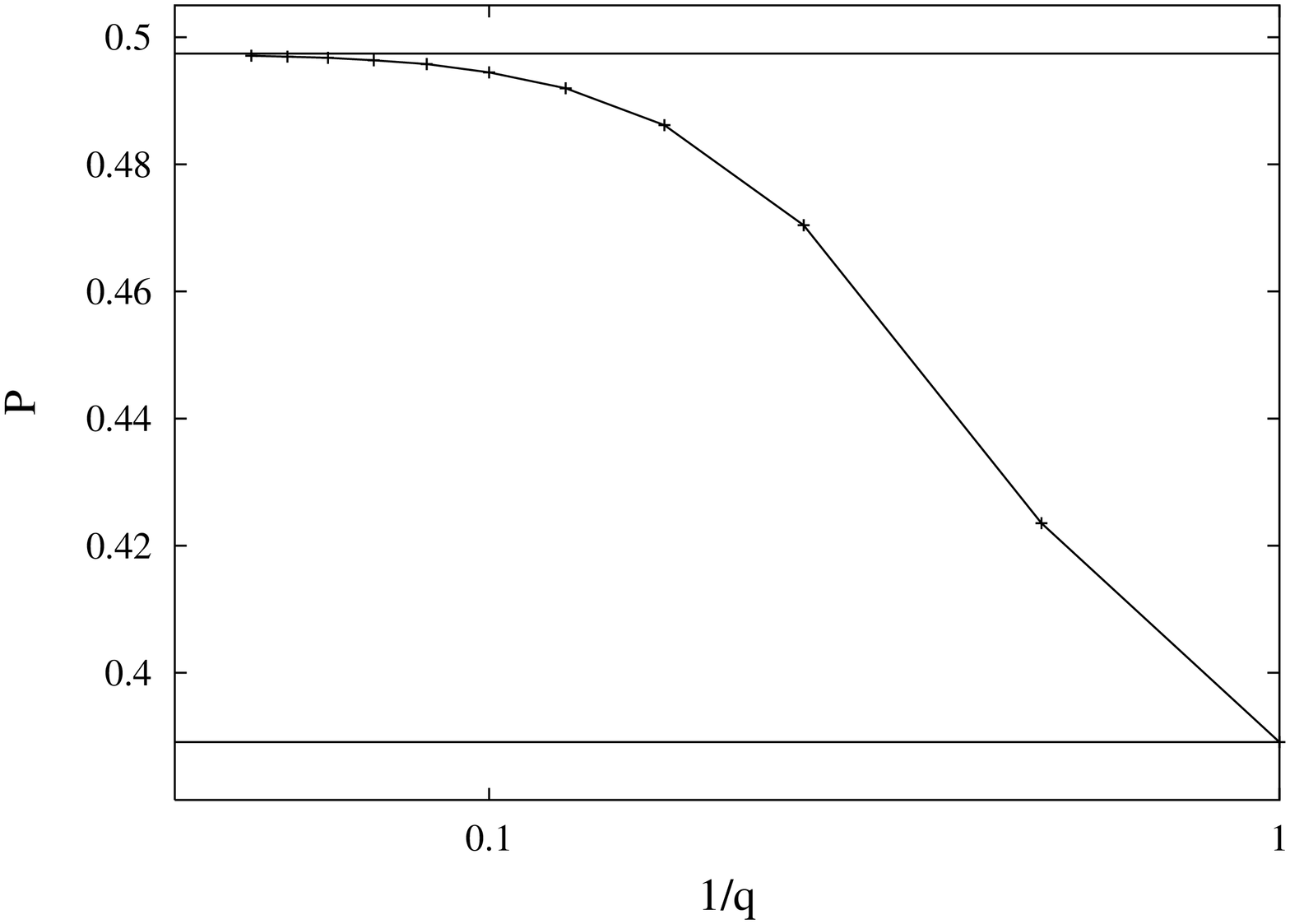}
\caption{Probability (as a function of $q$) of finding a
qubit-qutrit state which either has both positive relative
$q$-entropies and a positive partial transpose, or has a negative
relative $q$-entropy and a non positive partial transpose.} 
\label{fig8cond}
\end{center}
\end{figure}

We have performed for qubit-qutrit systems calculations similar to
the ones that we have already discussed for two-qubits systems.
The results are summarized in Figures \ref{fig6cond} and \ref{fig7cond}. 
Fig.\ref{fig6cond} depicts the probability of finding (for different values of $q$) a
qubit-qutrit state which, for a given degree of mixture $R= 1/Tr
\, (\rho^2)$, has positive conditional $q$-entropies. The solid line
in Fig.\ref{fig6cond} corresponds to the probability of finding, for a given 
degree of mixture $R= 1/Tr \, (\rho^2)$, a qubit-qutrit state with
a positive partial transpose. Fig.\ref{fig7cond} depicts the probability of 
finding, for different values of $q$, a qubit-qutrit state which
has, for a given degree of mixture $R= 1/Tr \, (\rho^2)$, either
(i) its two conditional $q$-entropies positive, as well as a positive
partial transpose, or (ii)  a negative conditional $q$-entropy and a
non positive partial transpose.  We have also computed the
probability (for different  values of $q$) that an arbitrary
qubit-qutrit state (regardless of its degree of mixture) be
correctly classified, either as entangled or as separable,
 on the basis of the sign of the conditional $q$-entropies. These
 probabilities are depicted in Fig.\ref{fig8cond}, for values of $q$ in the 
interval $q\in [2,20]$. As happens with two-qubits systems, this
 probability is an increasing function of $q$. For $q=1$ the
 probability is equal to $0.3891$ and approaches
 the (approximate) value $0.4974$ as $q\rightarrow \infty$. For
 a given value of $q$, the probability of coincidence between the
 Peres' and the entropic separability criteria are seen to be
 smaller in the case of qubit-qutrit systems than in the case of
 two-qubits systems.

\begin{figure}
\begin{center}
\includegraphics[angle=270,width=.65\textwidth]{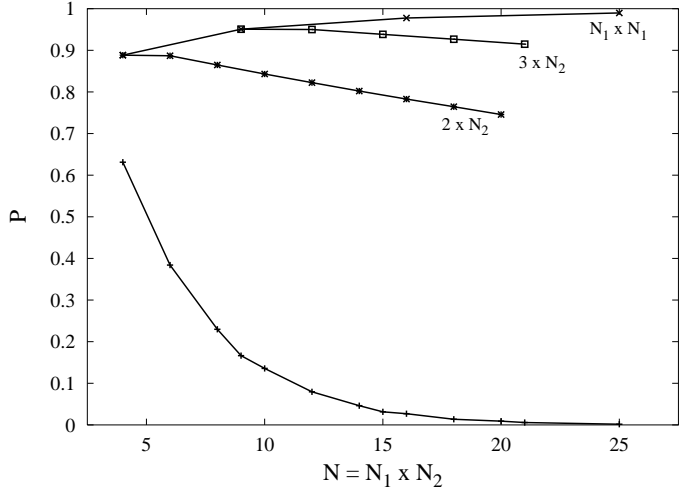}
\caption{Global probability of finding a state (pure or
mixed) of a bipartite quantum system with positive relative
$q$-entropies. $N_1$ and $N_2$ stand for the dimensions of the
Hilbert spaces associated with each subsystem, and $N=N_1\times
N_2 $ is the dimension of the Hilbert space associated with the
composite system as a whole. The three upper curves correspond (as
indicated in the figure) to composite systems of increasing
dimensionality, and with $N_1=2$, $N_1=3$, and $N_1=N_2$. The
probability of finding a state complying with the Peres partial
transpose separability ctiterium (lower curve) is also plotted.} 
\label{fig9cond}
\end{center}
\end{figure}

 Finally, we have computed the probabilities of finding states
 with positive conditional $q$-entropies (for the case
 $q=\infty$) for bipartite quantum systems described by Hilbert
 spaces of increasing dimensionality. Let $N_1$ and $N_2$ stand for
 the dimensions of the Hilbert spaces associated with each subsystem,
 and $N=N_1\times N_2 $ be the dimension of the Hilbert space
associated with the concomitant composite system. We have
considered three sets of systems: (i) systems with $N_1=2,3$ and
increasing values of $N_2$, and (ii) systems with $N_1=N_2$ and
increasing dimensionality. The computed probabilities  are
depicted in Fig.\ref{fig9cond}, as a function of the total dimension $N$. 
The three upper curves correspond (as indicated in the figure) to
composite systems with $N_1=2$, $N_1=3$, and $N_1=N_2$. For the
sake of comparison, the probability of finding  states complying
with the Peres partial transpose separability criterium (lower
curve) is also plotted. In order to obtain each point in Fig.\ref{fig9cond}, 
$10^8$ states\footnote{The error is then less than the size of the 
corresponding symbol in that Figure.} were randomly generated.

 Some interesting conclusions can be drawn from Fig.\ref{fig9cond}. 
In the case of composite systems with $N_1=N_2$, the probability
 of finding states complying with the classical ($q=\infty $)
 entropic inequalities (that is, having positive both conditional
 q-entropies) is an increasing function of the dimensionality.
 Furthermore, this probability seems to approach $1$, as 
 $N\rightarrow \infty$. In other words, Fig.\ref{fig9cond} provides 
numerical evidence that, in the limit of infinite dimension,
 two-qudits systems behave classically, as far as the signs of the
 conditional $q$-entropies are concerned.

 When considering composite systems with increasing dimensionality,
 but keeping the dimension of one of the subsystem constant
 ($N_1=2,3$), we obtained numerical evidence  that the probability
 of having positive conditional $q$-entropies (again, with $q=\infty $)
 behave in a  monotonous decreasing way with the total dimension $N$.

  It is interesting to notice that the probabilities of finding
  states with positive $q$-entropies are not just a function
  of the total dimension $N=N_1\times N_2$ (as happens,
  with good approximation, for the probability of having a
  positive partial transpose). On the contrary, they
  depend on the individual dimensions ($N_1$ and $N_2$) of both
  subsystems. Furthermore, the trend of the alluded to
  probabilities are clearly different if one considers composite
  systems of increasing dimension with either (i) increasing
  dimensions for both subsystems or (ii) increasing dimension
  for one of the subsystems and constant dimension for the other
  one.

\section{Maximally entangled mixed states (MEMS) viewed in the light of 
the entropic criterion}

Recourse to entanglement is required so as  to
implement quantum information processes \cite{W98} such as
quantum cryptographic key distribution \cite{E91}, quantum
teleportation \cite{BBCJPW93}, superdense coding \cite{BW93}, and
quantum computation \cite{BDMT98}. Indeed, production of
entanglement is a kind of elementary prerequisite for any quantum
computation. 

In practice, one will more often have to deal with mixed states than with pure ones.
From the point of view of entanglement-exploitation, one should
then be interested in maximally entangled mixed states (MEMS)
 $\rho_{MEMS}$ that have been studied, for example, in Refs. \cite{MJWK01,IH00} 
 for the two-qubits instance of two (one qubit-)subsystems $A$ and $B$. 
 These MEMS states have been recently achieved experimentally \cite{MEMSexp}. 
 We will focus attention on this kind of states here.
 
 MEMS for a given $R-$value have the following appearance in
 the computational basis ($|00\rangle,|01\rangle,|10\rangle,|11\rangle$) 
 \cite{MJWK01}.

\be \rho_{MEMS} = \left( \begin{array}{cccc}
g(x) & 0 & 0 & x/2\\
0 & 1 - 2g(x) & 0 & 0\\
0 & 0 & 0 & 0\\
x/2 & 0 & 0 & g(x) \end{array} \right), \label{MEMS} \ee with
$g(x)=1/3$ for $0\le x \le 2/3$, and  $g(x)=x/2$ for $2/3 \le x
\le 1$. The change of $g(x)-$regime ensues for $R=1.8$. {\it We
will reveal below some physical consequences of this
regime-change} . Of great importance are also mixed states whose
entanglement-degree cannot be increased by the action of logic
gates \cite{IH00} that, again in the same basis, are given by

\begin{equation} \label{IH}
\rho_{IH} = \left( \begin{array}{cccc}
p_2 & 0 & 0 & 0\\
0 & \frac{p_3+p_1}{2} & \frac{p_3-p_1}{2} & 0\\
0 & \frac{p_3-p_1}{2} & \frac{p_3+p_1}{2} & 0\\
0 & 0 & 0 & p_4 \end{array} \right),
\end{equation}
 whose eigenvalues are the $p_i;\,\,(i=1,\ldots,4)$ and   $p_1 \ge p_2 \ge p_3
 \ge p_4$. We call these states the Ishizaka and Hiroshima (IH) \cite{IH00} 
 ones and their concurrence $C_{IH}$ reads
\begin{equation} \label{CIH}
C_{IH}\,=\,p_1\,-\,p_3\,-\,2\,\sqrt{p_2\,p_4},
\end{equation}
a relation valid for ranks $\le 3$ that has numerical support also
if the rank is four \cite{IH00}. Of course, all MEMS belong to the
IH-class. Our goal is to uncover interesting correlations between
entanglement and mixedness that emerge when we study these states
from the view point of conditional entropies.

\subsection {Entropic inequalities and MEMS} 

We begin here with the presentation
of our results \cite{BCPP05}. A few of them are of an analytical nature. For
instance, in the case of all states of the forms (\ref{IH}) and/or
(\ref{MEMS}), the partial traces $\rho_{A/B}$ over one of the
subsystems $A$ or $B$ are equal, i.e., for the reduced density
matrices we have $\rho_A=\rho_B$, which entails
$S_{q}(A|B)=S_{q}(B|A)$\footnote{From now on let us replace $S^{(T)}_q$ and $S^{(R)}_q$ 
by $S_q$ and $R_q$, respectively.}. Notice that this is a particular feature of 
these states.

\nd As for the form (\ref{IH}), we establish a lower bound to its
states' concurrence for a considerable $R-$range 
(see Fig.\ref{fig3munro}),                                                                      
namely, \be
\label{lowerb} C_{IH;Min}=[\sqrt{3R(4-R)}-R]/(2R),\ee

\nd where $R$ is the participation ratio (\ref{partrad1}). In the case 
of MEMS and in the vicinity of $R=1$ we can
analytically relate entropic changes with concurrence-changes, in
the fashion (remember that for MEMS $C \equiv C_{Max}$)
\begin{equation} \label{relaS}
\Delta S_{q}(A|B)\,=\,-[2q/\{\ln(2)(q-1)\}]\,\Delta C.
\end{equation}
The case $q \rightarrow \infty$ is the strongest $q$-entropic
criterion \cite{batle}.
 Eq. (\ref{relaS}) expresses the fact that, for MEMS, small deviations
 from pure states (for which the $q$-entropic criteria are necessary and
sufficient separability conditions) do not change the criteria's
validity, that becomes then {\it extended} to a class of mixed
states.  

\begin{figure}
\begin{center}
\includegraphics[angle=270,width=.65\textwidth]{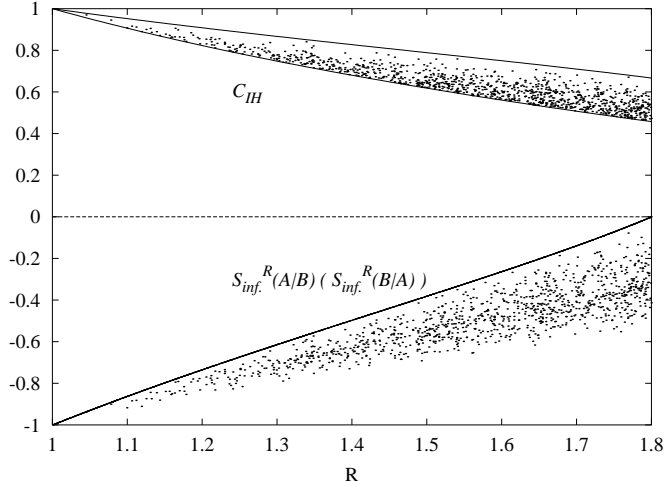}
\caption{Plot of the concurrence $C_{IH}$ for the two kinds
of maximally entangled states: Ishizaka and Hiroshima ones (with dots) and
MEMS vs. $R$ (upper solid curve) for a sample set. Their corresponding
$S_{\infty}(A|B)$-values are also shown. Contour lines can be found
analytically (see text for details). The entropic criteria always provide in
this range the correct answer in order to detect entanglement.} 
\label{fig1munro}
\end{center}
\end{figure}

As stated above, we deal with two kinds of maximally entangles states (MEMS and
Ishizaka and Hiroshima ones) \cite{BCPP05}. We call the class that comprises
both kinds the ME-one. Fig.\ref{fig1munro} depicts the overall situation. In 
the upper part we plot the ME-states' concurrence (\ref{CIH}) vs.
the participation ratio. $R$ ranges in the interval $1 < R < 1.8$
(the latter figure corresponds to the above mentioned transition
point for MEMS). (A): the upper line gives MEMS-states and the
lower one the lower bound (\ref{lowerb}). (B): the lower part
of the Figure gives the conditional entropy of the ME states
$S_{q}(A|B)$ for $q=\infty$ (the solid curve corresponds to the MEMS case).
It is always negative, so that here
the entropic inequalities for entangled states yield the expected
negative result.

\begin{figure}
\begin{center}
\includegraphics[angle=270,width=.65\textwidth]{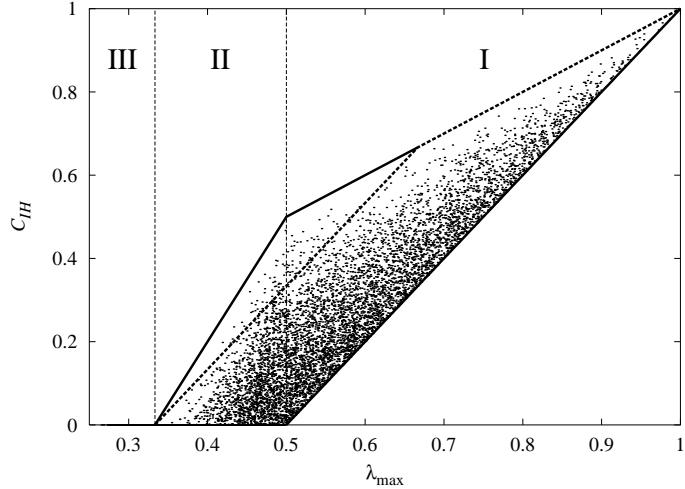}
\caption{Plot of the concurrence $C_{IH}$ for the class of maximally
entangled states vs. their maximum eigenvalue $\lambda_{max}$ for a sample set
of states. The dashed line corresponds to $\rho_{MEMS}$-states. Notice the
fact that these states are not maximally entangled if mixedness is not given
by $R$. Maximum and minimum contours of $C_{IH}$ are analytical.
See text for details.} 
\label{fig2munro}
\end{center}
\end{figure}

\begin{figure}
\begin{center}
\includegraphics[angle=270,width=.65\textwidth]{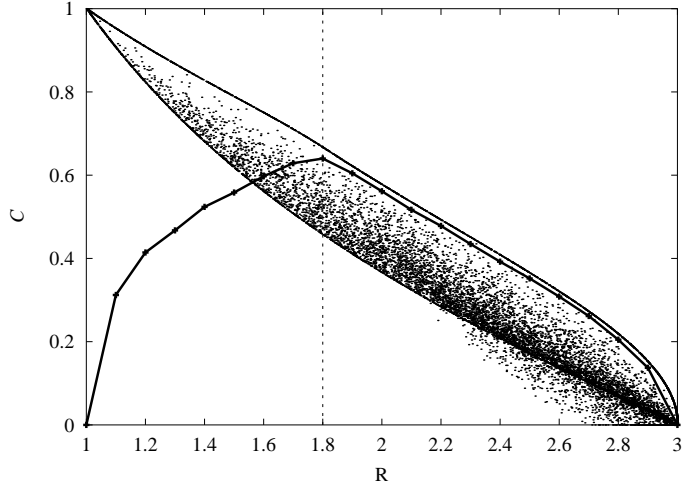}
\caption{Same as in Fig.\ref{fig1munro}, with an extended $R$-range.
The lower curve with crosses represents the maximum concurrence
for each $R$ for those states which observe the classical entropic
inequalities (both positive). The curve presents a maximum at
$R=1.8$ and it is null at $R=1, 3$ where the entropic criterion is
necessary and sufficient. This curve does not exactly match the
upper MEMS-curve in the range $[1.8,3)$. This is due to the
relative scarcity of the pertinent states (generated randomly
according to the ZHSL measure). See text for details.} 
\label{fig3munro}
\end{center}
\end{figure}

\nd Fig.\ref{fig2munro} is a plot of the concurrence $C_{IH}$ vs. 
$\lambda_{max}$, the maximum eigenvalue of our ME bipartite states
$\hat \rho$. The dashed line corresponds to MEMS. The graph confirms
the statement made in  \cite{MJWK01} that the latter are not
maximally entangled states if mixedness is measured according to a
criterion that is not the $R-$one. Three separate regions (I, II,
III) can be seen to emerge. The maximum and minimum (continuous)
contour lines are of an analytical character:
\begin{itemize} \item {\sf First zone:}
a) $C_{IH}^{max}= \lambda_{max}$ for $\lambda_{max} \in [1/2, 1]$
\item b) $C_{IH}^{min}= 2\lambda_{max}-1 $ for Bell diagonal states.

\item {\sf Second:} a) $C_{IH}^{max}= 3\lambda_{max}-1$ for $\lambda_{max} \in [1/3, 1/2]$
\item b) $C_{IH}^{min}= 0$
\item  {\sf Third:} All states are separable $C_{IH}= 0$.
\end{itemize}
\nd Our three zones  (I, II, III) can be characterized according
to strict geometrical criteria, as extensively discussed in
\cite{BCPP02b}. Fig.\ref{fig3munro} is a $C_{IH}$ vs. $R$ plot like that of                  
Fig.\ref{fig1munro}, but for an extended $R-$range ($1<R<3$). The pertinent                  
bipartite states (randomly generated according to the ZHSL-measure
of Ref. \cite{Z99,BCPP02a} fill a ``band". In Fig.\ref{fig3munro} we focus                   
attention on a special type of bipartite states: those that, being
entangled, do fulfill the inequalities (\ref{qsepar}).

{\it For these states}, let us call them entangled states with
classical conditional entropic behaviour (ESCRE) \cite{BCPP05}, 
the quasi-triangular solid line depicts, for each $R$, the maximum degree of
entanglement attainable. Interestingly enough, {\sf the maximum
degree of entanglement for ESCRE obtains at} $R=1.8$, which
signals the change of regime for MEMS (Cf. (\ref{MEMS}) and
commentaries immediately below that equation). {\it This fact
gives an entropic meaning to that $R-$value}.  We can state then
that i) whenever the entropic criterium turns out to
constitute a necessary and sufficient condition for separability (at
$R=1$ and $R=3$), the ESCRE-degree of entanglement is null, and
ii) the ESCRE-degree of entanglement is maximal at the Munro {\it et
al.} change-of-regime $R-$value of 1.8.

\section{Correlations between quantum entanglement and entropic measures}

It is our intention here to investigate the
  degree of correlation between (i) the amount of entanglement $E[\rho_{AB}]$
  exhibited by a two-qubits state $\rho_{AB}$, and (ii)
  the $q$-entropies (or $q$-information measures) of $\rho_{AB}$ (notice that we refer here
  to the total $q$-entropy of the density matrix $\rho_{AB}$ describing
  the composite system as a whole. We shall not consider conditional
  $q$-entropies). It is well known that the amount entanglement and the
  degree mixture, as measured by the $q$-entropies: $R_q$ (R\'{e}nyi's) or $S_q$ (Tsallis'), 
  of a state $\rho_{AB}$
  are independent quantities. However, there is a certain degree of
  correlation among them. States with an increasing degree of mixture tend to
  be less entangled. In point of fact, all two-qubits states with a large enough
  degree of mixture are separable. We want to explore to what extent does
  the strength of the alluded to correlation depend upon the parameter $q$
  characterizing the $q$-entropy used to measure the degree of mixture.
  In particular, we want to find out if there is a special value of $q$
  yielding a better entropy-entanglement correlation than the entropy-entanglement
  correlations associated with other values of $q$. 

Our investigations will be based upon a Monte Carlo exploration of  ${\cal S}$:
the set of {\it all states, pure and mixed} of a two-qubits
 system, exactly in the same way as previously done. In the present investigation we shall 
 deal with the case $N=4$.

  Most recent research efforts dealing with the relationship between the degree
  of mixture and the amount of entanglement focus on the behaviour, as a
  function of the degree of mixture, of the entanglement properties exhibited
  by the set of states endowed with a given amount of mixedness. For instance,
  they consider the behaviour, as a function of the degree of mixture
  (as measured, for instance, by $S_2$), of the average entanglement of
  those states characterized by a given value of $S_2$.
  Here we are going to adopt, in a sense, the reciprocal (and complementary)
  point of view. We are going to study the behaviour, as a function of $C^2$,
  of the entropic properties associated with the set of states characterized
  by a given value of $C^2$. This vantage point will enable us to clarify some
  aspects of the $q$-dependence of the entanglement-mixedness correlation. In
  particular, we want to asses, for different $q$-values, how sensitive are
  the average entropic properties to the value of the entanglement of
  formation (or, equivalently, to the value of the squared concurrence $C^2$).
      
\begin{figure}
\begin{center}
\includegraphics[angle=270,width=.65\textwidth]{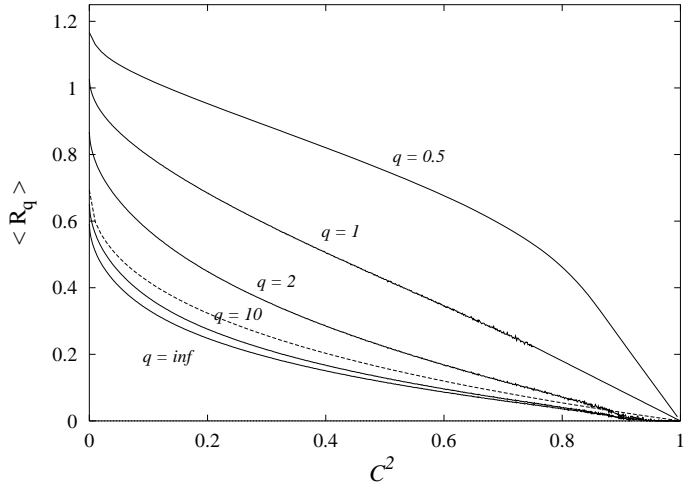}
\caption{Average value of the R\'enyi entropy $\langle R_q
\rangle$ of all states with a given squared concurrence $C^2$, as
a function of $C^2$, and for several $q$-values (solid lines). The
dashed line depicts the functional dependence of the $R_{\infty}$
R\'enyi entropy, as a function of $C^2$, for two-qubits states
diagonal in the Bell basis. All depicted quantities are
dimensionless.} 
\label{fig1Renyi}
\end{center}
\end{figure}

  We computed, as a function of $C^2$, the average value of
the R\'enyi entropy $R_q$ associated with the set of states
endowed with a given value of the squared concurrence $C^2$ \cite{xerec}. The
results are exhibited in Fig.\ref{fig1Renyi} (solid lines), where the mean 
value $\langle R_q \rangle$ is plotted against $C^2$, for $q
=0.5,\,1,\,2,\,10,$ and $\infty$. As stated, the averages are
taken over all the states $\hat \rho \in {\cal S}$ that are
characterized by a fixed concurrence-value. For all $q$ the
average entropies diminish as $C$ grows. This behaviour is
consistent with the fact that states of increasing entropy tend to
exhibit a decreasing amount of entanglement
\cite{ZHS98,BCPP02a,BCPP02b}. As $q$ grows, the average entropy
decreases, for any $C^2$, although the decreasing tendency slows
down for large $q$-values.
Many recent efforts dealing with the relationship between $q$-entropies and
 entanglement were restricted to states  $\rho_{\rm Bell}$ diagonal in the
 Bell basis. For such states, both the $R_q$ entropy and the squared
concurrence $C^2$ depend solely upon $\rho_{\rm Bell}$'s largest
eigenvalue, so that $R_q$ can be expressed as a function of $C^2$.
The dashed line in Fig.\ref{fig1Renyi} depicts the functional dependence of the 
$R_{\infty}$ R\'enyi entropy, as a function of $C^2$, for
two-qubits states diagonal in the Bell basis. It is instructive to
compare, in Fig.\ref{fig1Renyi}, the curve corresponding to states diagonal in 
the Bell basis with the curve corresponding (with $q=\infty $) to
all two-qubit states. It can can be appreciated that these two
curves, even if sharing the same qualitative appearance, differ to
a considerable extent.

\begin{figure}
\begin{center}
\includegraphics[angle=270,width=.65\textwidth]{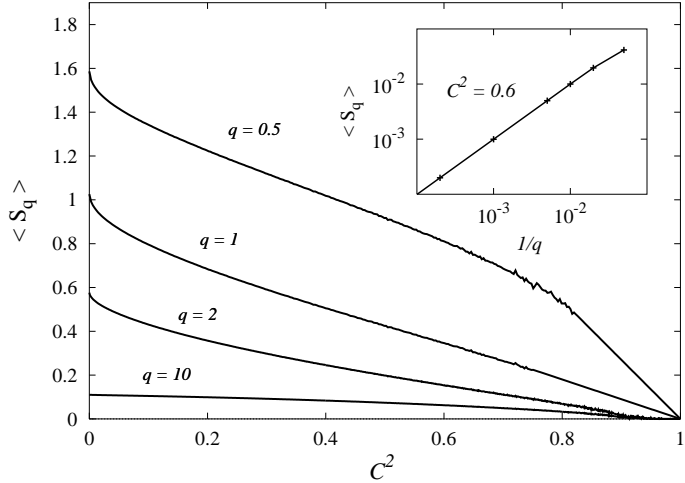}
\caption{Average value of the Tsallis' entropy $\langle
S_q \rangle$ of all states with a given squared concurrence $C^2$,
as a function of $C^2$, and for several $q$-values. The inset
shows $\langle S_q \rangle$ vs. $1/q$ for the particular value of
the squared concurrence $C^2 = 0.6$. All depicted quantities are
dimensionless.} 
\label{fig2Renyi}
\end{center}
\end{figure}

 For the sake of comparison, we plotted in Fig.\ref{fig2Renyi} the mean value         
 $\langle S_q \rangle$ of Tsallis' entropy, as a function of $C^2$, for
 $q =0.5,\,1,\,2,$ and $10$. Again, for each value of $C^2$, the entropy's
 average was computed over all those states characterized by
 that particular $C^2$-value. Notice that for large $q$-values, the Tsallis
entropy is approximately constant for all $C^2$ values, while the
R\'enyi one seems to be much more sensitive in this respect.
Entropies tend to vanish for $C^2 \to 1$, because only pure states
can reach the maximum concurrence value. In the inset of Fig.\ref{fig2Renyi} we      
depict the behaviour of $\langle S_q \rangle_{C^2}$ as a function
of $1/q$ for a given value of the concurrence ($C^2 =0.6$), thus
illustrating the fact that the mean entropy is a monotonically
decreasing function of $q$. For large $q$-values the Tsallis
entropy cannot discriminate between different degrees of
entanglement for states with $C^2 <1$, while R\'enyi's measure can
do it. This fact is related to an important difference between the
behaviours, as a function of the parameter $q$, of R\'enyi's $R_q$
and Tsallis' $S_q$ entropies. The maximum value $R_q^{\rm max}$
attainable by R\'enyi's entropy (corresponding to the
equi-probability distribution) is independent of $q$,

\be
R^{\rm max}_q \, = \,  -\ln N, \ee

\noindent where $N$ is the total number of accesible states. On the contrary,
the maximum value reachable by the Tsallis entropy $S_q$ does depend upon $q$,

\be
S^{\rm max}_q \, = \, \frac{1 - N^{1-q}}{ (q-1)}.
 \ee

\begin{figure}
\begin{center}
\includegraphics[angle=270,width=.65\textwidth]{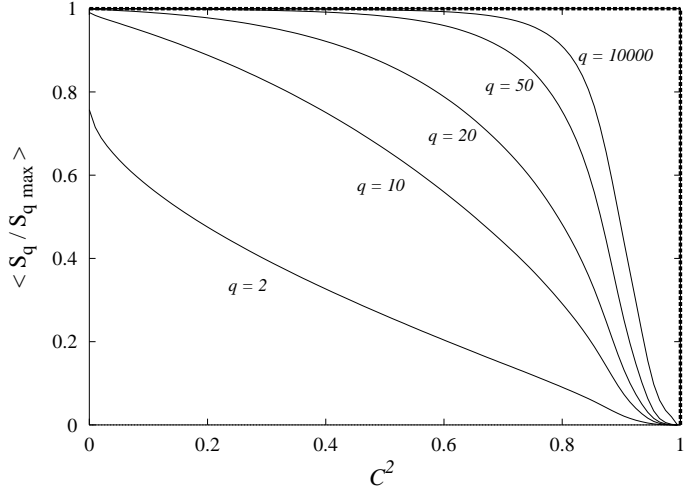}
\caption{Average value of the normalized Tsallis entropy
$\langle S_q \rangle/S_{q{\rm max}}$ vs. $C^2$, for several
$q$-values.  All depicted quantities are dimensionless.} 
\label{fig3Renyi}
\end{center}
\end{figure}

\noindent Clearly, $S^{\rm max}_q \rightarrow 0$ for $q \rightarrow \infty$.
One may think that the $q$-dependence of $S^{\rm max}_q$ may be appropriately
taken into account if one considers (instead of Tsallis' entropy itself), a
{\it normalized} Tsallis' entropy (see Fig.\ref{fig3Renyi}), 

\be \label{normal} S'_q = \frac{S_q}{S_q^{max}}, \ee

\noindent  For instance, in the case of two qubits one has,

\be \label{2q} S_q^{max} = \frac{1 - 4^{1-q}}{ (q-1)}, \ee

\noindent and we deal then with \be \label{normal1} S'_q = \frac{1-Tr [\hat
\rho^q]}{1-4^{1-q}} = \frac{1-\{[Tr (\hat \rho^q)]^{1/q}\}^q}{1-4^{1-q}}. \ee
Consider now the limit $q \rightarrow \infty$ for a density matrix $\hat \rho$
corresponding to a state of fixed concurrence $C$. In such a process one
immediately appreciates the fact that $[Tr (\hat \rho^q)]^{1/q} \rightarrow
\lambda_{max}$, where $\lambda_{max}$ is the largest eigenvalue of $\hat
\rho^q$. Thus, the limiting value we reach is \be \label{limite} S'_q
\rightarrow [1 - (\lambda_{max})^q], \ee and we see that this is always equal
to unity for all $C^2<1$ and vanishes exactly if $C^2=1$ (see Fig.\ref{fig3Renyi}). 
Consequently, even employing the normalized $S'_q$, the information concerning
the entropy-entanglement correlation tends to disappear in the $q\rightarrow
\infty$ limit.

  Returning to our discussion of the connection between entanglement and
(total) $q$-entropies of bipartite quantum systems, we have seen
that R\'enyi's entropy is particularly well suited for (i)
discussing the $q\rightarrow \infty$ limit and (ii) studying the
$q$-dependence of the entropy-entanglement correlations. For these
reasons, in the rest of the present contribution we will focus
upon R\'enyi entropy.

\begin{figure}
\begin{center}
\includegraphics[angle=270,width=.65\textwidth]{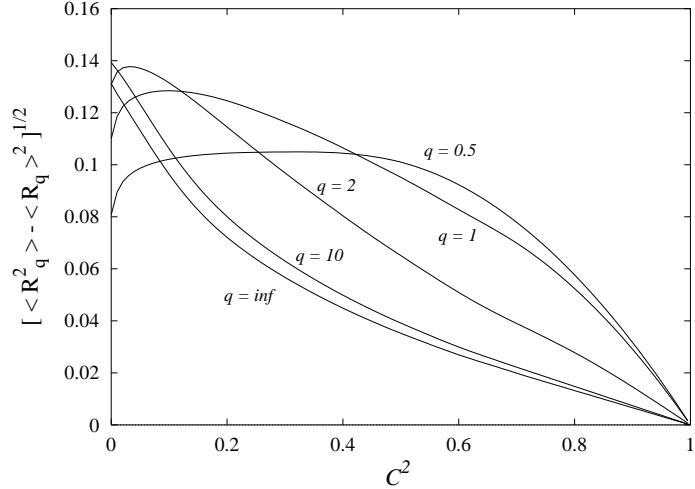}
\caption{Dispersion of the R\'enyi entropy
$\sigma^{(R)}_q=\left[\langle R_q^2 \rangle- \langle R_q \rangle^2
\right]^{1/2}$ for all qubits states with a given $C^2$, as a
function of $C^2$, and for several $q$-values.  All depicted
quantities are dimensionless.} 
\label{fig4Renyi}
\end{center}
\end{figure}

 We tackle now the question of the dispersion around these entropic averages.
Fig.\ref{fig4Renyi} is a graph of the dispersions 

\be \label{dispersion} \sigma^{(R)}_q \, = \,
 \left[\langle R_q^2 \rangle-
\langle R_q \rangle^2 \right]^{1/2}, \ee

\noindent as a function of $C^2$, for the same $q$-values of Fig.\ref{fig1Renyi}. We see 
that the size of the dispersions diminishes rather rapidly as $C^2$ increases
towards unity. Also, dispersions tend to become smaller as $q$ grows. This
suggests that, as $q$ increases, the correlation between $\langle R_q \rangle$
and entanglement improves. A similar tendency, but in the case of $S_q$, was
detected in \cite{CR02}.

  In order to estimate in a quantitative the sensitiveness of the average
$q$-entropy to changes in the value of $C^2$, we computed the
derivatives with respect to $C^2$  of the average value of
R\'enyi's entropy associated with states of given $C^2$,

  \be \label{derivadas} \frac{d\langle R_q \rangle}{d(C^2)} . \ee
                                                                             
\begin{figure}
\begin{center}
\includegraphics[angle=270,width=.65\textwidth]{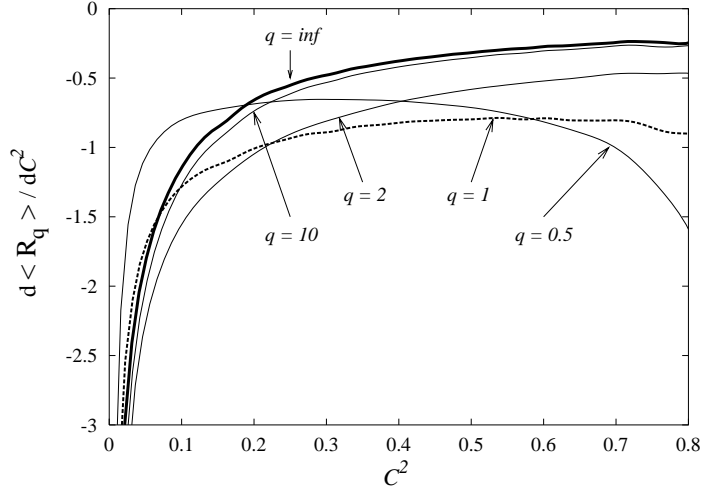}
\caption{The derivative $d\langle R_q \rangle/d(C^2)$, as
a function of the squared concurrence $C^2$, for several values of
the $q$-parameter. All depicted quantities are dimensionless.} 
\label{fig5Renyi}
\end{center}
\end{figure}

\noindent In Fig.\ref{fig5Renyi} we plot the above derivatives, against $C^2$, for $q=0.5,       
1, 2, 10$, and $\infty$. These derivatives fall abruptly to zero, in the
vicinity of the origin, as $C^2$ diminishes. As a counterpart, for all $q$, the
derivatives exhibit a rapid growth with $C^2$ for small values of the
concurrence. This growing tendency stabilizes itself and, for $q$ large enough,
saturation is reached.

Now let us assume that we know the value of the entropy $R_q[\rho]$ of certain
state $\rho$. How useful is this knowledge in order to infer the value of
$C^2$?. In other words, how good is $R_q$ as an ``indicator" of entanglement?
It has been suggested that $q=\infty $ provides a better ``indicator" of
entanglement than other values of $q$ \cite{CR02,GG01}. There are two
ingredients that must be taken into account in order to determine the $q$-value
yielding the best entropic ``indicator" of entanglement. On the one hand, the
sensitivity of the entropic mean value $\langle R_q \rangle$ to increments in
$C^2$, as measured by the derivative $d\langle R_q\rangle/d(C^2)$. On the other
hand, the dispersion $\sigma^{(R)}_q$, given by (\ref{dispersion}). A given
$q$-value would lead to a good entropic ``indicator" if it corresponds to (i) a
large value of $d\langle R_q\rangle/d(C^2)$, and (ii) a small value of
$\sigma^{(R)}_q$.  These two factors are appropriately taken into account if we
compute the ratio

\be \label{ratio} r \, = \, \left|\frac{\sigma^{(R)}_q}{d\langle R_q
\rangle/d(C^2)}\right|, \ee

\noindent
 between the dispersions depicted in Fig.\ref{fig4Renyi} and the derivative 
 of Fig.\ref{fig5Renyi}. The ratio $r$ provides a quantitative measure for the
strength of the entropic-entanglement correlations. The quantity $r$
constitutes an estimate of the smallest increment $\Delta C^2$ in the squared
concurrence which is associated with an appreciable change in $R_q$. In order
to clarify this last assertion, an analogy with the uncertainty associated with
the measurement of time in quantum mechanics can be established. Let us assume
that we can measure an observable $\hat A$. Then, the time uncertainty $\Delta
t$ depends upon two quantities, (i) the time derivative of the expectation
value of the observable, $d \langle \hat A \rangle/dt$, and (ii) the
uncertainty of the observable, $\Delta \hat A = [\langle \hat A^2 \rangle
-\langle \hat A \rangle^2]^{1/2}$. The time uncertainty is given by \cite{M61}

\be \label{deltime} \Delta t \, = \, \frac{\Delta \hat A}{d \langle \hat A
\rangle/dt}
 \ee

\begin{figure}
\begin{center}
\includegraphics[angle=270,width=.65\textwidth]{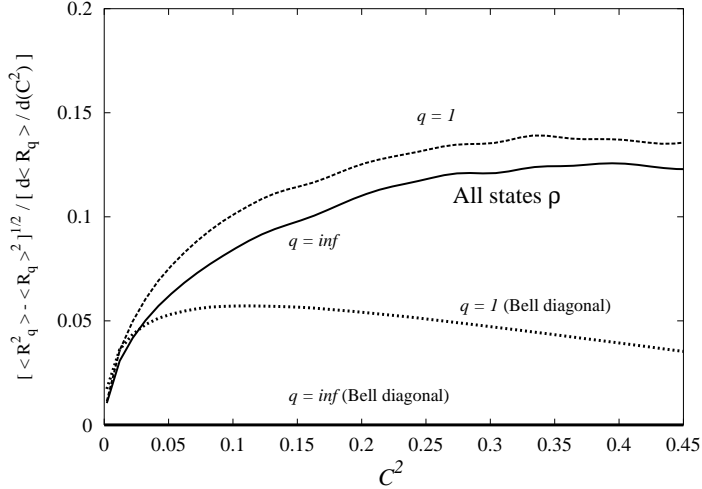}
\caption{The absolute value of the quotient $r =
\left|\frac{\sigma^{(R)}_q}{d\langle R_q \rangle/d(C^2)}\right|$,
as a function of the squared concurrence $C^2$, for $q=1$ and
$q=\infty$. The two upper curves correspond to all states in the
two-qubit state-space ${\cal S}$. The lower curves correspond to
states diagonal in the Bell basis. All depicted quantities are
dimensionless.} 
\label{fig6Renyi}
\end{center}
\end{figure}

\noindent The above expression for $\Delta t$ gives an estimation
of the smallest time interval that can be detected from
measurements of the observable $\hat A$. In the analogy we want to
establish, $C^2$ plays the role of $t$, and $R_q$ plays the role
of the observable $A$. The ratio $r$ is depicted in Fig.\ref{fig6Renyi}, 
as a function of $C^2$, for $q=1$ and $q=\infty$. The two upper curves
in Fig.\ref{fig6Renyi} correspond to the $r$-values obtained when all the 
states in the two-qubits state-space ${\cal S}$ are considered. On
the other hand, the lower curves are the ones obtained when the
computation of $r$ is restricted to states diagonal in the Bell
basis. When all states in ${\cal S}$ are considered, the values of
$r$ associated with $q=\infty$ are seen to be smaller than the
values corresponding to $q=1$, which can be construed as meaning
that the $q$-entropies with $q=\infty$ can indeed be regarded as
better ``indicators" of entanglement than the $q$-entropies
associated with finite values of $q$, as was previously suggested
in \cite{CR02,GG01}. Alas, the results depicted in Fig.\ref{fig6Renyi} 
indicate that this improvement of the entropy-entanglement correlation
associated with $q=\infty$ is not considerable. The usefulness of
$q$-entropies with $(q\rightarrow \infty)$ as an ``indicator" of
entanglement was proposed in \cite{GG01} on the basis of the
behaviour of states diagonal in the Bell basis. As already
mentioned, the squared concurrence $C^2$ of states $\rho_{\rm Bell
}$ diagonal in the Bell basis can be expressed as a function of
$R_{\infty}$, since both these quantities depend solely on the
largest eigenvalue $\lambda_m$ of $\rho_{\rm Bell }$ (in
particular, $R_{\infty}=-\ln \lambda_m$). This means that, as
pointed out in \cite{CR02,GG01}, for states diagonal on the Bell
basis there is a perfect correlation between $C^2$ and
$(q\!=\!\infty)$-entropies (and, consequently, $r$ vanishes). This
implies that, when restricting our considerations {\it only} to
states diagonal in the Bell basis, the entropy-entanglement
correlation is much more strong for $q=\infty$ than for other
values of $q$. States diagonal in the Bell basis are important for
many reasons, but their properties are by no means typical of the
totality of the state-space ${\cal S}$. See for instance, as
depicted in Fig.\ref{fig6Renyi}, the behaviour of $r$ (for $q=1$) 
associated with (i) all states in ${\cal S}$ and (ii) states diagonal 
in the Bell basis. There are remarkable differences between the two cases.

We thus find ourselves in a position to assert that the
relationship between the $q$-entropies and the amount of
entanglement exhibited by the family of states diagonal in the
Bell basis does not constitute a reliable guide to infer the
typical behaviour of states in the two-qubits state-space ${\cal
S}$ \cite{xerec}. When considering the complete state-space ${\cal S}$, the
${q\!=\!\infty}$-entropies turn out to be only a slightly better,
as entanglement ``indicators", than the entropies associated with
other values of $q$.

\section{Concluding remarks}

In this Chapter we have extensively explored all possible connections 
of the so called $q$-entropic information measures and their connection 
with entanglement. We have first explored in composite bipartite systems the 
features relevant to a family of these entropies, the conditional $q$-entropies, in 
terms of the total volume occupied by those states which do not violate the 
classical entropic inequalities. 
These inequalities constitute a necessary criterion for clearing up the 
separability of a given mixed state $\rho$, being also sufficient for pure states.
Secondly, the connection with entanglement and mixedness has been also studied 
along similar lines. Also, we focused our attention on the interesting properties 
that link a particular class of states, the so called Maximally Entangled Mixed 
States (MEMS), with the violation of the former entropic inequalities. Finally, 
we showed that there exist a direct correlation between average entropies for a given 
value of the concurrence and this measure of entanglement, for different values 
of the entropic parameter $q$. 

We then conclude that \cite{batle,JPhysA2,BCPP05,xerec}
\begin{itemize}

\item After a systematic survey of the space of pure and mixed states of
bipartite systems of arbitrary dimension, the
monotonicity with $q$ of both the Tsallis and R\'{e}nyi entropies
has been analyzed for two-qubits and a qubit-qutrit system, for
different values of the rank of the pertinent (mixed state)
statistical operator $\rho$. In spite of the fact that most states
have a Tsallis or R\'{e}nyi conditional entropy behaving in a
monotonic fashion with $q$, the proportion of these states always
diminishes as the rank of the state $\rho$ decreases, regardless
of the dimension of the system and the conditional entropy used.
The proportion of states with a monotonous conditional entropy is
larger for the case of the Tsallis information measure.

Concerning the volumes in state-space associated with states complying with the
``classical" entropic inequalities, we have presented results for states of
dimensions $2 \times 2$ up to $2 \times 10$ and for states ranging from $3
\times 3$ to $3 \times 7$. In general, the volume occupied by states with
positive conditional $q$-entropies (for a given $q$) is not a function solely
of the total dimension $N=N_1\times N_2$. Instead, it depends  on both
subsystems' dimensions, $N_1$ and $N_2$. For a given fixed value of $N_1=2,3$,
and for $q$-values above a special value $q^{*}$ (which itself depends upon
$N_1$), the alluded to volume decreases in a monotonous way with $N_2$.

In addition, the behaviour of two-qudits systems of dimension $3 \times 3$ and
$4 \times 4$ has also been taken into account. In all these cases, our
numerical results indicate that the probability of finding states endowed
either with (i) positive conditional $q$-entropies and a positive partial
transpose, or (ii) negative conditional $q$-entropies and a non positive
partial transpose, increase in a monotonic way with $q$. However, the largest
value of this probability (corresponding to $q=\infty$) diminishes in a very
fast fashion with $D$.

Finally, we computed the volumes (for composite systems with Hilbert space
dimensions $2\times N_2$ and $3\times N_2$) occupied by states complying with
the majorization separability criterion, and compared them with the volumes
corresponding to states endowed with positive $(q=\infty)$-conditional
entropies. The qualitative behaviour (as a function of $N_2$) of the volumes
associated with states complying (i) with the majorization condition and (ii)
with the classical, $(q=\infty)$-conditional entropic inequalities, turned out
to be qualitatively alike (and very close to each other in the case of systems
of dimension $3\times N_2$).

\item We have determined, as a
function of the degree of mixture, and for different values of the
entropic parameter $q$, the volume in state space occupied by
those states characterized by positive values of the conditional
$q$-entropy. We also computed, for different values of $q$, the
global probability of classifying correctly an arbitrary state of
a two-qubits system (either as separable or as entangled) on the
basis of the signs of its conditional $q$-entropies. This probability
exhibits a monotonous increasing behaviour with the entropic
parameter $q$. The approximate values of these probabilities are
$0.6428$ for $q=1$ and $0.7428$ in the limit $q\rightarrow
\infty$.

An interesting conclusion that can be drawn from the numerical
results reported here is that, notwithstanding the known non
monotonicity in $q$ of the conditional $q$-entropies \cite{VW02}, the
volume corresponding to states with positive conditional
$q$-entropies ($q>0$) is, for any degree of mixture, a monotonous
decreasing function of $q$.

Similar calculations were performed for qubit-qutrit systems and
for composite systems described by Hilbert spaces of larger
dimensionality. We pay particular attention to the limit case
$q\rightarrow \infty$. Our numerical results indicate that, for
composite systems consisting of two subsystems characterized by
Hilbert spaces of equal dimension $N_1$, the probability of
finding states with positive $q$-entropies tend to 1 as $N_1$
increases. In oder words, as $N_1\rightarrow \infty $ most states
seem to behave (as far as their conditional $q$-entropies are
concerned) classically.

\item The maximally entangled states of Munro, James, White, and
Kwiat \cite{MJWK01} are shown to exhibit
interesting features vis \`a vis conditional entropic measures. The
same happens with the Ishizaka and Hiroshima states \cite{IH00}, 
whose entanglement-degree can not be
increased by acting on them with logic gates. Maximally 
entangled states with classical entropic behaviour are seen to
exist in the space of two qubits. Special meaning can be assigned
to the Munro {\it et al.} special participation ratio of 1.8.
For entangled states with classical conditional entropic behaviour
(ESCRE),  the
maximum degree of entanglement attainable obtains at $R=1.8$.
Even though the entropic criteria are not universally valid for all
two-qubits states (yielding only a necessary condition for
separability), they have been shown here  to preserve their full
applicability for an important family of states, namely, those with
cannot increase their entanglement under the action of logic gates
for participation rations in the interval ($R\in[1,1.8]$). This in
turn, gives an entropic meaning to this special $R-$value
encountered by Munro {\it et al.} \cite{MJWK01}.
We find explicit ``boundaries" to $C_{IH}$ when we express the
degree of mixture using the maximum eigenvalue $\lambda_{max}$ of
$\rho^{IH}$. It would seem that the characterization of the
entanglement for these states, using the $\lambda_{max}$ criterion,
provides the best insight into the entanglement features of these
states.
Beyond a certain value for the participation ratio, namely, $R=1.8$, {\it
all states}, not necessarily the ones considered before, can be
correctly described by the entropic inequalities as far as this
criterion is concerned. One may argue that if the quantum
correlations are strong enough (greater than $C^{max}_{R=1.8}$ or
$C^{max}_{\lambda_{max}=\frac{2}{3}}$), there is still room for
entropic-based separability criteria to hold.

\item By recourse to the same Monte Carlo procedure as before, we have studied 
the $q$-dependence of 
the correlations exhibited by two-qubits states between (i) the
amount of entanglement and (ii) the $q$-entropies. It was previously conjectured by
 other researchers, on the basis of the study of states diagonal in the Bell
 basis, that the $q$-entropies associated with $q=\infty $ are better
 ``indicators" of entanglement than the entropies corresponding to finite
 values of $q$. In other words, it was suggested that the $q$-entropy with
 $q=\infty $ exhibits a stronger correlation with entanglement than the other
 $q$ entropies. By a comprehensive numerical survey of the complete (pure and
 mixed) state-space of two-qubits, we have shown  here that the alluded to
 conjecture is indeed correct. However, when globally considering the whole state-space
 the advantage, as an entanglement indicator, of
$(q\!=\!\infty)$-entropy turns out to be much smaller than what can be inferred from 
the sole study of
 states diagonal in the Bell basis. This constitutes an instructive example of
 the perils that entails trying to infer  typical properties of general two-qubits
 states from the study of just a particular family of states, such as those diagonal
 in the Bell basis.

\end{itemize}

\chapter{Entanglement, $q$-entropies and mixedness}

  The amount of entanglement and the purity of quantum states of composite
  systems exhibit a dualistic relationship. As the degree of
  mixture increases, quantum states tend to have a smaller
  amount of entanglement. In the case of two-qubits systems,
  states with a large enough degree of mixture are always
  separable \cite{ZHS98}. A detailed knowledge of the relation between the degree
  of mixture and the amount of entanglement is essential in order to
  understand the limitations that mixture imposes on quantum information
  processes such as quantum teleportation or quantum computing.
  To study the relationship between entanglement and mixture
  we need quantitative measures for these two quantities.
  The entanglement of formation provides a natural quantitative
  measure of entanglement with a clear physical motivation.
  As for mixedness, there are several measures of mixture that can
  be useful within the present context. The von Neumann measure

  \be \label{slog}
  S_1 \, = \, - Tr \left( \hat \rho \ln \hat \rho \right),
  \ee

  \noindent is important because of its relationship with the thermodynamic
  entropy. On the other hand, the so called participation ratio,

  \be \label{partrad}
  R(\hat \rho) \, = \, \frac{1}{Tr(\hat \rho^2)},
  \ee

  \noindent
  is particularly convenient for calculations \cite{ZHS98,MJWK01}.
  The $q$-entropies, introduced in Chapters 4 and 8, which are functions of 		   
  the quantity

  \be \label{trq}
  \omega_q \, = \, Tr \left( \hat \rho^q \right),
  \ee

  \noindent
  provide one with a whole family of measures for the degree of mixture.
  In the limit $q\rightarrow 1 $ these measures incorporate (\ref{slog})
  as a particular instance. On the other hand, when $q=2$ they are
  simply related to the participation ratio (\ref{partrad}). 

  Next in the present Chapter we study some aspects of the
  relationship between entanglement and purity, using the
  $q$-entropies as measures of mixture \cite{BCPP02b}. In particular, we 
  derive analytically the 
  probability (density) $F(R)$ to find a two qubit state with
  participation ratio $R$. Several distributions for higher 
  bipartite systems are obtained numerically, and some analytical 
  results are found. We shall also discuss in detail the
  limit case $q\rightarrow \infty $ and its connection with the
  use of the largest eigenvalue of $\hat \rho$ as a measure of
  the degree of mixture. For $q \in [2,\infty)$, we obtain 
  analytically the values of the $q$-entropies 
  above which all states are separable. Finally, 
  we derive the analytic probability distribution to find a 
  qubit-qutrit state endowed with a maximum eigenvalue $\lambda_m$ 
  and expose the general geometric framework for deriving 
  equivalent distributions in arbitrary bipartite systems 
  of dimension $N=N_A\times N_B$.

\section{Distribution of two-qubits states according to their mixture}


As described in Sec. (7.1), the space ${\cal S}$ of all (pure and mixed) states of a quantum
system described by an $N$-dimensional Hilbert space can be
regarded as a cartesian product space ${\cal S} = {\cal P} \times \Delta$, where 
$\cal P$ stands for the family of all
complete sets of ortonormal projectors $\{ \hat P_i\}_{i=1}^N$,
$\sum_i \hat P_i = I$ ($I$ being the identity matrix), and $\Delta$
is the set of all real $N$-tuples $\{\lambda_1, \ldots, \lambda_N
\}$, with $\lambda_i \ge 1$ and $\sum_i \lambda_i = 1$. A detailed description of 
the space ${\cal S}$ can be found in Appendix B. 
It suffices here to mention that the natural measure

\be \label{memu}
 \mu = \nu \times {\cal L}_{N-1}
 \ee

 \noindent is the one used in the random generation of two-qubits 
 states ($N=4$). Also, in order to study the distribution of two-qubit 
 states according
  to their degree of mixture, we identify the simplex $\Delta $ with 
  a regular tetrahedron	of side length 1, in ${\cal R}^3$, 
  centred at the origin. The mapping connecting the points
  of the simplex $\Delta $ (with coordinates $(p_1,\ldots, p_4)$)
  with the points $\bf r$ of the tetrahedron is given explicitly 
  in Appendix C. Next we consider two special $q$-cases that are 		 
  relevant in our study.
  
  \subsection{The case $q=2$}

  In this case the degree of mixture is characterized by the
  quantity $\omega_2 = Tr(\hat \rho^2)=\sum_i p_i^2$. 
  This quantity is related to the distance $r=\mid {\bf r} \mid$
  to the centre of the tetrahedron $T_{\Delta}$ by

  \be \label{tetra2}
  r^2 \, = \, -\frac{1}{8} \, + \, \frac{1}{2} \omega_2.
  \ee

 \noindent
 Thus, the states with a given degree of mixture lie on the
 surface of a sphere of radius $r$ concentric with the
 tetrahedron $T_{\Delta}$. See Appendix C for details.


The volume associated with states endowed with a value of
$\omega_2$ lying within a small interval $dw_2$ is clearly
associated with the volume $dV$ of the subset of points in
$T_{\Delta}$ whose distances to the centre of $T_{\Delta}$ are
between $r$ and $r+dr$, with $r dr = \omega_2 \, d\omega_2$. Let
$\Sigma_r$ denote the sphere of radius $r$ concentric with
$T_{\Delta}$. The volume $dV$ is then proportional to the area
$A(r)$ of the part of $\Sigma_r$ which lies within $T_{\Delta}$.
In order to compute the aforementioned area, it is convenient to
separately consider three different ranges for the radius $r$.

Let us first consider the range of values  $r \in [0, h_1]$, where
$h_1={1 \over 4 }\sqrt{2 \over {3}}$ is the radius of a sphere
tangent to the faces of the tetrahedron $T_{\Delta}$. In this case
the sphere $\Sigma_r$ lies completely within the tetrahedron
$T_{\Delta}$. Thus, the area we are interested in is just the area
of the sphere,

\begin{equation} \label{a1}
A_{I}(r) = 4 \pi r^{2}. \label{s1}
\end{equation}

We now consider a second range of values of the radius, $r \in
[h_1, h_2]$, where  $h_2 ={\sqrt{2}\over 4}$ denotes
 the radius of a sphere which is tangent to the sides of the tetrahedron
 $T_{\Delta}$. In this case, the area of the portion of $\Sigma_r$
 which lies within $T_{\Delta}$ is

\begin{equation} \label{a2}
A_{II}(r) = 4 \pi \left[r^{2}-2r(r-h_1)\right]. \label{s2}
\end{equation}

 Finally, we consider the range of values  $r \in [h_2, h_3]$, where
  $h_3={\sqrt{6}\over 4}$ is the radius of a sphere passing through the vertices
 of $T_{\Delta}$. In this case the area $A_{III}$ of the part of the sphere
 $\Sigma_r$ lying within $T_{\Delta}$ is

\begin{equation} \label{a3}
A_{III}(r) = 4 (S_{A}-3S_{B}), \label{s3}
\end{equation}

\noindent where

\ben S_A &=& r^2 (3 \alpha - \pi), \cr
 S_B &=& r^2 \left[ h (-\pi + 2 \sin^{-1} (C_1 C_2)) +
 2 \sin^{-1} \sqrt{1 - C^2_1 C^2_2 \over 1
+ C^2_2} \right]. \een

\noindent The quantities appearing in the right hand sides of the
above expressions are defined by

\be
 \alpha = \cos^{-1} \left[{\cos A - \cos^2 A \over \sin^2
A}\right]; \,\,\,\, A = 2 \sin^{-1} (D_1/r);  \,\,\,\, D_1 = {1
\over 2} \left({1\over 2} - \sqrt{r^2 - {1 \over 8}}\right),
 \ee

 \noindent
 and

\ben
h &=& h_1/r; \ \  C_1 ={h \over \sqrt{1 - h^2}}; \ \ C_2 = {C_B
\over \sqrt{1 - C^2_B} }; \cr
C_B &=& \sqrt{D^2_2 - D^2_1 \over r^2 - D^2_1}; \ \ D_2 = r \sqrt{1 - h^2}. 
\een

\noindent
 Using the relation between $r$ and the participation rate
$R = 1/ Tr(\rho^2)$,

\be \label{radiopart}
 r^2 = -  {1 \over 8} + {1 \over 2 R},
\ee

\noindent
 we analytically obtained the probability $F(R)$ of
finding a quantum state with a participation rate $R$,

\be
\label{ffrr} F(R) \, = \, f(r)   \left| \frac{dr}{dR} \right|,
 \ee

 \noindent

\begin{figure}
\begin{center}
\includegraphics[angle=270,width=0.65\textwidth]{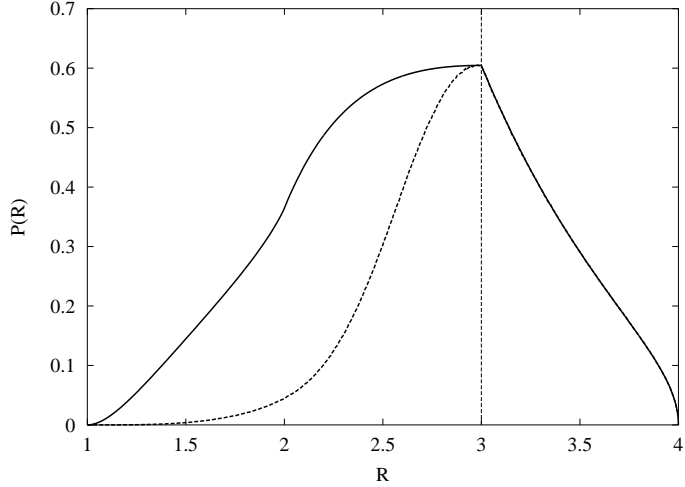}
\caption{Probability (density) $P(R)$ of finding two-qubits states with a given
amount of mixedness $R$. The solid line corresponds to all states and the 
dashed line to those ones which are separable {\it only}, 
given by $P_{sep}(R)$. This curve is normalized to the overall probability 
of finding a separable state so that both $P(R)$ and 
$P_{sep}(R)$ can be compared at the same scale. The maximum occurs at the 
separability border $R=3$.}
\label{fig1mix}
\end{center}
\end{figure}

 where $f(r) = A(r)/({\rm Volume}[T_{\Delta}])$, and $A(r)$ is given by
 equations (\ref{a1}-\ref{a3}). The plot of $F(R)$ is depicted in Fig.\ref{fig1mix}. 
The distribution $F(R)$ was first determined numerically by
 Zyczkowski et al. in \cite{ZHS98}. Here we compute $F(R)$
 analytically \cite{BCPP02b} and, as expected, the calculations coincide with the 
 concomitant numerical results and the ones reported in \cite{ZHS98}.
\newline
\newline
{\bf Information-theoretical approach to entanglement 
and $R$}
\newline

\begin{figure}
\begin{center}
\includegraphics[angle=270,width=0.65\textwidth]{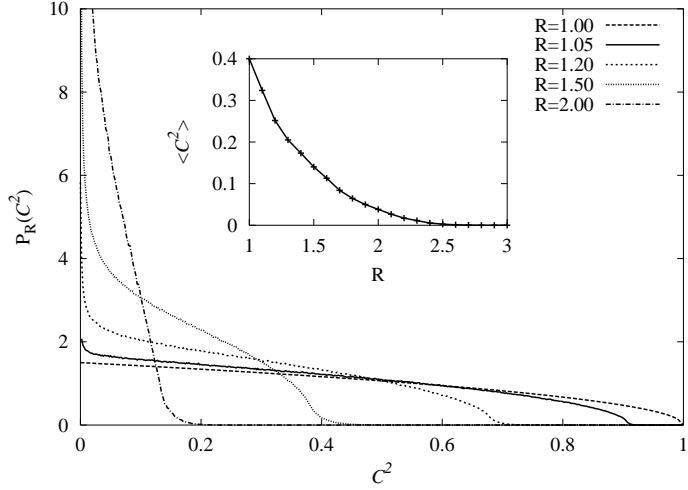}
\caption{Probability (density) $P_R(C^2)$ of finding two-qubits states 
(generated according to the $\mu_Z$-measure) with fixed degree of mixture $R$, 
and endowed with a given value of the 
concurrence squared $C^2$. The range of available values for $C^2$ 
continuously decreases as we approach the limiting value $R=3$. 
The inset shows the evolution of $C^2$ vs. $R$.}
\label{fig1bismix}
\end{center}
\end{figure}

\begin{figure}
\begin{center}
\includegraphics[angle=270,width=0.65\textwidth]{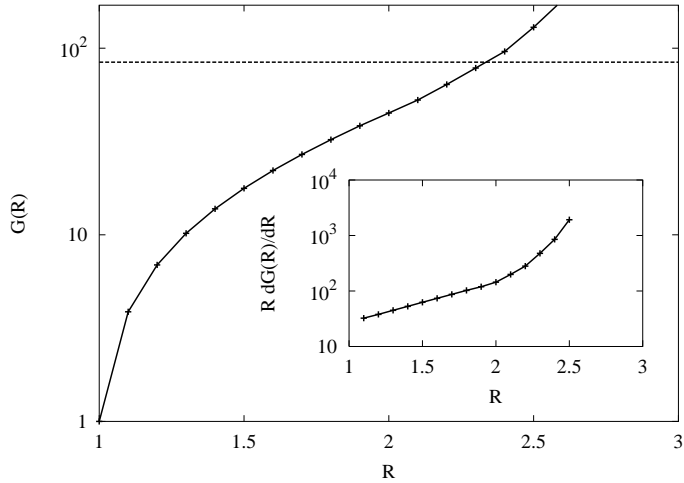}
\caption{$\ln{G}$ vs. $R-$plot, with
$G=[S_R-H_{max}]/S_{R=1}$  the information gain with respect to
the uniform distribution $H_{max}$ and $S_R$ the
information-amount contained in the distribution $P_R(C^2)$ of
Fig.\ref{fig1bismix}. The horizontal dashed line indicates the information gain
 corresponding to the knowledge of the probability $P(C^2)$ for all
 states, mixed or not. The inset shows (again in a logarithmic plot)
 the evolution of the analogue of a specific heat $C_R=  R \frac{d G}{d R}$ vs. $R$.
 A quadratic evolution is observed in the range $\in [1,2]$, which becomes 
steeper in $\in [2,3]$.}
\label{fig1bis2mix}
\end{center}
\end{figure}

According to the $R-$value, two zones are 
clearly to be distinguished. In Fig.\ref{fig1mix}, entanglement is to be 
encountered only in the zone lying at the ``west" ($R<3$) of the graph. 
As matter of fact, Fig.\ref{fig1bismix} depicts, for several values of the 
participation ratio $R$, the probability $P_R(C^2)$ of finding in our 15-dimensional 
space $\mathcal{S}$ of two-qubits a bipartite state of 
concurrence squared $C^2$ (\ref{concurrence}).  
The inset is a $\langle C^2 \rangle$ vs. $R$ plot.
 It is well known that i) no
 entanglement exists for $R > 3$ and ii) $R=1$ only for pure states.
 It is clear that $R$ plays an important role in pre-determining the
 possibility of finding entanglement. Obviously, to each curve 
$P_R(C^2)\,$vs.$\, C^2$ one can assign a Shannon information measure

 \be S_R= -\int_0^1\,dC^2\,P_R(C^2)\,\ln{[P_R(C^2)]}, \ee

 \noindent that measures the informational amount contained in the
 distribution $P_R(C^2)$. If we, in turn, plot now $S_R$ vs. $R$, we
 establish a correlation between the participation ratio and our
 knowledge regarding the entanglement-distribution (ED) in
 $\mathcal{S}$. This is done in Fig.\ref{fig1bis2mix}, that is a logarithmic plot. 
In the vertical axis we have the logarithm of the information gain
 $G=[S_R-H_{max}]/S_{R=1}$ (with reference
 to the uniform distribution, that always entails maximum ignorance).
   For pure states ($R=1$) $S_{R=1}$ is large, but it does not reach the
 uniform value  $H_{max}$. As stated, we calibrate the information (or
 ignorance) units so that $G=1$ corresponds to the ignorance
 concerning the ED for bipartite pure states. It is clearly seen that, as
 $R$ grows, $G$ steadily increases. We
 can assert, for instance, that if $R=1.5$, our information gain
 amounts to
 $\sim 10$ times the one that we possess
 at $R=1$.
 The upper, dashed horizontal line indicates the information gain
 corresponding to the knowledge, with reference to our 15-dimensional
  space, of the probability $P(C^2)$ for all
 states, mixed or not (See for instance \cite{BCPP02b}). Of course, we
 are gaining information because of the known fact that the
 number of entangled states diminishes with $R$.

  Let us recall here that for a continuous probability distribution (cpd) the
 entropy is defined only up to an arbitrary additive constant. Only
 entropy {\it differences} do make sense (the information gain is obtained 
with respect to $H_{max}$). Also, for
 cpd's, these differences can become infinite, because one needs an
 infinite informational amount to pick up a point in the continuum. 
 \newline
 \newline
 {\bf The general $F(R)$ in $N=N_A\times N_B$ dimensions}
 \newline

 We have just obtained the distribution $F(R)$ vs. $R$ for two-qubits 
 ($N=4$), under the assumption that they are distributed according 
 to measure (\ref{memu}). Through a useful analogy, we have mapped the 
 problem into a geometrical one in ${\cal R}^3$ regarding interior and 
 common sections of two geometrical bodies. In the previous case 
 we saw that the main difficulty lies in the third region, where 
 the region of the growing sphere inside the tetrahedron is not 
 described by a spherical triangle. The extension to higher 
 dimensions, however, requires a thorough account of the geometrical 
 tools required, but still it is in principle possible. 
 So, one can find the distribution of states according to 
 $R = 1/ Tr(\rho^2)$ basically by computing the surface area of a 
 growing ball of radius $r$ in $N-1$ dimensions ({\it sphere}) 
 that remains inside an outer regular $N$-polytope $T_{\Delta}$ 
 ({\it tetrahedron}) of unit length, excluding the common regions. 
 A $(N-1)$-dimensional sphere can be parameterized in cartesian 
 coordinates

\begin{figure}
\begin{center}
\includegraphics[angle=270,width=0.65\textwidth]{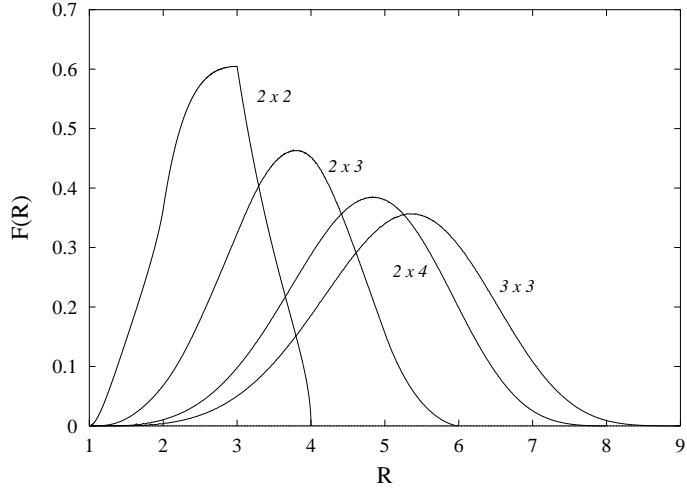}
\caption{Plot of the $F_N(R)$ distributions of mixed states $\rho$ numerically 
computed in arbitrary dimensions, generated according to (\ref{memu}). 
As we increase the total dimension $N$, the curves become smoother, 
in correspondence with our geometric interpretation. See text for details.} 
\label{figentR}
\end{center}
\end{figure}

\ben \label{vectSphere}
x_1 &=& r \sin(\phi_1) \sin(\phi_2) \sin(\phi_3)\,...\,\sin(\phi_{N-3}) 
\sin(\phi_{N-2}) \nonumber \\
x_2 &=& r \sin(\phi_1) \sin(\phi_2) \sin(\phi_3)\,...\,\sin(\phi_{N-3}) 
\cos(\phi_{N-2}) \nonumber \\
x_3 &=& r \sin(\phi_1) \sin(\phi_2) \sin(\phi_3)\,...\,\cos(\phi_{N-3}) 
\nonumber \\
...\nonumber \\
x_{N-2} &=& r \sin(\phi_1) \cos(\phi_2) \nonumber \\
x_{N-1} &=& r \cos(\phi_1),
\een

\noindent with the domains $0 \le \phi_j \le \pi$ for $1 \le j \le N-3$ and 
$0 \le \phi_{N-2} < 2\pi$. The definition of the 
$N$-polytope $T_{\Delta}$ then is required. This problem is not 
trivial at all, because new geometrical situations appear in the intersection 
of these two bodies. In point of fact there are $N-2$ intermediate 
regimes between $R=N$ and $R=1$ appearing at integer 
values of $R$ (recall the previous two-qubits case), where a change 
in the growth of interior hyper-surfaces occurs (at the values 
$r_i=\sqrt{(N-R_i)/2R_iN}, R_i=1,...,N$). In any case we can 
always generate random states $\rho$ in arbitrary dimensions (see Appendix B)
and compute the corresponding $F_N(R)$ distributions. This is done 
in Fig.\ref{figentR} for several cases.                    	                     
The relation (\ref{radiopart}) is generalized to $N$ dimensions in the form 

\be \label{rRN}
r^2 = -  {1 \over 2N} + {1 \over 2 R}.
\ee

\noindent The distribution $F_N(R), \, R \in [N-1,N]$ can be obtained 
analytically 

\be
F_N(R) \, \sim \, \frac{1}{R^2}\big[ \frac{1}{R}-\frac{1}{N} 
\big]^{\frac{N-3}{2}}
\ee

\noindent which has been numerically checked. 
The particular form of $F(R)$ for arbitrary $N$ is difficult 
to obtain, but nevertheless one can obtain quantitative results for 
asymptotic values of $N$. It may be interesting to know the position 
of the maximum of $F(R)$ or the mean value $\langle R \rangle$, which turns 
to be $\simeq N/2$ \cite{Zyckasimp} for states $\rho$ generated 
according to (\ref{memu}). There is the so called Borel lemma \cite{Borel} 
in discrete mathematics that asserts that (translated to our problem) 
when you grow a $(N-1)$-ball inside $T_{\Delta}$, from the moment that it 
swallows, say, 1/2 of the volume of it, then the area outside drops very 
quickly with further grow. So the maximum intersection with the sphere should 
be approximately for the radious $r^{*}$ where the volume of the ball $V_{N-1}$ 
{\it equals} that of the $T_{\Delta}$-polytope $V_T$\footnote{The usual formulas 
for the volumes of ($N-1$)-dimensional spheres and regular unit $N$-polytopes are 
$V_{N-1}=\frac{\pi^{(N-1)/2}}{\Gamma(\frac{N-1}{2}+1)}r^{N-1}$ and 
$V_T=\frac{1}{(N-1)!}\sqrt{\frac{N}{2^{N-1}}}$, respectively.}. 
It is then that we can assume 
that the position of $R(r^{*}) \simeq R^{\prime}$ such that $F(R=R^{\prime})$ 
is maximal. Substituting $r^{*}$ in (\ref{rRN}), and after some algebra, 
we obtain the beautiful result

\be \label{deltaR}
\lim_{N \to \infty} \, \frac{1/R(r^{*})}{1/N} \, = \, \frac{2\pi+e}{2\pi} 
\, \simeq \, 1.43.
\ee

\noindent In other words, $F_N(1/R) \sim \delta(1/R-1/N)$ for large $N$.

We must emphasize that this type of distributions $F_N(R)$ are 
``degenerated" in some cases, that is, different systems may 
present identical $F(R)$ distributions (for instance, there is nothing 
different from this perspective between $2\times 6$ and 
$3\times 4$ systems). We do not know to what extend these 
distributions are physically representative of such cases, as far 
as entanglement is concerned. What is certain is that all 
states $\rho$ with $R \in [N-1,N]$ possess a positive partial transpose. In 
point of fact, they are indeed separable, as shown in \cite{balls}.  
We merely mean by this that a state close enough to the maximally mixed state 
$\frac{1}{N}I_N$ is always separable. In other words, states lying on 
$(N-1)$-spheres with radius $r \le r_c\equiv 1/\sqrt{2N(N-1)}$ 
are always separable.

\subsection{The case $q= \infty $}

Coming back to two-qubits, the quantity $\omega_q$ is not appropriately 
suited to discuss the
limit $q\rightarrow \infty$. However, $\omega_q^{1/q}$ does
exhibit a nice behaviour when $q\rightarrow \infty$. Indeed, we
have

\be \label{limiqinf} \lim_{q\to \infty } \, \left( Tr \rho^q
\right)^{1/q} \, = \,\lim_{q\to \infty } \, \left( \sum_i
p_i^q\right)^{1/q} \, = \, \lambda_m, \ee

\noindent where

\be \lambda_m = \max_{i} \{ p_i \} \ee

\noindent is the maximum eigenvalue of the density matrix $\rho$.
Hence, in the limit $q\to \infty $, the $q$-entropies (when
properly behaving) depend only on the largest eigenvalue of the
density matrix. For example, in the limit $q\to \infty $, the 
R\'{e}nyi entropy reduces to

\be \label{reninf} S^{(R)}_{\infty} \, = \, -\ln \left( \lambda_m
\right).
 \ee

\noindent
 It is worth realizing that the largest eigenvalue itself
constitutes a legitimate measure of mixture. Its extreme values
correspond to (i) pure states ($\lambda_m =1$) and (ii) the
identity matrix ($\lambda_m = 1/4$). It is also interesting to
mention that, for states diagonal in the Bell basis, the
entanglement of formation is completely determined by $\lambda_m$ 
(This is not the case, however, for general states of
two-qubits systems).

 In terms of the geometric representation of the simplex $\Delta$,
 the set of states with a given value $\lambda_m$ of their maximum 
 eigenvalue is represented by (see Appendix C) the tetrahedron                 
determined by the four planes

 \be \label{geolambda}
  \lambda_m \, = \, 2({\bf r} \cdot {\bf r}_i) \, + \, \frac{1}{4},
  \,\,\,\,\,\, i=1,\ldots, 4.
 \ee

  \noindent
  The four vertices of this tetrahedron are given by the
  intersection points of each one of the four possible triplets
  of planes that can be selected among the four alluded to planes.

For $q \to \infty$ the accessible states with a given degree of
mixture are on the surface of a small tetrahedron $T_l$ concentric
with the tetrahedron $T_{\Delta}$. We are going to characterize
each tetrahedron $T_l$ (representing those states with a given
value of $\lambda_m$) by the distance $l$ between (i) the common
centre of $T_{\Delta} $ and $T_l$ and (ii) each vertex of $T_l$.
The volume associated with states with a value of $\lambda_m$
belonging to a given interval $\lambda_m$ is proportional to the
area $A(l)$ of the portion of $T_l $ lying within $T_{\Delta}$.

Following a similar line of reasoning as the one pursued in the
case $q=2$, we consider three ranges of values for
$l$. The first range of $l$-values is given by $l \in [0, h_1/3]$.
The particular value  $l = h_1/3$ corresponds to a tetrahedron
$T_l$ whose vertices are located at the centres of the faces of
$T_{\Delta}$. Within the  aforementioned range of $l$-values,
$T_l$ is lies completely  within $T_{\Delta}$. Consequently,
$A(l)$ coincides with the area of $T_l$,

\begin{equation}
A_{I}(l) = 24 \sqrt 3 l^2 \label{s12}.
\end{equation}

\noindent The second range of $l$-values corresponds to $l \in
[h_1/3, h_1]$. The area of the part of $T_l$ lying within
$T_{\Delta}$ is now

\begin{equation}
A_{II}(l) =3 \sqrt 3 \left[8l^2 - {3 \over 2} (3l-h_1)^2\right]
\label{s22}
\end{equation}

\noindent Finally, the third range of $l$-values we are going to
consider is $l \in [h_1, 3 h_1]$. In this case we have

\begin{equation}
A_{III}(l) ={3 \over 2} \sqrt 3(3h_1-l)^2 \label{s32}
\end{equation}

\noindent
 In a similar way as in the $q=2$ case, the above
expressions for $A(l)$ lead to the analytical form of the
probability (density) $F(\lambda_m)$ of finding a two-qubits state
with a given value of its greatest eigenvalue,

\be \label{largesteinge}
 F(\lambda_m) \, = \, \frac{A(l)}{{\rm Volume}[T_{\Delta}]} \,
 \left| \frac{dl}{d\lambda_m} \right|.
\ee

\begin{figure}
\begin{center}
\includegraphics[angle=270,width=0.65\textwidth]{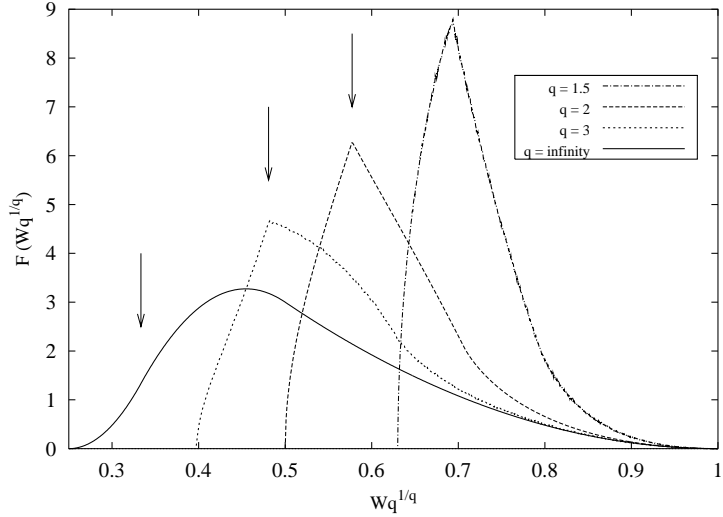}
\caption{Probability $F(w_q^{1/q})$  versus $w_q^{1/q}
= \left(\sum_i p_i^q \right)^{1/q}$ for several $q$ values. The
value of $w_q^{1/q}$ up to which all the states are separable is
also indicated.} 
\label{fig2mix}
\end{center}
\end{figure}

\noindent Regarding $\omega_q^{1/q}$ as a measure of mixture, the
probability of finding states with  given degrees of mixture is
depicted in Fig.\ref{fig2mix}, for different values of $q$. In the limit 
$q\rightarrow \infty$, this probability distribution is given by
$F(\lambda_m)$ (which is included in Fig.\ref{fig2mix}). 
Remarkably enough, as $q$ tends to infinity all discontinuities 
in the derivative of $F(\lambda_m)$ disappear. Is seems as in 
the $\lambda_m$-domain the distribution is completely smooth, 
as opposed to the $R$-domain.

\section{$q$-Entropies and the separability threshold}

  Quantum states of two-qubits are always separable (that is, their
 entanglement is equal to zero) if the degree of
 mixture is high enough. This fact was first demonstrated in \cite{ZHS98}. 
 With the aid of $q$-entropies it is possible
 to give this property of two-qubits states a quantitative expression \cite{BCPP02b}.
 For values of the parameter $q$ belonging to the interval $[2,\infty
 )$, a special value of the $q$-entropy exists such that states with an
 entropy greater than that particular value are always separable.

 Let us consider the subset $\Lambda (q;Z)$ of the simplex $\Delta $
 consisting of all the points of $\Delta $ such that

 \be \label{qesferoid}
 \omega_q \, = \, \sum_{i=1}^4 \, p_i^q \, \le z.
 \ee

 \noindent
 Clearly $\Lambda (q;z)$ is a convex set. In our geometric picture
 of the simplex, the set $\Lambda (q;z)$ corresponds to a subset
 of the tetrahedron. Incurring in a minor abuse of notation, we
 shall design this subset with the same symbol $\Lambda (q;z)$.

 It is clear that

 \be \label{inclu1}
 q>2 \,\, \Longrightarrow \,\, \Lambda (q;z) \, \subset \, \Lambda (2;z).
 \ee

 Let $z(q)$ denote the particular value of $z$ for which
 the set $\Lambda(q;z)$ becomes tangent to the faces of the
 tetrahedron. The points where $\Lambda(q;z(q))$ ``kisses"
 the faces of the tetrahedron are those points (within each of the
 four faces of the tetrahedron) where the quantity $\omega_q$
 adopts its minimum value. The minimum of $\omega_q$
 corresponds to a maximum of a $q$-entropy (for instance, to
 a maximum of Renyi entropy $S^{(R)}_q$). Now, within a face
 of the tetrahedron one of the $p_i$'s vanishes. The maximum of a
 $q$-entropy then corresponds to the centre of the face, where
 the three non vanishing $p_i$'s are equal to $1/3$. Consequently
 we have that

 \be \label{zq}
 z(q) \, = \, 3 \, (1/3)^q.
 \ee

 Now, for $q=2$, we have $ z(2) \, = \, \frac{1}{3}$.
 This value of $\omega_2$ corresponds to a participation
 ration $R=3$. Consequently, the points belonging to
 $\Lambda(2;z(2))$ are precisely those with $R \le 3$,
 which are always separable. Now, we have,

 \ben \label{inclu2}
 q>2 \,\,  & \Longrightarrow & \,\, z(q) \, < \, z(2) \cr
      \,\, & \Longrightarrow & \,\, \Lambda (q;z(q)) \, \subset \, \Lambda
      (q;z(2)),
 \een

 \noindent
 which combined with the inclusion relation (\ref{inclu1}) leads
 to

 \be \label{inclu3}
 q>2 \,\, \Longrightarrow \,\, \Lambda (q;z(q)) \, \subset \, \Lambda (2;z(2)).
 \ee

 \noindent
 The above equation implies that all states belonging to $\Lambda
 (q;z(q))$ are separable.
 Summing up, we have proved that, for $q \ge 2$, those
 states with

 \be
 w_q \, = \, Tr(\rho^q) \le \, 3 \, (1/3)^ q
 \ee

 \noindent
 are always separable. In terms of the Renyi entropies we
 then have that, for $q\ge 2$, those states with

 \be \label{srln3}
 S^{(R)}_q[\hat \rho] \, \ge \ln 3,
 \ee

 \noindent
 are always separable. In the limit $q\rightarrow 1$ (see Eq.(\ref{reninf})),
 the separability criterium given by equation (\ref{srln3}) implies
 that all states whose largest eigenvalue $\lambda_m$ is less than
 $1/3$ are separable.

 It is interesting that, expressed
 in terms of the Renyi entropies, the separability threshold
 does not depend (for $q\ge 2$) on the value of the parameter
 $q$. The problem of determining the limit values of the Renyi
 entropies $S^{(R)}_q$ such that states with entropies above
 them are always separable was first addressed by Zyczkowski
 {\it et al.} in \cite{ZHS98}, where this problem was studied
 numerically. On the basis of their numerical results,
 it was conjectured in the aforementioned paper that
 the limit value of $S^{(R)}_q$ is a decreasing function of $q$.
 Here we prove that, for $q\ge 2$, {\it this limit value is not
 dependent on $q$ and is equal to $\ln 3$}.


\section{Analytical distributions of arbitrary states vs. 
their maximum eigenvalue $\lambda_m$. The qubit-qutrit case}

 As we have seen in previous sections, regarding the maximum 
 eigenvalue $\lambda_m$ as a proper degree of mixture one is 
 able to find a geometrical picture analogue to the one of the 
 growing sphere. In that case a nested inverted tetrahedron 
 grows inside the outer tetrahedron representing the simplex 
 of eigenvalues $\Delta$. The generalization to higher bipartite 
 systems is similar to the $R$-case, but far much easier to 
 implement mathematically. As in that case, we have a high degree 
 of symmetry in the problem. The advantage is that one does not 
 deal with curved figures but perfectly flat and sharp surfaces instead. This 
 fact makes the general problem more approachable.

 We have seen that the problem of finding how the states of a bipartite quantum 
mechanical system are distributed according to their degree of mixedness 
can be translated to the realm of discrete mathematics. 
If we consider our measure of mixedness to be the maximum eigenvalue $\lambda_m$ 
of the density matrix $\hat \rho$ and the dimension of our problem to be 
$N=N_A \times N_B$, we compute the distribution of states in arbitrary dimensions 
by letting an inner regular $N$-polytope $T_l$ to grow inside an outer unit length 
$N$-polytope $T_{\Delta}$, the vertices of the former pointing towards the 
centre of the faces of the latter. In fact, it can be shown that the radius 
$l$ of the maximum hypersphere that can be inscribed inside the inner 
polytope is directly related to $\lambda_m$. 

By computing the surface area of $T_l$ 
strictly inside $T_{\Delta}$, we basically find the desired probability 
(density) $F_N(\lambda_m)$ of finding a state $\hat \rho$ with maximum 
eigenvalue $\lambda_m$ in $N$ dimensions. 

To fix ideas, it will prove useful first to define the vertices of $T_{\Delta}$ 
and $T_l$. In fact it is essential, because we need to deal with elements of 
cartesian geometry in $N$-dimensions. This vectors are given as

\begin{eqnarray}
\vec{r_1} &=& (-\frac{1}{2},-\frac{1}{2\sqrt{3}},-\frac{1}{4}\sqrt{\frac{2}{3}},...,-\frac{1}{N-1}\sqrt{\frac{N-1}{2N}}) \nonumber \\
\vec{r_2} &=& (\frac{1}{2},-\frac{1}{2\sqrt{3}},-\frac{1}{4}\sqrt{\frac{2}{3}},...,-\frac{1}{N-1}\sqrt{\frac{N-1}{2N}}) \nonumber \\
\vec{r_3} &=& (0,\frac{1}{\sqrt{3}},-\frac{1}{4}\sqrt{\frac{2}{3}},...,-\frac{1}{N-1}\sqrt{\frac{N-1}{2N}}) \nonumber \\
\vec{r_4} &=& (0,0,\frac{3}{4}\sqrt{\frac{2}{3}},..,-\frac{1}{N-1}\sqrt{\frac{N-1}{2N}}) \nonumber \\
...& & \nonumber \\
\vec{r_N} &=& (0,0,0,...,\sqrt{\frac{N-1}{2N}}),\label{vectr}
\end{eqnarray}

\noindent with $\sqrt{\frac{N-1}{2N}}$ being the distance from the center 
to any vertex of this regular $N$-polytope of unit length. One can easily check 
that $\sum_{i} \vec{r_i}=\sum_{i,j} \vec{r_i} \cdot \vec{r_j}=0$, as required. 
This particular choice for the position of the vertices of this $N$-simplex is 
such that it simplifies going from one dimension to the next by adding a new 
azimuthal axis each time. This vectors comply with the relations

\ben
\vec{r_i}\cdot\vec{r_j}\, &=& \, -\frac{1}{2N}+\frac{1}{2}\delta_{ij}, \cr
\lambda_m\, &=& \, 2(\vec{r}\cdot\vec{r_i}) + \frac{1}{N}, \,\,\,\,\,\,\,\, 
i=1...N,
\een 

\noindent where the last equation is the general form of (\ref{geolambda}).

Once we have a well defined $T_{\Delta}$, to know the coordinates of $T_l$ is 
straightforward. In fact, $T_l$ is the reciprocication (see \cite{Sommer}) 
of $T_{\Delta}$. This means that the coordinates of $T_l$ are obtained by 
reversing the sign of the ones of $T_{\Delta}$, multiplied by a suitable 
factor (which can be shown to be $\sqrt{2N(N-1)}l$, with $l$ defined as the 
length between the centre of $T_l$ to the centre of any of its faces, which 
in turn points towards the vertices of $T_{\Delta}$). Thus, we can 
relate $l$ with $\lambda_m$ through a general (\ref{geolambda})-relation 
$\lambda_m=2\,l\,\sqrt{\frac{N-1}{2N}}+\frac{1}{N}$, 
such that $\frac{dl}{d\lambda_m}=\sqrt{(2N)/(N-1)}/2$.

\begin{figure}
\begin{center}
\includegraphics[angle=270,width=0.65\textwidth]{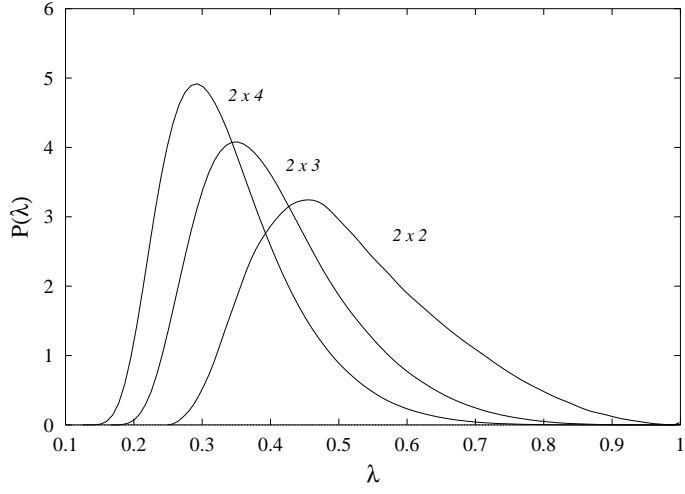}
\caption{Plot of the $F_N(\lambda_m)$ distributions of mixed states $\rho$ numerically 
computed in arbitrary dimensions, generated according to (\ref{memu}). 
As we increase the total dimension $N$, the curves tend to peak around 
$1/N$. See text for details.} 
\label{figentL}
\end{center}
\end{figure}

Several distributions $F_N(\lambda_m)$ are obtained numerically by generating 
random states $\rho$ according to (\ref{memu}) in Fig.\ref{figentL}. 
It becomes apparent that as $N$ grows, the distributions 
are biased towards $\lambda \simeq 1/N$, in absolute agreement with 
the result (\ref{deltaR}). 

As in the $R$-case, $F_N(\lambda_m)$ is distributed into $N-1$ regions 
separated at fixed values of $\lambda_m^{(i)}=\frac{1}{N-i}, \, i=1..(N-2)$. 
The general recipe for obtaining $F_N(\lambda_m)$ is tedious and long, but 
some nice general results are obtained. The $F_N(\lambda_m)$-distributions 
for the ranges a) $\lambda_m \in [\frac{1}{N},\frac{1}{N-1}]$ and 
b) $\lambda_m \in [\frac{1}{2},1]$ are general and read

\ben \label{region}
F_I(\lambda_m) \, &=& \, \kappa \frac{N}{(N-2)!}
\sqrt{\frac{N-1}{2^{N-2}}} \big[\sqrt{2N(N-1)}l(\lambda_m) \big]^{N-2}, \cr
F_{Last}(\lambda_m) \, &=& \, \kappa \frac{N}{(N-2)!}
\sqrt{\frac{N-1}{2^{N-2}}} \bigg[ \frac{\sqrt{\frac{N-1}{2N}}-l(\lambda_m)}
{\sqrt{ \frac{N}{2(N-1)} }} \bigg]^{N-2},
\een

\noindent respectively, 
where $\kappa\equiv\frac{dl}{d\lambda_m}/{\rm Volume}[T_{\Delta}]$ 
is introduced for convenience. 

For the sake of completeness, we briefly describe the physical case 
of a qubit-qutrit system $(N=6)$. Defining $r_i \equiv \sqrt{\frac{N-1}{2N}}$ 
and $y_i \equiv (l(\lambda_m)(N-1)-\frac{i}{N-i}r_i)/(\sqrt{N/2(N-1)})$, 
in addition to the previous regions (\ref{region}) we obtain 

\ben
F_{II}(\lambda_m) \, &=& \, \kappa \big[F_I(\lambda_m)/\kappa -
\frac{(N-1)N}{(N-2)!}\sqrt{\frac{N-1}{2^{N-2}}} [y_{i=1}]^{N-2} \big]; \cr
F_{III}(\lambda_m) \, &=& \, \kappa \big[F_{II}(\lambda_m)/\kappa+ 
\frac{2^9}{5^4}\frac{(N-1)N}{(N-2)!}\sqrt{\frac{N-1}{2^{N-2}}} [y_{i=2}]^{N-2} \big]; \cr
F_{IV}(\lambda_m) \, &=& \, \kappa \big[F_{III}(\lambda_m)/\kappa- 
2\frac{3^4}{5^4}\frac{(N-1)N}{(N-2)!}\sqrt{\frac{N-1}{2^{N-2}}} [y_{i=3}]^{N-2} \big],
\een

\noindent for $\lambda_m \in [\frac{1}{5},\frac{1}{4}],[\frac{1}{4},\frac{1}{3}]$, and 
$[\frac{1}{3},\frac{1}{2}]$, respectively. From the previous formulas one can 
infer a general induction procedure. 
Analytical results are in excellent agreement with numerical generations.

\section{Concluding remarks}

In this Chapter we have revisited the relationship between entanglement 
and purity of states of two-qubits systems, in the light of the 
$q$-entropies as measures of the degree of mixture \cite{BCPP02b}.

The probability $F$ of finding quantum states of 
two-qubits with a given degree of mixture (as measured by an 
appropriate function of $\omega_q$) is analytically found for $q=2$ 
and $q \to\infty$. In the latter case, the $q$-entropies become functions
of the statistical operator's largest eigenvalue $\lambda_m$. In
point of fact, $\lambda_m$ itself constitutes a legitimate measure
of mixture. During the derivation of the probability (density) distributions 
$F_N$ of finding a bipartite mixed state in arbitrary dimensions 
$N=N_A\times N_B$ with a given degree of mixture, we saw that it is more 
convenient to use $\lambda_m$ instead of $R$. In point of fact, 
we analytically entend 
by direct demonstration the separability threshold to $q>2$, when using 
the $q$-entropies as measures of the degree of mixture. 
In the limit case $q \to\infty$, 
we see that states $\rho$ with $\lambda_m \leq \frac{1}{3}$ are always 
separable or, alternatively, whenever the information content of the 
R\'{e}nyi entropy $S^{(R)}_q(\rho) \ge \ln 3$ the state is unentangled.  

In the case $q=2$, we saw that the amount of information concerning 
entanglement distribution is univocally fixed by the participation ratio $R$, 
which shows in a more direct way the intimate connexion existing between 
entanglement and mixedness. In other words, we ascertain the amount of 
information that, with regards to the distribution of entanglement 
among the states of our space, accompanies the knowledge of the degree 
of mixedness. 
Finally, we have derived explicitly the 
distribution $F_N(\lambda_m)$ vs. $\lambda_m$ for the physical meaningful 
case of a qubit-qutrit system ($N=6$).


\chapter{Structure of the space of two-qubit systems: 
metrics and entanglement}


 The two-qubits systems with which  we are going to be concerned
in this Chapter are the simplest quantum mechanical systems exhibiting the
entanglement phenomenon and play a fundamental role in quantum
information theory. The concomitant space ${\cal S}$ of {\it mixed
states} is 15-dimensional and its properties are not of a trivial
character. {\it Important features of this space, related to the
phenomenon of entanglement, have not yet been characterized in full
detail}, notwithstanding the existence of many interesting efforts
towards the systematic exploration of the space of arbitrary (pure
or mixed) states of composite quantum systems 
that have determined typical features with regards to the
phenomenon of quantum entanglement. It is mandatory
to stress the fact that i) the majority of states in this space are
{\it mixed} and ii) that most of the exciting proposals in quantum information 
theory address mainly pure states, contrary to the usual situation encountered 
by the experimentalist.

In the present Chapter we undertake a Monte Carlo exploration over the space 
${\cal S}$ of two-qubits mixed states in order to elucidate the features of the 
concomitant structure (in terms of different metrics) related to the issue 
of entanglement \cite{metricsXXX}. An {\it ab initio} detailed analytical description of the 
aforementioned structures is not possible because it requires a complete 
characterization of the ``geometry" of states which are invariant under PPT action 
(that is, $\rho \ge 0 \Longrightarrow \rho^{T_A} \ge  0$), or simply 
{\it separable}, which is not available up to date. In consequence, we 
carry out numerical computations by randomly generating states of two-qubits 
systems according to an appropriate measure, studying the ensuing entanglement 
properties.

We also focus our attention to the different ways that the space ${\cal S}$ 
can be generated, discussing the adequacy of several distributions for 
the simplex $\Delta$ of eigenvalues of $\rho$. Finally, the study of the 
features of real quantum mechanics is drawn in the context of {\it two-rebits} 
systems \cite{BCPP02a}.

\section{Metrics and entanglement}

 The protagonist of the following considerations is the maximally
mixed (MM) bipartite state $\rho_{MM}= I/4$. This state is 
surrounded by a separable ball \cite{balls}, where all states ``close 
enough" to it are separable. Therefore it becomes apparent that a way to 
characterize the space of two-qubits is through distances from a
given state $\rho$, to the MM-one and ask questions like, for instance,
starting from the MM, how far do we have to go to find an entangled state?
 The volume measure $\mu_Z \equiv \mu=\nu\times {\cal L}_{N-1}$ (\ref{memu}), thoroughly 
described in previous Chapters and in Appendix B, is not based        
 upon any distance-measure, and consequently has been criticized by Slater 
\cite{sla1,sla2,sla3,sla4} on two grounds: i) it is not associated
with the volume element of any monotonic metric and ii) it is
over-parameterized because the number of variables it needs to
parameterize de convex set of $N\times N$ density matrices $\rho$ is
$N^2+N-1$ rather than the theoretical minimum number $N^2-1$. {\sf
On the other hand} \cite{inclus}, these facts are to be weighted
against these other two: the  measure $\mu_Z$ allows for rapid
convergence and provides a simple procedure to investigate different
separability criteria for bipartite states, as seen in previous Chapters. 
Regarding ``overparameterization" by $\mu_Z$, it is possible to express this 
measure strictly using $N^2-1$ parameters. 

It is likely that answers to questions like the one posed before 
might presumably depend on the choice of volume measure. This is precisely 
what we investigate. A most appropriate alternative is to base
 our measure on the distance between density matrices. To such an end, 
let us survey different aspects that may provide some insight into the 
structure of two-qubits systems.
\newline
\newline
{\bf A. The Bures distance}
\newline

Let us describe this measure for the sake of completeness. 
Given two (not necessarily commuting) density matrices $\rho_1$
and $\rho_2$, H\"ubner \cite{Hub} has found an explicit form for the
distance $d_{Bures}$ between them, following well known Bures tenets
\cite{Bures} that apply for density operators. One has

\begin{equation} \label{d_Bures}
d_{Bures}( \rho_1, \rho_2) \, = \, \sqrt{2} \, \bigg[1 \, - \, Tr \,
\bigg( \big[ \rho_1^{1/2} \rho_2 \,
 \rho_1^{1/2}\big]^{1/2} \bigg) \bigg]^{1/2}.
\end{equation} Let

\begin{equation}
\hat \rho_1 |i\rangle \, = \, a_i |i\rangle,
\end{equation}
\noindent be the eigenvalue equation for the statistical operator
$\hat \rho_1$ whose associated density matrix is, of course,
$\rho_1$, so that

\begin{equation}
\hat \rho_1^{1/2} \, = \, \sum_{i} a_i^{1/2} |i\rangle \langle i|.
\end{equation}
\noindent In this basis, we also have

\begin{equation}
\hat \rho_2 \, = \, \sum_{k,j} b_{k,j} |k\rangle \langle j|,
\end{equation}
\noindent an then the triple product in (\ref{d_Bures}) yields an
operator $\hat T$

\begin{equation}
\hat T \, = \, \sum_{i,m} a_i^{1/2} b_{i,m} a_m^{1/2} |i\rangle
\langle m|.
\end{equation}
It is necessary now to diagonalize $T$ so as to get

\begin{equation}
\hat T^{1/2} \, = \, \sum_{\beta} t_{\beta,\beta}^{1/2}
|\beta\rangle \langle \beta|,
\end{equation}
\noindent and finally, be in a position to define the Bures distance

\begin{equation}
d_{Bures}( \rho_1, \rho_2) \, = \, \sqrt{2} \, \bigg[1 \, - \,
\sum_{\beta} t_{\beta,\beta}^{1/2} \bigg]^{1/2}.
\end{equation}
The Bures distance is a function of the so-called ${\it fidelity}$

\be
F(\rho_1,\rho_2)\,=\,\Bigl[ \mbox{Tr}\sqrt{\sqrt{{\rho}_1}
{\rho}_2\sqrt{{\rho}_1} } \Bigr]^2
\ee

\noindent between the two states
 $\rho_1$ and $\rho_2$ \cite{Fidelity}. The fidelity is a relevant quantity
for information processing purposes: It constitutes  a
generalization to mixed states of the overlap-concept between pure
states. Thus, whenever the Bures distance is employed, we also describe 
the fidelity-degree extant between the maximally mixed state $\rho_{MM}$ and an
arbitrary mixed state $\rho$. 

Lower and upper bounds for the function $d_{Bures}$ (and, in turn, for
the fidelity $F$) vs. $R$ are obtained in analytical fashion. A ``band" of 
Bures distances exist for each $R-$value, that 
ranges from a minimal up to a maximal $d_{Bures}(R)$. There are three
$R$ zones in our bipartite state-space, namely, i) $1 \le R \le 2$,
ii) $2 \le R \le 3$, and iii) $3 \le R \le 4$. For each 
of these zones we consider states that, in the product space 
$\mathcal{S}$ read, respectively,

\begin{enumerate}
\item $\rho_1 = diag(0,0,x,1-x)$,
\item $\rho_2 = diag(0,x,x,1-2x)$,
\item $\rho_3 = diag(x,x,x,1-3x)$.
\end{enumerate}
By setting the condition $Tr (\rho^2) =1/R$, in the case of $\rho_{1,\,2}$ 
we get only a physical root for $x$ that corresponds to the maximal distance. 
For $\rho_3$ we get two such roots that yield minimal and maximal values. It 
is important to stress that the aforementioned states are universal, that is, 
independent of the generation of ${\cal S}$.
\newline
\newline
{\bf B. Measures and distances}
\newline

One of our present purposes is to replace the volume measure
$\mu_Z$ by other volume measures based upon proper distance-definitions.
  One can then randomly generate states according to volume measures
  pertaining to these different metrics.
   Of course, within a given state-generation
   procedure (that uses a volume measure
   that may or may not be associated to a given metric) one can
   still determine distance between
   states according to different distance-recipes.

One can then generate the simplex $\Delta $ of eigenvalues (a
subset of a $(N-1)$-dimensional hyperplane of ${\cal R}^N$) 
using different measures \cite{Z01,Fidelity}, namely,
\begin{itemize}

\item $\mu_Z$

\item A Bures volume measure $\mu_B$ induced by the
Bures metric (see above) in the simplex of eigenvalues

\be 
\mu_B(\Delta) \, = \, \frac{2^{N^2-N}\Gamma(N^2/2)}{\pi^{N/2} \Gamma(1)..\Gamma(N+1)}
 \frac{\delta(\sum_{j=1}^{N}\lambda_j-1)}{\sqrt{\lambda_1 \lambda_2...\lambda_N}} 
\, \prod_{i<j}^{N} \frac{(\lambda_i-\lambda_j)^2}{\lambda_i+\lambda_j}.
\ee

\item A Hilbert-Schmidt (HS) volume  measure
(induced by the HS metric) in the $\Delta-$simplex of eigenvalues

\be 
\mu_{HS}(\Delta) \, = \,
\frac{\Gamma(N^2)\,\delta(\sum_{j=1}^{N} \lambda_j-1)}
{\prod_{j=0}^{N-1} \Gamma(N-j)\Gamma(N-j+1)} \, 
\prod_{i<j}^{N} (\lambda_i-\lambda_ j)^2. 
\ee

\end{itemize}

\nd Independently of the volume measure choice we can still speak of
distances between bipartite states measured according to either

\begin{itemize}

\item The Bures distance  (\ref{d_Bures}) or

\item The Hilbert-Schmidt distance

\begin{equation} \label{d_HS}
d_{HS}( \rho_1, \rho_2) \, = \, \sqrt{Tr[(\rho_1-\rho_2)^2]}.
\end{equation}

\end{itemize}
\noindent 
\newline
\newline
{\bf C. A random walk}
\newline

\begin{figure}
\includegraphics[angle=270,width=0.5\textwidth]{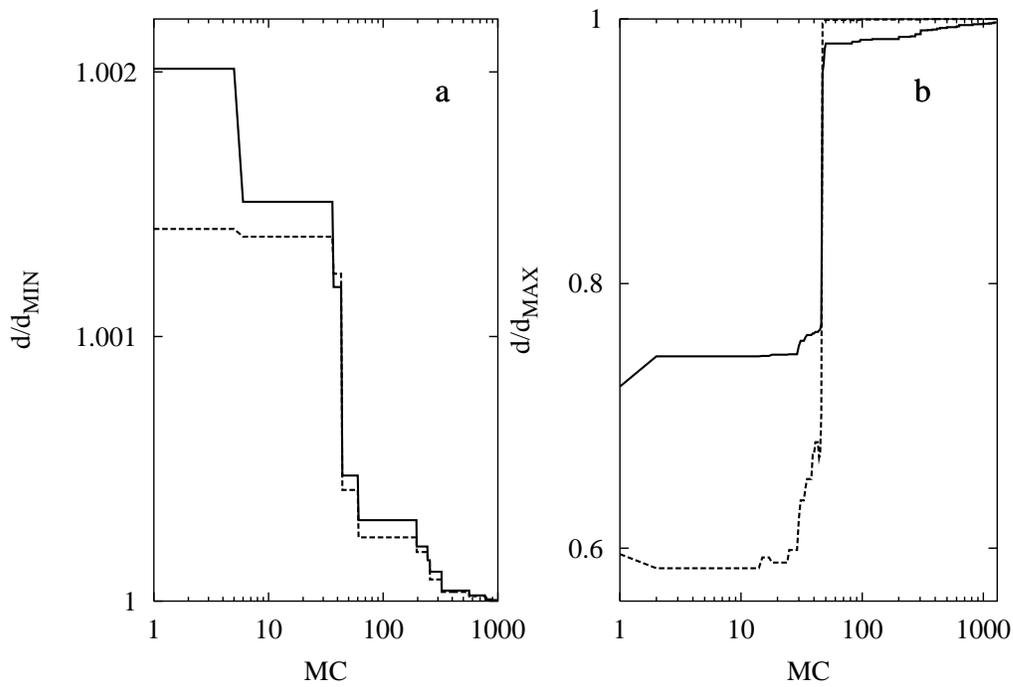}
\caption{a) Evolution of the ratio $d/d^{MIN}$ vs. the
number of Monte Carlo (MC) steps, as we approach the maximally
mixed state $\frac{1}{4}I$ from outside the set ${\mathcal
{S}}_{sep}$ of separable states, for the Bures (solid line) and
Hilbert-Schmidt (dashed line) distances (see text for details).
Convergence with the theoretical value is reached quickly. b)
Evolution of the ratio $d/d^{MAX}$ vs. the number of MC steps
throughout the interior of ${\mathcal {S}}_{sep}$, for the Bures
(solid line) and Hilbert-Schmidt (dashed line) distances.
Agreement with predicted values is excellent.}
\label{fig1metr}
\end{figure}

Assume that we are 
interested in reaching the MM bipartite state $\frac{1}{4} I$,
starting from an entangled initial bipartite state $\rho_0$. We call
${\mathcal{S}}_{sep}$ the set of separable states
${\mathcal{S}}_{sep}\subset \mathcal{S}$. The task can be performed
by following a simulated-annealing minimization-procedure in the
15-dimensional set ${\mathcal{S}}_{sep}$. The ensuing  Monte Carlo
recipe resembles a ``random walk". One takes advantage of the fact
that ${\mathcal{S}}_{sep}$ is a convex set. Thus, optimization is
equivalent to a random walk subject to constraints. Our search
begins then somewhere in exterior of ${\mathcal {S}}_{sep}$ and we
look for a minimal distance $d^{MIN}$ to $\rho_{MM}$. This minimal
distance can be either of the Bures ($d_{Bures}$)  or of the HS
($d_{HS}$) kind. In view of the fact that all states  $\in
\mathcal{S}$ are separable for $R >3$, our minimal
distance to $\rho_{MM}$ is obtained for a state characterized by a
participation ratio $R=3$. Interestingly enough, these minimal
distances can be obtained not only numerically but also in analytic
fashion. The pertinent results are depicted in Fig.\ref{fig1metr}a, 
where we plot, for the two distances,  the ratio $d/d^{MIN}$ vs. number of
Monte Carlo (MC) steps. By a MC step we mean a realization of the
set of variables that parameterize our two-qubits state $\rho$,
consisting in changing  the configuration of $\rho$ a reasonably
large number of times until ``thermalization" is reached. The
condition $R=3$ entails finding the appropriate $x$ value for a
state that, in the product space $\mathcal{S}$, is of the form $\rho
=diag(x,x,x,1-3x)$. 
Two solutions exist, $x=1/3,1/6$. Choosing the latter, one
encounters

 \ben d_{Bures}^{MIN}&=&d_{Bures}(I/4,\rho) \sim 0.26105 \cr
d_{HS}^{MIN}&=&d_{HS}(I/4,\rho)= \frac{1}{2\sqrt{3}}. \een Analytic
and numerical results are seen to eventually match each other. More
importantly, it is seen that which of the two distances one uses is
of no importance whatsoever.

\nd Another question that might be profitably asked is the
following: without leaving ${\mathcal{S}}_{sep}$, what is the
farthest from $\rho_{MM}$ that you can go? Let us call $d^{MAX}$ the
concomitant ``length". Again, we will have a Bures and an HS
quantity (see Fig.\ref{fig1metr}b). It is clear that the most dissimilar (to MM) 
density matrix is, in the product space $\mathcal{S}$, the state
$\rho_1=diag(1,0,0,0)$ (or any (diagonal) permutation thereof),
which will belong to ${\mathcal{S}}_{sep}$ if a suitable basis is
chosen. Our quantities are then distances to such a state from the
MM one. We immediately get

\ben d_{Bures}^{MAX}&=&d_{Bures}(I/4,\rho_1) =1 \cr
d_{HS}^{MAX}&=&d_{HS}(I/4,\rho_1)= \frac{\sqrt{3}}{2} . \een The
agreement between the above analytical results and numerical
simulations is excellent. 
\newline
\newline
{\bf D. Distribution of distances and CNOT gate}
\newline

Let us concern ourselves with one of the basic
constituents of any quantum processing device: {\it quantum logical
gates}, i.e., unitary evolution operators $\hat U$ that act on the
states of a certain number of qubits. If the number of such qubits
is $m$, the quantum gate is represented by a $2^m \times 2^m$ matrix in
the unitary group $U(2^m)$. These gates are reversible: one can
reverse the action, thereby recovering an initial quantum state from
a final one. We shall work here with $m=2$.
 The simplest nontrivial two-qubits
operation is the quantum controlled-NOT, or CNOT (equivalently, the
exclusive OR, or XOR) (See Sec.(2.6)). Its classical counterpart is a reversible
logical gate operating on two bits: $e_1$, the control bit,  and
$e_2$, the target bit. If $e_1=1$, the value of $e_2$ is negated.
Otherwise, it is left untouched. The quantum CNOT gate $C_{12}$ (the
first subscript denotes the control bit, the second the target one)
plays an important role in both experimental and theoretical efforts
that revolve around the quantum computer concept. In a given
ortonormal basis $\{\vert 0 \rangle,\,\vert 1 \rangle\}$, and if we
denote addition modulo 2 by the symbol $\oplus$, we have

\be \label{uno} \vert e_1 \rangle\, \vert e_2 \rangle \rightarrow
C_{12}\rightarrow \vert e_1 \rangle\, \vert e_1 \oplus e_2 \rangle.
\ee

In conjunction with simple single-qubit operations, the CNOT gate
constitutes a set of gates out of which {\it any quantum gate may be
built}. In other words, single qubit and CNOT gates
are universal for quantum computation. A more detailed account on 
quantum gates is given in Chapter 11.   

\begin{figure}
\includegraphics[angle=270,width=0.5\textwidth]{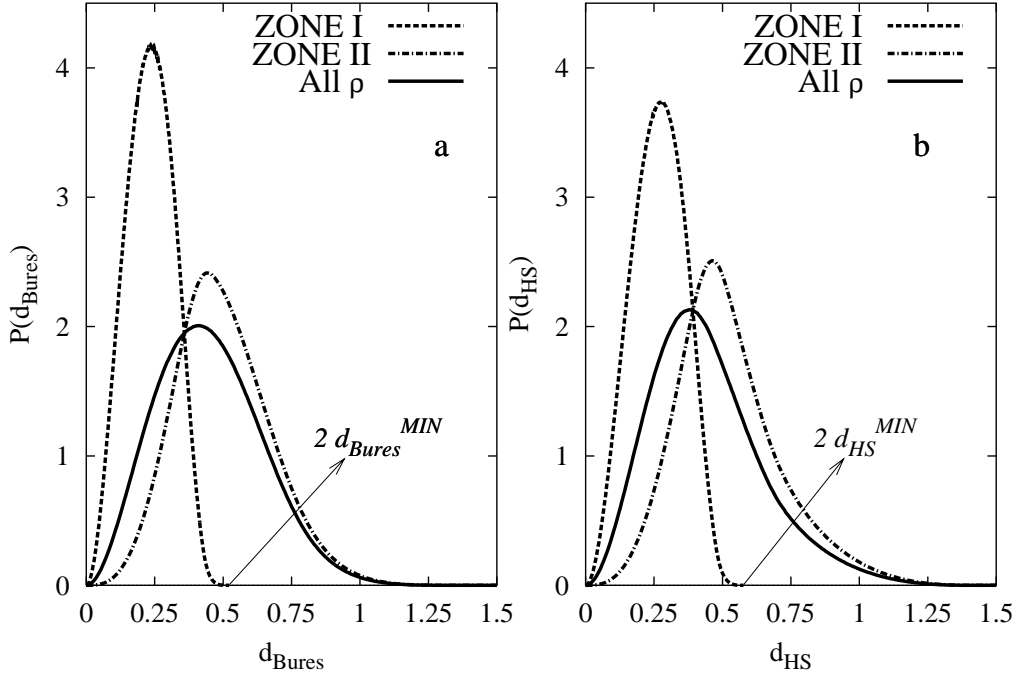}
\caption{a) Probability (density) distribution of changes 
in the Bures distance $d_{Bures}$ between states ($\mu_{Z}$-generated) 
after the application 
of a CNOT gate. The set of states upon which the gate acts is divided into 
i) ZONE I: states with a $d_{Bures}$ (from $\frac{1}{4}I$) $<$ $d_{Bures}^{MIN}$, 
ii) ZONE II: states with a $d_{Bures}$ (from $\frac{1}{4}I$) $>$ $d_{Bures}^{MIN}$, 
and iii) all states. These curves coincide at 
the $d_{Bures}=d_{Bures}^{MIN}$ boundary (see text for details). 
b) Same plot as before using the Hilbert-Schmidt distance.}
\label{fig2metr}
\end{figure}                              

In Fig.\ref{fig2metr} we depict the probability
$P(d)\,\,vs.\,\, d$ of encountering a giving distance $d$ between
initial and final states under the  CNOT-action, with $d$ given by
either the Bures (\ref{fig2metr}a) or the HS (\ref{fig2metr}b) definitions. 
Initial states are generated according to $\mu_Z$ and chosen in such a way 
that their distance to $\rho_{MM}$ is
\begin{enumerate}
\item always $< d^{MIN}$ (solid curve)
\item always $> d^{MIN}$ (dashed line)
\item no $d^{MIN}-$criterium is employed (dotted curve).
\end{enumerate}
Notice that if the selected initial state is located at any distance smaller 
than the $d^{MIN}$, by the application of any unitary transformation like 
the CNOT gate we always obtain a state inside of the sphere of radius 
$d^{MIN}$, and consequently $P(d)$ is only different from zero 
in the interval $\left[0, 2d^{MIN}\right]$. 
If the initial state is outside the sphere of radius $d^{MIN}$, $P(d)$ 
is non null in the interval  $\left[0, 2d^{MAX}\right]$ because the final state
obtained by using the CNOT-gate is not restricted to a given value of $d$.
When we choose random the initial state, the value of $P(d)$ slightly moves to 
the small $d$ values because the final state obtained using an initial state 
belonging to zone 1 are always restricted to $d< d^{MIN}$.
Finally, it is remarkable that the selection of Bures or Hilbert-Schmidt 
distances does not seem to produce great differences.
\newline
\newline
{\bf E. Effects of the $\mu-$choice}
\newline

\begin{figure}
\begin{center}
\includegraphics[angle=270,width=0.9\textwidth]{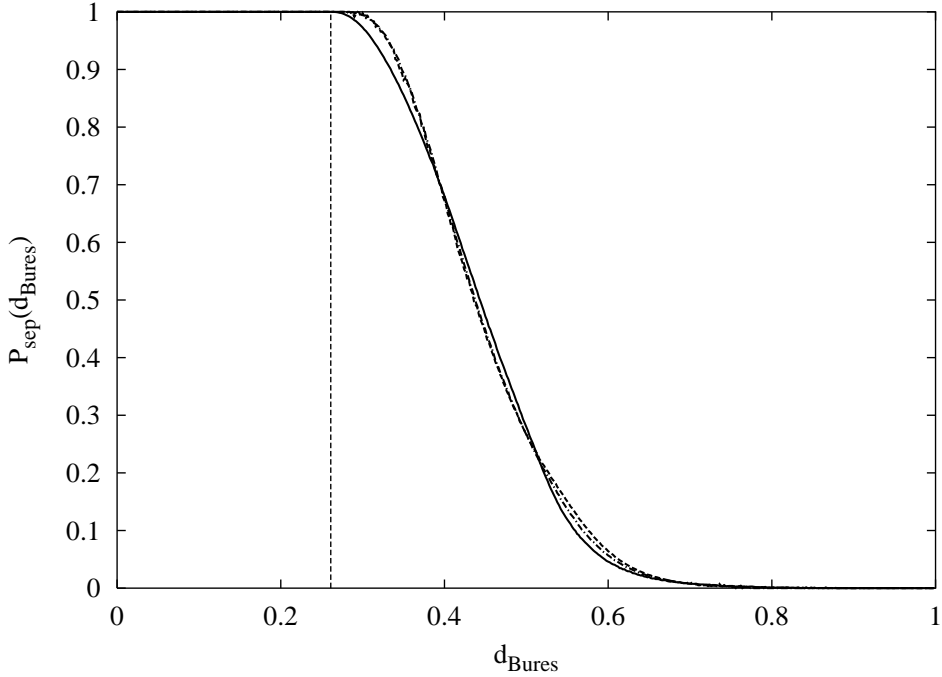}
\caption{Probability of finding a separable state in the
space of all two-qubits $\mathcal {H}$, as measured by the Bures
distance $d_{Bures}$ from the maximally mixed state
$\frac{1}{4}I$, for three different  measures of states in
$\mathcal {H}$: i) $\mu_{Z}$ (solid line), ii) $\mu_{Bures}$
(dashed line) and ii) $\mu_{Bures}$ (dot-dashed line). The three
curves are all alike, but ii) and iii) are closer (nearly coincide
at this scale) to one another (indeed they are metric-based
measures).}
\label{fig3metr}
\end{center}
\end{figure}    

Let us try to discern differences
 between randomly generating states ${\in \mathcal{S}}$ either according to
i) $\mu_Z$, ii) $\mu_{Bures}$, and  iii) $\mu_{HS}$. Again, $d$ is
the distance between a given state ${\in \mathcal{S}}$ and
$\rho_{MM}$ (Here we use only  $d_{Bures}$). We compute the
probability $P_{sep}$ of Monte Carlo-finding  a separable bipartite
state as a function of $d_{Bures}$. Our results are given in Fig.\ref{fig3metr}. 
{\it Notice  the important fact that  $P_{sep}$ does not strongly
depend on the $\mu-$choice}. In point of fact the curves generated 
according to the Bures or the Hilbert-Schmidt metrics does not differ 
from each other as much as the one obtained using $\mu_Z$. Nevertheless 
the differences are very tiny and the overall behaviour of the three curves are
quantitatively and qualitatively the same. This is probably the most important 
result of this Chapter: regardless of the nature of the measure employed 
in the generation of states, the entanglement properties of the set ${\cal S}$ 
are basically the same. In other words, the space ${\cal S}_{sep}$ of states 
which comply with the PPT separability criterion is not highly sensitive 
to the way the states $\rho$ are distributed in ${\cal S}$.


\section{Comment on the non uniqueness of the generation of bipartite 
mixed states. Examples}

In the papers by Zyczkowski {\it et al.} \cite{ZHS98,Z99}, 
a basic question regarding a natural measure $\mu$ for the set of mixed states 
$\rho$ was debated. As described in Secs. (7.1) and (9.1), it is know, the set 
of all states ${\cal S}$ can 
be regarded as the cartesian product ${\cal S} = {\cal P} \times \Delta$, 
where $\cal P$ stands for the family of all
complete sets of ortonormal projectors $\{ \hat P_i\}_{i=1}^N$,
$\sum_i \hat P_i = I$ ($I$ being the identity matrix), and $\Delta$
is the set of all real $N$-tuples $\{\lambda_1, \ldots, \lambda_N
\}$, with $\lambda_i \ge 1$ and $\sum_i \lambda_i = 1$. As discussed in 
those papers and in Appendix B, it is universally accepted to assume            
the Haar measure $\nu$ to be the one defined over ${\cal P}$, because of 
its rotationally-invariant properties. But when it turns to discuss an 
appropriate measure over the simplex $\Delta$, some controversy arises. 
In all previous considerations here, we have regarded the Leguesbe measure 
${\cal L}_{N-1}$ as being the ``natural" one. But one must mention that 
Slater has argued \cite{sla1,sla2} that, in analogy
to the classical use of the volume element of the Fisher information
metric as Jeffreys' prior \cite{jef} in Bayesian theory, a natural measure
on the quantum states would be the volume element of the Bures
metric. The problem lies on the fact that there is no unique 
probability distribution defined over the simplex of eigenvalues $\Delta$ of 
mixed states $\rho$. In point of fact, the debate was motivated by the fact 
that the volume occupied by separable two-qubits states (see Chapter 7) was     
found in \cite{ZHS98} to be greater than $50\%$ ($P_{sep}=0.6312$) using 
the measure $\mu$, something which is surprising.

One such probability distribution that is suitable for general 
considerations is the Dirichlet distribution \cite{Z99}

\be \label{dirich}
P_{\eta}(\lambda_1, \ldots, \lambda_N) \,=\, 
C_{\eta} \lambda_1^{\eta-1}\lambda_2^{\eta-1}...\lambda_N^{\eta-1},
\ee

\noindent with $\eta$ being a real parameter and 
$C_{\eta}=\frac{\Gamma[N\eta]}{\Gamma[\eta]^N}$ the normalization 
constant. This is a particular case of the more general Dirichlet distribution. 
The concomitant probability density for variables $(\lambda_1,...,\lambda_N)$ 
with parameters $(\eta_1,...,\eta_N)$ is defined by 

\be \label{dirich2}
P_{{\bf \eta}}(\lambda_1, \ldots, \lambda_N) \,=\, 
C_{\bf {\eta}} \lambda_1^{\eta_1-1}\lambda_2^{\eta_2-1}...\lambda_N^{\eta_N-1},
\ee 

\noindent with $\lambda_i \ge 0$, $\sum_{i=1}^{N} \lambda_i=1$ and 
$\eta_1,...,\eta_N > 0$, and $C_{\bf {\eta}}=\Gamma(\sum_{i=1}^N \eta_i)/
\prod_{i=1}^N \Gamma(\eta_i)$. Clearly, distribution\footnote{This distribution admits 
a clear interpretation. As known, the multinomial distribution provides a probability of 
choosing a given collection of $M$ items out of a set of $N$ items with repetitions, 
the probabilities being $(\lambda_1,...,\lambda_N)$. These probabilities are the parameters of 
the multinomial distribution. The Dirichlet distribution is the conjugate prior of the parameters 
of the multinomial distribution.} (\ref{dirich2}) generalizes 
(\ref{dirich}). A new measure then can be defined as 
$\mu_{\eta}=\nu\times \Delta_{\eta}$, where $\Delta_{\eta}$ denotes the simplex 
of eigenvalues distributed according to ({\ref{dirich}}) 
(The Haar measure $\nu$ remains untouched). Thus, one clearly recovers 
the Leguesbe measure ${\cal L}_{N-1}$ for $\eta=1$ (uniform distribution), 
and Slater's argumentation reduces to 
take $\eta=\frac{1}{2}$ in ({\ref{dirich}}). For $\eta \rightarrow 0$ one 
obtains a singular distribution
concentrated on the pure states only, while for $\eta \rightarrow \infty$, 
the distribution peaks on the maximally mixed state $\frac{1}{N}I$. We will 
see shortly that changing the continuous parameter $\eta$ indeed modificates 
the average purity (as expressed in terms of $R=1/Tr(\rho^2)$) 
of the generated mixed states. 

\begin{figure}
\begin{center}
\includegraphics[angle=270,width=0.65\textwidth]{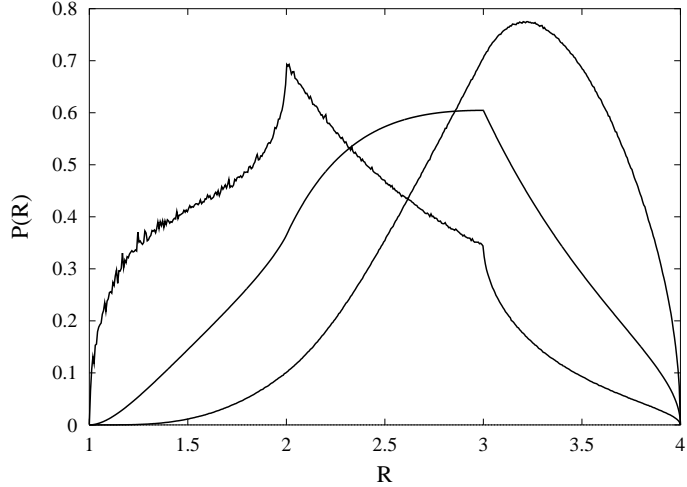}
\caption{$P(R)$ vs. $R$ distributions for two qubit systems, 
whose eigenvalues are distributed according to ({\ref{dirich}}), for the values 
$\eta=\frac{1}{2},1,2$ (from left to right in this order). It is plain 
from this figure that the uniform distribution ($\eta=1$) appears 
more balanced that the others. Also, the particularity of $R=2,3$ seems to disappear 
for $\eta > 1$.}
\label{figcomment01}
\end{center}
\end{figure}

\begin{figure}
\begin{center}
\includegraphics[angle=270,width=0.65\textwidth]{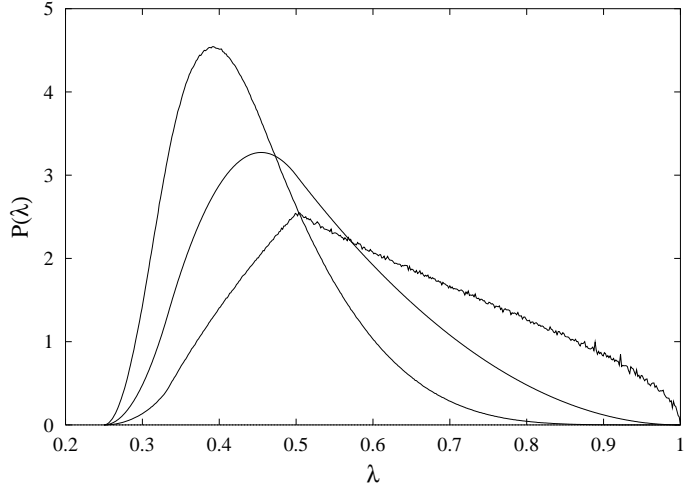}
\caption{Probability (density) distributions of the maximum eigenvalue $\lambda_m$ 
of two qubit systems, whose eigenvalues are distributed according to 
({\ref{dirich}}), for the values 
$\eta=\frac{1}{2},1,2$ (from right to left). When employing $\lambda_m$ as a 
degree of mixture, the derivative of these distributions is discontinuous at the 
special values $\lambda_m=\frac{1}{2},\frac{1}{3}$ for $\eta < 1$.}
\label{figcomment02}
\end{center}
\end{figure}

In what follows\footnote{J. Batle. Unpublished (2003).} 
we numerically generate mixed states whose eigenvalues are 
distributed following ({\ref{dirich}}). This is done is order to tackle 
the dependence of relevant quantities on the parameter $\eta$. Let us 
consider the way mixed states are distributed according to $R$. We focus our 
attention on the two-qubits instance, but similar studies can be extended 
to arbitrary bipartite dimensions. 
As shown in Fig.\ref{figcomment01}, the distributions $P(R)$ vs. $R$ are shown for 
$\eta=\frac{1}{2},1,2$ (from left to right in this order) while Fig.\ref{figcomment02} shows 
analogous distributions for the maximum eigenvalue 
$\lambda_m$ for $\eta=\frac{1}{2},1,2$ (from right to left). Notice the different shapes. 
We can no longer attribute a 
geometrical description (as done in Chapter 9) to $P(R)$ except for $\eta=1$. 
In \cite{Z99} $P(R)$ for $\eta=\frac{1}{2}$ was first derived. Here we can 
provide different distributions for arbitrary $\eta$-values.

\begin{figure}
\begin{center}
\includegraphics[angle=270,width=0.65\textwidth]{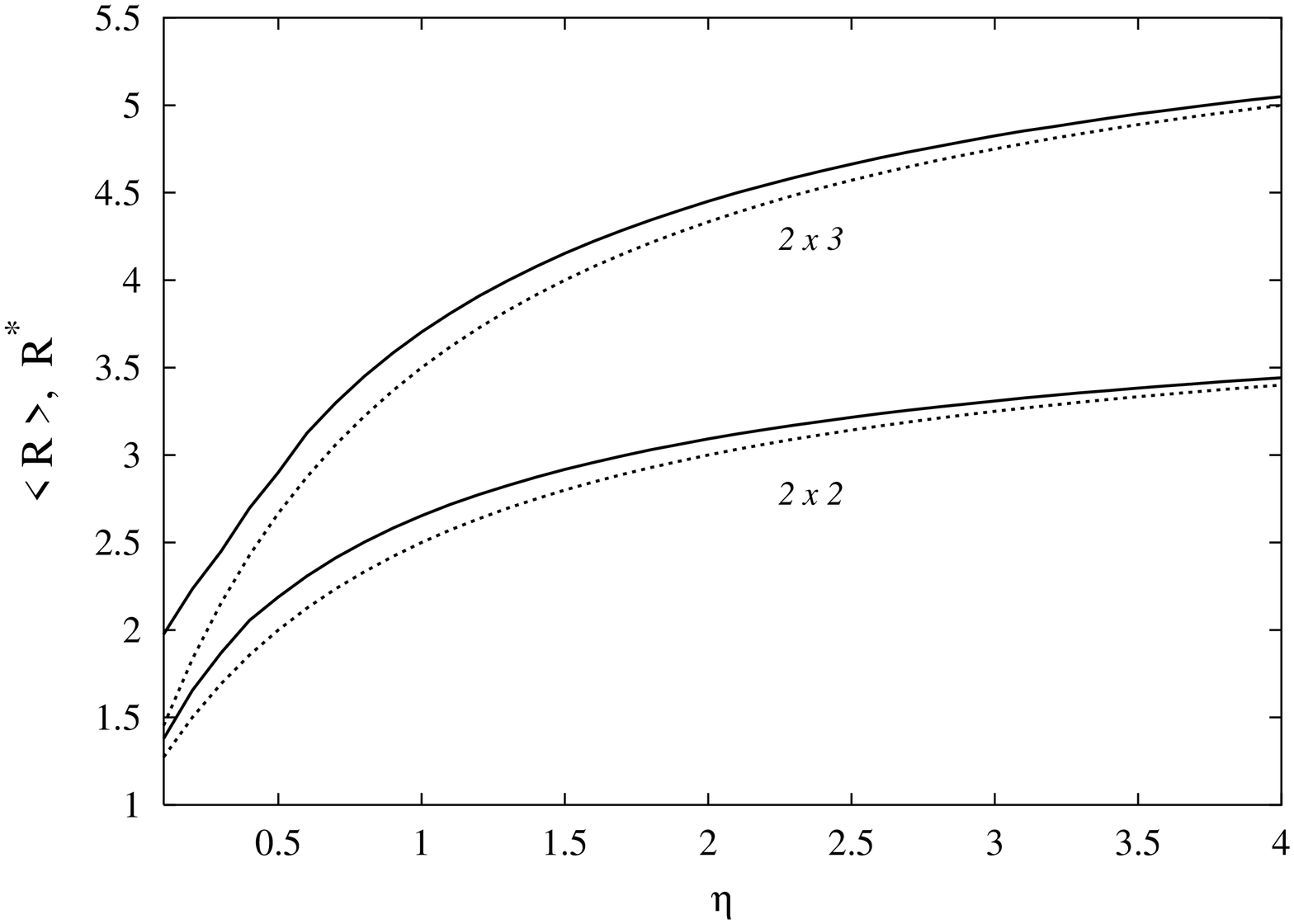}
\caption{Average $R$-value $\langle 1/Tr(\rho^2) \rangle$ (solid line) and 
$R^{*} \equiv 1/\langle Tr(\rho^2) \rangle$ (dashed line) for 
two qubit and one qubit-qutrit systems, plotted versus the Dirichlet 
parameter $\eta$.}
\label{figcomment1}
\end{center}
\end{figure}

A way to devise a certain range of reasonable $\eta$-values is to study the 
average $R$ induced for every $\eta$-distribution. This is performed in 
Fig.\ref{figcomment1}. The average $R$-value $\langle 1/Tr(\rho^2) \rangle$ and 
$R^{*} \equiv 1/\langle Tr(\rho^2) \rangle$ are plotted versus $\eta$. 
$\langle R \rangle$ (solid line) can only be computed numerically, 
but luckily $R^{*}$ (dashed line) is obtained in analytical fashion 
{\it for all N}

\ben \label{traceN}
\langle {\rm Tr} \rho^2 \rangle_N(\eta) &=& C_{\eta} 
\int_{0}^{1}d\lambda_1\lambda_1^{\eta-1}
\int_{0}^{1-\lambda_1}d\lambda_2\lambda_{2}^{\eta-1}...
\int_{0}^{1-\sum_{i=1}^{N-2}\lambda_i}d\lambda_{N-1}\lambda_{N-1}^{\eta-1} \cr 
& &(1-\sum_{i=1}^{N-1}\lambda_i)^{\eta-1}\,\big[\sum_{j=1}^{N} 
\lambda_{j}^{2}\big]\, = \, \bigg[N - \frac{N-1}{\eta+1} \bigg]^{-1}.
\een

\noindent The fact that $R^{*}$ matches exact results 
validates all our present generations. The actual value 
$\langle R \rangle$ is slightly larger than $R^{*}$ for all values 
of $\eta$, but both of them coincide for low and high values of the 
parameter $\eta$. It is obvious from Fig.\ref{figcomment1} that we cannot 
choose distributions that depart considerably from the uniform one $\eta=1$, 
becasue in that case we induce probability distributions that favor high or 
low $R$ already.

\begin{figure}
\begin{center}
\includegraphics[angle=270,width=0.65\textwidth]{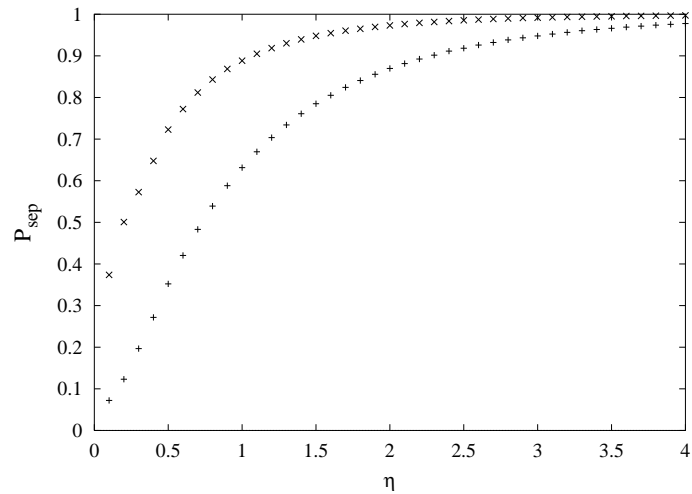}
\caption{Probability of finding a state $\rho$ of two-qubits being positive 
partial transposed (lower curve), and violating the strongest entropic 
criterion $q=\infty$ (upper curve). This figure illustrates the fact that 
one can arbitrarily choose any $P_{sep}$ by generating two qubit mixed states 
with different Dirichlet parameter $\eta$.}
\label{figcomment2}
\end{center}
\end{figure}  

Perhaps the best way is to go straight to the question that originated 
the controversy on the $\Delta$-measures: what is the dependency of 
the {\it a priori} probability $P_{sep}$ of finding a two-qubits mixed state 
being separable? In Fig.\ref{figcomment2} we depict $P_{sep}$ vs. $\eta$ for 
states complying with PPT (lower curve) and those which violate the $q=\infty$-entropic 
inequalities (upper curve). It seems reasonable to assume that a permissible range 
of $\eta$-distributions belong to the interval $[\frac{1}{2},2]$, within 
which $P_{sep}$ remains around the reference point $P_{sep}=0.5$. 

However, in view of the previous outcomes we still believe that the 
results obtained considering the uniform 
$\eta=1$-distribution for the simplex $\Delta$ remain the most natural 
choice possible, independent of any form that one may adopt for a generic 
probability distribution.


\section{Quantum mechanics defined over ${\cal R}$: two-rebits systems}

Pointed out by Caves, Fuchs, and Rungta \cite{CFR01}, 
real quantum mechanics (that is, quantum
mechanics defined over real vector spaces \cite{S60,GPRS61,E86,W02}) provides
an interesting foil theory whose study may shed some light on just which
particular aspects of quantum entanglement are unique to standard quantum
theory, and which ones are more generic over other physical theories endowed
with this
phenomenon. In the same spirit, let us explore numerically, as well
 as conceptually, the entanglement properties of two-rebits systems,
 as compared to the usual two-qubits ones, so as to detect  the
 differences between the two types of system \cite{BCPP02a}.

  For quantum mechanics defined over real vector spaces the
  simplest composite systems are two-rebits systems. 
Pure states of rebits-systems are described by
  normalized vectors in a two dimensional real vector space.
 The correspondent space of mixed two-rebits states is 9-dimensional 
(vis-\`a-vis 15 for two-qubits).

In the space of real quantum mechanics we can represent rebits on the Bloch
sphere. The poles  correspond to classical bits $|0\rangle$, $|1\rangle$, but
the sphere for a qubit reduces itself now to a maximum unit circle, 
described by just one parameter $\phi$. We thus have 
cos$\phi |0\rangle+e^{i\psi}$sin$\phi |1\rangle$
$\rightarrow$ cos$\phi |0\rangle+$sin$\phi |1\rangle$. Entanglement can also be
described in such a context with suitable modifications. Caves, Fuchs, and
Rungta's (CFR) formula for the entanglement of formation of a two-rebits state
$\rho$ changes considerably the new concurrence 
$C[\rho] \, = \, \mid \!
{\rm tr} (\tau) \! \mid \, = \,
 \mid \! {\rm tr} (\rho \, \sigma_y \otimes \sigma_y) \! \mid$, which has to
be evaluated using the matrix elements of $\rho$ computed  with respect to the
product basis, $\mid \! i,j \rangle = \mid i \rangle \! \mid j
\rangle, \,\, i,j=0,1$.

  For a two-rebits state the entanglement of formation is completely determined
  by the expectation value of one single observable,
  namely, $\sigma_y \otimes \sigma_y$, contrary to the two-qubits case. 
As shown in \cite{BCPP02a}, there are mixed
  states of two rebits with maximum entanglement
  (that is, with $C^2=1$) within the range $1\le R \le 2$.
        This is clearly in contrast to what happens with two-qubits states,
 because only pure states ($R=1$) have maximum entanglement.

The exploration of ${\cal S}_R$, the space of all two-rebits states, is 
analogous to the one for two-qubits (See Appendix B).                           
An arbitrary (pure and mixed) state $\rho$ of
a (real) quantum system described by an $N$-dimensional real
Hilbert space can always be expressed as the product of three
matrices,

\begin{equation}\label{odot} \rho \, = \, R D[\{\lambda_i\}] R^{T}. \end{equation}

\noindent Here $R$ is an $N\times N$ orthogonal matrix and
$D[\{\lambda_i\}]$ is an $N\times N$ diagonal  matrix whose
diagonal elements are $\{\lambda_1, \ldots, \lambda_N \}$, with $0
\le \lambda_i \le 1$, and $\sum_i \lambda_i = 1$.
   The group of orthogonal matrices $O(N)$ is
endowed with a unique, uniform measure $\nu$ \cite{PZK98}. On the
other hand, the simplex $\Delta$, consisting of all the real
$N$-tuples $\{\lambda_1, \ldots, \lambda_N \}$ appearing in
(\ref{odot}), is a subset of a $(N-1)$-dimensional hyperplane of
${\cal R}^N$. Consequently, the standard normalized Lebesgue
measure ${\cal L}_{N-1}$ on ${\cal R}^{N-1}$ provides a natural
measure for $\Delta$. The aforementioned measures on $O(N)$ and
$\Delta$ lead then to a natural measure $\mu = \nu \times {\cal L}_{N-1}$
on the set ${\cal S}_R$ of all the states of our (real) quantum system. 
In random matrix analysis, a state like (\ref{odot}) belongs to the 
Circular Orthogonal Ensemble (COE). See Appendix B for more details.

\begin{figure}
\hspace{2cm}
\includegraphics[angle=270,width=.4\textwidth]{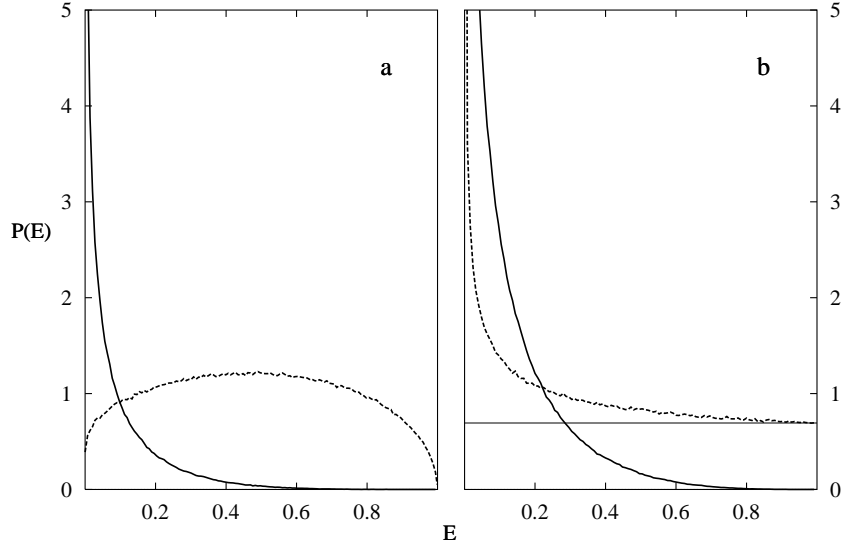}
\caption{a) Plot of the probability $P(E)$ of finding
two-qubits states endowed with a given entanglement $E$. The solid
line correspond to arbitrary states and the dashed line to pure states.	
b) Plot of the probability $P(E)$ of finding two-rebits states
endowed with a given entanglement $E$. The solid line correspond to arbitrary
states and the dashed line to pure states. The horizontal line corresponds to
the limit value $P(E=1)=\ln 2$ of the probability density  associated with pure
two-rebits states.} 
\label{fig4R}
\end{figure}

\begin{figure}
\begin{center}
\includegraphics[angle=270,width=.65\textwidth]{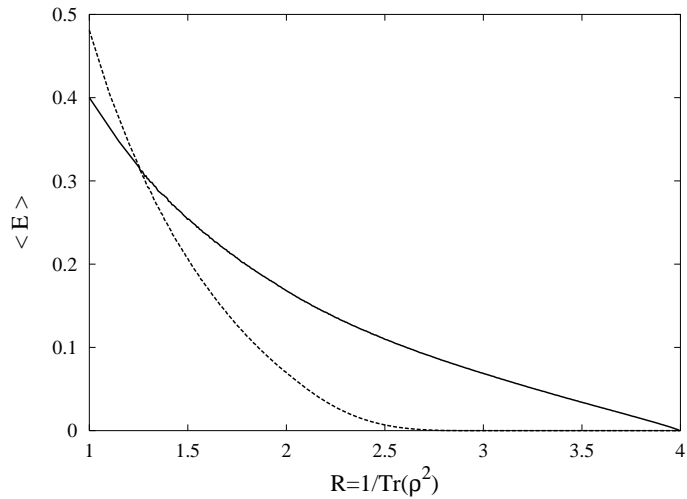}
\caption{The continuous line shows the behaviour of the mean
entanglement of formation $\langle E \rangle $  of real density matrices (given
by the CFR expression) as a function of the participation ratio $R$. The dashed
line shows the behaviour of the mean entanglement of formation $\langle E
\rangle $ of complex density matrices (given by Wootters' formula) as a function
of the participation ratio $R$.} 
\label{fig2R}
\end{center}
\end{figure}

The relationship between the amount of entanglement and the purity
of quantum states of composite systems has been discussed previously. 
As the degree of mixture increases, quantum states tend to have a smaller
  amount of entanglement. 
To study the relationship between entanglement and mixture in real
quantum mechanics, we compute numerically the probability $P(E)$ of finding
 a two-rebits state endowed with an amount of entanglement of formation $E$.
In Fig.\ref{fig4R} we compare (i) the distribution associated with two-rebits states 
with (ii) the one associated with two-qubits states. Fig.\ref{fig4R}a depicts 
the probability $P(E)$ of finding two-qubits states endowed with a given 
entanglement of formation $E$. In a similar way, Fig.\ref{fig4R}b exhibits a plot                        
of the probability $P(E)$ of finding two-rebits states endowed with a given
entanglement $E$ (as computed with the CFR formula). 
Comparing Figs.\ref{fig4R}a and \ref{fig4R}b we find that the 
distributions $P(E)$ describing arbitrary states (that is,
both pure and mixed states) exhibit the same qualitative shape for both
two-qubits and two-rebits states: in the two cases the distribution $P(E)$ is a
decreasing function of $E$. The mean entanglement also differs between 
standard and real quantum mechanics. 
The continuous line in Fig.\ref{fig2R} illustrates  
the behaviour of the mean entanglement of formation $E$ of
real density matrices (given by the CFR expression) as
a function of the participation ratio $R$. The dashed line
in Fig.\ref{fig2R} shows the behaviour of the mean entanglement 
of formation $E$ of complex density matrices (given
by Wootters' formula) as a function of the participation
ratio $R$. The two curves are quite different. In fact, if we were to generate 
states like (\ref{odot}) {\it only} and compute the ensuing mean entanglement 
by recourse of both formulas (Wootters' and CFR), we would notice that CFR 
constitutes an upper bound to the Wootters' one {\it for all R}. 
In the CFR case one can encounter entangled states for all $R$. Of course this 
is wrong, but completely consistent in the framework of real quantum mechanics.

  The distribution $P(E)$ or $P(C^2)$ for pure two-rebits
  states can be
  obtained analytically. Let us write a pure two-rebits state
  in the form

 \begin{equation}\label{repuro}
\mid \Psi \rangle \, = \,  \sum_{i=1}^4 \, c_i
  \mid \phi_{i}\rangle,
 \end{equation}

 \noindent
 where

 \begin{equation}\label{esfera4}
 \sum_{i=1}^4 \, c_i^2 \, = \, 1, \,\,\,\,\,\, c_i\in {\cal R}.
 \end{equation}

 \noindent
 The states $( \mid \phi_{i}\rangle, \,\,\, i=1,\ldots, 4$)
 are the eigenstates of the operator $\sigma_y \otimes \sigma_y$. The four real
 numbers $c_i$ constitute the coordinates of a point lying on
 the three dimensional unitary hyper-sphere $S_3$ (which is
 embedded in ${\cal R}^4$). We now introduce on $S_3$ three
 angular coordinates, $\phi_{1}$, $\phi_{2}$, and $\theta $,
 defined by

 \ben \label{triangle}
 c_1 \, &=& \, \cos \theta \cos \phi_1, \cr
 c_2 \, &=& \, \cos \theta \sin \phi_1, \cr
 c_3 \, &=& \, \sin \theta \cos \phi_2, \cr
 c_4 \, &=& \, \sin \theta \sin \phi_2, \,\,\,\,\,0\le \theta < \frac{\pi}{2},
 \,\, 0\le \phi_1,\phi_2 <2\pi.
 \een

\noindent In terms of the above angular coordinates, the
concurrence of the pure state $\mid \!\Psi \rangle $ is given by

\begin{equation} \label{cetita} C \, = \, \mid \! \langle \sigma_y \otimes
\sigma_y \rangle \! \mid \, = \, \mid \!\cos 2\theta \!\mid. \end{equation}

Using (\ref{triangle}) and (\ref{cetita}) one deduces that the probability
density $P(C^2)$ of finding a pure two-rebits state with a  squared concurrence
$C^2$ is given by \cite{BCPP02a}

\begin{equation}
 P(C^2) \, = \, \frac{1}{2\sqrt{C^2}}.
\end{equation}

\noindent The distribution is to be compared with the one obtained for pure
states of two-qubits systems, which is (analytically) found to be \cite{Z01}

\begin{equation}
 P(C^2) \, = \, \frac{3}{2}\sqrt{1-C^2}.
\end{equation}

 As well, we can determine analytically which is the maximum entanglement
 $E_m$ of a two-rebits state compatible with a given participation ratio $R$.
 Since $E$ is a monotonic increasing function of the
 concurrence $C$, we shall find the maximum value of $C$
 compatible with a given value of $R$. In order to solve the
 ensuing variational problem (and bearing in mind that
 $C = \mid \!\! \langle \sigma_y \otimes \sigma_y \rangle \!\! \mid$ ),
 let us {\it first} find the state that extremizes
  ${\rm Tr} (\rho^2)$ under the constraints associated with a
  given value of  $\langle \sigma_y \otimes \sigma_y \rangle $,
 and the normalization of $\rho $. This variational
 problem can be cast in the fashion

\begin{equation} \label{maxent1}
 \delta \Bigl[ {\rm Tr} (\rho^2) \, + \, \beta
\langle \sigma_y \otimes \sigma_y \rangle -\alpha {\rm Tr} (\rho)
\Bigr] \, = \, 0, \end{equation}

\noindent where $\alpha $ and $\beta $ are appropriate Lagrange
multipliers.

 After some algebra, and  expressing  the expectation value of
$\langle \sigma_y \otimes \sigma_y \rangle $ in terms of the parameter $\beta$,
 one finds that the maximum value of $C^2$ compatible with a given value
 of $R$ is given by

 \begin{equation} \label{cedoser}
 C^2_m \, = \, \left\{ 1 \,\,\,\,\,\,\,\,\,\,\,\,\,\, ; \,\,\, 1\le R \le 2
 \atop \frac{4}{R}-1 \,\,\,         ; \,\,\, 2\le R \le 4. \right.
 \end{equation}

\noindent 
\newline

The gathering of all these results concerning entanglement in the framework 
of real quantum theory can be considered as a complement towards a better 
understanding of which features of entanglement are unique and which
are more generic across various foil theories. The natural step to an 
extension of entanglement to quaternionic quantum mechanics formalism
\footnote{A current review on the experimental status of 
quaternionic quantum mechanics can be found in \cite{RevQuater}.} is sketched in 
\cite{MINSK2002}.


\section{Concluding remarks}

One of the goals of the present Chapter was to illuminate some 
further details concerning entanglement in the product space
$\mathcal{S}$ of two qubits, a 15-dimensional one. Distances between 
states can be calculated in diverse fashion. To what an extent does this 
fact influence the conclusions derived from bipartite entanglement 
explorations? In order to answer this question it has been 
shown that changing the measure $\mu$ that defines the way 
mixed states are distributed does not affect the
description of $\mathcal{S}-$ properties. Changing the way in which 
distances between states in $\mathcal{S}$ are evaluated does have 
sensible effects in {\it some} cases. Different metrics (Bures and 
Hilbert-Schmidt) have been used in the description of $\mathcal{S}$, 
but which of these distances one employs does not affect the description 
of the action of logic gates.

Pursuing the complete characterization of $\mathcal{S}$, we have 
described that the way of defining a measure over this set of states is not 
unique. By exploring the outcomes of different probability distributions 
over the simplex of eigenvalues $\Delta$, we end up in a position where 
the usual $\mu_Z$ measure so far used in all considerations can still 
be regarded as a natural measure, is spite of several criticisms that 
can be weighted against more practical grounds.

As well, we have explored numerically the entanglement properties of two-rebits 
systems. A systematic comparison has been established between many statistical
properties of two-qubits and two-rebits systems, paying particular attention
to the relationship between entanglement and purity in both quantum mechanical
frameworks. We also determined numerically the probability densities
$P(E)$ of finding (i) pure two-rebits states and (ii) arbitrary two-rebits
states, endowed with a given amount of entanglement $E$ or concurrence squared
$C^2$. In particular, we determined analytically the maximum possible value of
the concurrence squared $C^2$ of two-rebits states compatible with a given
value of mixedness $R$. Where all this real quantum mechanical approach will 
lead is uncertain, but it constitutes one step further in the understanding 
of the intriguing phenomenon of entanglement.

\chapter{Distribution of entanglement changes produced by unitary operations}

In this Chapter we shall investigate the changes of entanglement 
$\Delta E$ produced by  quantum logical gates acting
on composite quantum systems. Quantum gates, the quantum generalization of 
standard logical gates, play a fundamental role in quantum 
computation and other quantum information processes. 
Quantum gates are described by unitary transformations
$\hat U$ acting on the relevant Hilbert space describing 
the system under study (usually a multi-qubit system). In 
general, a quantum gate acting on a composite system changes 
the entanglement of the system's concomitant quantum state. It 
is them a matter of interest to obtain a detailed characterization of the
aforementioned entanglement changes.  Quite interesting work has recently been
performed to this effect (see, for instance,
\cite{Z01,batle,BPCP03,KC01,DVCLP01,WZ02}).

 A physically motivated measure of entanglement is
 provided by the entanglement of formation $E[\rho]$  \cite{BDSW96}.
 This measure quantifies the resources needed to create a 
given entangled state $\rho$. This will be the measure employed in 
the quantification of the entanglement change of a state induced by 
quantum gates.  

 One of the simplest nontrivial two-qubit
operation is the quantum controlled-NOT, or CNOT (equivalently, the exclusive
OR, or XOR). Its classical counterpart is a reversible logic gate operating on
two bits: $e_1$, the control bit,  and $e_2$, the target bit. If $e_1=1$, the
value of $e_2$ is negated. Otherwise, it is left untouched. The quantum CNOT
gate $C_{\rm CNOT}$ plays a paramount role in both experimental and theoretical
efforts that revolve around the quantum computer concept. In a given
ortonormal basis $\{\vert 0 \rangle,\,\vert 1 \rangle\}$, and if we denote
addition modulo 2 by the symbol $\oplus$, we have \cite{barenco}

\be  \vert e_1 \rangle\, \vert e_2 \rangle \rightarrow
C_{\rm CNOT}\rightarrow \vert e_1 \rangle\, \vert e_1 \oplus e_2 \rangle.   \ee
In conjunction with simple single-qubit operations, the CNOT gate constitutes a
set of gates out of which {\it any quantum gate may be built} \cite{barenco1}.
In other words, single qubit and CNOT gates are universal for quantum
computation \cite{barenco1}.

As stated, the CNOT gate  operates on quantum states of two qubits and is
represented by the $4\times 4$-matrix,

\be \label{CNOT}
U_{\rm CNOT} \, = \,
\left(
\begin{array}{cccc}
 1 & 0 & 0 & 0 \\
 0 & 1 & 0 & 0 \\
 0 & 0 & 0 & 1 \\
 0 & 0 & 1 & 0
\end{array}
\right)
\ee

We are also going to consider the parameterized
family of transformations $\hat U_{\theta}$ described
by the matrices

\be \label{tetransform}
U_{\theta} \, = \, \left(
\begin{array}{cccc}
 1 & 0 & 0 & 0 \\
 0 & 1 & 0 & 0 \\
 0 & 0 & \cos(\theta) & \sin(\theta) \\
 0 & 0 & -\sin(\theta) & \cos(\theta)
\end{array}
\right)
\ee

\noindent We have selected this family of unitary transformations because
we can explore for different $\theta$ values the changes
of entanglement on a two-qubit space.

As advanced, we study some aspects of the
entanglement changes generated by one of the basic
constituents of any quantum information processing device: 
unitary evolution operators $\hat U$ that act on the states of a 
composite quantum system. For instance, given an initial degree
of entanglement of formation $E$, what is the probability $P(\Delta E)$ of
encountering a change in entanglement $\Delta E$ upon the action of $\hat
U$?

To answer this type of questions we will perform a Monte Carlo exploration of
the quantum state-space ${\cal S}$, in the same fashion as previously done in previous 
chapters. Therefore all our present considerations are based on the assumption
 that the uniform distribution of states of the composite
 quantum system under study is the one determined by the 
 measure (\ref{memu}). Thus, in our numerical computations 
 we are going to randomly generate states of a the system 
 according to the measure (\ref{memu}) and investigate 
 the entanglement evolution of these states upon the action 
 of quantum logical gates $\hat U$.

During the generation of $P(\Delta E)-$distributions of different quantum gates, 
there appears a remarkable fact. The final numerical distribution $P(\Delta E)$ 
of entanglement changes for different gates are nearly identical. For instance, 
the ones corresponding to $CNOT$ and $U_{\pi/2}$ are characterized by
the same $P(\Delta E)-$distribution. Why is this so? Is it a coincidence than 
they look similar, or perhaps we should blame the numerical resolution? Let us  
discuss it in more detail. 

  Let us consider two quantum gates $U$ and $U_T$, which act on
  a two-qubits system, and are related by

\be \label{ut}
U_T \, = \, U_{LA} \, U \, U_{LB}, \ee

\noindent where

\be \label{uta} U_{LA} \, = \, U_{A1} \otimes U_{A2}, \ee

\be \label{utb} U_{LB} \, = \, U_{B1} \otimes U_{B2}, \ee

\noindent and the unitary transformations
 $ U_{Ai}, U_{Bi}, \,\,\, (i=1,2),$
act on the $i$-qubit. The unitary transformations $U_{LA}$ and $U_{LA}$ are
tensor products of unitary transformations acting
 locally on each qubit. That is, they are local transformations.
Local unitary transformations do not change the amount of entanglement of the
two-qubit quantum state. Now we are going to compare the distributions of
entanglement changes $P(\Delta E)$ generated, respectively, by the
transformations $U$ and $U_T$. In order to do this, it is convenient first to
consider separately the transformations

\be
U_{TA} \, = \, U_{LA} \, U, \ee

\noindent and

\be
U_{TB} \, = \, U \, U_{LB}. \ee

\noindent Given an initial state $\rho_i$ endowed with entanglement $E_i$, it
is plain that the entanglement change $\Delta E = E_f - E_i$, where $ E_f $ is
the entanglement of the state obtained after applying the transformation, is
the same for the transformations $U$ and $U_{TA}$. This is so because to apply
$U_{TA}$ is tantamount to apply first $U$, and them to apply $U_{LA}$. But
$U_{LA}$, being local, does not alter the entanglement of the final state.

Let us now discuss what happens with the entanglement changes generated by
$U_{TB}$. Consider an initial state $\rho_i$ and a small neighbourhood $d{\cal
S}$ of $\rho_i$ with volume $d\Omega$ (the volume evaluated according to the
product measure). The product measure is invariant under unitary
transformations. Consequently, the set $d{\cal S}^{\prime}$ obtained applying
the transformation $U_{LB}$ to each of the members of  $d{\cal S}$ has the same
volume $d\Omega$ as the set $d{\cal S}$. In particular, the state

\be
\rho_i^{\prime} \, = \, U^{\dagger}_{LB} \, \rho_i \, U_{LB} \ee

\noindent
belongs to $d{\cal S}^{\prime}$. Now, let us consider the state

\be
\rho_f \, = \, U^{\dagger} \, \rho^{\prime}_i U \, = \, U^{\dagger}_{TB} \,
\rho_i \, U_{TB}. \ee

\noindent Taking into account that both the entanglement of formation and
the product measure are invariant under $U_{LB}$, it follows that the
contribution of the states within $d{\cal S}$ to the probability distribution
$P(\Delta E)$ associated with $ U_{TB}$ is equal to the contribution of the
states in $d{\cal S}^{\prime}$ to the probability distribution $P(\Delta E)$
corresponding to the transformation $U$. Now, since this holds true for any
small element $d{\cal S}$ of the state space ${\cal S}$, and the local unitary
transformation $U_{LB}$ is a one-to-one map of the state space ${\cal S}$ into
itself, we can conclude that the unitary transformations $U$ and $U_{TB}$
exhibit the same probability distribution $P(\Delta E)$ of entanglement
changes.

 Summing up, we have that

 \begin{itemize}
 \item{(A) The unitary transformations $U$ and $U_{TA}$ have the same distribution $P(\Delta E)$.}
 \item{(A) The unitary transformations $U$ and $U_{TB}$ have the same distribution $P(\Delta E)$.}
 \end{itemize}

 \noindent
 Combining (A) and (B) we can conclude that the transformations $U$ and $U_{T}$
 (eq. \ref{ut}) share the same $P(\Delta E)$-distribution as well. It is important
to realize that the only property of the state-space volume measure which is relevant
for the above argument is that the measure must be invariant under unitary transformations. 
Consequently, the above argument is valid for the product measure 
(\ref{memu}), as well as for any other measure which is invariant under unitary 
trasnformations. Moreover, the above argument does not hold only for two-qubits
systems. It holds for general composite quantum systems.

 As an illustration, let us compare the CNOT logical gate with the gate $U_{\pi/2}$
 (see Eq. (\ref{tetransform})). Defining

\be
U_{LA} \, = \, \left(
\begin{array}{cc}
 1 & 0 \\
 0 & e^{i\pi/2}
\end{array}
\right) \otimes
 \left(
\begin{array}{cc}
e^{-i\pi/2} &  0 \\
   0        &  1
 \end{array}
 \right),
\ee

\noindent and

\be
U_{LB} \, = \, \left(
\begin{array}{cc}
 1 & 0 \\
 0 & 1
\end{array}
\right)
\otimes
 \left(
\begin{array}{cc}
e^{i\pi/2} &  0 \\
   0        &  1
 \end{array}
 \right),
\ee

\noindent we have,

\be
U_{\pi/2} \, = \, U_{LA} \, U_{\rm CNOT} \, U_{LB}. \ee

\noindent Consequently, the gates $CNOT$ and $U_{\pi/2}$ are characterized by
the same distribution $P(\Delta E)$ of entanglement changes, as viewed by 
numerical inspection. However, it is
worth while to mention that the gates $U_{CNOT}$ and $U_{\pi/2}$
{\it do not yield the same changes of entanglement when acting on individual
states}.

\section{The Hadamard-CNOT quantum circuit}

Let us discuss an interesting example of $P(\Delta E)$ distributions in 
the form of the Hadamard-CNOT circuit. The Hadamard-CNOT quantum circuit 
combines two gates: a single-qubit one (Hadamard's) with a two-qubits gate 
(CNOT). The simplest nontrivial two-qubit 
operation is the quantum controlled-NOT, or CNOT. Its classical counterpart is
a reversible logic gate operating on two bits: $e_1$, the control bit,  and
$e_2$, the target bit. If $e_1=1$, the value of $e_2$ is negated. Otherwise, it
is left untouched. The quantum CNOT gate $C_{12}$ (the first subscript denotes
the control bit, the second the target one)  plays a paramount role in both
experimental and theoretical efforts that revolve around the quantum computer
concept. In a given ortonormal basis $\{\vert 0 \rangle,\,\vert 1 \rangle\}$,
and if we denote addition modulo 2 by the symbol $\oplus$, we have
\cite{Galindo}, $C_{12}:\, \vert e_1 \rangle\, \vert e_2 \rangle
\rightarrow \vert e_1 \rangle\, \vert e_1 \oplus e_2 \rangle. $ In conjunction
with simple single-qubit operations, the CNOT gate constitutes a set of gates
out of which {\it any quantum gate may be built} \cite{barenco1}. This gate is
able to transform factorizable pure states into entangled ones,
 i.e., $  C_{12}: [c_1 \vert 0 \rangle + c_2 \vert
1 \rangle] \vert 0 \rangle \leftrightarrow c_1 \vert 0 \rangle \vert 0 \rangle
+ c_2 \vert 1 \rangle \vert 1 \rangle. $ This transformation can be reversed by
applying the CNOT operation once more.

The Hadamard transform $T_H$ ($T_H^2=1$) is given by $T_H= \frac{1}{\sqrt{2}}
[\sigma_1 + \sigma_3],$ and acts on the single qubit basis $\{ |0>,\,\, |1> \}$
in the following fashion, $T_H |0> = \frac{1}{\sqrt{2}}[|1> - |0>]$,  $T_H |1>
=\frac{1}{\sqrt{2}}[|0> + |1>]$. Consider now the two-qubits uncorrelated basis
$\{ |00>,\,  |01>,\,  |10>,\,  |11> \}$. If we act with $T_H$ on the members of
this basis we obtain $\frac{1}{\sqrt{2}}\,\,\big[|1> - |0>\big]\,|0>$,
$\frac{1}{\sqrt{2}}\,\,\big[|1> - |0>\big]\,|1>$,
$\frac{1}{\sqrt{2}}\,\,\big[|0> + |1>\big]\,|0>$,
$\frac{1}{\sqrt{2}}\,\,\big[|0> + |1>\big]\,|1>$. The posterior action of the
CNOT gate yields $\frac{1}{\sqrt{2}}[ |1>|1> - |0>|0> ]$,
$\frac{1}{\sqrt{2}}\,[ |1>|0> - |0>|1> ]$, $\frac{1}{\sqrt{2}}\,[ |0>|0> +
|1>|1> ]$, $\frac{1}{\sqrt{2}}\,[ |0>|1> + |1>|0> ]$
i.e., save for an irrelevant overall phase factor in two of the
kets, the maximally correlated Bell basis $\vert \phi^{\pm}
\rangle$, $\vert \psi^{\pm} \rangle$. We see then that the
$T_H$-CNOT combination transforms an uncorrelated basis
into
the maximally correlated one.

 The two-qubit systems are, as we know, the
simplest quantum mechanical systems exhibiting the entanglement
phenomenon and play a fundamental role in quantum information
theory. They also provide useful limit cases for testing the
behaviour of more involved systems \cite{gellmann}.

\begin{figure}
\hspace{2cm}
\includegraphics[angle=270,width=0.4\textwidth]{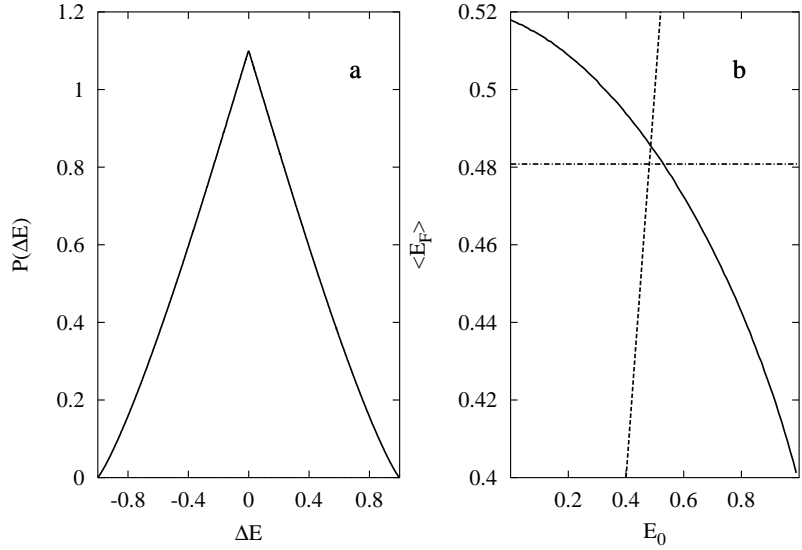}
\caption{a) $P(\Delta E)$ vs. $\Delta E$ for pure states. 
The change of entanglement $\Delta E$ arises as a result of the
action of the $T_H$-CNOT quantum circuit. b) Probability of
obtaining, via the $T_H$-CNOT transformation, a final state with mean
entanglement $\langle E_F \rangle$, when the initial state is endowed with a given
entanglement $E_0$ (solid line). The horizontal line depicts the
mean entanglement of all pure states. The diagonal (dashed line)
is drawn for visual reference.}
\label{had1}
\end{figure} 

We shall perform a systematic numerical survey of the action of  the $T_H$-CNOT
circuit on our 15-dimensional space \cite{circuit}. We will try to answer
the question: given an initial degree of entanglement of formation $E$, what is
the probability $P(\Delta E)$ of encountering a change in entanglement $\Delta
E$ upon the action of this circuit?

Our answer will arise from a Monte Carlo exploration of ${\it H}$ by randomly 
generating states of a two-qubit system according to the usual measure
 $\mu$ (\ref{memu}), studying the entanglement evolution of these states
upon the action of our $T_H$-CNOT circuit.

\begin{figure}
\hspace{2cm}
\includegraphics[angle=270,width=0.4\textwidth]{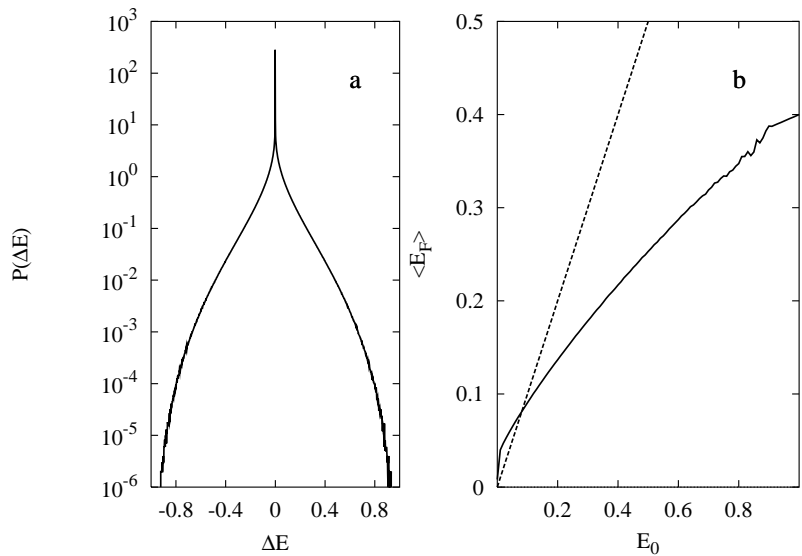}
\caption{Same as Fig.\ref{had1} for all states (pure and mixed).}
\label{had2}
\end{figure}

We deal with pure states only in Fig.\ref{had1}. Fig.\ref{had1}a plots the 
probability $P(\Delta E)$ of obtaining via the $T_H$-CNOT quantum
 circuit a final state with entanglement change $\Delta E=E_F-E_0$.
 In Fig.\ref{had1}b we are concerned with the average value $\langle E_F
 \rangle$ pertaining to final states that result from the gate-operation on initial
 ones of a given (fixed) entanglement $E_0$ (solid line). The horizontal line
 is plotted for the sake of reference. It corresponds to the
 average entanglement of two-qubits pure states, equal to
 $1/(3\ln{2})$. The diagonal line $\langle E_F \rangle$ = $E_0$ is also shown (dashed line).
$\langle E_F \rangle$ is a decreasing function of $E_0$ although
the quantum circuit considered increases the mean final
entanglement by amounts of up to 0.5 for states with $E_0$ lying 
in the interval $[0,0.5]$.

The same analysis, but involving now all states (pure and mixed),
 is summarized in Fig.\ref{had2}. The graph \ref{had2}a is the counterpart of 
 \ref{had1}a, while \ref{had2}b is that of \ref{had2}a.
 The dashed line of 2b, given for the sake of visual reference, if
 just the line $\langle E_F  \rangle=E_0$.
 The two Figs. allow one to appreciate the fact that it is quite unlikely 
 that we may generate, via the $T_H$-CNOT quantum circuit,
 a significant amount of entanglement if the initial state is
 separable. In Fig.\ref{had2} we see that the mean final entanglement
 $\langle E_F \rangle$ rises rapidly near the origin, from zero, with $E_0$ .
 The rate of entanglement-growth decreases steadily with $E_0$ and the interval
in which $\langle E_F \rangle$ is greater than $E_0$ is significantly smaller
that the one corresponding to pure states (Fig.\ref{had1}b). The $P(\Delta E)$ vs.
$\Delta E$ plots exhibit a  nitid peak at $\Delta E=0$. The peak is enormously
exaggerated if mixed states enter the picture (\ref{had2}a). Thus, if the initial state
has null entanglement, our survey indicates that the most probable circumstance
is that the circuit will leave its entanglement unchanged.

\section{Entanglement distribution and entangling power of quantum gates}

Quantum gates, the quantum generalization of the so-called standard 
logical gates, play a fundamental role in quantum 
computation and other quantum information processes, being 
described by unitary transformations $\hat U$ acting on 
the relevant Hilbert space (usually, that for a multi-qubit system). 
In general, a quantum gate acting on a composite system changes 
the entanglement of the system's concomitant quantum state. It 
is then a matter of interest to obtain a detailed characterization of the 
aforementioned entanglement changes (\cite{Z01,batle,KC01,DVCLP01}). 
To such an end, the study is greatly simplified if one is able to 
conveniently parameterize the gate's non-local features. We need 
$N^2-1$ parameters to describe a unitary transformation $U(N)$ in a system 
of $N = N_A \times N_B$ dimensions. In the case of two qubits ($N=2\times2$) 
one needs just three parameters ${\bf \lambda} 
\equiv (\lambda_1,\lambda_2,\lambda_3)$ (\cite{KC01}), since any 
two-qubits quantum gate can be decomposed in the form of a product of local 
unitarities, acting on both parties, and a ``nuclear" part $\tilde U$, 
which is completely non-local. Given a quantum gate $U$, the 
concomitant distribution of entanglement changes is equivalent, on average, 
to the one produced by $\tilde U$, and we need to know the vector 
${\bf \lambda}$.

In addition to studying  changes in the entanglement of a given state produced 
by quantum gates, we would like  to ascertain  entangling 
capabilities of  unitary operations or evolutions. In point of fact, 
the latter enterprise complements the former. By looking at the distribution 
of entanglement changes induced by several quantum gates, 
one can deduce a special formula that quantifies the 
``entangling power". To such an end we use the definition introduced 
by Zanardi {\it et al.} \cite{Zan00}, and introduce a new one as well, 
based exclusively on the shape of a particular probability (density) 
distribution: that for finding a state with a given entanglement change 
$\Delta E$, measured in terms of the so called entanglement of 
formation \cite{WO98}. We will see that the distribution obtained by 
randomly picking up two states measuring their relative entanglement change is 
optimal in the context of our new measure. Moreover, the two-qubits instance 
will be seen to be rather peculiar in comparison with its counterpart 
for larger dimensions (bipartite systems, 
like two-qudits $N_A \times N_A$, for $N_A$ = 3,4,5 and 6).

Extending the above considerations to mixed states requires the introduction 
of a measure for the simplex of eigenvalues of the matrix $\hat \rho$ instead 
of dealing with pure states distributed according to the invariant 
Haar measure. Rather than mimicking the aforementioned evaluation, which 
could be easily achieved by introducing a proper measure for the generation 
of mixed states, we will generate them in the fashion of 
Refs. (\cite{ZHS98,Z99,Z01,MJWK01,IH00,BCPP02a,BCPP02b}). In such a 
connection we discuss the action of the exclusive-OR or controlled-NOT 
gate (CNOT in what follows) in the 15-dimensional 
space $\cal S$ of mixed states and compare our results with those 
obtained using the well known Hilbert-Schmidt and Bures metrics \cite{ZyckJPA}.

Also we study numerically how the entanglement is distributed when more than 
two parties are involved (multipartite entanglement). By applying locally 
the CNOT gate to a given pair of two-qubits in a system of pure states 
composed by three or four qubits, we shall study the concomitant distributions 
of entanglement changes among different qubits, pointing out the 
differences between them. Great entanglement changes are appreciated 
as we increase the relevant number of qubits \cite{optics2005}.

\subsection{Optimal parameterization of quantum gates for two-qubits systems}

Two-qubits systems are the simplest quantum ones exhibiting 
the entanglement phenomenon. They play a fundamental role in 
quantum information theory. There remain still some features 
of these systems, related to the phenomenon of entanglement, that 
have not yet been characterized in enough detail, as for instance, 
the manner in which $P(\Delta E)$, the probability of generating a 
change $\Delta E$ associated to the action of these operators, 
is distributed under the action of certain quantum gates. In this 
vein it is also of interest to express the general quantum two-qubits 
gate in a way as compact as possible, i.e., to find an optimal parameterization.

Since any quantum logical gate acting on a two-qubits system can be 
expressed in the form \cite{VHC02},

\begin{equation} 
\label{lambdecomp1} 
\left( v_1 \otimes v_2 \right) \, \exp 
\left[-i
\sum_{i=1}^3 \lambda_k \sigma_k \otimes \sigma_k \right] \, 
\left( w_1 \otimes w_2  \right), 
\end{equation} 

\noindent where the transformations $v_{1,2}$ and $w_{1,2}$ act only on one of 
the two qubits, and $\sigma_k$ are the Pauli matrices. Note that it is always 
possible to chose the $\lambda$-parameters in such a way that

\begin{eqnarray}
\lambda_1 \ge \lambda_2 \ge |\lambda_3|, \cr \lambda_1, \lambda_2 \in
[0,\pi/4], \cr \lambda_3 \in (-\pi/4, \pi/4], 
\end{eqnarray}

\noindent and consider the parameterized unitary transformation

\begin{equation} \label{lambdecomp2} \tilde U_{(\lambda_1,\lambda_2,\lambda_3)} 
\, = \, \exp \left[-i \sum_{i=1}^3 \lambda_k \sigma_k \otimes \sigma_k \right]. 
\end{equation} 

From previous work \cite{VHC02,PhysicaApending} we know that the unitary 
transformations (\ref{lambdecomp1}) and (\ref{lambdecomp2}) share the same 
probability distribution $P(\Delta E)$ of entanglement changes. Consequently, 
the $P(\Delta E)$-distribution generated by any quantum logical gate 
acting on a two-qubits system coincides, for appropriate values of the 
$\lambda$-parameters, with the distribution of entanglement changes 
associated with a unitary transformation of the form (\ref{lambdecomp2}). 
This means that the set of all possible $P(\Delta E)$-distributions for 
two-qubits gates constitutes, in principle, a three-parameter family 
of distributions.

\begin{figure}
\begin{center}
\includegraphics[angle=270,width=0.95\textwidth]{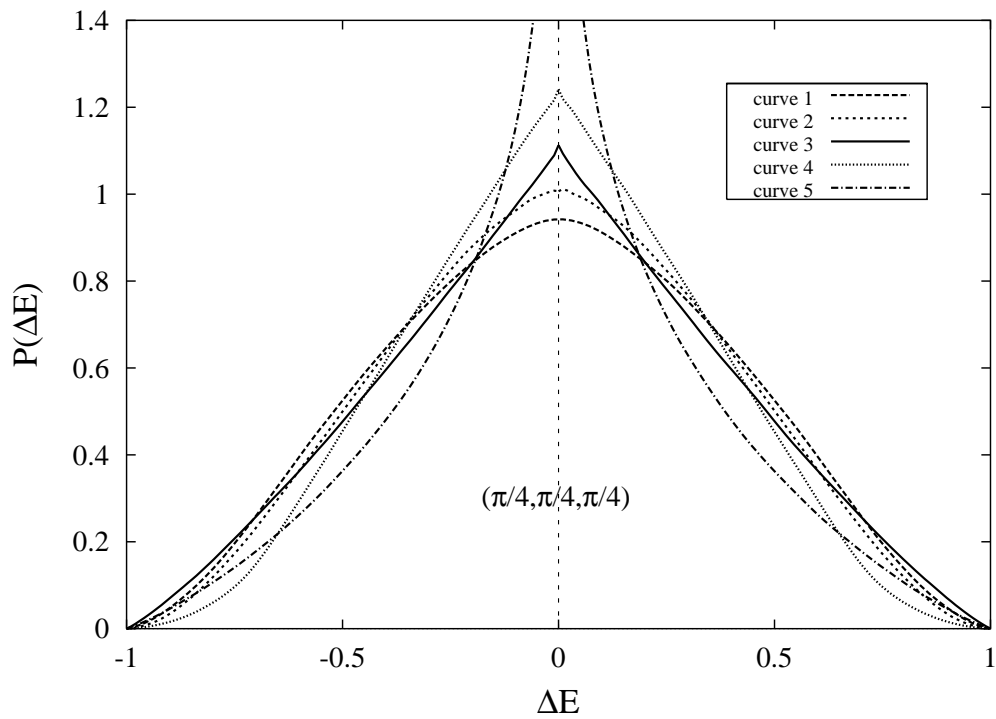}
\caption{$P(\Delta E)$-distributions generated by the two-qubits quantum gates, 
parametrized in an optimal way. Curve 1 corresponds 
to $\lambda$ = ($\pi/4$, $\pi/8$, 0), curve 2 to ($\pi/4$, $\pi/8$, $\pi/16$), 
curve 3 to ($\pi/4$, 0, 0) (or equivalently to the CNOT gate), 
curve 4 to ($\pi/4$, $\pi/8$, -$\pi/8$) and curve 5 to 
($\pi/8$, $\pi/8$, $\pi/8$). The vertical line represents any gate 
that can be mapped to the identity $\hat I$. All depicted quantities 
are dimensionless.}
\label{fig1U}
\end{center}
\end{figure}

We have explored the two-qubits space by means of a Monte Carlo simulation 
\cite{ZHS98,Z99,PZK98} and in Fig.\ref{fig1U} we depict the action of several gates 
acting on 
two-qubit pure states, as described by different values of the 
vector $(\lambda_1,\lambda_2,\lambda_3)$. We see how different 
the associated entanglement probability distributions are. In point of fact, 
the CNOT gate (solid line) is equivalent (on average) to $(\pi/4,0,0)$. 
Curve 1 corresponds to $\lambda$ = ($\pi/4$, $\pi/8$, 0), curve 2 
to ($\pi/4$, $\pi/8$, $\pi/16$), curve 3 to ($\pi/4$, 0, 0), 
curve 4 to ($\pi/4$, $\pi/8$, -$\pi/8$), and curve 5 to 
($\pi/8$, $\pi/8$, $\pi/8$). All these gates have the common property 
that they reach the extremum $|\Delta E|=1$ change if have given 
the appropriate ${\bf \lambda}$ vectors. This is not the case for other gates 
like the $U_{\pi/4}$ one \cite{batle}. The vertical dashed line represents 
any gate that can be mapped to the identity $\hat I$, so that no change in 
the entanglement occurs (we get a delta function 
$\delta(\Delta E)$)\footnote{In point of fact, in Ref. \cite{PhysicaApending} it is shown that 
these distributions can be well fitted using a simple Tsallis $q$-distribution 
with one or two constraints.}.

\subsection{Quantum gates' entangling power: qubits and qudits}

As stated, a quantum gate (QG), represented by a unitary 
transformation $\hat U$, changes the entanglement of a given state. As 
a matter of fact, we may think of the QG as an ``entangler". This 
particular transformation represents the abstraction of some physical 
interaction taking place between the different degrees of freedom of 
the pertinent system. A natural question then arises: how good 
a quantum gate is as an entangler?, or in other words, can we quantify 
the set of quantum gates in terms of a certain ``entanglement capacity"? 
The question is of some relevance in Quantum Information. A quantum gate 
robust against environmental influence becomes specially suitable 
in the case of networks of quantum gates (quantum circuits, quantum 
computer, etc) as described by Zanardi {\it et al.} \cite{Zan00}, 
where the so called ``entangling power" $\epsilon_P(\hat U)$ of a 
quantum gate $\hat U$ is defined as follows

\begin{equation} \label{eP}
  \epsilon_P(\hat U) \, \equiv \, \overline{E\big((\rho_A\otimes\rho_B) 
  \hat U (\rho_A\otimes\rho_B)^{\dag}\big)}^{\rho_A,\rho_B},
\end{equation}

\noindent where the bar indicates averaging over all (pure) product states 
in a bipartite quantum state described by 
$\rho_{AB}=\rho_A\otimes\rho_B \in \cal H=\cal H_A \otimes \cal H_B$ and 
$E$ represents a certain measure of entanglement, in our case the 
entanglement of formation, that, in the case of pure states becomes just 
the binary von Neumann entropy of either reduced state 
$E(\rho_{AB})=-$Tr$(\rho_A \log_2 \rho_A)=-$Tr$(\rho_B \log_2 \rho_B)$. 
It greatly simplifies the numerics of our study to assume 
that the separable states $\rho_{AB}$ are all equally likely. 
The corresponding (special) form of (\ref{eP}) exhibits the advantage 
that it can be generalized to any dimension for a bipartite system. 
In our case, we are mostly interested in  two-qubits systems 
(the $2 \times 2$ case). In \cite{Zan00} the concept of {\it optimal} gate 
is introduced, where by {\it optimal} one thinks of a gate that makes 
(\ref{eP}) maximal. It is shown there that the CNOT gate is an optimal gate.

\begin{figure}
\begin{center}
\includegraphics[angle=270,width=0.65\textwidth]{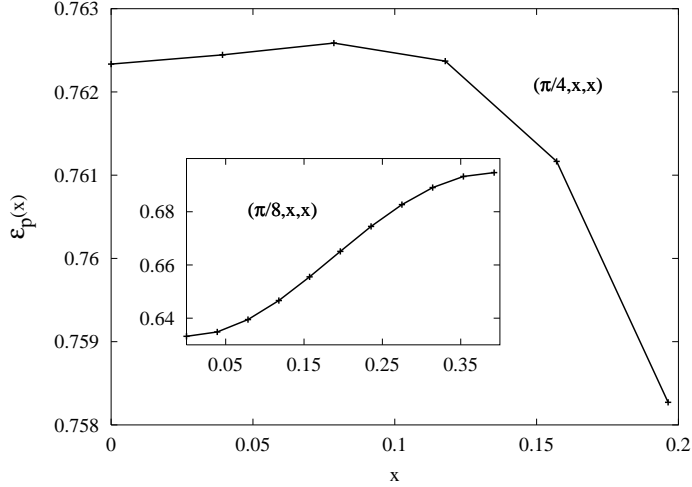}
\caption{Entangling power $\epsilon_{P}$ of the perturbed CNOT gate, expressed 
in the form of ($\pi/4, x, x$). Small perturbations around this optimal 
gate ($x=0$) find gates which are also optimal (greater $\epsilon_{P}$). 
Large deviations diminish the concomitant $\epsilon_{P}$. 
A perturbed non-optimal gate, like ($\pi/8, x, x$) shown in the inset, 
increases its $\epsilon_{P}$. See text for details. All depicted quantities 
are dimensionless.}
\label{fig2U}
\end{center}
\end{figure}

Let us suppose now that we make use of the special parameterization 
$\cal{P}$ (\ref{lambdecomp2}) for the unitary transformations $U(N)$. 
In the case of the CNOT gate, it was clear that $\cal{P}$ is 
equivalent (on average) to the $(\pi/4,0,0)$ gate. This fact allow us 
to see how the entangling power (\ref{eP}) evolves when we perturb 
the CNOT gate in the form  $(\pi/4,x,x)$, $x$ being a continuous parameter. 
To such an end we numerically generate {\it separable} 
\footnote{Or unentangled states. Let us remind the reader that by construction, 
product states are states with no quantum correlation between parties. 
A general necessary criterion for ascertaining when a state (pure or mixed) 
is entangled or not is given by the so called Positive Partial Transpose
criterion (PPT), first derived by Peres \cite{Peres}. Is is proven 
to be sufficient for $2 \times 2$ and $2 \times 3$ systems \cite{HoroPPT}.} 
states $\rho_A\otimes\rho_B$ according to the Haar measure on the group 
of unitary matrices $U(N)$ that induces a unique and uniform 
measure $\nu$ on the set of pure states of two-qubits ($N=4$) 
\cite{ZHS98,Z99,PZK98}. The corresponding results are shown in Fig.\ref{fig2U}. 
Every point has been obtained averaging a sampling of $10^9$ states, 
so that the associated error is of the order of the size of the symbol. 
It is clear from the plot that large deviations imply a smaller 
entangling power $\epsilon_P(CNOT_{pert.})$. Notice that a small 
perturbation around the origin (CNOT gate) {\it increases} 
the entangling power. This fact leads us to conclude that, in the 
space of quantum gates, and in  the vicinity of an optimal gate, there exists 
an infinite number of optimal gates. On the other hand, if we perturb a quantum 
gate which is not optimal, like $(\pi/8,x,x)$, any deviation, no matter how 
small, will lead to an increasing amount of the entangling power $\epsilon_P$. 
This latter case is depicted in the inset of Fig.\ref{fig2U}.

It is argued in \cite{Zan00} that the two-qubits case presents 
special statistical features, as far as the entangling power 
is concerned, when compared to $N_A \times N_A$ systems (two-qudits). 
We investigate this point next, not by making use of any 
quantum gate, or by recourse to Eq. (\ref{eP}). What we do instead 
might be regarded a ``no gate action": we look at the probability 
(density) distribution $P_{R}$ obtained by randomly picking up two 
pure states generated according to the Haar measure in $N_A \times N_A$ 
dimensions, and determine then the relative entanglement change 
$\Delta E$ in passing form one of these states to the other. 
The distribution $P_{R}$ is \cite{PhysicaApending}

\begin{equation} \label{randomint}
P_{R}(\Delta E) \, = \, \int^{1-|\Delta E|}_0 \, dE \, P(E) \, P(E+|\Delta E|).
\end{equation}

\begin{figure}
\begin{center}
\includegraphics[angle=270,width=0.65\textwidth]{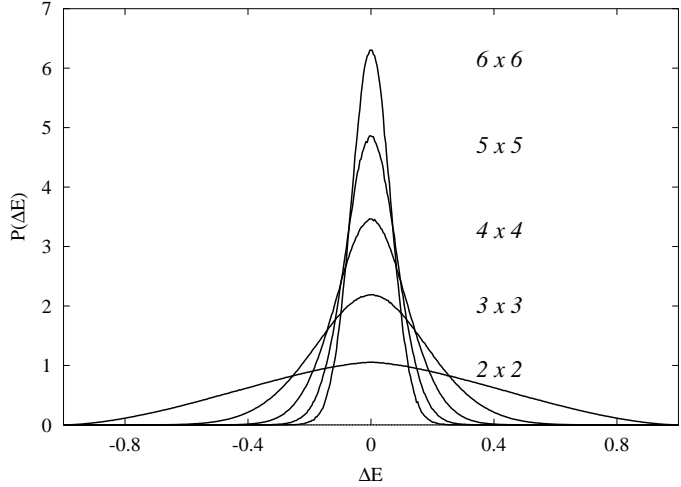}
\caption{$P(\Delta E)$-distributions generated ($\Delta E$ being the change 
in the entanglement of formation) by randomly choosing the initial and final 
pure two-qubits states ($2 \times 2$), and several two-qudits states 
($N_A \times N_A$, for $N_A=3,4,5,6$). The two-qubits instance appears to 
be a peculiar case. All depicted quantities are dimensionless.}
\label{fig3U}
\end{center}
\end{figure}

\noindent The distribution $P_R(\Delta E)$ is thus related to the probability 
density $P(E)$ of finding a quantum state with entanglement $E$. Notice that 
the above expression holds for any states space measure 
invariant under unitary transformations and for any bipartite quantum system
 consisting of two subsystems described by Hilbert spaces of the same 
dimensionality. We must point out that the entanglement is measured for every 
two-qudits in terms of $E=S(\rho_{A})/\log(N_{A})$, 
where $S$ is the von Neumann entropy, so that it ranges from 0 to 1 
($N_A$ is the dimension of subsystem $A$). The resulting distributions 
are depicted in Fig.\ref{fig3U}. The five curves represent the 
$2\times2, 3\times3, 4\times4, 5\times5$ and $6\times6$ systems. A first 
glance at the corresponding plot indicates a sudden change in the available 
range of $\Delta E$. The width of our probability distribution is rather large 
for two-qubits and it becomes narrower as we increase the dimensionality 
of the system. With this fact in mind, one may propose the {\it natural width} 
of these distributions as some measure of its entangling power. We choose 
the maximum spread of the distribution in $\Delta E$ at half its maximum 
height $P(0)$. If we use this definition of entangling power $W_{\Delta E}$, 
Fig.\ref{fig3U} provides numerical evidence for the peculiarity of the 
two-qubits instance. One may dare to conjecture, from inspection, 
that for large $N_A$, $W_{\Delta E}$ decays following a 
power law: $W_{\Delta E} \sim 1/N_A^{\alpha}$.

\subsection{Two-qubits space metrics and the entangling power of a quantum gate}

So far we have considered the QG ``entangling power" as applied to the case 
of pure states of two-qubits. In order to do so, it has been sufficient to 
generate pure states according to the invariant Haar measure. In passing to 
mixed two-qubits states, the situation becomes more involved. 
Mixed states appear naturally when we consider a pure state that is decomposed 
into an statistical mixture of different possible states by environmental 
influence (a common occurrence). It may seem somewhat obvious to extend 
to mixed states the previous study of the entangling power of a certain 
quantum gate by following the steps given by formula (\ref{eP}). Instead, 
we will consider a heuristic approach to the problem.

The space of mixed states $\cal S$ of two-qubits is 15-dimensional, 
which implies that it clearly possesses non-trivial properties. 
In the usual generation of states $\rho$, we compute at the same time 
distances between states, which can be evaluated by certain measures 
\cite{ZyckJPA}. The ones that are considered here are the Bures distance

\begin{equation} \label{dBures}
d_{Bures}(\hat \rho_1,\hat \rho_2) \, = \, \bigg(2-2\,Tr 
\sqrt{(\sqrt{\hat \rho_2} \hat \rho_1\sqrt{\hat \rho_2)}}\bigg)^{\frac{1}{2}},
\end{equation}

\noindent and the Hilbert-Schmidt distance

\begin{equation} \label{dHS}
d_{HS}(\hat \rho_1,\hat \rho_2) \, = \, \sqrt{|Tr [\hat \rho_1-\hat \rho_2]^2|}.
\end{equation}

\noindent Remember that these distances were carefully studied in Chapter 10 in 
order to grasp the features of the structure of two-qubit systems. The goal here 
is to generate unentangled 
states $\rho$ (according to (\ref{memu})) of two-qubits and 
to compute by means of measures (\ref{dBures},\ref{dHS}) the 
average distance reached in $\cal S$ by a final state 
$\rho^{\prime}$, once the CNOT gate 
(\ref{CNOT}) is applied. In other words, we quantify the action of the CNOT gate
acting on the set ${\cal S}^{\prime}$ of completely separable states. The 
several distances between final (after CNOT) and initial states are computed, 
and a probability (density) distribution is then obtained.

\begin{figure}
\begin{center}
\includegraphics[angle=270,width=0.65\textwidth]{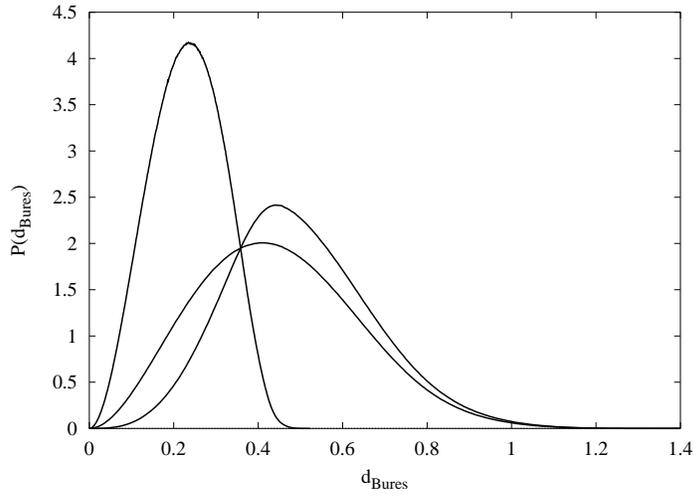}
\caption{Probability (density) distributions of finding a state of 
two-qubits (pure or mixed) being sent a distance $d_{Bures}$ away 
from the original state $\hat \rho$, after the action of the 
CNOT gate. All initial states belong to the set ${\cal S}^{\prime}$ 
of separable states. Two regions are defined. See text for details. 
All depicted quantities are dimensionless.}
\label{fig4U}
\end{center}
\end{figure}

\begin{figure}
\begin{center}
\includegraphics[angle=270,width=0.65\textwidth]{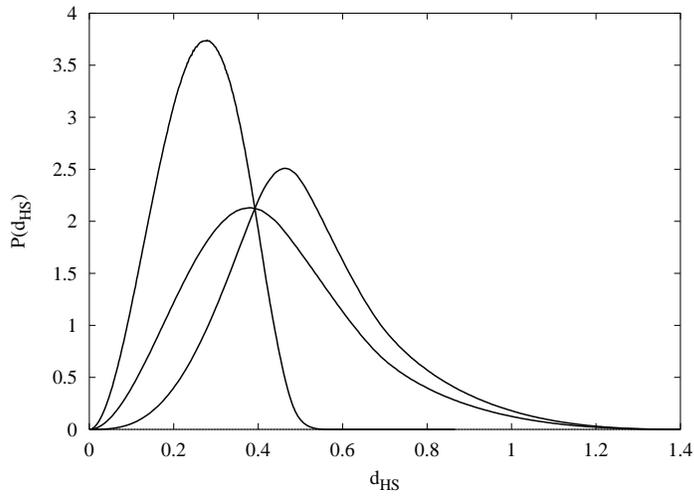}
\caption{Same as in Fig.\ref{fig4U}, using the Hilbert-Schmidt distance 
$d_{HS}$ between states. Both figures show similar qualitative features. 
See text for details. All depicted quantities are dimensionless.}
\label{fig5U}
\end{center}
\end{figure}

The probability distributions for the Bures and Hilbert-Schmidt distances 
are depicted in Fig.\ref{fig4U} and Fig.\ref{fig5U}, respectively. However, one has to 
bear in mind that these absolute distances between states do not take into account 
the fact that the set ${\cal S}^{\prime}$ may have (and indeed such is the case) 
a certain non-trivial geometry, which makes the shape of the convex set 
of separable states ${\cal S}^{\prime}$ highly anisotropic \cite{future work}. 
Therefore, in order to clarify the action of the CNOT gate, we separate the 
set ${\cal S}^{\prime}$ into two parts: I) ${\cal S}^{\prime}_{I}$, which is 
the set of unentangled states inside the minimal separable ball around 
$\frac{1}{4} \hat I$ of radius $d_{min}$, as measured with either 
(\ref{dBures}) or (\ref{dHS}), and II) ${\cal S}^{\prime}_{II}$, which is 
nothing but ${\cal S}^{\prime} - {\cal S}^{\prime}_{I}$. In point of fact, 
$d_{min}$ corresponds to the radius of a hypersphere in 15 dimensions 
whose interior points have Tr($\hat \rho^2$)$\le 1/3$ (\cite{BCPP02b}). 
As seen from Fig. \ref{fig4U} or Fig. \ref{fig5U}, the first case exhibits 
a well defined range. 
This is due to the fact that any unitary evolution (CNOT in our case) 
does not change Tr($\hat \rho^2$), so that the CNOT gate cannot 
produce entanglement at all or, in other words, cannot ``move" to any 
extent a state $\hat \rho$ out of ${\cal S}^{\prime}_{I}$. On the other hand, 
CNOT may entangle in ${\cal S}^{\prime}_{II}$ and displace the whole 
distribution to the right. Indeed, if we consider for both graphs 
the total set ${\cal S}^{\prime}$, the concomitant distributions look 
rather alike. The crossing point of the three curves in Fig.\ref{fig4U} and 
Fig.\ref{fig5U} corresponds to the border defined by $d_{min}^{Bures}$ 
and $d_{min}^{HS}$, respectively.

In view of these results, one may call a QG ``strong" if its entangling power, 
in acting on  a separable state, is great. Thus a semi-quantitative 
strength-measure could be the average value of the distance 
$\overline{d}^{{\cal S}^{\prime}}$ over the whole set of separable states. 
However, it should be pointed out that any definition of entangling power 
for mixed states would turn out to be metric-dependent, i.e., it 
depends on the set of eigenvalues $\Delta$ wherefrom $\hat \rho$ is generated.

\subsection{Entanglement distribution in multiple qubit systems}

So far we considered logical QGs acting on two-qubits systems. We pass now to 
multipartite ones (nothing strange: the environment can be regarded as a third 
party), composed of many subsystems \cite{optics2005}. We thus deal with a network of qubits, 
interacting with each other, and with a given configuration. More specifically, 
one could consider the set $\cal S$ of pure states 
$\hat \rho = |\Psi\rangle \langle \Psi|$ ``living" in a Hilbert space of 
$n$ parties (qubits) ${\cal H} = \otimes_{i=1}^{n}{\cal H}_i$.  

The usual three party, physically-motivated case, is the two-qubits system 
interacting with an environment which, as a first approximation, could be 
treated roughly as a qubit (two-level system). In any case, the issue of 
how the entanglement present in a given system is ${\it distributed}$ 
among its parties is interesting in its own right. Therefore, it should 
be of general interest to study the general case of multipartite 
networks of qubits on the one hand, while discussing, on the other one, 
how the dimensionality (the number of qubits) affects the distribution 
of the bipartite entanglement between pairs when we apply, locally, 
a certain quantum gate.

In what follows we consider the Coffman ${\it et \, al.}-$approach 
of \cite{CKW00} and consider firstly the case of three qubits in a pure 
state $\hat \rho_{ABC}$. An important inequality exists that refers to 
how the entanglement between qubits is pairwise distributed. 
The entanglement is measured by the concurrence squared $C^2$. 
Even though we handle pure states, once we have traced over 
the rest of qubits we end up with mixed states of two qubits, so that 
a measure for mixed states is needed. $C^2$ is related to the entanglement 
of formation \cite{WO98}. It ranges from 0 to 1. The concurrence is given 
by $C \, = \, max(0,\lambda_1-\lambda_2-\lambda_3-\lambda_4)$, 
$\lambda_i, \, (i=1, \ldots 4)$ being the square roots, in decreasing 
order, of the eigenvalues of the matrix $\rho \tilde \rho$, with 
$\tilde \rho \, = \, (\sigma_y \otimes \sigma_y) \rho^{*} 
(\sigma_y \otimes \sigma_y)$. The latter expression has to be 
evaluated by recourse to the matrix elements of $\rho$ computed 
with respect to the product basis. Considering the reduced 
density matrices $\hat \rho_{A}=Tr_{BC}\,(\hat \rho_{ABC})$, 
$\hat \rho_{AB}=Tr_{C}\,(\hat \rho_{ABC})$ and 
$\hat \rho_{AC}=Tr_{B}\,(\hat \rho_{ABC})$, the following 
elegant relation is derived:

\begin{equation} \label{d}
C^{2}_{AB} \,+\, C^{2}_{AC} \le 4 \, det \hat \rho_{A} \, (\equiv C^{2}_{A(BC)})
,
\end{equation}

\noindent where $C^{2}_{A(BC)}$ shall be regarded as the entanglement of qubit 
$A$ with the rest of the system. In fact, we are more concerned in quantifying 
$d_W \, \equiv \, C^{2}_{A(BC)} \, - \, C^{2}_{AB} \,-\, C^{2}_{AC}$. From 
inspection, $d_W$ ranges from 0 to 1 and can be regarded as a legitimate 
multipartite entanglement measure, endowed with certain 
properties \cite{CKW00}. 

\begin{figure}
\begin{center}
\includegraphics[angle=270,width=0.65\textwidth]{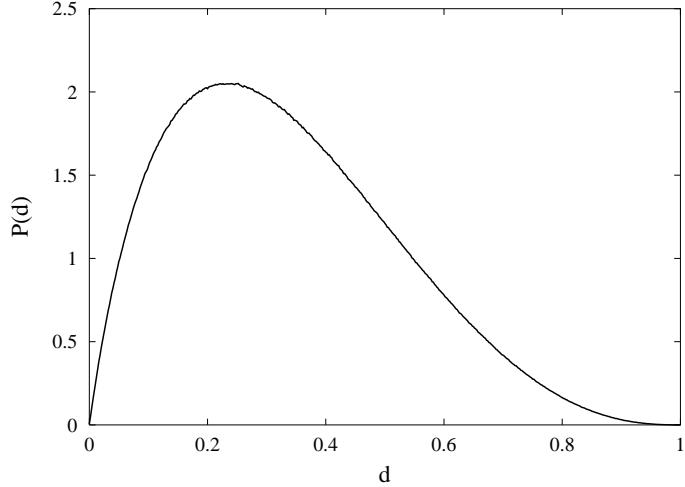}
\caption{Probability (density) distribution of finding a pure state 
of three-qubits with a given value of $d_W$ (\ref{d2r2}), a measure of 
the distribution of the pairwise entanglement in the system. The 
curve is biased to low values of $d_W$, 
and $\overline{d_W} \simeq 1/3$. All depicted quantities are dimensionless.}
\label{fig6U}
\end{center}
\end{figure} 

\begin{figure}					     
\begin{center}
\includegraphics[angle=270,width=0.75\textwidth]{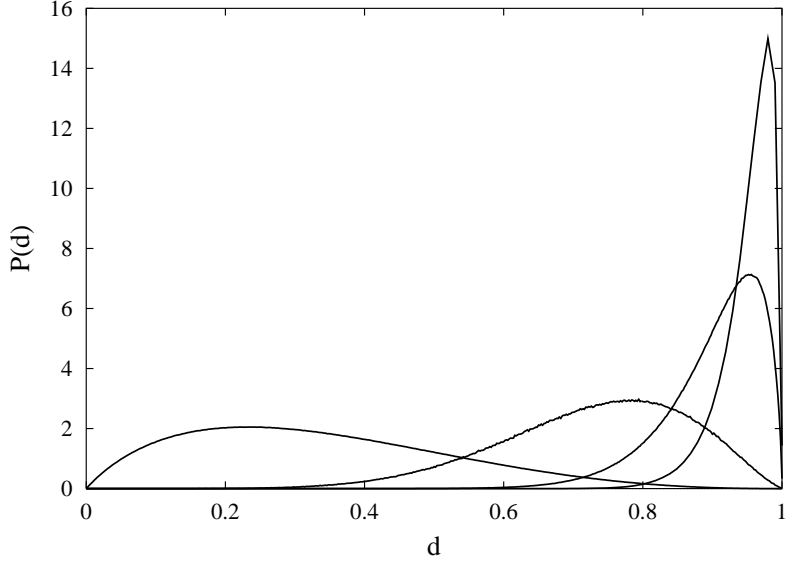}
\caption{Probability (density) distribution of finding a pure state 
of $n=3,4,5$ and 6 qubits (from left to right) with a given value 
of $d_W$ (\ref{d2r2}). As we increase 
the number of parties, measure (\ref{d2r2}) dramatically localizes entanglement 
on each of the qubits.}
\label{fig6bisU}
\end{center}
\end{figure}

In Fig.\ref{fig6U} the probability (density) function $P(d_W)$ is obtained by generating 
a sample of pure states of three qubits according to the invariant Haar measure, 
as we did for $n=2$. It is interesting to notice the bias 
of the distribution, and the remarkable fact that numerical evaluation 
indicates that $\overline{d_W} \simeq 1/3$. Also, we can increase the number of 
qubits forming the network. It is an exponentially time-consuming procedure to increase the 
total number of qubits, in consequence we limit ourselves to the additional cases $n=4,5$ 
and $6$. The concomitant probability (density) distributions $P(d_W)$ are depicted   
if Fig.\ref{fig6bisU}. First of all, our numerical calculations support the 	  
conjecture made in \cite{CKW00} that 

\be \label{d2r2}
0 \, \le \, d_W\equiv C^{2}_{1(2..n)}-\sum_{i=2}^{n} C^{2}_{1i} \, \le \, 1
\ee

\noindent holds for an arbitrary number $n$ of qubits in a pure state 
$\rho=|\Psi\rangle_{1..n}\langle \Psi|$. $C^{2}_{xy}$ stands for the 
concurrence squared between qubits $x,y$ and $C^{2}_{1(2..n)}=4\,$det$\rho_1$, 
with $\rho_1$=Tr$_{2..n}$($\rho$). It is apparent from Fig.\ref{fig6bisU} that the 
entanglement present in the system tends to become more and more concentrated 
on each qubit {\it individually}, then the ``residual" entanglement 
$\sum_{i=2}^{n} C^{2}_{1i}$ tends to zero.


\begin{figure}
\includegraphics[angle=270,width=0.5\textwidth]{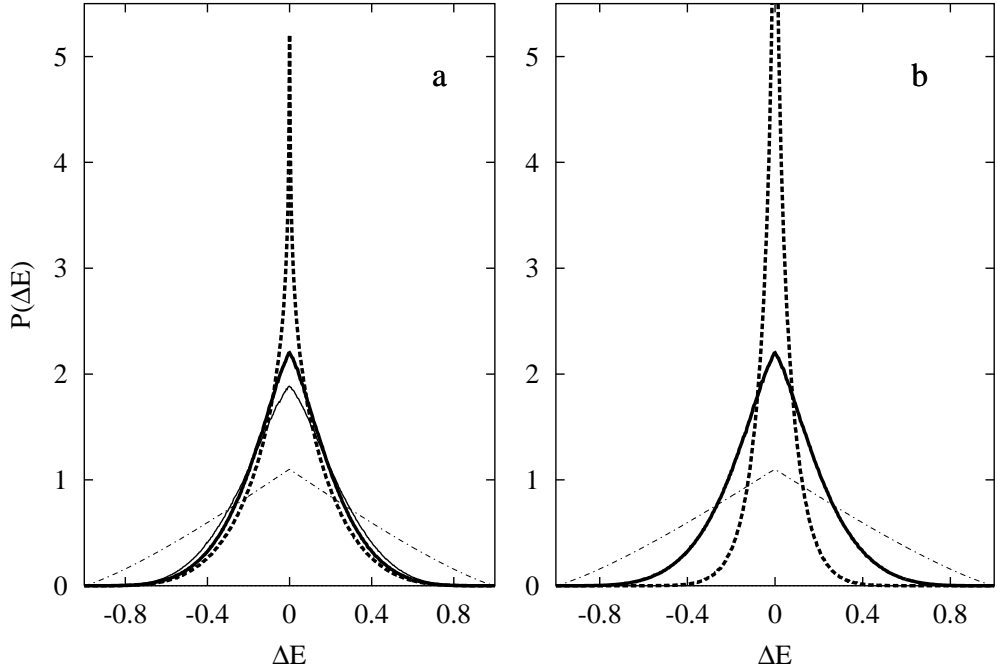}
\caption{a) $P(\Delta E)$-distributions generated by the CNOT 
quantum gate $\hat U^{CNOT}_{AB} \otimes \hat I_{C}$, acting on the pair 
$AB$ of a pure state of three-qubits. The resulting distribution 
(solid thick line) is to be compared with the one of the pairs $AC,BC$, 
equal on average (dashed line), the random case where no gate is applied 
(solid thin line) and the case of solely two-qubits CNOT gate 
$P(\Delta E)$ distribution (thin dot-dashed line). As compared to 
the three-qubit random instance, it possesses a width slightly 
inferior, being much narrower than in the two-qubits case. 
This fact indicates that the entanglement available to the pair 
$AB$ is diluted by the presence of a third party. 
b) These distributions result from the action of the CNOT gate 
$\hat U^{CNOT}_{AB}$ on two-qubits ($n=2$, dot-dashed line), 
$\hat U^{CNOT}_{AB} \otimes \hat I_{C}$ on three qubits ($n=3$, solid line), 
and $\hat U^{CNOT}_{AB} \otimes \hat I_{C} \otimes \hat I_{D}$ 
on four qubits ($n=4$, dashed line) pure states. The width of these distributions, 
or entangling power $W_{\Delta E}$ (see text), decreases exponentially 
as the number of qubits is increased. 
All depicted quantities are dimensionless.}
\label{fig7U}
\end{figure}

Now, suppose that we apply the CNOT gate to the pair of qubits $AB$. 
This means that the unitarity acting on the state $\hat \rho_{ABC}$ 
is described by $\hat U^{CNOT}_{AB} \otimes \hat I_{C}$, where 
$\hat I_X$ is the identity acting on qubit $X$. Making then 
a numerical survey of the action of this operator on the evolution 
of the system we show the concomitant, pairwise entanglement-change 
$\Delta E$ as the probability distributions plotted in Fig.\ref{fig7U}a 
(as measured by the entanglement of formation $E$). Two types of 
entanglement are present in the system, namely, the one between 
the pair $AB$, where the gate is applied, and the remaining possibilities 
$AC$ and $BC$, symmetric on average. The solid thick line depicts 
the first kind $AB$, while the second type $AC,BC$ exhibits a 
sharper distribution (dashed line). One is to compare this 
distributions to the one obtained by picking up two states at 
random (solid thin line), which resembles the case of Fig.\ref{fig3U}. 
Again, the random case exhibits a larger width for the 
distribution. When compared to the two-qubits CNOT case 
(thin dot-dashed line), we may think of the existence of a third 
party as a rough ``thermal bath" that somehow dilutes the 
entanglement available to the pair $AB$, as prescribed by 
the relation (\ref{d}). This is why the CNOT distribution 
for $n=3$ seems ``sharper" than that for $n=2$. As a matter of fact, 
if we continue increasing the number of qubits present in the system, 
we can numerically check that the generalization of (\ref{d}) 
still holds. In such a (new) instance, the action of the 
CNOT gate is equivalent to the evolution governed by 
$\hat U^{CNOT}_{AB} \otimes \hat I_{C} \otimes \hat I_{D}$. As it is shown in 
Fig.\ref{fig7U}b, the 
new distribution of the entanglement changes for $n=4$ in the 
$AB$ pair (dashed line, out of scale) and, as expected, is more peaked than 
for the $n=2$ (dot-dashed line) and, $n=3$ (solid line) cases, reinforcing 
our thermodynamical analogy \cite{PhysicaApending}. If we compute 
their entangling power (EP) with $W_{\Delta E}$, our new measure 
defined previously, we could conjecture that the EP decreases 
exponentially with the number of qubits $n$ 
($W_{\Delta E}^{n=2} \simeq 0.437, W_{\Delta E}^{n=3} \simeq 0.196, 
W_{\Delta E}^{n=4} \simeq 0.002$).

\section{Concluding remarks}


In the present work we have focused attention upon the action of quantum gates 
as applied to 
multipartite quantum systems and presented the results of a systematic 
numerical survey. 

Firstly, we proved that 
the $\Delta E$-distributions generated by quantum 
gates that can be obtained from each other by recourse 
to appropriate, local unitary transformation are the 
same, even if these gates (in general) produce different 
changes of entanglement on individual states. We also 
studied numerically some features of the probabilities of 
obtaining different values of $\Delta E$ \cite{PhysicaApending}. 

Secondly, we explored the entanglement changes associated with the
action of the $T_H$-CNOT circuit (two-qubit systems). We found that 
the probability
distribution of entanglement changes obtained when the circuit
acts on pure states is quite different from the distribution
obtained when the circuit acts on general mixed states. The
probability of entangling mixed states turns out to be rather
small. On average, the $T_H$-CNOT transformation is more
efficient, as entangler, when acting upon states with small
initial entanglement, specially in the case of pure states.

In addition, we investigated  
aspects of the quantum gate or unitary operation (acting on two-qubits 
states) as conveniently represented by a vector ${\bf \lambda} \equiv 
(\lambda_1,\lambda_2,\lambda_3)$, visualizing the ``entangling power" of unitary 
quantum evolution from two different perspectives.

\begin{itemize}
\item{The first one refers to pure 
states of a bipartite system. One has here a well defined formula that 
quantifies the ability of a given transformation $\hat U$ to entangle, 
on average, a given state that pertains to the set ${\cal S}^{\prime}$ 
of unentangled pure states. We have seen 
that the collective of all possible quantum gates, 
as defined by the vector ${\bf \lambda}$, possesses the following property: in 
the vicinity of an optimal gate there are infinite  quantum gates which are 
optimal as well. In addition, we introduced a measure of the entangling power 
above referered to: $W_{\Delta E}$, on the basis of the 
probability (density) distribution (associated with a quantum gate)  
of finding a state that experiences a given change $\Delta E$ in 
its entanglement $E$. A power-law decay is conjectured: 
$W_{\Delta E} \sim 1/N_A^{\alpha}$, $N_A$ being the dimension of the 
subsystem ($N = N_A \times N_A$).} 

\item{The second instance deals with mixed states and the metrics of the 
15-dimensional space $\cal S$ of mixed states of two-qubits. We introduce an 
heuristic measure based on an  average distance $\overline d$ obtained from  
the  distribution of distances between states in $\cal S$, as defined by the 
action of a definite quantum gate acting (again) on the set of unentangled 
states ${\cal S}^{\prime}$.} 
\end{itemize}

Finally, we have studied i) some basic properties of the 
distribution of entanglement in multipartite systems (MS) (network of qubits) 
and ii) the effects 
produced by two-qubits gates acting upon MS. The fact that the entanglement 
between pairs becomes diluted by the presence of third or fourth parties becomes 
apparent from the concomitant distribution of entanglement changes. 
Their natural width $W_{\Delta E}$ decreases with the number of parties $n$,  
in what seems to be an exponential fashion.


\chapter{Temporal evolution of states assisted by quantum entanglement}

Due to its essential connection both with 
 our basic understanding of quantum mechanics, 
 and with some of its most revolutionary (possible)
 technological applications, it is 
 imperative to investigate in detail the relationships 
 between entanglement and other aspects of quantum 
 theory. In particular, it is of clear interest to 
 explore the role played by entanglement in the 
 dynamical evolution of composite quantum systems.
 It was recently discovered by Giovannetti, Lloyd, and
 Maccone \cite{GLM03a,GLM03b} that, in certain cases,
 entanglement helps to ``speed up" the time evolution
 of composite systems, as measured by
the time a given initial state requires to evolve to an orthogonal
state. The problem of the ``speed" 
 of quantum evolution has aroused considerable
 interest recently, because of its relevance
 in connection with the physical limits imposed
 by the basic laws of quantum mechanics on the
 speed of information processing and information
 transmission \cite{ML98,CD94,L00}.

We provide here a systematic study of this effect for pure
states of bipartite systems of low dimensionality, considering both
distinguishable (two-qubits) subsystems, and systems constituted of
two indistinguishable particles. Therefore the aim of the present Chapter 
is to investigate in detail, for bipartite systems of low 
 dimensionality, the connection between entanglement and 
 the speed of quantum evolution. We are going to focus
 our attention on (i) two qubits (distinguishable) systems 
 and (ii) bosonic or  fermionic composite (bipartite) 
 systems of lowest dimensionality. The importance of the 
statistics of the particles will become apparent \cite{BatleTemp}.

\section{Two entangled distinguishable particles}

 We are going to investigate first the case of two equal but 
 distinguishable subsystems evolving under a local Hamiltonian.
 Let us then consider a two qubits system whose evolution is governed
 by a (local) Hamiltonian 
 
 \be \label{haloloco}
 H \, = \, H_A \otimes I_B + I_A \otimes H_B,
 \ee

 \noindent
 where $H_{A,B}$  have eigenstates $|0\rangle$ and $|1\rangle$ with 
 eigenvalues $0$ and $\epsilon$, respectively. That is, the eigenstates of
 $H$ are $|00\rangle$, $|01\rangle$, $|10\rangle$, and $|11\rangle$,
 with eigenvalues respectively equal to $0$, $\epsilon$ (twofold degenerate) 
 and $2\epsilon$. For pure states $|\Psi \rangle  $ of our composite system
 the natural measure of entanglement is the usual reduced von Neumann 
 entropy $S[\rho_{A,B}] = -Tr_{A,B} (\rho_{A,B} \ln \rho_{A,B}) $ 
 (of either particle $A$ or particle $B$) where $\rho_{A,B} = Tr_{B,A} 
 (|\Psi \rangle \langle \Psi |)$. It is convenient for our present purposes
 to use, instead of $S(\rho_{A,B})$ itself, the closely related 
 ${\it concurrence \,\, C}$, given by
 
 \be \label{concurre1}
 C^2=4 \det \rho_{A,B}. 
 \ee

\noindent
Both the entanglement entropy $S[\rho_{A,B}]$ 
and the concurrence $C$ are preserved under 
the time evolution determined by the local 
Hamiltonian (\ref{haloloco}). Given an 
initial state

\begin{equation} \label{instate}
|\Psi(t=0)\rangle = c_0 |00\rangle+c_1 |01\rangle+c_2 |10\rangle+c_3 |11\rangle,
\end{equation}

\noindent
its concurrence is,

\be \label{concurre2}
C^2 \, = \, 4|c_0c_3-c_1c_2|^2.
\ee

\noindent
The overlap between the initial state (\ref{instate}) 
and the state at time $t$ is given by

\begin{equation} \label{overlape}
\langle\Psi(t)|\Psi(t=0)\rangle = |c_0|^2+(|c_1|^2+|c_2|^2)z+|c_3|^2z^2,
\end{equation}

\noindent 
where
$z \equiv {\rm exp}(i \epsilon t/\hbar) \equiv {\rm exp}(i\alpha)$, that is, 
$\alpha=\frac{t\epsilon}{\hbar}$.

\noindent
Thus, the condition for the state at time $t$ to be orthogonal to the
initial state is,

\be
P(z) \, = \, |c_0|^2+(|c_1|^2+|c_2|^2)z+|c_3|^2z^2 = 0.
\ee

\noindent
The above polynomial equation  can be cast as,

\be
|c_3|^2 (z-z_1) (z-z_2) = 0,
\ee

\noindent
where $z_1$ and $z_2$ are the roots of $P(z)$. If the
initial state $(\ref{instate})$ is to evolve to an 
orthogonal state, then the two roots of $P(z)$
have to be two (complex conjugate) numbers of modulus
equal to one. That is $z_{1,2} = \exp(\pm i \alpha)$. 
In that case we shall have,

\begin{eqnarray}
|c_0|^2=|c_3|^2&=&\Gamma, \cr
|c_1|^2+|c_2|^2&=&-2\Gamma {\rm cos}\alpha. 
\end{eqnarray}

\noindent
Appropriate normalization of the initial
state also implies that the concomitant 
coefficients can be parameterized as,

\begin{eqnarray}
|c_0|^2&=&|c_3|^2=\Gamma, \cr
|c_1|^2&=&-2\delta\Gamma {\rm cos}\alpha \cr
|c_2|^2&=&-2(1-\delta)\Gamma {\rm cos}\alpha, 
\end{eqnarray}

\noindent with $\Gamma=\frac{1}{2(1-{\rm cos}\alpha)}$ and 
$\alpha \in [\frac{\pi}{2},\frac{3\pi}{2}]$, $\delta \in [0,1]$. 
In other words, we have $\alpha={\rm arccos}(\frac{2\Gamma-1}{2\Gamma})$.

\noindent
The initial state's energy mean value and energy 
uncertainty are, respectively,  

\begin{eqnarray} \label{medenerg}
E \, &=& \, \langle H \rangle \, = \, 
\epsilon(|c_1|^2+|c_2|^2)+2\epsilon |c_3|^2=\epsilon \cr
\Delta E \, &=& \, \sqrt{\langle H^2 \rangle - \langle H \rangle^2} 
\, = \, \big(\epsilon^2(|c_1|^2+|c_2|^2)+4\epsilon^2 
|c_3|^2  - \epsilon^2
\big)^{\frac{1}{2}}\cr
&=&\epsilon \sqrt{2\Gamma}. 
\end{eqnarray}

\noindent 
The time $\tau$ required to evolve into an orthogonal state admits
the lower bound \cite{GLM03a,GLM03b},

\begin{equation} \label{Tmin} 
T_{min}=max\bigg(\frac{\pi\hbar}{2E},\frac{\pi\hbar}{2\Delta E}\bigg),
\end{equation}

\noindent
which, together with equations (\ref{medenerg}), lead to

\begin{equation} \label{Tminina} 
T_{min}=
\frac{\pi\hbar}{2\epsilon\sqrt{2\Gamma}}.
\end{equation}

\noindent
The concurrence of the (pure) state under 
consideration, defined as $C^2= 4|c_0c_3-c_1c_2|^2$ 
(see equations (\ref{concurre1}-\ref{concurre2})) is 

\be \label{concurre3}
C^2=4 \Big|\Gamma-e^{i\phi}
\sqrt{\delta(1-\delta)}
2\Gamma{\rm cos}\alpha \Big|^2.
\ee

\noindent
The modulus of the coefficients $c_i$ are completely determined
by the two parameters $\alpha$ (or $\Gamma$) and $\delta$. 
The dependence of $C^2$ on the phases of the coefficients 
$c_i$ can be absorbed into one single phase $e^{i\phi}$, 
thus incorporating a new parameter $\phi$ into the expression 
(\ref{concurre3}) for $C^2$. 

\begin{figure}
\begin{center}
\includegraphics[angle=270,width=.65\textwidth]{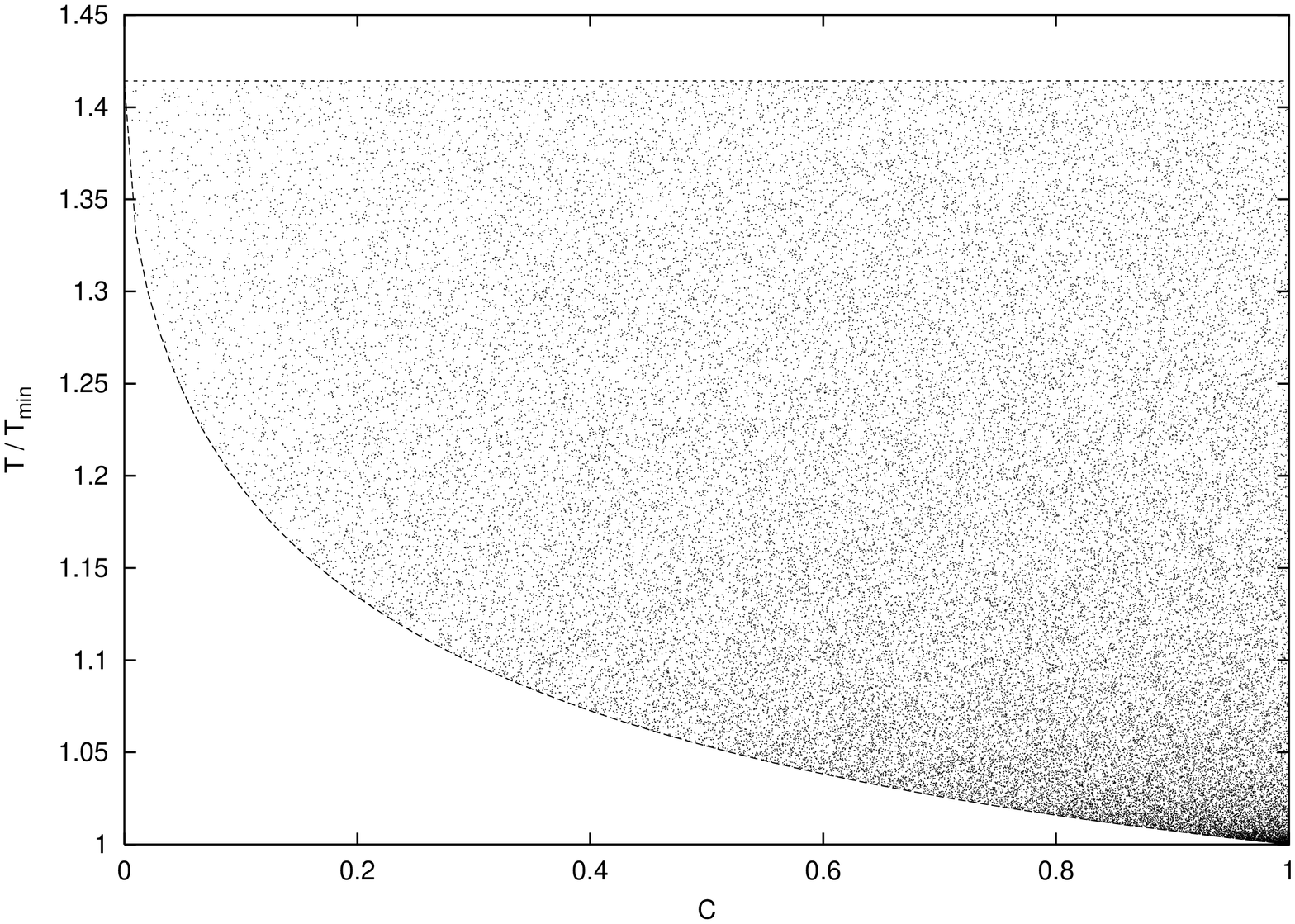}
\caption{Curves in the $(C,\tau/T_{min})$-plane corresponding, 
for each value of $C$, to the states of
two (distinguishable) qubits with maximum and 
minimun $\tau/T_{min}$. The points represent 
randomly generated individual states that evolve 
to an orthogonal state. All depicted quantities
are dimensionless.} 
\label{time1}
\end{center}
\end{figure}

After some algebra, the expressions for the minimum and 
maximum values for the evolution time $\tau$ that 
are actually realized for states of a given concurrence 
$C^2$, read

\begin{equation} \label{tautau}
\frac{\tau}{T_{min}(\Gamma)}=\frac{2}{\pi}\sqrt{2\Gamma}\,{\rm arccos}
\left(\frac{2\Gamma-1}
{2\Gamma} \right),
\end{equation}
  
\noindent
where the maximum evolution time for a fixed $C^2$ 
(or a fixed $C$) corresponds to $\Gamma=1/4$ (constant value), 
while the minimum one to $\Gamma=(1+\sqrt{C^2})/4$. 
The two curves in the $(C,\tau/T_{min})$-plane 
corresponding, for each value of $C$, to the states
with maximum and minimum $\tau/T_{min}$ are depicted
in Fig.\ref{time1}. All states that eventually evolve into 
an orthogonal state (that is, states characterized 
by different $\delta$'s and $\phi$'s) lie between 
these two curves. Some important features of the 
connection between entanglement and speed of 
evolution (for two qubits) transpire from 
Fig.\ref{time1}. First, we see that 
the minimum time required to reach an orthogonal 
state is a monotonously
decreasing function of the concurrence. 
Second, the lower bound for the evolution
time to an orthogonal state is saturated 
by (and only by) the maximally entangled
states ($C=1$). These features provide 
further support to the idea that 
entanglement tends to ``speed up" 
quantum evolution.
\newline

\begin{figure}
\begin{center}
\includegraphics[angle=0,width=.65\textwidth]{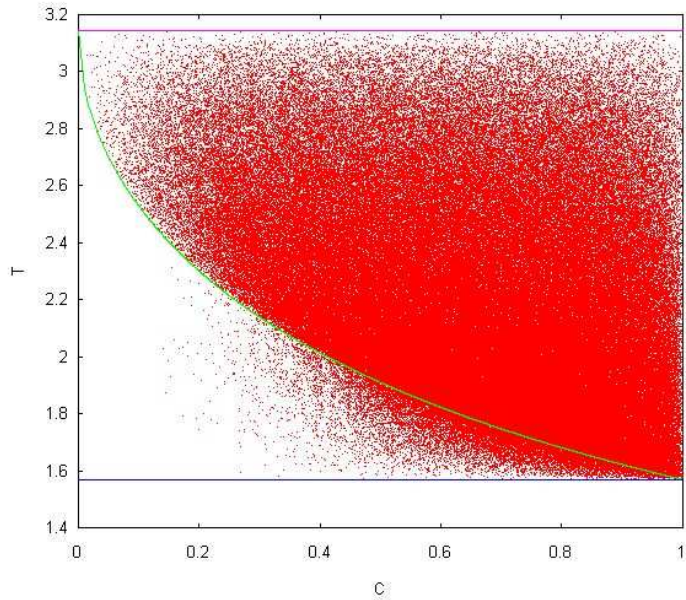}
\caption{Sample of one million points representing pure states 
generated according to the usual Haar measure. The time $\tau$ (in units 
of $\hbar/\epsilon$) needed to 
reach its minimum overlap vs. its entanglement is shown. We also 
plot the $C-$ min $\tau$ curve for comparison. It is plain from this 
figure that most of states with minimum overlap tend to stay 
above the $C-$ min $\tau$ curve. See text for details.} 
\label{timex}
\end{center}
\end{figure}

\begin{figure}
\begin{center}
\includegraphics[angle=270,width=.65\textwidth]{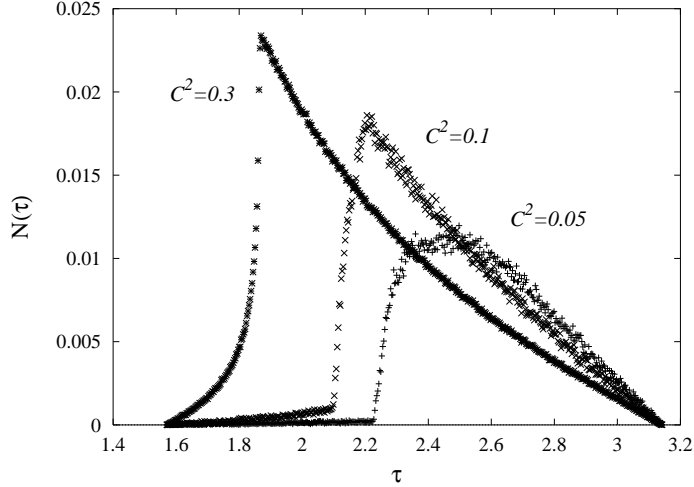}
\caption{Distributions $N(\tau)$ (not normalized) of the number of states in 
Fig.\ref{timex}, as a function of time $\tau$ (in units 
of $\hbar/\epsilon$), for different fixed amounts of entanglement. With further increase 
of entanglement, these distributions become 
more neatly peaked. See text for details.} 
\label{timexx}
\end{center}
\end{figure}

At this point, one may wonder if entanglement only ``speeds up" the evolution of 
a state towards an orthogonal one. So far the discussion has involved the zeros 
of Eq. (\ref{overlape}), but this is not always possible. Let us consider 
the general case where our state is a pure state, distributed 
according to the usual rotationally invariant Haar measure. During the time 
evolution of this state, the concomitant 
overlap (\ref{overlape}) evolves with time. A way to visualize if there exists 
any correlation between entanglement and time evolution in those cases where an 
orthogonal state cannot be reached, consists of computing the time needed for 
any initial separable state to reach its minimum overlap (\ref{overlape}) --under 
the action of a certain Hamiltonian-- as well as 
the entanglement of this final state. In the case of the simplest Hamiltonian 
(\ref{haloloco}), we compute a sample of one million states and follow the 
aforementioned procedure. In Fig.\ref{timex} we plot the time needed to reach a minimum 
overlap vs. the concurrence of the final state. The curve that appears is nothing but 
the minimum time evolution (towards an orthogonal state) compatible with a 
given concurrence, which arises from Eq. (\ref{tautau}). It is apparent from 
this Figure that all states with minimum overlap are likely to cover the upper 
part of the curve. In point of fact, en exhaustive exploration of all pure states 
shows that only 1.7 per cent of the space of pure, arbitrary states of two-qubit 
systems tend to evolve faster than the depicted $C-min\tau$ curve. Furthermore, 
if we fix the amount of entanglement $C$, the outgoing distributions $N(\tau)$ 
of time evolutions $\tau$ (Fig.\ref{timexx}) tend to peak around the concomitant time 
corresponding to that of an orthogonal evolution. As we increase the amount of 
entanglement, these distributions become more neatly peaked. This fact 
demonstrates that entanglement also influences those evolutions towards a minimum 
overlap (\ref{overlape}), and confirms that this tendency increases with the 
entanglement of the final state. 

\section{Two entangled indistinguishable particles}
 
 Here we are going to explore the connection between 
 entanglement and the speed of quantum evolution for 
 systems constituted by two indistinguishable particles.
 In this case the concept of entanglement exhibits some
 extra subtleties, as compared with the case of 
 distinguishable subsystems. When dealing with 
 indistinguishable particles, the correlations
 that arise purely from the concomitant
 statistics (either fermionic or bosonic) 
 do not constitute a useful resource and,
 consequently, must not be regarded as 
 contributing to the amount of entanglement 
 of the system's state \cite{identical}.
 A useful formalism to describe the entanglement of
 systems consisting of identical particles, that takes
 into account the above remarks, has been 
 advanced by  Eckert {\it et al.} in \cite{identical}. For 
 two identical bosons, the system of lowest 
 dimensionality exhibiting the phenomenon of entanglement 
 is a pair of bosons with a two dimensional single 
 particle Hilbert space. The simplest fermionic 
 system endowed with entanglement is a system 
 of two fermions with a three dimensional
 single particle Hilbert space.

\subsection{Bosons}

Using the second quantization formalism, 
the general (pure) state of two bosons 
(with a two-dimensional single particle 
Hilbert space) can be written under the 
guise \cite{identical}

\begin{equation} \label{sboson}
|V\rangle = \sum_{i,j=0}^{1} v_{ij} b_i^{\dagger}b_j^{\dagger} |0\rangle,
\end{equation}

\noindent  
where $b_i^{\dagger}$ and $b_i$ denote
bosonic creation and annihilation operators, 
the coefficients $v_{ij}$ constitute 
the symmetric matrix

\begin{equation}
\hat V= \left(
\begin{array}{cccc}
v_{00} & v_{01} \\
v_{10} & v_{11} \end{array} \right).
\end{equation}

\noindent That is, $v_{ij}=v_{ji}$. Normalization 
imposes the condition 
$2\sum_{i,j=0}^{1} |v_{ij}|^2=1$. 

The Hamiltonian associated with two 
non-interacting bosons is,

\begin{equation} \label{hamil}
\hat H = \sum_{k=0}^{1} \epsilon_k b_k b_k^{\dagger},
\end{equation}

\noindent where $b_0^{\dagger}|0\rangle$ is the single particle 
ground state with energy $\epsilon_0 = 0$, and  
$b_1^{\dagger}|0\rangle$ is the single particle excited 
state with energy $\epsilon_1 = \epsilon$. The state 
(\ref{sboson}) evolves according to the time-dependent 
Schr\"odinger equation, 

\begin{equation}
i\hbar \frac{d}{dt} |V(t)\rangle=\hat H |V(t)\rangle = 
\sum_{i,j=0}^{1} (\epsilon_i+\epsilon_j)
v_{ij}(t) b_i^{\dagger}b_j^{\dagger} 
|0\rangle,
\end{equation}

\noindent The general solution of this evolution equation
is given by the time dependent coefficients,

\begin{equation}
v_{ij}(t) = v_{ij}(0)\,e^{-i\frac{(\epsilon_i+\epsilon_j)}{\hbar}t}.
\end{equation}

\noindent The time $\tau $ required to evolve into an 
ortonormal state is

\begin{equation} \label{orthoboson}
\langle V(0)|V(\tau)\rangle=2 \sum_{i,j=0}^{1} |v_{ij}(0)|^2\,
e^{-i\frac{(\epsilon_i+\epsilon_j)}{\hbar}\tau}=0.
\end{equation}

\noindent Setting $z \equiv e^{-i\frac{\epsilon \tau}{\hbar}}
=e^{-i\alpha}$, the orthogonality condition 
(\ref{orthoboson}) can be recast as a 
polynomial equation in $z$, that has 
to admit roots of modulus equal to $1$.
From this last requirement, and taking into 
account the symmetries in the coefficients 
$v_{ij}$, it follows that the coefficients
can be parameterized as,

\begin{eqnarray}
|v_{00}|^2&=&\Gamma \cr
|v_{01}|^2&=&-\Gamma {\rm cos}\alpha \cr
|v_{11}|^2&=&\Gamma,
\end{eqnarray}

\noindent 
with $\Gamma > 0$ and $\alpha \in [\pi/2, 3\pi/2] $.
The normalization constraints also implies that 
$\Gamma=\frac{1}{4(1-{\rm cos}\alpha)}$. The 
expectation values of the energy and its square
read

\begin{eqnarray}
E&=&\langle H \rangle = 
2\sum_{i,j=0}^{1} |v_{ij}(0)|^2\,
(\epsilon_i+\epsilon_j) = \epsilon \cr
\langle H^2 \rangle&=& 
2\sum_{i,j=0}^{1} |v_{ij}(0)|^2\,
(\epsilon_i+\epsilon_j)^2=
(4 \Gamma + 1)\epsilon^2,
\end{eqnarray}

\noindent and consequently the minimum 
evolution time (\ref{Tmin}) is 

\be
T_{min} = \frac{\pi \hbar}{4 \epsilon \sqrt{\Gamma} }.
\ee

\noindent
The formula for the concurrence in 
the two-boson case is \cite{identical}

\be
C_{B}=4|v_{00}v_{11}-v_{01}^2|, 
\ee

\noindent
which is clearly time-independent. 

\begin{figure}
\begin{center}
\includegraphics[angle=270,width=.65\textwidth]{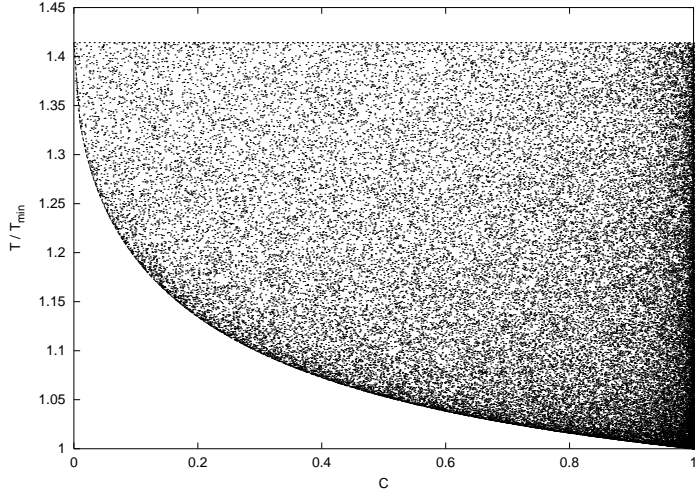}
\caption{Curves in the $(C,\tau/T_{min})$-plane 
corresponding, for each value of $C$, to the states 
of two bosons with maximum and minimum $\tau/T_{min}$.
The points represent randomly generated individual 
states that evolve to an orthogonal state.
All depicted quantities
are dimensionless.} 
\label{time2}
\end{center}
\end{figure}

For a given value of the concurrence, the minimum 
and maximum times for evolution to an orthogonal 
state can be obtained in the same way as in the
case of two distinguishable qubits. 
%
%
%
%
The concomitant curves 
are exhibited in Fig.\ref{time2}.

Comparing Fig.\ref{time2} with Fig.\ref{time1} we see 
that the same general trends exhibited by a
system of two distinguishable qubits are also 
observed in the case of two identical boson.

\subsection{Fermions}

Now we are going to study a system of two identical 
fermions with a three dimensional single particle
Hilbert space. In second quantization notation,
the general (pure) state of such a system is,

\begin{equation} \label{sfermion}
|W\rangle = \sum_{i,j=0}^{3} w_{ij} f_i^{\dagger}f_j^{\dagger} |0\rangle,
\end{equation}

\noindent where $f_i^{\dagger}$ and $f_i$ denote
fermionic creation and annihilation operators, 
respectively, and the coefficients $w_{ij}$ 
constitute the anti-symmetric matrix

\begin{equation}
\hat W= \left(
\begin{array}{cccc}
0 & w_{01} & w_{02} & w_{03} \\
w_{10} & 0 & w_{12} & w_{13} \\
w_{20} & w_{21} & 0 & w_{23} \\
w_{30} & w_{31} & w_{32} & 0 \end{array} \right).
\end{equation}

\noindent That is, $w_{ij}=-w_{ji}$. 
Normalization imposes the condition 
$\sum_{i,j=0}^{3} |w_{ij}|^2=1/2$. 
The Hamiltonian describing two non-interacting 
particles is given by,

\begin{equation} \label{hamilhamil}
\hat H = \sum_{k=0}^{1} \epsilon_k f_k f_k^{\dagger},
\end{equation}

\noindent
and the coefficients

\begin{equation}
w_{ij}(t) = w_{ij}(0)\,
e^{-i\frac{(\epsilon_i+\epsilon_j)}{\hbar}t},
\end{equation}

\noindent 
describe a general solution of the concomitant time depending
Schr\"odinger equation. Let $z \equiv 
e^{-i\frac{\epsilon \tau}{\hbar}}=
e^{-i\alpha}$. The evolution time to an orthogonal state 
follows from the condition 

\begin{eqnarray} \label{ferpolizet}
\langle W(0)|W(\tau)\rangle &=& 2 \sum_{i,j=0}^{3} |w_{ij}(0)|^2\,
e^{-i\frac{(\epsilon_i+\epsilon_j)}{\hbar} \tau} \cr
&=& 4 z \left( |w_{01}|^2+|w_{02}|^2\,z+M z^2+
|w_{13}|^2\,z^3+|w_{23}|^2\,z^4\right)\, \cr
&=& \, 0,  
\end{eqnarray}

\noindent with $M=|w_{03}|^2+|w_{12}|^2$. The polynomial equation (\ref{ferpolizet}) 
may have either (i) fourth real roots, (ii) two real roots 
and two complex (complex conjugated) roots, or 
(iii) two pairs of complex conjugated roots. 
Since we are interested in solutions of the type 
$e^{-i\frac{\epsilon}{\hbar}\tau}$, the most general 
case of interest is (iii). Consequently, the 
two solutions of (\ref{ferpolizet}) corresponding 
to (positive) times of evolution into an orthogonal 
state are of the form $z_1 \equiv e^{-i\alpha}$ and 
$z_2 \equiv e^{-i\beta}$. Taking into account the 
antisymmetric nature of $w_{ij}$, we get 
the following relations

\begin{eqnarray}
|w_{01}|^2 &=& x \\
|w_{02}|^2 &=& -2\,x({\rm{cos}}\alpha + {\rm{cos}}\beta)\\
|w_{03}|^2+|w_{12}|^2 &=& 2\,x(1+2{\rm{cos}}\alpha\,{\rm{cos}}\beta)\\
|w_{13}|^2 &=& -2\,x({\rm{cos}}\alpha + {\rm{cos}}\beta)\\
|w_{23}|^2 &=& x,
\end{eqnarray}

\noindent where the value of the parameter $x$ 
is determined by the normalization requirement. 
We want to find the fastest solution to the 
first orthogonal state. The time $\tau $ 
required to reach an orthogonal state is 

\be
\tau= \frac{\hbar}{\epsilon}  
\times min(\alpha,\beta).
\ee

Let us consider the case $\beta=\pi$. Then, the
time required to arrive to an orthogonal state
is equal to $\tau = \hbar \alpha/\epsilon$
and the coefficients characterizing the quantum
state are,

\begin{eqnarray}
|w_{01}|^2 &=& \frac{1}{32\,(1-{\rm{cos}}\alpha)} \\
|w_{02}|^2 &=& \frac{1}{16} \\
|w_{03}|^2+|w_{12}|^2 &=& \frac{1-2{\rm{cos}}\alpha}{16\,(1-{\rm{cos}}\alpha)} 
\\
|w_{13}|^2 &=& \frac{1}{16} \\
|w_{23}|^2 &=& \frac{1}{32\,(1-{\rm{cos}}\alpha)}, 
\end{eqnarray}

\noindent 
with the obvious condition $ \cos \alpha < 1/2$
(that is, $\alpha \in [\pi/3,\pi ]$). 
The energy and energy square expectation values 
read

\begin{eqnarray}
E&=& \langle H \rangle = 
2\sum_{i,j=0}^{3} 
|w_{ij}(0)|^2\,(\epsilon_i+\epsilon_j) = 
3 \epsilon \cr
\langle H^2 \rangle &=&
2\sum_{i,j=0}^{3} |w_{ij}(0)|^2\,
(\epsilon_i+\epsilon_j)^2 = \, \frac{\epsilon^2}{2} 
\left[\frac{21-19\cos \alpha}{1 - \cos \alpha} \right],
\end{eqnarray}

\noindent and, consequently, the minimum 
evolution time (\ref{Tmin}), after 
some calculation, is

\be
T_{min} \, = \, \frac{\pi \hbar}{2 \epsilon} \,
\sqrt{\frac{2(1 - \cos \alpha)}{3 - \cos \alpha}}.
\ee 

\noindent
The formula for the concurrence in the 
two-fermion case is \cite{identical}

\be \label{fermiconc}
C_{F}=8|w_{01}w_{23}-w_{02}w_{13}+w_{03}w_{12}|,
\ee

\noindent
which is clearly time-independent (for the Hamiltonian
(\ref{hamilhamil})).

One can check that the lowest value of 
$\tau/T_{min}$ corresponds to 
$\cos \alpha = 1/2$. That is, the 
state closest to saturate the 
lower bound for the time required
to reach an orthogonal state is
given by $\alpha=\pi/3$. In this
case the fermionic concurrence 
reads 

\begin{equation} \label{C_F}
C_F = \frac{|e^{i\phi_{01}+i\phi_{23}}-e^{i\phi_{02}+i\phi_{13}}|}{2}, 
\end{equation}

\noindent where $\phi_{ij}$ denotes the phase
associated with the coefficient $w_{ij}$.
Now, with an appropriate choice of the $\phi's$, 
we can make (\ref{C_F}) either $0$ or $1$. 
In other words, among those states that saturate
the lower bound on the time to evolve to an orthogonal
state, there are states of zero entanglement, 
as well as maximum entangled states.

\begin{figure}
\begin{center}
\includegraphics[angle=270,width=.65\textwidth]{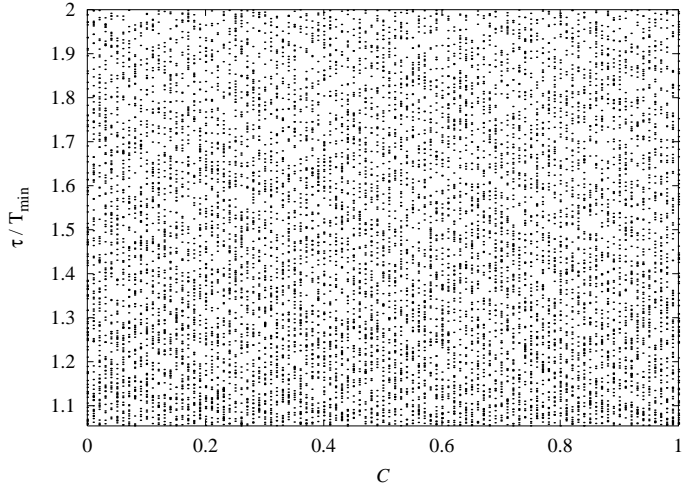}
\caption{Randomly generated states of two fermions that
evolve to an orthogonal state. Each point corresponds 
to one of those states, represented in the 
$(C,\tau/T_{min})$-plane. This curtain-like plot is obtained by fixing 
the concurrence at intervals of $0.01$. It transpires from the figure 
that for each value of the concurrence $C$ the time  
$\tau/T_{min}$ needed to reach an orthogonal state 
may adopt any value, from $\frac{1}{3}\sqrt{10}$
up to a maximum equal to $2$. All depicted quantities
are dimensionless.} 
\label{time3}
\end{center}
\end{figure}

Fig.\ref{time3} exhibits a plot in the $(C,\tau/T_{min})$-plane 
of a set of randomly generated states of two fermions 
that evolve to an orthogonal state. Each point represents 
one of those states. It transpires from the figure that 
for each value of the concurrence $C$ the time  
$\tau/T_{min}$ needed to reach an orthogonal state may 
adopt any value, from $\frac{1}{3} \sqrt{10}$ 
up to a maximum equal to $2$. 

We see that, as far as the connection between 
entanglement and the ``speed" of quantum evolution
is concerned, the behaviour of fermionic systems 
differs considerably  from the behaviour of
systems consisting either of bosons, or of
distinguishable particles.

\section{Concluding remarks}

 We have explored, for bipartite systems of low 
 dimensionality, some aspects of the connection 
 between entanglement and the speed of quantum 
 evolution. We considered (i) two qubits (distinguishable) 
 systems  and (ii) systems composed of two (bosonic or 
 fermionic) identical particles, with single
 particle Hilbert spaces of lowest dimensionality.

 These results corroborate that 
 there is a clear correlation between the
 amount of entanglement and the speed of 
 quantum evolution for systems of two-qubits 
 and systems of two identical bosons. On 
 the contrary, such a clear correlation is 
 lacking in the case of systems of 
 identical fermions \cite{BatleTemp}.

\chapter{Evolution of entanglement in a quantum algorithm: Grover's 
search algorithm}

Let us revisit the algorithm introduced by Lov Grover in 1996 for a faster 
search in a database than any classical computer can perform (See Chapter 2 
for more details). 
We shall pay special attention to the {\it evolution} of the entanglement 
present between 
the qubits in a given register because, as exhaustively mentioned throughout 
the present work, quantum entanglement is an essential ingredient for all 
these new revolutionary tasks. By tracing the evolution of entanglement 
during the search, we shall obtain a better insight into how this 
algorithm works. 

Suppose that we have a quantum circuit with an input register of $n$ qubits 
plus some auxiliary ancillas, which are not of our concern now. 
A key ingredient in the algorithm is the Hadamard gate

\be
H=\frac{1}{\sqrt{2}}\left [
\matrix{1 & 1\cr
1 & -1\cr
}
\right ],
\ee
\noindent which converts single qubit states into a coherent superposition of 
them. It is convenient to introduce the gate $W$ resulting of the action 
of $n$ times the application of the Hadamard gate in our quantum circuit 

\ben \label{Wg}
W &=& H\otimes H\otimes ... \otimes H \,(\equiv H^{\otimes n}) \, 
|0...0\rangle \,=\, \frac{1}{\sqrt{2^n}} \big(|00...00\rangle+\cr
& & |00...01\rangle+...+|11...11\rangle \big) \,=\, \frac{1}{\sqrt{2^n}} 
\sum_{i=0}^{2^n-1}|i\rangle
\een

\noindent on the initial register of $n$ qubits, set initially at 
${\bf x}=|{\bf 0}\rangle$. It is clear that what the $W$ gate does is to create 
a uniform superposition of all possible states of the register of $n$ qubits, 
starting from an initial state preparation of all states being reset to 
$|{\bf 0}\rangle$. 

We also need an operator ($2^n\times 2^n$ matrix) that flips only the first 
qubit ($|0\rangle \rightarrow -|0\rangle$), while leaving the remaining $n-1$ 
untouched ($|{\bf x}\rangle \rightarrow |{\bf x}\rangle$). We then have the 
gate

\be \label{I0}
I_0=\left [
\matrix{-1 & 0 & 0 & \cdots & 0\cr
0 & 1 & 0 & \cdots & 0\cr
0 & 0 & 1 & \cdots & 0\cr
\vdots & \vdots & \vdots & \ddots & \vdots \cr
0 & 0 & 0 & \cdots & 1\cr
}
\right ].
\ee

\noindent Now we wonder about the composite action $G = W I_0 W$ on an arbitrary 
state $\Psi=$($|a\rangle |b\rangle |c\rangle ... |z\rangle$), that is, we firstly 
apply in our quantum circuit (\ref{Wg}) to $\Psi$, followed by (\ref{I0}) 
and finally let us act (\ref{Wg}) again. The resulting state is given by 

\be \label{final}
\left [
\matrix{(-1+\frac{2}{2^n})a+\frac{2}{2^n}(b+c+...+z) \cr
\frac{2}{2^n}a + (-1+\frac{2}{2^n})b+\frac{2}{2^n}(c+d+...+z)\cr
\vdots \cr
\frac{2}{2^n}(a+b+...+y)+(-1+\frac{2}{2^n})z\cr
}
\right ]. 
\ee

\noindent The outcome of $G|{\bf 0}\rangle$ has a clear significance: 
{\it every element is inverted around its mean}, that is, 
every single element value $x_j$ of the $n$ qubits at stage or iteration $j$, 
turns into a new value $x_{j+1}=2\overline{\alpha}-x_j$. We have to wait for 
the last ingredient to give sense to $G|{\bf 0}\rangle$. During 
our search we need a ``black box" or oracle that identifies the element(s) we 
seek in the search. This is tantamount to assign the value $0$ to the element 
which does not accomplish the characteristics of the search, while to 
give the value $1$ to the element(s) that is(are) sought. To be more precise, 
every time we ask the oracle, we perform the operation

\be
F({\bf x}): x\Longrightarrow (-1)^{f(x)}|x\rangle,
\ee
\noindent where $f(x)$ is either zero or one for some $k$ values out 
of $2^n$ components of the general state $|\Psi_j({\bf x})\rangle$, 
at iteration $j$, of our register of $n$ qubits .

Now we can give sense to the action $-GF$ on the register $|{\bf x}\rangle$. 
Suppose that there is only one element out of $n$ qubits that is being sought. 
The initial state preparation is such that we have the following configuration
\newline
\newline
\newline
\vrule height 15pt $\_{}$$\_{}$$\_{}$$\_{}$$\_{}$$\_{}$$\_{}$\vrule height 15pt 
$\_{}$$\_{}$$\_{}$$\_{}$$\_{}$$\_{}$$\_{}$$\_{}$\vrule height 15pt 
$\_{}$$\_{}$$\_{}$$\_{}$$\_{}$$\_{}$$\_{}$$\_{}$\vrule height 15pt 
$\_{}$$\_{}$$\_{}$$\_{}$$\_{}$$\_{}$$\_{}$$\_{}$\vrule height 15pt 
$\_{}$$\_{}$$\_{}$$\_{}$$\_{}$$\_{}$$\_{}$$\_{}$\vrule height 15pt 
$\_{}$$\_{}$$\_{}$$\_{}$$\_{}$$\_{}$$\_{}$$\_{}$\vrule height 15pt 
$\_{}$$\_{}$$\_{}$$\_{}$$\_{}$$\_{}$$\_{}$$\_{}$\vrule height 15pt 
$\_{}$$\_{}$$\_{}$$\_{}$$\_{}$$\_{}$$\_{}$$\_{}$\vrule height 15pt 
$\_{}$$\_{}$$\_{}$$\_{}$$\_{}$$\_{}$$\_{}$$\_{}$\vrule height 15pt 
$\_{}$$\_{}$$\_{}$$\_{}$$\_{}$$\_{}$$\_{}$$\_{}$\vrule height 15pt 
$\_{}$$\_{}$$\_{}$$\_{}$$\_{}$$\_{}$$\_{}$$\_{}$
$\_{}$$\_{}$$\_{}$$\_{}$$\_{}$$\_{}$$\_{}$$\_{}$\vrule height 15pt 
\newline
\newline
$c_0\,\,\,\,\,\,\,\,\,c_1\,\,\,\,\,\,\,\,\,\,\,c_2
\,\,\,\,\,\,\,\,\,\,\,\,\,c_3\,\,\,\,\,\,\,\,\,\,\,\,c_4
\,\,\,\,\,\,\,\,\,\,\,\,c_5\,\,\,\,\,\,\,\,\,\,\,\,\,c_6
\,\,\,\,\,\,\,\,\,\,\,c_7\,\,\,\,\,\,\,\,\,\,\,\,\,c_8
\,\,\,\,\,\,\,\,\,\,\,\,c_9\,\,\,\,\,\,\,\,\,\,\,c_{10}
\,\,\,\,\,\,\,\,\,\,\,...\,\,\,\,\,\,\,\,c_{N-1}$
\newline
\newline
 
\noindent where $N=2^n$ and all the amplitudes are equal to 
$\frac{1}{\sqrt{2^n}}$. Suppose that the element $c_0$ is the one we 
are looking for. Let us now apply $F$ such that
\newline
\newline
\newline
$\_{}$$\_{}$$\_{}$$\_{}$$\_{}$$\_{}$$\_{}$\vrule height 15pt 
$\_{}$$\_{}$$\_{}$$\_{}$$\_{}$$\_{}$$\_{}$$\_{}$\vrule height 15pt 
$\_{}$$\_{}$$\_{}$$\_{}$$\_{}$$\_{}$$\_{}$$\_{}$\vrule height 15pt 
$\_{}$$\_{}$$\_{}$$\_{}$$\_{}$$\_{}$$\_{}$$\_{}$\vrule height 15pt 
$\_{}$$\_{}$$\_{}$$\_{}$$\_{}$$\_{}$$\_{}$$\_{}$\vrule height 15pt 
$\_{}$$\_{}$$\_{}$$\_{}$$\_{}$$\_{}$$\_{}$$\_{}$\vrule height 15pt 
$\_{}$$\_{}$$\_{}$$\_{}$$\_{}$$\_{}$$\_{}$$\_{}$\vrule height 15pt 
$\_{}$$\_{}$$\_{}$$\_{}$$\_{}$$\_{}$$\_{}$$\_{}$\vrule height 15pt 
$\_{}$$\_{}$$\_{}$$\_{}$$\_{}$$\_{}$$\_{}$$\_{}$\vrule height 15pt 
$\_{}$$\_{}$$\_{}$$\_{}$$\_{}$$\_{}$$\_{}$$\_{}$\vrule height 15pt 
$\_{}$$\_{}$$\_{}$$\_{}$$\_{}$$\_{}$$\_{}$$\_{}$
$\_{}$$\_{}$$\_{}$$\_{}$$\_{}$$\_{}$$\_{}$$\_{}$\vrule height 15pt 
\newline
\vrule height 15pt
\newline
$c_0\,\,\,\,\,\,\,\,\,c_1\,\,\,\,\,\,\,\,\,\,\,c_2
\,\,\,\,\,\,\,\,\,\,\,\,\,c_3\,\,\,\,\,\,\,\,\,\,\,\,c_4
\,\,\,\,\,\,\,\,\,\,\,\,c_5\,\,\,\,\,\,\,\,\,\,\,\,\,c_6
\,\,\,\,\,\,\,\,\,\,\,c_7\,\,\,\,\,\,\,\,\,\,\,\,\,c_8
\,\,\,\,\,\,\,\,\,\,\,\,c_9\,\,\,\,\,\,\,\,\,\,\,c_{10}
\,\,\,\,\,\,\,\,\,\,\,...\,\,\,\,\,\,\,\,c_{N-1}$
\newline
\newline
 
\noindent $c_0$ is being flipped\footnote{The order in which the elements appear 
is irrelevant; it could had been say $c_5$ instead.}. By reversing about the 
mean,
\newline
\newline
\newline
\vrule height 28pt $\_{}$$\_{}$$\_{}$$\_{}$$\_{}$$\_{}$$\_{}$\vrule height 10pt 
$\_{}$$\_{}$$\_{}$$\_{}$$\_{}$$\_{}$$\_{}$$\_{}$\vrule height 10pt 
$\_{}$$\_{}$$\_{}$$\_{}$$\_{}$$\_{}$$\_{}$$\_{}$\vrule height 10pt 
$\_{}$$\_{}$$\_{}$$\_{}$$\_{}$$\_{}$$\_{}$$\_{}$\vrule height 10pt 
$\_{}$$\_{}$$\_{}$$\_{}$$\_{}$$\_{}$$\_{}$$\_{}$\vrule height 10pt 
$\_{}$$\_{}$$\_{}$$\_{}$$\_{}$$\_{}$$\_{}$$\_{}$\vrule height 10pt 
$\_{}$$\_{}$$\_{}$$\_{}$$\_{}$$\_{}$$\_{}$$\_{}$\vrule height 10pt 
$\_{}$$\_{}$$\_{}$$\_{}$$\_{}$$\_{}$$\_{}$$\_{}$\vrule height 10pt 
$\_{}$$\_{}$$\_{}$$\_{}$$\_{}$$\_{}$$\_{}$$\_{}$\vrule height 10pt 
$\_{}$$\_{}$$\_{}$$\_{}$$\_{}$$\_{}$$\_{}$$\_{}$\vrule height 10pt 
$\_{}$$\_{}$$\_{}$$\_{}$$\_{}$$\_{}$$\_{}$$\_{}$
$\_{}$$\_{}$$\_{}$$\_{}$$\_{}$$\_{}$$\_{}$$\_{}$\vrule height 10pt 
\newline
\newline
$c_0\,\,\,\,\,\,\,\,\,c_1\,\,\,\,\,\,\,\,\,\,\,c_2
\,\,\,\,\,\,\,\,\,\,\,\,\,c_3\,\,\,\,\,\,\,\,\,\,\,\,c_4
\,\,\,\,\,\,\,\,\,\,\,\,c_5\,\,\,\,\,\,\,\,\,\,\,\,\,c_6
\,\,\,\,\,\,\,\,\,\,\,c_7\,\,\,\,\,\,\,\,\,\,\,\,\,c_8
\,\,\,\,\,\,\,\,\,\,\,\,c_9\,\,\,\,\,\,\,\,\,\,\,c_{10}
\,\,\,\,\,\,\,\,\,\,\,...\,\,\,\,\,\,\,\,c_{N-1}$
\newline
\newline

\noindent we obtain an enhancement of the amplitude of the element we are 
looking for. By repeating the action of $-GF$ several times, we arrive 
at the desired result with probability $p=c_0^2=1$.

How many steps do we have to do in order to find the solution? Or in other 
words, what is the efficiency of the algorithm? 
So far we have supposed that there is one only item to be found, 
but there can exist several of them. Suppose that according to 
this criterion, we represent the general state vector of 
the register at iteration $j$ by the wave function

\be\label{Psir}
|\Psi_j({\bf x})\rangle \, = \, s_j \sum_{{\bf x}\in S} |{\bf x}\rangle + 
c_j \sum_{{\bf x}\in NS} |{\bf x}\rangle,
\ee

\noindent where $S$ is the set of $k$ solutions of the 
oracle $f({\bf x})=1$ (number of items pursued), whereas there are $2^n-k$ 
terms of (\ref{Psir}) which are not (set $NS$). This decomposition proves to be 
extremely useful. We assume without loss of generality that the coefficients 
in (\ref{Psir}) are real. After the oracle, we have

\be\label{Psir2}
|\Psi_j^{\prime}({\bf x})\rangle \, = \, -s_j \sum_{{\bf x}\in S} |{\bf x}\rangle + 
c_j \sum_{{\bf x}\in NS} |{\bf x}\rangle.
\ee

\noindent Recall that the average of amplitudes of (\ref{Psir2}) at this stage 
is given by

\be
\overline{\alpha} \, = \, \frac{1}{2^n}\big( -s_j k + c_j (2^n-k)\big).
\ee

\noindent After application of operator $-GF$, we get the new state ($j+1$)

\be\label{Psir+1}
|\Psi_{j+1}({\bf x})\rangle \, = \, s_{j+1} \sum_{{\bf x}\in S} |{\bf x}\rangle + 
c_{j+1} \sum_{{\bf x}\in NS} |{\bf x}\rangle,
\ee

\noindent where we have the celebrated ``inversion around the mean" expressions 
$s_{j+1}=2\overline{\alpha}-s_j,\,c_{j+1}=2\overline{\alpha}-c_j$. 
Expanding coefficients we have two recursion relations between 
coeficients $s_{j+1}$ and $c_{j+1}$, that transforms ($s_j,c_j$) into 
($s_{j+1},c_{j+1}$) (the action of $-GF$). Because all operations done on the 
generic state $|\Psi_j({\bf x})\rangle$ are unitary, due to the fact that 
it is initially normalized to unity ($s_{0}=c_{0}=\frac{1}{\sqrt{2^n}}$), 
it must preserve its norm. This condition entails that

\be
|\langle \Psi_j({\bf x})|\Psi_j({\bf x})\rangle|^2\,=\, k\,s_{j}^2\,
+\,(2^n-k)c_{j}^2\,=\,1,
\ee
\noindent which is equivalent to an ellipse with coordinates 
$s_j=\frac{1}{\sqrt{k}}\sin\theta_j$, 
$c_j=\frac{1}{\sqrt{2^n-k}}\cos\theta_j$, for some angle $\theta_j$. After 
simplifying the aforementioned recursion relation, we obtain 

\ben
\sin\theta_{j+1} &=& \sin(\theta_{j}+\omega), \cr
\cos\theta_{j+1} &=& \cos(\theta_{j}+\omega),
\een

\noindent provided we identify $\cos\omega$ with $1-\frac{2k}{2^n}$. 
After imposing the initial conditions mentioned before, we arrive at 
the final expression for the $2^n$ coefficients of $|\Psi_j({\bf x})\rangle$ 
at step $j$:

\ben \label{evol}
s_{j} &=& \frac{1}{\sqrt{k}}\sin\big((2j+1)\nu\big), \cr
c_{j} &=& \frac{1}{\sqrt{2^n-k}}\cos\big((2j+1)\nu\big),
\een 

\noindent with $\sin^2 \nu = \frac{k}{2^n}$. 

Now we are in a position to answer the previous question. 
We have finished the search once we have absolute certainty about 
the result. In other words, $ks_{j}^2=\sin^2\big((2j+1)\nu\big)=1$ 
for some $j^{*}$. If the number of qubits $n$ is high enough, 
then $j^{*}$ is the closest integer value to 

\be \label{j}
\bigg[\frac{\pi}{4}\sqrt{\frac{2^n}{k}}\bigg]\,=\,O(\sqrt{2^n}).
\ee

\noindent With this analysis we show that Grover's algorithm 
is of $O(\sqrt{N})$, as opposed to the best classical result $N/2$. 
\newline

Where is entanglement in all this business? Whenever we apply $-GF$ on a 
reference state, we induce all qubits to interact between them. If we start 
with the state ${\bf x}=|{\bf 0}\rangle$ at zero iterations, we do not have 
any entanglement initially. But as soon as we make them interact, we create 
several superpositions of all possible states of the register, until a single 
state is reached, the solution to the search algorithm. Therefore we end up 
in a product state and no entanglement is present. 

Let us consider the measure of entanglement introduced in previous Chapters. 
It is based on the conjecture (numerically checked by us) that 
the inequality \cite{CKW00}

\be \label{dxd}
0 \, \le \, d_E\equiv C^{2}_{1(2..n)}-\sum_{i=2}^{n} C^{2}_{1i} \, \le \, 1
\ee

\noindent holds for an arbitrary number $n$ of qubits in a pure state 
$\rho=|\Psi\rangle_{1..n}\langle \Psi|$. $C^{2}_{xy}$ stands for the 
concurrence squared between qubits $x,y$ and $C^{2}_{1(2..n)}=4\,$det$\rho_1$, 
with $\rho_1$=Tr$_{2..n}$($\rho$). We regard the quantity $d_E$ between 
inequalities as a proper measure for multipartite entanglement, an so it 
is considered here.

\begin{figure}
\begin{center}
\includegraphics[angle=270,width=.65\textwidth]{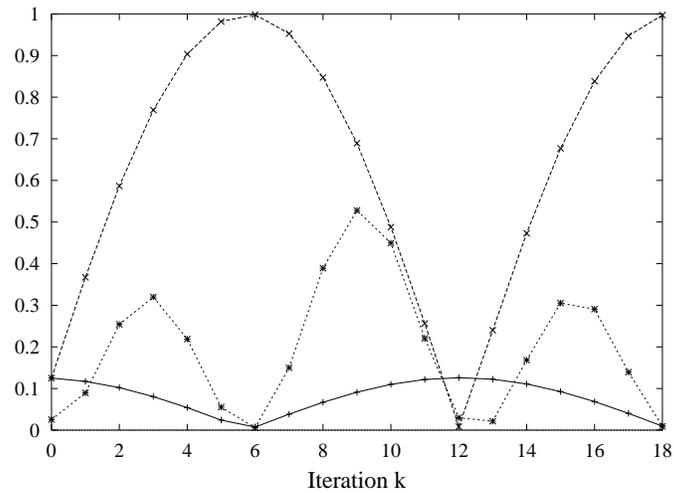}
\caption{Evolution for a 6-qubit register of the target amplitude or probability 
(curve reaching 1), and 
the amplitude or probability of the remaining non-target qubits (lowest curve), in 
absolute value. The curve with two maximums plots the evolution of \ref{dxd} during 
a whole period. See text for details.} 
\label{grover6}
\end{center}
\end{figure}	

\begin{figure}
\begin{center}
\includegraphics[angle=270,width=.65\textwidth]{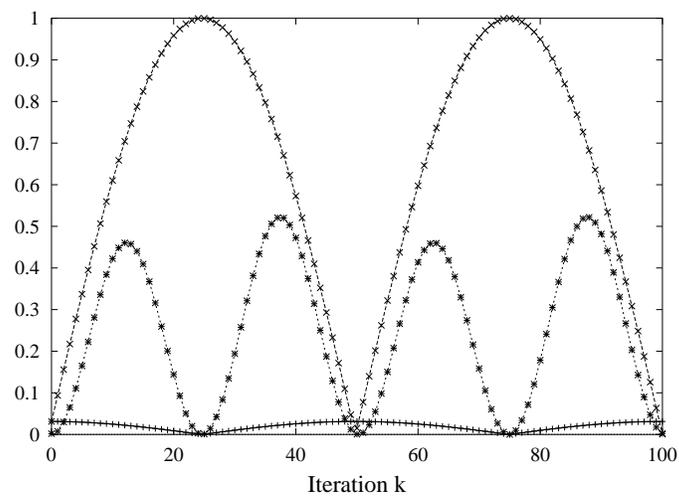}
\caption{Same as in Fig. \ref{grover6}, for a 10-qubit register. The perfect 
harmonic evolution of the amplitudes becomes apparent as the number of 
qubits increases. See text for details.} 
\label{grover10}
\end{center}
\end{figure}

Suppose the we have a quantum computer with a definite number of 
qubits upon which we want to apply our Grover's search algorithm. 
In Fig.\ref{grover6} we plot the evolution for a system with $n=6$ qubits of the 
absolute values of the amplitudes for the ``target" state (curve reaching 1) and 
all the remaining 
amplitudes (curve reaching 0) versus the number of iterations $j$. The periodic 
evolution described by (\ref{evol}) is apparent. Even when we have already 
found the solution (amplitude equal to one), we obtain it again and again. 
From inspection, we see that the critical value at which the solution is 
reached is $j_c=6$, which coincides with the predicted value $j^{*}(k=1)$ 
(\ref{j}). The evolution of entanglement $d_E$ is described by the curve 
in between. It is remarkable that the maximum value is obtained at exactly $j_c/2$ 
iterations, being null again when the solution is found. Fig.\ref{grover10} depicts 
the same quantities for $n=10$ qubits. In this plot the shape of a sinus or 
a cosinus for the amplitudes is more precise. Again we observe a periodical 
evolution for $d_E$, but there appears an interesting feature. Once the 
solution has been reached at $j=j_c$, there is a revival at $3 j_c/2$ in the 
entanglement which is {\it greater} that in the way to pursue the result at 
$j=j_c/2$. We do not quite understand the meaning of this feature, but there is 
no reason that prevents this fact. 
\newline

This example of quantum algorithm improves the speed of calculation by a factor 
of $O(\sqrt{N})$ with respect to the best classical result. Some may say 
that this is not the spectacular exponential result that computer scientists 
(induced by physicists) promised with the quantum computer business. Well, 
indeed it is a substantial improvement even nowadays. Let us see why.

Whenever we make a search in the 
internet, for instance we are looking for ``Perico de los Palotes Bush", there 
usually appears the result within few seconds or fractions of seconds. 
Notwithstanding the 
fact that it is certainly a useful achievement, it cannot be used against 
an eventual quantum computer. The tricky thing is that the searches are performed 
in a database arranged alphabetically, and it automatically updates during the night. 
So it is not the kind of problem of searching for a needle in a haystack like the 
one we are dealing with here. It resembles more the situation of finding a given number 
in the phone book. In point of fact, the algorithm we use in the real life is 
similar to a bisection method: we take half of the book, then one of the quarters, 
and so forth until we find the desired item. We could say that convergence is 
exponential. But what if we are given the telephone number and look for the owner? 
Suppose that we live in the Balearic Islands and it is high season for tourism
\footnote{During summer, mainly.}. Suppose 
that there is a population of one million inhabitants. If they were registered 
at random, to find someone would take of the order of 500000 steps {\it minimum}, 
while around 1000 using a quantum computer and the Grover's algorithm 
(only 10 qubits needed). Certainly not an exponential improvement, but considerable.
The paradigm of exponential speed-up is the Shor algorithm for factorizing large 
numbers which owes, in turn, much of its efficiency to the algorithm that 
performs the Quantum Fourier Transform (QFT) (see Chapter 2 for more details). Although 
it is a bit more complex than Grover's, the QFT algorithm relies also on registers 
of qubits that are collectively entangled during the different iterations that take 
place. If we had followed its study along the same line as Grover's, we would 
had seen that entanglement constitutes indeed an essential ingredient.

Summing up, two physical mechanisms make possible the astonishing predicted achievements 
of quantum computing: on the one hand {\bf quantum entanglement}, a purely quantum 
correlation, and on the other hand {\bf quantum parallelism}, the superposition principle.


\chapter{Entanglement and quantum phase transitions}

In Chapter 5 the concept of Generalized Entanglement and the $\fh$-purity measure 
\cite{barnum1} were introduced, a generalization which goes beyond the usual 
notion \cite{WO98}, where a subsystem
decomposition exists, and which is relative to a preferred set of observables.
In this definition a pure quantum state is {\it Generalized Unentangled} ({\it
Generalized Entangled}) if it induces a pure (mixed) reduced state on that set
of observables.

A measure of Generalized Entanglement is the relative purity or $\fh$-purity,
where $h=\{ O_j \}$ is an orthogonal and finite set of observables. The
$\fh$-purity of a pure quantum state $\ket{\psi}$ is defined as
\begin{equation} \label{puresa1}
P_\fh = {\sf K} \sum_j |\langle \psi | O_j \ket{\psi}|^2
\end{equation}
where ${\sf K}$ is a normalization factor chosen such that the maximum
relative purity is 1. In the cases analyzed in this work, a quantum
state will be generalized unentangled (GU) with respect to the set $\fh$,
whenever its $\fh$-purity is maximum and generalized entangled (GE) otherwise 
(\cite{somma1}).

Nice properties are obtained whenever the set $\fh$ is a simple Lie Algebra. In
this case, the GU states are the Generalized Coherent States (GCS) of $\fh$,
obtained by applying a group operation to some reference state, i.e., $\ket{\sf
GCS}=\exp [i\sum c_j O_j] \ket{\sf vac}$ ($c_j \in \mathcal{C}$). Also, the
relative purity is group-invariant.

The traditional notion of entanglement is 
recovered\footnote{We do not mean that the purity measure, or some function 
of it, reduces to the {\it reduced} von Neumann entropy for pure states. Conceptually 
they are {\it similar}, but not {\it equal}. Roughly speaking one may think of the 
purity measure as 1 minus the normalized (to 1) reduced von Neumann entropy, once a local set 
of observables is chosen, which is tantamount as partitioning the Hilbert space of the 
system in a preferred way.} when choosing the preferred set
of observables $\fh$ as the set of all local observables, corresponding to a
particular
subsystem decomposition (\cite{somma1}). For example, if the system studied
consists of $N$-spin 1/2 particles, a measure of the usual (traditional) entanglement is
obtained when  calculating the purity relative to the set 

\begin{equation}
\fh=\{ \sigma_x^1,
\sigma_y^1, \sigma_z^1, \cdots, \sigma_x^N,\sigma_y^N,\sigma_z^N\}, 
\end{equation}

\noindent with $\sigma_\alpha^j$ the Pauli spin-1/2 operators; that is,
\begin{equation}
P_{\fh} = \frac{1}{N} \sum\limits_{\alpha, j} |\langle
\psi |\sigma_\alpha^j
\ket{\psi}|^2 ,
\end{equation}
which reaches its maximum value ($P_\fh=1$)  only for a product state of the
form $\ket{\psi}= \ket{\phi_1}_1 \otimes \cdots \otimes \ket{\phi_N}_N$. These
states are the GCSs of $\fh$. The purity is then invariant under local
rotations, which are the group operations in this case.

Although the usual notion of entanglement is easily recovered in this
framework, here we analyze the generalized entanglement relative to sets of
observables other than the local ones, in order to capture the most important
quantum correlations that describe the physics of the models studied.
\newline

The main purpose of this Chapter is to relate quantum phase transitions and entanglement. But 
let us first remember the basics of a phase transition occurring at finite temperatures 
($T\neq 0$). 
Either if we are dealing with classical or quantum systems, they are equally 
described by Landau's theory \cite{geotopphys}, being the protagonist some broken 
symmetry of the system and some corresponding order parameter\footnote{However, 
non-broken-symmetry quantum phase transitions do exist. That is, they do not possess a local 
order parameter. Mermin and Wagner proved \cite{Mermin66} that there is no spontaneous 
magnetization in the two-dimensional isotropic Heisenberg model. On the other hand, there is 
evidence from high temperature expansions of the magnetic susceptibility \cite{duality} 
that this system undergoes a phase transition. 
Distinguishing a broken from a non-broken-symmetry QPT, remains an open question. In any case, 
we focus our interest in this Chapter in those transitions that do present a broken symmetry, 
and therefore detectable using the purity measure (\ref{puresa1}). 
For an interesting insight, see \cite{duality}.}. The idea is the 
following: given a Hamiltonian $H$ describing a physical system, we may assume 
$H$ to be invariant under a certain symmetry operation. However, the ground 
state of that system need not preserve that very same symmetry. We say then, 
in this case, that the system undergoes a spontaneous symmetry breakdown. Let us 
illustrate it with the Heisenberg Hamiltonian

\begin{equation} \label{Hprova}
H=-J\sum_{(i,j)} {\bf S_i}\cdot{\bf S_j} + {\bf h}\cdot\sum_i {\bf S_i},
\end{equation}

\noindent of $N$ spins and interaction mediated only through nearest-neighbours, 
with ${\bf h}$ being an external magnetic field. We recall that the average 
magnetisation per spin reads 

\begin{equation}
{\bf m} \equiv \frac{1}{N}\sum_i \langle {\bf S_i} \rangle.
\end{equation}

\noindent In the limit ${\bf h} \rightarrow 0$, (\ref{Hprova}) is still invariant 
under $SO(3)$ rotations of all spins, but ${\bf m}$ does not vanish, and consequently 
does not observe the $SO(3)$ symmetry. We then say that there appears an 
{\it spontaneous magnetisation}, with a \underbar{critical (maximum) temperature} such that 
${\bf m} \neq 0$. It is said that the vector ${\bf m}$ plays the role of an 
{\it order parameter}, that describes a phase transition between an ordered state 
(${\bf m} \neq 0$) and a disordered state (${\bf m}=0$). As we have seen, this 
mechanism is driven by the temperature: at low temperature, 
the term $TS$ in the free energy $F=\langle H \rangle-TS$ ($S$ being the entropy) may 
be negligible, so that minimum $F$ is equivalent to minimum $\langle H \rangle$,  
which happens when all spins are aligned in the same direction (ordered phase). 
As we increase the temperature, $TS$ dominates in $F$ and minimizing the free 
energy $F$ is tantamount as maximizing the entropy $S$, which is attained when 
all spins point at random (disordered phase).

With this example, we illustrate the tenets of Landau's theory of phase transitions. 
Now then, what is the mechanism of a phase transition that occurs at zero 
temperature?\footnote{We follow the description given by S. Sachdev in \cite{QPT}.} 
Could the ground state energy of the system at $T=0$ be the equivalent of 
the previous free energy $F$? Let us consider a Hamiltonian $H(g)$ in the context 
of condensed matter, whose degrees of freedom reside on the sites of a lattice. 
Let us further follow the evolution of the ground state energy of $H(g)$ as a 
function of the parameter $g$\footnote{The discussion could generalize the situation 
to many parameters.}. One possibility is that the ground state is a smooth function 
of $g$. On the other hand, $g$ can couple to a conserved quantity: $H(g)=H_0+gH_1, 
[H_0,H_1]=0$. Therefore we can diagonalize both $H_0$ and $H_1$ at the same time 
so that the eigenfunctions are independent of $g$ even if the eigenvalues vary with 
$g$. This situation implies that whenever there is a level crossing at some 
$g=g_c$, where an excited state becomes the ground state, we come across with 
a point of nonanalyticity in the ground state. However, even if there is no 
level-crossing, it can become sharper and sharper as we increase the number of 
sites. In the general case, we shall identify any point of non-analytic behaviour 
of the ground state energy as a quantum phase transition (QPT). We do not 
enter too much into all the details of a QPT. The reader is referred to \cite{QPT} for 
a comprehensive insight. Roughly speaking, the usual terminology employed in 
classical phase transitions ({\it critical exponents, characteristic lengths, ..}) 
also applies at zero temperature, where the basic 
difference is that transitions are induced by the change of a set of parameters 
$\{g_i\}$ of a given Hamiltonian. Even in this case, there exists two types of 
quantum phase transitions, namely, i) those that occur strictly a $T=0$, and ii) 
the ones that can take place at very low temperatures, remaining close enough to 
the critical point(s). In this Chapter we focus our interest in the former ones.

Summarizing, quantum phase transitions (QPTs) are qualitative changes
occurring in the properties of the ground state of a many-body
system due to modifications either in the interactions
among its constituents or in their interactions with an external
probe \cite{QPT}, while the system remains at zero temperature.
As explained, such changes occur as some parameter $g$ of a set $\{g_i\}$ 
of a given Hamiltonian vary across some critical point or manifold. 
It is plain from the nature of the QPTs that thermal fluctuations do not 
play a relevant role, as they do in the usual phase transitions. Instead, 
there appear fluctuations in the expectation value of some observable, implying 
that QPTs are purely quantum phenomena. Examples of QPTs are provided 
in \cite{QPT}, being a paradigm the quantum paramagnet to
ferromagnet transition occurring in Ising spin systems under
an external transverse magnetic field \cite{barouch1,barouch2}.
\newline

How do all these features relate to quantum entanglement? How does 
entanglement fit in the context of condensed matter physics? 
Entanglement is a property inherent to quantum
states, which lies at the core of generic ``truly" quantum correlations, 
as Schr\"odinger first pointed out \cite{Schro}. 
These correlations arise when particles interact, leaving the system in 
a state that has to be described as a whole, regardless of its subsystems. 
In view of this definition, one would expect some considerable 
change in the ground state of a given system as it undergoes a quantum phase 
transition. In other words, one could regard the entanglement present in a quantum 
system as a detectable property, on equal footing as several 
thermodynamic variables. Specifically, a quantum phase transition offers 
a unique framework where to study the intimate connection between entanglement 
and a physical system in the thermodynamic limit $N\rightarrow \infty$.

The first steps made towards this connections have focused on characterizing 
entanglement using information-theoretic
concepts, such as the entropy of entanglement \cite{BDSW96}, the concurrence \cite{WO98}, 
and other measures of entanglement, originally developed for bipartite systems. 
The arena for these studies has been those 
exactly solvable models that can support a quantum phase transition 
\cite{barnum3,Osborne,Osterloh,Arnesen,ERico,Verst}. 
These efforts have employed for instance the concurrence between neighbours, 
nearest-neighbours and so forh in the $XY$ model in a transverse magnetic 
field \cite{Osborne,Osterloh}, or have borrowed tools from renormalization group theory in the 
study of the entanglement between a block of nearby spins and the rest of the 
chain \cite{ERico}. These studies certainly constitute important 
contributions towards the study of entanglement and criticality (universality classes, 
critical exponents,..) in quantum systems, 
but one may agree with the that fact it is a limitation to use solely bipartite 
entanglement measures in this enterprise. It therefore seems plausible 
to employ a measure of entanglement that is capable of grasping the features of 
all constituents in a quantum systems {\it at once}, specially in the 
thermodynamic limit $N\rightarrow \infty$. Here is the point where the 
generalized entanglement (the purity measure) introduced in Chapter 5, and at the 
beginning of the present one, makes its appearance. It constitutes an extension of 
the essential properties of entanglement beyond the conventional subsystem-based 
framework. References \cite{barnum1,barnum3} provide a thorough 
account of it, and \cite{somma1} makes explicit use of the purity measure in the 
context of QPTs. In the latter context, the most critical step is
to determine which subset of observables may be relevant in
each system under study, since the $\fh$-purity must contain information about
the quantum correlations that play a dominant role in the 
quantum phase transition (QPT). In particular, if the ground state of the 
models we consider can be exactly calculated then the relevant quantum
correlations in the different phases are well understood, therefore 
choosing this subset of observables becomes relatively easy. This will be 
the case in this Chapter for the $XY$ model, and the $XY$ model with 
bond alternation. In a more general case where 
the ground state of the system cannot be exactly computed, the 
application of these concepts can be done in principle by following the same 
strategy (see Ref. \cite{somma1} for a comprehensive account of these 
features). 

As discussed in \cite{barnum1,barnum3}, the purity measures recovers the properties 
of bipartite (as well as multipartite) entanglement measures, which are of common use 
in quantum information processing (QIP). However, we do not pretend to discuss its utility in 
those areas of QIP where measures such as the concurrence or the entropy of entanglement 
suffices to explain the corresponding features described. It is not the intention of 
this Chapter to confront different positions regarding the nature of the measure of 
entanglement or its relativity, as explained in Chapter 5. The aim of the present 
Chapter is to expand the analysis initiated
in \cite{barnum1,barnum3,somma1} by focusing on the detection of QPTs due to a broken
symmetry. The behaviour of an appropriate relative purity of the ground state 
will prove its utility not only in detecting a QPT (in the cases of the $XY$ 
model, and the $XY$ model with bond alternation), but will also unfold 
several non-ergodic features in the dynamic evolution of a system during a QPT.

\section{Static case}

In this section we study the connection between quantum phase transitions (QPTs) 
and the concept of Generalized Entanglement, with no dynamics involved. 
In order to illustrate this feature, we could use the anisotropic $XY$ model, already 
employed in the work of the Los Alamos group (\cite{somma1}). However, 
it is more instructive to provide an additional 
example with a slightly modified model, namely, the anisotropic $XY$ model with bond 
alternation in a transverse magnetic field. Besides, we first provide here, 
to our knowledge, the analytical solution of this model. In any case, the original 
anisotropic $XY$ model is revisited in full detail in the section devoted to the 
dynamic case. For an account on QPTs and systems of spins the reader is referred to 
\cite{Bishop2005}.

With these examples, 
we claim to use the purity measure as a detector of a quantum phase transition in 
the condensed matter framework. 


\subsection{The anisotropic $XY$ model with bond alternation}

In this section we study the quantum phase transitions (QPTs)
of the anisotropic $XY$ model with bond alternation. The corresponding 
model Hamiltonian is
\begin{equation}
\label{Hamilt4}
H = -g \sum\limits_{j=1}^N (1+(-1)^j \delta) \left[ (1+\gamma)\sigma_x^j
\sigma_x^{j+1} + (1-\gamma) \sigma_y^j \sigma_y^{j+1} \right] +
\sum\limits_{j=1}^N \sigma_z^j,
\end{equation}
where $g$ is the coupling constant, $\gamma$ the anisotropy, and $\delta$ is the
bond-alternation constant. Making $\delta=0$ in (\ref{Hamilt4}), we recover 
the usual anisotropic $XY$ model with transverse magnetic field. As usual, 
periodic boundary conditions are considered
($\sigma_\alpha^j \equiv \sigma_\alpha^{j+N}$). This model was first 
studied in \cite{japanese} without including a transverse field. A 
similar model is studied in \cite{japanese2} at finite temperatures. 
In the present effort we also consider the action 
of a transverse uniform magnetic field, which modifies the concomitant ground state. 
To our knowledge, this is the first time that the QPTs associated to 
Hamiltonian (\ref{Hamilt4}) are studied in an analytic fashion. 

The ground state of system (\ref{Hamilt4}) can be exactly obtained by mapping the spin 1/2
Pauli operators into the fermionic algebra in the following way:
\begin{eqnarray}
a^\dagger_{2j-1} &=& \prod\limits_{k=1}^{2j-2} (-\sigma_z^k) \sigma^+_{2j-1} \cr
b^\dagger_{2j} &=& \prod\limits_{k=1}^{2j-1}(-\sigma_z^k) \sigma^+_{2j},
\end{eqnarray}
where $a^\dagger_{2j-1}$ and $b^\dagger_{2j}$ are the fermionic operators that
create a spinless fermion at site $2j-1$ (odd) and $2j$ (even), respectively.
Here, $j=1..M=\frac{N}{2}$ and $\sigma^+_l = \frac{\sigma_x^l + i
\sigma_y^l}{2}$. Defining the fermionic operators
\begin{eqnarray}
a^\dagger_k &=& \frac{1}{\sqrt{M}} \sum\limits_{j=1}^M e^{ik (2j-1)}
a^\dagger_{2j-1} \cr
b^\dagger_k &=& \frac{1}{\sqrt{M}} \sum_{j=1}^M e^{ik 2j} b^\dagger_{2j}
\end{eqnarray}
with $k = l \frac{2 \pi}{N}$, and $l =1 \cdots M$, the Hamiltonian of Eq.
\ref{Hamilt4} can be rewritten as $H= \sum_{k \in W} H_k$, with $W = \{
\frac{2\pi}{N}, \frac{4\pi}{N}, \cdots, \frac{\pi}{2}\}$, and

\begin{eqnarray}
\label{Hamilt5}
H_k &=& J_k a^\dagger_k b_k + J^*_k a^\dagger_{-k} b_{-k} + \Gamma_k a^\dagger_k
b^\dagger_{-k} + \Gamma^*_k a^\dagger_{-k} b^\dagger_k + 2(a^\dagger_k a_k +
b^\dagger_k b_k) + h.c., \cr
J_k &=& -4g (\cos k - i \delta \sin k), \cr
\Gamma_k &=& -4g \gamma (-\delta \cos k + i \sin k).
\end{eqnarray}

Defining the {\it vector} operators
\begin{equation}
\hat{A}_k = \pmatrix { a_k \cr a^\dagger_{-k} \cr b_k \cr b^\dagger_{-k}},
\end{equation}
and $\hat{A}^\dagger_k = (\hat{A}_k)^\dagger$, we obtain
\begin{equation}
\label{Hamilt6}
H_k = \hat{A}^\dagger_k \hat{H}_k \hat{A}_k,
\end{equation}
with $\hat{H}_k$ being the Hermitian  matrix
\begin{equation}
\label{Hamilt7}
\hat{H}_k = \pmatrix{ 2 & 0 & J_k & \Gamma_k \cr 0 & -2 & -\Gamma_k & -J_k \cr
J^*_k & -\Gamma^*_k & 2 & 0 \cr \Gamma^*_k & -J^*_k & 0 & -2}.
\end{equation}
Therefore, the problem reduces to the diagonalization of the matrices
$\hat{H}_k$ for each $k \in W$. The difference with the results obtained in \cite{japanese} 
is that the terms in (\ref{Hamilt7}) contain the interaction with an external magnetic field, 
which induces as we shall see a richer structure in the concomitant phase diagram. In 
this way, it can be proved that the ground
state energy in the thermodynamic limit 
is given by (up to an irrelevant global constant)
\begin{equation}
\label{energy}
E_g = \frac{1}{2\pi} \int_0^{\pi/2} dk \left[ \lambda_1(k) + \lambda_2(k)
\right],
\end{equation}
with 
\begin{eqnarray}
\lambda_1(k) =-\sqrt{4 -\xi_k + 16g^2 ((1+ \gamma^2 \delta^2) \cos^2k + (\gamma^2 +
\delta^2) \sin^2 k)} \\
\lambda_2(k) =-\sqrt{4 +\xi_k + 16g^2 ((1+ \gamma^2 \delta^2) \cos^2k + (\gamma^2 +
\delta^2) \sin^2 k)} ,
\end{eqnarray}
and $\xi_k =
2 \sqrt{16g^2 (2+ (2+16 g^2 \gamma^2) \delta^2 - 2(-1
+\delta^2) \cos 2k) }$. Obviously, $\lambda_1(k)$ and $\lambda_2(k)$ are the
negative eigenvalues of the matrices $\hat{H}_k$. Alternatively, we can diagonalize 
the matrix ${H}_k$ spanned by the eigenvectors 

\begin{eqnarray} \label{base}
{\cal D}&=&\{|0\rangle, a_{k}^{\dagger}a_{-k}^{\dagger}|0\rangle, 
b_{k}^{\dagger}b_{-k}^{\dagger}|0\rangle, a_{k}^{\dagger}b_{k}^{\dagger}|0\rangle, 
a_{k}^{\dagger}b_{-k}^{\dagger}|0\rangle, a_{-k}^{\dagger}b_{k}^{\dagger}|0\rangle,\cr 
& & a_{-k}^{\dagger}b_{-k}^{\dagger}|0\rangle, 
a_{k}^{\dagger}a_{-k}^{\dagger}b_{k}^{\dagger}b_{-k}^{\dagger}|0\rangle \}
\end{eqnarray}

\noindent which is an $8\times 8$ matrix, whose negative eigenvalues obviously coincide 
with the previous $\lambda_1(k), \lambda_2(k)$. Once we have found the eigenvalues, 
we find the concomitant ground state $|\Psi\rangle=\prod_{k>0}^{\pi/2} |\Psi_k\rangle$, 
with $|\Psi_k\rangle=u_1\, |0\rangle +
 u_1 \,a_{k}^{\dagger}a_{-k}^{\dagger}|0\rangle+...+ 
 u_8 \,a_{k}^{\dagger}a_{-k}^{\dagger}b_{k}^{\dagger}b_{-k}^{\dagger}|0\rangle$, 
 in terms of the coefficients $\{u_i\}$. The ground state will 
 be coherent\footnote{This is due to the fact that Hamiltonian (\ref{Hamilt6}) contains 
 biquadratic fermionic operators that form an $so(2N)$ Lie algebra.} with respect to 
 $so(2N)$, and therefore generalized unentangled with respect to this algebra. 
 As we saw, this is tantamount as possessing purity $P_{so(2N)}=1$, 
 which is of no interest. Depending on the subset of observables chosen, the $h$-purity 
 contains information about different $n$-body correlations present in the quantum
 state, allowing for a more general and complete characterization of entanglement. 

 Instead, if we consider the $u(N)$ algebra, we have a non-trival structure. We 
 need $N^2$ operators, which are found to be in the basis (\ref{base}):

\begin{eqnarray} \label{set}
& & \sqrt{2}(a_{k}^{\dagger}a_{k}-\frac{1}{2})\cr
& & \sqrt{2}(b_{k}^{\dagger}b_{k}-\frac{1}{2})\cr
& & (a_{k}^{\dagger}a_{k'}+a_{k'}^{\dagger}a_{k}) \,\,\,k\neq k'\cr
& & i(a_{k}^{\dagger}a_{k'}-a_{k'}^{\dagger}a_{k}) \,k\neq k'\cr
& & (b_{k}^{\dagger}b_{k'}+b_{k'}^{\dagger}b_{k}) \,\,\,k\neq k'\cr
& & i(b_{k}^{\dagger}b_{k'}-b_{k'}^{\dagger}b_{k}) \,k\neq k'\cr
& & (a_{k}^{\dagger}b_{k'}+b_{k'}^{\dagger}a_{k}) \,\,\, \forall \,k,k'\cr
& & i(a_{k}^{\dagger}b_{k'}-b_{k'}^{\dagger}a_{k}) \, \forall \,k,k',
\end{eqnarray}
 
\noindent with $-\frac{\pi}{2}<k<\frac{\pi}{2}$. If we are eager to obtain the purity 
$P_{\fu(N)}$ in the ground state, we take advantage of the symmetries of Hamiltonian 
(\ref{Hamilt5}) ($H_k$ mixes $k$ with $-k$). This means that few operators of the set 
(\ref{set}) will survive in the computation of $P_{\fu(N)}$\footnote{Remember: it is 
all about computing expectation values of a set of operators that form a given 
Lie algebra, $u(N)$ in this case.}. Therefore, what remains is

\begin{eqnarray} \label{bondbond}
P_{\fu(N)}&=& \frac{2}{N}\sum_{k>0}^{\pi/2} \bigg( 2[\langle a_{k}^{\dagger}a_{k}-\frac{1}{2}\rangle ^2+
\langle a_{-k}^{\dagger}a_{-k}-\frac{1}{2}\rangle ^2+\langle b_{k}^{\dagger}b_{k}-\frac{1}{2}\rangle ^2+\cr
& & \langle b_{-k}^{\dagger}b_{-k}-\frac{1}{2}\rangle ^2]+4|\langle a_{k}^{\dagger}a_{-k}\rangle |^2+4|\langle b_{k}^{\dagger}b_{-k}\rangle |^2+\cr
& & 4|\langle a_{k}^{\dagger}b_{k}\rangle |^2+4|\langle a_{k}^{\dagger}b_{-k}\rangle |^2+4|\langle a_{-k}^{\dagger}b_{k}\rangle |^2+\cr
& & 4|\langle a_{-k}^{\dagger}b_{-k}\rangle |^2 \bigg).
\end{eqnarray}

\noindent The expectation values are obtained in terms of coefficients $\{u_i\}$ of 
the ground state (for instance, $\langle a_{k}^{\dagger}a_{k}\rangle =|u_2|^2+|u_4|^2+|u_5|^2+|u_8|^2$). 
The numerical computation of the purity $P_{\fu(N)}$ as $N\rightarrow \infty$ is 
then carried out. In the case of no bond alternation ($\delta=0$) one recovers the 
study performed in \cite{somma1}. It is shown there that the corresponding $\fu(N)$ 
purity $P_{\fu(N)}$ reads  

\begin{equation}\label{puresaXY}
P_{\fu(N)}=\left\{
\begin{array}{cl}
\frac{1}{1-\gamma^2} \bigg(1\,-\,\frac{\gamma^2}{\sqrt{1-4g^2(1-\gamma^2)}} \bigg) & 
\mbox{  if } g\leq \frac{1}{2} \cr
\frac{1}{1+\gamma} & 
\mbox{  if } g > \frac{1}{2} 
\end{array}
\right. \;.
\end{equation}

\noindent Eq. (\ref{bondbond}) reduces to (\ref{puresaXY}) for $\delta=0$, which presents 
an analytical expression. In forthcoming sections we shall study the dynamic 
properties of the $XY$ model with no bond alternation, and eventually compare 
those features with the concomitant static case, 
whose purity is given by (\ref{puresaXY}).
\newline

\begin{figure}[htbp]
\begin{center}
\includegraphics[angle=0,width=.95\textwidth]{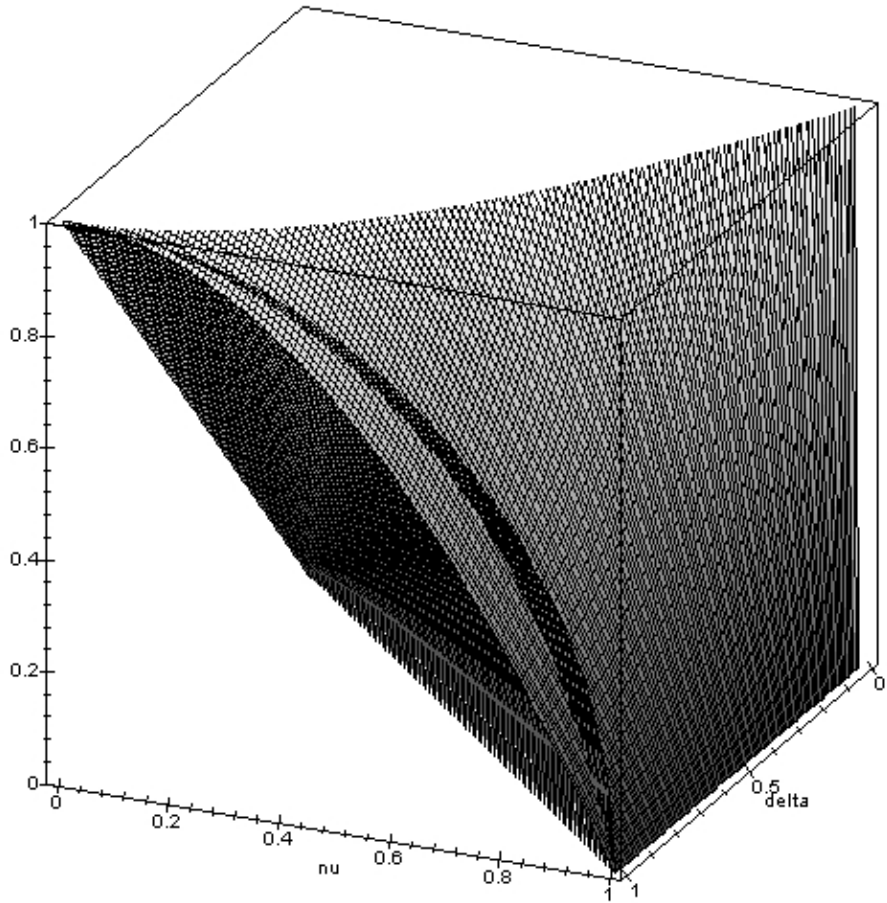}
\caption{Phase diagram of Hamiltonian (\ref{Hamilt4}) representing 
$\gamma^{(I)}$ and $\gamma^{(II)}$ in (\ref{critical}) vs. the bond 
alternating parameter 
$\delta$ and $\nu \equiv \frac{h}{2g}$ (see text). On the one hand, as we 
decrease the transverse magnetic field $h$ ($\nu \rightarrow 0$) in 
(\ref{Hamilt4}), the system presents a bond-order, so the surface 
collapses to the line $\gamma = \delta$, agreeing with the results in 
Ref. \cite{japanese}. On the other hand, by reducing the bond alternating parameter 
$\delta$, the surface reduces to the vertical line $\nu=1$, 
which is the critical point of the usual anisotropic $XY$ model with 
transverse magnetic field. It is apparent from this figure that the 
addition of a bond alternating component to the latter model induces a very rich 
structure in the phase diagram. See text for details.} 
\label{phase}
\end{center}
\end{figure}

The ground state of this bond alternating system undergoes several QPTs 
while changing the coupling constants $g$ and $\delta$. The quantum phases can be studied
by taking different limits in Eq. \ref{Hamilt4}. For example, if $g<0$,
$\gamma=1$, and $\delta=0$, there is no bond alternation and the model
transforms into  the antiferromagnetic Ising model in a transverse magnetic
field. Here, the ground state undergoes an antiferromagnetic-to
paramagnetic second order QPT at the critical point $g_c=-1/2$ (analogous to
the ferromagnetic case). In fact, when $g \rightarrow -\infty$, the ground
state of the system is twofold degenerate, i.e., $\ket{G} = 2^{-1/2}
\left[\ket{+ -
\cdots + -} \pm \ket{- + \cdots - +}\right]$, with
$\ket{+}=2^{-1/2}(\ket{\uparrow}+ \ket{\downarrow})$  and
$\ket{-}=2^{-1/2}(\ket{\uparrow}-\ket{\downarrow})$  the eigenvectors of the
$\sigma_x$ operators, while when $g \rightarrow 0 $ the spins align with the
external magnetic field, i.e., $\ket{G} = \ket{\downarrow
\downarrow \cdots \downarrow}$.

The limit $g \rightarrow - \infty$, with $1 > (\gamma, \delta) >0$ has been
studied in Ref. \cite{japanese}. Here, the external magnetic field is irrelevant. As
mentioned above, when $\delta \rightarrow 0$ and $\gamma \rightarrow 1$ the
phase is antiferromagnetic, while for $\delta \rightarrow 1$ (i.e., Eq.
\ref{Hamilt4} becomes a sum of isolated two-spin interactions) a bond-order
develops. In particular, it has been proved (\cite{japanese}) that an
antiferromagnetic-to-bond-order second order QPT occurs at the critical line
$\delta=\gamma$, in this limit.

\begin{figure}[htbp]
\begin{center}
\includegraphics[angle=270,width=.75\textwidth]{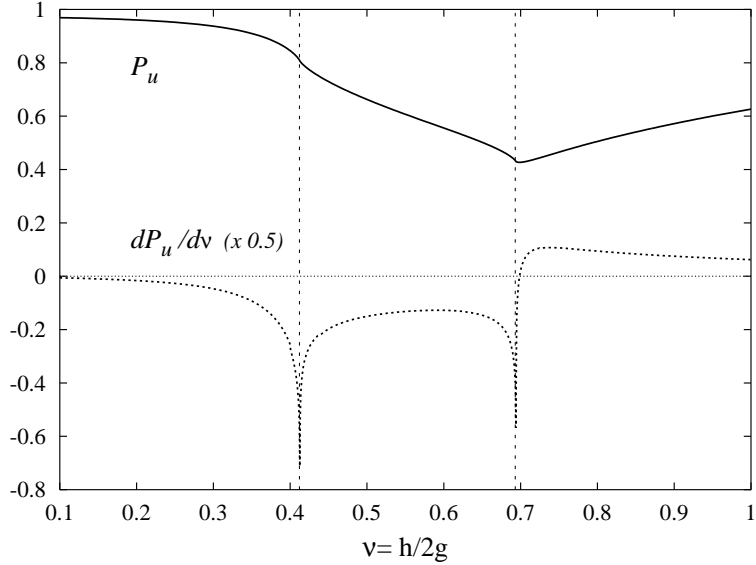}
\caption{Evolution of the purity measure $P_{\fu(N)}$ (solid line) as a function of the 
parameter $\nu \equiv \frac{h}{2g}$ for the bond alternating system (\ref{Hamilt4}). 
Its derivative (dashed lines) ``detects" the presence of critical points in phase 
space given by (\ref{critical}), which for the case of $\delta=0.9,\gamma=0.8$ and 
$g=1$ are given by $\nu=0.412$ (region (I) in (\ref{critical})) and $\nu=0.693$ 
(region (II) in (\ref{critical})). For this set of parameters, an antiferromagnetic phase 
developes from $\nu=0$ to $\nu=0.412$, a bond order one from $\nu=0.412$ to $\nu=0.693$, and 
a final paramagnetic phase with further $\nu$-grow.} 
\label{purPath}
\end{center}
\end{figure}

The phase diagram of the Hamiltonian (\ref{Hamilt4}) is, after some algebra, 
defined by the regions

\begin{eqnarray} \label{critical}
\gamma^{(I)} \,&=& \,  \sqrt{\delta^2-\nu^2} \cr 
\gamma^{(II)} \,&=& \, \frac{\sqrt{1-\nu^2}}{\delta}  
\end{eqnarray}

\noindent in the $\gamma,\delta$ and $\nu \equiv \frac{h}{2g}$ space of 
critical parameters ($h$ being the transverse magnetic field set to 1). 
The critical points can be obtained either numerically by finding the zeros
$(g_c,\gamma_c,\delta_c)$ of the functions $\lambda_1(k)$ or $\lambda_2(k)$, 
or analytically. In this way, we assure that the second
derivatives of the ground state energy with respect to these parameters 
diverge or present a discontinuity. In Fig.\ref{phase} we plot the regions of 
the phase diagram defined by (\ref{critical}). Also, in \cite{RoloiJo} 
we show the phase-diagram for this model in the antiferromagnetic region
($g<0$), which agrees with the above description for different limits.
One observes \cite{RoloiJo} that for small values of $g$ (high values of $\nu$) 
the phase is paramagnetic, therefore the purity in Fig.\ref{purPath} tends to 
one\footnote{Remember that this means disorder in the systems, therefore no 
entanglement whatsoever.}, 
and as we decrease $g$ (increase $\nu$) the phase becomes first antiferromagnetic 
(N\'{e}el order), and then it presents a bond-order. For small values of $\delta$ the 
antiferromagnetic region
is dominant but it dissapears when $\delta \rightarrow 1$. Moreover, when $g
\rightarrow - \infty$ ($\nu \rightarrow 0$) the critical line separating the 
antiferromagnetic and bond-order phases is located at $\gamma = \delta$, 
agreeing with the results in Ref. \cite{japanese}. As seen from Fig.\ref{phase}, 
the phase diagram indeed reduces to the line $\gamma = \delta$. Also, 
at zero anisotropy $\gamma=0$ we encounter a new critical line at $\nu=\delta$, 
which results from the competition of the bond alternating order between spins 
and their alignment with the external magnetic field $h$.

These 
analytic results are in excellent agreement with the numerical results 
computing the points where the derivative of the purity $P_{\fu(N)}$ (with 
respect to any element of the parameter space) diverge. This fact validates 
all our previous calculations regarding the features relative to the purity 
as a detector of a QPT. 

One way to check the validity of (\ref{critical}) is to trace the evolution 
of the purity $P_{\fu(N)}$ as a function of $\nu \equiv \frac{h}{2g}$. 
This is done in Fig.\ref{purPath}. Let us consider the $g=1$ (Ising case) throughout the 
following discussion. The solid line depicts the purity that corresponds 
to $\delta=0.9,\gamma=0.8$, 
while the dashed one plots its derivative $\frac{dP_{\fu(N)}}{d\nu}$. 
As $\nu\equiv \frac{h}{2g} \rightarrow 0$, which means zero magnetic field applied, 
the purity reaches the value of one, that is, 
the ground state of the system Hamiltonian (\ref{Hamilt4}) is unentangled. 
As we increase $\nu$, we cross two different regions: the first 
one corresponds to configuration (I) in (\ref{critical}) (vertical line 
at $\nu=0.412$), while the second case corresponds to (II)-instance in 
(\ref{critical}) (vertical line at $\nu=0.693$). For both cases, the purity tends 
to $1$ as $\nu$ increases (increasing magnetic field), as 
expected\footnote{Indeed, $\nu = \frac{h}{2g} = 0$ is equivalent to 
paramagnetic order: the spins align with the external magnetic field, i.e., 
$\ket{G} = \ket{\downarrow\downarrow \cdots \downarrow}$.}. 
Numerical computations are in full agreement with exact solutions (\ref{critical}).

\section{Dynamic case: non-ergodicity of entanglement}

The generalized entanglement present in the time-dependent anisotropic
spin-1/2 $XY$ model is
studied in this section by analyzing the behaviour of its relative purity. 
The obtained asymptotic values ($t \rightarrow \infty$) will strongly depend on 
the initial
($t=0$) conditions and the way interactions are turned on, indicating 
non-ergodicity phenomena. For an adiabatic
passage, the results of the static case are recovered and the
relative purity fully characterizes the paramagnetic-to-ferromagnetic quantum
phase transition present in this model, behaving as a disorder 
parameter \cite{somma1}. 

\subsection{The time-dependent anisotropic $XY$ model in the presence of an
external magnetic field}
\label{xymodel}
In this section we obtain the dynamical equations of the anisotropic $XY$ 
model in presence of an external magnetic field by using the symmetries of the
system. These results allow us to use the relative purity as a measure 
capable of distinguishing different orders (quantum correlations) in the
dynamics, as shown in the following section. The time-dependent Hamiltonian is
given by
\begin{equation}
\label{Hamilt1}
H(t)=-g(t) \sum\limits_{j=1}^N [(1+\gamma) \sigma_x^j \sigma_x^{j+1}
+ (1-\gamma) \sigma_y^j \sigma_y^{j+1}] + h(t) \sum\limits_{j=1}^N \sigma_z^j,
\end{equation}
where $g(t)$ is the nearest-neighbor time-dependent coupling,  $h(t)$ is the
time-dependent transverse magnetic field, $\gamma$ is  the anisotropy, and
$\sigma_\alpha^j$ are the Pauli spin-1/2 operators at site $j$. Periodic
boundary conditions are considered:  $\sigma_\alpha^j \equiv
\sigma_\alpha^{j+N}$.

The Hamiltonian of Eq. (\ref{Hamilt1}) can be exactly diagonalized \cite{barouch2}
by mapping the spin operators into the fermionic operators
using the Jordan-Wigner transformation \cite{jordan1}:
\begin{eqnarray}
\sigma^+_j &=& c^\dagger_j K_j, \\
\sigma^-_j &=& c_j K_j, \\
\sigma_z^j &=& 2n_j -1,
\end{eqnarray}
where $c^\dagger_j$ ($c_j$) are the creation (annihilation) fermionic operators
at site $j$, $\sigma^{\pm}_j = \frac{\sigma_x^j \pm i \sigma_y^j}{2}$, $K_j
= \prod_{k<j} (-\sigma_z^k)$, and $n_j=c^\dagger_j c_j$ is the fermionic
number operator at site $j$. Note that these fermions are spinless, i.e., at
most one fermion can occupy a single site.

In Ref. \cite{somma1} it was introduced the procedure for diagonalization. Since
translation invariance is considered, the first step is to perform a Fourier
transform of the fermionic operators
\begin{equation}
\label{ft}
c^\dagger_k = \frac{1}{\sqrt{N}} \sum\limits_{j=1}^N e^{-i k j } c^\dagger_j,
\end{equation}
where $k \in V+\{\pm \frac{\pi}{N}, \pm \frac{3\pi}{N}, \cdots, \pm
\frac{(N-1)\pi}{N}\}$ (the lattice constant is the unity), and $c_k =
(c^\dagger_k)^\dagger$. Then, Eq.~\ref{Hamilt1} can be written as:
\begin{eqnarray}
\label{Hamilt2}
H(t)= \sum_{k \in V^+} H_k(t) &=& \sum_{k \in V^+} \bigg( 
\frac{\beta_k(t)}{2} \left [c^\dagger_k c_k + c^\dagger_{-k} c_{-k}\right ] +\cr
& & \alpha_k(t) \left [c^\dagger_k c^\dagger_{-k} + c_k c_{-k} \right ] 
- 2h(t) \bigg),
\end{eqnarray}
with 
\begin{eqnarray}
\label{param0}
\alpha_k(t) &=& i4g(t) \gamma \sin k, \\
\label{param00}
\beta_k(t) &=& -4 \left [ 2g(t) \cos k - h(t)\right ],
\end{eqnarray}
and $V^+ = \{ +\frac{\pi}{N}, +\frac{3\pi}{N}, \cdots, +\frac{(N-1)\pi}{N}\}$.
Many symmetries of this model can be easily found from Eq.~\ref{Hamilt2}.  In 
particular, it is simple to observe that
\begin{equation}
\label{sim1}
\left[ H_k(t) , H_{k'}(t') \right ] = 0 \,for\,all \,t,t' \,and\, k \neq k',
\end{equation}
with $[A,B]=AB-BA$. Due to this symmetry, the non-degenerate  eigenstates of
the Hamiltonian at any fixed time $t$ can be written as a product of states
$\ket{\eta_k}$ ($k \in V^+$), each belonging to a Hilbert subspace
$h_k$ with corresponding  state basis $\mathcal{B}_k  = \{ \ket{0}
, c^\dagger_k\ket{0}, c^\dagger_{-k}\ket{0}, c^\dagger_k c^\dagger_{-k}
\ket{0} \}$, where $\ket{0}$ is the vacuum (no-fermions) state. 
Moreover, since the operators $H_k$ preserve the parity of the number of
fermions in each subspace $h_k$, the states $\ket{\eta_k}$ contain
either an even (zero and/or two) or an odd (one) number  of fermions.

In Ref. \cite{somma1} it was shown that the ground state $\ket{G}$ of the system
(Eq.~\ref{Hamilt1}) has even parity, taking the following form
\begin{equation}
\label{gs}
\ket{G}= \bigotimes\limits_{k \in V^+} \ket{\eta_k} = \prod\limits_{k \in V^+}
\left( u_k + i v_k c^\dagger_k c^\dagger_{-k} \right) \ket{0}.
\end{equation}
Considering that in the period $-\infty
< t \leq 0$ the interactions of the  system (e.g., coupling constant and
magnetic field) do not change, and defining the initial set of parameters as
$g(t=0)=g_0$ and $h(t=0)=h_0$, the parameters $u_k,v_k$ ($\in \mathcal{R}$)
are
\begin{eqnarray}
\label{param1}
u_k=\cos \left ( \frac{\phi_k}{2} \right), \\
\label{param2}
v_k=\sin \left ( \frac{\phi_k}{2} \right),
\end{eqnarray}
with
\begin{equation}
\label{param3}
\tan (\phi_k) = \frac{2 g_0 \gamma \sin k}{2 g_0 \cos k - h_0},
\end{equation}
taken such that $\cos (\phi_k) < (>) 0$ if $2g_0 \cos k > (<) h_0$.

Nevertheless, in this Chapter we are interested in studying the evolution
of the purity relative to some set of observables, while evolving the ground
state of the system (for a particular initial set of interactions
$g_0,h_0,\gamma$) by changing either $g(t)$ or $h(t)$.
In other words, we are interested in studying the generalized
entanglement of the evolved state ($t>0$)
\begin{equation}
\label{evol1}
\ket{\psi(t)} = U(t) \ket{G},
\end{equation}
where $U(t)=\hat{T}\exp[-i \int_0^t H(t') dt']$ is the unitary evolution given by the 
Schr\"odinger equation when changing the interactions of the system, $\hat{T}$ is the 
time-ordering operator and $\ket{G}$ is the ground state of the system at time $t=0$ (\ref{gs}). 

The time-dependent state $\ket{\psi(t)}$ (\ref{evol1}) can also be obtained
exactly (\cite{barouch1}). Since the different representations
corresponding to the subspaces $h_k$ do not mix in time 
(Eq.~\ref{sim1}), we can rewrite Eq.~\ref{evol1} as
\begin{equation}
\label{evol2}
\ket{\psi(t)} = \bigotimes\limits_{k \in V^+} \ket{\eta_k(t)} =
\bigotimes\limits_{k \in V^+} U_k(t) \ket{\eta_k},
\end{equation}
with $U_k(t)=\hat{T} \exp [-i \int_0^t H_k(t') dt']$ and $H_k$ defined in Eq.
\ref{Hamilt2}. Due to (even) parity invariance we obtain 
$\ket{\eta_k(t)} = [a_k(t) +  i b_k (t) c^\dagger_k c^\dagger_{-k}] \ket{0}$, 
with $a_k(t),b_k(t)  \in \mathcal{C}$, and $a_k(0)=u_k$ , $b_k(0)=v_k$, as
defined by Eqs.~\ref{param1}, \ref{param2}, and \ref{param3}.

Defining the reduced density operators $\rho_k(t) =  \ket{\eta_k(t)} \langle
\eta_k(t) |$, with $k \in V^+$, their matrix representations in the subspaces
$\tilde{h}_k$, with (reduced) basis states $\tilde{\mathcal{B}}_k = 
\{\ket{0}, c^\dagger_k c^\dagger_{-k} \ket{0} \}$, are
\begin{equation}
\label{reddensity}
\tilde{\rho}_k(t) = \pmatrix { 1-x_1^k(t) & x_2^k(t) \cr 
(x_2^k(t))^* & x_1^k(t)},
\end{equation}
with $x_1^k(t) = b_k(t) b_k^*(t) = 1-a_k(t) a_k^*(t) $; $x_1^k \in
\mathcal{R}$, and $x_2^k(t)=-i a_k(t) b_k(t)$; $x_2^k \in \mathcal{C}$. The
corresponding  evolution equations in the Schr\"odinger picture are given by
\begin{equation}
\label{evoleq1}
\frac{\partial \tilde{\rho}_k(t)} {\partial t} = -i \left [ \tilde{H}_k(t),
\tilde{\rho}_k(t) \right],
\end{equation}
where $\tilde{H}_k(t)$ is the representation of $H_k$ in the state basis
$\tilde{\mathcal{B}}_k$:
\begin{equation}
\tilde{H}_k(t) = \pmatrix {0 & -\alpha_k(t) \cr \alpha_k(t) & \beta_k(t)},
\end{equation}
with $\alpha_k(t),\beta_k(t)$ as defined in Eqs.~\ref{param0} and \ref{param00}.

A simple matrix calculation of Eqs.~\ref{evoleq1} yields to
\begin{equation}
\label{evoleq2}
\left \{ \matrix{ \dot{x}^k_1(t) = \tilde{\alpha}_k(t) \left[ (x^k_2(t))^* 
+ x^k_2(t)
\right], \cr \dot{x}^k_2(t) = \tilde{\alpha}_k(t)-2 \tilde{\alpha}_k(t) x^k_1(t)
+i \beta_k(t) x^k_2(t) ,} \right.
\end{equation}
with $\tilde{\alpha}_k(t)=-i \alpha_k(t)$. Obviously, the initial conditions
are given by the initial interaction parameters: $x^k_1(0)=
b_k(0)b_k^*(0)=(v_k)^2$ and $x^k_2(0)=-i a_k(0) b_k(0) = -i u_k v_k$,  or
equivalently, $x^k_1(0)= \sin^2 (\phi_k /2)$ and  $x^k_2(0)=-i
\sin(\phi_k)/2$, with $\phi_k$ as defined by Eq.~\ref{param3}, and $g_0,h_0$
the initial coupling interaction and magnetic field, respectively.

\subsection{$\fu(N)$-purity and the time-dependent anisotropic $XY$ model}
\label{unpurity}
In \cite{somma1} it was shown that the $\fu(N)$-purity was a good measure
of generalized entanglement capable of characterizing the
paramagnetic-to-ferromagnetic quantum phase transition (QPT) present in the
static  anisotropic $XY$ model in a transverse magnetic field. The (shifted)
$\fu(N)$-purity behaves as a disorder parameter in this case, vanishing in the
ferromagnetic phase and presenting the correct critical behaviour close to the critical
point ($g_c=h/2$).  Here, we study the properties of the $\fu(N)$-purity for
the time-dependent model (Eq.~\ref{Hamilt1}). Thus,
the evolved state of Eq.~\ref{evol1} is no longer the ground state of
the system (Sec.\ref{xymodel}).  We expect that this measure still captures
the relevant correlations of the system and give us information about the
physics underlying the evolution.

In Sec.~\ref{xymodel}, we showed that  the fermionic algebra provides a
natural language when solving this model. In fact, the Hamiltonian of
Eq.~\ref{Hamilt2} belongs to the $so(2N)$ Lie algebra composed of the
biquadratic fermionic operators  $c^\dagger_k c_{k'}$, $c^\dagger_k
c^\dagger_{k'}$, and $c_k c_{k'}$, with $k,k' \in V$. The corresponding time
evolution operator is obtained by exponentiating the time-dependent
Hamiltonian, and is a group operation of $SO(2N)$. The evolved quantum state
of Eq.~\ref{evol1} is then a GCS and generalized unentangled with respect to
the $so(2N)$ algebra, having  maximum $so(2N)$-purity (i.e.,
$P_{so(2N)}(t)=1 \ \forall t$).  Hence, the $so(2N)$-purity does not give
any information about the evolution of the evolved state, so it is necessary
to look into subalgebras of $so(2N)$, where the state becomes generalized
entangled (\cite{somma1}).

The $\fu(N)$ algebra  of fermionic observables can be
expressed as the linear span of the following orthogonal observables:
\begin{equation}
\fu(N)=\left\{
\begin{array}{cl}
(c^{\dagger}_k c^{\;}_{k'} + c^{\dagger}_{k'} c^{\;}_k) & 
\mbox{  with } 1\leq k<k' \leq N \cr
i(c^{\dagger}_k c^{\;}_{k'} - c^{\dagger}_{k'} c^{\;}_k) & 
\mbox{  with } 1\leq k<k' \leq N \cr 
\sqrt{2}(c^{\dagger}_k c^{\;}_k - 1/2 ) & \mbox{  with }1 \leq k \leq N
\end{array}
\right. \;,
\end{equation}
where the operators $c^\dagger_k$ ($c_k$) create (destroy) a fermion in  the
$k$-th mode and satisfy the anticommutation relations for spinless fermions.
Obviously, $\fu(N)$ is a Lie subalgebra of $so(2N)$, having Slater
determinants or fermionic product states as GCSs; thus, GU. In other
words, states like \begin{equation} \ket{\psi}= \prod_m c_m \ket{\sf vac},
\end{equation} with $m$ denoting the mode (e.g., a site on a lattice or a wave
vector) and $\ket{\sf vac}$ the vacuum or no-fermionic state, are the
generalized unentangled states relative to the $\fu(N)$ algebra, having
maximum $\fu(N)$-purity. Since the evolved state is not in general a fermionic
product state (i.e., GE with respect to $\fu(N)$), its $\fu(N)$-purity changes
as a function of time.

We now proceed to calculate the $\fu(N)$-purity for the evolved state 
$\ket{\psi(t)}$ (Eq.~\ref{evol1}). In order to calculate  its $\fu(N)$-purity,
the expectation values of every observable of the  $\fu(N)$ algebra must be
computed. However,  using the symmetries of the model we simplify the calculation,
obtaining $\langle \psi(t)|c^\dagger_k c_{k'} \ket{\psi(t)}= x_1^k(t)
\delta_{kk'}$.
Therefore,
\begin{equation}
\label{purity1}
P_{\fu(N)}(t)=\frac{4}{N} \sum_k [x_1^k(t)-1/2]^2,
\end{equation}
where the normalization constant ${\sf K}=4/N$ has been taken such that the
maximum of  $P_{\fu(N)}$ is 1 (for fermionic product states). Note that a
similar expression for the static case was obtained in \cite{somma1}.

The time-dependent function $x_1^k(t)$ depends on $g(t)$ and $h(t)$ as shown
by Eqs.~\ref{evoleq2}. In the following sections, we study the solution and
behaviour of Eqs.~\ref{evoleq2} for different time-dependent regimes. (Again,
we consider that for $-\infty < t \le 0$ the state of the system is the ground
state $\ket{G}$ for interaction parameters $g_0,h_0,\gamma$.) To complete the 
analysis, we also study the total time-dependent
$\hat{z}$-magnetization 
\begin{equation}
\label{magnetization}
M_z(t)=\sum_{k\in V} x_1^k(t)
-N/2.
\end{equation}

Obviously, a numerical computation of quantities (\ref{purity1}) 
and (\ref{magnetization}) involve a finite number of sites $N$. In all figures, 
the magnetization (\ref{magnetization}) is multiplied by $2/N$ so that it reaches the maximum 
value of one. Although several 
results can be obtained analytically, most of the calculations are numerical, employing 
one million sites. The accuracy in the approach of the thermodynamic limit has been checked 
when comparing numerical and analytical results, thus validating our calculations. 

From now on we omit (in general) the $k$ index but it must be understood that
such index should appear in every variable. 
\newline
\newline
\newline
{\bf Step function magnetic field}
\newline

In this case, the coupling constant $g(t)=g_0$ is time independent but the
magnetic field suffers a sudden change at $t=0^+$:
\begin{equation}
h(t) = \left \{ \matrix{ h_0, & t\le0 \cr h_f, & t>0} \right.
\end{equation}
Then, Eqs.~\ref{evoleq2} are ($t>0$)
\begin{equation}
\label{case1-1}
\left\{
\matrix{ \dot{x}_1(t)=2 \tilde{\alpha} x_2^R(t), \cr
\dot{x}_2^R(t) = \tilde{\alpha} -2\tilde{\alpha} x_1(t) - \beta x_2^I(t), \cr
\dot{x}_2^I(t) = \beta x_2^R(t), }
\right.
\end{equation}
with $\tilde{\alpha}= 4g_0 \gamma \sin k$, $\beta=-4[2g_0 \cos k - h_f]$
(which are time independent), and the indices $R,I$ denoting the real and
imaginary parts of $x_2$ ($x_2=x_2^R\,+i\,x_2^I$), respectively.

A simple replacement and a time derivative performed in Eqs.~\ref{case1-1}
yield to
\begin{equation}
\label{case1-2}
\ddot{x}^R_2(t) = -[4 \tilde{\alpha}^2 +\beta^2] x_2^R(t),
\end{equation}
which denotes the dynamics of a simple harmonic oscillator with solution (for $k
\in V^+$) $x^{R,k}_2(t) = A_k \sin (\omega_k t)$. Considering that
$x^{R,k}_2(0^+)=0$, and $\dot{x}^{R,k}_2(0^+)= \tilde{\alpha}_k -2
\tilde{\alpha}_k \sin^2 (\phi_k/2) + \beta_k \sin(\phi_k) /2$, we obtain for
the parameters of the solution:
\begin{equation}
\label{case1-3}
\matrix{
\omega_k &=\sqrt{4 \tilde{\alpha}_k^2 +\beta_k^2}, \cr
A_k&=\frac{1}{\omega_k} \left[ \tilde{\alpha}_k \cos (\phi_k)
+ \frac{\beta_k \sin(\phi_k)}{2}
\right].}
\end{equation}
Finally, since $x_1(t)= x_1(0) + \int\limits_0^t \dot{x}_1(t') dt'$, 
we obtain
\begin{equation}
\label{case1-4}
x_1^k(t) = \sin^2 (\phi_k/2) + 2 \frac{\tilde{\alpha}_k}{\omega_k} A_k
[1-\cos(\omega_k t )],
\end{equation}
with all the parameters defined in Eqs.~\ref{param0},\ref{param00},
\ref{param3}, and \ref{case1-3}.

\begin{figure}
\begin{center}
\includegraphics[angle=270,width=.65\textwidth]{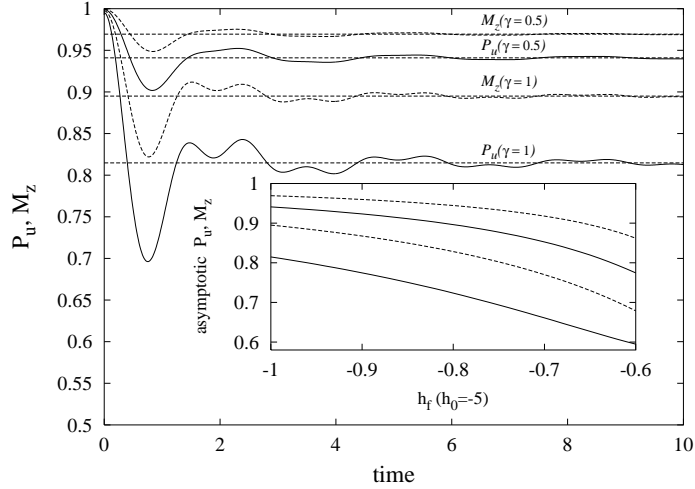}
\caption{Time-dependent $\fu(N)$-purity (solid lines)
and magnetization (dashed lines) of the evolved state
for the step-function case. The parameters are $g_0=-1/4$ (constant),
$h_0=-5$, and $h_f=-1$. The corresponding asymptotic values 
($t\rightarrow\infty$) as a function of $h_f$ are show in the 
inset of the figure.} 
\label{pur1}
\end{center}
\end{figure}

In Fig.\ref{pur1} we show the corresponding time-dependent $\fu(N)$-purity
(Eq. \ref{purity1}) and magnetization (Eq.~\ref{magnetization}) for 
$g_0=-1/4$, $h_0=-5$, and $h_f=-1$, and for different anisotropies 
$\gamma=0.5$ and $\gamma=1$.  The results are to be compared with  the ones of
the work of Barouch {\it et al.} \cite{barouch1} for finite  temperatures.
Note that the oscillations in $M_z(t)$ are present even at $T=0$. An
interesting point  is that for the step function case one can easily obtain
the  asymptotic values $P_{\fu(N)}(t=\infty)$ and $M_z(t=\infty)$ (shown by 
horizontal dashed lines) by time-averaging  the quantities appearing in their
definition. This is only possible  because the set of Eqs.~\ref{evoleq2} is
exactly solvable for  this case. The behaviour of the final values is shown in
the inset of  Fig.\ref{pur1}. As we increase the value of $h_f$ (with fixed
$h_0=-5$), the final  values decrease monotonically. 

From the results of Fig.\ref{pur1} one concludes that not only the
magnetization  $M_z(t)$ reaches a non-equilibrium value (as already pointed
out for finite temperatures  in \cite{barouch1}), but the $\fu(N)$-purity 
presents {\it non-ergodicity} features as well. This is a significant result,
which complements its relation with the detection of a QPT.

The computation of  either $P_{\fu(N)}(t)$ or $M_z(t)$ for finite times is
tedious and long, and not much  representative for the results we are concerned
here.  Nevertheless, the asymptotic ($t \rightarrow \infty$) behaviour for
$P_{\fu(N)}(t)$ is relevant and can be analytically  cast for the step-like case
in the same fashion as $M_z(t)$ is obtained in \cite{barouch1}. 
Basically, since the oscillations of these functions
disappear for $t \rightarrow \infty$ (Fig.\ref{pur1}), a time average of
expressions (\ref{purity1}) and (\ref{magnetization}) gives the desired result.
In particular, for $h_f=0$ and in the thermodynamic limit ($N\rightarrow
\infty$), the asymptotic values of $P_{\fu(N)}$ are given by
\begin{eqnarray}
\label{equiP}
 P_{\fu(N)}(\infty) & = & 8 \frac{1}{2 \pi} \int_{-1}^{1} dy
 \frac{1}{\sqrt {1- y^2}} \times \cr 
 & & \bigg( \frac{1}{4} \frac{( 2 g_0y- h_0 )^2} { h_0^2 +4
g_0^2 \gamma^2 - 4 g_0 h_0 y + 4 g_0^2 (1- \gamma^2 ) y^2} \\
\nonumber
& & \frac{3}{2} \frac{\gamma^4  (1- y^2 )^2 h_0^2}{4 (\gamma^2+ (1-\gamma^2 )
y^2 )^2 ( h_0^2 + 4 g_0^2 \gamma^2- 4 g_0 h_0 y + 4 \ g_0^2 (1- \gamma^2)y^2 )} +\\
\nonumber
& & \frac{\gamma^2  (1- y^2)  h_0  (2 g_0y- h_0 )}{2 ( \gamma^2 + (1-\gamma^2 ) 
y^2 )( h_0^2 + 4 g_0^2 \gamma^2 - 4 g_0 h_0 y+4 g_0^2 (1- \gamma^2) y^2)} \bigg),
\end{eqnarray}
where $y = \cos k$. 
Similarly, we obtain for the magnetization ($h_f=0$)
\begin{eqnarray} 
\label{equiM}
M_z(\infty) &=& 2\frac{1}{2\pi} \int_{-1}^{1}  dy
\frac{1}{\sqrt {1- y^2}} \times \cr
& & \bigg( \frac {2 g_0 y-  h_0}{\sqrt {h_0^2 + 4 g_0^2  
\gamma^2 - 4 g_0 h_0 y + 4 g_0^2 (1- \gamma^2) y^2 }} + \\
\nonumber
& & \frac { \gamma^2 (1-y^2) h_0}{ ( \gamma^2 + (1- \gamma^2 )y^2)
\sqrt{ h_0^2 +4 g_0^2 \gamma^2 - 4 g_0 h_0 y+ 4 g_0^2 (1- \gamma^2) y^2}} \bigg).
\end{eqnarray} 

The analytical expression of Eq.~\ref{equiP} can be given for any value of
$\gamma$ but for simplicity  we only consider the case $\gamma=1.0$, obtaining 
\begin{eqnarray}
P_{\fu(N)}(t \rightarrow \infty) \vert_{\gamma=1,h_f=0} &=&
- \frac{3 h_0^4}{512 g_0^4} + \frac {h_0^2}{128 g_0^2} 
+ \frac {17}{32} - \frac{3 g_0^2}{8 h_0^2} + \\
\nonumber
& &\frac{1}{\sqrt{ \frac{h_0^2}{g_0^2} -8+ 16 \frac{g_0^2}{h_0^2} }} \times \cr
& &\left[ \frac{3 h_0^5}{512 g_0^5} - \frac{h_0^3}{32 g_0^3} + \frac{h_0}{16 g_0}
- \frac{g_0}{2h_0} + \frac{3 g_0^3}{2 h_0^3} \right].
\end{eqnarray}
Clearly, the relative purity depends on the initial conditions given by
$h_0,g_0$. In fact, in \cite{barouch1} the authors show that the
asymptotic magnetization for $\gamma=1.0$ and $h_f=0$ does not reach its
equilibrium value and strongly depends on the initial conditions, too. This is a
non-ergodic process which can be captured by the relative purity.


\begin{figure}
\begin{center}
\includegraphics[angle=0,width=.65\textwidth]{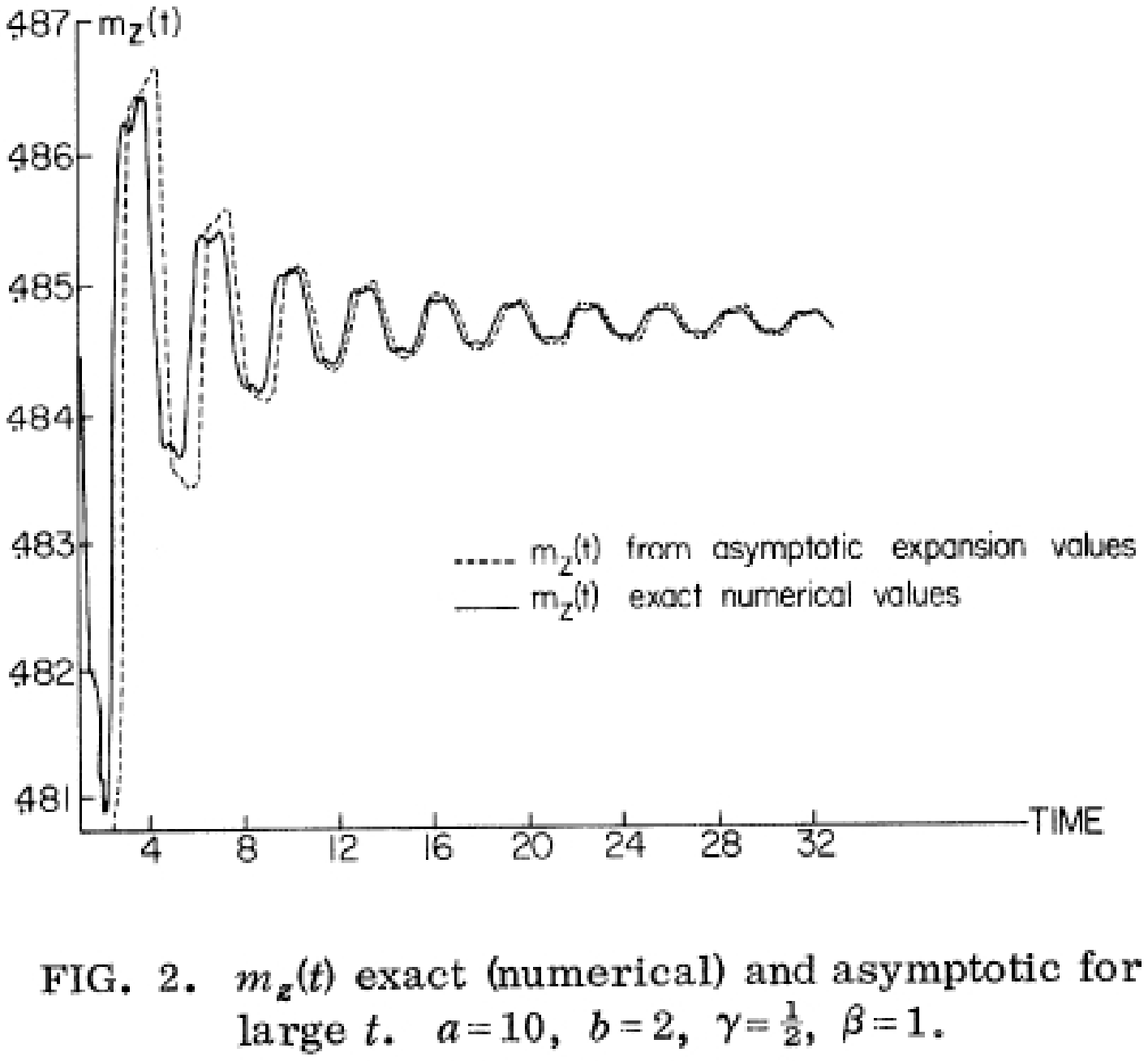}
\caption{Typical result for the magnetization in the step function magnetic 
field evolution, taken from the work of Barouch {\it et al.} \cite{barouch1}. 
The qualitative behaviour in the case of finite temperatures ($\beta\equiv 1/k_BT=1$) 
is similar to the quantum case at zero temperature studied here ($\beta=\infty$).} 
\label{pur56}
\end{center}
\end{figure}

\begin{figure}
\begin{center}
\includegraphics[angle=0,width=.75\textwidth]{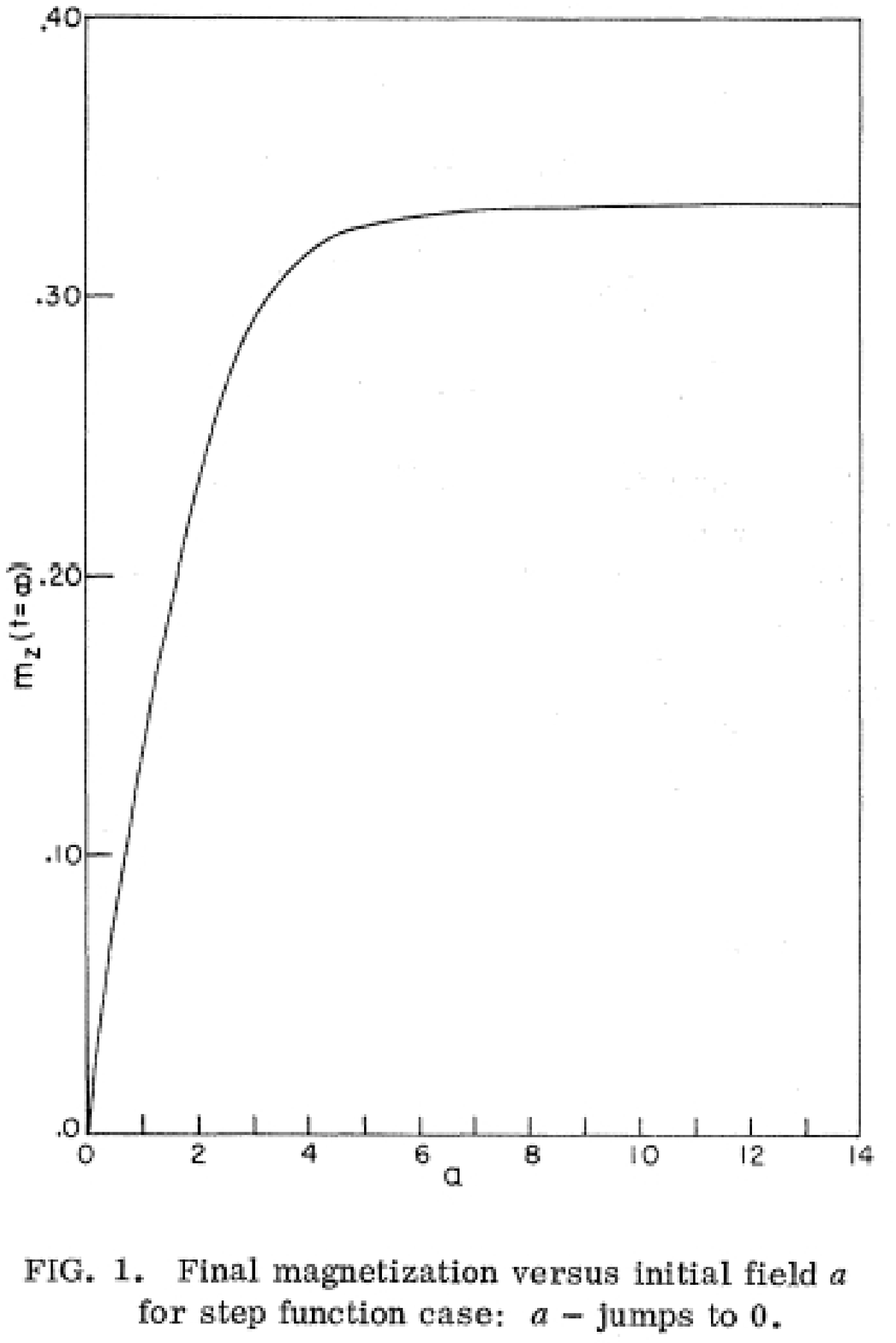}
\caption{Final magnetization in the case of final magnetic field $h_f=0$ 
(step-function magnetic field), taken from the work of 
Barouch {\it et al.} \cite{barouch1}. As seen in Fig.\ref{assimpt}, 
our $M_z(\infty)$ at $T=0$ presents a similar shape.} 
\label{pur57}
\end{center}
\end{figure}


Although only asymptotic values can be obtained in analytical fashion, we provide 
here for the sake of completeness a basic strategy to pursue in order to describe 
the ``transient" regime. This is done in order to justify the oscillations seen in 
Fig.\ref{pur1}, and following the steps of Barouch {\it et al.} \cite{barouch1}. 
As we can observe, all the time dependence of the purity (\ref{purity1}) is put 
in the coefficients $x_1^k(t)$ (\ref{case1-4}). In order to simplify things, 
let us consider the {\it fluctuations} of the purity $P_{\fu(N)}(t)$ minus 
its final value $P_{\fu(N)}(t \rightarrow \infty)$ in the form  $\Delta
P_{\fu(N)}(t)$ (there is an additional {\it static} term $\Delta P_{0}$  which
is not considered). Simplifying a bit more, let us consider the case 
$\gamma=1$. We thus have
\begin{equation} 
\label{time0}
\Delta P_{\fu(N)}(t)=\frac{4}{N} \sum_k [A(k)+B(k)\cos(\omega_k t)] 
\cos(\omega_k t),
\end{equation}  
\noindent 
where $A(k)$ and $B(k)$ are some functions of $k$ and 
the frequency (\ref{case1-3})
\begin{equation} 
\omega_k=\sqrt{4(4g\sin k)^2+(4\big(2g\cos k-h_f)\big)^2}
\end{equation}
\noindent 
is expressed in terms 
of the final magnetic field $h_f$, the interaction strength $g$ and 
the momenta $k$. Equation (\ref{time0}) suggests that we can consider 
$\Delta P_{\fu(N)}(t)$ in the thermodynamic limit $N\rightarrow \infty$ 
(sums become integrals) as the real contribution of a more general equation, 
namely
\begin{equation} 
\label{time}
\Delta P^{{\rm th.}}_{\fu(N)}(t)\,=\,4\,{\rm Re} 
\bigg( \frac{1}{2\pi} \int_{0}^{\pi} dk 
[A(k)+B(k)\cos(\omega_k t)] \,\,e^{i\,t\,\omega_k} \bigg).
\end{equation} 

Following the steps in \cite{barouch1}, further changes  of variables may lead
to an integration in the complex plane following  a given path, depending on
the values of the poles, which in turn  depend on the relative values of final
and initial magnetic fields and  the strength $g$. Although the solution to
(\ref{time}) is rather complex, it is plausible that the solution for either
$P_{\fu(N)}(t)$ or $M_z(t)$ drawn in Fig.\ref{pur1} represents a special combination
of time-dependent functions,  but with one definite frequency. As a matter of 
fact, one can check from inspection that the overall period of oscillation is
pretty much the same for all times.  In the discussion by Barouch {\it et al.}
\cite{barouch1}, several frequencies are involved. At zero temperature, we
pretty much obtain the same results. See Figs. \ref{pur56} and \ref{pur57}.

\begin{figure}
\begin{center}
\includegraphics[angle=270,width=.65\textwidth]{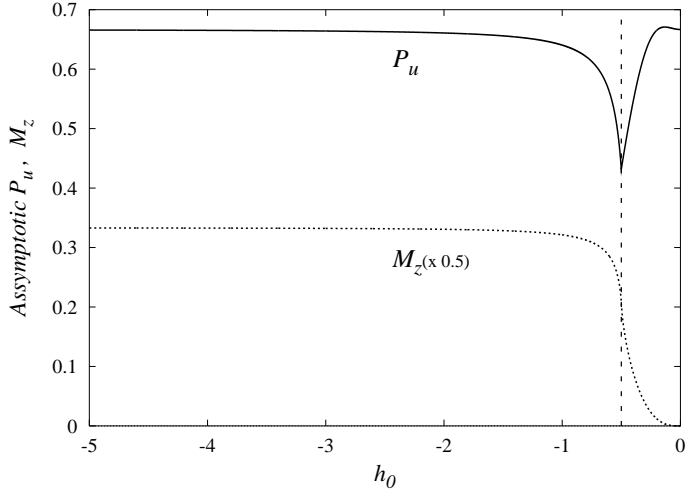}
\caption{Asymptotic (or stationary) values for $P_{\fu(N)}(t=\infty)$ 
and $M_z(t=\infty)$ for the step-function magnetic field $h(t)$ instance, 
plotted as a function of the initial $h_0$, with $h_f=0$, 
$\gamma=\frac{1}{2}$ and $g=-\frac{1}{4}$. $M_z(t=\infty)$ smoothly tends 
to zero, while $P_{\fu(N)}(t=\infty)$ still detects the critical point 
$h_0=2g=-0.5$ (vertical line). See text for details.} 
\label{assimpt}
\end{center}
\end{figure}

It is likely that the best way to discuss the equilibrium state is to evaluate 
the {\it time average} of expressions (\ref{purity1}) and (\ref{magnetization}). 
These curves are shown in Fig.\ref{assimpt}. We plot the asymptotic equilibrium 
($t=\infty$) values for the purity (\ref{purity1}) and 
magnetization (\ref{magnetization}) with $\gamma=\frac{1}{2}$, $g=-\frac{1}{4}$, 
versus the initial magnetic field $h_0$ and zero final magnetic field $h_f$ (step 
function magnetic field). We notice from inspection that 
the equilibrium purity also detects a change of regime at $h_0=2g=-0.5$, where 
the function has no well defined derivative. As we approach zero magnetic 
field, we recover the expected value 
$P_{\fu(N)}(t=\infty)\,\vert_{h_f=0}=1/(1+\gamma)$. Also, regarding 
the magnetization, this is the asymptotic value reached when 
$h_0\rightarrow -\infty$. The magnetization approaches a zero value as 
$h_0$ goes to zero, as it should be. 
The $\gamma=1$ case is analytic (other cases most probably not). The purity versus 
$h_0$ for a zero final magnetic field $h_f$ reads 
\begin{eqnarray}
P_{\fu(N)}(h_0;t=\infty)\,\vert_{\gamma=1}&=&
-{\frac {3}{512}}\,{\frac {{h_0}^{4}}{{g}^{4}}}+{\frac {1}{128}}\,{
\frac {{h_0}^{2}}{{g}^{2}}}+{\frac {17}{32}}+{\frac {3}{512}}\,{h_0}^{5}{g
}^{-5}\cr 
& & {\frac {1}{\sqrt {{\frac {{h_0}^{2}}{{g}^{2}}}-8+16\,{\frac {{g}^{
2}}{{h_0}^{2}}}}}}-1/32\,{h_0}^{3}{g}^{-3}{\frac {1}{\sqrt {{\frac {{h_0}^{2
}}{{g}^{2}}}-8+16\,{\frac {{g}^{2}}{{h_0}^{2}}}}}}+\cr
& & 1/16\,h_0{g}^{-1}{\frac {1}{\sqrt {{\frac {{h_0}^{2}}{{g}^{2}}}-8+16\,
{\frac {{g}^{2}}{{h_0}^{2}}}}}}-\cr
& & 1/2\,g{h_0}^{-1}{\frac {1}{\sqrt {{\frac {{h_0}^{2}}{{g}^{2}}}-8
+16\,{\frac {{g}^{2}}{{h_0}^{2}}}}}}-3/8\,{\frac {{g}^{2}}{{h_0}^{2}}}+\cr 
& & 3/2\,{g}^{3}{h_0}^{-3}{\frac {1}{\sqrt {{\frac {{h_0}^{2}}{{g}^{2}}}-8+16\,
{\frac {{g}^{2}}{{h_0}^{2}}}}}}.
\end{eqnarray}
\noindent Its derivative has a finite value at the critical point 
$\frac{h_0}{2g}$, though it is discontinuous
\begin{equation}
\frac{dP_{\fu(N)}(h_0;t=\infty)}{dh_0}\vert_{h_0/2g\rightarrow 1^{\pm}}=
-\frac{1}{16}\frac{\mp 4 + (\frac{1}{g^2})^{\frac{3}{2}}g^3}
{(\frac{1}{g^2})^{\frac{3}{2}}g^4}.
\end{equation}

\noindent These features transpire from Fig.\ref{assimpt}, computed for the 
$\gamma=\frac{1}{2}$ case. In fact, the magnetization 
$M_{z}(t=\infty)\,\vert_{h_f=0}$ possesses an analytic expression as well, 
but it is not given due to its complex form.

It is clear from the previous formulas that, although we reach an equilibrium 
value for $M_{z}(t=\infty)\,\vert_{h_f=0}$, this is different from zero and 
depends on the initial magnetic field $h_0$ as prescribed by relation 
(\ref{equiM}). This is the conclusion reached by Barouch 
{\it et al.} in \cite{barouch1} at finite temperatures, extended here to the 
case of zero temperature. What is also surprising is the fact that the own 
measure of entanglement $P_{\fu(N)}$ reaches an equilibrium value, which implies
 that not only the magnetization presents a non-ergodic behaviour, but 
{\bf entanglement itself} too.
\newline
\newline
\newline
{\bf Exponential decay}
\newline

In this case, $g(t)=g_0$ and
\begin{equation}
h(t)=\left\{
\matrix{ h_0 & t\le0 \cr h_f + (h_0 - h_f) e^{- \kappa t} & t>0 }
\right. ,
\end{equation}
where $\kappa$ plays the role of a {\it knob} adjusting the speed of the
passage.
Then, Eqs.~\ref{evoleq2} are ($t>0$)
\begin{equation}
\label{case2-1}
\left\{
\matrix{ \dot{x}_1(t)=2 \tilde{\alpha} x_2^R(t) \cr
\dot{x}_2^R(t) = \tilde{\alpha} -2\tilde{\alpha} x_1(t) - \beta(t) x_2^I(t) \cr
\dot{x}_2^I(t) = \beta(t) x_2^R(t) }
\right.
\end{equation}
where $\tilde{\alpha}= 4g_0 \gamma \sin k$, $\beta(t)=\beta_a+\beta_b z$, with
$\beta_a= -4(2g_0 \cos k - h_0)$, $\beta_b=4(h_f -h_0)$, and $z=1-e^{-\kappa
t}$. (Again, the indices $R,I$ denoting the real and imaginary parts of $x_2$,
respectively.) In terms of the variable $z$, Eqs.~\ref{case2-1}  read
($\partial_t= \kappa (1-z) \partial_z$)
\begin{equation}
\label{case2-2}
\left\{
\matrix{ \kappa (1-z) \dot{x}_1(z)=2 \tilde{\alpha} x_2^R(z) \cr
\kappa (1-z)\dot{x}_2^R(z) = \tilde{\alpha} -2\tilde{\alpha} x_1(z) - 
\beta(z) x_2^I(z) \cr
\kappa (1-z) \dot{x}_2^I(z) = \beta(z) x_2^R(z) }
\right.
\end{equation}
where the derivatives are now with respect to the variable $z$. The solution
to Eqs.~\ref{case2-2} can be obtained proposing an ansatz function of the form
$x_2^R =\sum_m a_m z^m$, where the coefficients $a_m$ are related by a 
recurrence relation, obtained when inserting the ansatz in Eqs.~\ref{case2-2} 
and keeping the same order terms in both hands of the Eqs. However, this
recurrence relation is not simple and a straight computation of the
$\fh$-purity and magnetization using numerical methods was performed.

\begin{figure}
\begin{center}
\includegraphics[angle=270,width=.65\textwidth]{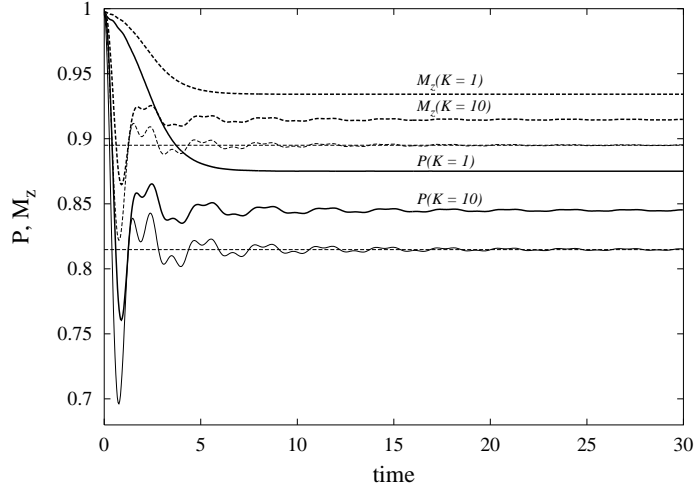}
\caption{Time-dependent $\fu(N)$-purity (solid lines) 
and magnetization (dashed lines) of the evolved state
for the exponential-decay case for different values of
the exponent $\kappa$. The initial and final magnetic fields are $h_0=-5$
and $h_f=-1$, respectively.} 
\label{pur2}
\end{center}
\end{figure}

In Fig.\ref{pur2} we show the time-dependent $\fu(N)$-purity and magnetization 
(Eqs.~\ref{purity1} and \ref{magnetization}, respectively) for $g_0=-1/4$,
anisotropy $\gamma=1$, and for $\kappa=1$, $\kappa=10$, and $\kappa=150$.  For
a {\it slow} passage ($\kappa=1$),  an equilibrium value is steadily and
monotonically reached.  No oscillations appear in this case.  On the contrary,
as we increase $\kappa$, $P_{\fu(N)}(t)$ and $M_z(t)$ present  oscillations
around their limiting values  $P_{\fu(N)}(t=\infty)$, $M_z(t=\infty)$. A  very
fast passage ($\kappa=150$) virtually coincides with a step function 
behaviour in $h(t)$,  and the result is equivalent
to the one showed in Fig.\ref{pur1}. 

As far as non-ergodicity is concerned, we appreciate also in Fig.\ref{pur2} that both 
$P_{\fu(N)}(t)$ and $M_z(t)$ tend to a stationary equilibrium which is non-ergodic, 
independently of the parameter $\kappa$. It is plausible to assume then that 
this is a special feature independent on the specific time evolution employed.
\newline
\newline
\newline
{\bf Hyperbolic magnetic field}
\newline

In this case, $g(t)=g_0$ and the time-dependent magnetic field is
\begin{equation}
h(t) = \left \{ \matrix{ h_0, & t\le0 \cr h_f + \frac{(h_0-h_f)}{1+t}, & t>0} 
\right ..
\end{equation}
The solutions to Eqs.~\ref{evoleq2} are obtained similarly to the
exponential-decay case by proposing an ansatz of the
form $x_2^R(t)=\sum_m a_m t^m$ and obtaining a recurrence relation for the
coefficients $a_m$. As in the exponential-decay case,  we did not obtain the
exact solution since numerical methods allow us to compute the time-dependent
$\fh$-purity  and magnetization directly, efficiently, and with high
accuracy.

\begin{figure}
\begin{center}
\includegraphics[angle=270,width=.65\textwidth]{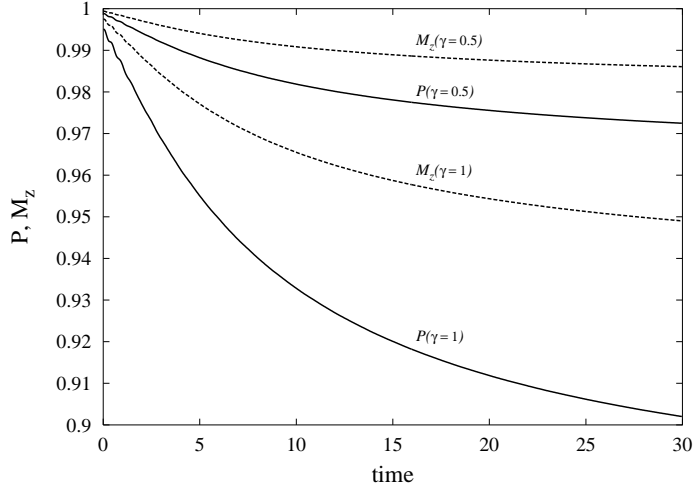}
\caption{Time-dependent $\fu(N)$-purity (solid lines) 
and magnetization (dashed lines) of the evolved state
for the hyperbolic-function case. Similarly to the step-function case, 
the interaction coefficients are $g_0=-1/4, h_0=-5, h_f=-1$.} 
\label{pur3}
\end{center}
\end{figure}

In Fig.\ref{pur3} we show the time-dependent $\fu(N)$-purity and magnetization for
$g_0=-1/4$, $h_0=-5$, and $h_f=-1$ (the same values used in the step function 
case), and with anisotropies $\gamma=0.5, 1$.
It is remarkable from inspection that both the purity and 
magnetization do not present oscillations, and the tendency to reach a stable 
value is slowed down. Again, and with a different time dependence in  
$h$, the non-ergodic features of $P_{\fu(N)}$ and $M_z$ are apparent 
as $t \rightarrow \infty$. We see that low $\gamma$-values is tantamount as 
high values of $P_{\fh}$ and $M_z$.

\subsection{Adiabatic evolution: recovering the static case}

As mentioned above, in \cite{somma1} the $\fu(N)$-purity was computed for 
the static anisotropic $XY$ model in a transverse magnetic field as a
function of the coupling constant $g$ and for $h=1$. Remarkably, the
$\fu(N)$-purity  characterizes the QPT present in this model, changing
drastically at the  critical point $g_c=1/2$. Now, a question arises: Do we
observe a QPT when a time dependence is involved? In other words, does the
speed of passage  through the critical point influence the very existence of a
QPT?  One would expect then to answer these questions by studying the behaviour
of the relative purity $P_{\fu(N)} (t)$.

An interesting issue appears when one considers slow evolutions. The adiabatic 
theorem states that if the time-dependent Hamiltonian $H(t)$ of a system 
evolves slow enough and no level crossing with excited states exists, 
the ground state remains as such with time. In fact, this is the case in the
time-dependent anisotropic $XY$ model in a transverse magnetic field.
Therefore, one should expect to recover the static case for slow evolutions
of the time-dependent parameters in $H(t)$.

For the sake of simplicity and without loss of generality, let us  focus
our attention in the Ising model  in a transverse {\it constant} magnetic 
field; i.e., $\gamma=1$ and $h(t)=1$ in Eq.~\ref{Hamilt1}.  The
time-dependence now is in the coupling parameter $g(t)$. Although
experimentally is more feasible to vary the external magnetic field,  the
conclusions in both cases are the same.  Let us consider then the following
time evolution for $g(t)$:
\begin{equation}
g(t) = \left \{ \matrix{ g_0, & t\le0 \cr g_f + (g_0-g_f)\exp(-\kappa t), & t>0} 
\right .,
\end{equation}
with $g_0=0$ and $g_f=1$.
In this way, the critical point $g_c=1/2$ is crossed at a speed given by
$\kappa$. Again, the time dependent $\fu(N)$-purity can be
numerically computed by inserting $g(t)$ in Eqs.~\ref{evoleq2} and then
inserting the corresponding solutions $x_1^k(t)$ in Eq.~\ref{purity1}.

\begin{figure}
\hspace*{-1.3cm}
\includegraphics[angle=270,width=1.2\textwidth]{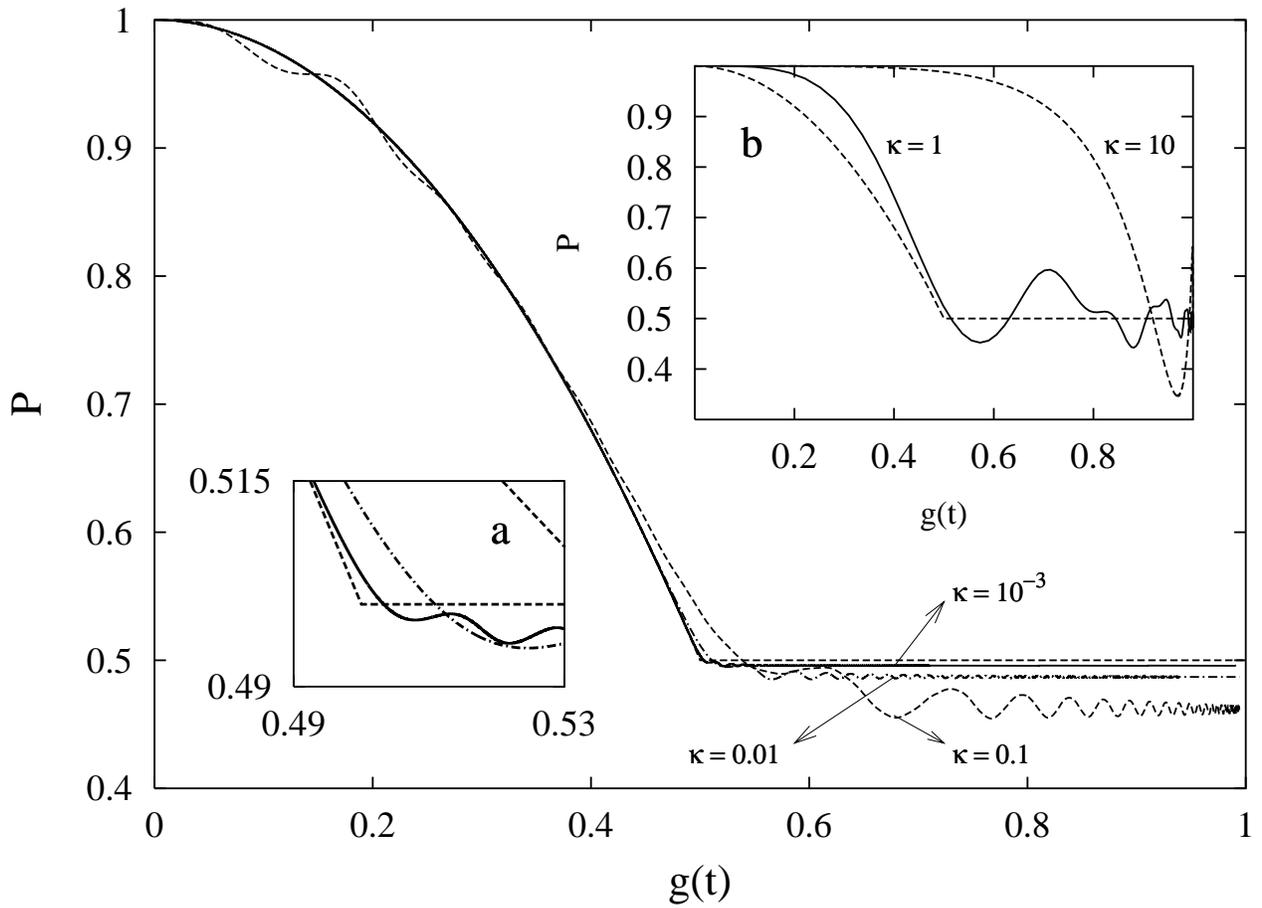}
\caption{Time-dependent $\fu(N)$-purity of the evolved state for an
exponential-like passage through the critical point $g_c=1/2$. For slow
passages $\kappa \rightarrow 0$ the static case (dashed lines) is recovered and
the $\fu(N)$-purity behaves as a disorder parameter. Inset {\it a} shows the 
details around the critical point $g_c=\frac{1}{2}$. As we slow down 
the passage through the critical point, we get closer and closer to 
the static case, given by Eq. (\ref{puresaXY}). For fast passages (inset {\it b}) 
we no longer agree with the static case. Here $h(t)=1$ and 
$\gamma=1$. $\kappa=10^{-3}$ (solid thick line) is very close to the 
static case, which is in full agreement with the adiabatic theorem.} 
\label{pur4}
\end{figure}

In Fig.\ref{pur4} we show $P_{\fu(N)}(t)$ as a function of $g$ and for $\kappa=10^{-3}$,
0.01, 0.1, 1, and $10$. The static case corresponds to the dashed thin line
(see \cite{somma1}).  We observe that the curves corresponding to the slow-speed
cases with $\kappa=10^{-3}$  (solid thick line) and $\kappa=0.01$ (dot dashed
thick line) are very close to the static case, as stated by the adiabatic
theorem. In the inset a) the region around the critical point is enhanced.
The average final value for the $\kappa=10^{-3}$ case is 0.496, remaining
extremely close to $g_c=1/2$. On the other hand, fast-speed  evolutions are
shown in the inset b), where no obvious critical point is detected through 
the $\fu(N)$-purity.

\section{Concluding remarks}

This Chapter has been devoted to the application of the notion of 
Generalized Entanglement (GE) to broken-symmetry QPTs. As we focused on a 
situation where the physically relevant observables form a
Lie algebra, a natural GE measure provided by the relative
purity of a state relative to the algebra has been used in order to 
identify and to characterize these transitions. Therefore the 
measure employed in order to detect a QPT was the purity $P_h$ 
relative to a set of observables $h$, which form a Lie algebra. 
We recall that the concept of $h$-purity encompasses the usual notion of
entanglement if the family of all local observables is distinguished. 
In addition, the possibility to directly apply the GE
notion to arbitrary quantum systems, {\it including indistinguishable 
particles}, becomes apparent when using a fermionic system as a relevant 
case study, as it is the case here.

The study performed in \cite{somma1} dealt with different exact 
lattice models in the application of $P_h$ in a context of QPT. In this Chapter 
we have first focused on a new lattice model which, 
to our knowledge, had not been solved analytically to date: the 
anisotropic $XY$ model with bond alternation in a transverse magnetic 
field. The analytical detailed resolution of this system is done 
for pedagogical reasons, because in the end we provide a formula 
for the evaluation of the purity $P_h$ relative to the $h=u(N)$ algebra 
of observables. The calculation of $P_h$ is performed numerically, 
but one must not rule out the possibility that it might admit 
an analytical expression in the thermodynamic limit $N\rightarrow \infty$, 
where $N$ is the number of lattices\footnote{In these systems, the 
usual notion of entanglement cannot be straightforwardly applied.}. 
However the concomitant phase diagram of this bond-alternating model is obtained 
in analytical 
fashion, and one can easily recover the usual anisotropic $XY$ model with a transverse 
magnetic field studied in \cite{somma1}. The derivative of $P_h$ 
with respect to some parameter in the phase diagram ``signals" the 
presence of a QPT, in complete agreement with analytic results.

In a second stage, we abandon the static description of QPT and GE through 
the study of the purity $P_h$ relative to the $h=u(N)$ and we let the time 
run. The goal was to study the dynamic entanglement features of the usual 
anisotropic $XY$ model when we let the coupling constant $g$ and the external, 
transverse magnetic field depend on time ($g(t),\,h(t)$). First, 
we found the general time dependent differential equations for the coefficients 
describing the ground state $|\Psi(t)\rangle$ of the system, for 
arbitrary $g(t),\,h(t)$. Then we calculated the purity $P_{u(N)}(t)$ and the 
magnetization $M_z(t)$ setting $g(t)=g_0$ and considering different 
cases for the time dependent magnetic field $h(t)$, namely, i) step function, 
ii) exponential and iii) hyperbolic. In the step function case, the equations 
for the ground state of the system can be found analytically, but the 
time dependent $P_{u(N)}(t)$ and $M_z(t)$ ought to be calculated numerically. 
However, for certain values of the parameters of the system Hamiltonian, 
the equilibrium values $P_{u(N)}(t\rightarrow \infty)$ 
and $M_z(t\rightarrow \infty)$ can be expressed in analytical fashion only 
in the step-function magnetic field $h(t)$ instance. Yet, at zero temperature, 
the magnetization reaches a final non-zero value for all $h(t)$ cases considered, 
which means that $M_z(t)$ is non-ergodic: it strongly depends on the initial 
values. This fact was already known to Barouch {\it et al.} in 
\cite{barouch1,barouch2} back in the 70s for the same system at finite 
temperatures. Here we show that it is indeed the same case for $T=0$.

But the magnetization is not the only non-ergodic quantity here. Entanglement itself 
($P_{u(N)}(t)$) pretty much behaves as the magnetization $M_z(t)$, that is, 
it is also non-ergodic. This fact is certainly relevant, for entanglement 
--understood in the framework of the theory of GE-- do behaves as a property of 
the system, similarly to the magnetization {\it on equal footing}.

Finally, we wanted to study the counterpart of previous sudden changes in the 
parameters $g(t),\,h(t)$. An interesting issue appeared when we considered slow 
evolutions. The adiabatic theorem states that if the time-dependent Hamiltonian 
$H(t)$ of a system 
evolves slow enough and no level crossing with exited states exists, 
the ground state remains as such with time, and in fact this was the case in the
time-dependent anisotropic $XY$ model in a transverse magnetic field.
Therefore, one expected to recover the static case for slow evolutions
of the time-dependent parameters in the Hamiltonian $H(t)$. By varying the coupling 
constant $g(t)$ exponentially, with $h(t)=1$ we compared the different results for the 
purity $P_{u(N)}(t)$ vs. $g(t)$. For different speed passages trough 
the critical point $g_c=\frac{1}{2}$, we show in the ``frozen" plot 
$P_{u(N)}(t)$ vs. $g(t)$ how it approached the static result for extremely 
slow passages, agreeing with the adiabatic theorem.

Although not discussed in this thesis it is worth commenting on the following important issue. 
As pointed out in \cite{somma1}, in general terms to determine in a systematic way the minimal subset
of observables $h$ whose purity is able to signal and characterize
the QPT, and thereby providing the relevant correlations,
requires an elaborate analysis. Going one step further, it is perhaps 
even more interesting to study the open question of finding the minimal number of
GE measures, possibly including measures of GE relative to
different observable sets, needed to unambiguously characterize a QPT, 
in terms of critical exponents, university classes, etc. However, 
this may constitute a subject of future study. A motivation to study the model 
Hamiltonian of Eq. (\ref{Hamilt4}) was precisely the investigation of a problem 
with various order parameters that would help us figuring out the systematic 
choice of minimal algebras.\newline

Note added.— After completion of this work, it was 
brought to our attention the fact that non-ergodicity 
was discovered also in \cite{losqueseavanzaron} for finite 
temperatures in the bipartite entanglement (the {\it concurrence} to be more precise) 
present between two sites in the $XY$ model.

%% file: conclusionsTesi.tex
\part{Conclusions}

Throughout the present Thesis we have emphasized the role played by information or 
entropic measures in the characterization of quantum entangled states. The correct 
use of MaxEnt procedures, subject to certain requirements, turned out to be a powerful 
and safe inference tool (safe in the sense of not ``faking" the entanglement present 
in a given state $\rho$). Thus Jaynes' principle is universally valid as long as 
we identify the correct input information. Also, the classic $q$-entropic inequalities 
offer a necessary criterion for separability which has interesting echoes in 
the volume of states occupied by those which do not violate them, as well as in 
their connection --chain of implications-- with other existing criteria. We have also seen 
that MEMS states play a distinguished role with regards to the violation of the aforementioned 
classical entropic inequalities. 
Besides, we have seen that it is possible to correlate total entropies (its mean value) 
of states of two qubits with their entanglement. The relationship between entanglement 
and mixedness has been established, and we have reviewed the pros and cons of different 
measures in the space of mixed two-qubit states. In this vein, the structure of two-qubit 
mixed states in connection with mixedness and entanglement appears slightly different 
when a different generation of states is performed. In clear 
connection with quantum computation, the entanglement distribution of quantum gates and their 
entangling power acting on pure or mixed two-qubit states has been investigated as well.\newline

Regarding the characterization of entanglement, we have seen that it constitutes an essential 
ingredient in certain quantum algorithms such as Grover's and, among other properties, 
we explicitly have shown that entanglement-assisted time evolution of states towards orthogonal ones 
does not occur in the case of indistinguishable particles, at least for fermions. 
Entanglement, in the particular form the so called {\it purity}, can also describe condensed 
matter systems. Particularly, we have seen that there exists a connection between 
quantum phase transitions (QPTs) and entanglement in the $XY$ bond alternating anisotropic model 
with a transverse magnetic field. By studying the dynamical evolution of the 
$XY$ anisotropic model, we have discovered interesting non-ergodic properties of the 
purity measure, which reinforces our perception of entanglement as a property that 
characterizes the quantum system, specially in broken-symmetry QPTs, signalling the position 
of the existing critical points.\newline

Summing up, the most important results of this Thesis appear as follows:

\begin{itemize}

\item Jaynes' principle does not ``fake" entanglement. We have exhaustively 
investigated Horodecki's ``fake" inferred entanglement problem, related with the use 
of the maximum entropy principle,
with reference to distinct inference schemes, and advanced a 
new one, reminiscent of Horodecki's, for dealing with general 
observables.

\item Inclusion relations between several separability have been checked numerically. 
We have explored the application of different separability criteria 
by recourse to an exhaustive Monte Carlo exploration involving the pertinent 
state-space of pure and mixed states. The corresponding  chain of implications 
of different criteria has been in such a way numerically elucidated. Besides, 
we have quantified for a bipartite system of arbitrary dimension, the proportion 
of states $\rho$ that can be distilled according to a definite criterion. 

\item We have extensively explored all possible connections 
of the so called $q$-entropic information measures and their connection 
with entanglement. Secondly, the connection with entanglement and mixedness 
has been studied too. Also, we focused our attention on the interesting properties 
that link a particular class of states, the so called Maximally Entangled Mixed 
States (MEMS), with the violation of the usual entropic inequalities.

\item The relationship between entanglement and purity of states of two-qubit 
systems has been revisited in the light of the $q$-entropies as measures of 
the degree of mixture. Probability distributions of finding quantum states of 
two-qubits with a given degree of mixture have been analytically found for $q=2$ 
and $q \to\infty$. We claim that $\lambda_m$ itself constitutes a legitimate measure
of mixture. 

\item The space $\mathcal{S}$ of mixed states ot two-qubits is not very sensitive, 
with regards to the concomitant entanglement-mixedness properties of states, 
to the measure used to generate them.

\item We have focused our attention on the action of quantum gates as applied to 
multipartite quantum systems and presented the results of a systematic 
numerical survey. We also studied numerically some features of the probabilities 
of obtaining different values of $\Delta E$. Quantum gates are more
efficient, as entanglers, when acting upon states with small
initial entanglement, specially in the case of pure states.

\item There is a clear correlation between the
 amount of entanglement and the speed of 
 quantum evolution for systems of two-qubits 
 and systems of two identical bosons. On 
 the contrary, such a clear correlation is 
 lacking in the case of systems of 
 identical fermions.

 \item There exists a clear connection between quantum phase transitions and entanglement, 
 as expressed by the so called {\it purity} measure, a generalized entanglement measure. 
 We have obtained --to our knowledge for the first time-- 
 the phase diagram of a bond alternating $XY$ model and checked that the purity, or its derivative 
 to be more precise, detects too the critical points in a richer phase diagram. Finally, the dynamic 
 evolution of the $XY$ anisotropic model in a transverse magnetic field reveals that entanglement 
 itself, applied in a condensed  matter scenario, can present non-ergodic features.

\end{itemize}

\pagebreak

{\bf OPEN QUESTIONS AND FUTURE WORK}
\newline							    
\newline

The field of quantum information theory and quantum computation is 
growing extremely fast, so quickly that it is difficult to compile 
all the relevant results that constantly appear in the vast literature 
of this newborn branch of theoretical and experimental physics. 
It is likely that no other field of physics had experienced before such a rapid 
theoretical and experimental development as it is happening with 
quantum information theory. Quantum communication, quantum teleportation and 
quantum computation are the main focus of experimental efforts. The promise 
of a secure communication and ultrafast computations is a subject 
that many governments\footnote{Just wondering if we should include Spain..} 
around the world have taken seriously into account. Therefore, lots of 
related difficulties naturally arise both in the theoretical and experimental 
implementations of these items, what makes the research in this field more 
exciting.

However, in the meantime, there have appeared skeptical voices of relevant 
scientists like Gerard 't Hooft \cite{Hooft} that conceive the impossibility to overcome the 
technical difficulties that appear. These criticisms are in the line that no matter 
how better we improve our technology, there will always appear some quantum 
limit that cannot be avoided. These limits translate into the fact that 
{\it quantum} information, as we have studied here, will end up being ultimate 
{\it classical} 
information (the usual bits 0 and 1). Nevertheless, proof-of-principle NMR and 
ion-trap quantum computing has been observed. So the basis for quantum computation 
exists. However, serious drawbacks are present when the 
experimentalist wants to isolate the system of qubits he/she is interested in 
from the rest of the universe. Decoherence times ought to be great enough so that 
a quantum gate operation can be performed. Tailoring interactions between subsystems, which is 
tantamount as engineering entanglement, constitutes a difficult issue. That is why 
new proposals for quantum computing appear from time to time. Physicists seek out 
potential proposals (physical systems) for quantum bits, such as the combined system of 
the spin of a nucleous and the one of the electrons surrounding it, that fulfils 
the requirements of DiVincenzo's criteria 
(see Chapter 2). Such systems do exist, but we are still unable to handle them in 
the way which is required. Difficulties are of technical nature. 
Another kind of criticism is the one against the object of studying algorithms and protocols 
in quantum computation and quantum communication, respectively. One could argue what is 
the purpose of studying processes that should still wait to be observed and controlled. 
This remark is, in fact, 
a confirmation that theoretical progress grows at a different rate as compared 
to the experimental stuff. It confirms, in turn, the huge interest that is growing 
in the physics community about these extraordinary quantum features. 

But again, would not it be easier to still consider classical information and build classical 
computers with one electron transistors, instead of complex, strange proposals 
that may not in the end work? Perhaps this is a short-term solution. 
It would make no use of entanglement whatsoever. Entanglement, what for? 
Well, it makes no reference to the inherent randomness of quantum mechanics. This fact means 
that it does not support any device that provides one with absolutely secure communication, as 
quantum cryptography is experimentally giving nowadays. Quantum teleportation? No pink!

Quantum computers are the definitive, once-for-all, long standing solutions to simulate 
quantum physics. Would not it be marvellous that one could map a quantum physical 
system like a cuprate superconductor into {\it another} quantum physical one like 
a quantum computer? If this were possible, one could simulate the intriguing 
features of non-understood phenomena so that we would get a better insight: to calculate the 
theoretical mass of the proton, to make precise meteorological predictions nearly 
in real time, and the list continues.. 

Finally, here come the one-million-dollar\footnote{I would personally prefer euros at the 
moment} questions: what if a quantum computer can not be built? What if 
we cannot control undesired interactions or instabilities? What if none of 
the proposals for quantum computing scales the number of qubits up to no more than 
a certain small number? What if we end up with a system that, yes, can perform better that a 
supercomputer, but not as much as we expected? Have we wasted precious time for 
nothing? 
Well, it is a common belief, and I shared that impression, that to 
build a quantum computer is not the ultimate goal. Never in the past a pure 
theoretical concept such as non-locality --crystallized into entanglement-- 
gathered different disciplines altogether (computer science, information 
theory, quantum mechanics and its foundations..) around this strange quantum 
correlation. As pointed out by Schr\"odinger \cite{Schro}, quantum entanglement 
is not a feature of quantum mechanics, but rather {\bf the} characteristic 
feature that makes quantum mechanics different from classical physics. As a 
consequence, it deserves to be studied in all possible ways. It constitutes, 
in the modern context of science, a clear example that fundamental or 
pure research need not has to find an application. But if it eventually does, 
the results become superb.
\newline
\newline

Now then, let us return from the future.. In this final Chapter we will not be 
dealing with the challenges that appear in 
the previous fields, mainly because we have not studied them in any way. Instead, 
we expose those problems that appear in basic research that we have dealt with 
in this Thesis. It's about a better understanding of the physical nature of 
entanglement, its detection and characterization in different contexts.
\newline

{\bf Detection of entanglement}
\newline
\newline

In the context of the orthodox view of unentangled states of $N$ parties,

\begin{equation} \label{sepaNN}
\rho \,=\, \sum_{k} p_k \, \rho^{(k)}_1 \otimes ... \otimes \rho^{(k)}_N
\end{equation}

\noindent with $0\le p_k \le 1$ and $\sum_k p_k =1$, there is no definitive 
criterion yet which can asserts if a given state $\rho$ can be written in the form 
(\ref{sepaNN}) or not, even in the general case of bipartite systems of arbitrary 
dimensions. The application in this case of the theory of positive maps advanced by the 
Horodecki family \cite{AB01}, turns out to be a hard problem to solve, and possibly 
may not have a solution. Also entanglement witnesses, which have proved to be 
quite successful, find it hard to go up to high dimensions. Indeed, much effort is 
done by many authors in order to find a solution to the {\it separability problem} 
as we present it here, but a conclusive 
answer still remains, at least for mixed states. In practice, what has been done 
in order to tackle the problem is to induce {\it partitions} in the system, that is, 
to consider bipartite entanglement --which is well understood-- between clusters 
of particles. Certainly this may not be a definite solution, because the problem 
of characterizing genuine multipartite entanglement still resists, in spite of many 
efforts \cite{ee1,ee2,ee3}. 

Motivated by these facts, one may wonder if the entropic criteria could shed some light 
on the problem. These simple, information-theoretical based criteria is endowed with 
the high physical and intuitive notion that the entropy of the total system, as described 
by the density matrix $\rho_{1,..,N}$, has to be greater that any of its subsystems for 
{\it classical} systems, but entangled states are so particular that this may happen or not. 
Therefore, we consider a subject of future study 
the derivation of general conditions for positivity using the entropic inequalities. This 
procedure should involve the characterization of the positivity of $2^N-2$ inequalities, 
with $N$ being the number of parties.
\newline

{\bf Characterization of entanglement}
\newline
\newline

As we have seen, the parameterization of the two-qubit space ${\cal S}$ of all pure 
and mixed states (see also the Appendix) has been made upon the assumption that states 
are distributed according to a definite measure. In the case of pure states, the 
only possible measure happens to be the Haar measure for unitary matrices. In this case, 
many average 
quantities such as the mean degree of mixture $R= 1/Tr \, (\rho^2)$, or even the 
distribution of concurrence squared $C^2$ ($P(C^2)$) can be found in analytical fashion. 
But when it comes to mixed 
states, the situation is more involved. The computation of the aforementioned (average) 
quantities characteristic of the space ${\cal S}$ is three-fold complex: firstly, i) 
there is no unique procedure to generate those mixed states, because one can choose 
different distributions for the simplex $\Delta$ describing the eigenvalues of the 
state $\rho$; secondly, ii) the final chosen distribution for $\Delta$ may or may not 
induce a real metric in the 15-dimensional space of two-qubits ${\cal S}$; and finally, 
iii) in either cases the computation of several properties has to be performed in 
numerical fashion.

If we were able to describe analytically several properties in this 
15-dimensional space 
of two-qubits ${\cal S}$, which is the simplest one exhibiting the feature of 
entanglement we could gain, on the one hand, more physical insight into features 
such as the geometrical meaning of positive partial transposition\footnote{The partial 
transpose condition could be used to find the set of separable and entangled
states by finding the regions for which the density matrix is positive semidefinite.} of 
$\rho$, and on the other hand, simplicity in the numerical generation of states. 
In the latter case, 
a clear example of this situation is encountered whenever we want to calculate distance 
measures as in the case of the minimum distance\footnote{There exist algorithms that minimize 
this quantity. In our case we performed (see Chapter 10) a stimulated annealing minimization 
taking advantage of the fact that the space of unentangled states ${\cal S}_{sep}$ is convex.} 
to a completely separable matrix
as a measure of separability. This is indeed the case for the so called robustness of 
entanglement or the relative entropy of entanglement introduced in Chapter 5, which do not 
possess an analytical formula to date. 

In order to overcome this facts for two-qubit systems, an interesting parameterization of 
bipartite systems based on SU(4) Euler angles has been introduced recently by 
T. Tilma {\it et al.} 
in \cite{Tilma}. Such a parameterization should be very useful for many calculations, 
especially numerical, concerning entanglement. This
parameterization would also allow for an in-depth analysis of the convex sets, subsets, 
and overall set boundaries of separable and entangled two-qubit systems. This simplification 
in the calculations, which could be done analytically, is certainly of interest for us and 
will be explored in the future.
\newline

The measure $d$ described in Chapter 12 in order to analyse genuine multipartite 
pure state entanglement 

\be \label{dmeasure}
0 \, \le \, d\equiv C^{2}_{1(2..N)}-\sum_{i=2}^{N} C^{2}_{1i} \, \le \, 1,
\ee

$N$ being the number of qubits in a pure state 
$\rho=|\Psi\rangle_{1..N}\langle \Psi|$, and $C^{2}_{xy}$ stands for the 
concurrence squared between qubits $x,y$, opens a new window to the study of the 
evolution of entanglement during the application of a given quantum algorithm. 

Along this line, it is of interest also to study the time evolution of pure states 
of $N$ parties, in similar fashion as done in Chapter 12 for two-qubit systems. 
Several efforts have been already done in systems of two-qutrits and three 
qubits\footnote{J. Batle {\it et al.} (2005). Unpublished.}. Although 
some results can be obtained analytically, it is clear than a direct relation between 
measure (\ref{dmeasure}) and the time evolution towards an orthogonal state will not be of 
trivial nature at all. Besides, one has to bear in mind that the results reported in the 
literature \cite{GLM03a,GLM03b} deal with special forms of states, which allow an 
analytical study. The relation between global properties of the set of pure and mixed 
states of $N$ parties, on the one hand, and the role of entanglement in the concomitant 
time evolution, one the other hand, will certainly deserve further exploration.
\newline

It is implicit from (\ref{sepaN}) that the $N$ parties of this composite system, whose state 
$\rho$ belongs to the Hilbert space 
${\cal H}={\cal H}_1 \otimes ... \otimes {\cal H}_N$, are {\it distinguishable} 
or, on the contrary, are indeed identical but can be addressed individually 
because the individual wavefunctions do not overlap. This is certainly the case for quantum 
communication, teleportation, and quantum cryptography, because communication only 
occurs between two parties\footnote{This situation will change in the future. There have been 
recent experiments reporting entanglement between four photons 
(group of Anton Zeilinger in Vienna) and even five photons (Jian-Wei Pan in Hefei, China).
Novel crystals and better laser systems will boost the number of entangled photons 
further and allow such systems be used for multiparty communication.} 
that are far apart from each other. 
Therefore we do not have to worry about the statistics of 
the particles involved (either bosons or fermions). In Chapter 14 we saw that entanglement 
plays a decisive role in a faster evolution of a pure state to an orthogonal one. In view of 
this fact, one is naturally led to encounter an experiment which, by indirect means, could 
detect the presence of entanglement if the time evolution of a given state to its 
orthogonal one is speeded-up. In the same spirit, we encountered in Chapter 14 that 
this pattern did not work for fermions. We did not find a direct correlation whatsoever 
between entanglement\footnote{We used an extension of the usual concurrence to 
identical particles --fermions in this case-- alone. No other measure was employed.} 
and time evolution. We will look in the future for the solution to this puzzle.
\newline

%% file: Appendices.tex
\chapter*{Appendices}
\addcontentsline{toc}{chapter}{Appendices}

\section{A. Landmarks in classical and quantum information theory}

 Here is a brief account of the most significant discoveries occurred in the fields 
of Information Theory, Computer Science and Quantum Information Theory. The list 
of the modern achievements in quantum information (a field growing rapidly and 
in constant change) presented here is far for complete. Thus, the following 
references\footnote{There is no need to provide a full bibliographic record. We provide 
these notes so as to serve as a guide of the historical evolution of the basic 
grounds of Quantum Information Theory.} could serve as guideline where the 
correlations between several disciplines 
become apparent with time. Those facts specially related to 
entanglement-separability appear \underbar{underlined}.

\begin{itemize}

\item {\bf 1870} J. C. Maxwell: public appearance of Maxwell's demon in the 
{\it Theory of Heat}. Probably the resolution of a paradox in physics had never 
been so fruitful before as Maxwell's demon

\vskip 0.5cm
\noindent {\it ``Now let us suppose that such a vessel is divided into 
two portions, A and B, by a division in which there is a small hole, and that 
a being... opens and closes this hole, so as to allow only swifter molecules 
to pass from A to B... He will thus, without expenditure of work, raise the 
temperature of B and lower that of A, in contradiction to the second law 
of thermodynamics"}
\vskip 0.5cm

\item {\bf 1929} L. Szilard: {\it \"Uber die Entropieverminderung in einem 
thermodynamischen System bei Eingriffen intelligenter Wesen}. 
Seminal paper relating the Maxwell paradox to entropy and information. Besides, 
the point stressed by Szilard was that information (not yet defined as such) 
is linked to a physical representation, pioneering future disciplines 
describing the intimate connection between physics 
(classical and quantum), information and computation

\vskip 0.5cm
\noindent {\it ``If we do not wish to admit that the Second Law has been 
violated, we must conclude that the intervention which establishes the 
coupling (the measuring instrument and the thermodynamic system) must be 
accompained by a production of entropy"}
\vskip 0.5cm

\item {\bf 1932} J. von Neumann: {\it Mathematische Grundlagen der 
Quantenmechanik}. 
Seminal work in quantum mechanics, revisits his introduction of the 
thermodynamical entropy $S(\rho)$ through the formalism of density matrices. 
Introduction of the measurement theory

\item{\underbar{\bf 1935}} A. Einstein, B. Podolsky and N. Rosen: 
{\it Can Quantum-Mechanical Description of Physical Reality Be 
Considered Complete?} Keystone paper in the foundations of quantum mechanics. 
Based in 
position-momentum arguments of two distant particles, it 
raised the question of whether Nature can be regarded as locally realistic as  
opposed to non-local or ``incomplete". The modern version with spins is due to 
D. Bohm (1951) 

\item{\underbar{\bf 1935}} E. Schr\"odinger: {\it Die gegenw\"artige Situation in 
der Quantenmechanik}. Seminal paper where {\it Verschr\"ankung} 
(German word for ``folding arms") or 
entanglement is introduced as ``the characteristic trait of quantum mechanics" 

\vskip 0.5cm
\noindent {\it ``When two systems, of which we know the states by their respective 
representatives, enter into temporary physical interaction due to known forces 
between them, and when after a time of mutual influence the systems separate 
again, then they can no longer be described in the same way as before... By 
the interaction, the two representatives (or $\Psi$-functions) have become 
entangled"}
\vskip 0.5cm

\item {\bf 1936} A. M. Turing, {\it On Computable Numbers, with an Application 
to the Entsheidungsproblem}. Keystone paper in computation science. He poses 
the basic operating principles (further developed by von Neumann) of the 
ordinary computers (birth of the Turing machine). Together with A. Church they 
formulate what is known as 
the ``Church-Turing hypothesis": every physically reasonable model of 
computation can be efficiently simulated on a universal Turing machine

\item {\bf 1949} C. E. Shannon and W. Weaver: {\it The Mathematical Theory of 
Communication}. Seminal paper in information theory 

\item {\bf 1957} E. T. Jaynes: {\it Information Theory and Statistical 
Mechanics}. 
The principle of maximum (informational) entropy is advanced as the basis 
of statistical mechanics 

\vskip 0.5cm
\noindent {\it ``Information theory provides a constructive criterion... which is 
called the maximum-entropy estimate... If one considers statistical mechanics as 
a form of statistical inference rather than a physical theory... the usual rules 
are justified independently of any physical argument, and in particular 
independently of experimental verification"}
\vskip 0.5cm

\item {\bf 1959} R. P. Feynman: {\it There is Plenty of Room at the Bottom}. 
Briefly exposes the fact that there is nothing in the physical laws that prevents 
from building computer elements enormously smaller than they are (or were). 
Constitutes the first wink to the physical limits of computation 

\item {\bf 1961} R. Landauer: {\it Irreversibility and Heat Generation in 
the Computing Process}. R. Landauer formulated his celebrated principle, 
stating that in order to erase one bit of information it is necessary to 
dissipate an amount of energy equal to $\ln 2 \,k_B T$, where $k_B$ is Boltzmann's 
constant, and $T$ is the temperature at which the computing device is working.

\item {\bf 1964} J. S. Bell: {\it On the Einstein-Podolsky-Rosen paradox}. 
Milestone paper in quantum mechanics. J. Bell proposes several inequalities 
in order to test whether Nature admits local realism or follows the tenets 
of quantum mechanics. The most famous inequality, the Clauser-Horne-
Shimony-Holt (CHSH) inequality follows from Bell's work. In the forthcoming 
decades experimental setups will support Bell's view

\item {\bf 1973, 1982} C. H. Bennett. Inspired by R. Landauer, Bennett 
demostrates that reversible, logically and thermodynamically (avoiding erasure), 
computation (classical) is possible. Reversibility in computation is first 
considered. It naturally arises 
when quantum gates are considered (unitary operations). Later on Maxwell's 
demon is exorcised using his memory erasure (the demon must store the 
information obtained)

\item {\bf 1982} R. P. Feynman: {\it Simulating Physics with Computers}. 
Quantum mechanical phenomena are extremely difficult (if not impossible) to
simulate on a digital (or classical) computer

\item {\bf 1982} W. K. Wootters and W. H. Zurek: {A single quantum cannot be 
cloned}. Simple but extremely important demonstration of 
the Non-Cloning Theorem. Its consequences range the whole quantum information 
field

\item {\bf 1984} C. H. Bennett and G. Brassard: {\it Quantum cryptography: 
Public key distribution and coin tossing}. Seminal paper, one of the first 
comprehensive protocols for quantum cryptography. Work followed by a series 
of fascinating experiments demonstrating quantum cryptography 
(entanglement is essential) at very long distances (Gisin group in Geneva)

\item {\bf 1985} D. Deutsch: {\it Quantum theory, the Church-Turing principle 
and the universal quantum computer}. Milestone paper, it constitutes the first 
formal bridge between quantum mechanics and computation science. Machines 
rely on characteristically 
quantum phenomena to perform computations, or in other words, the abstract 
mathematical idea of logical action during computation depends on the 
physical support. Recall, in a similar analogy, that geometry turned out to be 
falsable only in a physical context (General Relativity Theory) 

\item {\underbar{\bf 1989}} R. F. Werner: {\it Quantum states with 
Einstein-Podolsky-Rosen correlations admitting a hidden-variable model}. Entangled 
states are strange: they can be entangled and satisfy general 
Bell inequalities

\item {\bf 1992} C. H. Bennett and S. J. Wiesner: {\it Communication via one 
and two-particle operators on Einstein-Podolsky-Rosen states}. Paper 
where the issue of quantum dense coding is discovered and explained

\item {\bf 1993} C. H. Bennett, G. Brassard, C. Cr\'{e}peau, R. Jozsa, A. Peres 
and W. K. Wootters: {\it Teleporting an unknown quantum state via dual 
classical and EPR channels}. Striking paper where quantum teleportation is 
discovered and explained. Its experimental realization was carried out in 1996 
by the group of Anton Zeilenger.

\item {\bf 1994} P. W. Shor: {\it Algorithms for quantum computation, 
discrete logarithms and factorizing}. The field of quantum computing
blossomed into a new era. It was shown that on a (hypothetical) quantum 
computer there are polynomial time algorithms for factorization and 
discrete logarithms, impossible to achieve on a universal (classical) 
Turing machine. This 
fact implied that all the encryptation codes (based so far on the 
difficulty of factoring large integers) could be broken easily with the 
advent of a quantum computer. Factorization is reduced to period-finding

\item {\bf 1995} D. Deutsch, A. Barenco and A. Ekert: {\it Universality 
in quantum computation}. Fundamental paper where quantum computation is shown 
to be univeral on almost every two-qubits gate. Generalizes the result by 
DiVicenzo (1995) that the CNOT plus single qubit gates suffices for universal 
quantum computation

\item {\bf 1995} J. I. Cirac and  P. Zoller: {\it Quantum Computations with 
Cold-Trapped Ions}. Theoretical proposal for reliable quantum computation 
based on ions confined in a linear trap. Paper of enormous impact in the 
experimental realization and study of ion-trapped quantum computers 

\item {\underbar{\bf 1996}} A. Peres: {\it Separability Criterion for 
Density Matrices}. Formulation of the necessary condition for separability 
for bipartite general 
states $\rho$ based on the Positive Partial Transposition (PPT) of $\rho$. 
Peres also gives some examples showing that the PPT criterion is more 
restrictive than Bell's inequality, or than the $q$-entropic inequalities 
provided the same year 
by the Horodecki family. The same authors showed by the same dates that 
PPT was a necessary and sufficient separability criterion for $2 \times 2$ and 
$2 \times 3$ systems

\item {\bf 1997} D. Bouwmeester, J. W. Pan, K. Mattel, M. Eibl, H. Weinfurter 
and A. Zeilinger: {\it Experimental quantum teleportation}. The state of 
polarization of a photon is experimentally teleported. Milestone paper in 
all fields of physics, that soon will be followed by other groups in the 
world, using several different techniques (optical, NMR and ``squeezed" states 
of light) 

\item {\bf 1997} L. K. Grover: {\it Quantum mechanics helps in searching 
for a needle in a haystack}. Development of the quantum search algorithm. 
The improvement versus the classical case is of order $O(\sqrt{N})$. The 
original work dates back to 1996 

\item {\bf 1998} D. Loss and D. P. DiVincenzo: {\it Quantum computation with 
quantum dots}. Theoretical proposal for quantum computing using spin-based 
coupled quantum dots

\item {\bf 2004} M. Riebe, H. H\"affner, C. F. Roos,  W. H\"ansel, J. Benhelm, 
G. P. T. Lancaster, T. W. K\"orber, C. Becher, F. Schmidt-Kaler, 
D. F. V. James and R. Blatt: {\it Deterministic quantum 
teleportation with atoms}. Striking paper reporting for the first time 
the teleportation of the quantum state of a trapped calcium ion to 
another calcium ion. Constitutes the first time that teleportation has 
been achieved with atomic particles, as opposed to photons 

\end{itemize}

\section{B. The Haar measure and the concomitant generation of 
arbitrary states. Ensembles of random matrices} 

The applications that have appeared so far in quantum information theory, 
in the form of dense coding, teleportation, quantum cryptography and specially 
in algorithms for quantum computing (quantum error correction codes for instance), 
deal with finite numbers of qubits. A quantum gate which acts upon these qubits 
or even the evolution of 
that system is represented by a unitary matrix $U(N)$, with $N=2^n$ being the 
dimension of the associated Hilbert space ${\cal H}_N$. The state $\rho$ describing 
a system of $n$ qubits is given by a hermitian, positive-semidefinite ($N \times N$) 
matrix, with unit trace. In view of these facts, it is natural 
to think that an interest has appeared in the {\it quantification} of 
certain properties of these systems, most of the 
times in the form of the characterization of a certain state $\rho$, 
described by $N \times N$ matrices of finite size. Natural applications arise 
when one tries to simulate certain processes through random matrices, 
whose probability distribution ought to be described accordingly. In the work 
described in previous chapters, it was of great interest to study, for instance, 
volumes occupied by states $\rho$ complying with a given property.   
\newline
\newline
{\bf Pure states}
\newline

This enterprise requires a quantitative measure $\mu$ on a given set of 
matrices. Once we have chosen such a measure, averages over the aforementioned 
set will provide expectation values of the quantities under study. 
In the space of pure states, with $|\Psi\rangle \in {\cal H}_N$, 
there is a natural candidate measure, induced by 
the {\bf Haar measure} on the group ${\cal U}(N)$ of unitary matrices. 
In mathematical analysis, the Haar\footnote{This measure is named after 
Alfr\'{e}d Haar, a Hungarian 
mathematician who introduced this measure in 1933.} measure \cite{Haar33} 
is known to assign an ``invariant 
volume" to what is known as subsets of locally compact topological groups. 
In origin, the main objective was to construct a measure invariant under the 
action of a topological group \cite{Mehta90}. Here we present the formal 
definition \cite{Conway90}: given a locally compact topological group 
$G$ (multiplication is the group operation), consider a $\sigma$-algebra $Y$ 
generated by all compact subsets of $G$. 
If $a$ is an element of $G$ and $S$ is a set in $Y$, then the set 
$aS = $ $\{$ $as : s \in S$ $\}$ also belongs to $Y$. A measure $\mu$ on $Y$ will be 
letf-invariant if $\mu(aS)=\mu(S)$ for all $a$ and $S$. Such an invariant measure 
is the Haar measure $\mu$ on $G$ (it happens to be both left and 
right invariant). In other words \cite{Haarsimetria}, the Haar measure defines 
the unique invariant integration measure for Lie groups. It implies that a 
volume element d$\mu(g)$ is identified by defining the integral of a function 
$f$ over $G$ as $\int_G f(g) d\mu(g)$, being left and right invariant 

\begin{equation}
\int_G f(g^{-1}x) d\mu(x)\,=\,\int_G f(x g^{-1}) d\mu(x)\,=\,\int_G f(x) d\mu(x).
\end{equation}

\noindent The invariance of the integral follows from the concomitant invariance 
of the volume element d$\mu(g)$. It is plain, then, that once d$\mu(g)$ is fixed 
at a given point, say the unit element $g=e$, we can move performing a 
left or right translation. Suppose that the map $x \rightarrow g(x)$ defines the 
action of a left translation. We have $x^i \rightarrow y^i(x^j)$, with $x^i$ being 
the coordinates in the vicinity of $e$. Assume, also, that d$x^1 ...$d$x^n$ defines 
the volume element spanned by the differentials d$x^1$, d$x^2$, ..., d$x^n$ at point 
$e$. It follows then that the volume element at point $g$ is given by 
d$\mu(g)=|J|^{-1}$d$x^1 ...$d$x^n$, where $J$ is the Jacobian of the previous 
map evaluated at the unit element $e$: $J=\frac{\delta(y^1...y^n)}{\delta(x^1...x^n)}$. 
In a right or left translation, both d$x^1 ...$d$x^n$ and $|J|$ are multiplied by the 
same Jacobian determinant, preserving invariance of d$\mu(g)$. The Lie groups 
also allow an invariant metric and d$\mu(g)$ is just the volume element 
$\sqrt{g}$d$x^1 ...$d$x^n$. Let us provide an example. Consider the volume element 
of the group $SU(2)$. The elements of $SU(2)$ are expressed by the $2\times 2$ matrices

\begin{equation}
x\,=\,\sum_{\mu} x^{\mu}\tilde \sigma ^{\mu} \,\,\,\,\,\,\,\, \sum_{\mu} x^{\mu}x^{\mu}=1,
\end{equation}

\noindent with $\sigma_0=1$, $\tilde \sigma _i=-i\sigma_i$ and 
$\tilde \sigma _i \tilde \sigma _j=-\delta_{ij}+\epsilon_{ijk}\tilde \sigma _k$, 
with $\sigma_i$ being the usual Pauli matrices. The coordinates of $SU(2)$ are taken as 
$x^i, i=1,2,3$ and $x^0=\pm\sqrt{1-r^2}$, $r\equiv \sqrt{\sum_i x^i x^i}$. In this 
form, the $SU(2)$ group manifold can be regarded as a 3D sphere of unit radius 
($\sum_{\mu=1}^4 x^{\mu} x^{\mu}=1$) in euclidian space of four dimensions 
$E_4$. The unit element $e$ corresponds to the origin, and the left action of $x$ on $y$ 
can be written as $z=xy=\sum_{\mu} z^{\mu}\sigma^{\mu}$, with the coordinates 
$z^{i}=(x^0y^i+x^iy^0)+\epsilon_{ijk}x^jy^k, z^{0}=\sqrt{1-\sum_i z^i z^i}$. Therefore we 
obtain the Jacobian matrix and its determinant. The final invariant integration measure 
reads as 

\begin{equation}
d\mu\,=\, \frac{1}{\sqrt{1-r^2}}dx^1 dx^2 dx^3.
\end{equation}
\pagebreak

We do not gain much physical insight with these definitions of the Haar measure and its 
invariance, unless we identify 
$G$ with the group of unitary matrices ${\cal U}(N)$, the element $a$ with a 
unitary matrix $U$ and $S$ with subsets of the group of unitary matrices ${\cal U}(N)$, 
so that given a reference state $|\Psi_0\rangle$ and a unitary matrix $U \in {\cal U}(N)$, 
we can associate a state $|\Psi\rangle_0=U|\Psi_0\rangle$ to $|\Psi_0\rangle$.
Physically what is required is a probability measure $\mu$ invariant under 
unitary changes of basis in the space of pure states, that is, 

\begin{equation}
P^{(N)}_{Haar}(U\,|\Psi\rangle)\,=\,P^{(N)}_{Haar}(|\Psi\rangle).
\end{equation}

\noindent These requirements can only be met by the Haar measure, which is 
rotationally invariant.
\newline

Now that we have justified what measure we need, we should be able to generate  
random pure states according to such a measure in arbitrary dimensions. 
The theory of random matrices \cite{Mehta90} specifies different 
{\it ensembles} of matrices, classified according to their different 
properties. In particular, the Circular Unitary Ensemble (CUE) consists 
of all matrices with the (normalized) Haar measure on the unitary 
group ${\cal U}(N)$. The Circular Orthogonal Ensemble (COE) is described 
in similar terms using orthogonal matrices, and it was useful in order 
to describe the entanglement features of two-{\it rebits} systems. Given 
a $N \times N$ unitary matrix $U$, the minimum number of independent entries 
is $N^2$. This number should match those elements that need to describe 
the Haar measure on ${\cal U}(N)$. This is best seen from the following 
reasoning. Suppose that a matrix $U$ is decomposed as a product of two 
(also unitary) matrices $U = X Y$. In the vicinity of $Y$, we have 
\cite{Mehta90} $U + dU = X(1 + idK)Y$, where $dK$ is a hermitian matrix 
with elements $dK_{ij} = dK^{R}_{ij} + idK^{I}_{ij}$. Then the probability 
measure nearby $dU$ is $P(dU) \sim \prod_{i \le j}dK^{R}_{ij} 
\prod_{i < j}dK^{I}_{ij}$, which accounts for the number of independent 
variables. Such measure for CUE is invariant \cite{Mehta90} and therefore 
proportional to the Haar measure. 

Yet, the aforementioned description is not useful for pratical purposes. We 
need to parameterize the unitary matrices according to the Haar measure. 
Following the work by Po\'{z}niak {\it et al}. \cite{PZK98}, the 
parameterization 
for CUE dates back to Hurwitz \cite{Hur1887} using Euler angles. The basic 
assumption is that an arbitrary unitary matrix can be decomposed into 
elementary two-dimensional transformations, denoted by 
$E^{i,j}(\phi,\psi,\chi)$: 

\begin{eqnarray} \label{Eij} 
E^{i,j}_{kk} &=& 1 \,\,\,\,\,\,\,\,\,\,\,\,\,\,\,\,\,\,\,\, 
k=1, .., N; \,\,\,\,\,\,\,\,\,\, k \neq i,j \cr
E^{i,j}_{ii} &=& \cos \phi \, e^{i\psi}, \cr
E^{i,j}_{ij} &=& \sin \phi \, e^{i\chi}, \cr
E^{i,j}_{ji} &=& -\sin \phi \, e^{-i\chi}, \cr
E^{i,j}_{jj} &=& \cos \phi \, e^{-i\psi}. 
\end{eqnarray} 

\noindent Using these elementary rotations we define the composite 
transformations

\begin{eqnarray} \label{Es}
E_{1} &=& E^{N-1,N}(\phi_{01},\psi_{01},\chi_{1}), \cr
E_{2} &=& E^{N-2,N-1}(\phi_{12},\psi_{12},0) 
E^{N-1,N}(\phi_{02},\psi_{02},\chi_{2}), \cr
E_{3} &=& E^{N-3,N-2}(\phi_{23},\psi_{23},0) 
E^{N-2,N-1}(\phi_{13},\psi_{13},0) E^{N-1,N}(\phi_{03},\psi_{03},\chi_{3}), \cr
... &=& ... \cr
E_{N-1} &=& E^{1,2}(\phi_{N-2,N-1},\psi_{N-2,N-1},0) 
E^{2,3}(\phi_{N-3,N-1},\psi_{N-3,N-1},0)... \cr
&...& E^{N-1,N}(\phi_{0,N-1},\psi_{0,N-1},\chi_{N-1}),
\end{eqnarray} 

\noindent we finally form the matrix 

\begin{equation} \label{U}
U \,=\, e^{i\alpha}\,E_1 E_2 E_3 ... E_{N-1}
\end{equation}

\noindent with the angles parameterizing the rotations 

\begin{equation} \label{angles}
0 \le \phi_{rs} \le \frac{\pi}{2} \,\,\,\, 0 \le \psi_{rs} < 2\pi 
\,\,\,\, 0 \le \chi_{1s} < 2\pi \,\,\,\, 0 \le \alpha < 2\pi.
\end{equation}
  
\noindent The ensuing (normalized) Haar measure \cite{Girko90}

\begin{equation} \label{PHaar}
P_{Haar}(dU)\,=\,\sqrt{N!2^{N(N-1)}}d\alpha 
\prod_{1 \le r < s \le N} \frac{1}{2r} d [(\sin \phi_{rs})^{2r}] d\psi_{rs} 
\prod_{1 < s \le N} d\chi_{1s}
\end{equation}

\noindent provides us with a random matrix belonging to CUE.

Now given the set of pure states $\{$$|\Psi\rangle$$\}$ in a certain 
Hilbert space ${\cal H}_N$ of dimension $N$, the computation of some 
quantity\footnote{Bear in mind that we do not say ``observables".} 
${\cal A}_{|\Psi\rangle}$ (its mean value, to be more precise) 
should be nothing but

\begin{equation} \label{int}
\langle {\cal A} \rangle \, = \, \frac{1}{V_{N}} 
\int_{metric\,\,space} {\cal A}_{|\Psi\rangle} \, dV,
\end{equation}
  
\noindent where $dV = dV(|\Psi\rangle)$ is the volume element defined by the 
angles (\ref{angles}) according to (\ref{PHaar}). In those cases where 
${\cal A}_{|\Psi\rangle}$ represents a magnitude that is a explicit function 
of $|\Psi\rangle$, we say that we are computing an average value 
$\langle {\cal A} \rangle$. On the other hand, if we wish to compute the portion 
of states that verify a certain property ${\cal P}$, we then have that

\begin{eqnarray} 
{\cal A}_{|\Psi\rangle} &=& 1, \,\,\,\, {\rm if}\,\,{\cal P}\,\,{\rm is}\,\,{\rm verified} \cr
{\cal A}_{|\Psi\rangle} &=& 0, \,\,\,\, {\rm if}\,\,{\cal P}\,\,{\rm is}\,\,{\rm not}\,\,{\rm verified}.
\end{eqnarray} 

\noindent Performing (\ref{int}) 
{\it analytically} may imply a gigantic task unless we work at 
low $N$-dimensions or ${\cal A}_{|\Psi\rangle}$ presents a simple 
parametrization in terms of (\ref{angles}). Thus we are naturally ``invited" 
to explore most of the properties concerning the set of pure 
states $\{$$|\Psi\rangle$$\}$ numerically. In doing so, we randomly generate 
the angles (\ref{angles}) {\it uniformly} and finally get the desired random 
matrix $U$ (\ref{U}). The numerical recipe consists of a Monte Carlo 
integration of (\ref{int}): randomly generating the states according to 
the Haar measure, we keep those ones complying with the properties defined 
by ${\cal A}_{|\Psi\rangle}$. By doing the ratio of the latter number 
to the total number of generated states we get, within a definite 
precision, an estimation of the integral (\ref{int}). 
\newline
\newline
{\bf Mixed states}
\newline

So far we have discussed the generation of pure states according to a natural 
rotationally invariant measure called the Haar measure on the the group 
of unitary matrices ${\cal U}(N)$. Now we face the following questions: 
i) how do we generate random mixed states appropriately?, and 
ii) is there a ``universal" measure also in the general case? 

Consider a positive semi-definite hermitian density matrix $\rho$, 
with Tr[$\rho$]$=1$. It is well known that such an arbitrary mixed state of a 
quantum system described by an $N$-dimensional Hilbert space can always be 
expressed as the product of three matrices,

\begin{equation}
\rho \,=\, U \, D[\lambda_i] \, U^{\dag}.
\end{equation}

\noindent Here $U$ is the usual $N \times N$ unitary matrix and 
$D[$$\{$$\lambda_i$$\}$$]$  is an $N \times N$ diagonal matrix whose elements 
are $\{$$\lambda_1,...,\lambda_N$$\}$, with $0\le \lambda_i \le 1$, and 
$\sum_i \lambda_i=1$. Recall that for a single non-zero value of $\lambda_i$, 
the pure state case $\rho=|\Psi\rangle\langle\Psi|$ is recovered. 
More generically, the space ${\cal S}$ of mixed states can be
regarded as a product space ${\cal S} = {\cal P} \times \Delta$
\cite{ZHS98,Z99}, where $\cal P$ stands for the family of all
complete sets of ortonormal projectors $\{ \hat P_i\}_{i=1}^N$,
$\sum_i \hat P_i = I$ ($I$ being the identity matrix), and $\Delta$
is the set of all real $N$-tuples $\{\lambda_1, \ldots, \lambda_N
\}$, with $\lambda_i \ge 1$ and $\sum_i \lambda_i = 1$. From the fact that 
${\cal S}$ is the product of two spaces, it is obsvious that we require 
a {\it product} measure in order to describe a general mixed state $\rho$. 
We know already that the Haar
measure on the group of unitary matrices ${\cal U}(N)$ induces a unique,
uniform measure $\nu$ on the set ${\cal P}$
\cite{PZK98}. On the other hand, since the simplex
$\Delta $ is a subset of a $(N-1)$-dimensional hyperplane of
${\cal R}^N$, the standard normalized Lebesgue\footnote{The Leguesbe measure 
is an extension of classical notions such as length or area to more 
complicated sets. For example, the Leguesbe measure ${\cal L}$ of an open set 
$W \equiv \sum_i (x_i,y_i)$ of disjoint intervals is equal to 
${\cal L}(W)=\sum_i(y_i-x_i)$.} measure ${\cal
L}_{N-1}$ on ${\cal R}^{N-1}$ provides a reasonable measure for
$\Delta$. The aforementioned measures on $\cal P$ and $\Delta$
lead to a natural measure $\nu $ on the set $\cal S$ of quantum
states \cite{ZHS98,Z99},

\begin{equation} \label{nu}
 \nu = \mu \times {\cal L}_{N-1}.
\end{equation}

\noindent Therefore, answering i), to generate mixed states according 
to (\ref{nu}) is tantamount to generate random unitary matrices according to 
CUE (previous pure case) and random points on the $\Delta$-simplex 
(which gives the eigenvalues of the matrix $\rho$). 
Unfortunately, there is no universality in the way of generating the simplex. 
The adequacy of this product measure was already discussed in Chapter 10, 
where some controversy around the choice of the measure for the simplex 
$\Delta$ is exposed. There is some criticism on the fact that the volume 
element associated with the measure (\ref{nu}) does not arise from a metric 
in the state space ${\cal S}$. The reader is invited to follow that discussion 
for a further insight.     

The computation of the mean value of some quantity ${\cal A}$ or 
property is performed in the same way as it was conceived in (\ref{int}) 
for pure states.

\section{C. Generation of two-qubits states with a fixed value of 
the participation ratio $R$}

The two-qubits case ($N=2 \times 2$) is the simplest quantum mechanical 
system that exhibits the feature of quantum entanglement. The relationship 
between entanglement and mixedness is described in 	      
Chapter 9. One given aspect is that as we increase the degree               
of mixture, as measured by the so called participation ratio 
$R=1/$Tr[$\rho^2$], the entanglement diminishes (on average). 
As a matter of fact, if the state is mixed enough, that state will have 
no entanglement at all. This is fully consistent with the fact 
that there exists a special class of mixed states which have maximum 
entanglement for a given $R$ \cite{MJWK01} 
(the maximum entangled mixed states MEMS). 
These states have been recently reported to be achieved in the 
laboratory \cite{MEMSexp} using pairs of entangled photons. 
Thus for practical or purely theoretical purposes, it may happen 
to be relevant to generate mixed states of two-qubits with 
a given participation ratio $R$. It may represent an excellent tool 
in the simulation of algorithms in a given quantum circuit: 
as the input pure states go through the quantum gates, they interact with the 
environment, so that they become mixed with some $R$. This 
degree of mixture $R$, which varies with the number of iterations, can be used 
as a probe for the evolution of the degradation of the entanglement present 
between any two qubits in the circuit. Different evolutions of the degree of mixture on 
the output would shed some light on the optimal architecture of the circuit 
that has to perform a given algorithm.

Here we describe a numerical recipe to randomly generate two-qubit states, according 
to a definite measure, and with a given, fixed value of $R$. Suppose 
that the states $\rho$ are generated according to the product measure 
$\nu = \mu \times {\cal L}_{N-1}$ (\ref{nu}), where $\mu$ is the Haar measure 
on the group of unitary matrices ${\cal U}(N)$ and the Leguesbe measure 
${\cal L}_{N-1}$ on ${\cal R}^{N-1}$ provides a reasonable measure for
the simplex of eigenvalues of $\rho$. In this case, the numerical procedure 
we are about to explain owes its efficiency 
to the following {\it geometrical picture} which is {\it valid only 
if the states are supposed to be distributed according to 
measure} (\ref{nu}). We shall identify the simplex $\Delta $ with a regular 
tetrahedron of side length 1, in ${\cal R}^3$, centred at the origin. Let
  ${\bf r}_i$ stand for the vector positions of the tetrahedron's
  vertices. The tetrahedron is oriented in such a way that the vector
  ${\bf r}_4$ points towards the positive $z$-axis and the vector
  ${\bf r_2}$ is contained in the $(x,z)$-semiplane corresponding to
  positive $x$-values. The positions of the tetrahedron's vertices correspond to 
  the vectors

\begin{eqnarray} 
\bf{r_1} &=& (-\frac{1}{2\sqrt{3}},-\frac{1}{2},-\frac{1}{4}\sqrt{\frac{2}{3}}) \nonumber \\
\bf{r_2} &=& (\frac{1}{\sqrt{3}},0,-\frac{1}{4}\sqrt{\frac{2}{3}}) \nonumber \\
\bf{r_3} &=& (-\frac{1}{2\sqrt{3}},\frac{1}{2},-\frac{1}{4}\sqrt{\frac{2}{3}}) \nonumber \\
\bf{r_4} &=& (0,0,\frac{3}{4}\sqrt{\frac{2}{3}}).
\end{eqnarray}

\noindent The mapping connecting the points
  of the simplex $\Delta $ (with coordinates $(\lambda_1,\ldots, \lambda_4)$)
  with the points $\bf r$ within tetrahedron is given by the equations

  \begin{eqnarray} \label{tetra1}
  \lambda_i \, &=& \, 2({\bf r}\cdot {\bf r}_i ) \, + \, \frac{1}{4}
  \,\,\,\, i=1, \dots, 4, \cr
  {\bf r} \, &=& \, \sum_{i=1}^4 \lambda_i {\bf r}_i
  \end{eqnarray}

  \noindent The degree of mixture is characterized by the
  quantity $R^{-1} \equiv Tr(\rho^2) = \sum_i \lambda_i^2$. This
  quantity is related to the distance $r=\mid {\bf r} \mid$
  to the centre of the tetrahedron $T_{\Delta}$ by

  \begin{equation} \label{tetra3}
  r^2 \, = \, -\frac{1}{8} \, + \, \frac{1}{2} \sum_{i=1}^4 \lambda_i^2.
  \end{equation}

 \noindent Thus, the states with a given degree of mixture lie on the
 surface of a sphere $\Sigma_r$ of radius $r$ concentric with the
 tetrahedron $T_{\Delta}$. To choose a given $R$ is tantamount to define 
a given radious of the sphere. There exist three different possible regions 
(see Fig.\ref{tetra}): 

\begin{figure}
\begin{center}
\includegraphics[angle=0,width=0.8\textwidth,clip=true]{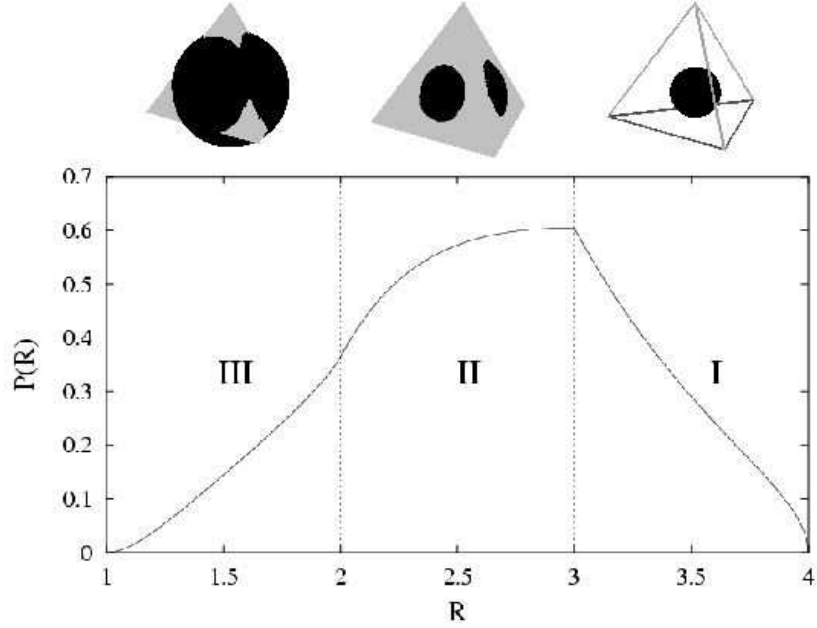}
\caption{Probability (density) distribution of stated $\rho$ generated 
according to the measure (\ref{nu}). The three regions described in the text are 
also shown.} 
\label{tetra}
\end{center}
\end{figure}

\begin{itemize}

\item region I: $r \in [0, h_1]$ ($R \in [4,3]$), where
$h_1 \equiv h_c={1 \over 4 }\sqrt{2 \over {3}}$ is the radius of a sphere
tangent to the faces of the tetrahedron $T_{\Delta}$. In this case
the sphere $\Sigma_r$ lies completely within the tetrahedron
$T_{\Delta}$. Therefore we only need to generate at random points over its 
surface. The cartesian coordinates for the sphere are given by

\begin{eqnarray} \label{spher}
x_1 &=& r \, \sin\theta\, \cos\phi \nonumber \\
x_2 &=& r \, \sin\theta\, \sin\phi \nonumber \\ 
x_3 &=& r \, \cos\theta, 
\end{eqnarray}

\noindent Denoting {\sf rand\_{}u()} a random number uniformly distributed 
between 0 an 1, the random numbers $\phi=2\pi${\sf rand\_{}u()} and 
$\theta=\arccos(2${\sf rand\_{}u()}$-1)$ (its probability distribution being 
$P(\theta)=\frac{1}{2}\sin(\theta)$) define an arbitrary state $\rho$ on the 
surface {\it inside} $T_{\Delta}$. The angle $\theta$ is defined between the 
centre of the tetrahedron (the origin) and the vector ${\bf r_4}$, and any point 
aligned with the origin. Substitution of ${\bf r}=(x_1,x_2,x_3)$ 
in (\ref{tetra1}) provides us with 
the eigenvalues $\{$$\lambda_i$$\}$ of $\rho$, with the desired $R$ as 
prescribed 
by the relationship (\ref{tetra3}). With the subsequent application of the 
unitary matrices $U$ we obtain a random state $\rho = U D(\Delta) U^{\dag}$ 
distributed according to the usual measure $\nu = \mu \times {\cal L}_{N-1}$.

\item region II: $r \in [h_1, h_2]$ ($R \in [3,2]$), where 
$h_2 \equiv \sqrt{h^{2}_{c}+(\frac{D}{2})^2}={\sqrt{2}\over 4}$ denotes
 the radius of a sphere which is tangent to the sides of the tetrahedron
 $T_{\Delta}$. Contrary to the previous case, part of the surface of the 
sphere lies outside the tetrahedron. This fact means that we are able to 
still generate the states $\rho$ as before, provided we reject those ones with 
negative weights $\lambda_i$.
									      
\item region III: $r \in [h_2, h_3]$ ($R \in [2,1]$), where
  $h_3 \equiv \sqrt{h^{2}_{c}+D^2}={\sqrt{6}\over 4}$ is the radius of 
a sphere passing through the vertices of $T_{\Delta}$. The generation 
of states is a bit more involved in this case. Again 
$\phi=2\pi${\sf rand\_{}u()}, but the available 
angles $\theta$ now range from $\theta_c(r)$ to $\pi$. It can be shown that 
$w\equiv\cos(\theta_c)$ results from solving the equation 
$3r^2 w^2 - \sqrt{\frac{3}{2}}r w + \frac{3}{8}-2r^2 = 0$. Thus, 
$\theta(r)=\arccos(w(r))$, with $w(r)=\cos\theta_c(r) + 
(1-\cos\theta_c(r))${\sf rand\_{}u()}. 
Some states may be unacceptable ($\lambda_i<0$) still, but the vast majority 
are accepted.

\end{itemize}

Combining these three previous regions, we are able to generate arbitrary 
mixed states $\rho$ endowed with a given participation ratio $R$.

%% file: cvEnglish.tex
%
\newlength{\defaultparindent}
\setlength{\parindent}{0cm}                 
 
\chapter*{}
\begin{center}%
\setlength{\parskip}{12pt}{\LARGE {\bf CURRICULUM VITAE}}

\end{center}\section*{Personal Data}

{\bf Name:} Josep Batle-Vallespir

\setlength{\parskip}{0pt}

{\bf D.N.I.:} 43.101.564-R

{\bf Birthdate:} June 8 1976

{\bf Birthplace:} Sa Pobla, Mallorca

\section*{Academic Data}        

Master Degree in Physics, University of Balearic Islands (1994-1998)\newline

University Professor Assistant, 2000-2001. Subject ``Fonaments 
F\'{\i}sics de l'Enginyeria", Enginyeria T\`{e}cnica Industrial especialitat 
Electr\`{o}nica. University of Balearic Islands\newline

PhD Student in Physics under FEDER funding (European Union), 2001-2002\newline

PhD Student in Physics under grant by the Conselleria d'Innovaci\'{o} i Energia, 
Government of Balearic Islands, 2002-2005\newline

High School Professor of Mathematics, 2005-
\newline
\newline

University of Balearic Islands

Physics Department

Mateu Orfila building

Carretera de Valldemossa, km 7.5

E-07122, Palma de Mallorca

Fax: +34 971173426

E-mail: vdfsjbv4@uib.es

\pagebreak

\section*{Publications}

\setlength{\parskip}{6pt}%

\begin{itemize}


\item J. Batle, M. Casas, A. Plastino, and A. R. Plastino, 
{\it On the Connection Between Entanglement and the Speed of Quantum Evolution}, 
Physical Review A {\bf 72}, 032337 (2005).

\item J. Batle, M. Casas, A. Plastino, and A. R. Plastino, 
{\it Werner States and the Two-Spinors Heisenberg Anti-ferromagnet}, 
Physics Letters A {\bf 343}, 12 (2005).
								       
\item J. Batle, A. R. Plastino, M. Casas, and A. Plastino, 
{\it Maximally Entangled Mixed States and Conditional Entropies}, 
Physical Review A {\bf 71}, 024301 (2005).

\item J. Batle, A. R. Plastino, M. Casas, and A. Plastino, 
{\it Some Statistical Features of the Entanglement Changes Associated 
with Quantum Logical Gates}, Physica A {\bf 356}, 385 (2005).

\item J. Batle, M. Casas, A. Plastino, and A. R. Plastino, 
{\it Entanglement Distribution and Entangling Power of Quantum Gates}, 
Optics and Spectroscopy {\bf 99}, 371 (2005). 

\item J. Batle, M. Casas, A. R. Plastino, and A. Plastino, 
{\it Quantum Entropies and Entanglement}, 
International Journal of Quantum Information {\bf 3}, 99 (2005).

\item J. Batle, M. Casas, A. Plastino, and A. R. Plastino, 
{\it Metrics, Entanglement, and Mixedness in the Space of Two-qubits}, 
Physics Letters A (2006), in press.

\item J. Batle, A. R. Plastino, M. Casas, and A. Plastino, 
{\it Inclusion Relations Among Separability Criteria}, 
Journal of Physics A {\bf 37}, 895 (2004).

\item J. Batle, A. R. Plastino, M. Casas, and A. Plastino, 
{\it Some Features of the Conditional q-Entropies of Composite Quantum Systems}, 
European Physical Journal B {\bf 35}, 391 (2003).

\item J. Batle, M. Casas, A. Plastino and A. R. Plastino, 
{\it On the Correlations Between Quantum Entanglement and q-Information Measures}, 
Physics Letters A {\bf 318}, 506 (2003).

\item J. Batle, M. Casas, A. Plastino, and A. R. Plastino, 
{\it The Statistics of the Entanglement Generated by the Hadamard-CNOT Quantum Circuit}, 
Physica A {\bf 327}, 140 (2003).

\item J. Batle,  M. Casas, M. de Llano, M. Fortes, and F. J. Sevilla, 
{\it Generalized BCS-BOSE Crossover Picture of Superconductivity}, 
International Journal of Modern Physics B {\bf 17}, 3271 (2003).

\item J. Batle, A. R. Plastino, M. Casas, and A. Plastino, 
{\it Understanding Quantum Entanglement: Qubits, Rebits and the 
Quaternionic Approach}, Optics and Spectroscopy {\bf 94}, 700 (2003).

\item J. Batle, A. R. Plastino, M. Casas, and A. Plastino, 
{\it On the Distribution of Entanglement Changes Produced by Unitary Operations}, 
Physics Letters A {\bf 307}, 253 (2003).

\item J. Batle, A. R. Plastino, M. Casas, and A. Plastino, 
{\it Conditional q-Entropies and Quantum Separability: a Numerical Exploration}, 
Journal of Physics A {\bf 35}, 10311 (2002).

\item J. Batle, M. Casas, A. R. Plastino and A. Plastino, 
{\it Inference Schemes and Entanglement Determination}, 
Physical Review A {\bf 65}, 024304 (2002); {\bf 65}, 049902 (2002).

\item J. Batle, A. R. Plastino, M. Casas, and A. Plastino, 
{\it On the Entanglement Properties of Two Rebits Systems}, 
Physics Letters A {\bf 298}, 301 (2002).

\item J. Batle, M. Casas, A. R. Plastino, and A. Plastino, 
{\it Entanglement, Mixedness and q-Entropies}, 
Physics Letters A {\bf 296}, 251 (2002).

\item J. Batle, A. R. Plastino, M. Casas and A. Plastino, 
{\it Quantum Evolution of Power-law Mixed States}, Physica A {\bf 308}, 233 (2002).

\item J. Batle, M. Casas, M. Fortes,  M. de Llano, and V. V. Tolmachev, 
{\it Generalizing BCS for Exotic Superconductors}, Journal of 
Superconductivity {\bf 15}, 655 (2002).

\item J. Batle, M. Casas, A. R. Plastino, and A. Plastino, 
{\it Supersymmetry and the q-Maxent Treatment}, Physica A {\bf 305}, 316 (2002).

\item J. Batle, M. Casas, A. R. Plastino, and A. Plastino, 
{\it On the ``Fake" Inferred Entanglement Associated with 
the Maximum Entropy Inference of Quantum States}, 
Journal of Physics A {\bf 34}, 6443 (2001).

\item J. Batle, M. Casas, M. Fortes, M.  A. Sol\'{\i}s, 
M. de  Llano, A. A. Valladares, and O. Rojo, 
{\it Bose-Einstein Condensation of Nonzero-Center-of-Mass-Momentum 
Cooper Pairs}, Physica C {\bf 364-365}, 161 (2001).

\item J. Batle, M. Casas, A. R. Plastino, and A. Plastino, 
{\it Tsallis Based Variational Method for Finding Approximate Wave 
Functions}, Physica A {\bf 294}, 165 (2001).


\end{itemize}

\setlength{\parskip}{6pt}%

\section*{Book Chapters}

\setlength{\parskip}{6pt}%

M. Casas, A. R. Plastino, A. Plastino, and J. Batle, \\
{\it Approximate density matrices for metal clusters and entangled states,}\\
Condensed Matter Theories Vol. 16, Nova (New York) ed. S. Hern\'andez (2001).

J. Batle, M. Casas, A. R. Plastino, and A. Plastino, \\
{\it Inference of quantum states: Maximum entropy and fake 
inferred entanglement,}\\
AIP Proceedings Maxent 2001, Baltimore (USA), ed. R. Fry (2001).

J. Batle, M. Casas, A. Plastino, and A. R. Plastino, \\
{\it A Survey of Entanglement Changes Associated with Quantum Gates Acting 
on Two-Qubits,}\\
Trends in Quantum Physics, Nova (New York) eds. V. Krasnoholovets and F. Columbus (2004).

M. Casas, J. Batle, A. Plastino, and A. R. Plastino, \\
{\it A Systematic Numerical Survey of the Separability Criteria for Bipartite Quantum Systems,}\\
Condensed Matter Theories Vol. 19, Nova (New York) eds. M. Belkacem and P. M. Dinh (2005).

\setlength{\parskip}{6pt}%

\section*{Communications \& Posters}

\setlength{\parskip}{6pt}%
 
J. Batle, M. Casas, A. R. Plastino and A. Plastino,\\
{\it Inference of quantum states: Maximum entropy and fake 
inferred entanglement}.\\
``$21^{ST}$  International Workshop on Bayesian Inference and Maximum Entropy Methods 
in Science and Engineering (MAXENT 2001)", Johns Hopkins University, 
Baltimore (USA), 4-9 August 2001. \\
Oral presentation (30 min).

J. Batle, M. Casas, A. R. Plastino and A. Plastino,\\
{\it Entanglement, maximum entropy and separability criteria}.\\
Reuni\'on Anual Grupo Especializado de F\'{\i}sica Nuclear, 
Universitat de Val\`{e}ncia, 22-23 February 2002. \\
Oral presentation (15 min).

J. Batle, M. Casas, A. R. Plastino and A. Plastino,\\
{\it Maximum entropy and fake 
inferred entanglement}.\\
``Second European Summer School on Microscopic 
Quantum Many-Body Theories and Their Applications", the Abdus Salam 
International Centre for Theoretical Physics (Miramare-Trieste, Italy), 
3-14 September 2001. \\
Poster presentation.

J. Batle, A. R. Plastino, M. Casas and A. Plastino,\\
{\it Understanding quantum entanglement: qubits, rebits and the quaternionic 
approach}.\\
``IX International Conference 
on Quantum Optics", Raubichi, Belarus, 14-17 May 2002.\\
Oral presentation as invited speaker (20 min).

J. Batle, M. Casas, A. Plastino and A. R. Plastino,\\ 
{\it On the Correlations Between Quantum Entanglement and q-Information Measures}.\\
``36th Course of the International School of Quantum Electronics on 
Advances in quantum Information Processing: From Theory to Experiment", 
Erice, Sicily, 15-22 March 2003.\\
Poster presentation.

J. Batle, A. R. Plastino, M. Casas, and A. Plastino,\\ 
{\it Spectral decomposition based separability criteria: a numerical survey}.\\
Centennial Anniversary of the Spanish Royal Society of Physics and Chemistry, 
Grupo expecializado en Informaci\'{o}n Cu\'{a}ntica (SQUIN), July 2003.\\
Oral presentation (30 min).

J. Batle, A. R. Plastino, M. Casas, and A. Plastino,\\ 
{\it Spectral decomposition based separability criteria: a numerical survey}.\\
4th European QIPC Workshop, Oxford, United Kingdom, July 13-17 2003.\\
Poster presentation.

J. Batle, M. Casas, A. R. Plastino and A. Plastino,\\
{\it Quantum Entropies and Entanglement}. \\
``International Meeting on Quantum Information Science", University of Camerino, 
Camerino, Italy, 22-19 April 2004.\\
Poster presentation.

J. Batle, M. Casas, A. Plastino, and A. R. Plastino,\\ 
{\it Entanglement Distribution and Entangling Power of Quantum Gates}.\\
``X International Conferece on Quantum Optics ICQO 2004", Minsk, Belarus, 
May 30 - June 4.\\
Oral presentation as invited speaker (30 min).

\setlength{\parskip}{6pt}%

\section*{Courses, Meetings \& Workshops}

\setlength{\parskip}{6pt}%

JUAS 2000 (Joint Universities Accelerator School) courses on 
``Accelerator Physics and Applications", sponsored by ESI 
(European Scientific Institute) and organized by several european universities 
in collaboration with CERN. Archamps (France), and Geneva (Switzerland), 
January - March 2000.

\setlength{\parskip}{6pt}%

Participation in the ``Workshop on Quantum Information and Quantum Computation", 
held at the Abdus Salam International Centre for Theoretical Physics (Miramare-Trieste, Italy), 
from 14 to 25 October 2002.

\setlength{\parskip}{6pt}%

Participation in the ``Joint ICTP-INFM School-Workshop on Entanglement at the Nanoscale", 
held at the Abdus Salam International Centre for Theoretical Physics (Miramare-Trieste, Italy), 
from October 28 to November 8 2002.

\setlength{\parskip}{6pt}%

Participation in the course "Informaci\'{o}n y Computaci\'{o}n Cu\'{a}nticas", 
delivered by Prof. Ignacio Cirac, held at the Theoretical Physics Department of the 
Universidad Aut\'{o}noma de Madrid, February 2-6 2004.

\setlength{\parskip}{6pt}%


\section*{Visits to Other Institutions}

\setlength{\parskip}{6pt}%

Collaboration, under the supervision of Prof. G. Ortiz, in the 
Theoretical (T-11) Division of Los Alamos National Laboratory, New Mexico, USA, 
regarding the issue of Entanglement and Quantum Phase Transitions, from September 
to December 2003. 

\setlength{\parskip}{6pt}%

\section*{Current Research Interests}

\setlength{\parskip}{6pt}%

The characterization of multipartite entangled states has not been by far completely 
achieved. This description of entanglement, in the orthodox view of Hilbet space 
partitioning, offers an arena where the application of the so called entropic inequalities 
could be of some relevance. Therefore the study of entanglement in multipartite 
quantum systems plays a paramount role in our current research interest.\newline

Other aspects of quantum entanglement, such as its connection with speed of the 
quantum evolution of a quantum state, among many other aspects of quantum mechanics, 
are also under study.\newline 

In a more recent status, the characterization of the dynamics of entanglement in quantum 
phase transitions is a subject that focuses our immediate attention.